\documentclass[reqno,12pt,oneside]{report} 

\usepackage{rac}         
\usepackage{aas_macros}  
\usepackage[intlimits]{amsmath} 
\usepackage{amsxtra}     
\usepackage{amsthm}
\usepackage{amssymb}
\usepackage{amsfonts}
\usepackage{graphicx}    
\usepackage{rotating}
\usepackage{color}
\usepackage{epsfig}
\usepackage{xspace}	
\usepackage{subfigure}   
\usepackage{verbatim}
\usepackage{tikz}
\usepackage{notoccite}
\def\checkmark{\tikz\fill[scale=0.4](0,.35) -- (.25,0) -- (1,.7) -- (.25,.15) -- cycle;} 

\usepackage[printonlyused]{acronym} 
\usepackage{setspace}    
\theoremstyle{plain}

\theoremstyle{definition}

\theoremstyle{remark}

\numberwithin{theorem}{chapter}     

\makeatletter
\def\cleardoublepage{\clearpage\if@twoside \ifodd\c@page\else
\hbox{}
\thispagestyle{empty}
\newpage
\if@twocolumn\hbox{}\newpage\fi\fi\fi}
\makeatother

\newcommand{\pythia}{\mbox{\sc{pythia}}\xspace}
\newcommand{\pp}{\mbox{$p$$+$$p$}\xspace}
\newcommand{\pau}{\mbox{$p$$+$Au}\xspace}
\newcommand{\pal}{\mbox{$p$$+$Al}\xspace}
\newcommand{\pa}{\mbox{$p$$+$A}\xspace}
\newcommand{\sqs}{\mbox{$\sqrt{s}$}\xspace}
\newcommand{\sqsn}{\mbox{$\sqrt{s_{_{NN}}}$}\xspace}
\newcommand{\pion}{\mbox{$\pi^0$}\xspace}
\newcommand{\h}{\mbox{h$^{\pm}$}\xspace}
\newcommand{\pt}{\mbox{$p_T$}\xspace}
\newcommand{\gammajet}{\mbox{$\gamma$-jet}\xspace}
\newcommand{\epsilonniso}{\mbox{$\epsilon_{\rm dec}^{\rm niso}$}\xspace}
\newcommand{\epsilondec}{\mbox{$\epsilon_{\rm tag}^{\rm dec}$}\xspace}
\newcommand{\alphamiss}{\mbox{$\alpha_{\rm inc}^{\rm miss, iso}$}\xspace}
\newcommand{\ntrig}{\mbox{$N_{\rm trig}$}\xspace}
\newcommand{\decpty}{\mbox{$Y_{\rm decay}^{\rm iso}$}\xspace}
\newcommand{\incpty}{\mbox{$Y_{\rm inclusive}^{\rm iso}$}\xspace}
\newcommand{\dirpty}{\mbox{$Y_{\rm direct}^{\rm iso}$}\xspace}
\newcommand{\rgammaprime}{\mbox{$R_\gamma^{\rm iso}$}\xspace}
\newcommand{\mapfxn}{\mbox{$P(p_T^{\pi^0},p_T^\gamma)$}\xspace}
\newcommand{\pout}{\mbox{$p_{\rm out}$}\xspace}
\newcommand{\dphi}{\mbox{$\Delta\phi$}\xspace}
\newcommand{\ptassoc}{\mbox{$p_{T}^{\rm assoc}$}\xspace}
\newcommand{\pttrig}{\mbox{$p_T^{\rm trig}$}\xspace}
\newcommand{\rgamma}{\mbox{$R_\gamma$}\xspace}
\newcommand{\rmspout}{\mbox{$\sqrt{\langle p_{\rm out}^2\rangle}$}\xspace}
\newcommand{\zt}{\mbox{$\langle z_T^{\pi^{0}}\rangle$}\xspace}
\newcommand{\rmsjt}{\mbox{$\sqrt{\langle j_T^2\rangle}$}\xspace}
\newcommand{\rmskt}{\mbox{$\sqrt{\langle k_T^2\rangle}$}\xspace}
\newcommand{\jt}{\mbox{$j_T$}\xspace}
\newcommand{\kt}{\mbox{$k_T$}\xspace}
\newcommand{\jpsi}{\mbox{$J/\psi$}\xspace}
\newcommand{\gev}{\mbox{GeV/$c$}\xspace}
\newcommand{\prob}{\mbox{$\mathcal{P}$}}
\newcommand{\probniso}{\mbox{$\mathcal{P}^{niso}$}\xspace}
\newcommand{\probiso}{\mbox{$\mathcal{P}^{iso}$}\xspace}
\newcommand{\photon}{\mbox{$\gamma$}\xspace}
\newcommand{\alphas}{\mbox{$\alpha_s$}\xspace}
\newcommand{\qsq}{\mbox{$Q^2$}\xspace}
\newcommand{\eplus}{\mbox{$e^+$}\xspace}
\newcommand{\eminus}{\mbox{$e^-$}\xspace}
\newcommand{\xe}{\mbox{$x_E$}\xspace}
\newcommand{\ups}{\mbox{$\Upsilon$(nS)}\xspace}
\newcommand{\gevc}{\mbox{GeV/$c$}\xspace}
\newcommand{\xt}{\mbox{$x_T$}\xspace}
\newcommand{\vtwo}{\mbox{$v_2$}\xspace}
\newcommand{\vthree}{\mbox{$v_3$}\xspace}
\newcommand{\qhat}{\mbox{$\hat{q}$}\xspace}
\newcommand{\ncoll}{\mbox{$N_{coll}$}\xspace}
\def\fig#1{Figure~\ref{#1}}

\begin{document}


\titlepage{Nonperturbative Factorization Breaking and Color Entanglement Effects in Dihadron and Direct Photon-Hadron Angular Correlations in $p$$+$$p$ and $p$$+$A Collisions}{Joseph Daniel Osborn}{Doctor of Philosophy}
{Physics}{2018}
{Associate Professor Christine Aidala, Chair\\
 Professor Wolfgang Lorenzon\\
 Associate Professor Abhijit Majumder, Wayne State University\\
 Professor Sara Pozzi \\
 Assistant Professor Joshua Spitz}
\initializefrontsections


\copyrightpage{Joseph Daniel Osborn}

\makeatletter
\if@twoside \setcounter{page}{4} \else \setcounter{page}{1} \fi
\makeatother


\dedicationpage{To Alyssa Osborn, without whom this work would not exist}

\startacknowledgementspage

As I sit here attempting to come up with a list of all the people who deserve acknowledgement over the course of the last five years, I recognize that I will almost surely fail to construct a complete list. Nonetheless, there are many people I would like to personally thank and recognize.

Firstly, I must thank my wife, Alyssa Osborn. Her unending and dedicated love, support, and patience with me over the last five years has not gone unnoticed and unappreciated. I would not be writing this if it had not been for her support for me to achieve a dream I have had since I could remember; her own sacrifices have allowed me to pursue this. For this I am and always will be eternally grateful.

Secondly, I'd like to thank my advisor Christine Aidala. The scientific environment that she has developed is unparalleled, and for that I am incredibly grateful. Her guidance and support as an advisor would rival the best of anyone in the country. In addition, her willingness to let me pursue my own scientific interests independently was an integral part of my own development as a scientist.

My academic curiosity began in high school where I attended the Math Science and Technology Center magnet program at Paul Laurence Dunbar high school. In particular, Karen Young deserves recognition for first introducing me to physics and getting me interested in understanding how nature works. Additionally, the many teachers at the MSTC program deserve recognition for their dedication to exciting the next generation of young scientists.

Previously to my time at the University of Michigan, I feel it is necessary to thank Renee Fatemi at the University of Kentucky. Allowing me to work in her group on the STAR experiment and the FGT project in addition to introducing me to the field of nucleon structure is what allowed me to find a field of study that would never satisfy my curiosity. 

At the University of Michigan, there are far too many people that need thanking. Starting from the beginning I must thank my first year classmates for propelling me, and sometimes dragging me, through E\&M and QFT amongst other classes. It is also necessary to thank Aaron White for setting up my computing environment on PHENIX, so that I was able to seamlessly join and start analysis work right away. I also need to thank Anthony Charles and Sebastian Ellis for never turning me away when I had phenomenological and/or theoretical questions to ask. When the weather was warm, research walks with Dan Marley were also an important part of my graduate training and taught me when your eyes hurt from writing code for too long, it was time to discuss college basketball and/or football. I'd also like to thank Ezra Lesser for the constant source of entertainment via his seemingly infinite appetite for free pizza, in addition to his willingness to listen to my rants about philosophical shortcomings of particle physics. I am also thankful to everyone that I have encountered in Christine's group, Bryan Ramson, Isaac Mooney, Catherine Ayuso, amongst others, for appropriately giving me the nickname ``Dad'' or ``Earl of Dad'' despite (at the time of writing) not having any children.

Amongst the PHENIX and sPHENIX collaborations, there are also many people who will not receive the appreciation that they deserve just from a single acknowledgement sentence. First I would like to thank my computing colleagues and their patience with my likely endless barrage of stupid questions: Chris Pinkenburg, Jin Huang, and Hubert Van Hecke. I would also like to thank Craig Woody and John Haggerty for their expertise in the sPHENIX EMCal and allowing me to get my hands dirty with hardware. My analysis efforts would have been incredibly unguided had it not been for the help of Justin Frantz, Megan Connors, and Nathan Grau. I am additionally thankful for the many stimulating physics discussions with Ron Belmont, John Lajoie, Paul Stankus, and Mike Tannenbaum, in addition to many of the aforementioned people. I am also grateful to Sasha Bazilevsky for fruitful discussions and always being available to help, guide, and mentor me, amongst many other students, while analyzing PHENIX data.

Outside of work, I must thank my family, including multiple sets of parents and siblings, who have always supported me in my endeavors despite never feeling like they could adequately say what I was doing with my life. Each one of them has served an important role and been an inspiration in a unique way. Adam Timaji must also receive adequate thanks for the many nights of video games which allowed me to leave the worries of graduate school behind even if just for half an hour. Ben Anderson, Alex Riggs, and Noah Troutman also deserve thanks for the visits (and my visits to them) to remind me how important good friends are in life. I am also grateful to my many friends in Michigan who have supported me and given me an outlet to leave the worries of graduate school behind.

\label{Acknowledgements}


\tableofcontents     
\begin{singlespace}
\listoffigures       

\listoftables        
\end{singlespace}


\startabstractpage

New predictions regarding the role of color flow in high energy Quantum Chromodynamics (QCD) processes have emerged in the last decade. Novel effects due to the non-Abelian nature of QCD have been predicted and are just now accessible experimentally due to significantly improved facilities that are able to measure multidifferential observables. High energy proton-proton collisions provide a testing ground to study nonperturbative QCD in a regime where perturbative calculations should be applicable; thus theoretical tools within a perturbative framework can be used to probe and constrain nonperturbative functions and effects in QCD. In particular, the role of color flow is now being explored through many different observables throughout various subfields of QCD; one such observable is nearly back-to-back hadron correlations in proton-proton collisions which are predicted to be sensitive to states that are entangled via their QCD color charge. 

The PHENIX detector at the Relativistic Heavy Ion Collider (RHIC) is well suited to study potential effects from color flow. In 2013 and 2015 the PHENIX experiment recorded data from proton-proton and proton-nucleus collisions. Angular correlations between two nearly back-to-back hadrons or a direct photon and hadron are measured to study the prediction of color entanglement; this refers to a novel entangled state of the two hard-scattering partons across the colliding hadronic system. These correlations can be treated in a transverse-momentum-dependent framework where sensitivity to these non-Abelian effects from color are predicted. The measurements presented here are the first ever to search for experimental evidence of these entangled states and furthermore will help establish color flow in hadronic interactions as a new area of focus within QCD research.

Results are presented for proton-proton collisions at center-of-mass energies of 200 and 510 GeV and proton-nucleus collisions at nucleon-nucleon center-of-mass energies of 200 GeV. World measurements of processes where factorization is predicted to hold are also compiled and analyzed to compare to the new experimental results presented here. The measured results, which include the first measurements of nonperturbative momentum widths in processes predicted to break factorization, do not indicate any obvious qualitative differences from observables where factorization is predicted to hold. This indicates that quantitative comparisons with phenomenological calculations will be necessary to identify the magnitude of effects from color entanglement. Future calculations will therefore have the opportunity to establish the magnitudes of non-Abelian color effects in hadronic collisions with comparisons to these results. In addition, future measurements of similar observables have the potential to further identify nontrivial effects from color interactions and color entangled states in hadronic collisions. As QCD is the only non-Abelian quantum field theory known to exist in nature that admits bound states, it will be essential to continue exploring unique QCD phenomena due to color interactions in controlled ways in the coming years.

\label{Abstract}


\startthechapters 

 \chapter{Introduction}
 \label{chap:Intro}
 \section{Quantum Chromodynamics}\label{sec:qcd}

Quantum Chromodynamics (QCD), the fundamental theory of the strong force, is the non-Abelian gauge invariant quantum field theory which describes the interactions between quarks and gluons. The QCD Lagrangian is shown in Eq.~\ref{eq:qcdlagrangian} and is analogous to the Quantum Electrodynamics (QED) Lagrangian

\begin{equation}\label{eq:qcdlagrangian}
	\mathcal{L}_{\rm QCD} = \bar{\psi}(i\gamma^\mu D_\mu-m)\psi~-~ \frac{1}{4}G_{\mu\nu}^aG_a^{\mu\nu}\,.
\end{equation} 
Here $\psi$ is a spin 1/2 fermion quark field, $D_\mu$ is a covariant derivative defined by $D_\mu = \partial_\mu + it^\alpha A^\alpha_\mu$, and $G_{\mu\nu}$ is the spin 1 boson gluon field tensor~\cite{Collins_book}. The gluon field tensor is expressed in terms of the gluon field $A_\mu$, defined as

\begin{equation}\label{eq:gluontensor}
G^\alpha_{\mu\nu} = \partial_\mu A^\alpha_\nu - \partial_\nu A^\alpha_\mu - f_{\alpha\beta\gamma}A^\beta_\mu A^\gamma_\nu\,,
\end{equation}
where $f_{\alpha\beta\gamma}$ are the structure constants of the SU(3) gauge group. While the Lagrangian as written in Eq.~\ref{eq:qcdlagrangian} looks identical to the QED Lagrangian, the final term in the gluon field tensor is the reason that QCD is fundamentally so different from QED; the gauge boson which mediates the strong nuclear force can interact with itself in addition to the fermions of the theory, unlike the photon in QED. \par

The symmetry structure of QCD is described mathematically with the SU(3) gauge group in which there are 8 generators, in this case 3x3 matrices, denoted by $t^\alpha$ in the covariant derivative definition above. The generators satisfy the commutation relations of the group in conjunction with the structure constants, $[t^\alpha, t^\beta] = if^{\alpha\beta\gamma}t^{\gamma}$. Physically this group corresponds to a theory in which there is an additional quantum number called color. Color charge is the QCD analogue to electric charge in QED; in SU(3) this quantum number can take on three different values, cleverly named red, green, and blue. However, it is important to emphasize that these have no relation to the actual visible color spectrum. Interestingly in QCD the gauge boson also carries this quantum number, while in QED the photon does not carry any electric charge. Based on the generator structure, there are eight linearly independent color combinations gluons can carry, while quarks carry one of the three color charges. The eight generators of the group, or physically the color combinations that gluons carry, encode the fact that a gluon's interaction with a quark rotates the quark's color in SU(3) space.

There is also an additional SU(3) flavor symmetry which arises due to the small mass of the lightest quarks (up, down, and strange). The quarks are classified into 3 generations, where up and down quarks comprise generation 1, charm and strange quarks comprise generation 2, and top and bottom quarks comprise generation 3. The unique features described above in QCD, such as the gluon self coupling and color charges, which differentiate it from QED led to the concepts of confinement and asymptotic freedom, for which the Nobel prize was awarded in 2004~\cite{Gross:1973id, Politzer:1973fx}. This was the symbolic closure of the initial development period of QCD, and cemented the theory as the correct theory of the strong force.  \par

\subsection{Asymptotic Freedom}

\begin{figure}[h]
	\centering
	\includegraphics[width=0.7\textwidth]{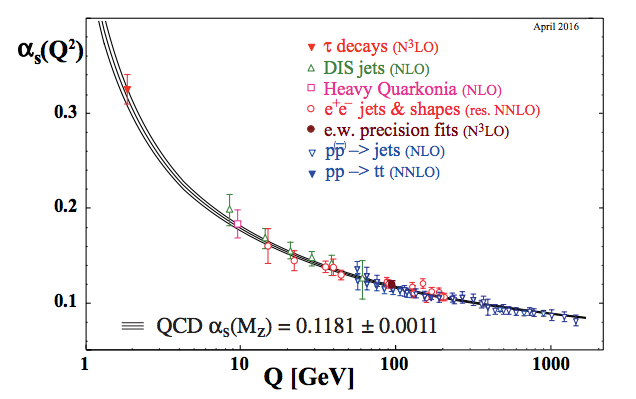}
	\caption{World measurements of the strong coupling constant \alphas are shown over a wide range of momentum transfers, taken from Ref.~\cite{Patrignani:2016xqp}. Perturbative calculations agree well with data.}
	\label{fig:alphas_fig}
\end{figure}

Both QCD and QED are characterized by a scale dependent coupling constant which depends on the energy scale of the interaction. Running coupling constants can be defined by a beta function, which is a differential equation that must be solved within the perturbative field theory based on a renormalization scale for which the theory breaks down. The one-loop running coupling constant in QCD has been calculated to be~\cite{Patrignani:2016xqp}

\begin{equation}\label{eq:alphas_eq}
\alpha_s(Q^2) = \frac{\alpha_s(\mu^2)}{1+\beta_0\alpha_s(\mu^2)\ln(Q^2/\mu^2)}\,,
\end{equation}
where $\mu^2$ defines the scale at which perturbative techniques break down and has been determined experimentally to be several hundred MeV, and 
\[
	\beta_0 = \frac{11N_c -2 n_f}{12\pi}\,.
\]
Higher order terms have been calculated~\cite{Patrignani:2016xqp}, but to see the dominant behavior only the one-loop term is necessary. In the standard model, where $N_c$, the number of colors, is three, and $n_f$, the number of quark flavors, is six, the quantity $\beta_0$ is dominated by the color term which comes from the gluon self coupling. This results in a strong decrease of $\alpha_s$ with the momentum transfer \qsq, which is referred to as the running of the strong coupling constant, or asymptotic freedom. The running of \alphas confirms the name of the strong force, namely that it is very strong at small energies but small enough to apply perturbative techniques at large energies. Figure~\ref{fig:alphas_fig} shows world measurements of \alphas as a function of momentum transfer, which agree excellently with perturbative calculations~\cite{Patrignani:2016xqp}. At low energies less than several GeV \alphas becomes large and thus perturbative expansions in \alphas break down.

Asymptotic freedom allows the use of perturbative techniques when processes are ``hard,'' or have a large momentum transfer such that the coupling constant \alphas is small. When the coupling constant is large, nonperturbative or ``soft'' processes dominate and quarks and gluons, collectively referred to as partons, can no longer be treated as free particles within the field theory. At this point partons hadronize into color neutral mesons and baryons, which are a quark-antiquark and a three quark state, respectively. These are the stable QCD bound states that are observed in nature. This occurs because as the distance between two partons becomes large, the energy carried by the gluon fields exceeds the threshold necessary to spontaneously create matter; thus, partons are created which hadronize into bound states. This is referred to as confinement and is why free partons cannot be observed. Ultimately the properties of the baryons and mesons, collectively referred to as hadrons, are generated nonperturbatively and depend on the long distance behavior of QCD. Thus, it is necessary to construct a formalism which relates the degrees of freedom of the gauge invariant field theory to the long range nonperturbative degrees of freedom of the hadrons within QCD.  \par

\subsection{Confinement and Factorization}
As perturbative techniques cannot be used to describe the bound state structure of hadrons, another method must be used to accurately characterize the nonperturbative behavior that leads to color neutral hadrons that are observed in nature. To probe the partonic structure of matter, high energy collisions are typically used so that perturbative QCD can be applied. Some typical interactions that are studied are $\eplus+\eminus\rightarrow q+\bar{q}+X$, $\eminus+p\rightarrow\eminus+X$, and $p+p\rightarrow X$, where $X$ refers to an arbitrary final-state that may or may not include QCD bound states. In any of these interactions, the cross section, which is the actual physical observable, must be written in a way that includes both the perturbative interaction as well as the nonperturbative long range dynamics in both the initial and final states which results in the measured cross section. \par

In $\eplus\eminus\rightarrow q\bar{q}$ production, the only nonperturbative QCD behavior is in the final fragmenting state. The long-range dynamics are written as a parametrization called the fragmentation function (FF), often denoted $D_q^h(z,\qsq)$. At leading order (LO) this is defined as the probability for a given parton $q$ to hadronize into a particular hadron $h$, and is to first approximation only a function of the longitudinal momentum fraction $z = p_h/p_q$ that the hadron carries with respect to the initial parton's momentum and the \qsq of the interaction. In $\eminus p\rightarrow\eminus X$, or deep-inelastic scattering (DIS), the hadron in the initial-state must also be parametrized in a similar way. Here the long-range dynamics are characterized by parton distribution functions (PDFs), and are often denoted $f_{q/h}(x,Q^2)$. At LO these are defined as the probability for a certain flavor of parton $q$ from hadron $h$ to interact in the hard scattering, and is to first approximation only a function of the longitudinal momentum fraction of the parton within the hadron $x = p_q/p_h$ and the momentum transfer of the process \qsq. Both PDFs and FFs can be intuitively thought of as probability density functions at LO, and are used to relate the partonic and hadronic degrees of freedom. \par

Therefore a generic cross section can be written as the convolution of PDFs, FFs, and a partonic hard scattering cross section which is perturbatively calculable: 

\begin{equation}\label{eq:factorization_theorem}
\frac{d\sigma}{d\qsq} = \sum_{ij} \int_0^1 \int_0^1 \int_0^1 dz_1  dx_1  dx_2 f_{1/h_1}(x_1,Q^2)f_{2/h_2}(x_2,Q^2)\frac{d\hat{\sigma}(x_1,x_2,z_1,i,j)}{d\qsq} D_1^{h_3}(z_1)
\end{equation}
where the $i$,$j$ indices are a sum over parton flavors in the hadrons. This is known as a factorization definition, since the long and short distance physics factorizes from one another, and the generic definition shown here necessarily depends on what kind of collision is being studied. For example in \eplus\eminus annihilation to hadrons there would only be FFs and no PDFs as there are no initial-state hadrons in this process. Factorization necessarily relies on a scale for which one can resolve the PDFs, and this is called a factorization scale and is generally set to the interaction hard scale \qsq. It can be conceptually thought of as a definition of what is included in the nonperturbative functions and what is included in the partonic hard function, as this is in a sense somewhat arbitrary. This scale is often adjusted as an estimate of theoretical uncertainties which might depend on the choice of what is included in the hard function and nonperturbative functions. The definition of collinear factorization, shown here as a function of only the longitudinal momentum fractions, for a particular process is often assumed, although it has only been rigorously proven in the processes $\ell^+\ell^-\rightarrow$~hadrons, $\ell p\rightarrow\ell+X$, and $pp\rightarrow\ell^+\ell^-+X$~\cite{Collins_book}. Collinear factorization has not been rigorously proven to all orders for hadronic collisions where final-state hadrons are measured, and in fact recent studies have shown that it is broken at the multi-loop level~\cite{Catani:2011st,Forshaw:2012bi}; however, it is generally assumed to be true and cross section calculations have been shown to match data with the precision of tens of percent (see e.g. Ref.~\cite{ppg186}). \par

It is important to emphasize that PDFs and FFs are not perturbatively calculable as they are nonperturbative functions; they require data which can then be used to constrain fitting procedures that extract the dependence of the functions on $x$ and \qsq. Lattice QCD offers an alternative method to perturbative techniques, where partons are placed on a three dimensional discrete lattice and interactions between these partons may be calculated. While many computations still require several caveats, recent lattice QCD calculations have made significant strides in calculating the full $x$ dependence of certain PDFs~\cite{Musch:2010ka,Musch:2011er,Ji:2013dva}. This is in contrast to previous studies which could only calculate moments of PDFs. Very recently, calculations have even been able to determine certain PDFs at the physical pion mass~\cite{Alexandrou:2018pbm}; this is a major step forward for lattice calculations which were previously only possible at unphysical pion masses of approximately 300 MeV/$c^2$. While lattice calculations are still at an early stage these recent studies show promising possibilities for future work as computational power limitations become less of a barrier. \par

The nonperturbative functions which describe the partonic structure of hadrons are also taken to be universal functions. This means that a function could be constrained with data from one process and then could be used in calculating a cross section for an entirely different process. For example, FFs could be constrained with data from \eplus\eminus annihilation to hadrons; since there are no initial-state hadrons this makes the determination of the FFs cleaner. These FFs could then be used in a cross section calculation for a different process, say $\eminus p\rightarrow\eminus h+X$, also known as semi-inclusive deep-inelastic scattering (SIDIS). The universality of these functions has allowed the beyond the Standard Model (BSM) community, for example, to make predictions for new physics processes at the Large Hadron Collider (LHC). Data from the HERA electron-proton collider facility has significantly constrained the collinear PDFs~\cite{AAron:2009aa} due to the cleaner nature of the DIS interaction, and these PDFs can then be used for predictions of BSM processes which have been searched for at the LHC. The universality of certain PDFs and factorization theorems for certain processes will be one of the focal points of this thesis.

\section{Nucleon Structure and Spin}

\subsection{Unpolarized Structure}
In the previous discussion, the spin of the quarks and gluons is averaged and the partons are assumed to be moving collinearly with the parent hadron. This is, by definition, an oversimplification since hadrons are bound states of partons; the uncertainty principle dictates that there must also be additional transverse degrees of freedom. As discussed and referenced above, the unpolarized collinear PDFs are quite well known; recent extractions have well controlled uncertainties down to $x\sim\mathcal{O}(10^{-3})$ (see e.g. Ref.~\cite{Dulat:2015mca}). When the transverse momentum of the partons is explicitly included in the definition of the PDF, transverse-momentum-dependent (TMD) PDFs can be defined. Thus, the collinear PDF definition described above can be extended to an unpolarized TMD PDF $f_{q/h}(x,k_T,Q^2)$ which is dependent on both the partons longitudinal and transverse momentum degrees of freedom. When transverse momentum degrees of freedom are integrated over, the collinear unpolarized PDFs can be recovered; thus TMD PDFs inherently contain more information about the nonperturbative structure of partons within hadrons.

\subsection{Polarized Structure}
\begin{figure}[tbh]
	\centering
	\includegraphics[width=0.6\textwidth]{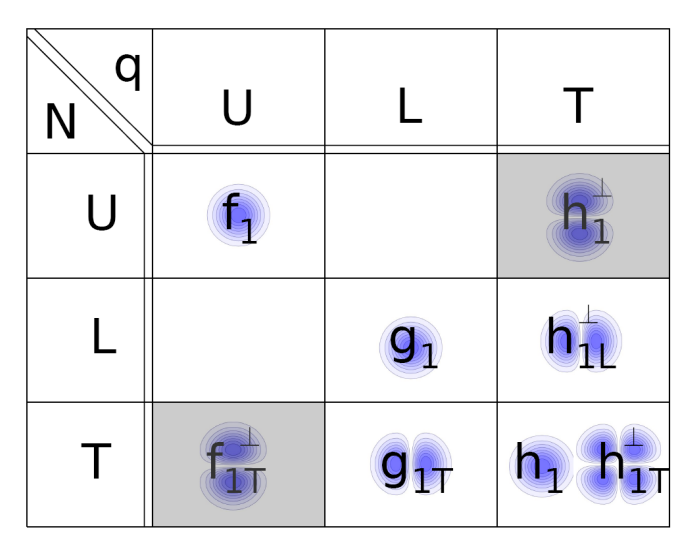}
	\caption{A table showing the eight possible TMD PDFs at twist-2 order, taken from a talk by Alexei Prokudin at the SPIN 2016 conference. The top row indicates the polarization of the quark within the nucleon, while the left column indicates the polarization of the nucleon. The unpolarized, helicity, and transversity distributions ($f_1$, $g_1$, and $h_1$, respectively) are the only functions that survive integration over partonic transverse momentum.}
	\label{fig:tmd_zoo}
\end{figure}

When the spin of the partons and nucleons is considered, there are two other TMD distributions that also survive integration over transverse momentum; these are referred to as the helicity distributions for longitudinally polarized partons and nucleons and the transversity distribution for transversely polarized partons and nucleons. Additionally, when partonic transverse momentum degrees of freedom are explicitly included in the functions, several new spin-dependent PDFs may be considered. In sum, at twist-2, or at leading power expansion in the hard scale $Q$, there are eight TMD PDFs which may depend on spin and partonic longitudinal and transverse momentum. These functions are shown in a table in Fig.~\ref{fig:tmd_zoo}, where the three PDFs without a $\perp$ superscript or T subscript on the diagonal are the only functions that survive integration over transverse momentum and include the unpolarized TMD PDF ($f_1$), the helicity PDF ($g_1$) and the transversity PDF ($h_1$). The other TMD PDFs manifest themselves as azimuthal modulations, which is why after integration over $k_T$ they become zero by symmetry considerations. It can also be considered that in the rigorous definition of each TMD PDF, there is a triple-product term that looks like $p\cdot(\vec{S}\times \vec{k}_T)$. Here, $p$ is the boost momentum of the hadron, $\vec{S}$ is the spin direction of the parton or nucleon, and $k_T$ is the transverse momentum degree of freedom. Thus, when integrating over $k_T$, this cross product becomes zero for those TMD PDFs that explicitly depend on $k_T$. \par

\subsubsection{Longitudinally Polarized Structure}

The helicity distribution functions represent the difference in probability of scattering off of a parton with its spin vector parallel vs.~antiparallel with the nucleon longitudinal spin. The helicity distributions are determined from global analyses of world data, similarly to the unpolarized collinear PDFs, and can be probed in a variety of high energy collisions. These distributions provide the spin contributions to the total spin of the nucleon, where the quark contribution is

\begin{equation}\label{eq:quark_spin_contribution}
\Delta\Sigma = \frac{1}{2}\sum_{q,\bar{q}}\int_0^1 dxx\Delta q_i(x) = \frac{1}{2}\int_0^1 dxx[\Delta q_i(x) + \Delta\bar{q}_i(x)]
\end{equation}
and the gluon contribution is
\begin{equation}\label{eq:gluon_spin_contribution}
\Delta G = \int_0^1 dxx\Delta g(x)\,.
\end{equation}

\begin{figure}
	\centering
	\includegraphics[width=0.7\textwidth]{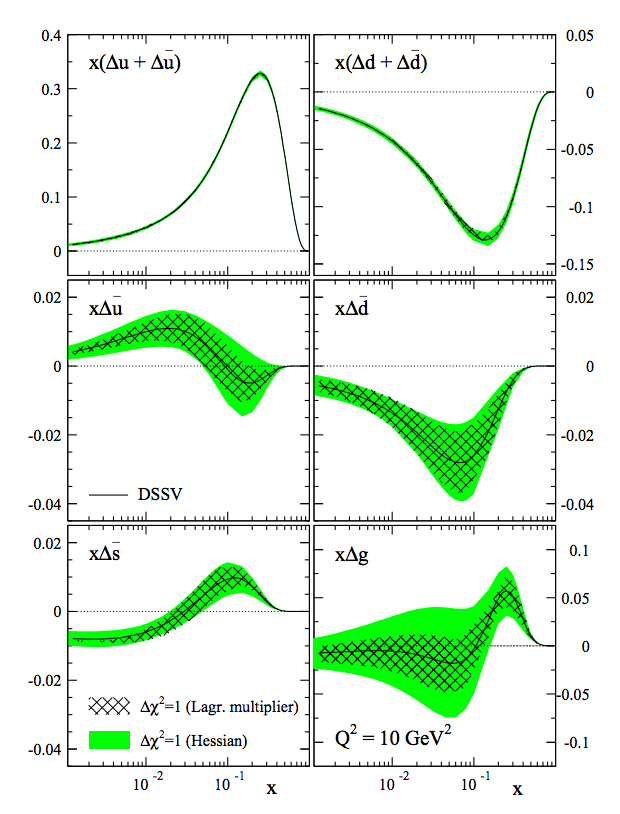}
	\caption{\label{fig:helicity_pdfs}The helicity PDFs for several quark flavors and gluons are shown, taken from Ref.~\cite{deFlorian:2009vb}.}

\end{figure}

A global analysis from 2008 shows the helicity distributions for various quark flavors, as well as the gluon in Fig.~\ref{fig:helicity_pdfs}~\cite{deFlorian:2009vb}. The up and down quark distributions are very well constrained, since in DIS the valence quarks can be accessed at LO. The anti quark helicity distributions have larger uncertainties as they are grouped together with the quark distributions since the probing lepton in DIS cannot differentiate between quarks and antiquarks; however, recent results from the PHENIX and STAR collaborations at the Relativistic Heavy Ion Collider (RHIC) have shown that the uncertainties on these distributions will be significantly reduced from $W^\pm$ boson longitudinal single spin asymmetry measurements~\cite{Aschenauer:2016our}. This process constrains antiquark distributions better than DIS since it tags a particular flavor antiquark explicitly in the creation of the W boson, most notably anti-up and anti-down quarks since high $x$ valence quarks are likely to be probed at RHIC center-of-mass energies.

\begin{figure}
	\centering
	\includegraphics[width=0.6\textwidth]{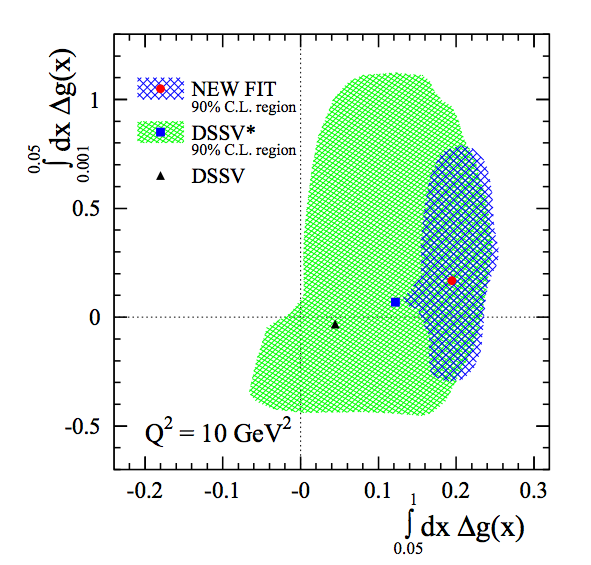}
	\caption{A recent global analysis of the gluon contribution to the proton spin has shown that at moderate and large $x$ the gluon contributes roughly 20\% of the total proton spin~\cite{deFlorian:2014yva}.}
	\label{fig:gluon_spin}
\end{figure}

In Ref.~\cite{deFlorian:2009vb} the gluon helicity distribution is also highly unconstrained since the gluon cannot be accessed at LO in DIS. The gluon helicity function has been of immense interest within the nucleon structure community due to the so called ``spin crisis'' (see e.g. Ref.~\cite{Aidala:2012mv} for a review). It was largely assumed that the valence quarks must carry the majority of the nucleon spin $\hbar/2$; however, it is now well established that they only contribute roughly 30\% of the total proton spin. For this reason determining the gluon spin contribution to the proton has been a top priority at the RHIC facility as gluons interact at LO in hadronic collisions, and RHIC is the world's only hadronic collider facility capable of polarizing proton beams in both the longitudinal and transverse directions. A recent global analysis has indicated that gluons may contribute up to roughly 20\% of the total proton spin at large $x$~\cite{deFlorian:2014yva}. Figure~\ref{fig:gluon_spin} shows that this is only in the region of $0.05<x<1$; there are still significant uncertainties at smaller momentum fractions $x$ where the unpolarized collinear gluon distribution is known to be quite large. Recent inclusive jet and dijet measurements from STAR which probe smaller $x$ and have not been included in the most recent global analyses will almost certainly provide a significant reduction in these uncertainties.

\subsubsection{Transversely Polarized Structure}

When transverse polarization and partonic transverse momentum degrees of freedom are considered, additional PDFs may be defined as described above and shown in Fig.~\ref{fig:tmd_zoo}. In this figure the subscript 1 refers to leading twist, while the L or T refers to the polarization of the nucleon. The TMD PDFs are explicitly dependent on both the partonic longitudinal and transverse momentum, where the initial partonic transverse momentum is often denoted $k_T$. Each TMD PDF is uniquely defined by a particular configuration of nucleon and partonic spin. Note that there are also TMD FFs that can be defined in addition to the collinear FFs, where in the final fragmentation state the function is explicitly dependent on the longitudinal and transverse momentum, $z$ and $j_T$ respectively, of the hadron with respect to the outgoing parton. For an observable to have sensitivity to TMD PDFs and/or TMD FFs, the observable must be sensitive to two scales $Q^2$ and $q_T$, where \qsq is the hard scale of the partonic interaction and $q_T$ is a soft scale on the order of $\Lambda_{QCD}$ such that $\Lambda_{QCD}\lesssim q_T\ll Q$. Observables sensitive to TMDs are therefore particularly interesting since they probe multiple scales.

\begin{figure}[tbh]
	\centering
	\includegraphics[width=1\textwidth]{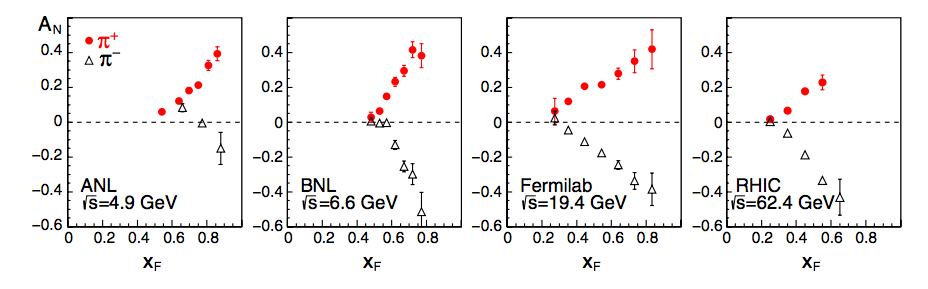}
	\caption{A collection of several transverse single-spin asymmetry measurements showing the surprisingly large dependence on $x_F$ and large asymmetries at large \sqs~\cite{Aidala:2012mv}.}
	\label{fig:tssa_measurements} 
\end{figure}

Transversely polarized proton structure came to the forefront of nucleon structure physics in the 1970's when the first measurement of the transverse single spin asymmetry (TSSA) was made by Ref.~\cite{Klem:1976ui} in collisions of a transversely polarized proton with an unpolarized proton. The TSSA is defined by 
\begin{equation}
A_N(\phi) = \frac{\sigma^\uparrow(\phi)-\sigma^\downarrow(\phi)}{\sigma^\uparrow(\phi)+\sigma^\downarrow(\phi)}
\end{equation}
and is a left-right asymmetry measurement as a function of $\phi$ with respect to the transverse spin of the polarized proton. Purely perturbative calculations predicted that the asymmetry should be very small such that $A_N\sim m_q/Q$ where $m_q$ is a mass scale of the interacting quark and $Q^2$ is the hard scale of the interaction~\cite{Kane:1978nd}. The measurement in Ref.~\cite{Klem:1976ui} was performed at center-of-mass energies of approximately \sqs$\approx$~5 GeV, and found that asymmetries rose to tens of percent at large $x_F=2p_z/\sqs$; this was orders of magnitude larger than the perturbative prediction. This discovery prompted many future measurements to understand if this result was potentially just due to the small center-of-mass energies where perturbative calculations were known to be unrealiable. Nonetheless, a flurry of additional measurements in the last several decades have shown that the asymmetries persist up to center-of-mass energies of 500 GeV and \pt of 8 GeV/$c$, well beyond the limit where perturbative calculations have been successful in describing inclusive hadron cross sections~\cite{Arsene:2008aa, Abelev:2008af, Adare:2013ekj,Adare:2014qzo, Dilks:2016ufy}. Figure~\ref{fig:tssa_measurements} shows a collection of several of these measurements. Nonetheless, neutral pion asymmetries have been found to be consistent with zero at midrapidity up to $p_T\sim$~15 \gev with excellent statistical precision~\cite{Boer:2017ycl}, where perturbative calculations would be expected to hold. See e.g. Ref.~\cite{Aidala:2012mv} for an extensive list of previous TSSA measurements. It is also surprising that the TSSA measurements show no obvious dependence on \sqs; these measurements have indicated that the asymmetries must be nonperturbatively generated in the initial and/or final hadronic states. They also appear to be highly dependent on the rapidity of the final-state hadron, regardless of the \sqs or hadron \pt. \par

The measurements in the 1970's and 1980's prompted theoretical work on transverse partonic dynamics in the early 1990's to attempt to explain these surprising results~\cite{Mulders:1995dh}, leading to the birth of the era focusing on parton dynamics in QCD. Ultimately, two TMD functions were proposed as explanations for the asymmetries: the Sivers function~\cite{Sivers:1989cc, Sivers:1990fh} in the initial state and the Collins function~\cite{Collins:1992kk} in the final state. The Sivers function corresponds to a spin-momentum correlation between the intial-state nucleon transverse spin and the partonic transverse momentum $k_T$, while the Collins function corresponds to a correlation between the final-state partonic transverse spin and the hadronic transverse momentum $j_T$. Measurements of these functions in SIDIS~\cite{Airapetian:2004tw, Adolph:2014zba, Qian:2011py, Avakian:2010ae, Airapetian:2010ds} and \eplus\eminus annihilation~\cite{Abe:2005zx,TheBABAR:2013yha} have shown that both the Sivers and Collins functions give rise to sizeable asymmetries, up to $\sim$15\% in the case of the Collins asymmetries in \eplus\eminus annihilation. Measurements in the Drell-Yan (DY) process, \pp$\rightarrow\ell^+\ell^-$, have shown additional TMD PDFs, such as the Boer-Mulders function, are nonzero as well~\cite{Guanziroli:1987rp,Conway:1989fs,Zhu:2008sj}.

In addition to the TMD framework, the collinear twist-3 framework has also been used to describe the large TSSA measurements. TMD functions are twist-2 functions, indicating that only the hard scattering of two partons is considered. In the collinear twist-3 framework, the nonperturbative functions remain dependent on only the longitudinal momentum fractions; while observables sensitive to TMD functions are two scale problems such that $\Lambda_{QCD}\lesssim q_T\ll Q$, observables sensitive to twist-3 collinear functions have only one scale such that $\Lambda_{QCD}\ll q_T\sim Q$. The theoretical calculation now considers additional partonic scatterings which can be correlated and can generate a large TSSA. In this case, a quark-gluon correlation function can be defined in either the initial state~\cite{Qiu:1998ia} or the final state~\cite{Kanazawa:2000hz}. The twist-3 framework has been used to describe TSSA measurements at RHIC~\cite{Kanazawa:2011bg,Metz:2012ct, Kanazawa:2014dca}; however, there is ultimately more information within the TMD framework since the collinear functions lack explicit dependence on partonic transverse momentum scales. The twist-3 approach has been related to $k_T$ moments of TMD PDFs and FFs (such as the Sivers and Collins functions)~\cite{Ji:2006ub}, and verifying the relation between these two frameworks with multi-differential measurements will be an important achievement in nucleon structure research.

\section{Modified Universality of Parity-Time-Odd TMD PDFs}

The recent focus on multidimensional structure of the nucleon has not only offered richer information about confined parton dynamics but has also brought to light fundamental predictions about QCD as a gauge invariant field theory. In particular, the role of color interactions between partons and remnants of the hard scattering have become clearer. Since QCD is a non-Abelian gauge theory which exhibits the characteristics of confinement, when partons interact via a hard scattering they must also be exciting the gluon field surrounding them; this can allow for soft gluon exchanges, and thus color exchanges, with the spectator partons from the remaining hadrons. These types of interactions have led to new predictions regarding the universality of certain TMD PDFs, which is in general regarded as a fundamental pillar of the treatment of nonperturbative functions.

In particular, certain TMD PDFs have been shown to exhibit ``modified'' universality due to color interactions. Recalling from Sec.~\ref{sec:qcd}, collinear PDFs are taken to be universal functions which can be measured in one process, e.g. SIDIS, and then used for predictions in another process, e.g. \pp scattering. It was first suggested that parity and time (PT) odd TMD PDFs could be nonzero due to phase interference effects from soft gluon exchanges in SIDIS~\cite{Brodsky:2002cx,Belitsky:2002sm}. Shortly after this conclusion, due to the gauge invariant nature of QCD and PT odd nature of certain TMD PDFs, it was predicted that these TMD PDFs will be the same magnitude and shape but have opposite sign between SIDIS and DY interactions~\cite{Collins:2002kn}. Twist-2 TMD PDFs that involve only one polarization vector are odd under a PT transformation, which is one simple explanation for the effect.

\begin{figure}[tbh]
	\centering
	\includegraphics[width=0.49\textwidth]{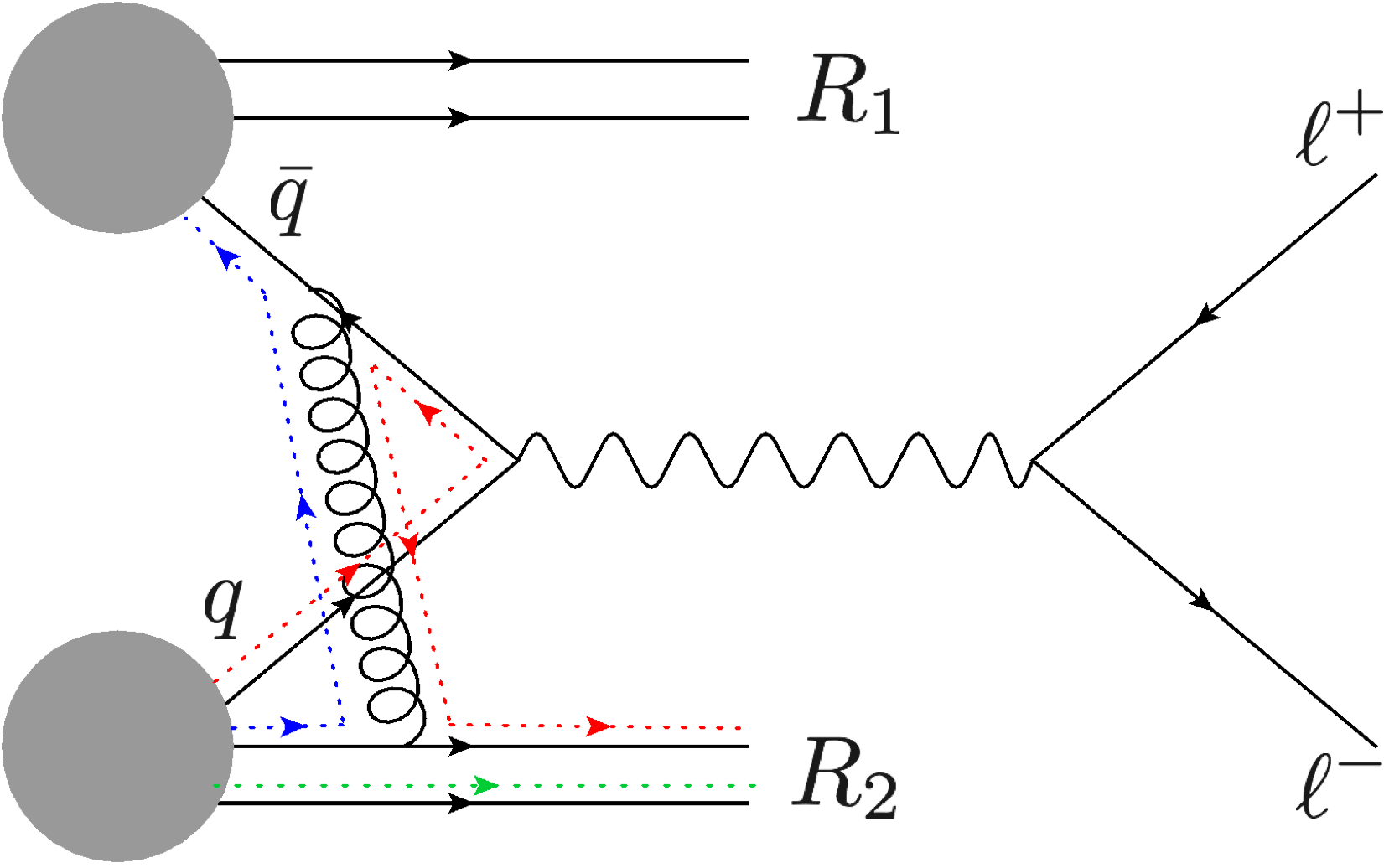}
	\includegraphics[width=0.49\textwidth]{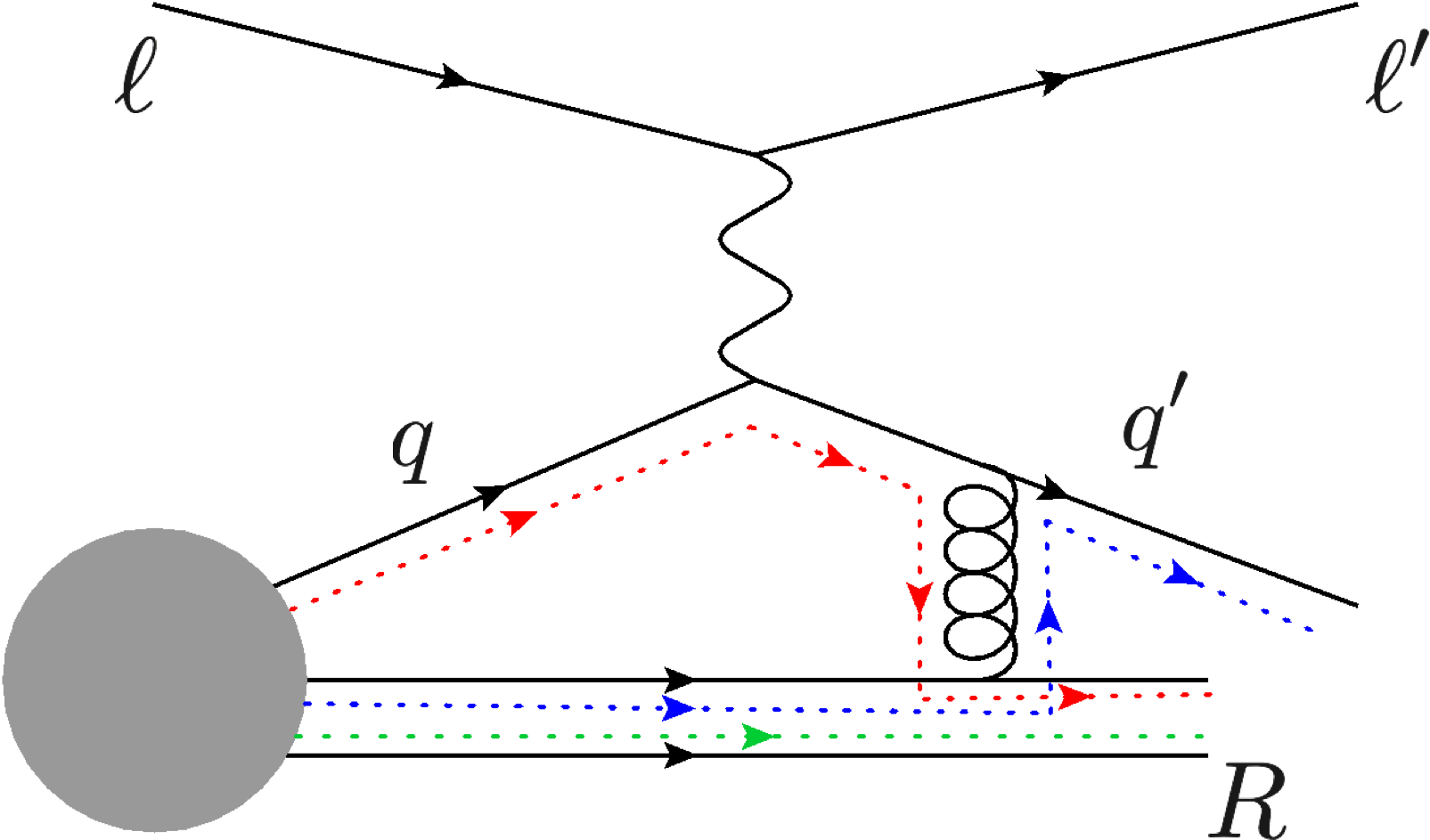}
	\caption{The color flow in DY (left) and SIDIS (right) is shown when a soft gluon attaches between a hard scattered parton and spectator in the initial state and final state, respectively. In the figures, $\ell$ refers to a lepton, $q$ or $\bar{q}$ refer to a quark or anti-quark, respectively, and $R$ refers to the remnants of a nucleon after a hard scattering.}
	\label{fig:color_flows}
\end{figure}

The sign difference arises due to different color flows that are possible in the initial vs. final state in these two processes. Figure~\ref{fig:color_flows} shows two Feynman diagrams of DY and SIDIS, each with a soft gluon attachment between a hard scattered parton and spectators of the collision. Since the DY process has no final-state colored particles, soft gluons can only attach in the initial state. This leads to the color flow shown; note that due to the gluon exchange, and thus color rotation, the red and anti-blue color lines on the interacting quark and remnant are pointing in the opposite direction. In SIDIS there are only final-state spectators, thus soft gluons can only attach with remnants after the hard scattering. In this case, the gluon exchange flips the color of the red quark, however due to the nature of the interaction the color lines of the outgoing quark and remnant are pointing in the \emph{same} direction. This difference is another qualitative feature that would lead us to expect modified universality as it is a feature of the color fields; in the case of DY the gluon fields destructively interfere while in the case of SIDIS they constructively interfere. These interferences lead to the predicted modified universality of PT odd TMD PDFs, which are the Sivers function~\cite{Sivers:1989cc,Sivers:1990fh} and the Boer-Mulders function~\cite{Boer:1997nt}. \par

\begin{figure}[tbh]
	\centering
	\includegraphics[width=0.8\textwidth]{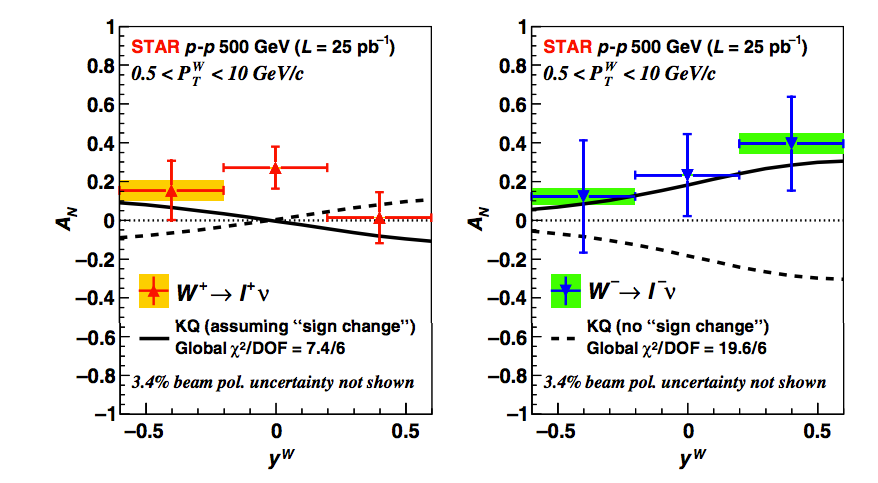}
	\caption{The $W$ boson TSSA as measured by the STAR collaboration~\cite{Adamczyk:2015gyk}. Calculations show that the measurement is more compatible with the prediction of modified universality; however, these calculations are highly dependent on TMD evolution effects which are still largely unknown.}
	\label{fig:star_w_an}
\end{figure}

\begin{figure}[tbh]
 	\centering
	\includegraphics[width=0.7\textwidth]{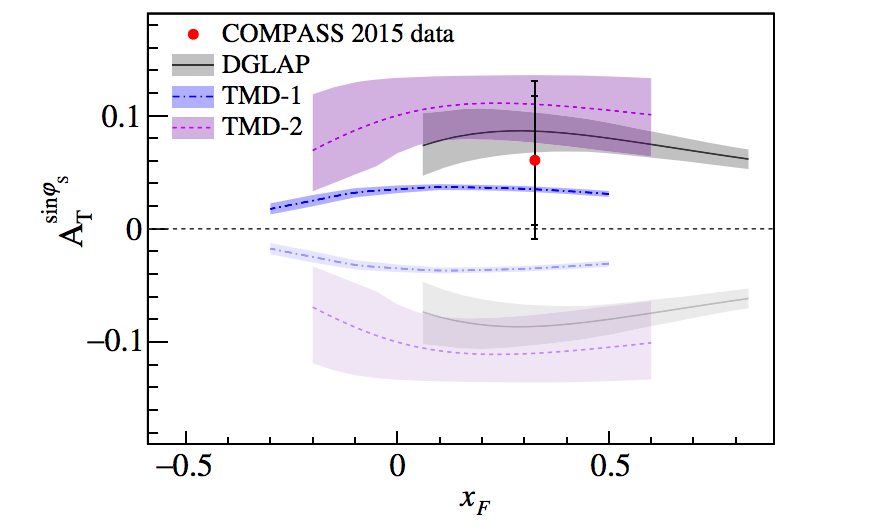}
	\caption{The first measurement of the DY TSSA, shown with calculations assuming the sign change (darkly shaded bands) and without the sign change (lightly shaded bands)~\cite{Aghasyan:2017jop}. The measurement is more compatible with the prediction of modified universality of the Sivers function; however, the uncertainties are still large enough that a conclusive statement cannot be made.}
	\label{fig:compass_dy_an}
\end{figure}

The Sivers function, rather than the Boer-Mulders function, has received the most attention in attempting to verify this prediction; this is like because it may contribute to the large TSSA measurements which have puzzled the nucleon structure community for decades. Many measurements from SIDIS exist~\cite{Airapetian:2004tw, Adolph:2014zba} which have shown that the Sivers function is nonzero. Despite this prediction being made 15 years ago, it was only very recently that the first measurements of DY and DY-like processes were reported which could be compared to the SIDIS measurements~\cite{Adamczyk:2015gyk,Aghasyan:2017jop}. Note that Ref.~\cite{Adamczyk:2015gyk} measured the $W^\pm$ and $Z^0$ boson TSSA, however the $W$ measurement should behave exactly like the DY process since it is a $q\bar{q}$ interaction in which the color annihilates to produce a colorless electroweak final state, therefore soft gluon exchanges between hard scattered partons and remnants are only possible in the initial-state. These measurements are shown in Figs.~\ref{fig:star_w_an} and~\ref{fig:compass_dy_an}, which show that they are still statistically limited. One recent theoretical examination of the data also shows compatibility with the predicted sign change~\cite{Anselmino:2016uie}. Nonetheless, significantly more data needs to be collected to draw a complete conclusion; as the prediction was made for the Sivers function the sign change must hold true for the function across all $x$ and $k_T$. Thus, more measurements are necessary before stronger conclusions can be drawn; however, these are important first steps towards verifying this important prediction of the QCD TMD factorization framework. The STAR collaboration collected nearly 20 times more data in 2017 than what was published in Ref.~\cite{Adamczyk:2015gyk}, and this will result in a sizable reduction in uncertainties in the W and Z boson TSSA measurements~\cite{Aschenauer:2016our}.

It is important to emphasize the significant qualitative shift in thinking that the predicted sign change of PT odd TMD PDFs has brought about. Previously, QCD interactions were treated as $2\rightarrow2$ hard scatterings, where the nucleonic structure is parameterized in the form of PDFs and FFs. The prediction of modified universality shows the importance of interactions between the hard scattered parton and the remnants of the collision. In particular, it emphasizes that non-negligble interactions take place in hard scatterings with the remnants of the collision and takes into account more thoroughly the bound state nature of the nucleon. This qualitative shift in thinking is significant as the predicted sign change shows nonzero effects, calculated rigorously within perturbative QCD, that are due to soft interactions of partons with other colored objects. 

\section{Factorization Breaking}

The predicted modified universality of the Sivers TMD PDF ultimately arises due to the difference in color field interference in the initial vs. final state that is possible in DY and SIDIS respectively. At tree level DY and SIDIS are both electroweak processes in which an electroweak boson is exchanged at the interaction vertex, and they represent only a small phase space of processes which can be used to probe QCD interactions. In hadronic collisions where a final-state hadron is measured, soft gluons can be exchanged in both the initial and final states since color is exchanged at the interaction vertex, and is therefore necessarily present in both the initial \emph{and} final state. For observables sensitive to a small transverse momentum scale in hadronic collisions where a final-state hadron is measured, factorization breaking has been predicted in a TMD framework in both polarized and unpolarized interactions~\cite{Bomhof:2006dp, Collins:2007nk, Collins:2007jp,Rogers:2010dm}. The nonperturbative objects in the calculation of a cross section become correlated with one another, such that a convolution of individual TMD PDFs and TMD FFs cannot be defined; this intuitively means that the partons are correlated across the initial-state colliding protons and final-state hadrons. There are no current theoretical claims that the perturbatively calculated partonic cross section does not factorize from the nonperturbative correlation function. Factorization breaking has also recently been explored in the twist-3 framework~\cite{Zhou:2017mpw}, however it is more established in the TMD framework and further studies within the collinear twist-3 framework will be important in establishing the connection between the TMD and twist-3 frameworks in the intermediate region of \qsq where they describe similar physics. It should be stressed that this result is due to the non-Abelian nature of QCD~\cite{Rogers:2010dm}, thus by investigating factorization breaking effects we are probing fundamental physics about non-Abelian gauge invariant quantum field theories and, in this case, manifestations of the SU(3) gauge group in nature. \par

\begin{figure}[tbh]
	\centering
	\includegraphics[width=0.8\textwidth]{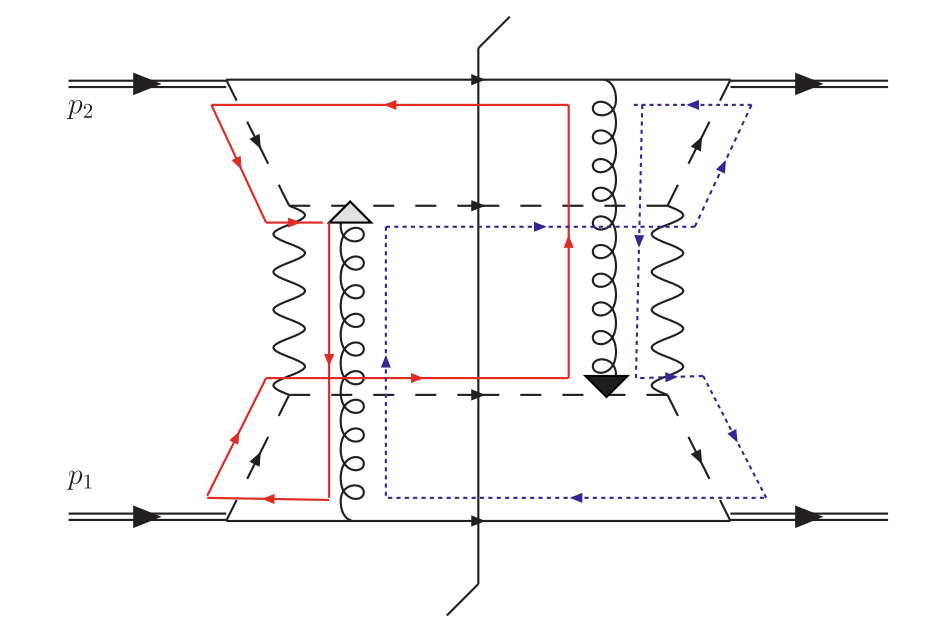}
	\caption{A diagram showing the color flow through the process $\pp\rightarrow h_1+h_2$, taken from Ref.~\cite{Rogers:2010dm}. The color flows connect the partons from one proton with the remnants of the other proton.}
	\label{fig:color_entanglement}
\end{figure} 

Because of the complicated color flows that are possible in processes which break factorization, it is commonly referred to as ``color entanglement'' since the color flow through the hard scattering is what is responsible for the nonperturbative TMD functions becoming correlated with one another. Throughout this work, the terms factorization breaking and color entanglement will refer to the same physical phenomena. Figure~\ref{fig:color_entanglement} shows the most simple example of these color flows in a diagram where the partons exchange soft gluons with the spectators of the collision in both the initial and final state. The process diagrammed is $\pp\rightarrow h_1+h_2$, where $h_1$ and $h_2$ are two nearly back-to-back hadrons. When the hadrons have large \pt and are nearly back-to-back they can be treated in a TMD framework since the acoplanarity in the transverse plane gives a small transverse momentum scale. The diagram shows that the color through the parton from hadron $p_1$ (red) can flow through the hard scattering and the remnant of hadron $p_2$, connecting the parton with the other hadron via its color. Similarly in the final-state, the color of the parton from hadron $p_2$ (blue) can flow through the hard scattering and the remnant of hadron $p_1$. Physically, the color which flows throughout the scattering is what causes the nonperturbative functions to become correlated, and thus no longer factorizable. Note that this is only possible when at least two gluons are exchanged, which is why factorization is predicted to hold in both DY and SIDIS in a TMD framework.

It is important to emphasize that the same physics which motivates the predicted sign change of PT odd TMD PDFs is responsible for the result of factorization breaking in more complicated QCD processes. Soft gluon exchanges which can occur in the initial and final states with spectators of the hard interaction cannot in general be eliminated via a gauge transformation. Thus, their color fields create interference effects that necessarily affect the partons and remnants involved in the hard scattering. The ideas behind these predictions reflect major qualitative shifts in the way calculations are performed. Previous perturbative calculations only considered the hard scattering and additional radiations at NLO and subsequent orders, but did not consider the interaction of these gluons with spectators of the collision. The departure from these simplistic ideas represents an important step towards treating the nucleon as a composite and complicated object rather than just a collection of partons which are independent from one another. \par

Factorization is ultimately a tool that is used in order to make theoretical QCD computations easier; it is a prescription for which the nonperturbative behavior of QCD can be straightforwardly handled. Nonetheless, it is an assumption that has only been proven to all orders for a select few processes that always have an electroweak vector boson which mediates the scattering. Even at the collinear level factorization has been shown to break down at third loop order for processes with hadrons in the initial and final states~\cite{Catani:2011st,Forshaw:2012bi}, and this ultimately points towards a more accurate picture of nucleon structure which requires that partons are dynamic objects within a complicated quantum mechanical bound state. Results such as these challenge the conventional picture of two partons interacting via a single hard scattering and should push QCD into new research areas which investigate how to describe correlated partons. In particular, TMD PDFs and TMD FFs would be replaced by a correlation function that must be interaction dependent; however, no such theoretical studies have been performed to explore what this correlation function might look like. \par

\section{The Role of Color in Hadronic Collisions}

The predictions of modified universality of PT odd TMD PDFs and factorization breaking represent a major qualitative shift in the way QCD interactions are approached. In particular, the role of color has been shown to have nonzero and predictable qualitative and quantitative effects due to the non-Abelian nature of QCD. Within QCD a major paradigm shift has begun in the way color is treated. The modified universality of PT odd TMD PDFs and factorization breaking are just one example where differences in color flow lead to a predictable and experimentally measurable, in the case of the Sivers function, nonzero effect. In addition to effects in the TMD framework, other worldwide measurements are probing effects from color and correlations due to nontrivial color flows in hadronic interactions.

\subsection{Color Coherence}

Historically, color has always been an integral part of QCD; for the theory to obey SU(3) group symmetry this must be the case. However, efforts to quantify effects from color correlations have not always been at the forefront of QCD research. Some of the earliest studies of color effects known as color coherence were measured in $e^+e^-$ annihilation to three jet final states~\cite{Bartel:1983ij,Althoff:1985wt,Akrawy:1990ha}. Color coherence is a phenomenon predicted within QCD that soft radiation should be suppressed between color correlated partons. In the case of $e^+e^-\rightarrow\bar{q}q$ the quarks are necessarily color correlated, and when a third jet is measured, necessarily from a gluon radiation, measurements show that soft hadron production is depleted in certain regions of phase space. This depletion is due to destructive interference from gluon radiation; also known as the ``drag'' effect, the radiated gluon is said to ``drag'' color away from the partons leading to the interference effect~\cite{dokshitzer_basicspqcd}.

Before color coherence in QCD is discussed, coherence phenomena should be generally introduced. For an extensive discussion, see in particular chapter four of Ref.~\cite{dokshitzer_basicspqcd}. Coherence phenomena can be found in any gauge theory, and because of this we can first consider QED which is simpler than QCD. The crux of coherence is to what extent, for example in QED, a relativistic \eplus\eminus pair radiate photons independently from one another. For the example of bremsstrahlung, it can be shown that if the \eplus\eminus pair radiate photons independently, then the condition $\theta_{\gamma\eminus}<\theta_{\eplus\eminus}$ or $\theta_{\gamma\eplus} < \theta_{\eplus\eminus}$ must be met; here the angles $\theta$ are the respective angle between the particles noted in the subscripts. This result is referred to as ``angular ordering,'' and dictates the radiation pattern of relativistic particles. In the case of QCD, this results in a uniform decrease of opening angles in the partonic cascade; in other words, partons radiate gluons at large angles initially and these angles successively become smaller and more collimated. Angular ordered parton cascades are implemented in many Monte Carlo event generators, and are often attributed to the description of color coherence measurements (see e.g.~\cite{Chatrchyan:2013fha}).

\begin{figure}[tbh]
	\centering
	\includegraphics[width=0.7\textwidth]{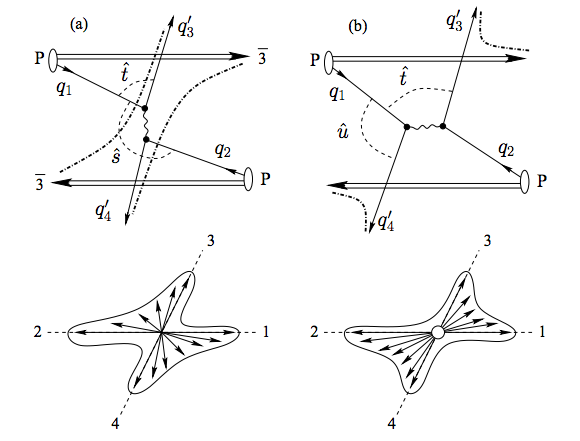}
	\caption{Hadron production regions for two high $p_T$ qq scattering processes with different color topologies, from Ref.~\cite{dokshitzer_basicspqcd}.} 
	\label{fig:color_topologies}
\end{figure}

In hadronic collisions, coherence effects are more complicated because color can connect both the initial and final states whereas in \eplus\eminus annihilation color correlations can only be present in the final state. For an extensive discussion of color coherence in hadronic collisions, see chapter ten of Ref.~\cite{dokshitzer_basicspqcd}. In these cases the color topology of a particular partonic hard process results in different regions of hadron depletion, as depicted in Fig.~\ref{fig:color_topologies}. For a generic $qq\rightarrow qq$ scattering, depending on the particular color exchange at the interaction vertex, gluons may destructively interfere in different areas of phase space which results in different regions of hadron depletion. Ultimately color connections manifest themselves in global event observables, and inclusive QCD measurements are not sufficient to have sensitivity to these types of color phenomena. This is likely why factorization generally holds for inclusive collinear observables in hadronic collisions; however, when additional, more realistic, considerations are made (e.g. in the case of dijet measurements or multi-loop processes), factorization breaks down due to complex color flows.  

It is particularly interesting to point out that in Fig.~\ref{fig:color_topologies}, the regions of depletion are determined via the dashed-dotted lines which connect hard scattered partons from one proton to the remnants of the other nucleon. These lines dictate the regions where hadron production will be largest; thus, they also dictate where gluons will destructively interfere leading to hadron depletions. This is reminiscent of color entanglement, where hard scattered partons interact with remnants and lead to complex color correlations. It is likely that color coherence and color entanglement are referring to the same physical phenomenon; namely that complex color flows connect all of the colored objects in a particular hadronic interaction.

\begin{figure}[tbh]
	\centering
	\includegraphics[width=0.49\textwidth]{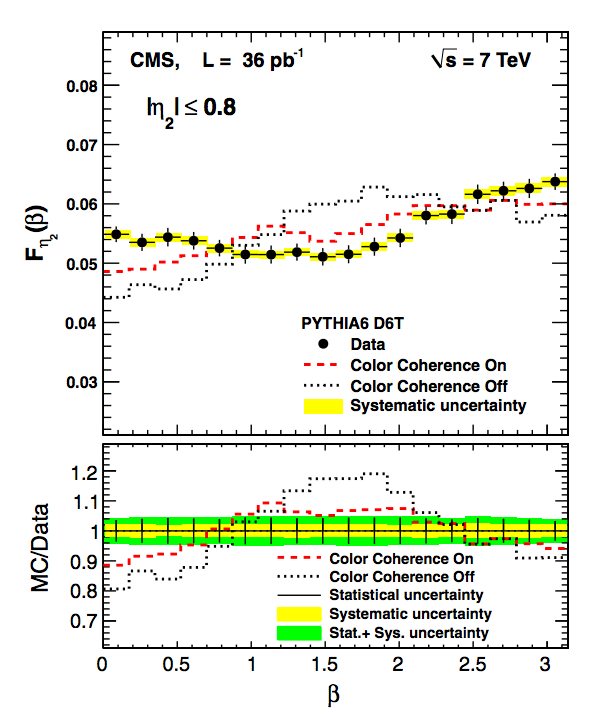}
	\includegraphics[width=0.49\textwidth]{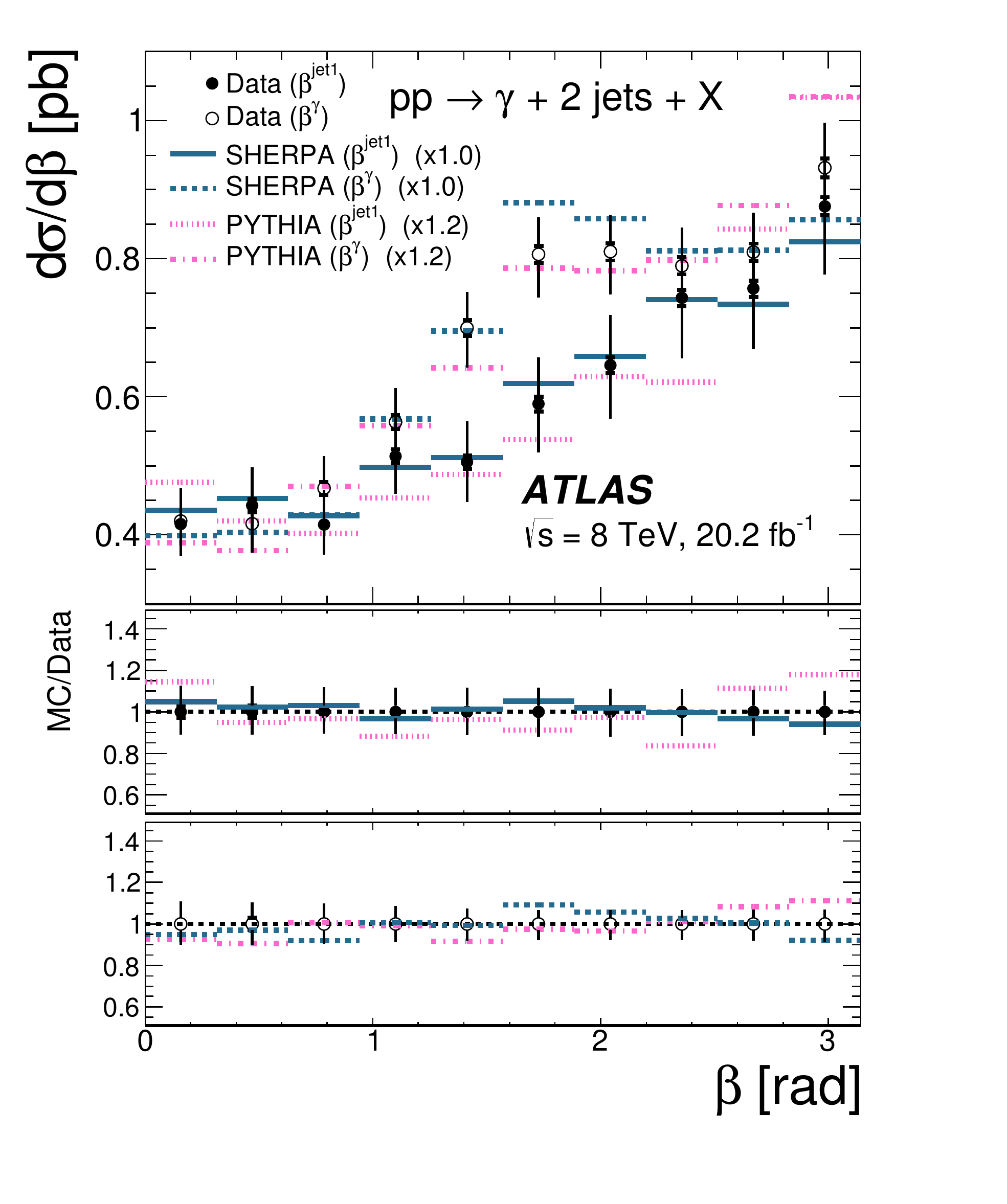}
	\caption{An observable which quantifies the angle between a subleading jet and sub-subleading jet in $\phi-\eta$ space shows evidence for color coherence in \pp collisions in both dijet (left) and $\gamma$-jet (right) events~\cite{Chatrchyan:2013fha,Aaboud:2016sdm}.}
	\label{fig:cdf_cc}
\end{figure}

Experimentally, color coherence effects can connect the initial and final states in hadronic collisions where at least one final-state jet is measured; however, modern observables require multijet final-states to be observed. These were first studied by the Tevatron experiments nearly 10 years after the first \eplus\eminus color coherence publication~\cite{Abe:1994nj,Abbott:1997bk, Abbott:1999cu}. Since the detectors at RHIC were not designed to be robust jet detectors, no RHIC experiments attempted to study color coherence effects; this left a dearth of studies until the LHC began collecting data. The topic was revived by the CMS collaboration in three-jet final states~\cite{Chatrchyan:2013fha} and has since been studied by the ATLAS collaboration in the photon+jets final state~\cite{Aaboud:2016sdm}, where the main results from each analysis are shown in Fig.~\ref{fig:cdf_cc}. These measurements have largely been used to constrain parton showering algorithms in Monte Carlo event generators (see e.g. Ref.~\cite{Ritzmann:2012ca}). Nonetheless, color coherence effects have been treated rigorously within QCD in the past (see e.g. Ref.~\cite{dokshitzer_basicspqcd} and references within) and could be used to constrain the magnitude of color correlations in future calculations. Future phenomenological calculations within a TMD framework may be able to use this data to constrain the magnitude of color entanglement effects, since each of these measurements requires a nearly back-to-back dijet or photon-jet pair.

\subsection{High Multiplicity Effects}

Since the LHC experiments began collecting data, novel phenomena in high multiplicity final states have been a subject of intense study in hadronic collisions. In particular, with the larger center-of-mass energies available at the LHC, events containing large final-state hadron multiplicities are produced significantly more frequently than at RHIC. These types of events have shown unexpected and striking similarities across many different collision systems, ranging from \pp to \pa to heavy nucleus-nucleus (A+A) collisions. The role of color in these interactions has also recently begun to be explored, for example in \jpsi and $\psi^\prime$ production in \pa collsions~\cite{Ma:2017rsu}.

One of the main goals set for the RHIC physics program was to measure the strongly interacting quark gluon plasma (QGP) and its properties. The QGP is formed in heavy nucleus-nucleus interactions, where the energy density of strongly interacting particles becomes so large that quarks and gluons become deconfined. Measuring the properties and characteristics of the QGP is a highly active field within QCD research, and various measurements have been proposed as potential signatures for the QGP. One such measurement is ``jet quenching,'' which refers to the suppression of high $p_T$ hadrons or jets in A+A collisions relative to \pp collisions. Jet quenching was predicted since the high $p_T$ partons should interact with the QGP in A+A collisions and not \pp collisions, thus losing energy from gluon bremsstrahlung amongst other interactions~\cite{Adcox:2001jp}. Another signature, the ``ridge,'' refers to a near-side in $\dphi\sim0$ long range in rapidity $\Delta\eta\gtrsim2$ two-particle correlation that was predicted to arise from a longitudinally expanding medium~\cite{Adams:2004pa}. A third signature predicted that bound states with strange quark content would be enhanced in A+A collisions relative to \pp collisions~\cite{Koch:1986ud}. In each of these cases, there are many measurements which show each of these predictions to be true in A+A collisions; for examples see Refs.~\cite{Adcox:2001jp,Adams:2004pa,Abelev:2007xp} and references and citations within. Many of these measurements were used to establish the existence of the QGP in A+A collisions.

These proposed signatures of the QGP came into question when, in 2010, the CMS collaboration reported measurements of the long range ridge structure in $\Delta\eta$ in high multiplicity \pp collisions~\cite{CMS_pp_collectivity}, where high multiplicity refers to events where the final-state track multiplicity was greater than 110. Since then, a surge of measurements has shown these signatures in \pp, proton-nucleus, deuteron-nucleus, and helium-nucleus collisions, even down to track multiplicities of 50 in \pp collisions~\cite{ATLAS_pp_collectivity,CMS_pp_collectivity2,Aidala:2016vgl,ATLAS_pPb_collectivity,CMS_pPb_collectivity,Aidala:2017ajz,Adare:2015ctn}. While there is little disagreement that a QGP is formed in A+A collisions, there is serious debate as to whether or not these results imply there is also a QGP being formed in smaller collision systems where the energy densities achieved are not nearly as large. Moreover, strangeness enhancement in high multiplicity \pp and \pa collisions was observed by the ALICE experiment~\cite{ALICE_strangeness}, meaning that two of the three proposed ``signatures'' for the QGP listed above have been measured in high multiplicity \pp and \pa collisions. However, jet quenching of high \pt jets or hadrons has yet to be measured in these small systems~\cite{Adler:2006wg,Adare:2015gla,Aad:2016zif,Khachatryan:2016xdg}, which calls into question the interpretation that a QGP is formed in \pa and \pp collisions. 

\begin{figure}[tbh]
	\centering
	\includegraphics[width=0.4\textwidth]{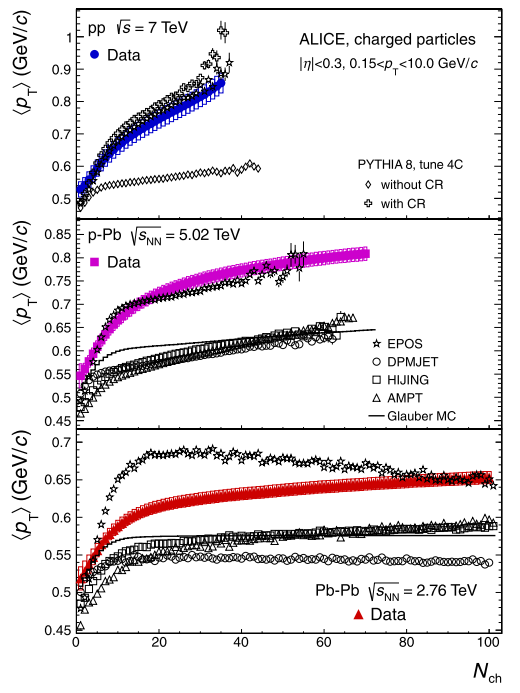}
	\caption{The average \pt of charged tracks as a function of final-state track multiplicity is shown in several hadronic collision systems~\cite{ALICE_pp_mult}. In particular the role of color reconnections is demonstrated in comparisons between the \pp data and the PYTHIA8 Monte Carlo generator.}
	\label{fig:avgpt_alice}
\end{figure}
 
These results may imply that there is an underlying QCD mechanism that is responsible for the surprising measurements in \pp and \pa collisions; color is an obvious candidate for these phenomena. For example, the average \pt of charged hadrons as a function of final-state multiplicity has been shown to be highly dependent on color connections and displays a similar shape across collision systems~\cite{ALICE_pp_mult,Skands_ColoredStrings}. As shown in Fig.~\ref{fig:avgpt_alice}, the average \pt in \pp collisions is not described at all by Monte Carlo event generators without color reconnections; these correlations have been tuned in event generators by measurements. Similarly the role of color correlations has also been explored in the long-range $\Delta\eta$ correlations that were surprisingly found in \pp collisions. In particular, it was found that color correlations between multiple partonic interactions could generate a nonzero second Fourier harmonic in the azimuthal angular distribution of particles with respect to the event plane of the nucleus-nucleus collision, the observable which is often attributed to collective behavior expected to be found only in A+A collisions~\cite{Ortiz:2013yxa}. Initial-state correlations that form instantaneously after the collision or must pre-exist in the incoming projectiles have also been proposed as explanations for the long range correlations~\cite{Dumitru_pp_ridge,Dusling:2017aot,Dusling:2017dqg}; additionally connections are now being made between the long-range phenomena and the three dimensional structure of the proton, including possible sensitivity to the protons', potentially fluctuating, color charge density~\cite{Dusling:2018hsg}. Interestingly, recent preliminary results from the ALEPH and ZEUS collaborations have shown that no long range pseudorapidity correlations have been measured in $e^+e^-$ annihilation and semi-inclusive DIS, respectively. This could suggest that initial-state color correlations are important in the measured effects in \pp and \pa collisions.

\subsection{Jet Substructure Techniques}

With the advent of robust jet finding algorithms in the last decade that are both infrared and collinear safe, a significant amount of study has since been invested in jet substructure techniques. In particular, the goal of jet substructure techniques is to systematically identify products from the hard scattered parton, rather than also grouping soft radiation that may have come from remnant interactions into the jet definition. These algorithms have been extensively developed in the search for BSM physics, and are just recently starting to be applied to more traditional QCD physics~\cite{Aad:2013gja}. So called jet grooming algorithms~\cite{Krohn:2009th,Larkoski:2014wba}, which have primarily been developed to remove physics related to remnant interactions, are also being used to probe color flows and interactions through high energy \pp scatterings. \par

\begin{figure}[tbh]
	\centering
	\includegraphics[width=0.6\textwidth]{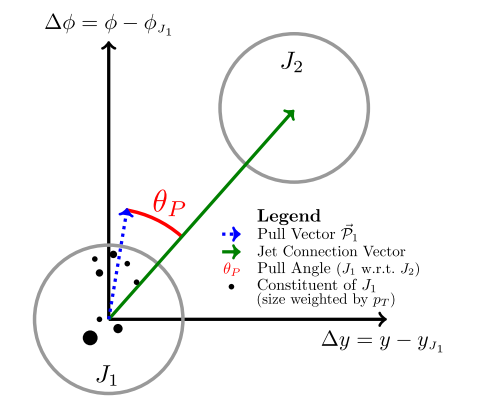}
	\caption{A diagram showing the definition of the jet pull vector for a generic dijet system, taken from Ref.~\cite{ATLAS:2017iaz}.}
	\label{fig:jet_pull}
\end{figure}

\begin{figure}[tbh]
	\centering
	\includegraphics[width=0.5\textwidth]{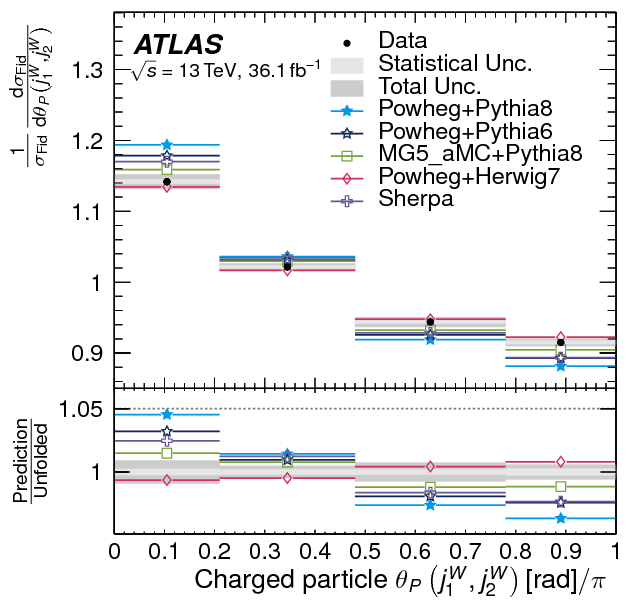}
	\caption{The ATLAS collaboration has measured the charged particle pull angle in $t\bar{t}$ events, showing that it is sensitive to the color flow in the process~\cite{ATLAS:2017iaz}.}
	\label{fig:atlas_ttbar} 
\end{figure}

For example, a recent study from the ATLAS collaboration has shown that an observable called the jet pull vector is sensitive to color connected top anti-top quark production in \pp collisions~\cite{ATLAS:2017iaz,Gallicchio:2010sw,Aad:2015lxa}. This vector is defined as a \pt weighted radial moment of the jet, and its direction in a dijet system is shown in Fig.~\ref{fig:jet_pull} as taken from Ref.~\cite{ATLAS:2017iaz}. For the case of a color connected system, the jet pull angle is expected to be aligned with the jet connection axis, or $\theta_{jp}\sim0$. If the jets are produced without color connection, the angle should be distributed uniformly. Both measurements in Ref.~\cite{Aad:2015lxa,ATLAS:2017iaz} find that the observable is strongly correlated to 0, and monotonically decreases towards large pull angles as shown in Fig.~\ref{fig:atlas_ttbar}; this indicates that the observable is indeed sensitive to the color flow of the $t\bar{t}$ quark events. There are also theoretical studies of the role of color flow in $t\bar{t}$ events, specifically within the TMD framework~\cite{Cao:2018ntd}, potentially drawing more connections between the largely separate fields of QCD and BSM physics.

While the previous measurements can be used to constrain QCD color connections, the motivation of the studies was largely to demonstrate that the observable could discern between various color flows and thus could be used to search for BSM model physics processes~\cite{Aad:2015lxa}. However, recent theoretical efforts in Soft Collinear Effective Theory (SCET) have indicated that jet grooming algorithms could potentially remove factorization breaking components from jet observables (see e.g. Refs.~\cite{Larkoski:2014wba,Dasgupta:2013via,Makris:2017arq,Schwartz:2018obd} and references within). If true, rather than removing them for the purpose of searching for BSM physics the contributions could in principle be isolated and perhaps studied further. Regardless, the substructure techniques have already been shown to be sensitive to color flow and the relation between the color correlations between partons and factorization breaking effects could potentially be shown with these new observables and techniques.

\section{Using Collins-Soper-Sterman Evolution to Search For Color Entanglement}

A nominally straightforward method to search for factorization breaking effects is to compare a measurement from a process predicted to break factorization to a calculation which assumes that factorization holds. From the comparison, one could then immediately conclude if the calculation replicates the data, and to what quantitative degree they match. This could in principle also provide insight into the magnitude of factorization breaking effects; there is general consensus in the literature that factorization breaking occurs in hadronic collisions where a final-state hadron is measured and a TMD framework is applicable, however there is still significant disagreement as to how large these effects are and to what extent they might modify the cross section. It may be that, for example, effects from factorization breaking are significantly smaller than the current level of uncertainties in theoretical calculations from the factorization scale, and thus they are a negligible systematic uncertainty to a particular calculation compared to current uncertainties. \par

Unfortunately this particular method is not effective due to the current uncertainties on TMD calculations. Even for unpolarized cross sections, TMD calculations still have large uncertainties due to the nonperturbative functions being generally unconstrained. The first ever global fit of TMD data was just recently published~\cite{Bacchetta:2017gcc}, and while this is an excellent step in the right direction it is not precise enough to constrain TMD functions to the level that would be necessary for comparing a calculation to measurements. The global fit reproduces transverse momentum dependent widths generally well but still has normalization problems. There is still not enough worldwide data to constrain TMD functions to the level that collinear functions have been constrained; even with the significant amount of collinear data available current collinear cross section uncertainties are on the order of tens of percent (see e.g. Ref.~\cite{ppg186}). Thus, a different method to attempt to observe modified behavior from factorization breaking must be pursued. \par

In calculations of TMD processes where factorization is predicted to hold, the hard scale evolution of the interaction is known to be governed by the Collins-Soper-Sterman (CSS) evolution equation~\cite{Collins:1981uk, Collins:1981uw}. In contrast to the Dokshitzer-Gribov-Lipatov-Altarelli-Parisi (DGLAP) evolution equations~\cite{Altarelli:1977zs,Dokshitzer:1977sg,Gribov:1972ri}, which are used to evolve collinear nonperturbative functions from one hard scale to another,  CSS evolution explicitly considers a small transverse momentum scale. Additionally, in contrast to DGLAP evolution, CSS evolution depends on nonperturbative contributions. While the DGLAP equation is purely perturbative, the kernel for the CSS evolution equation involves the CSS soft factor~\cite{Collins:1984kg}, which generally contains nonperturbative contributions. Because of this TMD evolution differs greatly from collinear evolution in that it cannot simply be calculated; there are contributions that must be constrained by data. As TMD data has only recently begun to be collected in large amounts, quantitative TMD evolution predictions can vary drastically depending on assumptions and inputs for the nonperturbative contributions~\cite{Collins:2014jpa}. 

However, CSS evolution qualitatively predicts that any momentum width sensitive to nonperturbative transverse momentum would increase as the hard scale of the interaction increases. This prediction does not depend on the nonperturbative input and can be understood intuitively as a broadening of the phase space for gluon radiation with the increasing hard scale. For example, if the typical DY invariant mass ranges of 4-9 GeV/$c^2$ are compared to the Z boson scale, 91 GeV/$c^2$, significantly harder gluons may be radiated from the Z than for the DY pair simply due to the larger energy in the interaction. Because of this, the Z can have more nonperturbatively generated $k_T$ contributing to its total $p_T$, and thus the nonperturbative momentum width is larger than at the typical DY scale. This has been phenomenologically studied in several DY and Z boson analyses~\cite{Landry:2002ix,Konychev:2005iy,Schweitzer:2010tt}, as well as in SIDIS~\cite{Schweitzer:2010tt,Nadolsky:1999kb,Aidala:2014hva} where factorization is also predicted to hold.  It is important to emphasize that the CSS evolution equation comes directly out of the derivation of TMD factorization~\cite{Collins:2012ss}. Thus, it follows that a promising avenue to investigate factorization breaking effects is to look for qualitative differences from CSS evolution in processes that are predicted to break factorization. This additionally requires less information about the functional form of distributions from non-factorizable processes since observing the rate of change of a distribution inherently requires less information than the distribution itself.

\section{Two-Particle Correlations}

To have sensitivity to factorization breaking effects and CSS evolution, an observable must meet three requirements:

\begin{enumerate}
	\item It must be sensitive to a small transverse momentum scale such that it can be treated in a TMD framework.
	\item It must have colored partons in both the initial and final states of the hard process.
	\item It must be measured over a range of hard scales to observe effects from CSS evolution. 
\end{enumerate}

Nearly back-to-back dijet production in \pp collisions satisfies these requirements, when the process is measured over a range of hard scales. The hadronization of the two partons also allows dihadron production to be used as a proxy for the dijets. Two-particle correlation measurements were first proposed as a way to measure initial-state transverse momentum during the early development of QCD~\cite{Feynman:1978dt}. Correlation measurements have since been used to measure the partonic transverse momentum $k_T$ over a large range of center-of-mass energies~\cite{Apanasevich:1997hm,ppg029,ppg095,Adam:2015xea}. The correlations are sensitive to the initial-state $k_T$ of the colliding partons since at LO, they should emerge exactly back-to-back simply due to transverse momentum conservation; however, NLO $k_T$ effects can cause the jets to be acoplanar in the transverse plane.

\begin{figure}[tbh]
	\centering
	\includegraphics[width=0.6\textwidth]{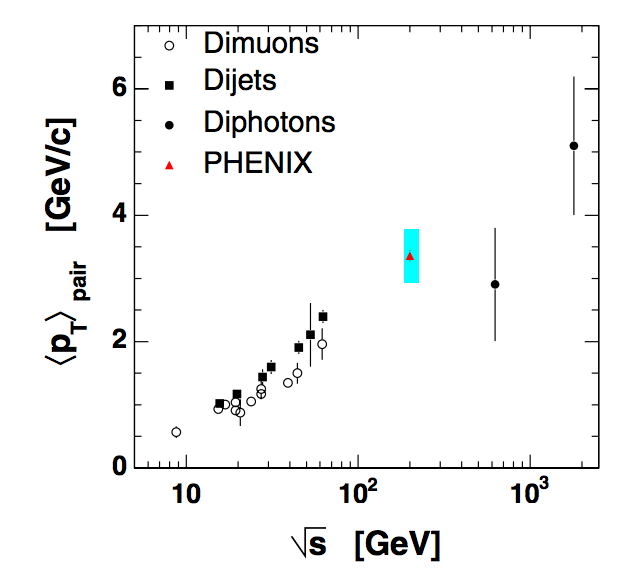}
	\caption{The \pt of a two-particle pair is shown as a function of \sqs for a variety of two-particle correlations processes~\cite{ppg029}.}
	\label{fig:ktmeasurements}
\end{figure}

Figure~\ref{fig:ktmeasurements} shows world measurements of the total \pt of two-particle correlations in various channels: dijets, diphotons, DY events, and dihadrons. Because of confinement, the uncertainty principle dictates that partons must contain a small transverse momentum on the order of several hundred MeV~\cite{Feynman:1978dt}. Historically it was expected that \kt effects would be small and limited to the nonperturbative behavior due to confinement. However measurements such as those shown in Fig.~\ref{fig:ktmeasurements} show that there are significant perturbative contributions to \kt at the \sqs probed by modern hadronic colliders. This figure is an indication of expectations from CSS evolution - namely that as the energy of the interaction increases, and thus the hard scale, the small transverse momentum scale from \kt should also increase.

\begin{figure}[tbh]
	\centering
	\includegraphics[width=0.6\textwidth]{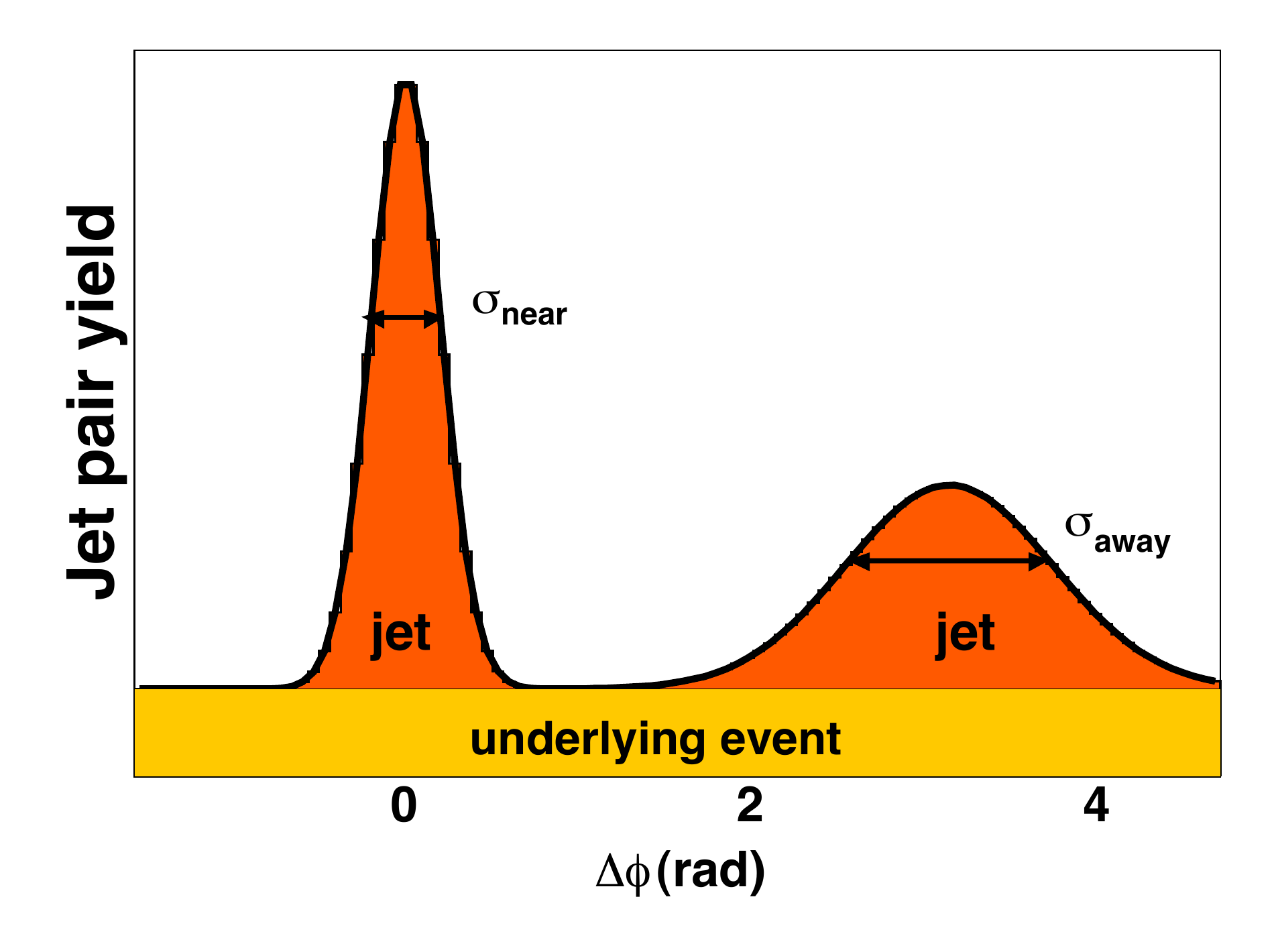}
	\caption{A cartoon illustrating the basic features of a dijet or dihadron two-particle correlation measurement.}
	\label{fig:dijet_cartoon}
\end{figure}

Rather than using jets, small transverse momentum scales can also be accessed with nearly back-to-back two-particle correlations, where the two particles are proxies for the jets. The particles are still sensitive to the small transverse momentum scale of the initial-state, but have an additional contribution from final-state fragmentation transverse momentum $j_T$. A cartoon diagram showing the basic features of nearly back-to-back correlations is shown in Fig.~\ref{fig:dijet_cartoon}. The two hadrons are separated by appromixately $\pi$ radians, and there is some smearing to the near-side jet due to final-state transverse momentum $j_T$. The away-side jet is smeared out even more due to the sensitivity to both \jt and \kt. The underlying event, which refers to particles that are produced uncorrelated with the hard scattering, is shown in yellow and is usually assumed to be flat as a function of \dphi in \pp collisions. 

\begin{figure}[tbh]
	\centering
	\includegraphics[width=0.8\textwidth]{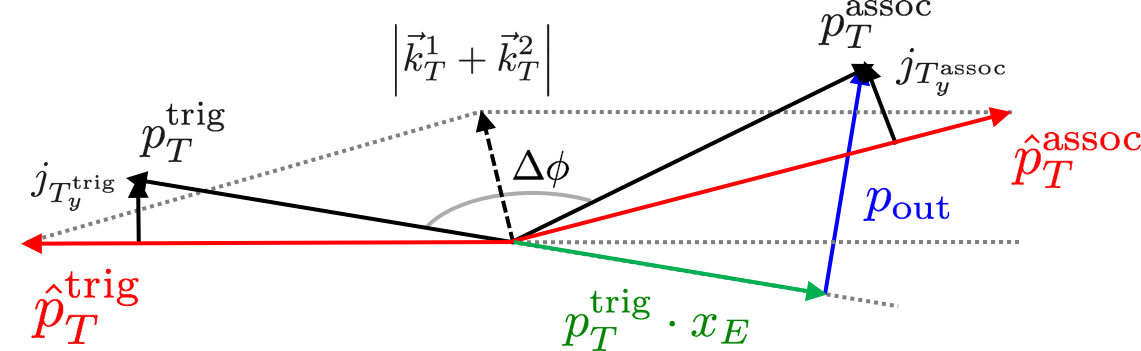}

	\caption{A diagram which shows the hard-scattering kinematics of a dihadron event in the transverse plane, taken from Ref.~\cite{ppg217}. The various vectors are described in more detail in the text.}
	\label{fig:ktkinematics}
\end{figure}

Figure~\ref{fig:ktkinematics} shows a diagram of a dihadron correlation in the transverse plane. In the figure, two hard scattered partons with transverse momentum $\hat{p}_T^{\rm trig}$ and $\hat{p}_T^{\rm assoc}$ are initially acoplanar due to the initial-state $k_T$ of the colliding partons. Note that any quantity with a hat, for example $\hat{p}_T^{\rm assoc}$, refers to a partonic quantity. The partonic vectors are shown in red in the diagram, and are acoplanar due to the vector sum $\vec{k}_T^1+\vec{k}_T^2$ of the partonic $k_T$. When the partons fragment, two hadrons can be measured, shown as the black vectors and labeled \pttrig and \ptassoc. These now include a transverse momentum component $j_{T_y^{\rm trig}}$ and $j_{T_y^{\rm assoc}}$ perpendicular to the jet axes in the transverse plane; these are assumed to be Gaussian such that \rmsjt = $\sqrt{2\langle j_{T_y^{\rm trig}}^2\rangle} = \sqrt{2\langle j_{T_y^{\rm assoc}}^2\rangle}$.

\begin{figure}[tbh]
	\centering
	\includegraphics[width=0.5\textwidth]{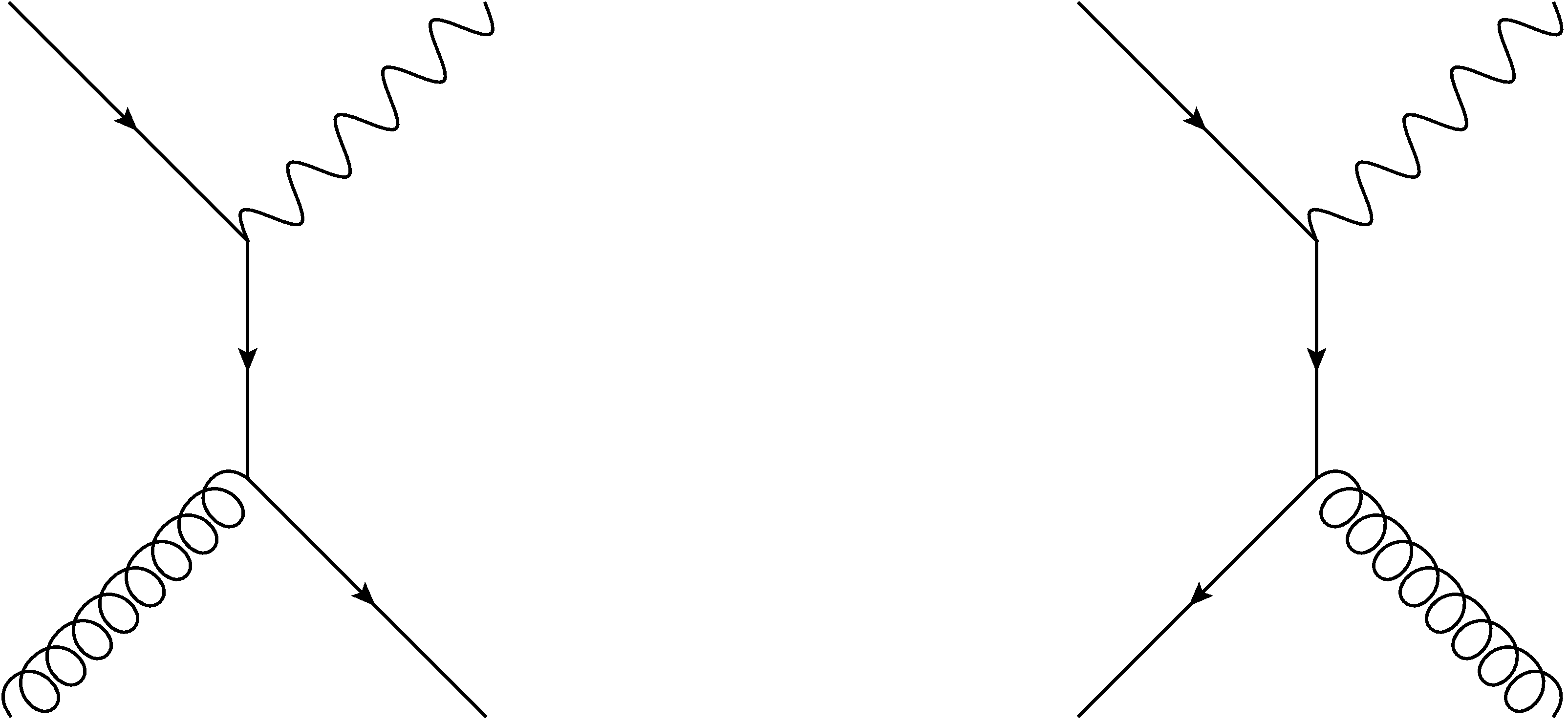}
	\caption{The leading order diagrams for direct photon production are shown. The two processes are QCD Compton scattering ($qg\rightarrow\gamma q$) and quark-antiquark annihilation ($q\bar{q}\rightarrow\gamma g$). Each process has an additional crossed diagram not shown here.}
	\label{fig:dp_lo_diagrams}
\end{figure}

Dihadron events are a useful probe for factorization breaking effects because of their relatively large cross section; dijet events are produced via LO partonic scatterings in \pp collisions. However, they are the most complicated two-particle correlation measurements to interpret because they are dependent on two TMD PDFs in the initial-state as well as two TMD FFs in the final-state assuming a factorizable picture, and thus offer more avenues for gluon exchanges. An alternative correlation measurement is the direct photon-hadron process, whose LO diagrams are shown in Fig.~\ref{fig:dp_lo_diagrams}. Often referred to as the ``golden channel,'' direct photon processes are one of the most ideal processes to study in hadronic collisions because the photon is emitted directly from the hard scattering; therefore it contains information about the partonic process. This provides direct access to the parton dynamics and is thus useful for studying nucleon structure.

Additionally direct photons have the added benefit that they do not interact via the strong force. This means that they do not suffer from final-state interaction effects via the strong force and thus, assuming a factorizable picture for photon-hadron correlations, only one final-state fragmentation function is necessary in a cross calculation rather than two; for this reason they are a widely sought after observable in nucleus-nucleus collisions because the photon is not modified by the strongly interacting QGP. For studying processes predicted to break factorization, they offer a potentially interesting comparison to dihadron correlations because there is one less avenue through which gluons may be exchanged since the photon is colorless. This may lead to modified behavior when comparing the two processes.

\begin{figure}[tbh]
	\centering
	\includegraphics[width=0.8\textwidth]{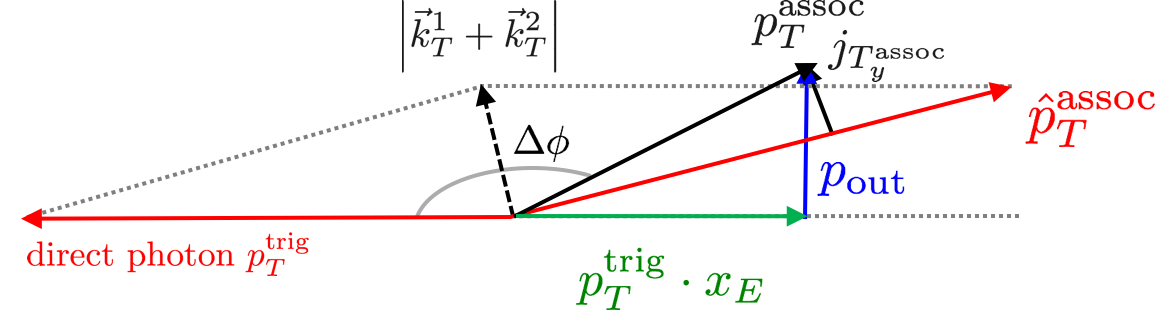}
	\caption{A diagram which shows the hard-scattering kinematics of a direct photon-hadron event in the transverse plane, taken from Ref.~\cite{ppg217}. The various vectors are described in more detail in the text.}
	\label{fig:dpktkinematics}
\end{figure}

Kinematically, direct photon-hadron events can be described in a similar way to dihadron events as shown in Fig.~\ref{fig:dpktkinematics}. Since the direct photon comes from the hard scattering, the red hard scattering vector \pttrig is the direct photon in Fig.~\ref{fig:dpktkinematics} whereas in dihadron events a jet fragment would be measured on the near-side. The away-side is constructed similarly, since the recoiled jet will still produce hadrons with some final-state transverse momentum component due to fragmentation.

\begin{figure}[tbh]
	\centering
	\includegraphics[width=0.6\textwidth]{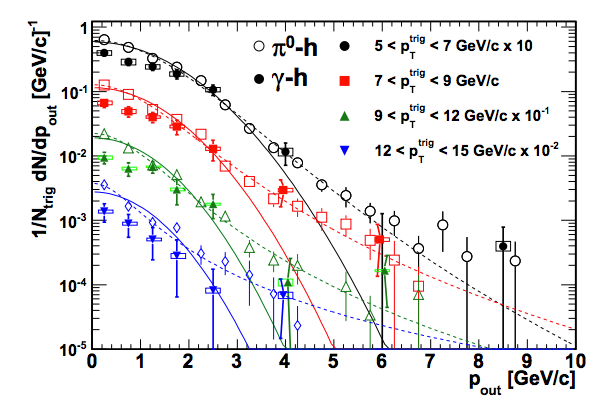}
	\caption{The \pout distributions measured in \pp collisions at \sqs=~200 GeV are shown from Ref.~\cite{ppg095}. The transition from nonperturbative to perturbative behavior is indicated by the failure of the Gaussian fit in describing the data at large \pout.}
	\label{fig:ppg095_pouts}
\end{figure}

To have sensitivity to small \kt and \jt transverse momentum scales, the two-particle correlations must be nearly back-to-back such that $\dphi\sim\pi$. As shown in Figs.~\ref{fig:ktkinematics} and~\ref{fig:dpktkinematics}, a vector \pout can be defined as a momentum space complement to the angular separation \dphi. The quantity \pout is the out of plane momentum component with respect to the near-side trigger particle, and thus is a transverse momentum dependent observable which could be used to study CSS evolution in processes predicted to break factorization. Mathematically it is defined as

\begin{equation}
	\pout = \ptassoc\sin\dphi
\end{equation}

\noindent and thus \pout is 0 when the two-particle pair is exactly back-to-back. Small deviations from 0 indicate sensitivity to nonperturbative \kt and \jt, while large deviations indicate sensitivity to perturbatively generated \kt and \jt. This quantity has already been shown to discriminate between perturbative and nonperturbative contributions, shown by the Gaussian and Kaplan fits in Fig.~\ref{fig:ppg095_pouts}~\cite{ppg095}. The Gaussian fits clearly fail at describing the data at large values of \pout, indicating that the large \pout values are generated by perturbative gluon radiation. The Kaplan fits, which describe both Gaussian behavior at small values of \pout and power law behavior at large \pout, describe the full functional form of \pout significantly better than the Gaussian fit alone. To quantify the nonperturbative behavior as a function of the hard scattering scale, the \pout distributions can be fit to Gaussian functions and the evolution of these widths with the hard scale can be studied. Since the Gaussian widths are sensitive to only nonperturbative \kt and \jt, if they do not follow the CSS evolution expectation then the processes could be exhibiting factorization breaking effects. 

The hard scattering quantity \xe is also shown in Figs.~\ref{fig:ktkinematics} and~\ref{fig:dpktkinematics} as a green vector antiparallel to the trigger particle. Mathematically \xe is defined as
\begin{equation}
	\xe = -\frac{\pttrig\cdot\ptassoc}{|\pttrig|^2} = -\frac{|\ptassoc|}{|\pttrig|}\cos\dphi
\end{equation}
and can be used as a modified proxy for the fragmentation variable $z$ where jet reconstruction is not possible. At LO for direct photon-hadron production, \xe is exactly $z$; however, NLO effects from \kt or \jt can smear this interpretation. Nonetheless, for the \pout distributions in the nearly back-to-back region, it is a good proxy for $z$ since the trigger and associated particles are nearly coplanar. This means the only deviation that \xe has from $z$ comes from approximating the away-side jet \pt as \pttrig. With the definitions of \pout and \xe, two-particle correlations can be used to study TMD observables multidifferentially as a function of both \pout and \xe, which can be taken as proxies for \kt and $z$ respectively. 

RHIC is an ideal facility to study effects from factorization breaking because the predicted effects can only be found in hadronic collisions where at least one final-state hadron is measured. The observable that is of interest must have a large \pt such that a hard scale can be defined, but it also must be sensitive to a small transverse momentum scale such that a TMD framework can be applied. At RHIC energies, the \pt reach for direct photons and neutral pions is sufficiently large to establish a hard scale while also being sensitive to a small transverse momentum scale. Additionally the detectors at RHIC have the necessary resolution in both azimuth and \pt at small \pt to be able to resolve small values of \pout. While jet processes at the LHC could be particularly interesting to study as well, the azimuthal and \pt resolution of the back-to-back jets must be small enough such that the small \pout region can be identified. Due to experimental capabilities at the RHIC experiments, \pion-hadron and direct photon-hadron are thus the ideal choices of processes to study potential factorization breaking effects.

\section{Motivation and Contributions of this Thesis}

The motivation for this thesis is to search for effects from the theoretically predicted color entangled states in hadronic collisions where at least one final-state hadron is measured. To perform this search, angular correlations of dihadrons and direct photon-hadrons are measured at the PHENIX experiment in a variety of collision systems. The correlations are collected at midrapidity where high \pt particles are most frequently produced. Observables sensitive to nonperturbative transverse momentum are collected and studied as a function of the leading \pt of the correlation to search for modifications from CSS evolution which comes directly out of the derivation for TMD factorization.

Specifically, the results of this thesis include measurements from \pp collisions at two center-of-mass energies, \sqs=~200 and 510 GeV, as well as \pal and \pau collisions at nucleon-nucleon center-of-mass energies of \sqsn=~200 GeV. These measurements represent the first dedicated studies to searching for these theoretically predicted color entangled states, and in particular are the first to identify an observable which more cleanly separates the nonperturbative from the perturbative contributions to the correlations. The analysis also allows for the nonperturbative observables to be probed with varying values of the longitudinal momentum fraction $x$ in \pp collisions since the hard scales are approximately the same at the two different center-of-mass energies. Measurements of correlations in \pa collisions allow for a nuclear dependence to be studied, which may lead to modified effects since there are more hadronic remnants which can lead to potential color entanglement effects.

To better interpret the results where color entanglement is predicted, phenomenological studies are also performed to compare to the results measured for this thesis. In particular, published data from the DY and SIDIS processes, where factorization is predicted to hold, are collected and analyzed to compile results that can be compared to those from the measured dihadron and direct photon-hadron correlations. The measurements dedicated to searching for evidence of factorization breaking, in addition to the compilation and analysis of data from processes where factorization is predicted to hold, will serve as a foundation for future TMD phenomenological efforts. In addition, several future observables and the prospects of measuring them are discussed in the context of the upcoming RHIC and LHC high luminosity running periods.

The rest of the thesis is organized as follows. Chapter 2 will introduce the RHIC experimental facility and the PHENIX apparatus with which the angular correlations are measured. Chapter 3 outlines the analysis methods used to measure both dihadron and direct photon-hadron correlations in the PHENIX spectrometer. Chapter 4 presents the \sqs=~510 GeV \pp results, while Chapter 5 presents both the \sqs=~200 GeV \pp results and the \sqsn=~200 GeV \pal and \pau results. Chapter 6 includes the phenomenological analysis of world data for several different processes predicted to factorize or break factorization, and Chapter 7 is a discussion of future measurements at the proposed sPHENIX experiment at RHIC as well as at the ATLAS and CMS experiments at the LHC. Finally, Chapter 8 includes a summary and conclusions concerning the results presented as well as the prospect for future constraints on color entanglement effects.

 \chapter{Experimental Setup}
 \label{chap:experiment}
 


\section{Relativistic Heavy Ion Collider}
The Relativistic Heavy Ion Collider (RHIC) at Brookhaven National Laboratory (BNL) is one of two operating hadronic colliders in the world, with the other collider being the LHC at CERN. RHIC is unique as a hadronic collider complex due to its versatility in collision species. Because of this versatility, RHIC allows for various asymmetric as well as symmetric systems to be collided such that the initial-state geometry of nuclear collisions can be studied. Additionally, it is the only polarized proton-proton collider in the world, making it an excellent facility to study spin-spin and spin-momentum correlations within the proton. \par

The RHIC ring is approximately 2 miles in circumference and was built to collide protons and heavy ions. The accelerator complex collides bunches of protons or ions in 106 ns intervals at 6 main interaction points on the RHIC ring. Four major detectors were built at the RHIC complex: PHENIX, STAR, BRAHMS, and PHOBOS. Of these four, only STAR is actively taking data, although there are future run plans for STAR and a proposed successor experiment to PHENIX called sPHENIX. The design luminosity was on the order of 10$^{26}$ cm$^{-2}$s$^{-1}$ and 10$^{31}$ cm$^{-2}$s$^{-1}$ for heavy ion collisions and proton-proton collisions, respectively~\cite{RHIC_overview1,RHIC_overview2}. \par

\begin{figure}[thb]
	\centering
	\includegraphics[width=0.7\textwidth]{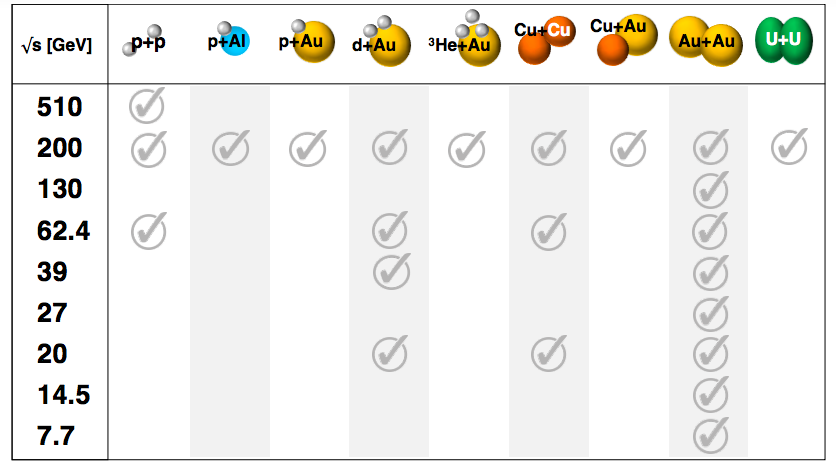}
	\caption{A pictorial representation of the various collision species that RHIC has provided for data collection. In addition to the wide range of light and heavy nuclei, both proton beams may be longitudinally or transversely polarized.}
	\label{fig:rhic_collision_species}
\end{figure}

RHIC is able to accelerate proton beams up to 255 GeV per proton and ion beams up to 100 GeV per nucleon. Protons and ions are accelerated from either a proton LINAC or a tandem Van-de-Graaff generator, respectively, into a booster ring. The booster ring subsequently injects the beams into the Alternating Gradient Synchotron (AGS) ring, which increases the energy of the beam. After the AGS the beams are injected into the RHIC ring, at which point they are accelerated to their maximum beam energies. The versatility of RHIC is demonstrated in \fig{fig:rhic_collision_species}, which shows all of the collision species and center-of-mass energies that RHIC has delivered to the PHENIX experiment. In addition to the various light and heavy nuclei that have been collided, the proton beams can be longitudinally or transversely polarized. \par

\section{The PHENIX Experiment}

The PHENIX experiment is one of two large collaborations at RHIC, with several hundred collaborators from countries across the globe. The spectrometer was designed to sacrifice acceptance for the ability to have excellent mass resolution and a high rate trigger system; therefore, the detector is designed to study rare and high \pt processes. The significant luminosity increases that RHIC was able to deliver in the last years of PHENIX operation allowed the experiment to collect what are referred to as ``golden data sets," which are PHENIX data sets in several different collision systems that have more integrated luminosity than all of the previous similar data sets combined. \par

\begin{figure}[htbp]
\begin{center}
	\includegraphics[width=0.7\textwidth]{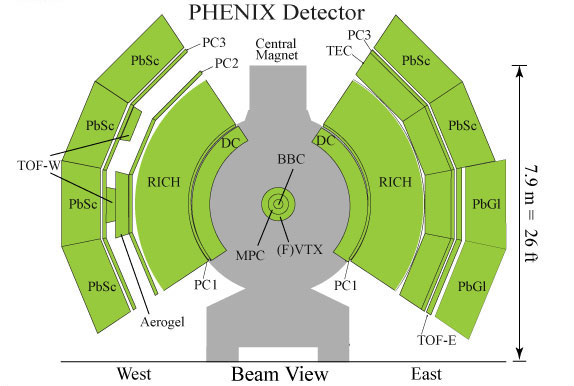}
\caption{A schematic diagram showing the PHENIX central arm spectrometers in the transverse plane. The picture is displayed with the beam pipe running out of the page.}
\label{fig:phenix_drawing}
\end{center}
\end{figure}

The PHENIX detector is comprised of two central arms which each span an azimuthal angle of \dphi$\sim\pi/2$ and are nearly back-to-back in azimuth. The central arms also cover a pseudorapidity region of $|\eta|<0.35$. Despite its limited acceptance, the subdetectors in the central arm are highly segmented, allowing for excellent spatial resolution. The central arms are used primarily for identifying charged hadrons and electromagnetic probes, but also have particle ID capabilities. Additionally there are two forward arms covering the full azimuth and pseudorapidity region $1.2<|\eta|<2.4$; these spectrometers specialize in the measurement of muons and decays from heavy flavor. In 2012 and 2013 two silicon vertex detectors were installed; the FVTX is located at forward rapidities $1<|\eta|<3$ and the VTX covers the barrel region surrounding the interaction vertex. These detectors were installed largely for the heavy flavor program at RHIC, which needs excellent z vertex resolution.  Reference~\cite{phenix_overview} gives an overview of the PHENIX detector. A schematic drawing of the PHENIX central arms after the 2012 running period is shown in \fig{fig:phenix_drawing}.

\subsection{Global Detectors}

To determine global event characteristics, PHENIX utilizes two forward detectors: the Beam Beam Counter (BBC) and the Zero Degree Calorimeter (ZDC). These detectors are used to determine the z-vertex position of the hard scattering collision as well as the centrality in collisions involving a nucleus; centrality is a proxy for the impact parameter between two nuclei in a relativistic nucleus collision. The centrality of the collisions ranges from 0-100\%, with 0\% denoting an impact parameter of 0 and 100\% denoting an impact parameter the size of the nucleus. \par

\begin{figure}[htbp]
\begin{center}
	\includegraphics[width=0.5\textwidth]{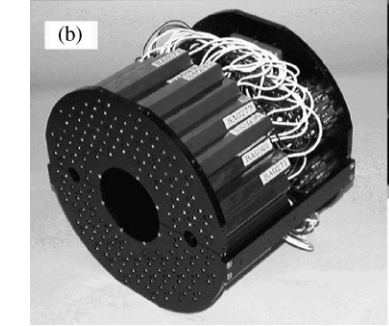}
\caption{A picture of one of the 64 element array BBCs. Each element is comprised of a 3 cm quartz radiator attached to a photomultiplier tube. The picture is taken from Ref.~\cite{BBC}.}
\label{fig:bbc_array}
\end{center}
\end{figure}

The BBCs~\cite{BBC} are situated approximately 144 cm from the nominal interaction point in PHENIX and cover a pseudorapidity of $3<|\eta|<3.9$ over the full azimuth. The north and south BBC consist of arrays of photomultiplier tubes (PMT) with a quartz $\check{\rm C}$erenkov radiator mounted to the head of the PMT. A picture of the 64 element array is shown in \fig{fig:bbc_array}. Since the BBCs are quite far forward with respect to the central arms, correlations with midrapidity particles are minimized and thus the BBCs permit detection of charged particles to determine the nominal collision vertex. \par

An important role of the BBC includes providing the timing measurement for the central arm time of flight detector as well as the for the PHENIX Level1 trigger system. The intrinsic resolution of each element in the BBC is roughly 50 ps; using the timing information from charged particles detected in both BBCs the vertex position can be determined with a resolution of roughly two cm in \pp collisions and 1 cm in A+A collisions. Additionally the centrality in collisions with a nucleus can be determined using the charge sum within the BBC in the nucleus going direction since there is a monotonic relationship between the BBC charge sum and the definition of centrality. \par

\begin{figure}[htbp]
\begin{center}
	\includegraphics[width=0.7\textwidth]{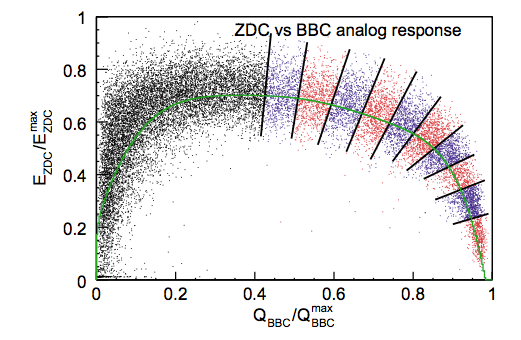}
\caption{The correlation between the energy deposited in the ZDC and the charge sum of the BBC is shown for Au+Au collisions~\cite{BBC}. The straight black lines indicate the centrality bins that are used in Au+Au collisions.}
\label{fig:cent_auau}
\end{center}
\end{figure}

The ZDC~\cite{ZDC} is another global detector which is a hadronic calorimeter located approximately 18 m from the nominal collision point in PHENIX. The ZDCs cover the full azimuth, lie at roughly $|\eta|\sim 6$, and are made of tungsten plates which are sandwiched with optical fibers read out by PMTs. The ZDCs are used primarily for the detection of very forward neutrons for the categorization of the collision centrality and/or the tagging of highly diffractive events. \fig{fig:cent_auau} shows the correlation between the total charge measured in the BBC and the energy deposited in the ZDC. The black lines denote the centrality categorizations in Au+Au collisions, where the far most right bin is the 0-5\% most central collisions. \par

In small collision systems like $p$+Au or d+Au, the collision centrality is determined specifically in the nucleus going direction since the multiplicities in the proton or deuteron going direction are not large enough to categorically determine the collision centrality. Here the centrality is determined with a Glauber-Monte Carlo calculation coupled with a simulation of the charge deposited in the BBC of the nucleus going direction. Geometric quantites associated with the different centrality selections can be determined with this method, and they show good agreement with measured data in d+Au collisions. A full description of the method used to determine the centrality in small systems can be found in~\cite{dAu_cent_determ}. It is important to note however that in small system measurements, e.g. $p$+A collisions, the interpretation of centrality as a proxy for impact parameter no longer holds. In these asymmetric measurements the centrality is an indication of final-state particle multiplicity rather than impact parameter, as an interaction with b = 0 fm and b = 1 fm could produce very similar responses of integrated charge in the BBC despite being very different impact parameters. \par




\subsection{Central Arm Spectrometers}

A diagram of the central arm spectrometers is shown in \fig{fig:phenix_drawing}. The central arms consist of two nearly back-to-back arms that cover roughly $\pi/2$ in azimuth each and $|\eta|<0.35$. In two-particle angular correlation analyses, the primary subdetectors used are the electromagnetic calorimeter (EMCal), the Drift Chamber (DC), and Pad Chamber (PC) tracking systems. A Ring Imaging  $\check{\rm C}$erenkov (RICH) detector is used for particle identification as well as the high energy cluster EMCal RICH Trigger (ERT) which is used to trigger on rare processes with a high \pt photon. \par

\subsubsection{EMCal}

The EMCal~\cite{EMCal} is located on the outermost edge of the central arms and measures the position and energy of photons and electrons. It is composed of eight sectors, six of which are lead scintillating (PbSc) calorimeters and two of which are lead glass (PbGl) $\check{\rm C}$erenkov calorimeters. The two types of calorimeter tower respond differently and thus care is taken in the analysis to appropriately treat each sector. The intrinsic resolution of the calorimeter in energy, $\eta$, and $\phi$ allows for high energy single photons to be reconstructed up to 25 GeV and neutral pions to be reconstructed via their two photon decay up to $\sim$17 GeV/c in \pt. \par

\begin{figure}[htbp]
\begin{center}
\includegraphics[width=0.6\textwidth]{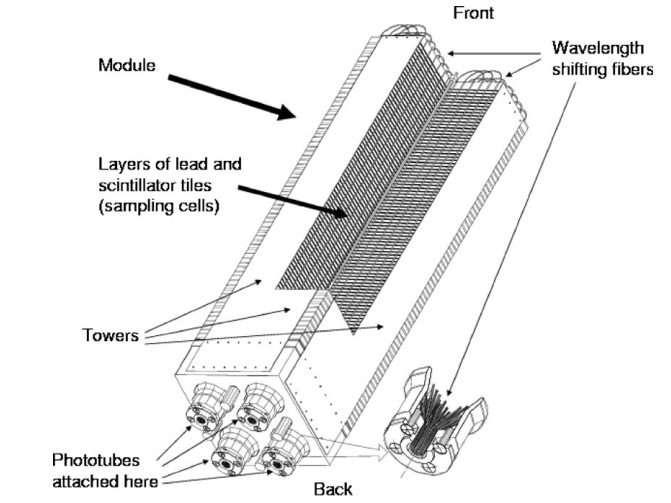}\\
\includegraphics[width=0.75\textwidth]{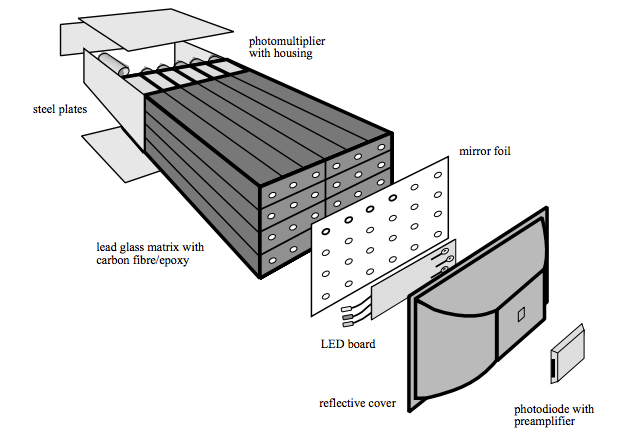}
\caption{Diagrams showing the composition of the lead scintillating (top) and lead glass (bottom) electromagnetic calorimeter towers.}
\label{fig:emcal_towers}
\end{center}
\end{figure}

The PbSc calorimeters are composed of alternating tiles of lead and plastic scintillator. In total the six sectors contain 15,552 individual towers. A diagram of one of the PbSc towers is shown in \fig{fig:emcal_towers}. The tower contains wavelength shifting fibers at the front which lead to phototubes in the back to read out the electromagnetic shower. The towers have a radiation length of 18$X_0$ and were shown to have an energy resolution in test beam data of 8.1\%/$\sqrt{E}~\oplus~$2.1\% in units of GeV. They additionally have an intrinsic timing resolution of 200 ps for electromagnetic showers. The inherent spatial resolution of the towers is $\Delta\eta\times\Delta\phi\sim$0.011$\times$0.011, where $\Delta\eta$ and \dphi refer to the pseudorapidity and azimuthal angular segmentation. \par

The PbGl calorimeter modules are comprised of a group of 24 towers bonded together as shown in \fig{fig:emcal_towers}. Photomultipliers at the front of the module measure photons and the lead glass matrix showers towards the back of the module where a photodiode measures the resulting energy of the shower. In total there are 9216 towers in the two sectors. In contrast to the PbSc towers, the PbGl towers have a radiation length of about 14$X_0$ and have an energy resolution from test beam data of 5.9\%/$\sqrt{E}~\oplus~$0.8\% in units of GeV. The spatial resolution of the PbGl towers is also finer than the PbSc towers at $\Delta\eta\times\Delta\phi\sim$0.008$\times$0.008. \par

\subsubsection{DC and PC}

\begin{figure}[htbp]
\begin{center}
\includegraphics[width=0.3\textwidth]{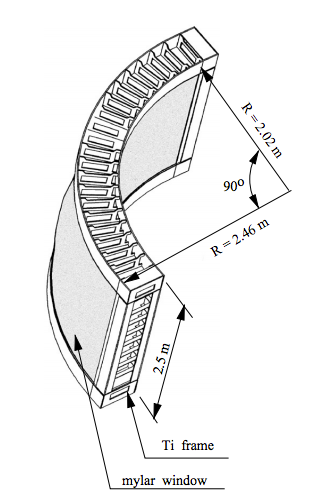}
\caption{A diagram showing one drift chamber frame. Each arm in the PHENIX spectrometer has two frames.}
\label{fig:dc_xsec}
\end{center}
\end{figure}

The DC~\cite{DC} is the primary subsystem for tracking charged particles in PHENIX. The two central arms each house a cylindrically shaped DC which lies between 2-2.4 meters from the beam pipe. A schematic diagram of one half of one DC arm is shown in \fig{fig:dc_xsec}. The DC measures charged particle trajectories in the $r$-$\phi$ plane and uses the information to determine the \pt of the track. These tracks can then be used to determine the invariant masses of parent particles if they are decay products; if not they can be matched to hits in the PC3 and associated with light in the RICH to determine if they are hadrons or electrons. The cylindrical frames are filled with a gas mixture of 50\% Argon and 50\% ethane which is ionized when a charged particle passes through it. The resulting charge is collected by sets of wire modules and used to reconstruct the track. \par

The DC consists of a frame which contains six types of wire modules stacked radially in each sector and labeled X1, U1, V1, X2, U2, and V2. The X1 and X2 wires run parallel to the beam pipe to perform tracking measurements in the $r$-$\phi$ plane. The other four sets of wires are angled by about 6$^\circ$ relative to the X1 and X2 wires to measure the $z$ coordinate of the track. The wires hold a voltage which allows the ionized electrons to drift towards them. Track position information is then reconstructed based on the drift time of the electrons. The PC improves both the track \pt resolution as well as the $z$ position of the track within the PHENIX detector volume. \par

\begin{figure}[htbp]
\begin{center}
\includegraphics[width=0.5\textwidth]{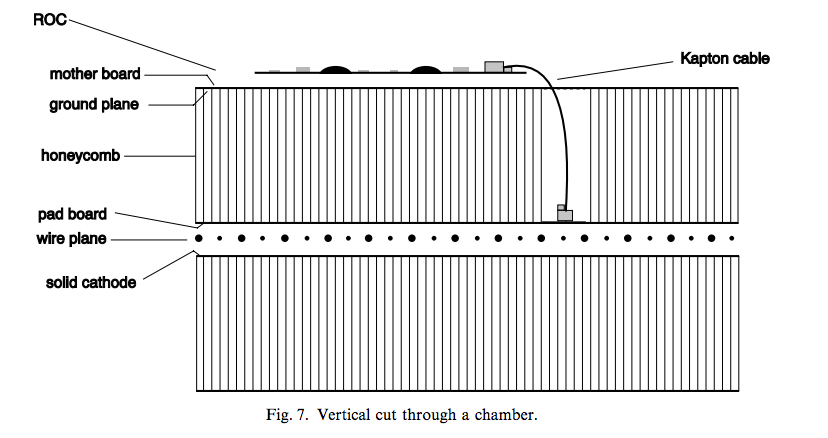}
\caption{A drawing showing the cross section of one of the pad chambers. Anode wires are enclosed by two cathode plates, and accumulated charge is read out by electronics at the top of the figure.}
\label{fig:pc_xsec}
\end{center}
\end{figure}

The PC tracking system is a set of three multiwire proportional chambers that are labelled as PC1, PC2, and PC3 in \fig{fig:phenix_drawing}. Each detector contains a plane of anode wires inside a gas volume bounded by two cathode planes. Charge is induced on one of the cathode planes when a particle passes through the gas volume, and this charge is then read out with readout electronics designed for the PCs. A diagram showing a cross section of the pad chamber design is shown in \fig{fig:pc_xsec}. The PC1 is located directly behind the DC, and with the z vertex resolution of the BBCs the DC and PC1 tracking system determines the momentum of charged tracks with resolution $\Delta p/p\sim 0.7$\%$~\oplus~1$\%$p$ in units of GeV/c. The PC1 position resolution was measured to be $\pm$1.7 mm in the $z$ direction, and thus with the DC the full momentum vector of charged tracks can be reconstructed. The two other pad chambers, PC2 and PC3, are located at further radial positions from the DC and PC1. Hits in these pad chambers are required in order to reject secondary tracks produced from decays, conversions, or interactions with the detector volume. \par

\subsubsection{RICH}

Since the goal of this analysis is to identify away-side charged hadrons, care must be taken to reject electrons that get reconstructed as charged hadrons. To do this, a RICH detector is used to identify, and thus reject, electrons. The RICH is located directly behind the DC and PC1 tracking systems, as can be seen in \fig{fig:phenix_drawing}. Spherical mirrors reflect $\check{\rm C}$herenkov light produced by electrons passing through radiator gas in the RICH onto PMTs. Electrons are identified and rejected based on the pattern of light collected in a ring around the projected track onto the PMT plane. Since high momentum pions also produce radiation and a signal in the RICH above $\sim$~5 GeV/c, the electron veto used in the analysis is not applied for charged tracks with \pt$>$~5 GeV/c. \par

\subsubsection{ERT Trigger}

The ERT trigger is the high-energy photon or electron trigger used to identify rare processes in PHENIX. For photon triggers, groups of EMCal towers called trigger tiles are iteratively scanned and the energy sum in the trigger tile is determined. If the energy sum is larger than a predetermined threshold, then the event is written out and kept for offline analysis. In conjunction with the RICH, the ERT can be used to trigger on electrons. These trigger tiles are nonoverlapping 2x2 tower regions. In this analysis, a 4x4 overlapping region was used as the trigger tile which is generically used for the high energy photon processes. There are three 4x4 triggers available for analysis (labeled ERT A, B, or C), with each trigger having a different energy threshold depending on the center-of-mass energy of the collision system. The trigger with the lowest energy threshold is the ERTC, while the trigger with the highest threshold is the ERTB trigger (while this is perhaps nonintuitive it is the nomenclature used within PHENIX for the trigger thresholds). \par

\subsection{Data Set Summaries}

In 2012, 2013, and 2015 the PHENIX experiment collected two of the best data sets that RHIC has provided in its 17 years of operation due to the significantly larger luminosity that the RHIC collider delivered to the PHENIX experiment. In 2012 and 2013, data from \pp collisions at \sqs=~510 GeV was collected. In the two years, approximately 30 and 150 pb$^{-1}$ of integrated luminosity was accumulated and used in the present analysis of dihadron and direct photon-hadron correlations. In 2015, PHENIX collected approximately 60 pb$^{-1}$ of \pp data at \sqs=~200 GeV. Additionally, for the first time ever, RHIC delivered \pau and \pal collisions at \sqsn=~200 GeV. PHENIX accumulated approximately 200 and 700 nb$^{-1}$ integrated luminosity for the \pau and \pal running, respectively; this corresponds to approximately 15.8 and 9.1 pb$^{-1}$, respectively, of equivalent \pp luminosity. These data sets are the focus of the analysis presented in this thesis. \par

\chapter{Analysis Details}
\label{chap:analysis}

\section{Particle Identification}

\subsection{Photons}\label{particle_cuts}
\begin{figure}[htbp]
\begin{center}
\includegraphics[width=0.6\textwidth]{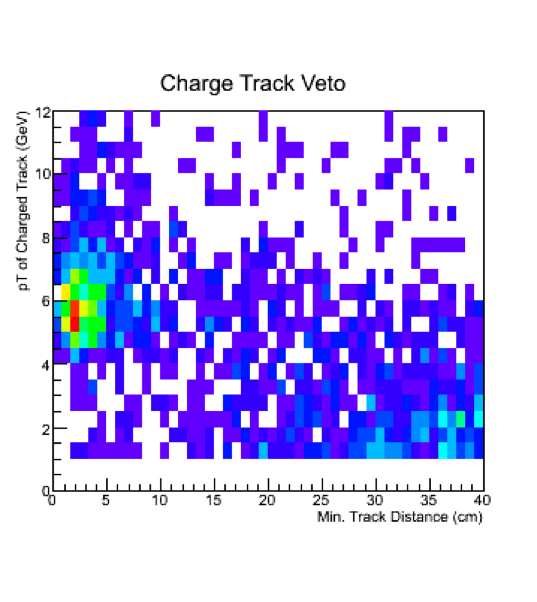}
\caption{The track \pt is shown as a function of the minimum distance to a shower in the EMCal. High \pt tracks which are traced to a shower in the EMCal are removed from the photon sample.}
\label{fig:chgtrack}
\end{center}
\end{figure}

Photon clusters are identified in the EMCal with the following criteria:
\begin{itemize}
	\item photon energy $1.0<e_{core}<20.0$ GeV
	\item Shower shape cut $\chi^2<3$ for PbSc clusters, or a dispersion cut for PbGl clusters
	\item Hot and dead tower map
	\item ERT supermodule not masked
	\item 2 tower edge cut around each sector
	\item $|z_{emc}|<155$ cm to reduce decay contributions from missed jets due to PHENIX acceptance
	\item ERT trigger tile matches EMCal tower check
	\item Track based elliptical charged hadron veto
\end{itemize}
The tower edge and fiducial cuts are implemented so that the single photon acceptance better matches the measured \pion acceptance. The edge cuts are additionally beneficial for the implementation of the isolation cone algorithm, which will be discussed later. The shower shape or dispersion cut was used to identify real electromagnetic showers in the EMCal rather than showers initiated by hadrons; it additionally suppressed background from merged clusters at high \pt from \pion decays. While this cut is effective for the majority of hadronic background, a non-negligible amount from hadronic showers remains. To further suppress this contribution a track based elliptical charged hadron cut is used with the addition of track information from the DC and PC3. Reconstructed track information is traced back to the EMCal tower, and if the reconstructed track points to the same tower where a shower was reconstructed in the EMCal, the cluster is rejected as a hadronic shower. Figure~\ref{fig:chgtrack} shows that there is a significant portion of high \pt tracks which can be traced to a showered cluster within several centimeters.  \par

\begin{figure}[tbh]
	\centering
	\includegraphics[width=0.7\textwidth]{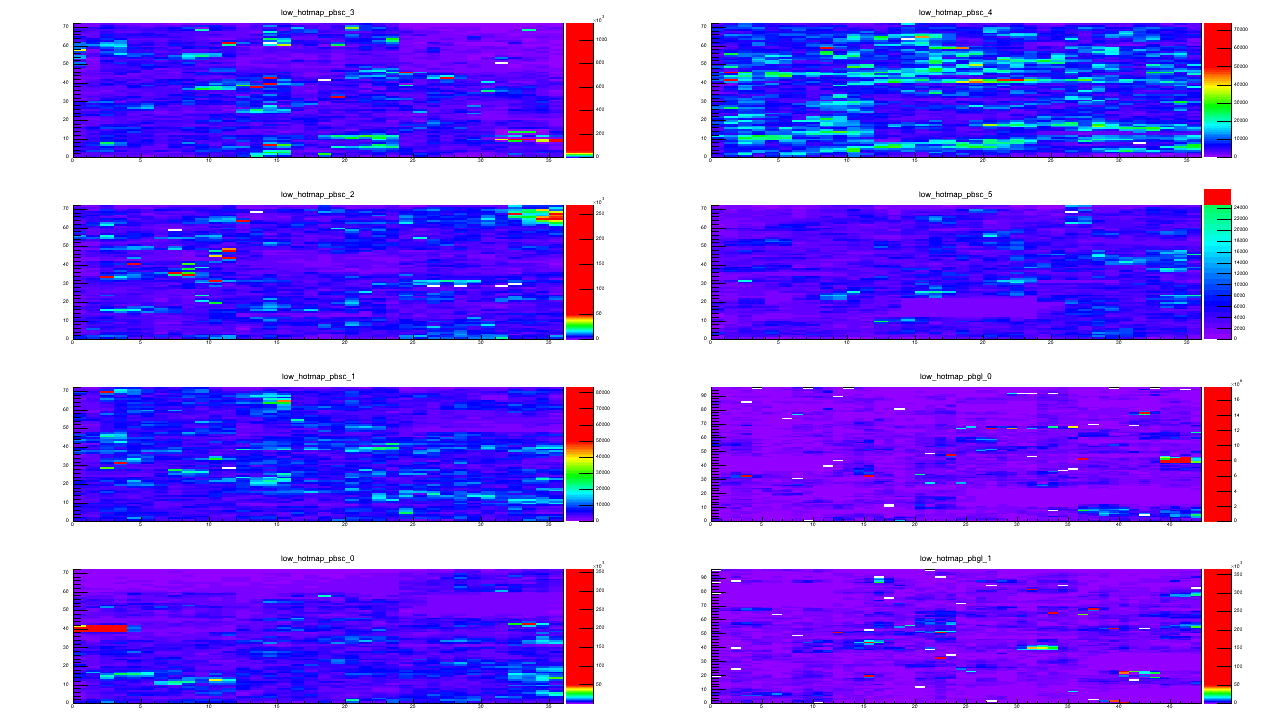} 
	\caption{Sector-by-sector comparisons for hits per-tower before hot towers are removed. The figure shows the low energy $1<e_{core}<5$ GeV tower hits. The x and y axes are \textit{iz} vs. \textit{iy} tower labels used to identify individual towers within the PHENIX EMCal. The contour levels are the same to show the towers with a large difference in hits from the majority.}
	\label{fig:EMCalnomaps1}
\end{figure}

\begin{figure}[tbh]
	\centering
	\includegraphics[width=0.7\textwidth]{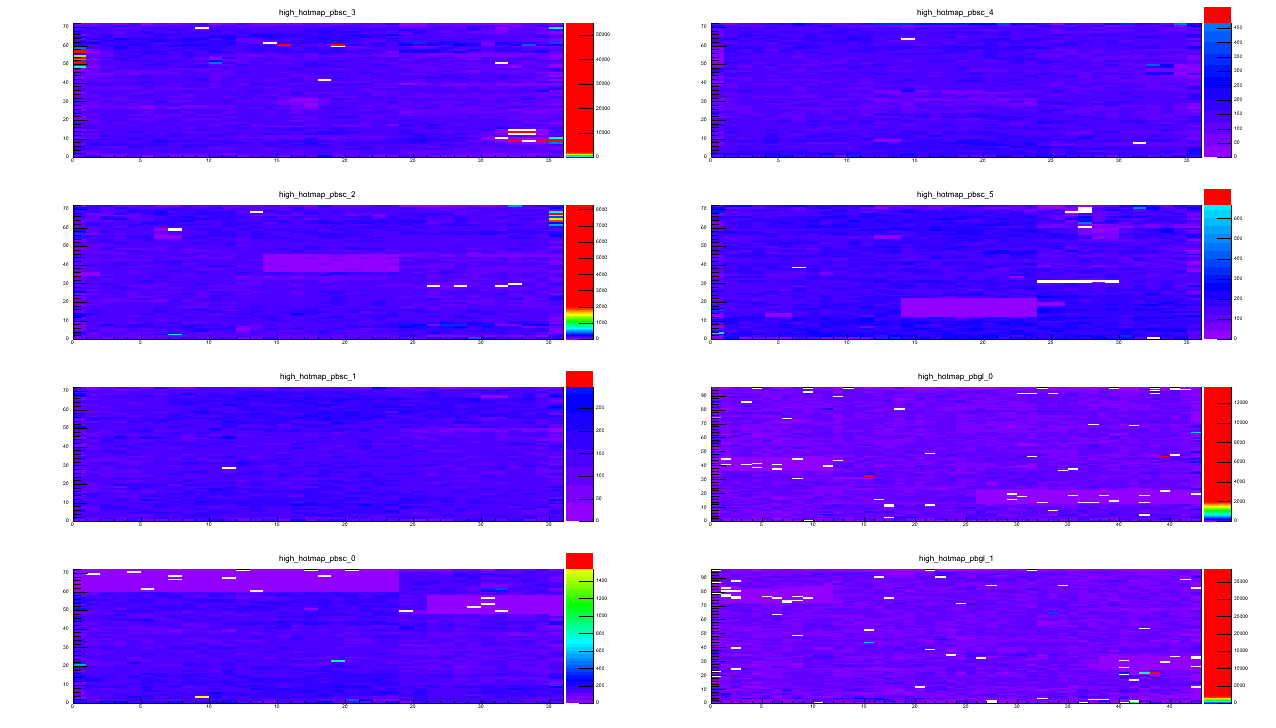}
	\caption{Sector-by-sector comparisons for hits per-tower before hot towers are removed. The figure shows the low energy $5<e_{core}<20$ GeV tower hits. The x and y axes are \textit{iz} vs. \textit{iy} tower labels used to identify individual towers within the PHENIX EMCal. The contour levels are the same to show the towers with a large difference in hits from the majority.}
	\label{fig:EMCalnomaps2}
\end{figure}

\begin{figure}[tbh]
	\centering
	\includegraphics[width=0.7\textwidth]{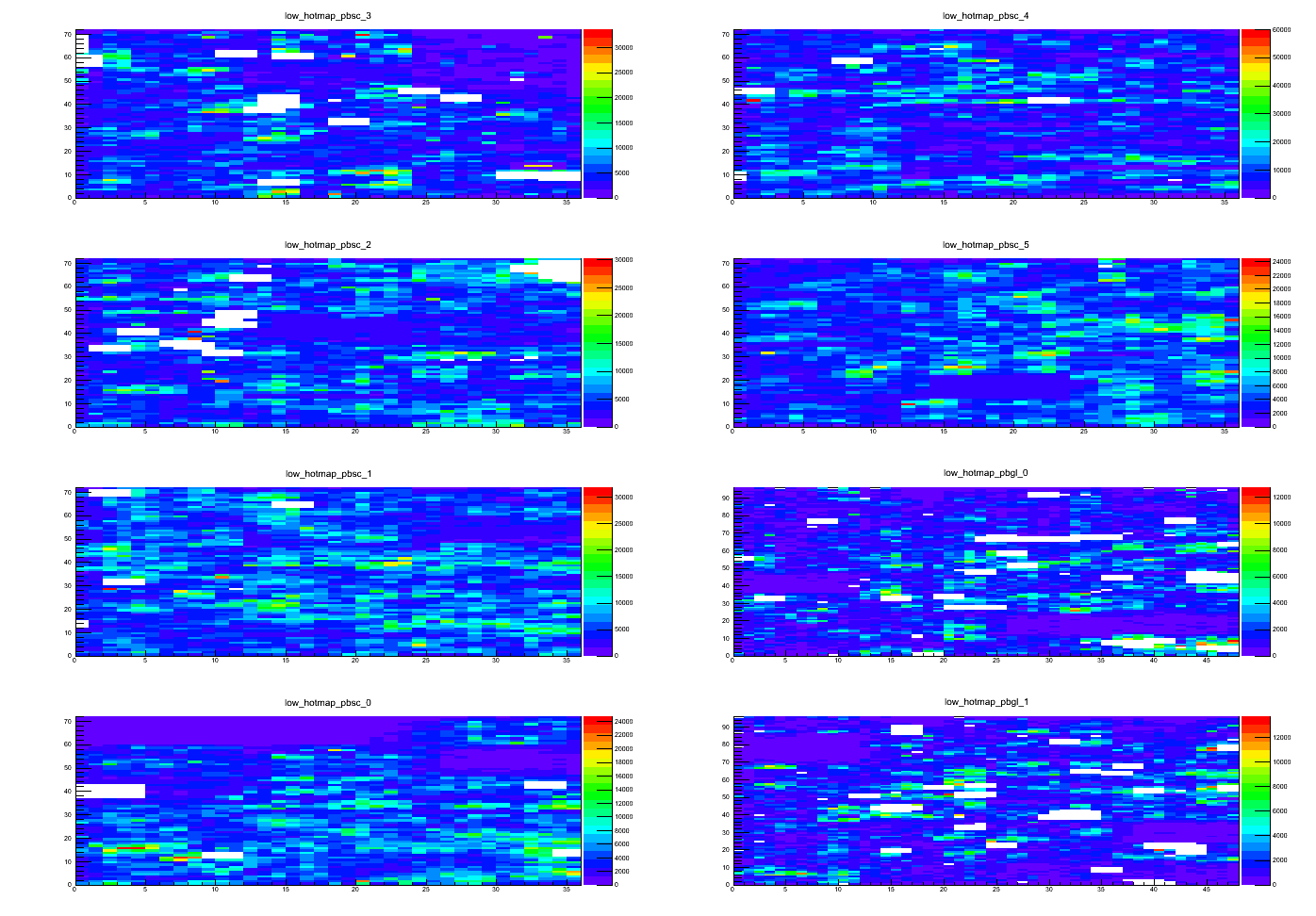} 

	\caption{Hot tower maps after the towers and surrounding towers identified as ``hot" were removed. The figure is for the low energy $1<e_{core}<5$ GeV region. The x and y axes are \textit{iz} vs. \textit{iy} tower labels.}
	\label{fig:EMCalwithmaps1}
\end{figure}

\begin{figure}[tbh]
	\centering
	\includegraphics[width=0.7\textwidth]{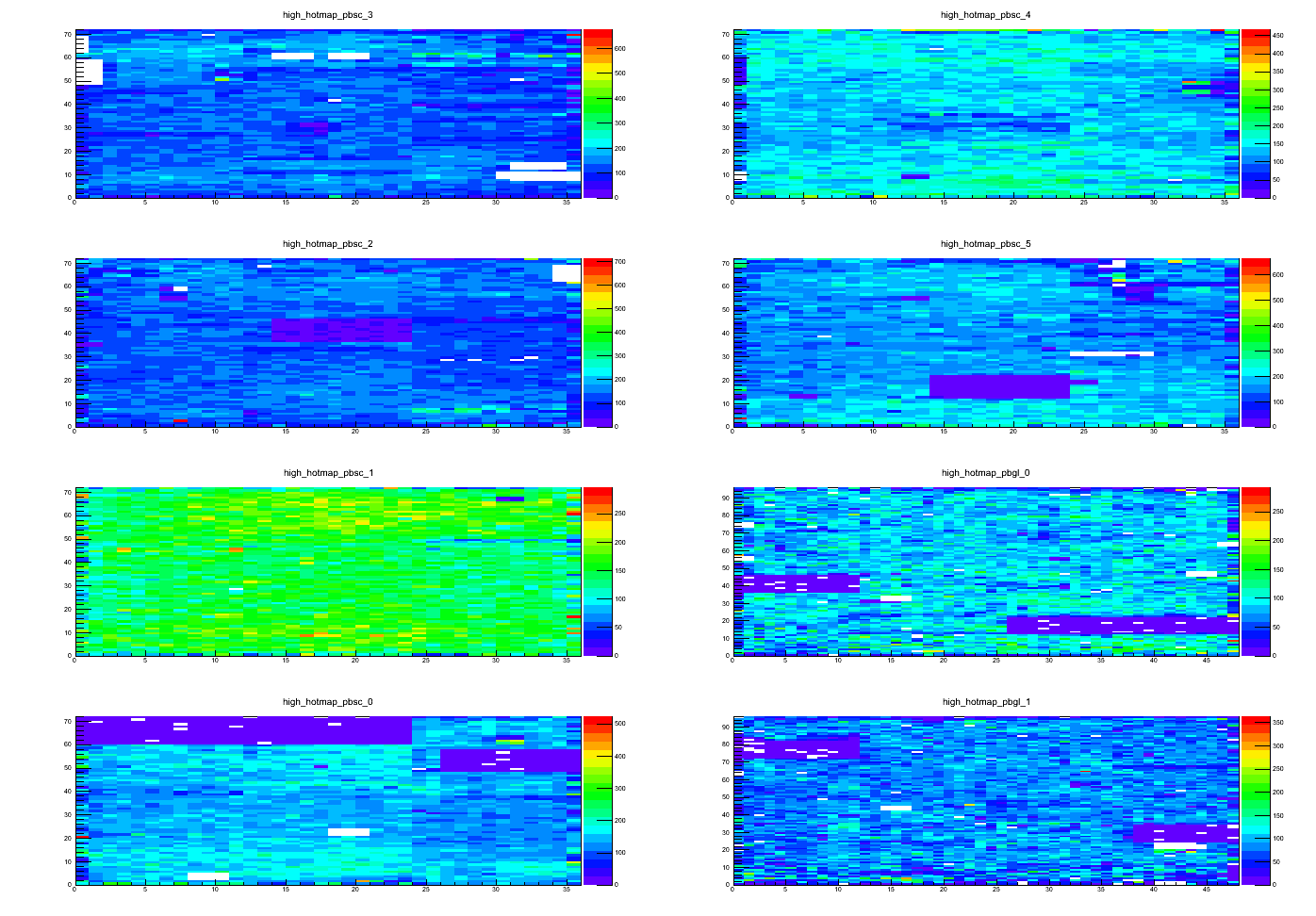}

	\caption{Hot tower maps after the towers and surrounding towers identified as ``hot" were removed. The figure is for the high energy $5<e_{core}<20$ GeV region. The x and y axes are \textit{iz} vs. \textit{iy} tower labels.}
	\label{fig:EMCalwithmaps2}
\end{figure}

To identify problematic towers within the EMCal, hot tower maps were constructed. Towers were classified as hot if the number of hits in a given tower was more than the sector's mean number of hits by 6 times the root mean square; thus hot tower maps are constructed on a sector-by-sector basis. Maps were constructed with the set of inclusive photons that satisfy the energy and shower shape cuts, which are the minimum number of cuts to identify photons. Since clusters can be found in adjacent towers, if a tower is identified as hot the 8 surrounding towers are additionally excluded from analysis. Maps were constructed in two photon energy bins, $1<E<5$ GeV and $5<E<20$ GeV. As an example, Figs.~\ref{fig:EMCalnomaps1} and~\ref{fig:EMCalnomaps2} show the sector-by-sector number of hits in run-15 \pau running. After the hot tower algorithm is applied, the towers that fail the criteria are excluded; these maps are shown in Figs.~~\ref{fig:EMCalwithmaps1} and~\ref{fig:EMCalwithmaps2}. The number of hits per tower are significantly more uniform after the hot towers are removed as would be expected. \par

\subsection{\pion and $\eta$}
\begin{figure}[tbh]
\begin{center}
\includegraphics[width=0.7\textwidth]{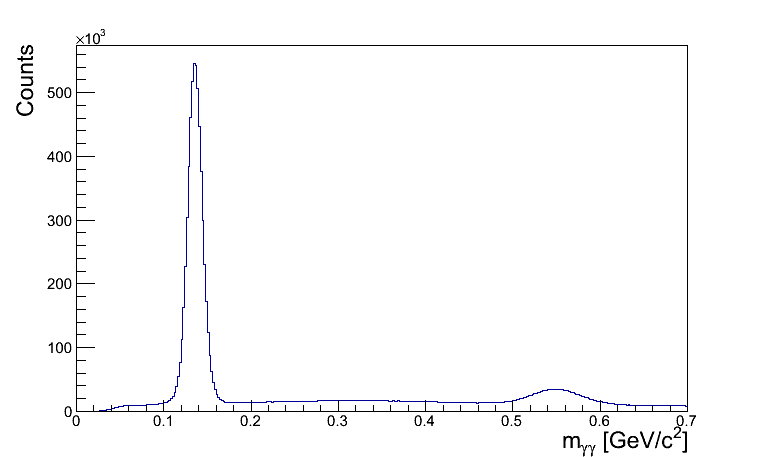}
\caption{The two-photon invariant mass spectrum is shown for run-15 \pp collisions. Peaks are apparent for both the \pion and $\eta$ meson.}
\label{fig:invmass}
\end{center}
\end{figure}

\pion and $\eta$ mesons are identified via their two-photon decay channels, using the aforementioned analysis cuts. The invariant mass is reconstructed and \pion mesons are identified in the range $120<m_{\gamma\gamma}<160$ MeV/$c^2$, while $\eta$ mesons are identified in the range $500<m_{\gamma\gamma}<600$ MeV/$c^2$. Subleading photons are required to have an energy larger than 1 GeV to suppress the combinatorial background. To avoid double counting the highest energy photon is always taken as the leading photon. Figure~\ref{fig:invmass} shows an example invariant mass spectrum, indicating the \pion and $\eta$ peaks near their respective nominal invariant masses of 135 and 547 MeV/$c^2$.

\subsection{Charged Hadrons}\label{charged_hadron_id}
DC tracks are selected as nonidentified charged hadrons, meaning not explicitly identified as $\pion^\pm$, $K^\pm$, or $p(\bar{p})$ hadrons, according to the following criteria:
\begin{itemize}
	\item Quality selection of 63 or 31
	\item PC3 track matching of 2$\sigma$
	\item RICH ring veto for \pt$>$5 \gev
\end{itemize}

The selection of tracks with a quality of 63 or 31 is the highest track selection quality available in tracking analyses at PHENIX. A quality of 63 indicates unique X1,X2, and UV wire hits found in the DC as well as unique PC1 hits associated to the track. A quality of 31 is only slightly different, where the PC1 hits are not required to be unique. Background from tracks not originating from the primary event vertex and from decays or conversions are reduced with the PC3 track matching cut. Tracks found in the DC and PC1 are projected to the PC3 and matched with hits in the PC3. If the matched hit falls outside 2$\sigma$ of the track projection in both the d$\phi$ and dz directions, the track is rejected. The RICH ring veto is used to exclude electrons that leave a $\check{\rm C}$erenkov ring in the detector.

\subsection{Run Quality Assurance}
To first determine what runs are available are good for analysis, run quality assurance (QA) was performed for both charged hadrons and photons. Since multiple subsystems are used, each must be checked on a run-by-run basis to assure that any runs where subsystems were performing sub-optimally are excluded. In run-13 and 15 this is particularly important due to the significantly higher luminosity that RHIC was able to deliver to the PHENIX experiment. \par

\begin{figure}[htbp]
\begin{center}
\includegraphics[width=0.7\textwidth]{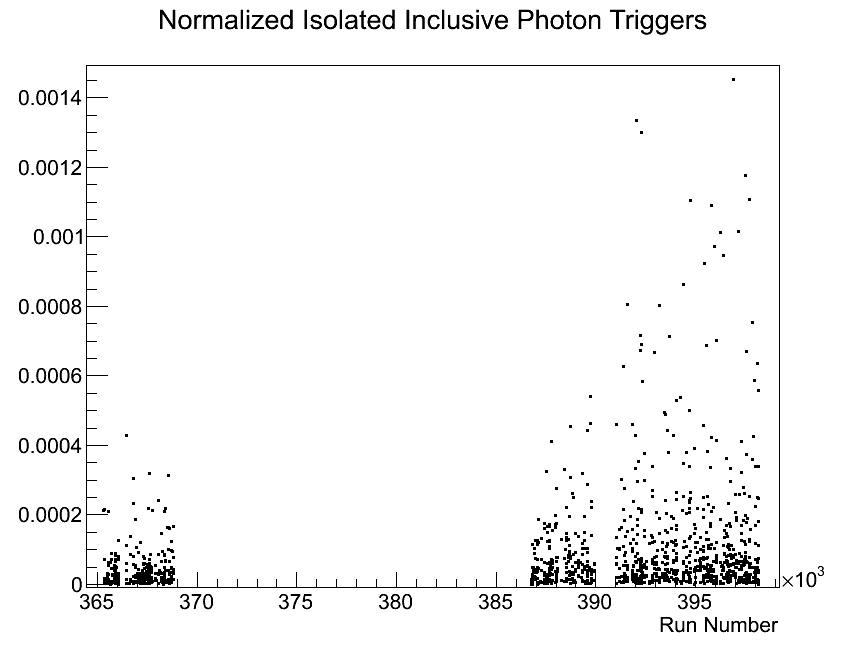}
\caption{Isolated inclusive triggers normalized by number of events in the run. Runs with yields above 0.0003 were excluded.}
\label{fig:run13_qa1}
\end{center}
\end{figure}

\begin{figure}[htbp]
\begin{center}
\includegraphics[width=0.6\textwidth]{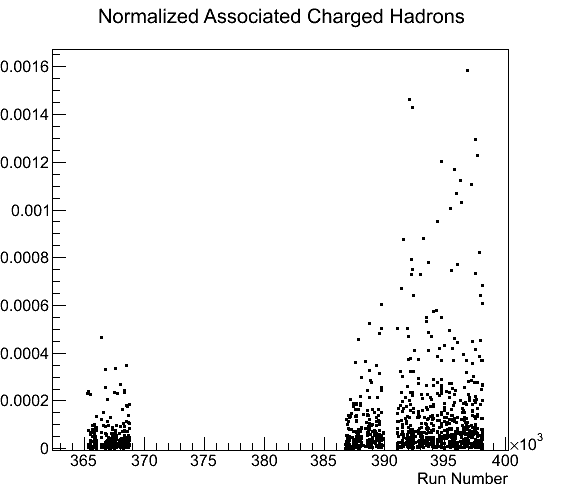}
	\caption{Associated charged hadrons normalized by number of events in the run.}
	\label{fig:run13_qa2}
\end{center}
\end{figure}

For the \sqs=~510 GeV \pp run, the number of isolated inclusive photon triggers and the number of associated charged hadrons normalized by the number of events in a run was determined on a run-by-run basis. Any runs that show abnormally high or low yields per event should be excluded from subsequent analysis. Figures~\ref{fig:run13_qa1} and~\ref{fig:run13_qa2} show the normalized isolated inclusive photon and associated charged hadrons as a function of the run number. The cluster of runs on the left side of the plots is from the short \sqs=~510 GeV data collection at the end of the run-12 period; the larger cluster on the right side of the plots shows the longer run-13 \sqs=~510 GeV run. Immediately it is clear that several runs should be excluded due to the high yields per run when compared to the average. The mean plus $5\sigma$ value of isolated trigger photons per event was found to be 3$\times$10$^{-4}$, so all runs above this were excluded in further analysis. Unsurprisingly, these runs were the same runs that showed abnormally large yields for the associated charged hadrons. \par

\begin{figure}[tbh]
	\centering
	\includegraphics[width=\textwidth]{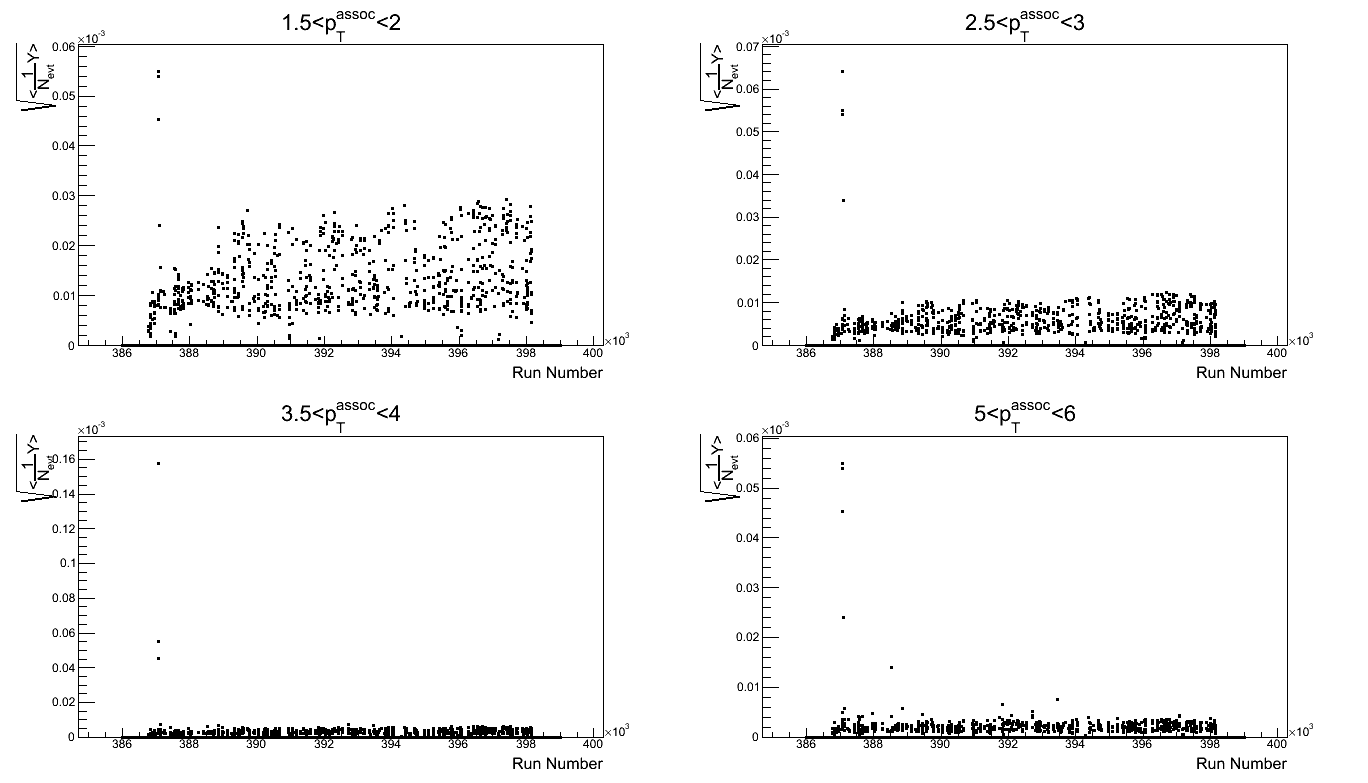}
	\caption{The RMS of the normalized minimum bias charged hadron yields as a function of the run number in run-13. An additional 5 runs were excluded from subsequent analysis.}
	\label{fig:run13_dcqa}
\end{figure}

Due to this correlation, run QA was also performed for the charged hadrons from the DC alone. This QA also accounts for any bad runs that should be excluded from the mixing correction, discussed further in Section~\ref{acceptance_correction}. Minimum bias charged hadrons were collected with the same cuts used in the triggered sample. The root mean square of the yields per run normalized by the number of events for several \pt bins is shown in Fig.~\ref{fig:run13_dcqa}. There were five runs that were not accounted for by the EMCal photon QA, so these runs were excluded in addition to the runs noted above. \par

\begin{figure}[tbh]
	\centering
	\includegraphics[width=0.8\textwidth]{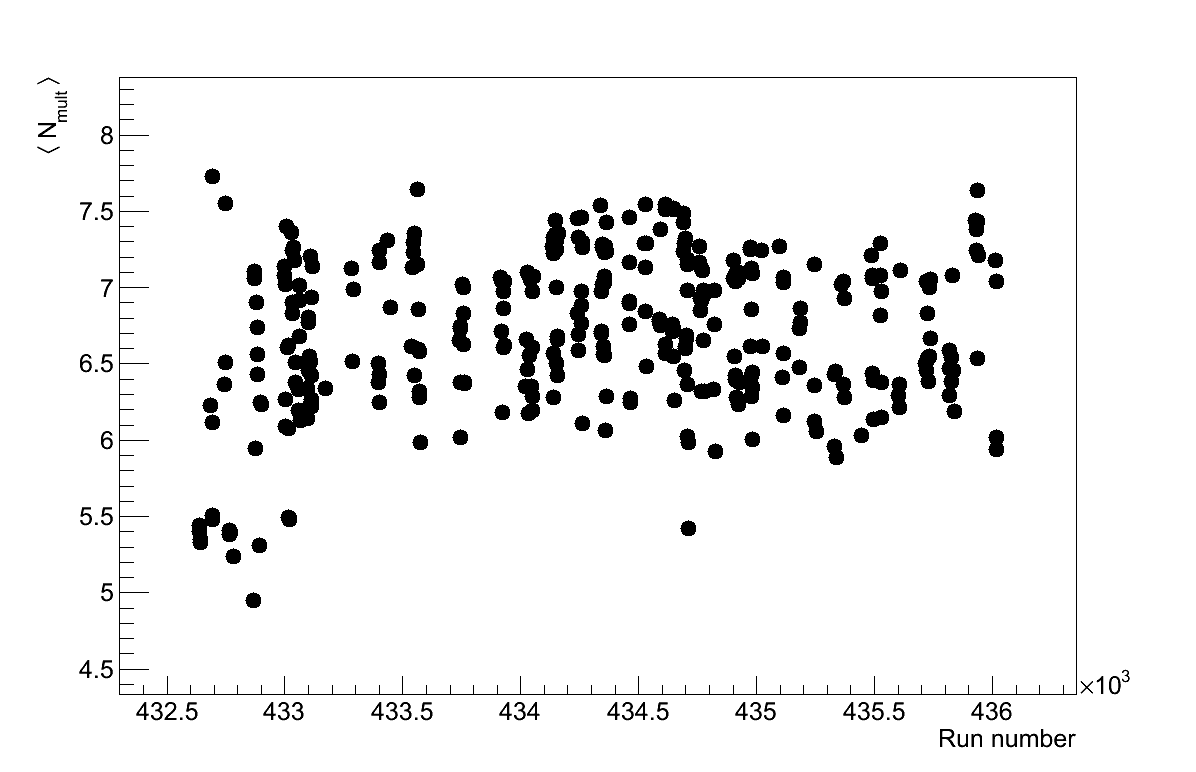}
	\caption{The number of minimum bias charged hadron tracks as a function of the run number is shown. No runs appear to be abnormal.}
	\label{fig:mbchghadqa_pau}
\end{figure}

Run QA was also performed for the 2015 \pp and \pa run period. The average multiplicity per run number was plotted to determine if any bad runs were to be removed, shown in Fig.~\ref{fig:mbchghadqa_pau}. The average multiplicities include tracks that pass all of the standard cuts used for the hadrons, described in more detail in Section~\ref{particle_cuts}. No runs appear to be abnormal from the minimum bias hadron sample, so no runs are removed from this QA. 

\begin{figure}[tbh]
	\centering
	\includegraphics[width=0.8\textwidth]{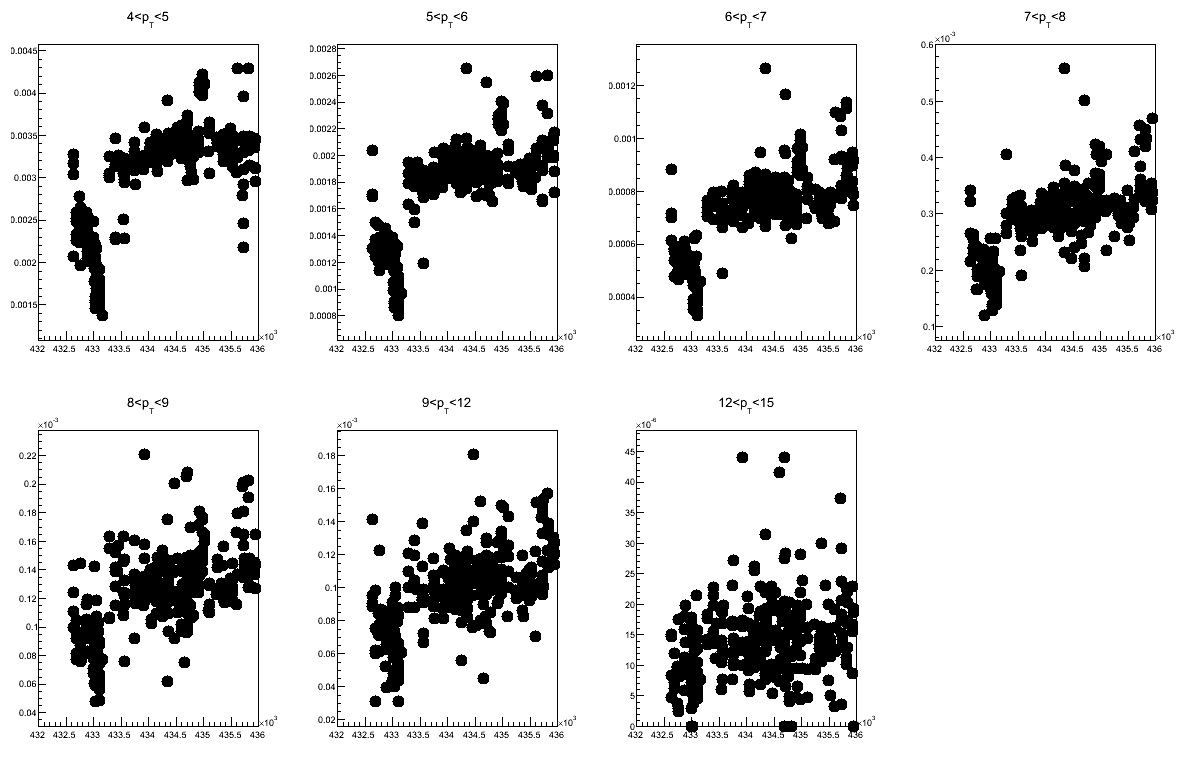}
	\caption{The number of neutral pions per event in several $p_T$ bins is shown as a function of run number for $p$+Au running.}
	\label{fig:ertpi0qa_pau}
\end{figure}

EMCal run QA was determined with the number of $\pion$s collected normalized for total number of events, shown in Fig.~\ref{fig:ertpi0qa_pau}. These $\pion$s are from ERT triggered events since for the final analysis we are interested in high $p_T$ $\pion$s and photons. No runs are obviously bad. Note that the consistently lower yields in the lower run numbers matches other \pion analyses in PHENIX and is due to the ERT trigger not performing at 100\% efficiency in this subset of runs.  

Several runs were excluded by virtue of no photons passing the ERT trigger bits. Investigating further, it was determined that six runs in run-15 had special notes that the ERT trigger was not working, due to the live time of the ERT trigger bits being 0. Therefore, these runs are excluded simply because there are no measured high \pt photons that fired the trigger bit. \par

\begin{figure}[tbh]
	\centering
	\includegraphics[width=0.7\textwidth]{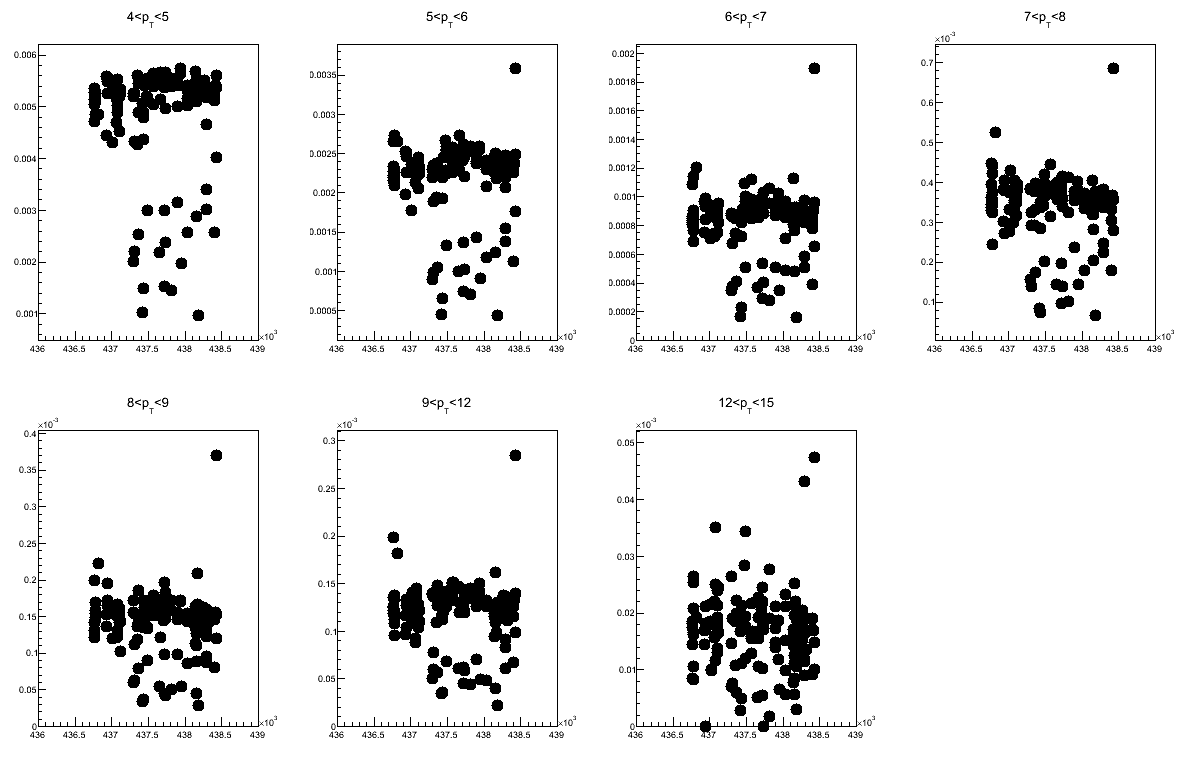}
	\caption{The number of neutral pions per event in several $p_T$ bins is shown as a function of run number for $p$+Al running.}
	\label{fig:palpi0qa}
\end{figure}

\begin{figure}[tbh]
	\centering
	\includegraphics[width=0.7\textwidth]{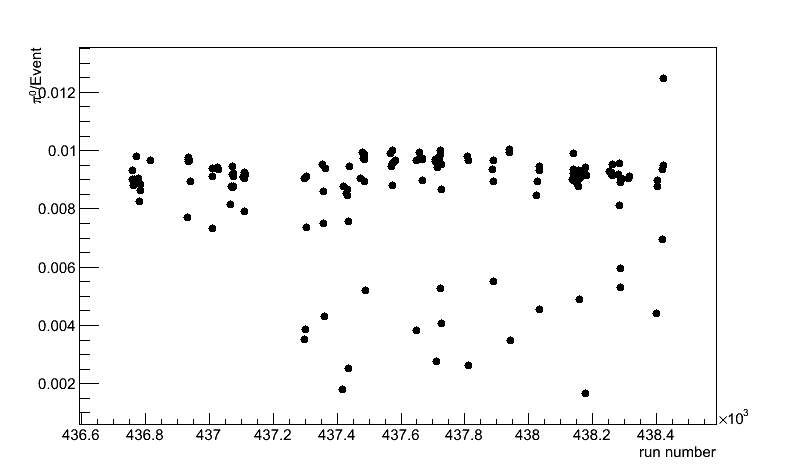}
	\caption{The number of neutral pions per event integrated over $p_T$ is shown as a function of run number for $p$+Al running.}
	\label{fig:palpi0qaint}
\end{figure}

The same tests were done for the p+Al running. Run QA for the p+Al is shown below. The ERT $\pion$ triggered QA can be found in Figs.~\ref{fig:palpi0qa} and \ref{fig:palpi0qaint}. There are no runs that stand out as particularly bad, despite several that show lower yields in general. The p+Au running shows this feature as well in some runs where the ERT was not functioning at maximum efficiency. \par

\begin{figure}[tbh]
	\centering
	\includegraphics[width=0.7\textwidth]{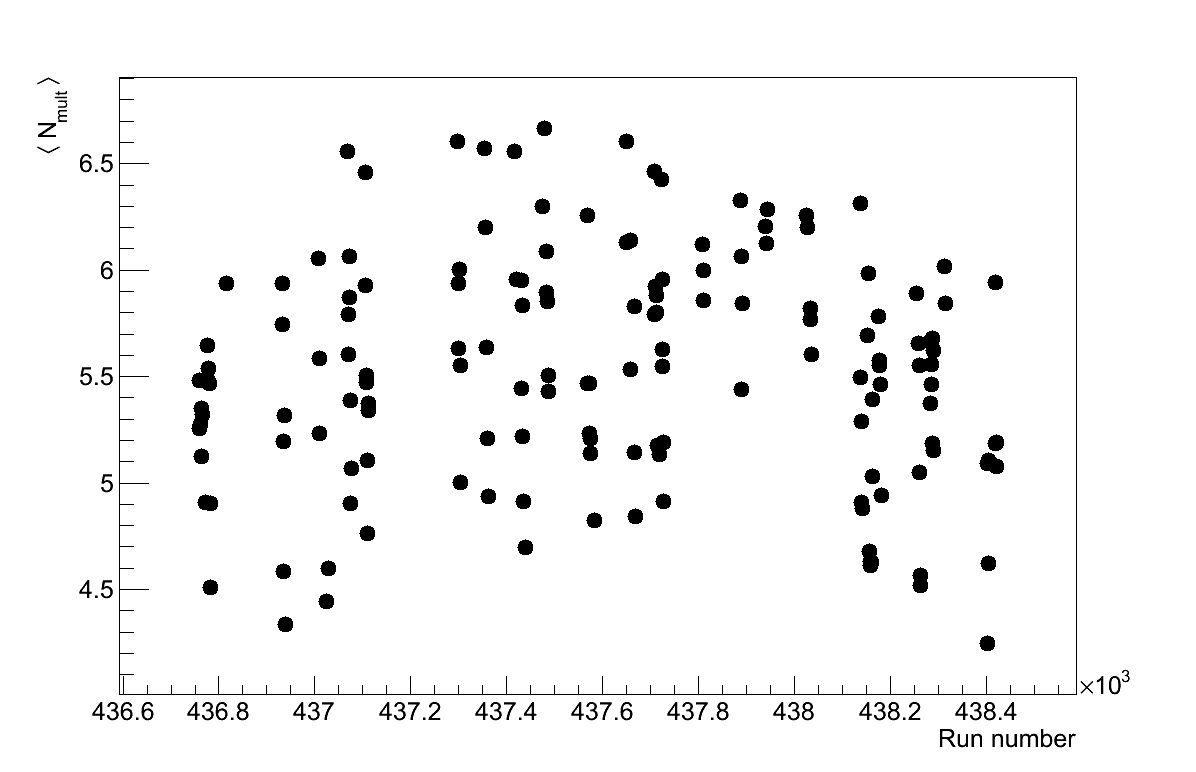}
	\caption{The average multiplicity for nonidentified charged hadrons in run-15 \pal running as a function of run number is shown.}
	\label{fig:palmbhadqa}
\end{figure}

Similar minimum bias hadron QA was done for the background determination in $p$+Al. The average multiplicity of minimum bias hadrons is shown in Fig.~\ref{fig:palmbhadqa}. There are no obviously bad runs in the minimum bias hadrons either, so all available runs for the $p$+Al run period are used. 

\section{Two-Particle Correlations}

\subsection{Correlation Functions}

Two-particle correlations are a powerful observable that can be used to probe a variety of physics mechanisms at collider facilities. Unlike an inclusive analysis, here we primarily concern ourselves with the production of associated charged hadrons \textit{given} a certain species of trigger particle. Rather than quoting a yield as one would in an inclusive analysis, we instead quote per-trigger yields (PTY), which are defined as the number of associated charged hadrons per trigger particle measured and written as

\begin{equation}
Y=\frac{N^{pair}}{N^{triggers}}\,.
\end{equation}

The correlation function can then be constructed as a function of a given observable between the trigger and associated pair; this is typically the variable \dphi which quantifies the angular separation between the two particles in azimuthal space. This quantity most obviously identifies the jet structure between the two particles. To construct the correlation function, we divide the number of associated charged hadrons by the number of trigger particles and two correction factors. The correction factors, a charged hadron acceptance and a background acceptance, quantify inefficiencies in the PHENIX detector. Mathematically the correlation function looks like

\begin{equation}\label{eq:corrfxn}
	Y = \frac{1}{\ntrig}\frac{dN}{d\dphi} = \frac{1}{\ntrig}\frac{dN/d\dphi_{\rm raw}}{dN/d\dphi_{\rm mix}\epsilon_{\rm had}(p_T)}\,.
\end{equation}
Here \ntrig is the number of trigger particles measured, regardless of if there was or was not a correlated hadron measured as well, $dN/d\dphi_{\rm raw}$ is the raw correlated charged hadron yield as a function of \dphi, $dN/d\dphi_{\rm mix}$ is the mixed background acceptance correction, and $\epsilon_{\rm had}(p_T)$ is the single particle charged hadron efficiency. The mixed event background correction accounts for the shape of the distributions, while the other factors in the denominator of Eq.~\ref{eq:corrfxn} are normalization quantities. This defines the correlation function and it can be equivalently defined for any observable that one could construct in a two-particle correlation; for example the observable \pout can be straightforwardly inserted into Eq.~\ref{eq:corrfxn} for \dphi. This method which utilizes this definition of the correlation function has been used in several previous PHENIX analyses~\cite{ppg029,ppg089,ppg090,ppg095}. It has also been extended to include $\Delta\eta$ correlations at the LHC where the pseudorapidity coverage of ATLAS and CMS is significantly larger than PHENIX~\cite{CMS_pp_collectivity,ATLAS_pp_collectivity}. Here we do not account for the $\eta$ range as in the PHENIX detector it is quite small, so any yields correspond to $|\eta|<0.35$.  \par




\subsection{Charged Hadron Efficiency}

Previous PHENIX publications have determined the charged hadron efficiency by bootstrapping the measured nonidentified charged hadron cross sections to per event yields and comparing to the measured yields in the analysis. Unfortunately there are no previous cross section measurements for nonidentified charged hadrons in \pp at \sqs=~510 GeV and \pa at \sqs=~200 GeV. To determine the charged hadron efficiency used in the correlation functions, single particle simulations were used with a full GEANT3 description of the PHENIX detector, referred to as PISA. \par

\begin{figure}[htbp]
\begin{center}
\includegraphics[width=0.45\textwidth]{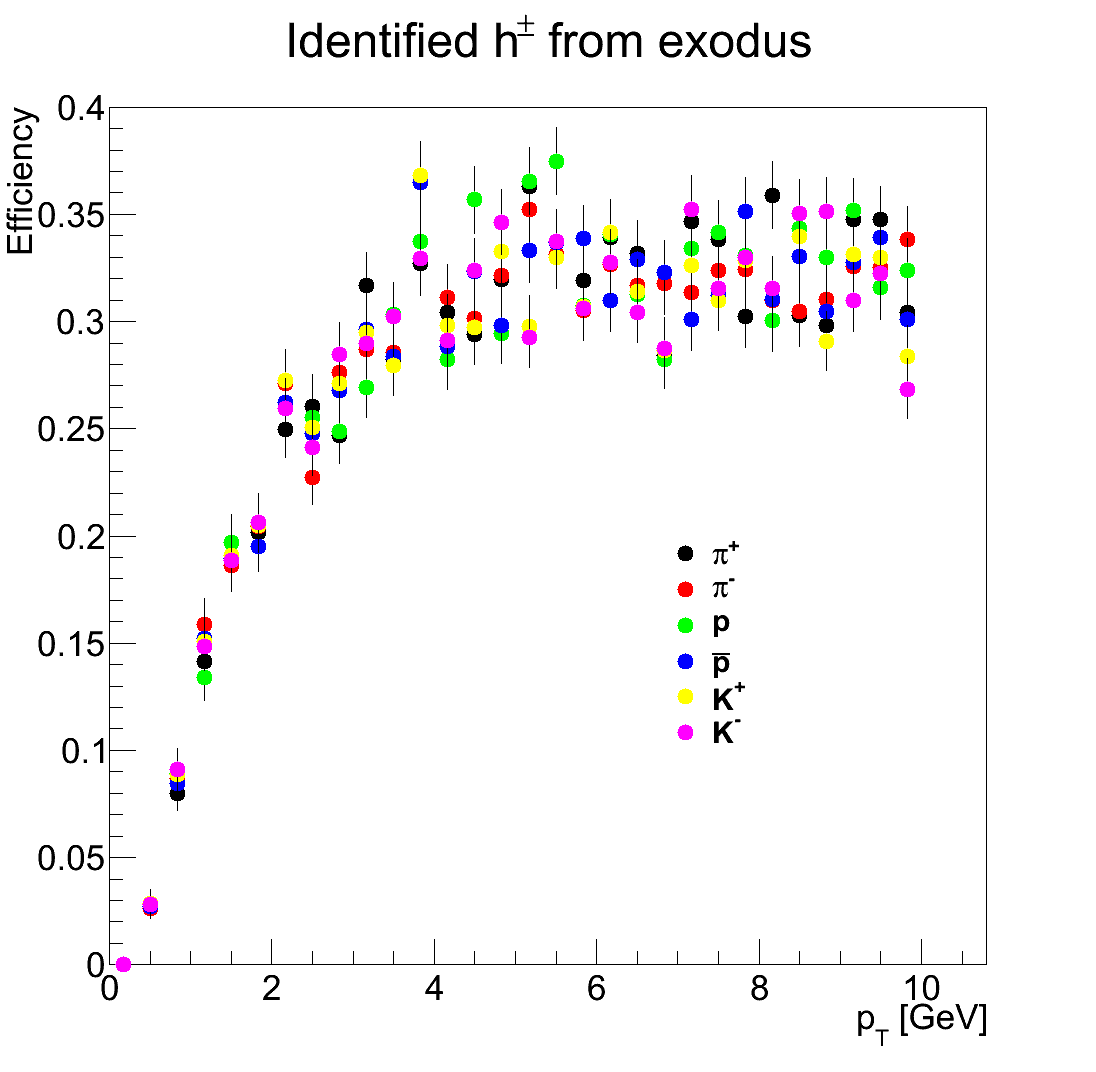}
\caption{The identified charged hadron efficiencies are shown with the run-13 PHENIX configuration using a single particle generator and a complete GEANT3 simulation of the detector.}
\label{fig:chg_had_eff_run13}
\end{center}
\end{figure}

EXODUS, which is the PHENIX single particle Monte Carlo generator, was used to throw 50,000 single $\pi^\pm$, $K^\pm$, $p$ and $\bar{p}$ each. Each set of particles was independently run through PISA to determine the detector response and thus the acceptance and efficiency. Single particle ntuples were generated from a flat \pt distribution with $0<\pt<11$, $0<\phi<2\pi$, and $|\eta|<0.5$. Particles were thrown in a wider pseudorapidity and \pt range than what is actually measured to avoid any possible edge effects from the simulation. The resulting output from PISA was then subjected to the same charged hadron cuts that are applied in the actual data analysis, which are described in Section~\ref{charged_hadron_id}. An additional cut was required that the particles have a generation of 1 to avoid any feed down or decay hadrons. The efficiencies were constructed by dividing the number of reconstructed hadrons by the number of truth hadrons thrown within the pseudorapidity of $|\eta|<0.35$ since this is the only region where PHENIX can in principle detect hadrons. The single particle efficiencies in run-13 \pp are shown in Fig.~\ref{fig:chg_had_eff_run13}. Efficiencies were found to be smaller than in previous PHENIX runs which is expected due to the degradation of the DC over time. \par

\begin{figure}[htbp]
\begin{center}
\includegraphics[width=0.6\textwidth]{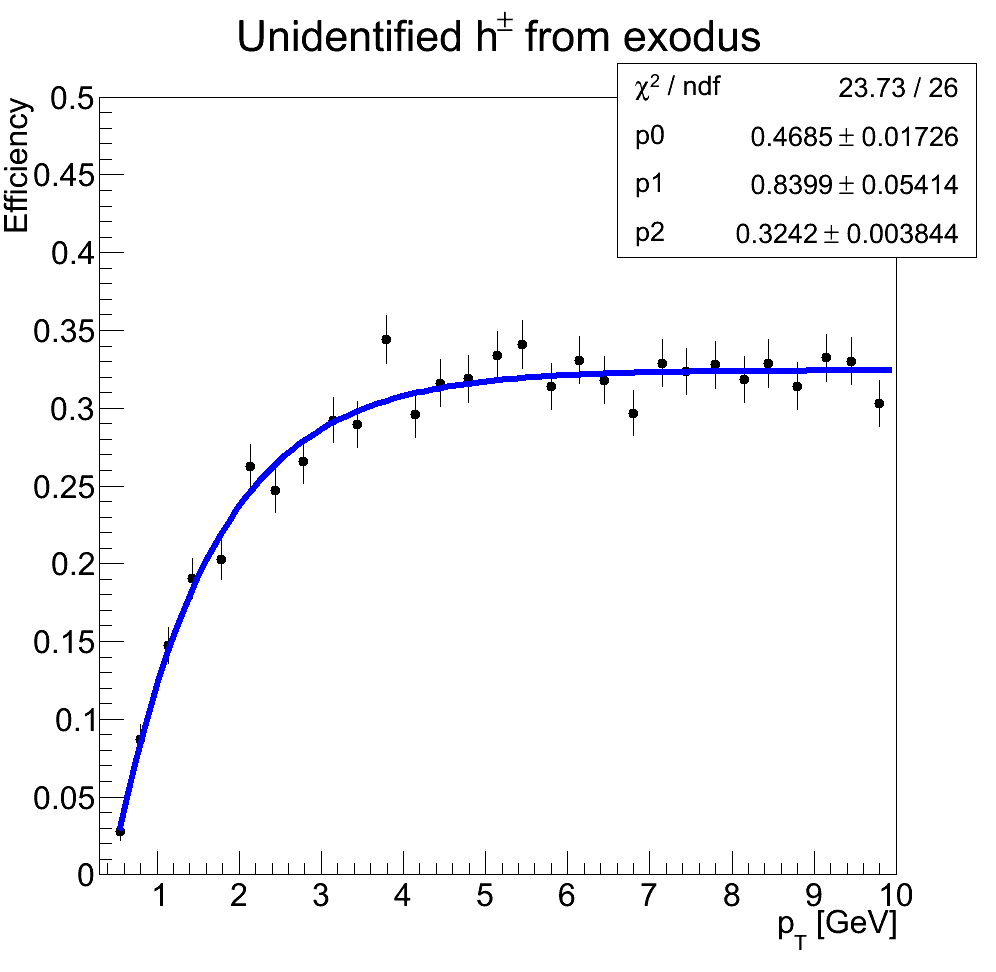}
\caption{The averaged nonidentified charged hadron efficiency is shown for run-13.}
\label{fig:nonchg_had_eff_run13}
\end{center}
\end{figure}

Since the efficiencies were found to be similar to each other, they were averaged to get a nonidentified charged hadron efficiency. Additionally since there are no particle identification (PID) cross sections from PHENIX at \sqs=~510 GeV, it would not be possible to average them based on their respective PID ratios. This will be discussed as a systematic uncertainty later. The nonidentified charged hadron efficiency for run-13 is shown in Fig.~\ref{fig:nonchg_had_eff_run13}. To account for the flat \pt spectrum thrown and to give a more realistic fit to data, the \pt bins were calculated with a weighted average from the actual charged hadron spectrum found in the analysis data. The efficiencies are fit with a saturated exponential of the form $\epsilon = -Ae^{-Bp_T}+C$, which is also shown in the figure with the fit constants.  \par

\begin{figure}[htbp]
	\centering
	\includegraphics[width=0.6\textwidth]{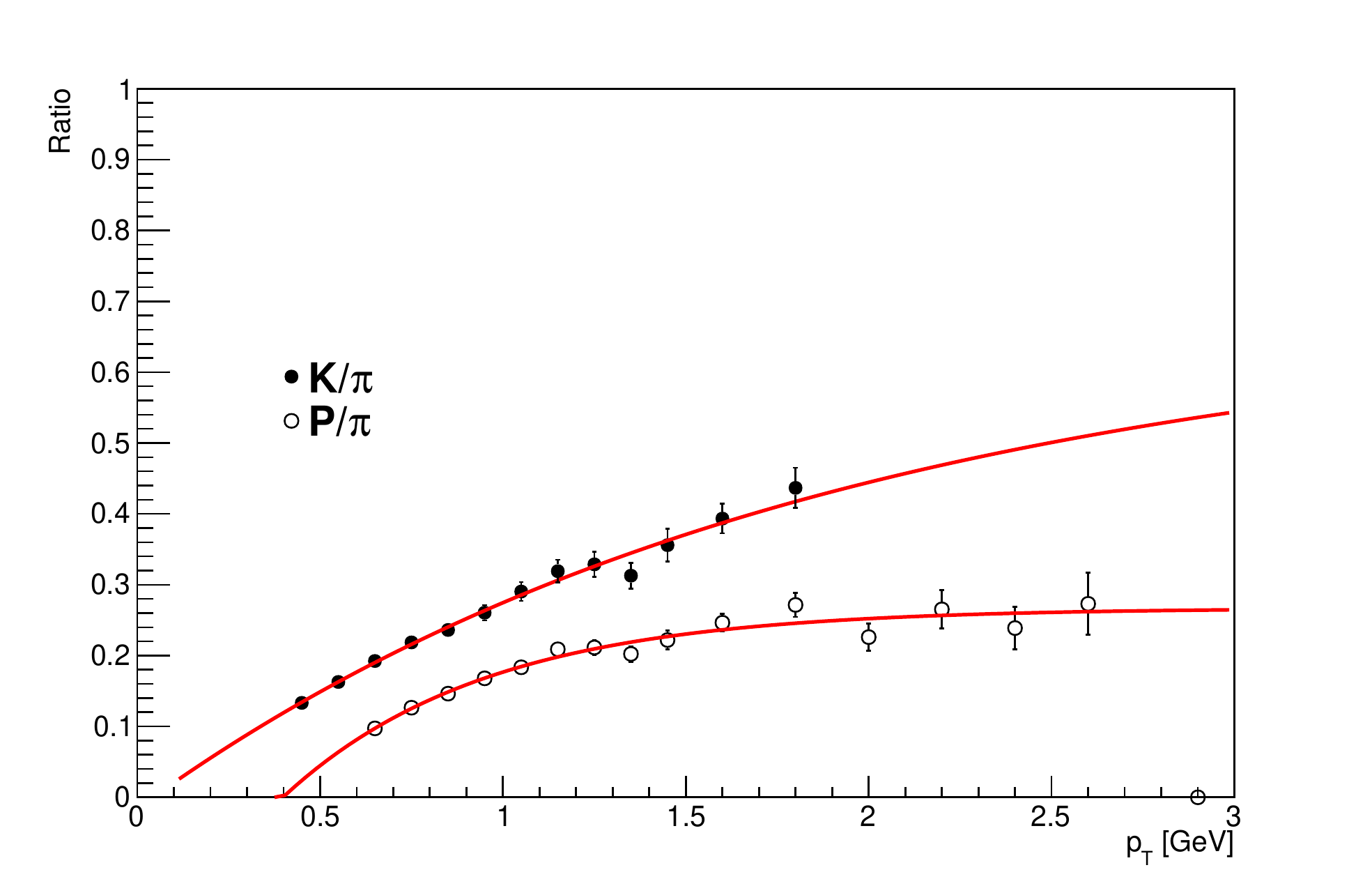}
	\caption{Measured identified particle cross section ratios are shown from PHENIX~\cite{PPG030}.}
	\label{fig:pid_ratios}
\end{figure}

\begin{figure}[htbp]
\begin{center}
\includegraphics[width=0.6\textwidth]{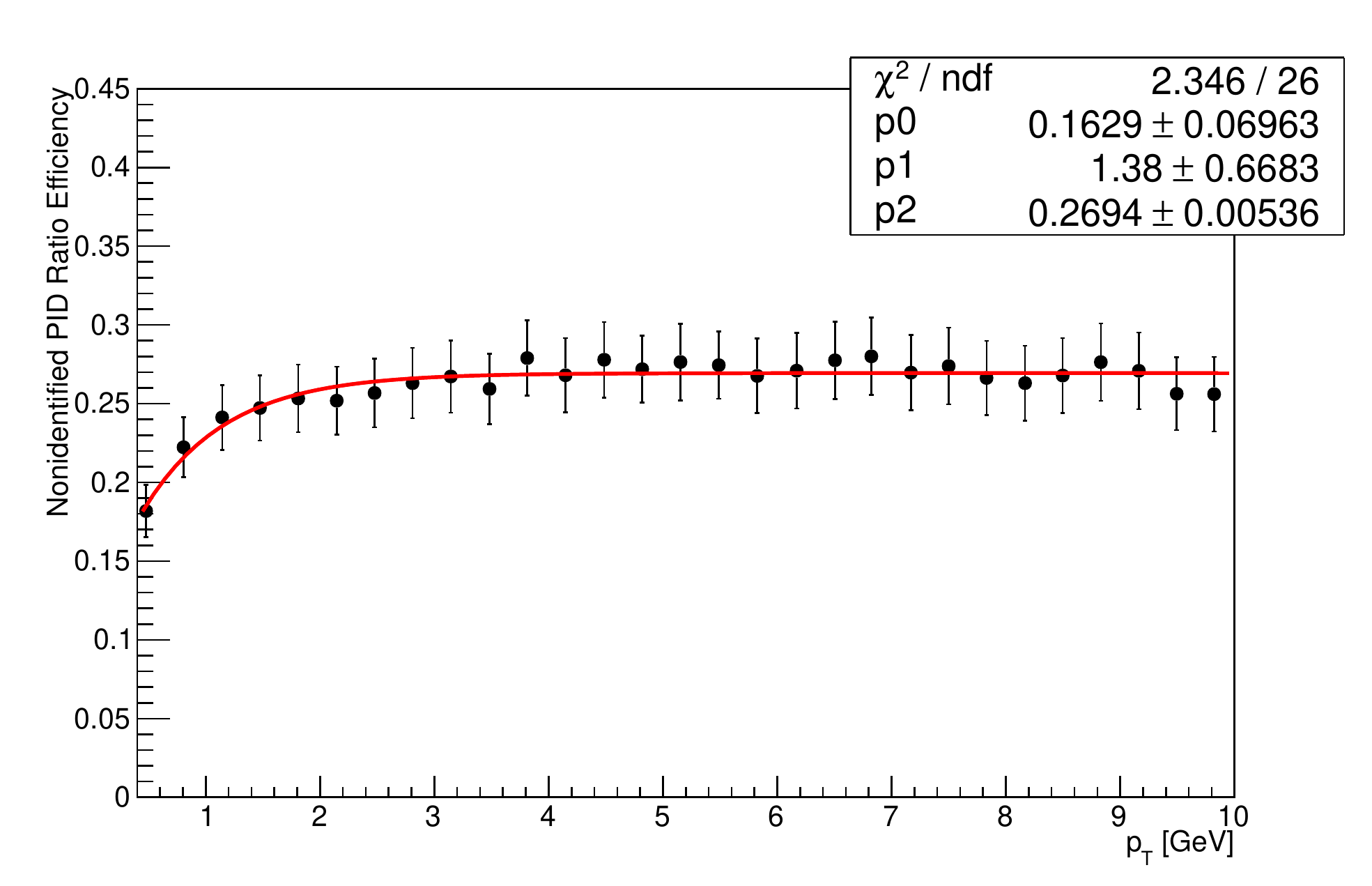}
\caption{The nonidentified charged hadron efficiency for run-15 is shown after appropriately weighting the particle species.}
\label{fig:nonchg_had_eff_run15}
\end{center}
\end{figure}

The run-15 charged hadron efficiency could be constructed with the additional benefit of the measured \pp PHENIX identified hadron cross sections at \sqs=~200 GeV~\cite{PPG030}. The identified hadron cross sections were plotted as ratios $\pi/K$ and $\pi/p$ and fit with saturated exponential functions as shown in figure~\ref{fig:pid_ratios}. At first the fact that the cross sections have only been measured up to 2.5 GeV/$c$ in \pt is concerning as the fit function is unconstrained at high \pt; however, we have several reasons to trust this functional choice. Firstly, there are measurements from the STAR collaboration which show that the particle ratios are largely \pt independent at high \pt~\cite{Agakishiev:2011dc}. There is also an additional physical reason to expect this because at high \pt kaon in-flight decays become a progressively smaller effect. To estimate an uncertainty on the fit functional form a linear fit is also used to produce a conservative upper limit on the particle ratios. This is by no means intended to be physical, but rather just to form an upper limit. Using this fit changes the hadron efficiency by at most 3\% from the saturated exponential functional form, thus this is the systematic uncertainty ascribed to the PID ratios. The run-15 identified efficiencies are found with a similar method to the run-13 efficiencies described above; averaging the identified hadron efficiencies from run-15 with the PID ratios given by the saturated exponential functions yields the nonidentified charged hadron efficiency shown in Fig.~\ref{fig:nonchg_had_eff_run15}. The plateau  of the efficiency is smaller in Fig.~\ref{fig:nonchg_had_eff_run15} than in Fig.~\ref{fig:nonchg_had_eff_run13} due to the physical degradation of the DC with time.

\subsection{Acceptance Correction}\label{acceptance_correction}

Since PHENIX is composed of two nearly back-to-back central arms, even isotropic particle production will appear to be back-to-back in azimuth. The acceptance correction, or mixed background correction, is applied to the correlation functions to account for the non-uniform PHENIX acceptance in azimuthal space. The acceptance effects of the detector are determined by constructing a mixed event background distribution, defined as

\begin{equation}\label{eq:mixedbckgd}
	A(\Delta\phi)=\frac{1}{C}\frac{dN_{mix}}{d\Delta\phi},\hspace{15pt} \text{where} \hspace{15pt} C=\frac{1}{2\pi}\int_0^{2\pi} \frac{dN_{mix}}{d\Delta\phi}d\Delta\phi\,.
\end{equation}

The intent of the acceptance correction is to correct the entire geometry to the same average efficiency, so that the charged hadron efficiency can then be applied as only a function of $p_T^{\rm hadron}$ and not also of $\phi$. In this way the $\phi$ dependence of the charged hadrons (and the correlation) is completely encapsulated in the mixed event correction. Note that this correction also takes into account possible warm towers in the EMCal, as more charged hadrons will then be mixed with the towers that were more likely to fire. \par

The integral is in the infinite bin limit, so in practice the mixed event distribution is made in bins of (\dphi, \pttrig, \ptassoc), then the area under the curve is summed and divided by the number of bins to get the normalizing factor $C$. The mixing is performed by embedding trigger particles into minimum bias events with charged hadrons. A particular trigger particle, when found, is mixed with minimum bias hadrons from a different event in the same run. ERT triggered hadrons cannot be used due to the presence of the high energy photon; this skews the distribution of hadrons in the event which in turn affects what background is being accounted for. It is important that the minimum bias charged hadrons come from the same run (or a similar run) due to the significantly increased luminosity delivered by the RHIC accelerator. This resulted in varying yields in both the DC and EMCal run by run. Additionally the acceptance can change run by run due to towers in the EMCal that were turned off or areas in the DC that were damaged during the run. \par

\begin{figure}[tbh]
	\centering
	\includegraphics[scale=0.5]{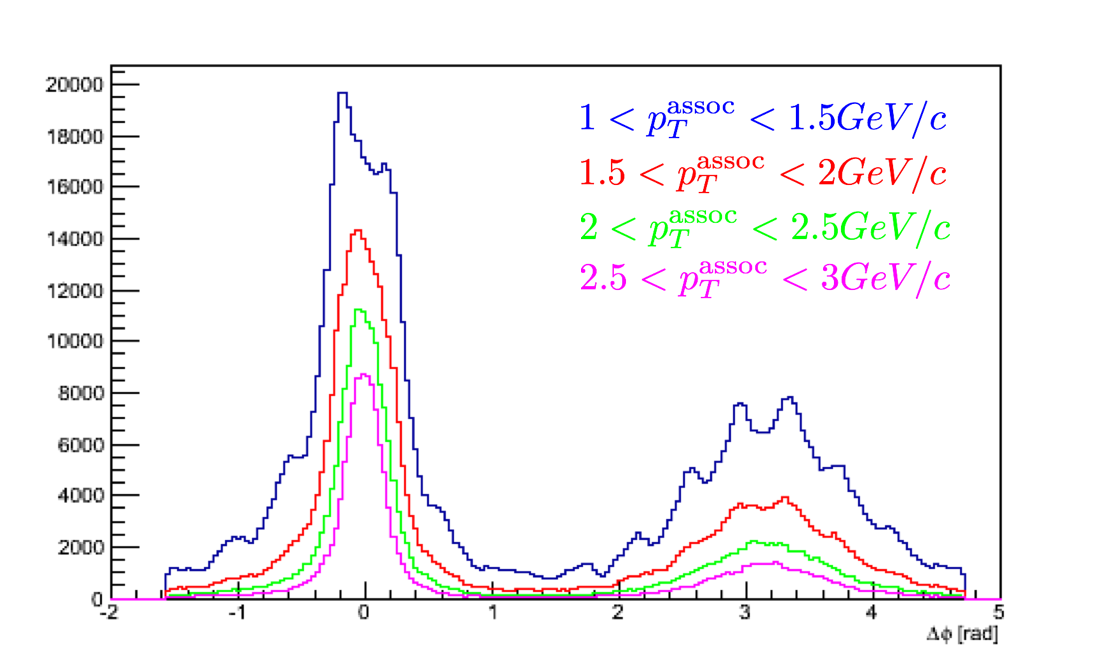}
	\caption{Uncorrected $\pion$ $\dphi$ correlations in $2\pi$ show an asymmetry about 0 and $\pi$ due to the PHENIX acceptance. The $\pttrig$ bin is the same for all colors, $5<\pttrig<6$ \gevc.} 
	\label{fig:uncorrectedpi0dphipt}
\end{figure}	

\begin{figure}[tbh]
	\centering
	\includegraphics[scale=0.6]{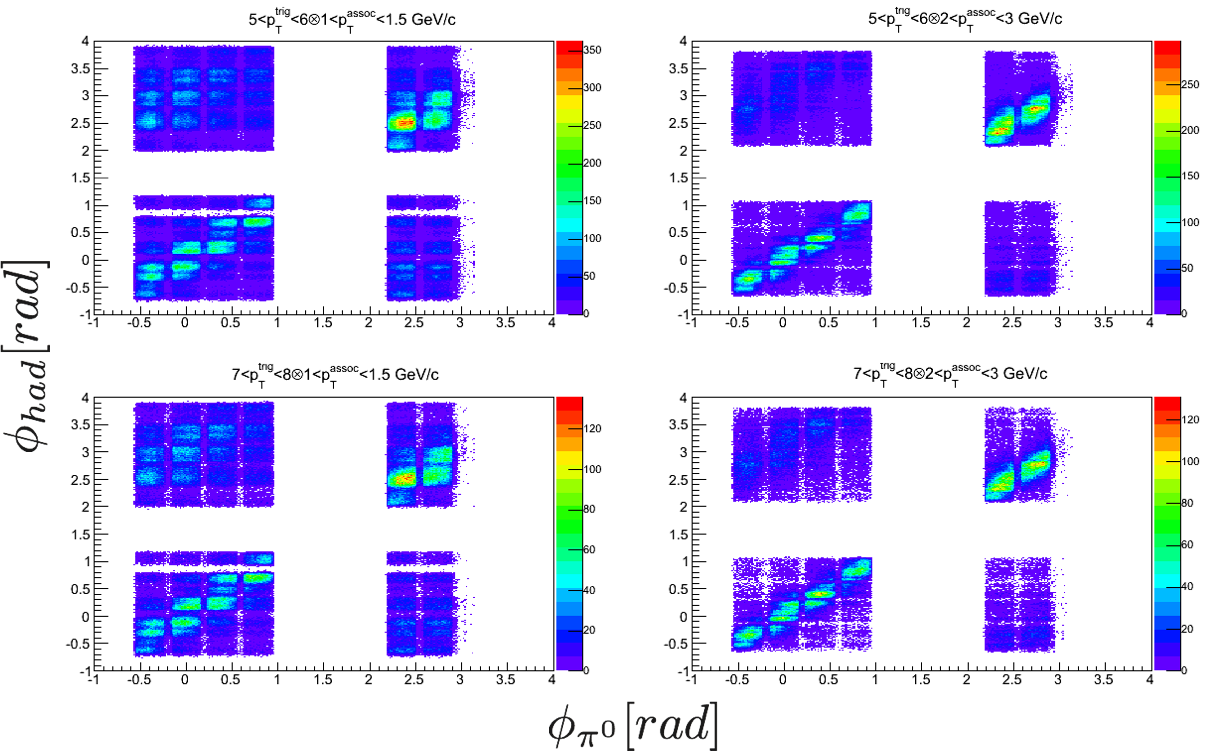}
	\caption{Uncorrected $\phi_{hadron}$ vs. $\phi_{\pion}$ for various bins of $\pttrig$ and $\ptassoc$ display the acceptance dependence. The ``hot spot" areas change with $\ptassoc$ but not with $\pttrig$. The same is true for the isolated inclusive photons. Therefore it is very important for the mixed corrections to be finely binned in $\ptassoc$ in order to capture this $\ptassoc$ dependence of the detector acceptance. This dependence is also shown explicitly as a function of $\dphi$ in figure \ref{fig:uncorrectedpi0dphipt}.}
	\label{fig:piphihadphi_pttrigptassoc}
\end{figure}

Ideally it would be best to have a continuous function (similar to the charged hadron efficiency), but in practice this is unrealistic for statistical reasons. Therefore we make the binning as finely as possible before running out of adequate statistical power. To determine in which direction(s) the binning should be finer, raw \pion-\h \dphi distributions for several bins of charged hadron \pt are shown in Fig.~\ref{fig:uncorrectedpi0dphipt}. The raw distributions clearly show highly asymmetric and non-jet-like structures as a function of both \ptassoc and \dphi; therefore the mixed background binning was chosen to be very fine in these two variables. The dependence on \pttrig was found to be small as shown in Fig.~\ref{fig:piphihadphi_pttrigptassoc}. While the statistical precision is reduced with increasing \pttrig, it is clear that the ``hot spots" which display the asymmetric nature of the PHENIX detector response are in each bin of \pttrig for varying \ptassoc. \par

The mixing also checks for a similar $z_{vtx}$ between events in order to check that the event characteristics are similar. In the \pa mixing the algorithm also checks for a similar centrality between the events as the distribution of hadrons can vary significantly with different centralities. To weight each minimum bias hadron appropriately a poolsize of 200 is used in the mixing. This way no single hadron is mixed more or less than 10\% from any other given hadron. To accomplish this, if a trigger particle is identified, a set of 200 minimum bias hadrons is mixed with the trigger particle, regardless of if the hadron passes the necessary $z_{vtx}$, \pt, or centrality criterion. The next trigger particle of the same kind would then mix with the next block of 200 minbias hadrons in the same run, and so on. In order to fully account for the asymmetric nature of the detector, the mixing is performed in the azimuthal range [0,2$\pi$] to get the appropriate correction value for a given \pttrig, \ptassoc, and \dphi triplet which completely describes the correlation phase space for a particular two-particle pair. \par

In the case of identifying PTYs of isolated quantities, great care must be taken in the event mixing to properly account for the background. Since the isolation cut is a cut which acts on the phase space of the identified particles, it is possible that it can alter the background associated with the measured quantities. For this reason, in the identification of isolated direct photon PTYs, the mixing algorithm rechecks that the trigger particle is isolated with the minimum bias charged hadrons from the mixed events. The same isolation cut criterion is used in the mixed event with the minimum bias tracks. Note that this applies to both isolated photons and \pion mesons in the application of the statistical subtraction method, detailed in Section~\ref{decay_sub_method}. \par

\begin{figure}[tbh]
	\centering
	\includegraphics[scale=0.4]{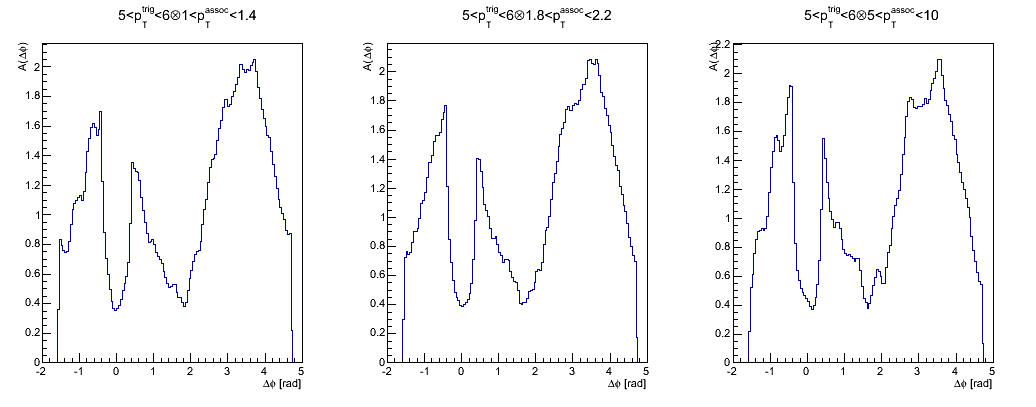}
	\caption{Example mixed event distributions for isolated inclusive photons are shown. The effect of rechecking the isolation cut is seen in the suppression of the near side jet structure. The isolated pion mixed event distributions look very similar.}
	\label{fig:mix_isophot}
\end{figure}

\begin{figure}[tbh]
	\centering
	\includegraphics[scale=0.4]{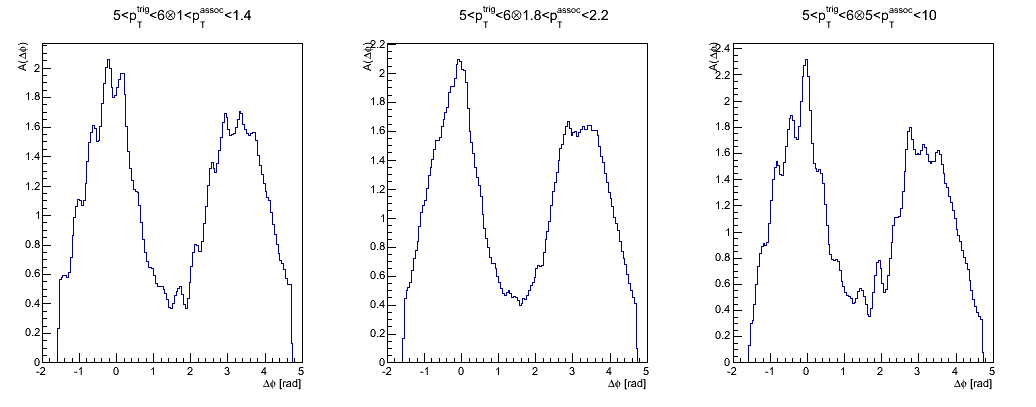}
	\caption{Several example mixed event distributions for $\pion\h$ correlations are shown.}
	\label{fig:mix_pi}
\end{figure}

Figures \ref{fig:mix_isophot} and \ref{fig:mix_pi} show a few examples of the mixed distributions for both isolated inclusive photons and $\pion$s, respectively. Since the binning is significantly finer than the final results, not all of the distributions are placed here; these figures just serve as examples. The figures show that the extremely fine binning of the mixed distributions show evidence of asymmetric areas in the detector acceptance. These could be caused from several effects; for example hot/cold areas in the DC and/or the EMCal. This encapsulates any bias that the detector(s) may have, and correctly accounts for these uneven areas. In the case of the isolated mixed distributions, the effect of rechecking the isolation algorithm is clear in the dearth of particles within the isolation cone. \par

\section{Decay Statistical Subtraction Method}\label{decay_sub_method}
The identification of direct photons on an event-by-event basis is difficult due to the significantly higher cross section for dijet events, and thus \pion production. Roughly 99\% of neutral pions decay to two photons, which introduces significant background that must be properly removed and accounted for when identifying direct photons. To determine the direct photon PTY, a statistical subtraction method is implemented that was first used in Refs.~\cite{ppg090,ppg113} and later extended with the use of isolation and tagging cuts in Ref.~\cite{ppg095}.

\subsection{Statistical Subtraction Method}
To identify direct photons via a statistical subtraction method, all decay photons must be accounted for and subtracted from an inclusive sample. Following the methods of Ref.~\cite{ppg090}, the total number of photons can be divided into two categories,

\begin{equation}
N^\gamma_{\rm inclusive} = N^\gamma_{\rm decay}+N^\gamma_{\rm direct}\,,
\end{equation}
where $N^\gamma_{\rm inclusive}$ is the total number of photons, $N^\gamma_{\rm decay}$ is the total number of decay photons, and $N^\gamma_{\rm direct}$ are all other photons. With this definition, $N^\gamma_{\rm direct}$ is defined as all photons that come directly from the hard scattering and includes a contribution from NLO fragmentation photons. The above definition can be extended to PTYs with the inclusion of a weighting factor

\begin{equation}\label{eq:statsub1}
Y_{\rm inc} = \frac{N_{\rm direct}^\gamma}{N^\gamma_{\rm inclusive}}Y_{\rm direct} + \frac{N^\gamma_{\rm decay}}{N^\gamma_{\rm inclusive}}Y_{\rm decay}\,.
\end{equation}
The weighting factor can be expressed in terms of a more familiar term in the literature \rgamma which is defined as
\begin{equation}\label{eq:rgamma}
	\rgamma(p_T^\gamma) = \frac{N^\gamma_{\rm inclusive}}{N^\gamma_{\rm decay}} = 1+\frac{N^\gamma_{\rm direct}}{N^\gamma_{\rm decay}}\,.
\end{equation}
From Eq.~\ref{eq:rgamma} it is clear that if there is any direct photon production, \rgamma must be greater than 1. Including this in the subtraction from Eq.~\ref{eq:statsub1} and rearranging, the PTY of direct photon production is given by
\begin{equation}\label{eq:ppg090statsub}
Y_{\rm direct} = \frac{1}{\rgamma-1}\left(\rgamma Y_{\rm inclusive}-Y_{\rm decay}\right)\,.
\end{equation}

\subsection{Isolated Statistical Subtraction Method}
Since the method used in Ref.~\cite{ppg090} includes contribution from NLO fragmentation photons, we would ideally like to impose additional cuts to reduce this component. Reference~\cite{ppg095} pioneered this method with the inclusion of an isolation cut and tagging cuts. Since direct photons are produced directly from the hard scattering, they should be produced with little activity in the near vicinity of $\eta$ and $\phi$ space. Isolation cuts have been used for decades across many collider experiments to better identify direct photons (see e.g.~\cite{ATLAS_isophotons,CDF_isophotons}); the cut requires that the candidate photon have little electromagnetic and hadronic activity in a cone of some size surrounding the photon. Tagging cuts aim to reduce the background from \pion and $\eta$ decays by removing photons on an event-by-event basis that can be tagged as decay photons from these sources. The specific details of these cuts used in this analysis are described in Section~\ref{isolation_tagging_cuts}. \par

To use these cuts we must apply them to Eq.~\ref{eq:ppg090statsub} as now the background we wish to statistically subtract has changed from all decay photon PTYs to decay photon PTYs which are isolated and unable to be tagged as decay products. This background is primarily due to large longitudinal momentum fraction $z$ neutral pions whose daughter decays either merge in the EMCal or which decay asymmetrically such that one of the decay photons is not detected by PHENIX. The statistical subtraction from equation~\ref{eq:ppg090statsub} now becomes
\begin{equation}\label{eq:isostatsub}
	\dirpty = \frac{1}{\rgammaprime-1}\left(\rgammaprime\incpty-\decpty\right)\,.
\end{equation}
In equation~\ref{eq:isostatsub} \rgammaprime is now a modified version of \rgamma that accounts for the isolation and tagging cuts, and is defined as
\begin{equation}\label{eq:rgammaprimedef}
	\rgammaprime(\pttrig) =\frac{N^{\rm iso}_{\rm inclusive}}{N^{\rm iso}_{\rm decay}} = \frac{N_{\rm inclusive}-N^{\rm tag}_{\rm decay}-N^{\rm niso}_{\rm inclusive}}{N_{\rm decay}-N^{\rm tag}_{\rm decay}-N^{\rm niso}_{\rm decay}}\,.
\end{equation}
In the previous equations ``niso" refers to ``not isolated" and ``iso" refers to ``isolated." In each of these equations, the inclusive sample is trivially determined as it is just the number of photons that passes the cuts; for example the numerator of Eq.~\ref{eq:rgammaprimedef} is just the total number of isolated inclusive photons that pass the tagging cuts. On the other hand, the decay quantities are not \textit{a priori} known since one cannot distinguish between a true isolated direct photon and an isolated decay photon. Therefore we must determine the number of isolated untagged decay photons in the denominator of Eq.~\ref{eq:rgammaprimedef} as well as the isolated decay photon PTY in Eq.~\ref{eq:isostatsub}. To do this we utilize a decay probability mapping function that accounts for the removal of photons tagged as coming from \pion decays. \par

\subsection{Determination of the Decay Per-Trigger Yield}

\begin{figure}[tbh]
	\centering
	\includegraphics[width=0.8\textwidth]{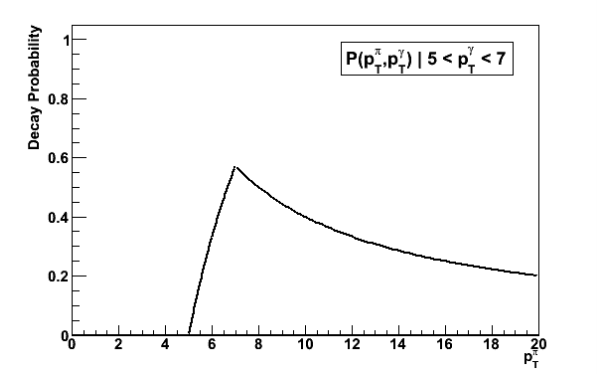}
	\caption{The probability for a $\pion$ of some $p_T$ to decay to a photon of $p_T$ between 5 and 7 \gev is shown~\cite{matt_thesis}. The normalization is arbitrary as it cancels when the per-trigger yields are calculated.}
	\label{fig:shark_notag}
\end{figure}

In the following discussion we restrict ourselves to the dominant source of background decay photons, the 2$\gamma$ decay of mesons such as the \pion or $\eta$. The solution to the problem discussed above is to determine a decay mapping function that quantifies the probability for a \pion of some $p_T^{\pion}$ to decay into a photon of some $p_T^\gamma$. Mathematically this function is a Green's function and can be described with the following equation, which relates the decay photon yields to the \pion yields
\begin{equation}\label{eq:green_fxn}
	\frac{dN^\gamma}{dp_T^\gamma} = \int dp_T^{\pion}\mapfxn\frac{dN^{\pion}}{dp_T^{\pion}}\,.
\end{equation}
This equation must be applied to both the \pion triggers as well as the \pion-\h pairs since we are interested in PTYs and thus the ratio of these two quantities. Conveniently the normalization of the function cancels in the PTY ratio, so we need not normalize these to unity to be true probability distributions. These functions are colloquially referred to as ``sharkfin" functions due to their shape. An example of a sharkfin function without tagging cuts applied is shown in Fig.~\ref{fig:shark_notag}. This shows the probability for a \pion of some \pt on the $x$ axis to decay to a photon with 5$<\pt<$7 \gev.

\begin{figure}[tbh]
	\centering
	\includegraphics[width=0.7\textwidth]{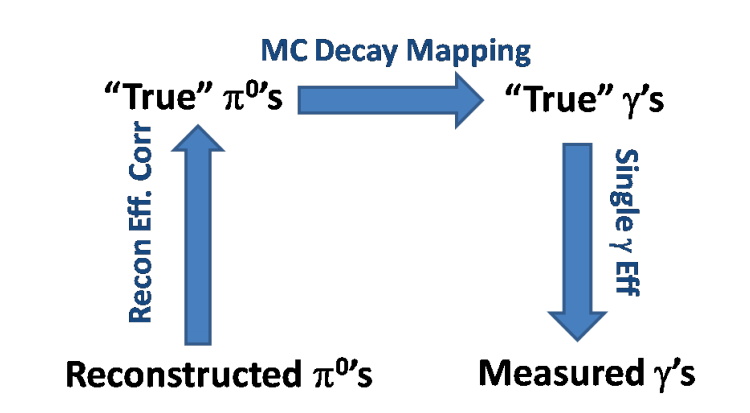}
	\caption{A diagram showing the logic flow of the Monte Carlo mapping procedure to estimate the decay photon PTYs~\cite{matt_thesis}. The single \photon efficiency is included in the Monte Carlo decay mapping functions.}
	\label{fig:decay_mapping_logic}
\end{figure}

In reality this function is also dependent on the z position of the \pion decay because in the PHENIX acceptance we are less likely to reconstruct photons that originate from pions close to the edge of the detector. In the continuous limit of Eq.~\ref{eq:green_fxn}, the decay photon yields look like

\begin{equation}\label{eq:green_fxn_contlimit}
N^{\gamma,meas} = \int_0^\infty dp_T^{\pi^0}\int dp_T^\gamma\int dz^{\pi^0}\int dz^\gamma \epsilon_{\pi^0}^{-1}(p_T^{\pi^0},z^{\pi^0})\frac{dN^{\pi^0,meas}}{dp_T^{\pi^0}}\mapfxn\epsilon_\gamma(p_T^{\pi^0},p_T^\gamma,z^\gamma)\,.
\end{equation}
Included in this more detailed equation are a \pion reconstruction efficiency as well as a single photon reconstruction efficiency, which are both dependent on the z position due to the acceptance of the PHENIX detector. The z position dependence of the \pion and the entire photon reconstruction efficiency are absorbed into the Monte Carlo determined mapping functions. The \pt dependent \pion efficiency is also applied to map the \pion yields found in this analysis to the measured \pion cross section. A schematic diagram shown in Fig.~\ref{fig:decay_mapping_logic} shows the logic flow to map the \pion-\h correlations to decay \photon-\h correlations. Using all of this the final decay photon PTYs are calculated by Eq.\ref{eq:decay_ptys}. Here $N$ refers to the raw yield of either a \pion-\h correlation or a \pion trigger, depending on whether the numerator or denominator of the PTY is being calculated.

\begin{equation}\label{eq:decay_ptys}
Y_{\rm decay}=\frac{N^{\gamma-h}}{N^\gamma}=\frac{\sum_i^{\pi^0-h}\epsilon_{\pi^0}^{-1}(p_T^{\pi^0})P(p_{T_i}^{\pi^0},z_{EMC_i})N^{\pi^0-h}}{\sum_i^{\pi^0}\epsilon^{-1}_{\pi^0}(p_T^{\pi^0})P(p_{T_i}^{\pi^0},z_{EMC_i})N^{\pi^0}}
\end{equation}

\begin{figure}[tbh]
	\centering
	\includegraphics[scale=0.5]{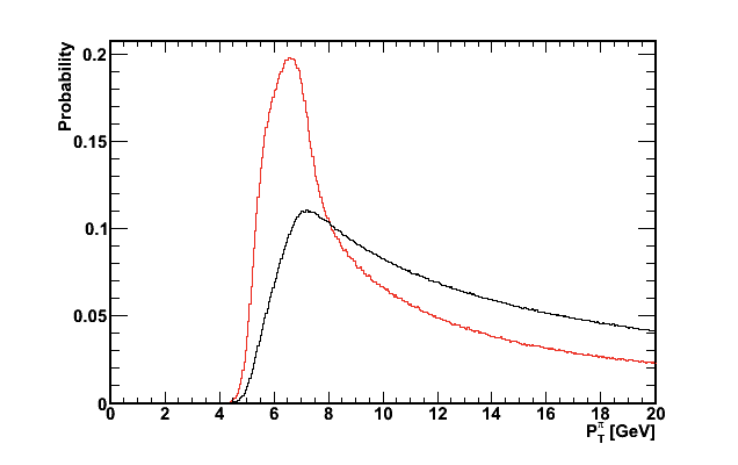}
	\caption{Decay probability functions for a $\pion$ to decay to a photon with $5<p_T^\gamma<7$ GeV with tagging (red) and without tagging (black)~\cite{matt_thesis}.}
	\label{fig:mod_sharkfins}
\end{figure}

We can now define an analogous expression after including the tagging and isolation cuts. The isolation cut alters the background from all decay photons to isolated decay photons; to estimate this contribution we measure isolated \pion mesons. The tagging cuts alter the shape of the sharkfin functions due to the higher likelihood of tagging high \pt \pion mesons; this is due to their higher probability of decaying symmetrically and thus detecting both photons in the PHENIX acceptance. To determine the isolated decay photon PTY we then have

\begin{equation}\label{eq:isodecpty}
\decpty =  \frac{\sum_i P^{tag}(p_T^i,p_T^\gamma)N^{iso}_{i-h}(p_T^i)}{\sum_i P^{tag}(p_T^i, p_T^\gamma)N^{iso}_i(p_T^i)}\,.
\end{equation}
Since the number of isolated decay photons comes primarily from isolated neutral pions, we utilize the mapping procedure to quantify this background. The Monte Carlo mapping functions $P^{tag}$ are just the decay probability where photon pairs that can be tagged as coming from a \pion decay are removed from the Monte Carlo calculation in the same way they are in the data. Figure~\ref{fig:mod_sharkfins} shows an example of the modified sharkfin for photons that decay from neutral pions in the range $5<p_T^\gamma<7$ \gev, both with and without the tagging included.

The shape of the modified sharkfin after including tagging cuts is altered due to the higher likelihood of being unable to tag asymmetric decays, or in other words, being unable to tag a \pion with a similar \pt to the observed isolated photon decay \pt. For example, if a \pion has a \pt of 6 \gev and it decays to a photon with \pt=~5 \gev, the other decay photon will be emitted highly asymmetrically from the higher energy photon in the boosted \pion frame. Therefore, it is more likely that the PHENIX acceptance misses this low energy photon and thus we cannot tag the 5 \gev photon as a decay product and remove it from the simulation. It follows that the probability that this \pion leads to an isolated decay photon is larger than, for example, a \pion with \pt=12 \gev where a 5 \gev decay photon requires a more symmetric two-photon decay. \par

\subsection{\pion Trigger Efficiency}

\begin{figure}[htbp]
\begin{center}
\includegraphics[width=0.7\textwidth]{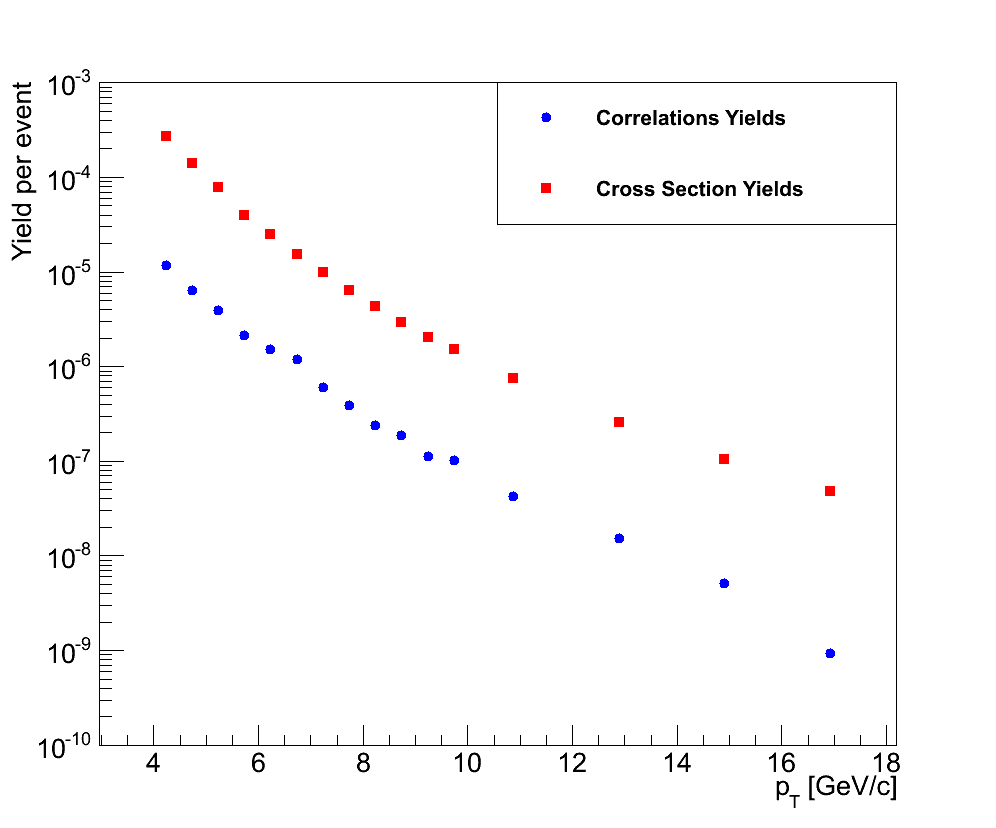}
\caption{Yields per event of both the run-13 \sqs=~510 GeV $\pion$ cross section and this analysis in minimum bias data are shown.}
\label{fig:pi0_xsec_yields}
\end{center}
\end{figure}

\begin{figure}[tbh]
	\centering
	\includegraphics[scale=0.27]{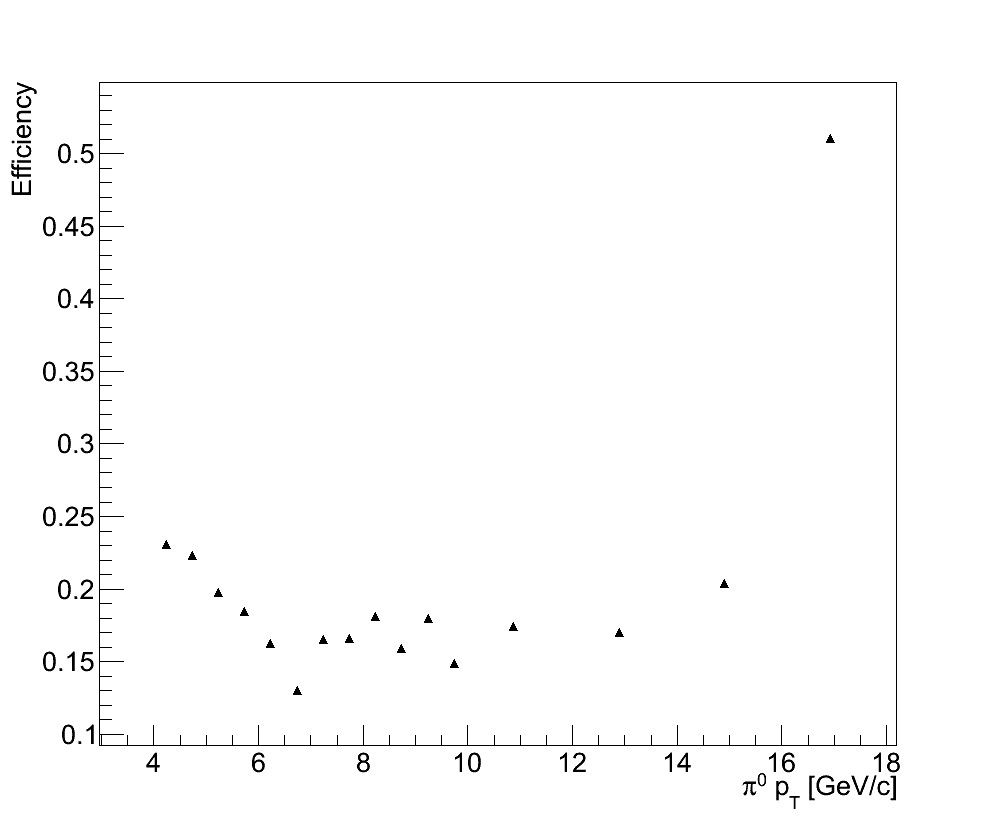}
	\caption{The $\pion$ trigger efficiency correction is found by taking the ratio of the cross section to correlations yields per event in \sqs=~510 GeV \pp collisions. The very poor efficiency correction at the highest \pt is due to two-photon merging effects becoming dominant in the PbSc EMCal sectors.}
	\label{fig:run13_pi0eff}
\end{figure}

\begin{figure}[tbh]
	\centering
	\includegraphics[scale=0.27]{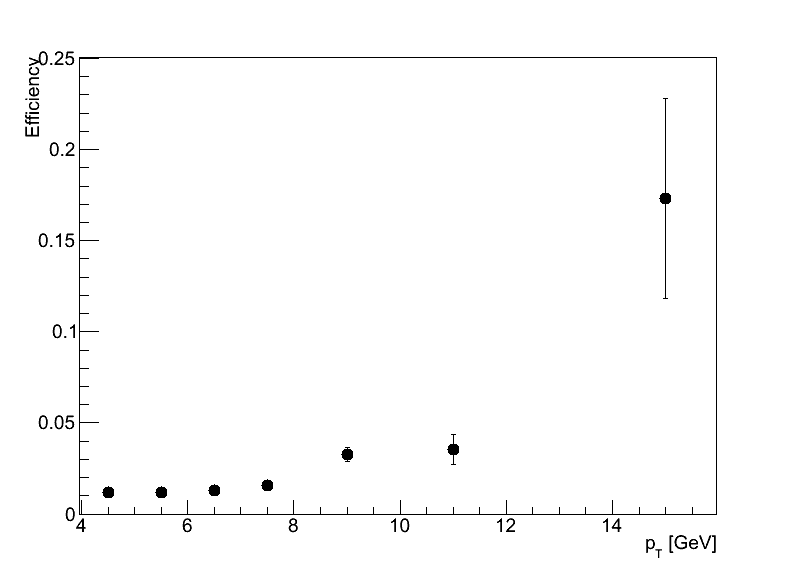}
	\caption{The $\pion$ trigger efficiency correction is found by taking the ratio of the cross section to correlations yields per event in \sqs=~200 GeV \pp collisions.}
	\label{fig:run15pp_pi0eff}
\end{figure}

As input for determining the decay photon PTYs, the \pion efficiency must be determined as outlined in Fig.~\ref{fig:decay_mapping_logic} and in Eq.~\ref{eq:decay_ptys} since this is not included in the Monte Carlo mapping. Since this analysis explores a new center-of-mass energy, this efficiency has yet to be determined. The \pt dependent efficiency is determined by bootstrapping the \pion yields in this analysis to the published \pion cross section from PHENIX at \sqs=~510 GeV~\cite{ppg186}. The cross section is converted to yields per event so that it can be directly compared to the yields found in this analysis. This is done via the following:

\begin{equation}\label{eq:bootstrap_xsec}
E\frac{d^3\sigma_{\pion}^{unbiased}}{d^3p}\cdot\frac{2\pi p_T}{\sigma_{BBC}C_{p+p}^{BBC Bias}} = \frac{1}{N_{BBC}^{total}}\frac{d^2N_{\pion}^{Total Biased}}{dp_Tdy}\,.
\end{equation}
This gives the yields differential in rapidity and \pt, normalized by the total number of BBC events. The $C_{p+p}^{BBC Bias}$ factor accounts for the fact that the BBC only measures a fraction of the total inelastic \pp cross section. The raw analysis yields are found in minimum bias data to remove any bias that may arise from the ERT trigger, but the exact same analysis cuts are used to reconstruct the \pion yields as in the ERT triggered data. Figure~\ref{fig:pi0_xsec_yields} shows the yields per event in this analysis from minimum bias data and the converted run-13 \pp cross section. The efficiency function is found by taking the ratio of the yields per event from the cross section to the yields per event from this analysis, shown in Fig.~\ref{fig:run13_pi0eff} for \sqs=~510 GeV and Fig.~\ref{fig:run15pp_pi0eff} for \sqs=~200 GeV \pp. The normalization of the function is not important as it cancels in the determination of the decay PTYs, so only the shape is relevant. The efficiency correction is larger at both low and high \pt, due to the effects of missing a decay photon from asymmetric decays from the PHENIX acceptance and merging \pion clusters, respectively. \par

\begin{figure}[tbh]
	\centering
	\includegraphics[width=0.6\textwidth]{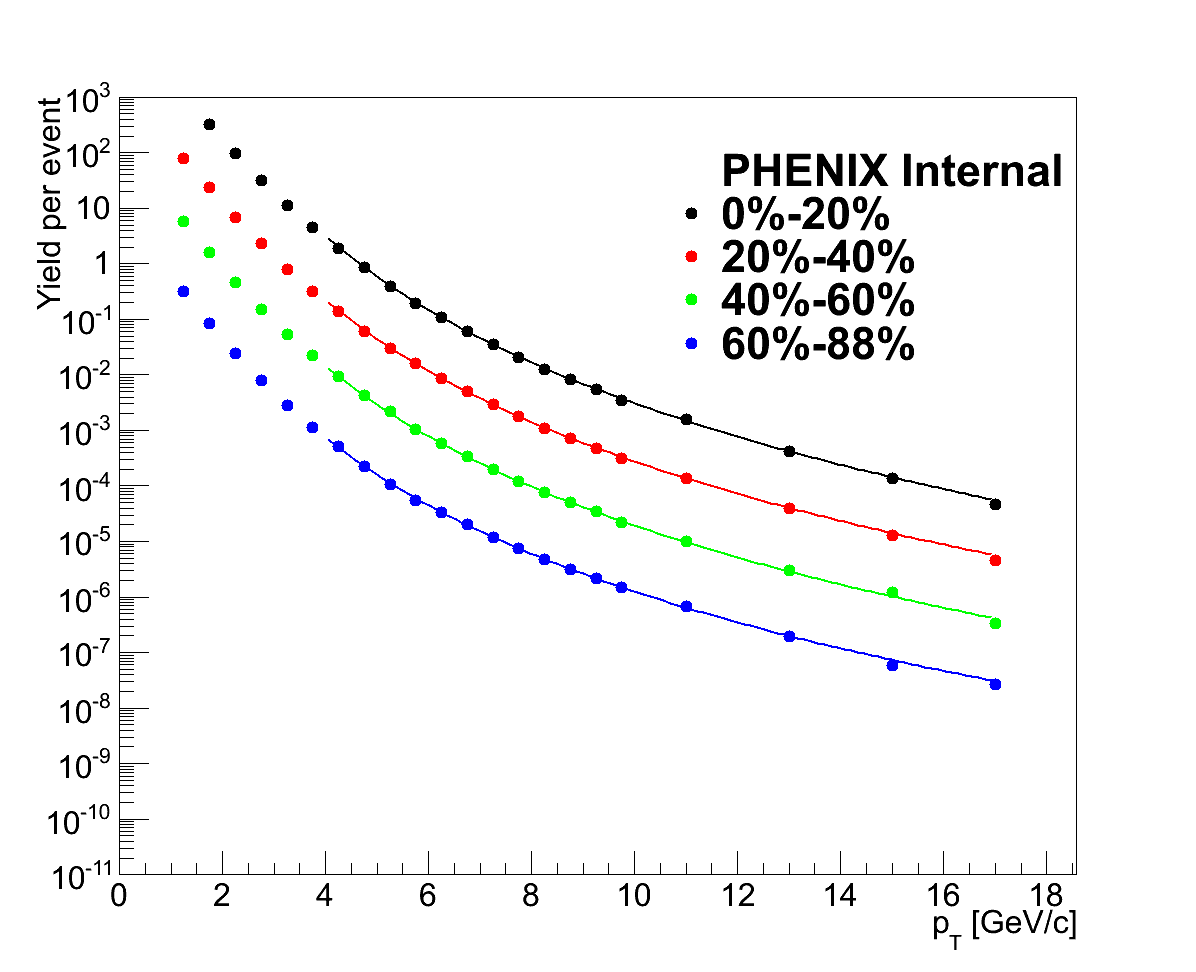}
	\caption{The fully corrected \pau $\pion$ cross sections from internal PHENIX analysis are shown with power law fits.}
	\label{fig:pa_centdep_pi0xsec}
\end{figure}

\begin{figure}[tbh]
	\centering
	\includegraphics[width=0.8\textwidth]{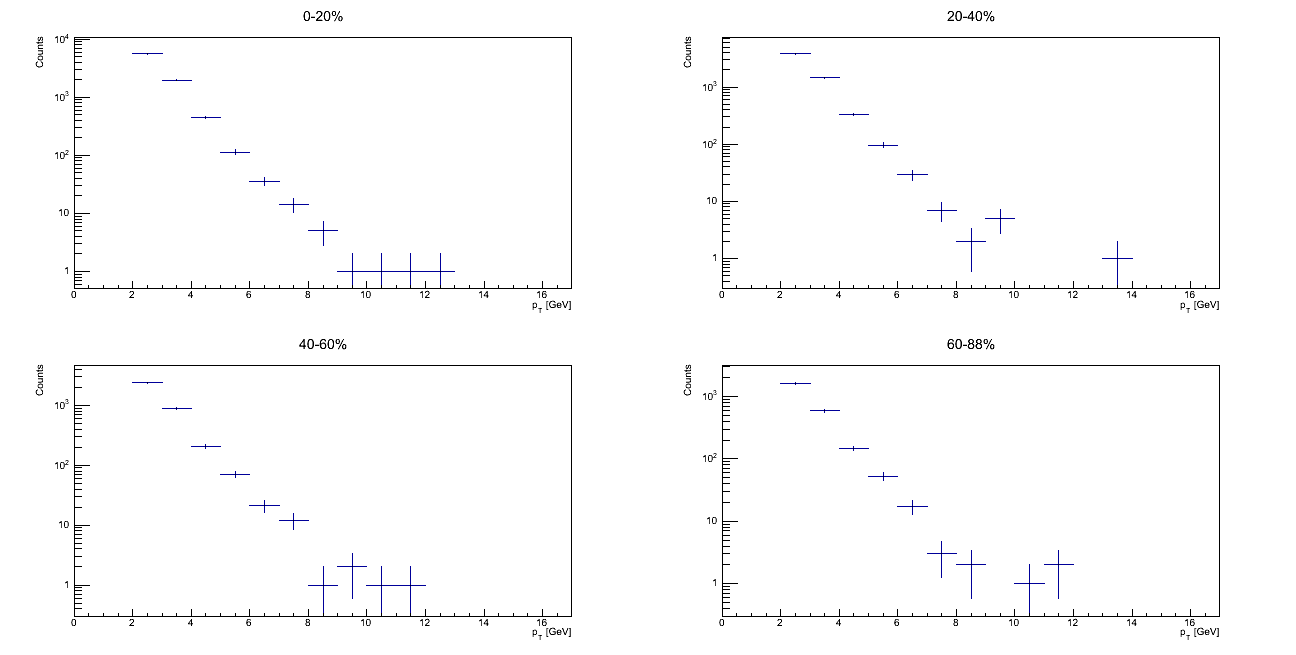}
	\caption{The raw minimum bias $\pion$ yields with the cuts used in the \pau analysis are shown in four different centrality bins.}
	\label{fig:raw_minbias_pi0_payields}
\end{figure}

\begin{figure}[tbh]
	\centering
	\includegraphics[width=0.6\textwidth]{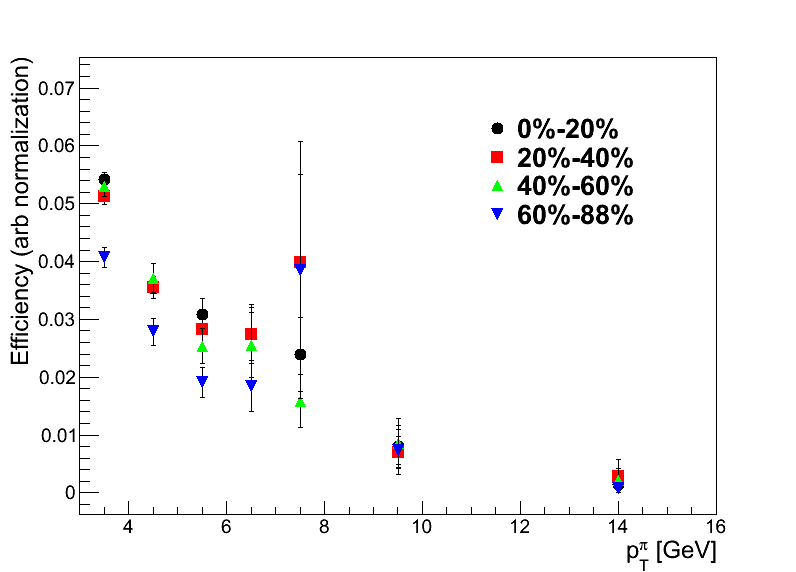}
	\caption{The run-15 $\pion$ efficiency correction is shown as a function of $\pion$ $p_T$. The efficiency correction takes the usual shape of larger at  low $p_T$ due to asymmetric $\pion$ decays.}
	\label{fig:pa_pi0eff}
\end{figure}

In \pa collisions, the collision centrality must also be considered in the \pion efficiency since the \pion production as a function of \pt could be dependent on the centrality. The same procedure described above was used except with the $p$+Au centrality dependent cross sections, shown in Fig.~\ref{fig:pa_centdep_pi0xsec}. The raw minimum bias yields of \pion production are shown in Fig.~\ref{fig:raw_minbias_pi0_payields}, and the resulting ratio of the minimum bias yields per event and cross section yields per event is shown in Fig.~\ref{fig:pa_pi0eff}. Here there is a clear centrality dependence; the high \pt behavior is less constrained due to the limited statistical precision of the minimum bias counts.

\subsection{Isolation and Tagging Cuts}\label{isolation_tagging_cuts}

True direct photons that emerge directly from the hard scattering are difficult to identify on an event-by-event basis even with the use of the statistical subtraction method outlined in section~\ref{decay_sub_method}. This is because at NLO, fragmentation photons from standard dijet events also have a nonzero contribution to the final ``direct photon'' PTYs. To reduce this contribution, isolation cuts are often used since true direct photons should be produced with little hadronic activity in the vicinity of the photon. Tagging cuts also help identify and remove backgrounds from decay photons; both of these cuts help boost the signal to background ratio of the direct photons to NLO fragmentation photons. \par

First, tagging cuts are applied on an event-by-event basis to remove any photons that can be tagged as coming from decay sources. Candidate direct photons are paired with all other photons in the event, satisfying very loose photon identification criteria. The sister photon is only required to have an energy greater than 1 GeV and pass basic shower shape and hot/dead tower cuts. If the invariant mass of the pair of photons is in the region $118<m_{\gamma\gamma}<162$ MeV/$c^2$ for \pion mesons or $500<m_{\gamma\gamma}<600$ MeV/$c^2$ for $\eta$ mesons, the candidate direct photon is removed from the sample. Note that the invariant mass range for the tagging of \pion decays is slightly larger than for the identification of \pion-\h pairs; this is because we should be more aggressive with the tagging cuts to eliminate as much decay background as possible. \par

The isolation cut requires that the sum of energy in the EMCal and the \pt of charged tracks measured in the DC/PC tracking system in a cone of radius 0.4 radians around the trigger photon be less than 10\% of the energy of the candidate photon. In other words the photon must pass
\begin{equation}
\sum\left(E^{\rm EMCal}+\pt^{DC}\right)\leq 0.1\times E_{\gamma}^{\rm direct}\,.
\end{equation}
This type of isolation cut is fairly standard throughout the literature~\cite{ppg095,CDF_isophotons}; isolation cuts have become more aggressive at the LHC due to the higher luminosities and pile up fractions available~\cite{Aaboud:2016sdm}. In determining the contributions to the isolation cone energy sum, the cuts placed on EMCal clusters and charged tracks are less stringent in order to err on the side of including more in the cone. Due to the large datasets available in 2013 and 2015 one can afford to be more aggressive with these cuts, similar to the tagging cuts. Note that for the determination of the isolated decay photon PTYs, \pion mesons undergo the same isolation cut algorithm that the candidate direct photons are subjected to.\par

\subsection{\rgamma and \rgammaprime Determination}\label{rgammasection}

Since there is no published data for \rgamma at \sqs=~510 GeV as there was for Ref.~\cite{ppg095} at \sqs=~200 GeV, a different method was used to determine its \pt dependence. Under the assumption that the decay photon and \pion spectra are given by a pure power law, the number of photons coming from $\pion$ decays at a given \pt is given by

\begin{equation}\label{eq:rgammadataeq}
\rgamma = \frac{N_{\rm inc}}{N_{\rm decay}} = 1 + \frac{n-1}{2} \frac{1}{1.23} \frac{dN/dp_{T}(\rm direct)}{dN/dp_{T}(\pion)}\,.
\end{equation}
In this equation, $(n-1)/2$ is a correction for $\pion\to\gamma\gamma$ decays and $1/1.23$ is a correction for decay photon contributions from other mesons, e.g. the $\eta$, $\omega$, etc. The validity of these correction factors was studied and found to be reasonably accurate in PYTHIA simulations in the associated work for~\cite{ppg095}. Since there was no published direct photon cross section at \sqs=~510 GeV, two methods were used to calculate \rgamma as a systematic cross check. \par

The first way to determine \rgamma was data driven by finding the fraction $N_{\rm inc}/N_{\pion}$ and then applying the above corrections which take into account the effects from decays. This ratio was also corrected by a single photon efficiency and a \pion efficiency to account for the different and non-canceling efficiencies for which PHENIX detects single photons vs. two-photon decays. To include these efficiencies, the above Eq.~\ref{eq:rgammadataeq} is modified to be
\begin{equation}
 	\rgamma = \frac{N_{\rm inc}}{N_{\pi^0}} \frac{1}{1.23} \frac{n-1}{2} \frac{\epsilon_{\pi^0}}{\epsilon_\gamma}\,.
\end{equation}
Here the efficiencies are defined as $\epsilon = N_{\rm reco}/N_{\rm truth}$, which is why they appear ``flipped" with respect to the number of inclusive photons and neutral pions.  \par

\begin{figure}[tbh]
	\centering
	\includegraphics[width=0.6\textwidth]{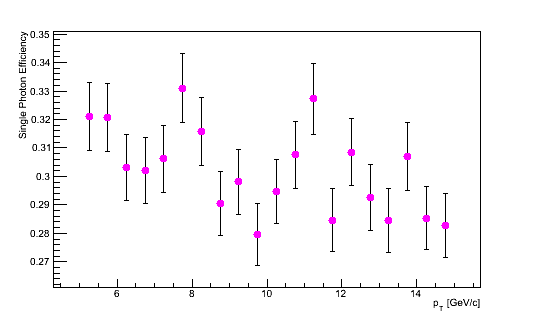}
	\caption{The run-13 single photon efficiency is shown as a function of \pt.}
	\label{fig:singlephotoneff}
\end{figure}

\begin{figure}[tbh]
	\centering
	\includegraphics[width=0.6\textwidth]{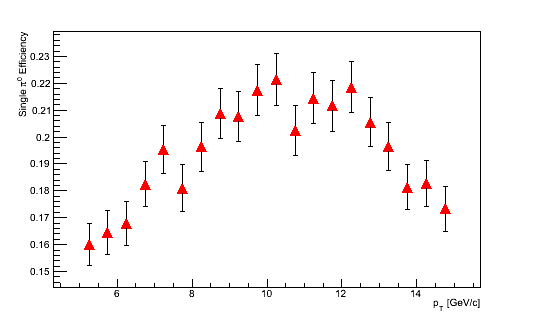}
	\caption{The run-13 single \pion efficiency is shown as a function of \pt.}
	\label{fig:singlepioneff}
\end{figure}

The two efficiencies were found similarly to the identified charged hadron efficiencies. Single photons and single neutral pions were generated independently with the EXODUS Monte Carlo generator. Particles were thrown with a flat \pt distribution in $2\pi$ and in $|\eta|<0.5$, and then simulated through the run-13 PISA configuration. Efficiencies were determined by dividing the number of reconstructed particles by the number of truth particles thrown in a particular \pt bin. Only truth particles within $|\eta|<0.35$ are considered in determining the efficiencies as these are the only particles that could in principle be detected. In reconstructing the single photons and neutral pions the same cuts as used in the data analysis are used here. The determined efficiencies are shown in Figs.~\ref{fig:singlephotoneff} and~\ref{fig:singlepioneff}. The shape of the single \pion efficiency is similar to the \pion trigger efficiency discussed above; the loss in neutral pions at low and high \pt is due to asymmetric decays and merged photons, respectively. Note that the exact shape is actually the inverse of the \pion trigger efficiency due to the different definition used for the calculation of \rgamma. \par

The single photon efficiency required additional study due to the modest downward slope as a function of \pt. In principle this efficiency should not be dependent on \pt. The cause of the downward slope was determined to be due to internal conversions with the VTX, a silicon tracking vertex detector that was installed surrounding the PHENIX interaction point for run-13. The VTX was installed for the PHENIX heavy flavor program; unfortunately here it is causing additional internal conversions from the single photons. Several short studies were performed to come to this conclusion.

In the single photon simulation, there were many reconstructed charged tracks from the DC. These tracks were found to be equally split between positively and negatively charged tracks, as determined by the DC tracking algorithm. Additionally these tracks nearly all had a RICH parameter $n0<$0, indicating that they were also reconstructed in the RICH as electron tracks. One can also look through the ancestry tree in these single Monte Carlo events, and the parent \pt of the electron or positron always matched the truth \pt of the thrown photon. This additional high rate of conversion is because the VTX adds about 16\% of a radiation length that the photon must traverse from the nominal interaction point. \par

\begin{figure}[tbh]
	\centering
	\includegraphics[width=0.6\textwidth]{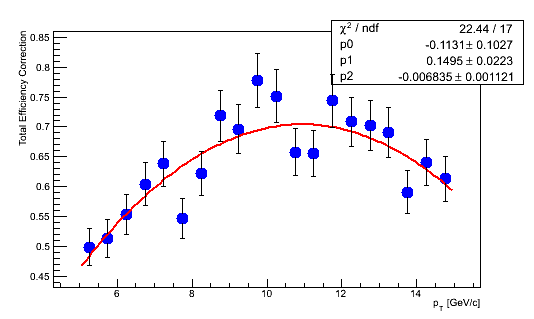}
	\caption{The total efficiency given by the single \pion efficiency divided by the single photon efficiency is shown, which is applied as a correction factor for determining \rgamma.}
	\label{fig:rgammaeff}
\end{figure}

The single \pion efficiency was then divided by the single photon efficiency and fit with a quadratic function to determine the total efficiency correction $\epsilon = \epsilon_{\pi^0}/\epsilon_\gamma$. The resulting fraction of inclusive photons to neutral pions from data was then corrected by this polynomial, shown in Fig.~\ref{fig:rgammaeff} as well as the other corrections from Eq.~\ref{eq:rgammadataeq}. Note that, as anticipated, the efficiencies do not  cancel in the ratio as they are not equal to each other. \par

\begin{figure}[thb]
	\centering
	\includegraphics[width=0.6\textwidth]{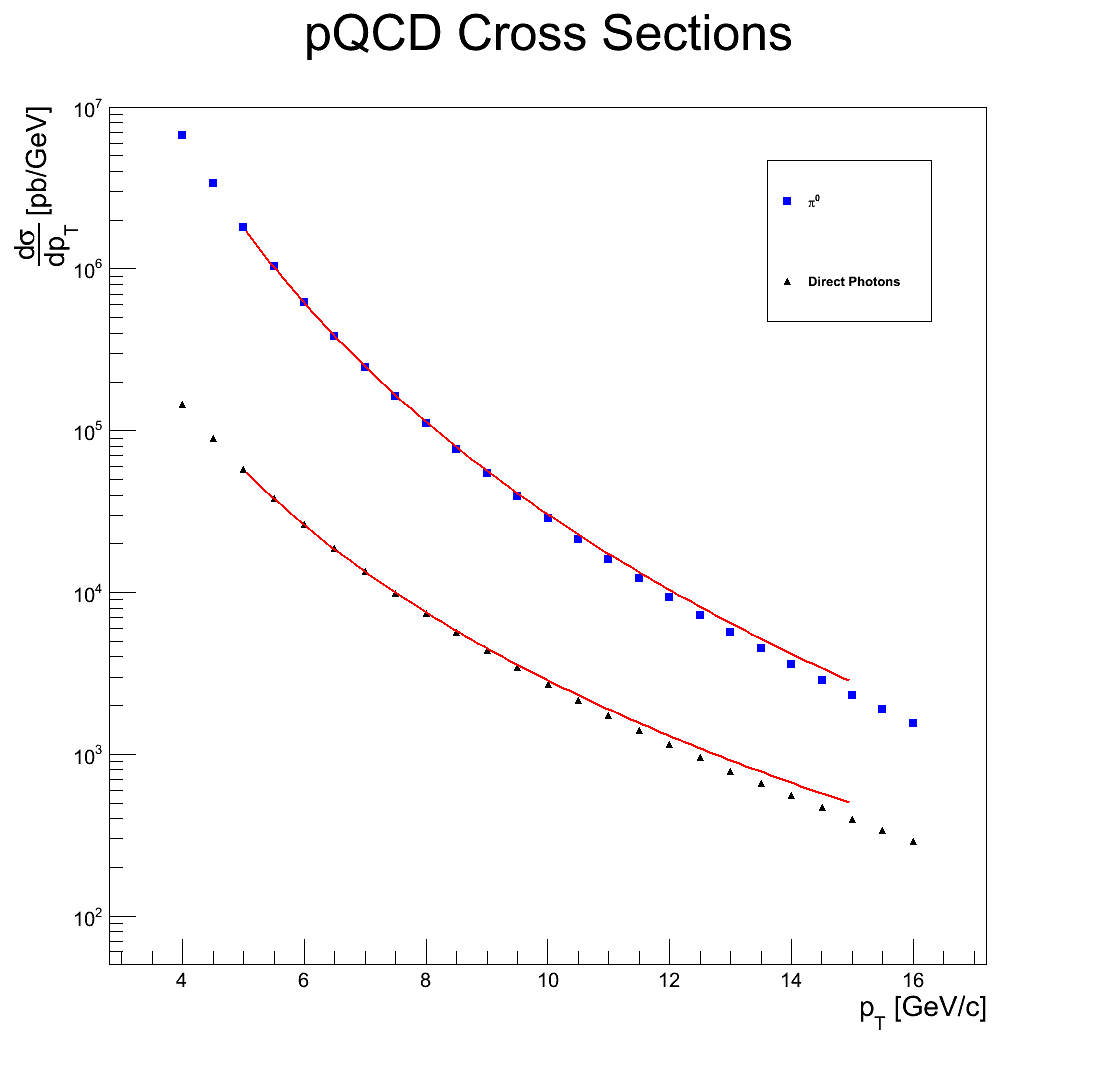}
	\caption{Perturbative QCD cross sections are shown with power law fits for both \pion and direct photon production at \sqs=~510 GeV. The calculations were performed with the CT10 PDFs~\cite{Lai:2010vv} and the DSS14 FFs~\cite{deFlorian:2014xna}.}
	\label{fig:pqcd_pi0_photons}
\end{figure}

\begin{figure}[thb]
	\centering
	\includegraphics[width=0.6\textwidth]{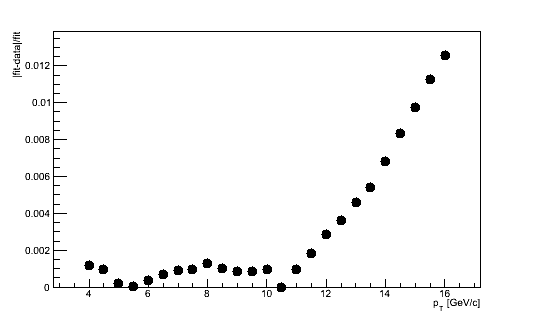}
	\caption{ \rgamma was calculated via the power law fits and directly from the pQCD cross sections and the value $|$fit-data$|$/fit was calculated to estimate the error in the power law fits. At most the fits give a 1\% error at high \pt which is negligible compared to the other assigned systematic uncertainties.}
	\label{fig:rgamma_dev_from_pqcd}
\end{figure}

The second method to calculate \rgamma was by using pQCD calculations for both the direct photon and \pion cross section at \sqs=~510 GeV, and then using Eq.~\ref{eq:rgammadataeq} to calculate \rgamma. The calculations were provided by Tom Kaufmann using the CT10 PDFs~\cite{Lai:2010vv} and DSS14 FFs~\cite{deFlorian:2014xna} for the neutral pions. The cross sections with power law fits are shown in Fig.~\ref{fig:pqcd_pi0_photons}. Since the power law fits slightly deviate from the calculation at high \pt, the effect on \rgamma from using the fits or data was determined. This effect is shown to be smaller than 1\% in Fig.~\ref{fig:rgamma_dev_from_pqcd}, so the fits were used as they are continuous functions. Additionally this error is quite small with respect to the systematic error taken at high \pt, so a 1\% difference is negligible. 

\begin{figure}[thb]
	\centering
	\includegraphics[width=0.6\textwidth]{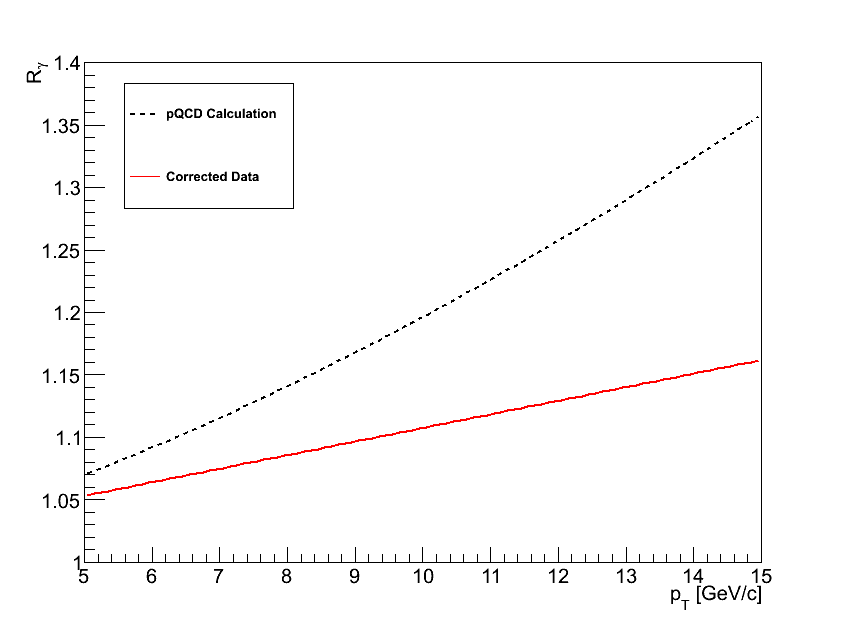}
	\caption{The \pt dependence of \rgamma is shown for both methods, as determined from a fit to data and as determined from perturbative calculations. The difference in the two methods is taken as systematic uncertainty.}
	\label{fig:rgamma_calcs}
\end{figure}

The calculations from the data driven method and from the perturbative calculations are shown in Fig.~\ref{fig:rgamma_calcs}. The resulting \rgamma data points were fit with a line at high \pt to determine the \pt dependence as shown in the figure. Systematic uncertainty in the determination of \rgamma is taken as the difference between the two methods, which is less than 20\% at the highest \pt observed in this analysis. This is consistent with percent errors that were taken for previous \rgamma results in PHENIX analysis, for example in Refs.~\cite{ppg090,ppg095}. \par

\begin{figure}[thb]
	\centering
	\includegraphics[width=0.7\textwidth]{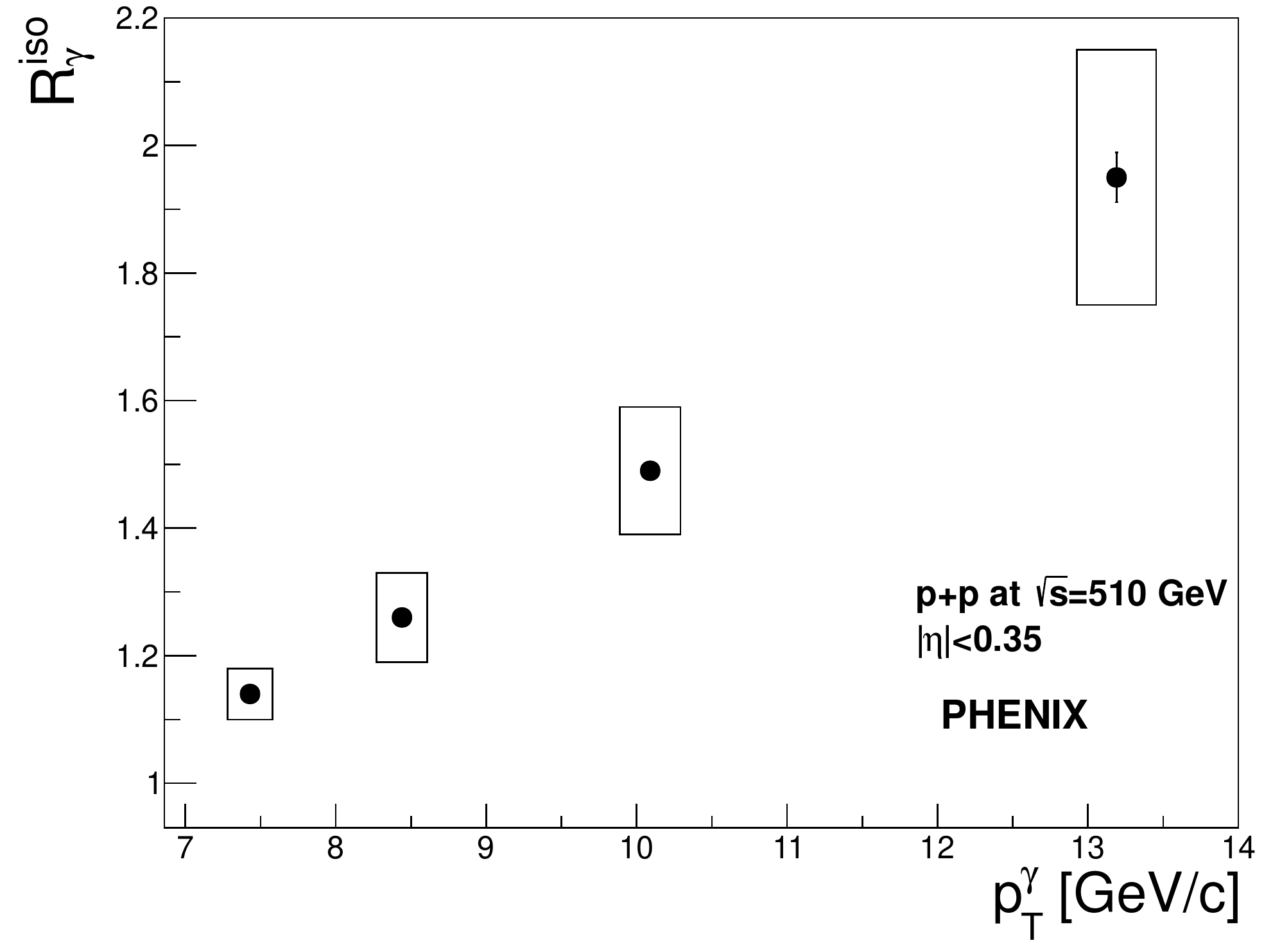}
	\caption{The values of \rgammaprime are shown in four \pt bins for \pp at \sqs=~510 GeV.}
	\label{fig:rgammaprime_pp}
\end{figure}

To determine the modified \rgammaprime due to the tagging and isolation cuts, the values of \rgamma must be corrected for tagging and isolation efficiencies as described in Section~\ref{decay_sub_method}. To determine \rgammaprime, one can derive an expression from the definition given in Eq.~\ref{eq:rgammaprimedef} that can be determined with tagging and isolation efficiencies. This derivation is shown in full in Appendix~\ref{app:rgammaprime_derivation} and was originally determined in the analysis corresponding to Ref.~\cite{ppg095}. The derivation shows that \rgammaprime can be determined with the following equation, where each value is dependent only on the \pttrig of the photon:

\begin{equation}
	\rgammaprime = \rgamma\frac{\alpha_{\rm inc}^{\rm miss, iso}}{(1-\epsilon^{\rm niso}_{\rm dec})(1-\epsilon^{\rm tag}_{\rm dec})}\,.
\end{equation}
Here the values in the fraction are the associated corrections that account for the tagging and isolation cuts and their associated efficiencies. They are defined as
\begin{equation}
	\alpha_{\rm inc}^{\rm miss, iso} = \frac{N_{\rm inc}-N_{\rm dec}^{\rm tag}-N_{\rm inc}^{\rm niso}}{N_{\rm inc}}\,,
\end{equation}
\begin{equation}
	\epsilon_{\rm dec}^{\rm niso} = \left(1+\frac{\sum_{\pi^0} \mapfxn\times N_{\pi^0}^{\rm iso}}{\sum_{\pi^0}\mapfxn\times N_{\pi^0}^{\rm niso}}\right)^{-1}\,,
\end{equation}
and
\begin{equation}
	\epsilondec = \frac{N_{\rm dec}^{\rm tag}}{N_{\rm inc}}\rgamma\,.
\end{equation}
Here \alphamiss is simply the number of photons that pass the isolation and tagging cuts divided by the total number of inclusive photons. \epsilonniso is the isolation efficiency and quantifies the efficiency with which the isolation cut removes decay photons. The Monte Carlo probability functions are used to map the effect from the parent mesons to the daughter photons. \epsilondec is the efficiency with which the tagging cut removes tagged \pion decays. Each efficiency correction is found by counting the number of photons that pass the various cuts. In the case of the isolation efficiency the neutral pions are counted rather than photons, since the probability functions map the parent meson to the daughter photon kinematics. The \rgamma values used are taken from data, with an uncertainty assigned by the difference in the data and pQCD values.  The resulting \rgammaprime values are shown in Fig.~\ref{fig:rgammaprime_pp}. Table~\ref{tab:rgammaprime13} shows the values of \rgammaprime in addition to the inputs used to calculate them. It is clear from the table that the isolation and tagging cuts raise the signal-to-background of direct photons to decay photons as the \rgammaprime values for \pttrig$>$7 \gev are all larger than the corresponding \rgamma values.  \par

\begin{table}
\begin{center}
\begin{tabular}{ |c|c|c|c|c|c|c|c| }
	\hline
	$p_T$ [GeV] & \rgamma & $\Delta\rgamma$ & \alphamiss & \epsilondec & \epsilonniso & $\rgammaprime$ & $\Delta\rgammaprime$ \\
	\hline
	5-6 & 1.06 & 0.02 & 0.268 & 0.294 & 0.589 & 0.975 & 0.02 \\
	6-7 & 1.07 & 0.03 & 0.264 & 0.321 & 0.604 & 1.04 & 0.03\\
	7-8 & 1.08 & 0.04 & 0.262 & 0.361 & 0.618 & 1.14 & 0.04 \\
	8-9 & 1.09 & 0.06 & 0.263 & 0.392 & 0.633 & 1.26 & 0.07 \\
	9-12 & 1.11 & 0.08 & 0.276 & 0.423 & 0.655 & 1.49 & 0.11 \\
	12-15 & 1.14 & 0.12 & 0.322 & 0.43 & 0.686 & 1.95  & 0.21\\
	\hline

\end{tabular}
\caption{$R_\gamma$ values from data and associated corrections for calculating $\rgammaprime$ in smaller $\pttrig$ bins.}
\label{tab:rgammaprime13}
\end{center}
\end{table}

One comment that should be made is that, in the 5-6 GeV/c \pttrig bin, \rgammaprime was determined to be less than 1. This is, of course, unphysical, and there is some additional discussion concerning this in the systematic uncertainties. The reason that this occurs is due to the tagging and isolation efficiencies being particularly small. With this method, at \sqs=~510 GeV, the ability of PHENIX to tag neutral pions is not sufficient to boost the signal of direct photons. For this reason this bin is excluded from any further analysis. Although the 6-7 GeV/c \pttrig bin is greater than 1, it is only 1$\sigma$ larger than unity. This bin is also excluded from analysis since there is a singularity in determining the final isolated direct photon PTYs at \rgammaprime=1 as seen in Eq.~\ref{eq:isostatsub}. Since this particular bin is quite close to this singularity, it is also removed from subsequent analysis. \par

In the run-15 \pa data, direct photon-hadron correlations were only collected for \pau collisions due to lack of statistical precision in \pal. The \rgammaprime used in these collisions was taken from the measured \rgammaprime from a previous preliminary \sqs=~200 GeV d+Au analysis. While the collision system is different, the difference in the production of isolated direct photons in d+Au and \pau collisions should be quite small, and certainly within the already large uncertainties assigned to \rgammaprime. \par

\begin{figure}[tbh]
	\centering
	\includegraphics[width=0.6\textwidth]{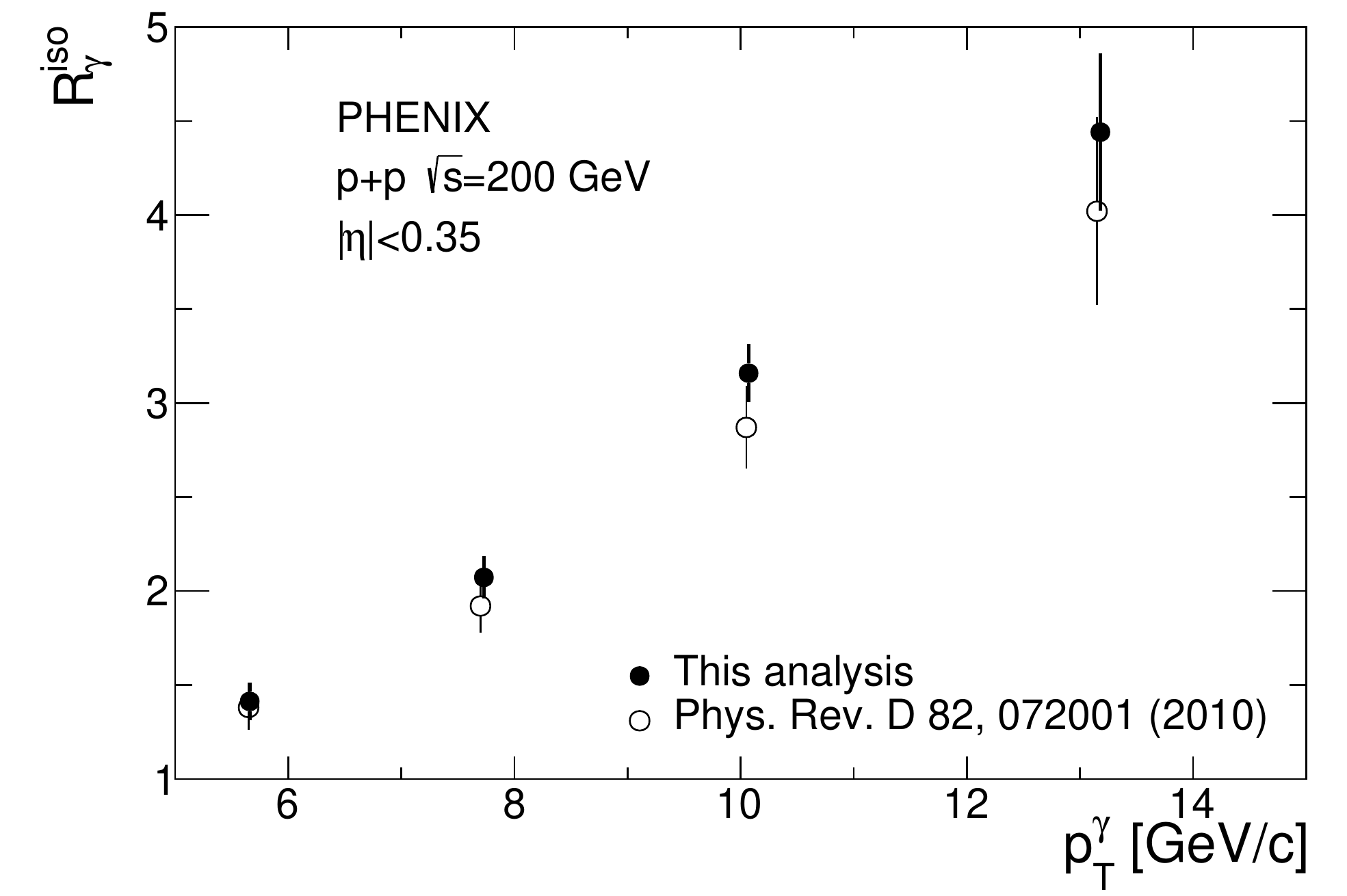}
	\caption{A comparison of the values of \rgammaprime measured in run-15 compared to the published values in Ref.~\cite{ppg095}.}
	\label{fig:rgammaprime_run15}
\end{figure}

The values of \rgammaprime in run-15 \pp could be computed directly from the published values of \rgamma~\cite{ppg095} and the corresponding efficiencies in run-15. This value is a physics quantity comparing the isolated direct photon to isolated decay photon cross sections, so we can also check that the efficiencies in run-15 are correct by comparing the values of \rgammaprime in run-15 to the previous measurement in Ref.~\cite{ppg095}. Figure~\ref{fig:rgammaprime_run15} shows this comparison between the previously published values and those measured in run-15. The values show good agreement, with slightly reduced uncertainties from this analysis due to the larger statistical sample taken in run-15 compared to runs 5 and 6. \par

\section{Systematic Checks and Uncertainties}

In this analysis there are three sources of systematic error on the per-trigger yields: the charged hadron normalization, the statistical subtraction decay method, and $\rgammaprime$. There is an additional uncertainty due to the underlying event subtraction that only applies to the \pout per-trigger yields. The charged hadron efficiency systematic applies to the normalization of the PTYs, and the statistical subtraction and \rgammaprime uncertainties apply only to the direct photon-hadron PTYs. For the direct photons, the uncertainty on \rgammaprime dominates due to the methods in which it was estimated in addition to the size of the uncertainties on \rgamma. The charged hadron, \rgammaprime, and underlying event uncertainties are described in greater detail in several forthcoming sections, while the uncertainty on the statistical subtraction decay method is described below. A brief table summarizing the various uncertainties and their relative contributions is shown in Table~\ref{tab:sysunc}. \par

\begin{center}
\begin{table}[tbh]
\begin{tabular}{|c|c|c|c|c|}
\hline
Sys. Unc. Type & \pion-\h & $\gamma$-\h & Relative Contribution [\%] & Type \\
\hline
Hadron Norm & \checkmark & \checkmark & 9-11 & C \\
Underlying Event Subtraction & \checkmark & \checkmark & 0.1-9 & B \\
Decay Photon Subtraction & $\times$ & \checkmark & 4 & C \\
\rgammaprime & $\times$ & \checkmark & 5-22 & B \\
\hline
\end{tabular}
\caption{A summary table of the systematic uncertainties. The relative contributions are shown for each of the four types of uncertainties, which are in percent of the total per-trigger yield and may depend on the data set. The table also identifies whether or not the per-trigger yields apply to the dihadron and/or direct photon-hadron correlations. Type C uncertainties are global normalization uncertainties, while type B uncertainties are point-to-point.}
\label{tab:sysunc}
\end{table}
\end{center}

Previous analyses quote a 3\% uncertainty on the statistical subtraction decay method as a whole~\cite{ppg090,ppg095}. This includes efficiencies used to determine \rgammaprime. This uncertainty largely accounts for any other decay photons that were not statistically subtracted out. Since this analysis does not include any subtraction of $\eta$-\h pairs, this uncertainty must be altered to reflect the extra contribution from isolated decay eta meson-hadron pairs. To account for this additional uncertainty, a 4\% uncertainty is ascribed to the method. This was estimated with the following: the branching ratio of $\eta\rightarrow\gamma\gamma$ is 40\%, and the ratio of $\eta$ mesons to $\pion$ mesons is about 40\%~\cite{Adare:2010cy}, which is reasonably constant as a function of $\sqrt{s}$. Therefore, 40\% of 40\% is 16\%, and 16\% of the original 3\% uncertainty ascribed to the statistical method is about 0.5\%. To be conservative, since this analysis only considers the $\eta\rightarrow\gamma\gamma$ decay, the uncertainty on the method is increased to 4\%. \par

The systematic uncertainty on the actual values of \pout and \dphi from the bin widths is different for each of the observables due to the various inputs required for their calculation. For $\pout$, the points in each bin are found by a weighted average of the inclusive $\pout$ distribution. No correction was applied for momentum smearing from the detector, so a 4\% systematic is assigned to account for this. This accounts for PHENIX's resolution in $\pout$, which is due to the inherent detector resolution of $\ptassoc$ and $\dphi$. In $\dphi$, the only systematic that would apply is a resolution on $\dphi$ and an error for the bin width. The resolution on $\dphi$ is estimated by the resolution of the EMCal and DC $\phi$, which is 1\% and 2\% respectively. The points are plotted at the center of the bin with a 3\% systematic uncertainty on \dphi from the inherent $\phi$ resolution of the EMCal and DC. Since the yield is not logarithmically changing as a function of \dphi, a weighted average was not used to determine the placement of the \dphi values like what was used for \pout. \par

\subsection{Charged Hadron Normalization Uncertainty}

Previous PHENIX charged hadron uncertainties on cross sections range from 8-12\%, depending on the collision system, DC performance, and other factors.  The uncertainty on the charged hadrons is due to the momentum scale, the averaging of all particle identified efficiencies into one unidentified efficiency, and the PC3 track matching, which is by far the dominant source of uncertainty. The momentum scale has a 1\% uncertainty as determined by the measurement of the proton and J$/\psi$ mass in the PHENIX tracking system, with the aid of the time-of-flight detectors. In Ref.~\cite{ppg095} the unidentified hadron efficiency was determined by averaging the identified hadron efficiencies weighted by their respective production cross sections. For the $\sqrt{s}=$510 GeV data no production ratios of $K/\pi$ or $p/\pi$ were available, so the identified hadron efficiencies were averaged without weights. Ref.~\cite{ppg095} cited a 3\% conservative uncertainty for the ratios, so here I assign a 4\% uncertainty. The run-15 \pp and \pa analysis also cite a 3\% uncertainty on the ratios since the particle ID dependence was evaluated in the same way as Ref.~\cite{ppg095}. \par

\begin{figure}[thb]
	\centering
	\includegraphics[width=0.7\textwidth]{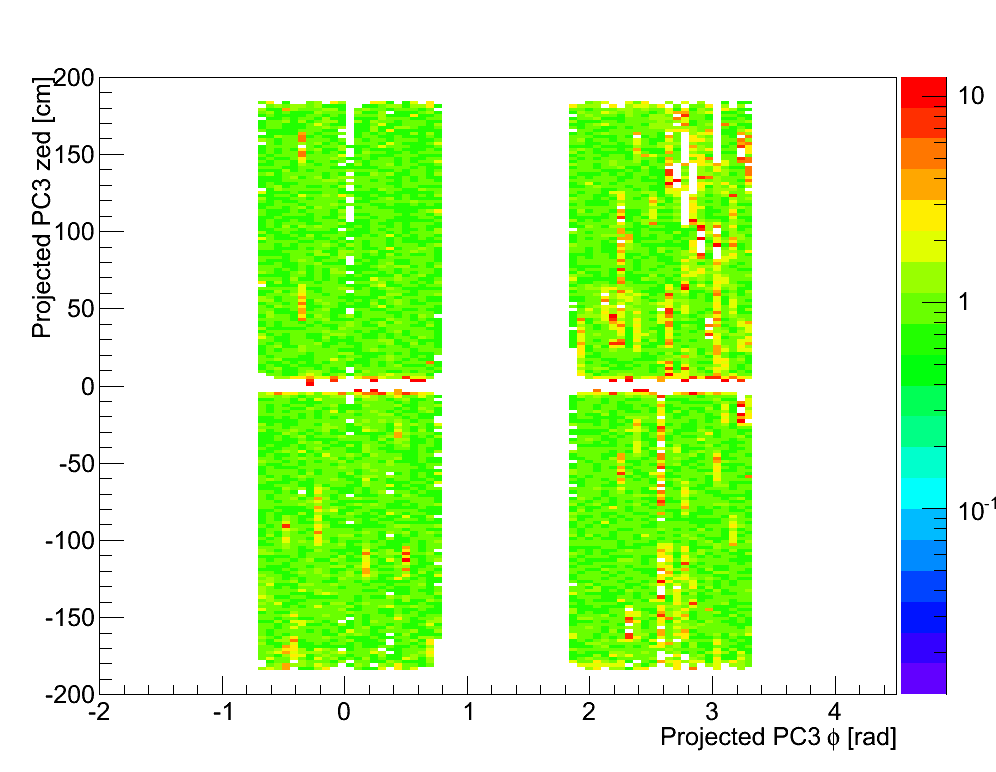}
	\caption{The track matching uncertainty from the PC3 is estimated as described in the text. 8 regions, 2 for each quadrant of the PC3, were selected to characterize possible inefficiencies of the track matching. A 7\% systematic uncertainty is assigned, similar to~\cite{ppg095}.}
	\label{fig:pc3matchtrack}
\end{figure}

The dominant source of the charged hadron uncertainty is the track matching uncertainty, which arises due to the matching of DC tracks to the outermost pad chamber, the PC3. Ref~\cite{ppg095} cited a 7\% uncertainty for the track matching, and to check that this was still valid the method for which this uncertainty was estimated was replicated. The example shown here is from the \pa running since this was the later run period, thus it in principle would show the largest deviation due to additional dead channels produced over the course of the many PHENIX data taking periods. The track matching $2\sigma$ cut was lifted from the reconstruction and the ratio of the matched tracks to unmatched tracks was taken as a function of the projected PC3 $\phi$ and zed. Note that each histogram was normalized by the number of tracks to make an apples-to-apples comparison. Figure~\ref{fig:pc3matchtrack} shows the resulting divided histograms. Naturally the majority of the PC3 is at 1, i.e. the track matching cut does not make a substantive difference. To estimate an error, one region in each quadrant was chosen to compare to the entire PC3. Each region was chosen to be in an area of the quadrant that is relatively stable for that quadrant, i.e. the region does not contain large fluctuating areas. The average difference was calculated in each of the 8 regions, and this number was compared to the average calculated from the entire PC3 matching. These ratios were calculated to be (one for each of the 8 regions) 0.90, 0.93, 0.93, 0.92, 0.93, 0.92, 0.93, 1.01. These values were averaged to get 7\%, which is the same value assigned to the track matching uncertainty in~\cite{ppg095}. The similar track matching efficiency is not surprising as the averages are taken from stable regions in the PC3, so the only differences between run-15 and previous PHENIX runs would be any new dead areas in the PC3. Ultimately this uncertainty is based on the efficiency of the PC3 track matching algorithm, so in fact the consistency with previous runs is a good check that the uncertainty is correct given that the track matching algorithm should not have changed significantly over time. \par

In total for the run-13 \pp data an uncertainty of 9\% was assigned on the charged hadron yields, and the run-15 \pp and \pa data was assigned an 8\% systematic uncertainty. The reduced uncertainty in run-15 is largely due to the availability of the particle ID ratios to determine the charged hadron efficiency. This uncertainty is an overall normalization uncertainty, so it affects each point on the PTYs equally.

\subsection{\rgamma and \rgammaprime}

For the run-13 \pp collisions, the error from \rgamma is propagated to the error on \rgammaprime, where the error on \rgamma is estimated by taking the average difference of the two methods described in section~\ref{rgammasection}. The actual value of \rgamma is taken as measured from data, and then the error on \rgamma is taken as the upper bound given by the difference between the measurement and the pQCD calculation. The final values of \rgammaprime as shown in Fig.~\ref{fig:rgammaprime_pp} are tabulated in Tab.~\ref{tab:rgammaprime13}. As commented above, the value of \rgammaprime in the 5-6 GeV/c bin is unphysical, and thus it is not included in any further analysis. The value of \rgammaprime in the 6-7 GeV/c bin is only $\sim$1.5$\sigma$ away from a singularity, and thus any further results with this bin were also discarded. As the value of \rgammaprime is greater than 3$\sigma$ away from the singularity in the 7-8 GeV/c bin, this bin and subsequently larger \pt bins were analyzed further. \par

\begin{figure}[thb]
	\centering
	\includegraphics[scale=0.4]{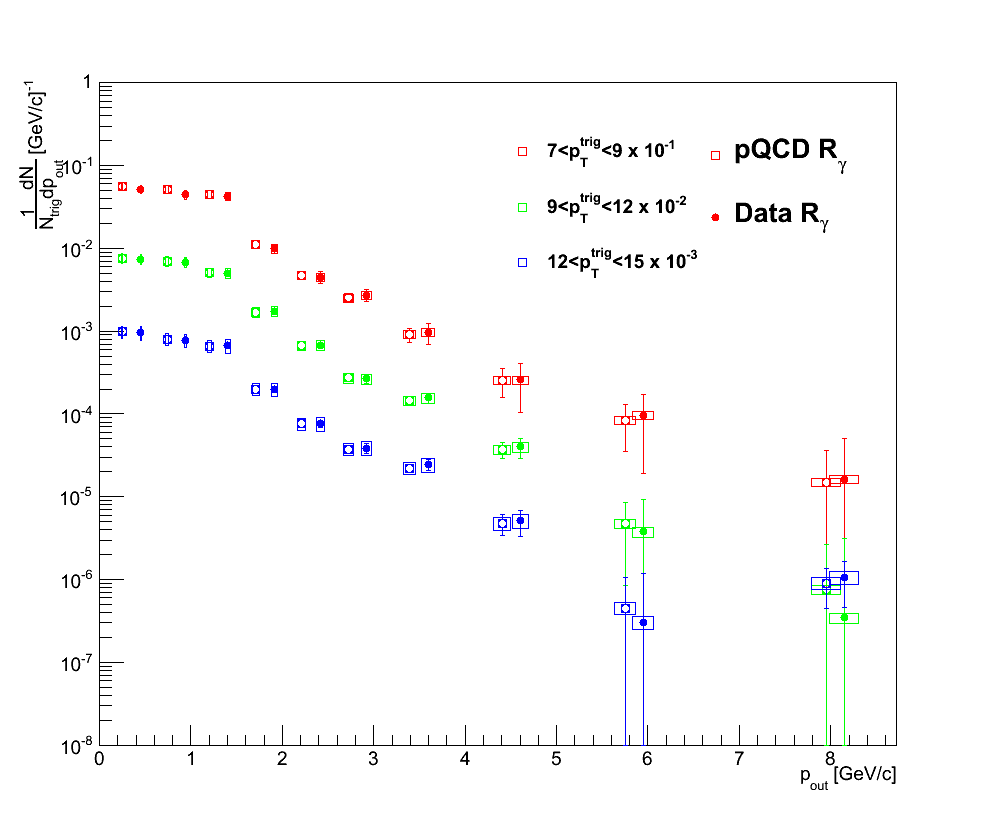}
	\caption{A comparison of the isolated direct photon \pout distributions with both the pQCD calculated \rgamma (filled circles) and the data calculated \rgamma (open circles) is shown. The filled circles are shifted by \pout=~0.2 for visibility sake.}
	\label{fig:rgammacomppout}
\end{figure}

\begin{figure}[thb]
	\centering
	\includegraphics[scale=0.4]{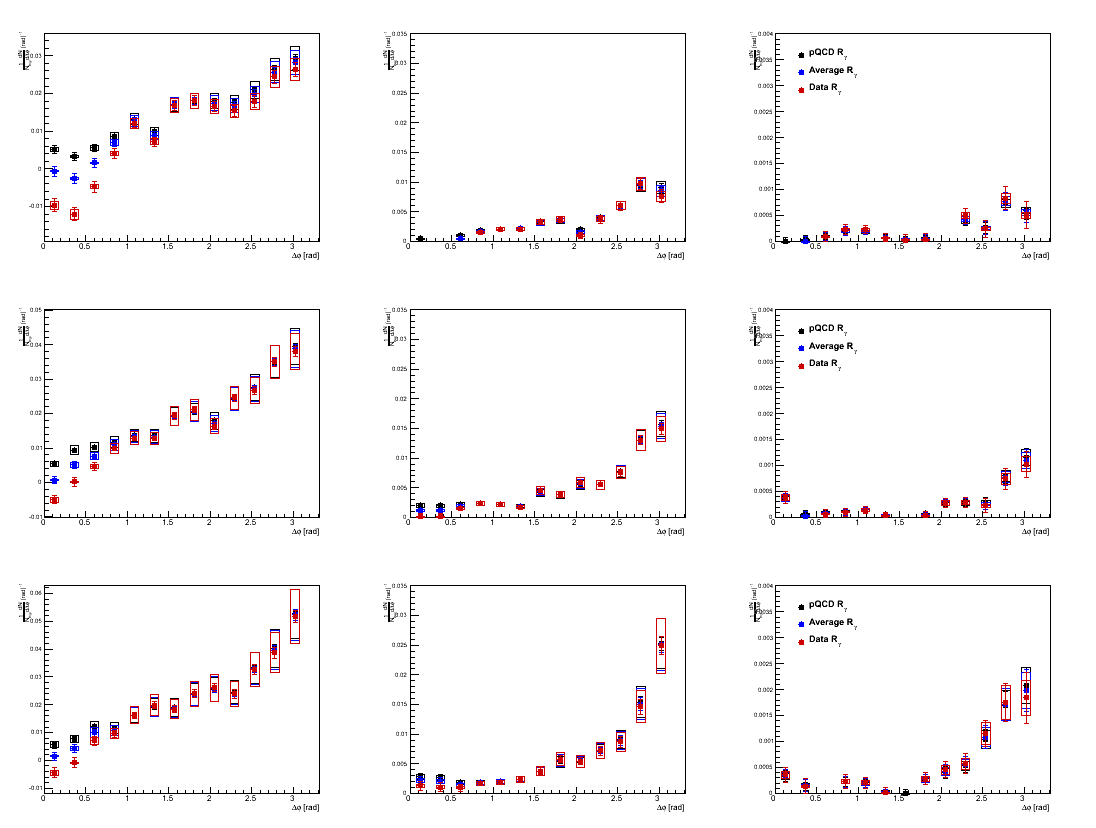}
	\caption{A comparison of the isolated direct photon \dphi distributions with all three sets of \rgamma is shown.}
	\label{fig:rgammacompdphi}
\end{figure}

To show that taking the uncertainty in \rgamma is a conservative estimate, the \pout and \dphi distributions for isolated direct photons were made for three different sets of \rgamma: \rgamma from the pQCD calculation, the average \rgamma between the pQCD calculation and data, and the data determined \rgamma. The observables are plotted together in Figs.~\ref{fig:rgammacomppout} and~\ref{fig:rgammacompdphi} to show that the different sets of \rgamma do not change the observables within the quoted systematic uncertainties. The largest difference can be seen in the low \ptassoc bins, at small \dphi. This is not a cause for concern as these differences are within the isolation cone and it is already difficult to physically interpret this region of \dphi space. In addition the near side points are omitted from the final plots anyway as is commonly done in the literature, due to this region being physically uninterpretable. The important point is that the away side yields do not change significantly within the assigned systematic uncertainties.


The decay photon tagging cut introduces another uncertainty into $\rgammaprime$ due to the possibility of the cut altering both $\epsilon_{niso}^{dec}$ and $\epsilon_{tag}^{dec}$. There exists the possibility that a direct photon could be falsely tagged as a decay photon, which would thus cause $\rgammaprime$ to be overestimated due to $\epsilon_{tag}^{dec}$ being overestimated. $\epsilon_{niso}^{dec}$ could be over- or under-estimated depending on if the direct photon was isolated or not. In Ref.~\cite{ppg095} this added systematic uncertainty was calculated and found to be very small in size relative to the uncertainty due to $\rgamma$. Therefore this is neglected here since the systematic uncertainty from the estimate of $\rgamma$ is already so large. At most this effect altered the systematic uncertainties by 1.5\% at high $\pttrig$ in Ref.~\cite{ppg095}, so this will be negligible here compared to the uncertainty assigned on the determination of \rgamma and \rgammaprime. \par

The uncertainty in \rgammaprime for \sqs=~200 GeV is propagated from the uncertainty in \rgamma from Ref.~\cite{ppg095} in addition to the uncertainties in the various efficiency factors used to calculate \rgammaprime. Because of the improved statistical significance of run-15 compared to runs 5 and 6, these uncertainties are reduced; however, the dominant uncertainty comes from the determination of \rgamma in Ref.~\cite{ppg095}. This uncertainty, when propagated, contributes the most; however, the reduced uncertainties in the various efficiency factors does shrink the overall uncertainties on \rgammaprime at \sqs=~200 GeV slightly when compared to Ref.~\cite{ppg095}.

\subsection{Underlying Event Statistical Subtraction for $\pout$}

\begin{figure}[tbh]
	\centering
	\includegraphics[scale=0.4]{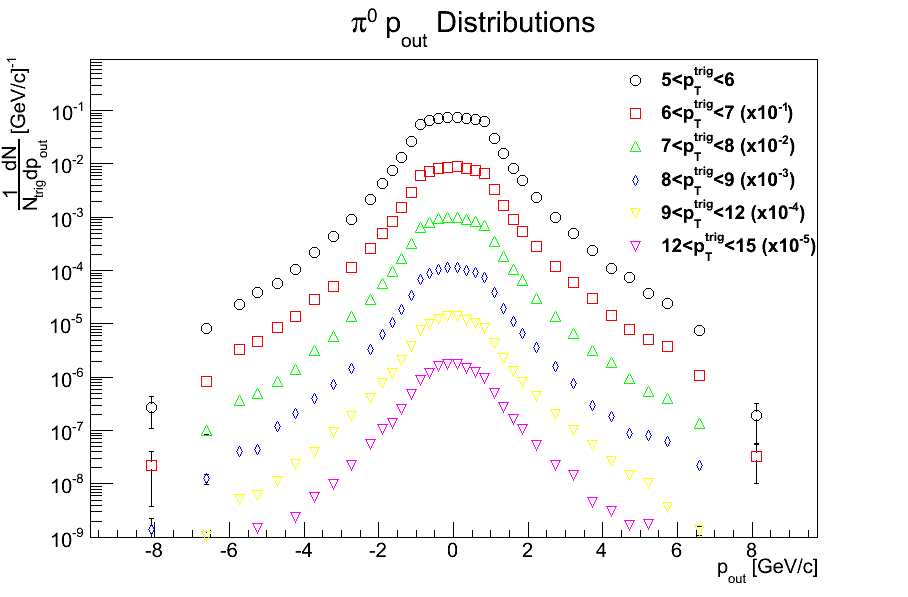}
	\caption{The $\pion$ $\pout$ distributions in run-13 are shown before the underlying event background subtraction. The underlying event contribution is evident in the ``corners'' that appear at small \pout.}
	\label{fig:pi0poutscorners}
\end{figure}

\begin{figure}[tbh]
	\centering
	\includegraphics[width=0.49\textwidth]{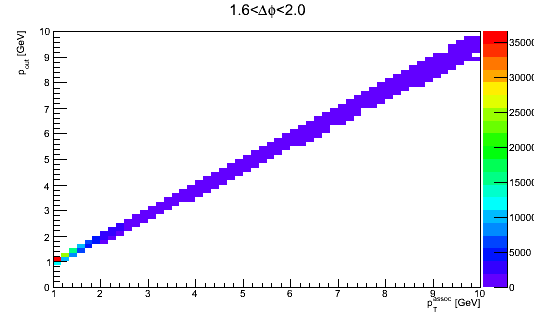}
	\includegraphics[scale=0.4]{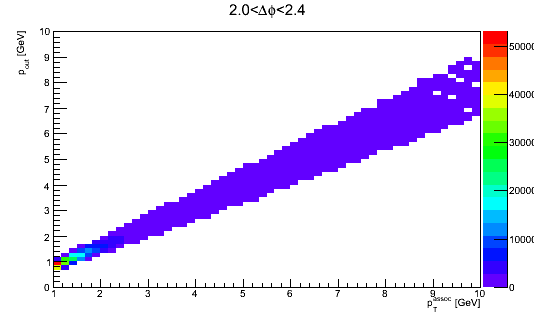}
	\caption{$\pout$ as a function of $\ptassoc$ in the region $1.6<\dphi<2.0$ radians (left) and $2<\dphi<2.4$ radians (right). The perturbative tail of $\pout$ is generated in this region from hard gluon radiation, but there is an excess of counts at 1 \gevc of $\pout$ and $\ptassoc$ from the underlying event.}
	\label{fig:pionpout1}
\end{figure}

\begin{figure}[tbh]
	\centering
	\includegraphics[width=0.49\textwidth]{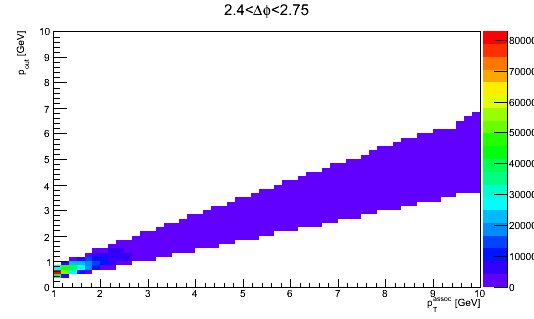}
	\includegraphics[width=0.49\textwidth]{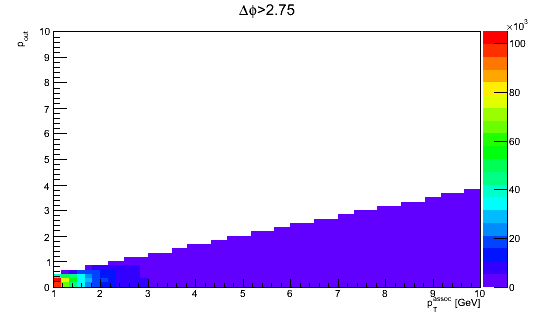}
	\caption{$\pout$ as a function of $\ptassoc$ in the region $2.4<\dphi<2.8$ radians (left) and $2.8<\dphi<\pi$ radians (right). }
	\label{fig:pionpout3}
\end{figure}

It was discovered that the \pout distributions suffer from background due to low \pt uncorrelated hadrons that are mostly in the underlying event region, i.e. close to \dphi$\sim\pi/2$. This can be observed in Fig.~\ref{fig:pi0poutscorners}, where there appear to be ``corners" in the distributions rather than a Gaussian behavior. Note that this behavior seems to disappear as the \pion \pt increases; this is because the signal to underlying event background ratio rises as the \pion \pt increases. Since this analysis is focused on characterizing the Gaussian behavior of $\pout$ at small values of $\pout$, this background needed to be investigated further. To determine the cause of this behavior, $\pout$ was plotted as a function of $\ptassoc$ in bins of $\dphi$ for $\pion$ triggers. The resulting plots are shown in Figs~\ref{fig:pionpout1} and~\ref{fig:pionpout3}, where the folded $\dphi$ bins are noted in the title of each plot. The plots show that the nonperturbatively generated $\pout$ from $k_T$ is mostly in the nearly back to back region, while in the underlying event region ($1.6<\dphi< 2.3$) there is a perturbatively generated tail at large $\pout$ but also an excess of counts at $\pout\approx 1$ from $\ptassoc$ hadrons of $p_T\approx 1-2$. This indicates that there is an excess of counts in the $\pout$ distributions around 1 GeV due to the underlying event, as it is most likely that the majority of 1-2 GeV hadrons produced in this $\dphi$ region are largely from the underlying event. For this reason, in order to isolate the Gaussian nature of $\pout$, a statistical subtraction of the underlying event yield was performed in constructing the $\pout$ distributions. \par

\begin{figure}[tbh]
	\centering
	\includegraphics[scale=0.6]{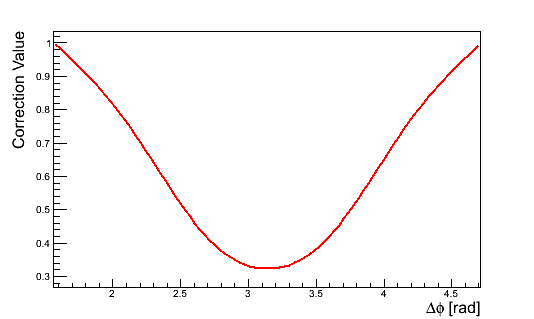}
	\caption{An example of a $\pion$ underlying event correction. These were made for all $\pttrig$ bins and all trigger types, i.e. the $\pion$ $\pout$ distributions were corrected by fits to the $\pion$ $\dphi$ correlation functions, the isolated inclusive photon $\pout$ were corrected by fits to the isolated inclusive photon $\dphi$ correlations, and so on. This example shows that the underlying event background is approximately 100\% at $\dphi\sim\pi/2$, while it is approximately 30\% at $\dphi\sim\pi$.}
\label{fig:uecorrfunc}
\end{figure}

The statistical subtraction of the underlying event was performed on the $\pout$ distributions using the fit function from Eq.~\ref{eq:rmspout} described in Section~\ref{rmspoutdetermination} to extract \rmspout. In this equation, $C_0$ parameterizes the underlying event as a constant function of $\dphi$, and the rest of the fit function characterizes the Gaussian nature of the away side jet. To do the underlying event statistical subtraction, the $\dphi$ correlations were fit with this function and the correction value was defined as the underlying event parameter divided by the total fit function. This was done in each $\pttrig\otimes\ptassoc$ bin. An example of the correction function for \pion mesons is shown in Fig.~\ref{fig:uecorrfunc}. The function shows that in the $5<\pttrig<6\otimes1<\ptassoc<2$ \pion-\h bin, at around $\pi$ radians, $\sim$30\% of the counts are from hard scattered jet structure and 50\% are from the underlying event. Similarly, at $\sim\pi/2$ radians, nearly 100\% of the jet structure is from underlying event in this particular bin. Using these functions, the yield in $\pout$ was modified as follows; the yield for $\pout$ in a given bin was altered by statistically subtracting the percentage of the yield that is from the underlying event if the associated hadron was in the range $0.7<\ptassoc<4$, since above this the yields in the underlying event are negligible. This correction was applied for $\pion$, inclusive, and decay yields and the correction function was made for each type of trigger in all $\pttrig$ bins in order to be consistent. This is effectively a Zero Yield At Minimum (ZYAM) type statistical subtraction of the underlying event commonly performed in heavy ion analyses, used only for the $\pout$ distributions. \par

\begin{figure}[tbh]
	\centering
	\includegraphics[scale=0.6]{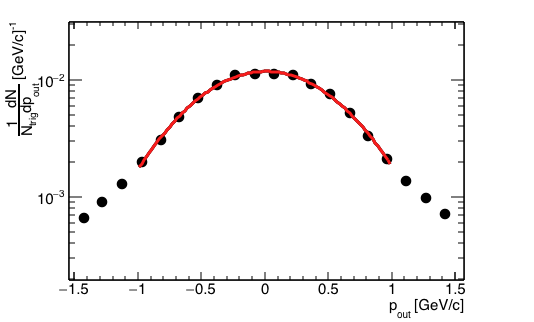}
	\caption{An example of the $\pion$ $\pout$ distributions corrected by the underlying event. For $5<\pttrig<6$ this effect was most prominent since the underlying event is a larger percentage of the overall jet structure; the correction cleans up the ``flatness" and ``corners" seen in Fig.~\ref{fig:pi0poutscorners}.}
	\label{fig:uecorrectedpi0pout}
\end{figure}

An example of the result of the underlying event subtraction on the $\pion$ $\pout$ distributions is shown in Fig.~\ref{fig:uecorrectedpi0pout}. The Gaussian behavior can be identified much better after the underlying event statistical subtraction. To be clear, the subtraction was done individually for each of the inclusive, decay, and $\pion$ triggers. This way the amount of underlying event from each trigger particle was correctly estimated. \par

Since this analysis was the first to subtract the underlying event from the \pout distributions, a systematic uncertainty needed to be assigned to take into account the method for which the underlying event was subtracted. Systematic uncertainties on the underlying event subtraction were evaluated by varying the parameter $C_0$ in the fit function by $\pm1\sigma$ and observing the resulting effect on the $\pout$ distributions. The $\pout$ distributions were constructed with the three underlying event parameters, $C_0$, $C_0+\sigma$ and $C_0-\sigma$, and then the ratio of the $\pout$ yields was taken to determine a systematic uncertainty. In other words, the ratio of the distributions is just the distribution with the underlying event varied $C_0\pm\sigma$ divided by the distribution with the underlying event at its nominal value from the fit $C_0$. The uncertainty in the $C_0$ parameter was taken from the fitting procedure. The resulting ratios are shown below for both $\pion$ and direct photon triggered correlations in run-13 \pp collisions. The difference in the distributions is on the order of tenths of a percent, which is unsurprising since the underlying event in the $\dphi$ correlations is determined to high accuracy with the statistical precision available in the \sqs=~510 GeV data. The irregularity of the direct photon ratios is due to the fact that the underlying event statistical subtraction is not applied directly to the direct photon distributions, because we do not have access to event by event direct photons since these are obtained by a statistical subtraction of decay photons. Rather, the underlying event subtraction is applied to both the isolated inclusive and isolated decay distributions, and then the isolated decay photon PTY statistical subtraction is applied to obtain the isolated direct photon PTYs. The direct photon systematic ratios do follow the expectation that they are inversely related between the $+\sigma$ and $-\sigma$ cases, i.e. the ratios should just be the inverse since the same amount of background is being added or subtracted from the nominal value $C_0$. The same behavior is seen in the $\pion$ ratios. The various systematic uncertainties here were added in quadrature point-to-point to the \pout PTYs with the other uncertainties assigned. \par

\begin{figure}[tbh]
	\centering
	\includegraphics[width=0.49\textwidth]{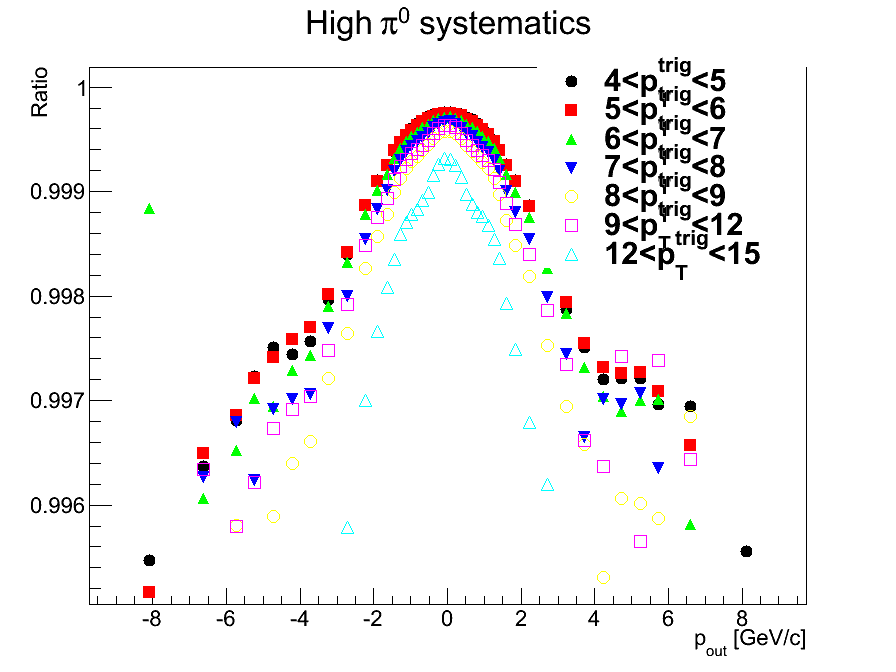}
	\includegraphics[width=0.49\textwidth]{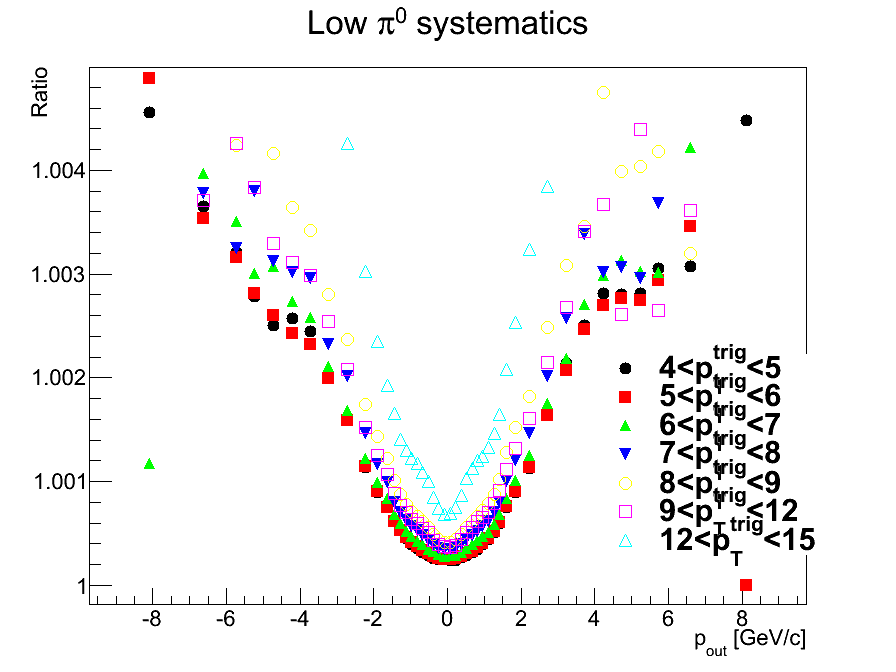}
	\caption{Systematic uncertainties on the underlying event background subtraction for $\pion$-h$^\pm$ correlations are shown for \sqs=~510 GeV \pp collisions.}
\end{figure}
\begin{figure}[tbh]
	\centering
	\includegraphics[width=0.49\textwidth]{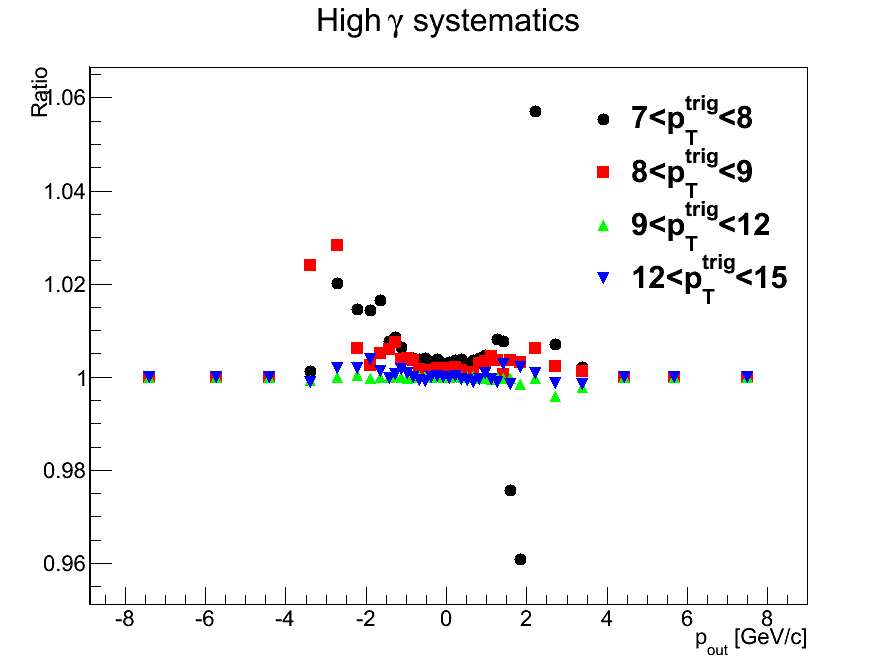}
	\includegraphics[width=0.49\textwidth]{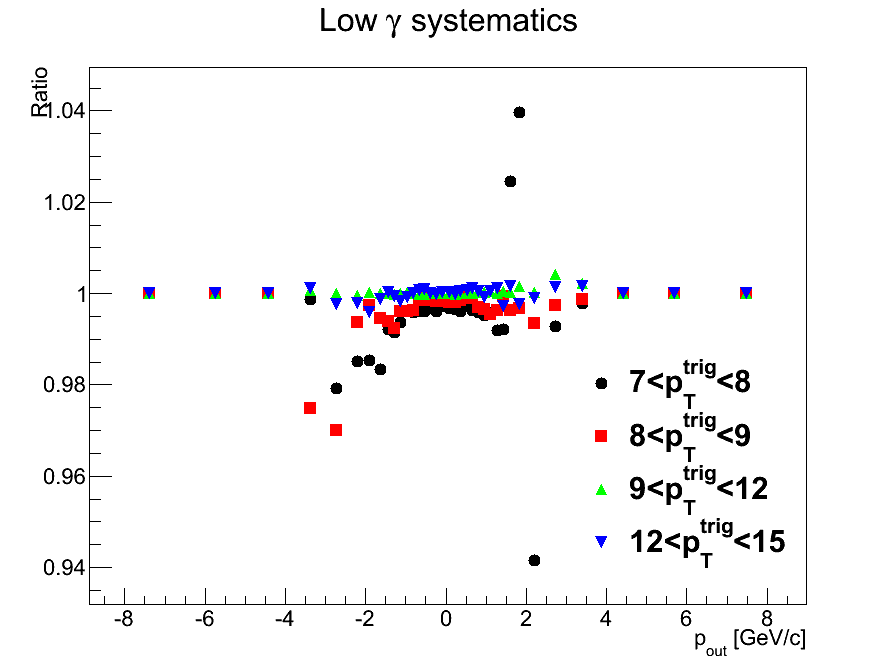}
	\caption{Systematic uncertainties on the underlying event background subtraction were evaluated as described in the text for direct photon-hadron correlations. The irregularity of the direct photon ratios is due to the fact that the underlying event statistical subtraction is not applied directly to the direct photon distributions; rather, it is applied to the inclusive and decay distributions and then the decay photon statistical subtraction is implemented to obtain the direct photon distributions. In the region of interest of small $\pout$ for this analysis, the uncertainty is on the order of tenths of a percent.}
\end{figure}

\begin{figure}[tbh]
	\centering
	\includegraphics[width=0.49\textwidth]{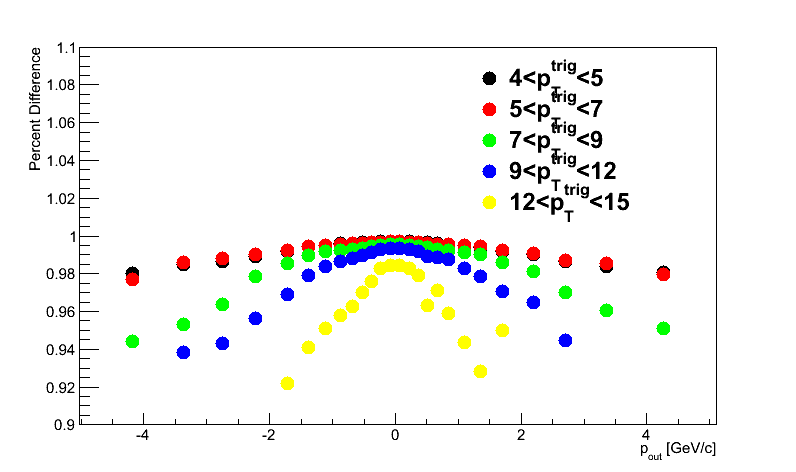}
	\includegraphics[width=0.49\textwidth]{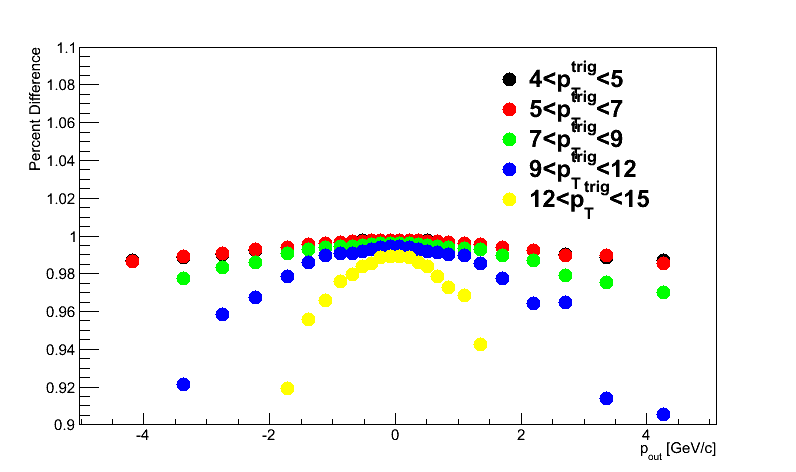}
	\caption{Systematic uncertainty for the subtraction of the underlying event for $p$+Al (left) and \pau (right) $\pout$ distributions in each $\pttrig$ bin for \pion-\h correlations.}
	\label{fig:pi0papoutuesys}
\end{figure}

\begin{figure}[tbh]
	\centering
	\includegraphics[width=0.5\textwidth]{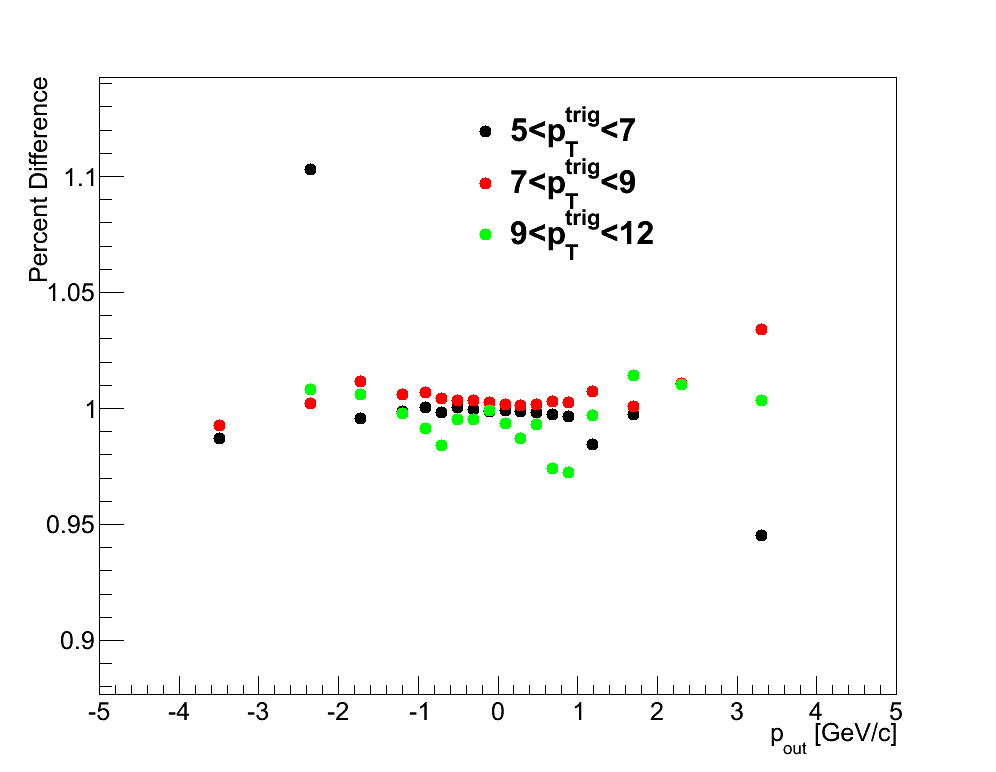}
	\caption{Systematic uncertainty applied point-to-point for the statistical subtraction of the underlying event in the isolated direct $\gamma$-h correlations in \pau. The not smooth shape is due to the statistical subtraction of decay photon-hadron pairs, and is discussed further in the text.}
	\label{fig:dppapoutuesys}
\end{figure}

The underlying event subtraction in the run-15 data was performed in the same way as described above. The systematic uncertainties were also estimated in the same way, and the corresponding uncertainties for the $p$+A systems are shown in Figs.~\ref{fig:pi0papoutuesys} and~\ref{fig:dppapoutuesys}. The systematic errors in the \pa data are larger than in the run-13 \pp data due to the data set being significantly smaller, thus statistical fluctuations in the PTYs translate to a larger uncertainty in the underlying event value, and thus a larger systematic uncertainty in the underlying event subtraction from the \pout PTYs. However, the smaller center-of-mass energy and smaller integrated luminosities in run-15 also lead to larger statistical uncertainties at large \pout in these distributions, so the larger systematic uncertainties from the underlying event subtraction are in general smaller than the statistical uncertainties at large \pout.

\chapter{\sqs=~510 GeV \pp Results}
\label{chap:ppresults}
\section{\dphi Correlation Functions}\label{ppresults}

\begin{figure}[tbh]
	\centering
	\includegraphics[width=0.49\textwidth]{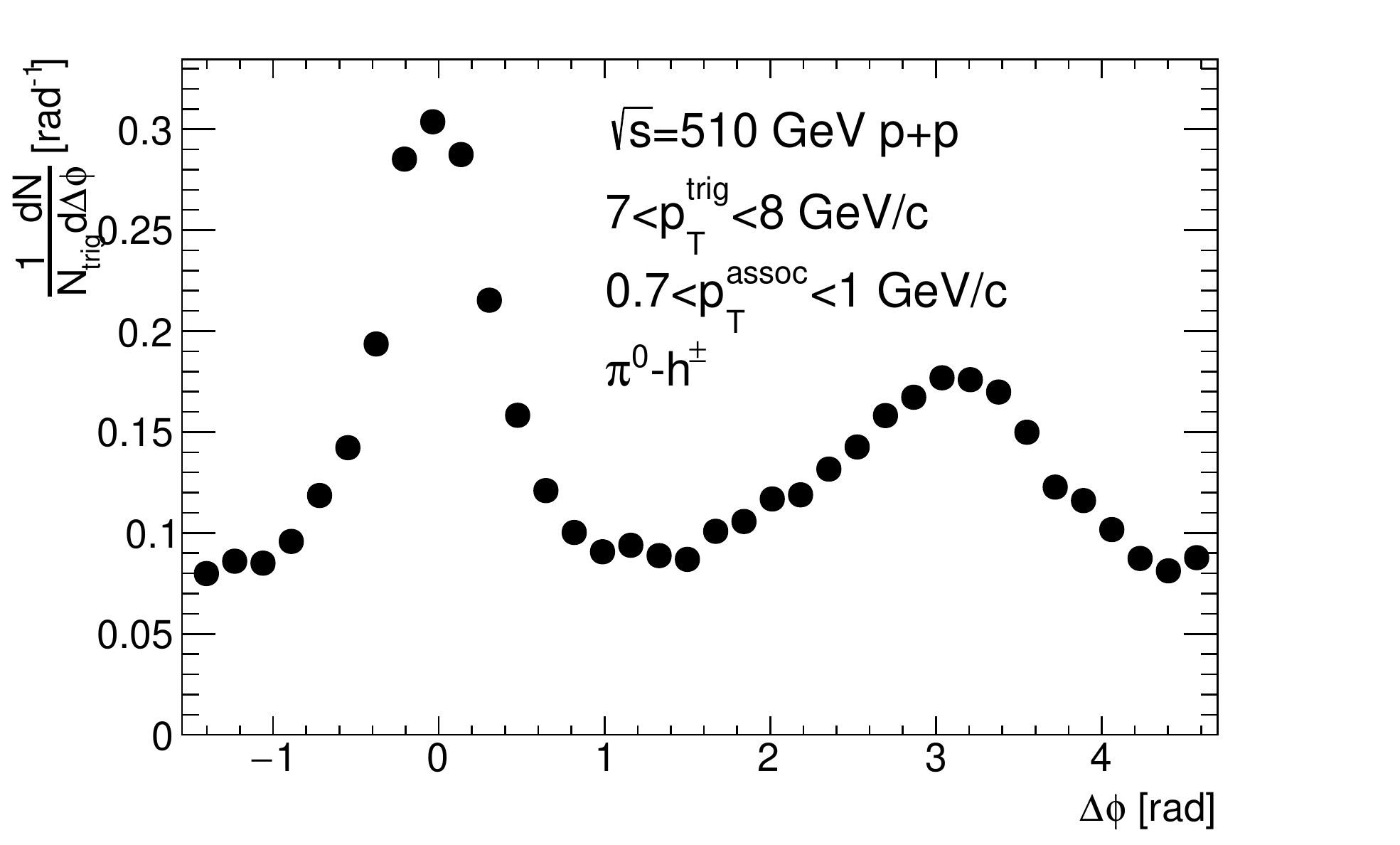}
	\includegraphics[width=0.49\textwidth]{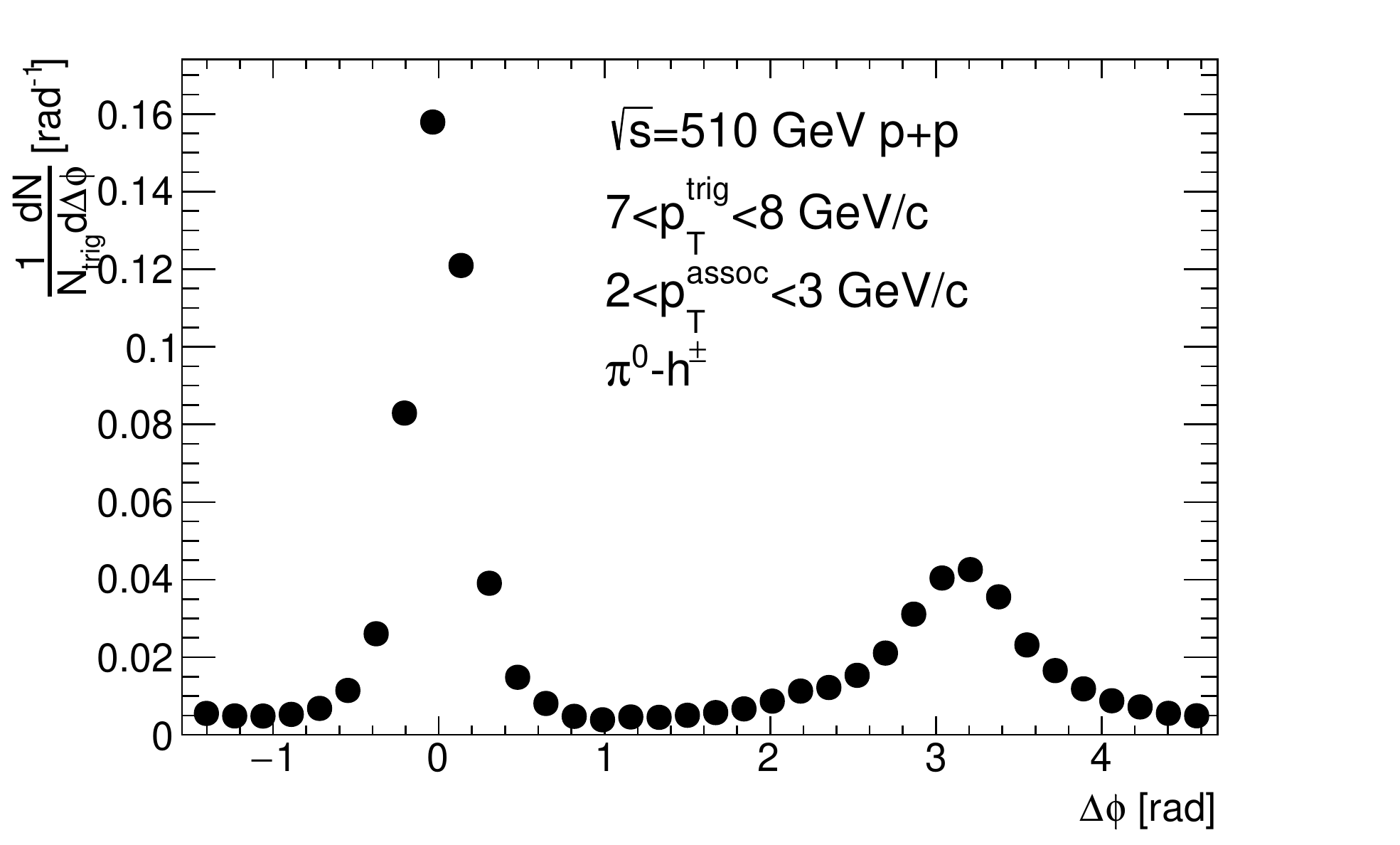}
	\includegraphics[width=0.49\textwidth]{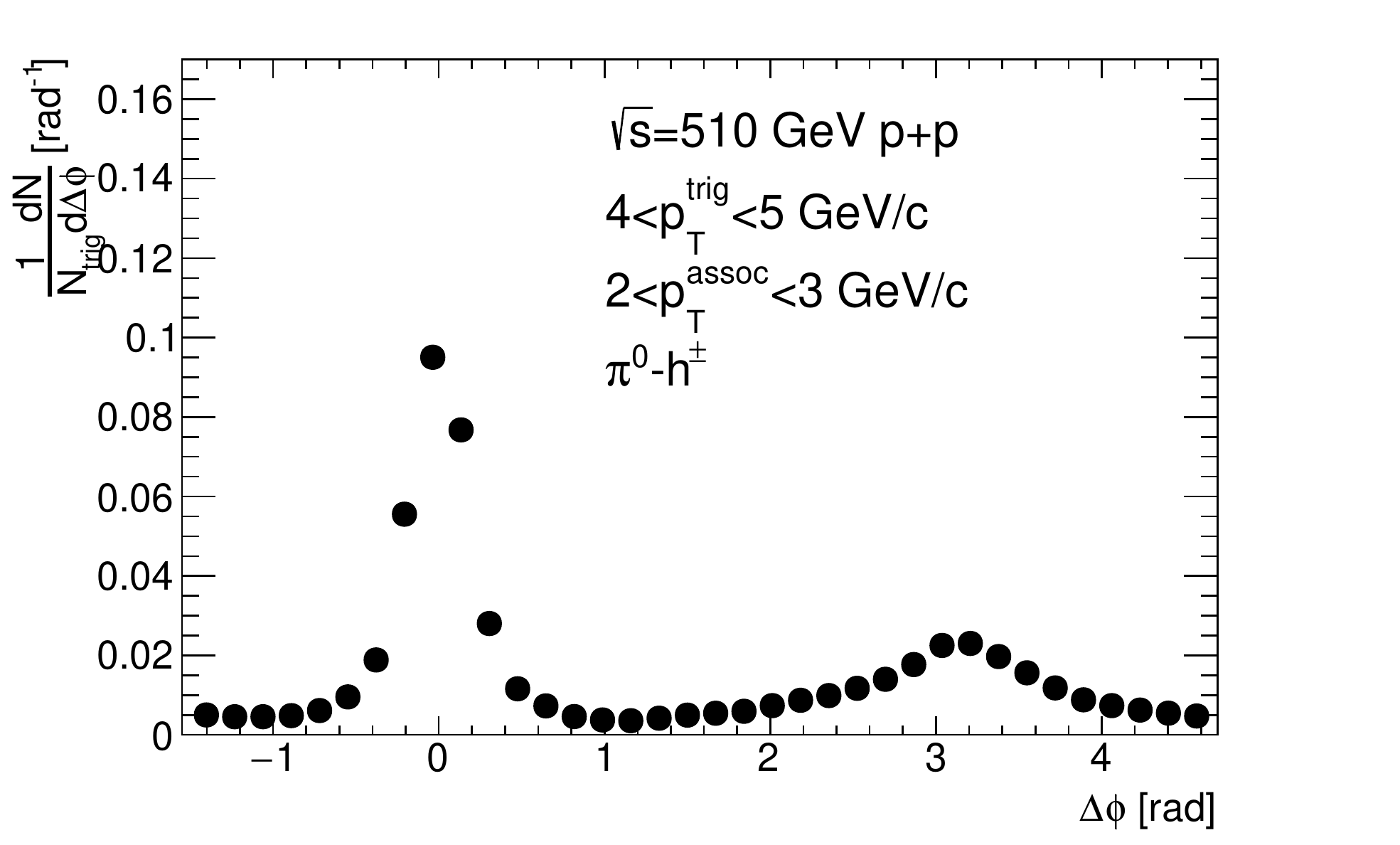}
	\caption{Some example correlation functions for \pion-\h per-trigger yields in bins of \pttrig, \ptassoc, and \dphi. The correlation functions show the expected dijet structure, with a peak at \dphi$\sim$~0 and $\pi$ radians.}
	\label{fig:pi0h_dphi_ptys}
\end{figure}

The \dphi correlation functions are visually instructive since they demonstrate the dijet and direct photon-jet structure one expects to see as a function of azimuth. Figure~\ref{fig:pi0h_dphi_ptys} shows two examples of a \pion-\h correlation function as a function of \pttrig, \ptassoc, and \dphi. The dijet structure is immediately obvious, with jet peaks at \dphi$\sim$~0 and $\pi$ radians. The underlying event region is considerably larger at small \ptassoc as would be expected from energy conservation. Additionally, the underlying event yield is largely constant as a function of \pttrig for a fixed \ptassoc bin as would be expected kinematically if the \pion is truly approximating the hard scale of the interaction and the underlying event is not associated with the hard interaction. The away-side peaks are smeared out in \dphi when compared to the near-side, indicating sensitivity to $k_T$ broadening effects. This additional significant smearing is because \kt has been measured to be significantly larger than \jt~\cite{ppg029,ppg095}, thus leading to a greater acoplanarity. \par

One comment that should be made is in regard to the size of the near-side peaks with respect to the away-side peaks. The near-side peaks are significantly larger than the away-side peaks in a given \pttrig$\otimes$\ptassoc bin due to the effect of ``trigger-bias," a poorly chosen term which refers to the inherent bias resulting from triggering on high \pt particles. When triggering, there is an intrinsic bias which results in the trigger selectively choosing very high momentum fraction $z$ hadrons within jets. This results in a higher likelihood to trigger on the higher $p_T$ jet of the dijet pair, or the jet which has the larger $k_{T_x}$ momentum imparted to it. Thus, the near-side actually samples larger $p_T$ jets on average, and thus has more yield associated with it. This is not a trigger-bias in the usual usage of the term; rather it refers to the inherent selection of high $z$ hadrons within the highest $p_T$ jet in a particular event. \par

\begin{figure}[thb]
	\centering
	\includegraphics[width=0.6\textwidth]{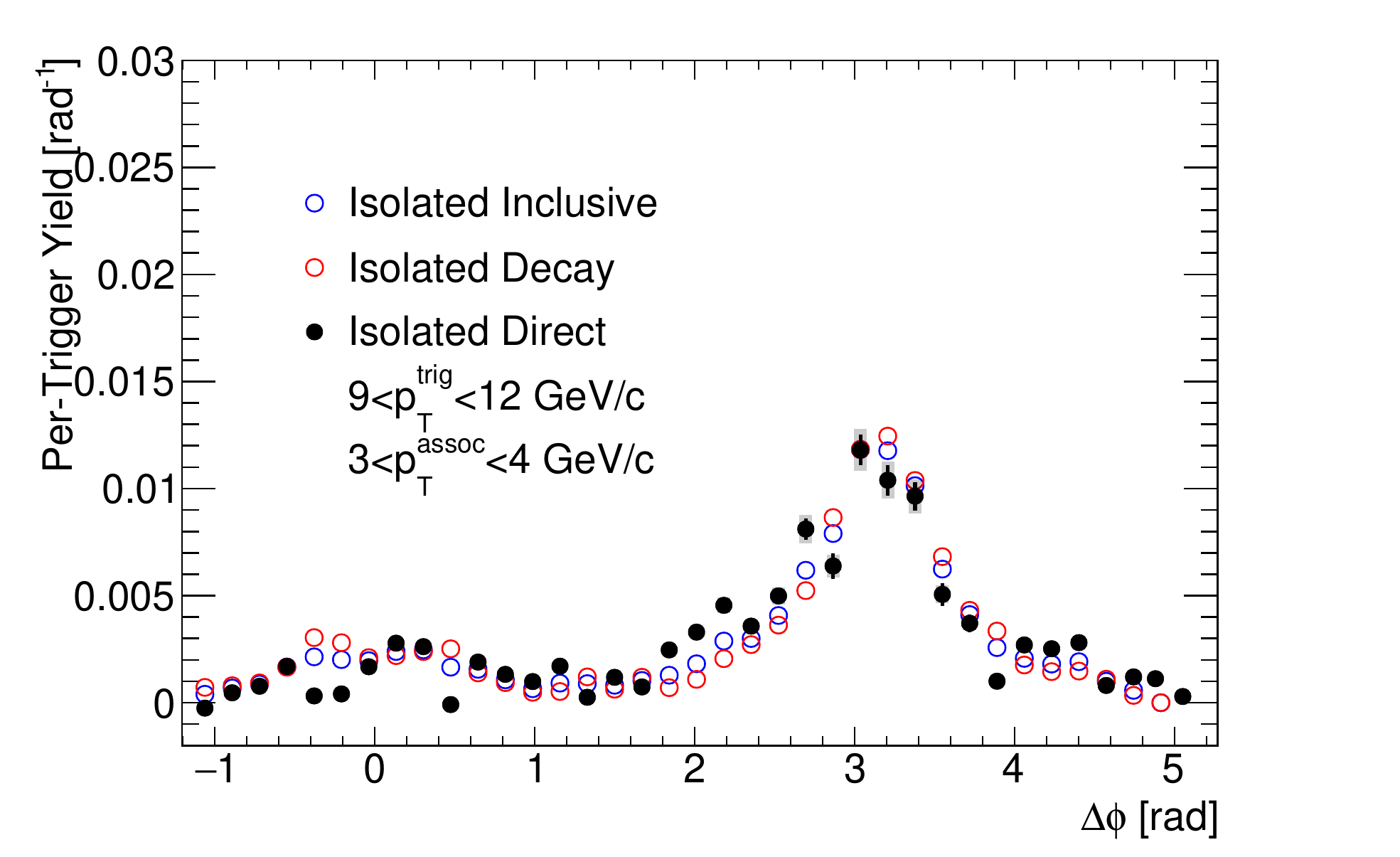}
	\caption{One example of the isolated direct photon per-trigger yields and the ingredients which were used to construct the PTY are shown. The isolated inclusive and decay per-trigger yields are used to obtain the isolated direct photon per-trigger yields as discussed in Chapter~\ref{chap:analysis}.}
	\label{fig:ex_dp_dphi_pty}
\end{figure}

An example of the isolated direct photon per-trigger yields and the corresponding isolated inclusive and decay per-trigger yields that were used in the statistical subtraction is shown in Fig.~\ref{fig:ex_dp_dphi_pty}. The near-side structures in the isolated inclusive and decay distributions are a result of re-checking for isolation when the photons are embedded into minimum bias events. The near-side of the isolated direct photon per-trigger yield is clearly suppressed and is roughly constant across \dphi, indicating that the leftover yield is due to contributions from the underlying event. There is a single away-side jet structure that is smeared in \dphi, indicating that the isolation cut has effectively identified direct photon-hadron events that are sensitive to initial-state $k_T$ and final-state $j_T$. \par

\begin{figure}[thb]
	\centering
	\includegraphics[width=0.75\textwidth]{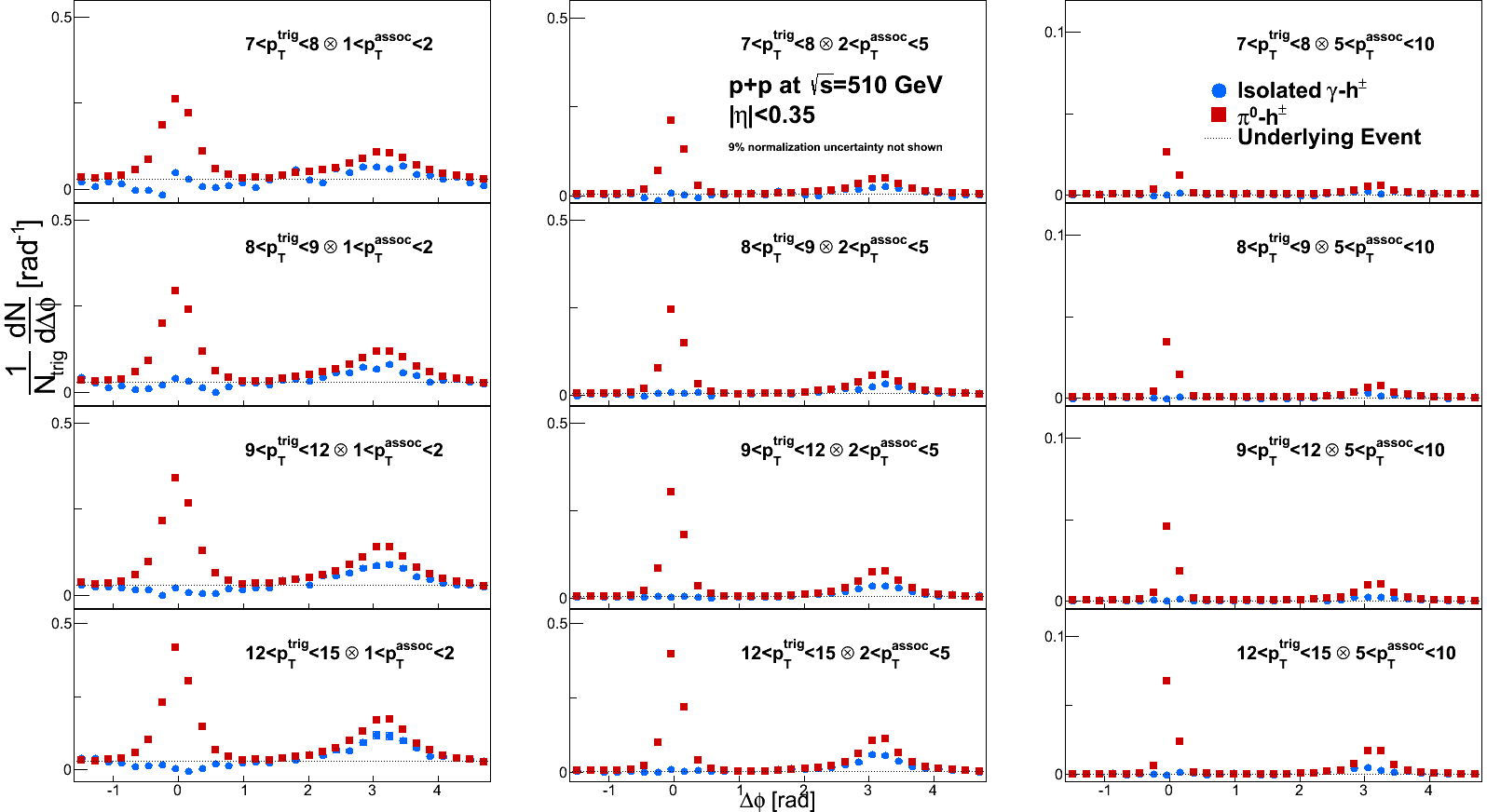}
	\caption{Examples of the isolated direct photon-hadron and \pion-hadron per-trigger yields as a function of \pttrig$\otimes~$\ptassoc. The absence of any near-side yield for the direct photon-hadrons indicates that the statistical subtraction adequately removes correlated near-side hadrons. In some bins the yield is negative, which is unphysical and is due to the combination of the $\gamma-\h$ isolation cut and statistical subtraction. Note that the near-side yield is not used for further analysis.}
	\label{fig:isodppiptys_iso}
\end{figure}

\begin{figure}[thb]
	\centering
	\includegraphics[width=0.75\textwidth]{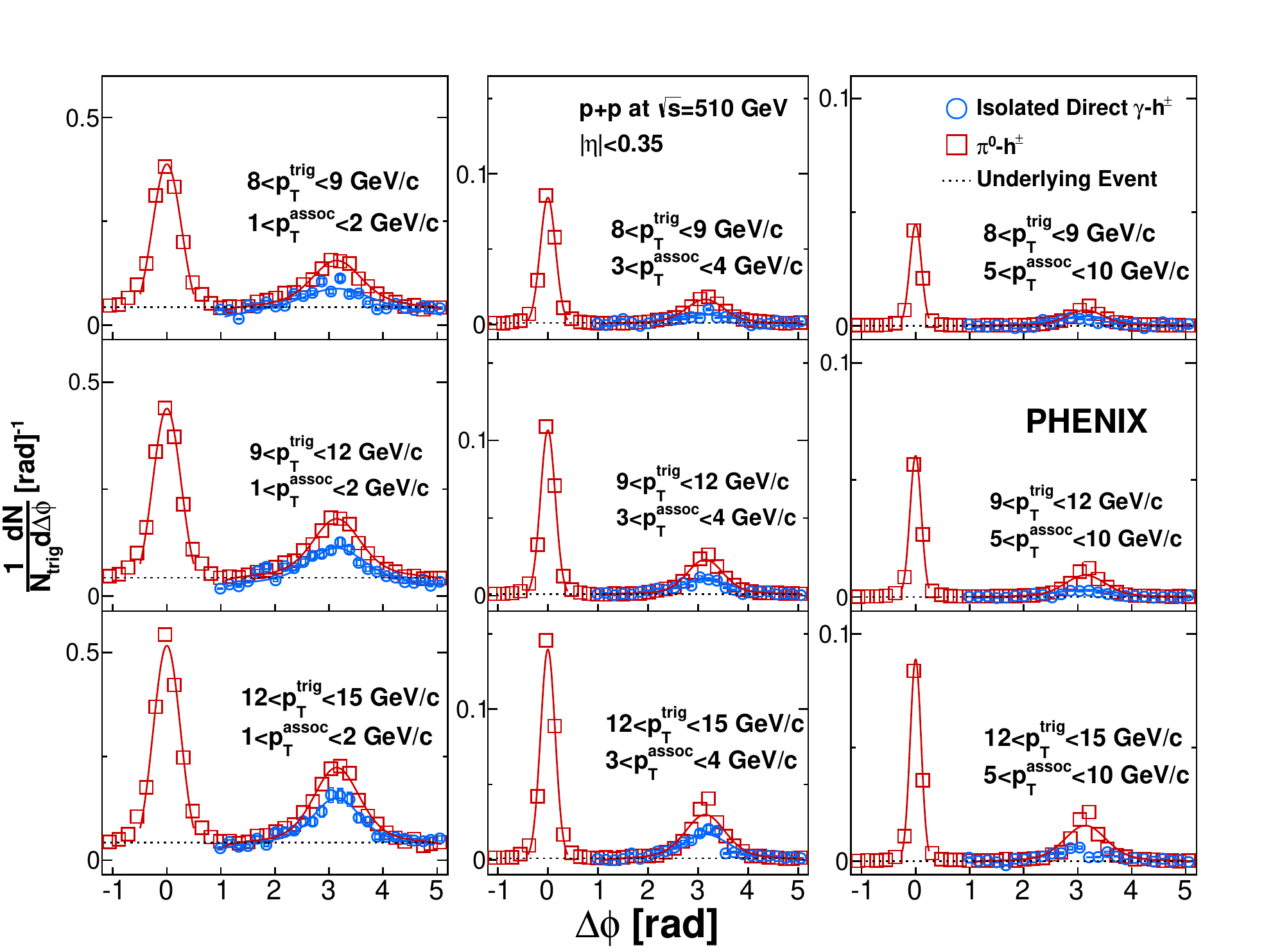}
	\caption{Published examples of the isolated direct photon-hadron and \pion-hadron per-trigger yields as a function of \pttrig$\otimes~$\ptassoc, allowing comparisons between dijet and $\gamma$-jet events.}
	\label{fig:isodppiptys}
\end{figure}

It is also beneficial to compare the direct photon-hadron correlations with the \pion-hadron correlations. Figure~\ref{fig:isodppiptys_iso} shows examples of the isolated direct photon-hadron per-trigger yields with the corresponding \pion-hadron per-trigger yields. Immediately the effect of the isolation cut is obvious; the near-side yield for the direct photons is significantly reduced compared to that of the near-side neutral pions. In the lower \ptassoc bins there are some cases where the yield is actually negative as a result of the statistical subtraction; this is again a result of the statistical subtraction procedure as well as the isolation cone significantly modifying the yield within and around the cone radius. Figure~\ref{fig:isodppiptys} shows the final correlation functions from Ref.~\cite{ppg195} where the yield in the region of the isolation cut is artificially removed since it is not physically interpretable and nonetheless is not dependent on both \kt and \jt. On the away-side, it is clear that in all of the bins the yield of the photon-hadrons is smaller than the $\pion$-hadrons. This is a result of the direct photons probing smaller hard scales than the neutral pions at the same \pttrig; since the $\pion$ is a fragment of a parent parton it is actually some fraction of the hard scale, whereas at leading order the direct photon approximates the hard scale since it is not a fragment. This is additional support that the correlation functions in question are actually probing direct photon-hadron events. \par

\begin{figure}[thb]
	\centering
	\includegraphics[width=0.7\textwidth]{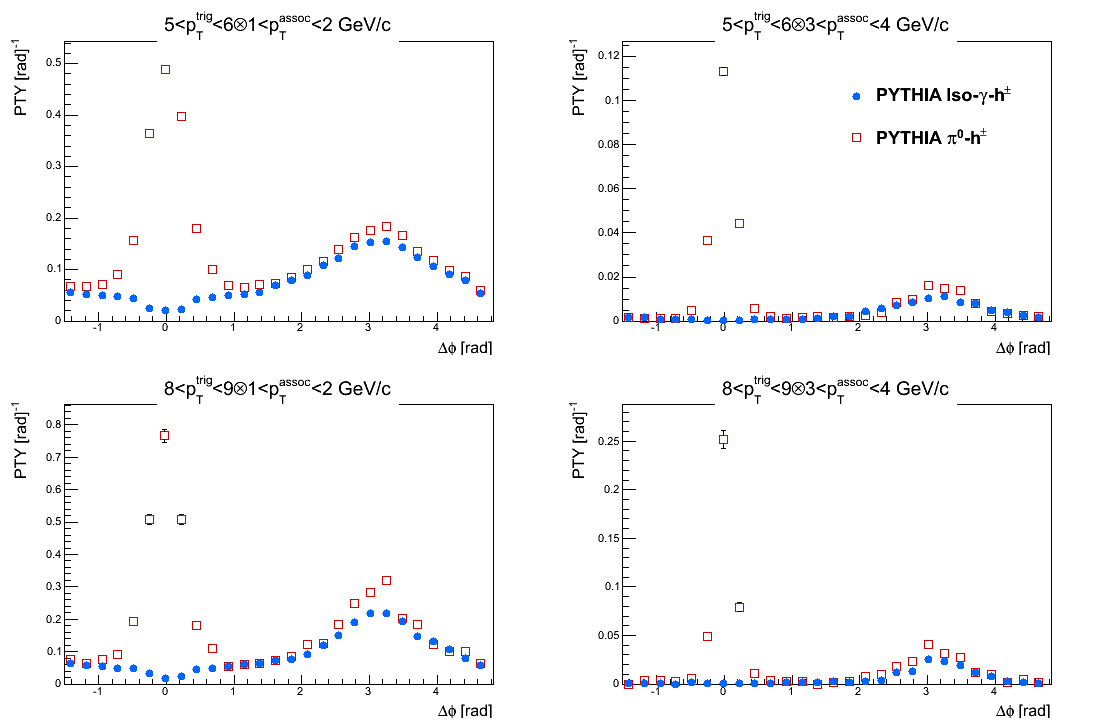}
	\caption{Comparisons are made between \pythia generated dihadron and direct \photon-hadron correlations. \pythia reproduces the kinematic effect of away-side jet yields being smaller for direct photons when compared to neutral pions.}
	\label{fig:pythia_ptys}
\end{figure}

If this effect is truly a kinematic effect, the \pythia~\cite{Sjostrand:2006za} event generator should be able to qualitatively reproduce this behavior. \pythia dijet and \photon-jet events were generated, and they were analyzed at the Monte Carlo truth level. Correlations were constructed similarly in \pythia as they were in data, with near-side neutral pions and near-side isolated direct photons. Charged hadrons were collected in the full azimuth, where this refers to pions, kaons, and (anti)protons. Since these events were analyzed at truth level, there was no need for an acceptance correction in \dphi; correlations were constructed within $|\eta|<~$0.35. Figure~\ref{fig:pythia_ptys} shows the results of the short study; the per-trigger yields are shown in several \pt bins for both \pion-\h and direct photon-\h correlations. Similarly to previous figures, the \pttrig bin increases moving top to bottom and the \ptassoc bin increases moving left to right.  The \pythia generated per-trigger yields replicate the qualitative behavior of the direct photon per-trigger yields having smaller yield than the \pion per-trigger yields in the same $\pttrig\otimes\ptassoc$ bin. \par

The complete published set of \dphi correlation functions for both \pion-\h and \photon-\h per-trigger yields can be found in the supplementary material of Ref.~\cite{ppg195}. While the correlation functions display the nearly back-to-back jet structure, the observable that will display the transverse-momentum-dependence is \pout as this is a momentum space observable which should be able to delineate the boundary between nonperturbative and perturbative contributions to the correlations. For measuring potential effects from TMD factorization breaking, this is necessary since a nonperturbative scale must explicitly be measured.

\section{\pout Distributions}

\begin{figure}[tbh]
	\centering
	\includegraphics[width=0.8\textwidth]{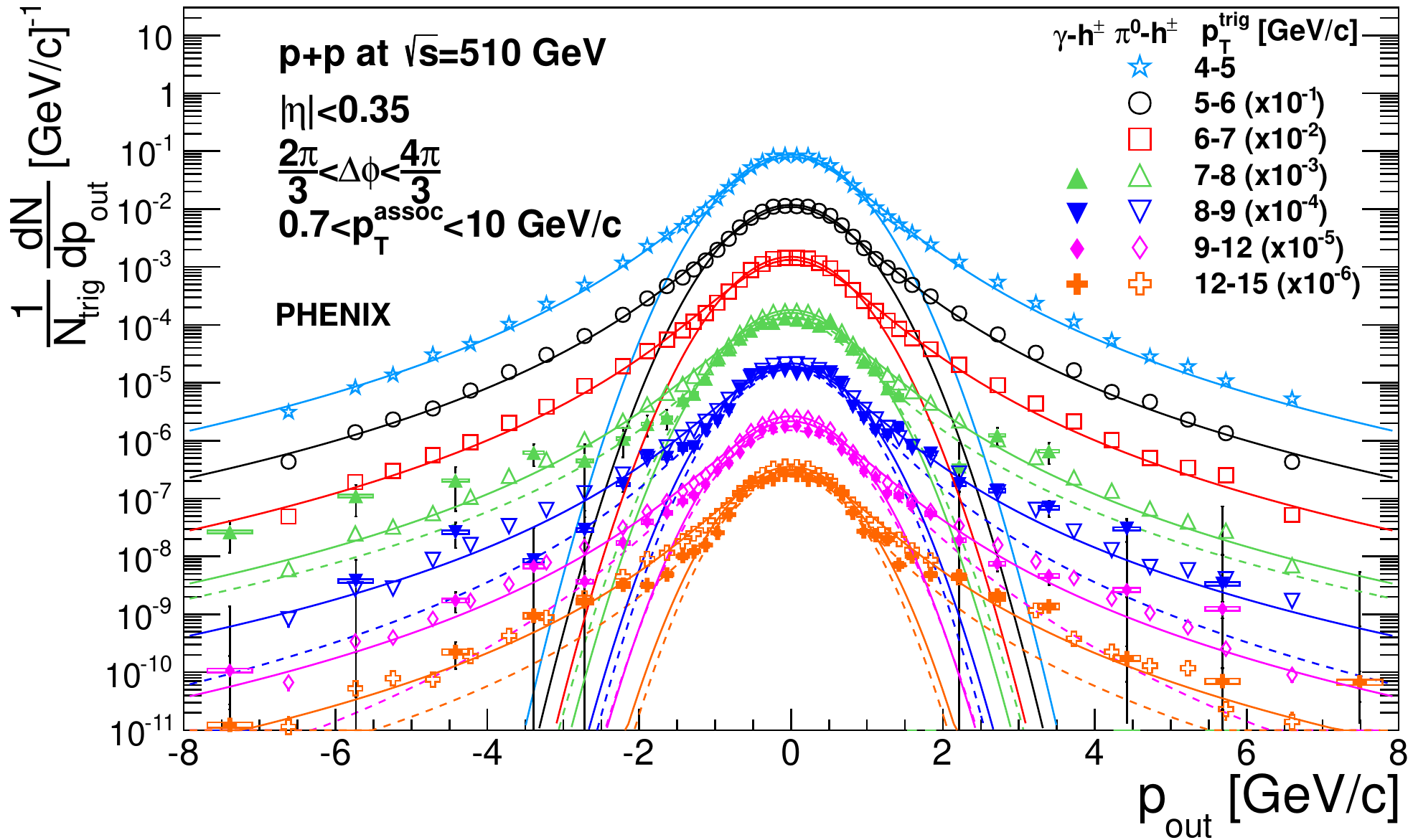}
	\caption{The per-trigger yields as a function of \pout are shown for both \pion-\h and \photon-\h correlations. The per-trigger yields are fit to Gaussian functions at small \pout and Kaplan functions over the entire region. The solid (dashed) lines are fits to the \pion-\h ($\gamma-\h$) per-trigger yields.}
	\label{fig:pout_ptys_run13}

\end{figure}

The \pout distributions are constructed for both \pion-\h and \photon-\h correlations to observe TMD effects in momentum space rather than azimuthal space. Figure~\ref{fig:pout_ptys_run13} shows the per-trigger yields for the correlations as a function of \pout for both \pion-\h and \photon-\h correlations. The open points show the \pion-\h per-trigger yields in various bins of \pttrig, while the filled points show the \photon-\h correlations in four \pttrig bins. Only away-side charged hadrons are used to construct the distributions, with the requirement that the correlated pair satisfy $2\pi/3<\dphi<4\pi/3$. This cut is enforced to ensure that only hadrons associated with the away-side jet are considered, since these are the correlated pairs sensitive to both $k_T$ and $j_T$. The distributions are fit to a Gaussian function in the region [-1.1,1.1] GeV/c, as well as a Kaplan function of the form $a(1+p_{out}^2/b)^{-c}$ where $a$,$b$, and $c$ are all free parameters of the fit. Solid line fits correspond to the dihadron correlations, while dotted line fits correspond to the \photon-hadron correlations. The fits show the change from nonperturbative to perturbative physics; the Gaussian functions clearly do not describe the \pout distributions after roughly $\pout\sim1.2$ GeV/c where there is a transition to power-law behavior. This indicates that in the small \pout region, the correlations are sensitive to nonperturbatively generated \kt and \jt transverse momentum, while in the large \pout region the correlations are sensitive to perturbatively generated \kt and \jt transverse momentum. \par

Dihadron and direct photon-hadron correlations are ideal processes to study in order to measure the \pout distributions accurately because of the good resolution on both \ptassoc and \dphi. Since \pout is dependent on both of these quantities, the inherent resolution of the detector on both \ptassoc and \dphi will determine the resolution on \pout. The \pt and $\phi$ resolution on hadrons is much better than, for example, jets, so \pout will have sensitivity to these modified TMD effects from soft transverse momentum. While measuring dijet and \photon-jet correlations are of interest as well, for other reasons which will be discussed in Chapter~\ref{chap:future}, the sensitivity to nonperturbative physics is essential for making TMD measurements. Jets have significantly worse resolution in \pt and $\phi$ when compared to hadrons, so using the hadrons as proxies for the jets allows much better resolution on \pout to be sensitive to TMD effects. This is additionally motivated by the fact that the original prediction of factorization breaking was for dihadron production~\cite{Rogers:2010dm}; however, more recent studies have shown that dijet and photon-jet correlations should exhibit factorization breaking effects as well~\cite{Rogers:2013zha,Schafer:2014xpa,Zhou:2017mpw}\par

\begin{figure}[tbh]
	\centering
	\includegraphics[width=0.85\textwidth]{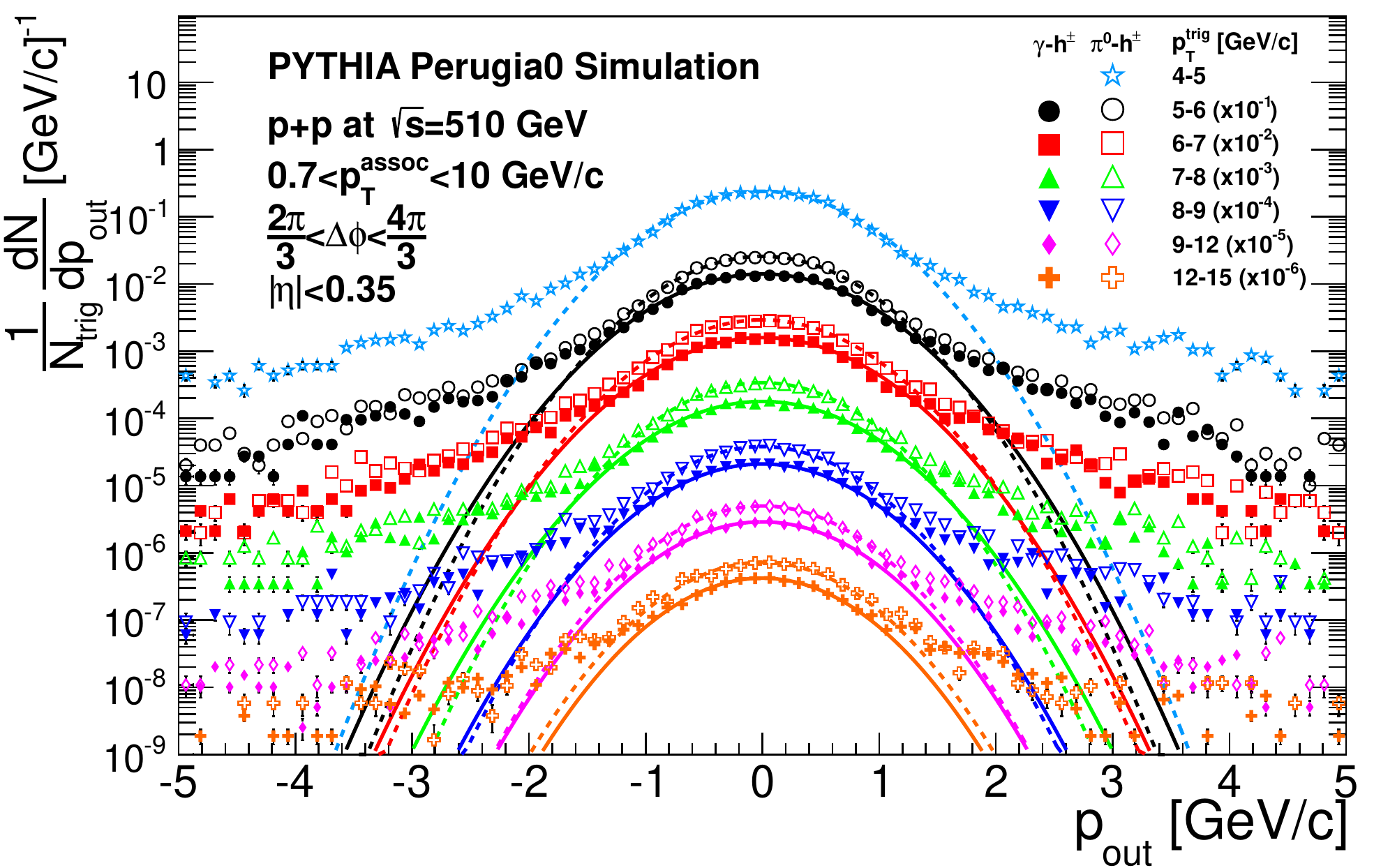}
	\caption{\pout per-trigger yields are constructed in \pythia for $2\pi$ in azimuth and $|\eta|<0.35$. The \pythia yields display similar characteristics to those measured in data.}
	\label{fig:pythia_pouts}
\end{figure}

Correlated \pout distributions, shown in Fig.~\ref{fig:pythia_pouts}, are also constructed in the \pythia model framework, in order to make kinematic comparisons. Here the underlying event distributions were statistically subtracted similarly to the procedure performed in data. \pythia shows similar characteristics to the data; namely a Gaussian distribution at small \pout that transitions to a power law tail at large \pout. The direct photon and \pion perturbative tails match reasonably well in \pythia, which would be expected since this region is generated by charged hadrons at large \dphi due to large angle hard gluon radiations. \pythia also replicates the data distributions in that the \pion-\h yields are larger than the corresponding \photon-\h yields, due to the different hard scales being probed. \par

The complete published set of \pout correlation functions for both \pion-\h and \photon-\h per-trigger yields can be found in the supplementary material of Ref.~\cite{ppg195}. The correlation functions display the relevant per-trigger yields, and in order to probe potential TMD factorization breaking effects the relevant nonperturbative momentum widths must be extracted from these correlation functions.

\section{\rmspout Determination}\label{rmspoutdetermination}

\begin{figure}[tbh]
	\centering
	\includegraphics[width=0.6\textwidth]{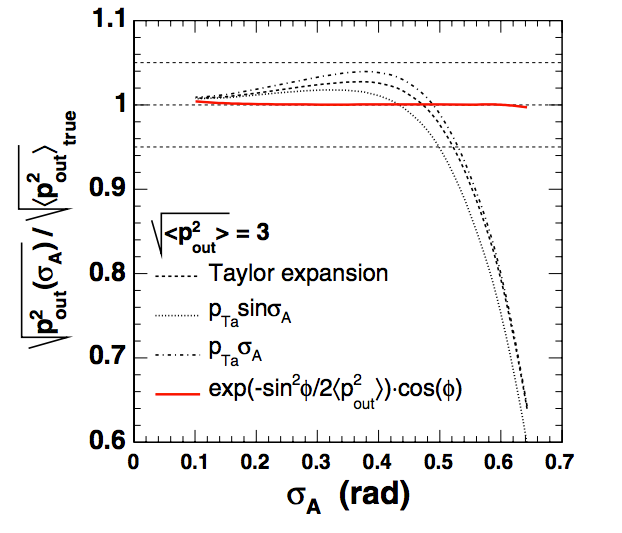}
	\caption{The relative error on \rmspout=3 GeV/c is shown as a red line when the quantity is extracted from the fit function in Eq.~\ref{eq:rmspout}~\cite{ppg029}. The additional lines correspond to alternative assumptions in extracting \rmspout; for example, with the assumption that \rmspout=\ptassoc$\sin\sigma_A$ where $\sigma_A$ is extracted from a Gaussian fit to the \dphi correlation function.}
	\label{fig:rmspout_validity}
\end{figure}

The quantity \rmspout can be extracted with a fit function to the away-side correlation function in \dphi. This quantity is useful as it can be extracted as a function of both \pttrig and \ptassoc, while the \pout distributions must be constructed integrated over \ptassoc due to the constraint imposed on \pout=~\ptassoc$\sin$\dphi for a limited range of \ptassoc. It additionally considers contributions across the entire away-side, which therefore includes both perturbatively and nonperturbatively generated charged hadrons. Therefore, useful information can be gained by extracting fit quantities from both the \dphi distributions and the \pout distributions. The fit function can be found in Eq.~\ref{eq:rmspout}
\begin{equation}\label{eq:rmspout}
\frac{dN}{d\Delta\phi} = C_0+C_1\frac{-p_T^{assoc}\cos\Delta\phi}{\sqrt{2\pi\langle\pout^2\rangle}Erf(p_{T}^{assoc}/\sqrt{2\langle\pout^2\rangle})}e^{-\frac{p_T^{assoc}p_T^{assoc}\sin^2\Delta\phi}{2\langle p_{out}^2\rangle}}\,,
\end{equation}
\noindent where the quantities $C_0$, $C_1$, and \rmspout are free parameters in the fit. $C_0$ quantifies the underlying event level, $C_1$ quantifies the away-side yield, and \rmspout characterizes the away-side jet width. In Eq.~\ref{eq:rmspout} the \ptassoc value used is the weighted average of the actual \ptassoc spectrum for a particular \ptassoc bin. The accuracy of the fit function in determining \rmspout was demonstrated in Ref.~\cite{ppg029}, and is shown in Fig.~\ref{fig:rmspout_validity}. As long as the away-side jet width is within [0.2,0.6] radians, the value of \rmspout from the fit function is nearly identical to the actual value of \rmspout from a direct calculation. \par

\begin{figure}[tbh]
	\centering
	\includegraphics[width=0.7\textwidth]{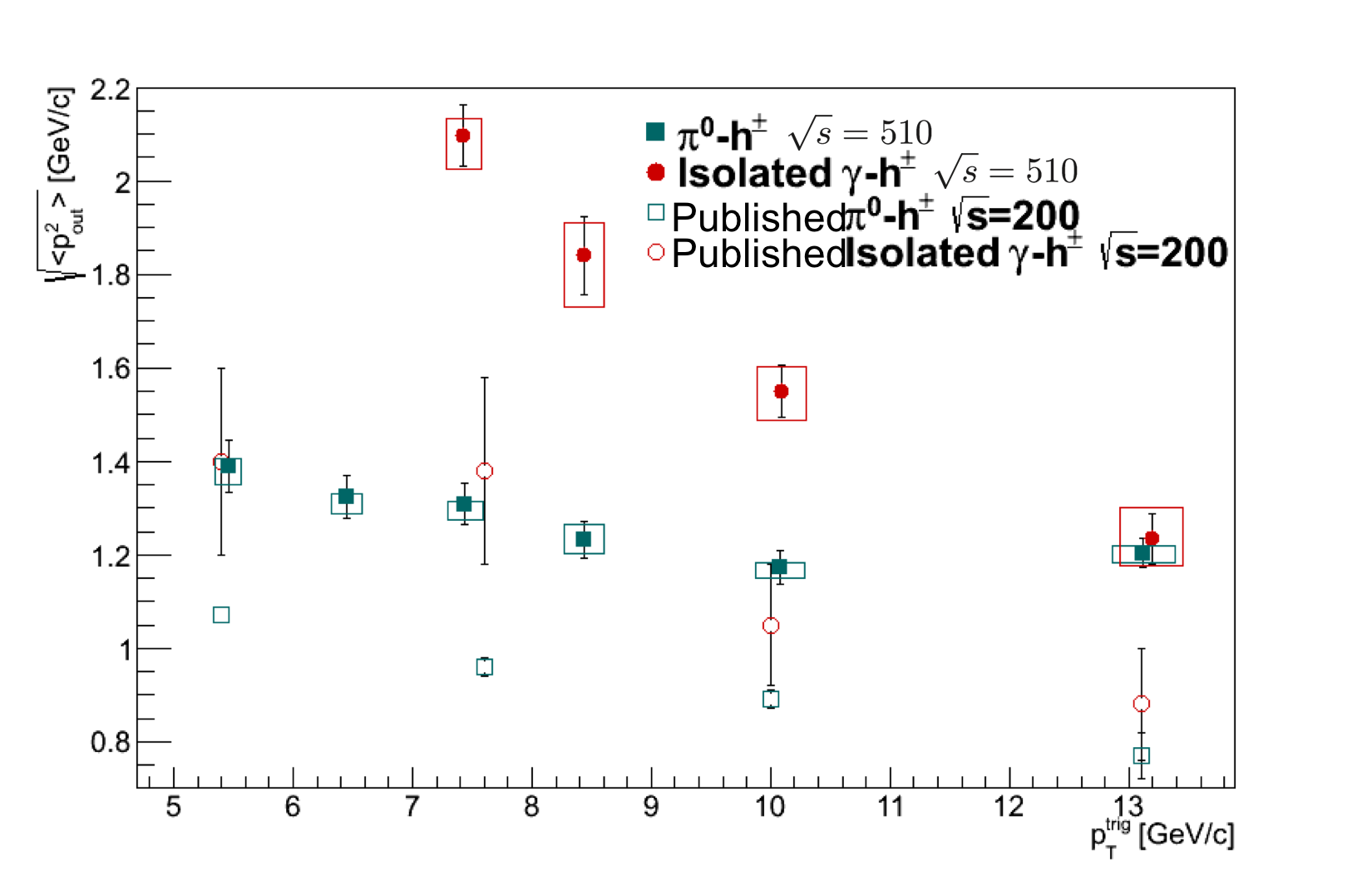}
	\caption{The \rmspout quantity at \sqs=~510 GeV is compared to the previous published values at \sqs=~200 GeV from Ref.~\cite{ppg095}. In the same bins there is a clear difference, indicating both the dependence of $k_T$ on \sqs as well as the momentum fraction $x$ that is probed in the two different data sets.}
	\label{fig:rmspout_comp_ppg095}
\end{figure}

To compare directly to Ref.~\cite{ppg095}, the fit was performed for the \ptassoc bin $2<\ptassoc<5$ GeV/c as was previously done. The results of the fit are shown in Fig.~\ref{fig:rmspout_comp_ppg095} with a comparison to the values from the previously published data. Systematic errors were assigned to \rmspout by varying the fit region by $\pm$ 0.2 radians in the underlying event region. They were assigned this way since \rmspout is the value taken from a fit function, so the systematic error assigned to it should reflect the uncertainty in the fit itself. Unsurprisingly the systematic errors for the \pion-\h correlations are small since the underlying event region is well defined; for the \photon-\h correlations the underlying event is determined from the statistical subtraction so there is a higher degree of variability in the result of the fit. The \pttrig values are the weighted averages of the actual \pion or isolated inclusive \photon spectrum. A 2\% systematic uncertainty is assigned to \pttrig to account for the inherent \pt resolution of the PbSc and PbGl electromagnetic calorimeters. \par

The comparison of \rmspout between \sqs=~510 and 200 GeV indicates several different physics effects. First, the value of the initial-state $k_T$ has been shown to increase with \sqs~\cite{ppg029}, thus since \rmspout is sensitive to initial-state $k_T$ it should be expected that it would be larger at \sqs=~510 GeV when compared to \sqs=~200 GeV in Ref.~\cite{ppg095} for the same \ptassoc bin. Additionally, the change in \sqs causes the average partonic momentum fraction $x$ that is probed to be different between the two data sets. Thus, if there is any correlation between the partonic $x$ and $k_T$, it should be reflected when comparing the two different center-of-mass energies. This discussion will be revisited when comparing the updated results from the run-15 \sqs=~200 GeV run in Chapter~\ref{chap:paresults}. \par

\begin{figure}[thb]
	\centering
	\includegraphics[width=0.7\textwidth]{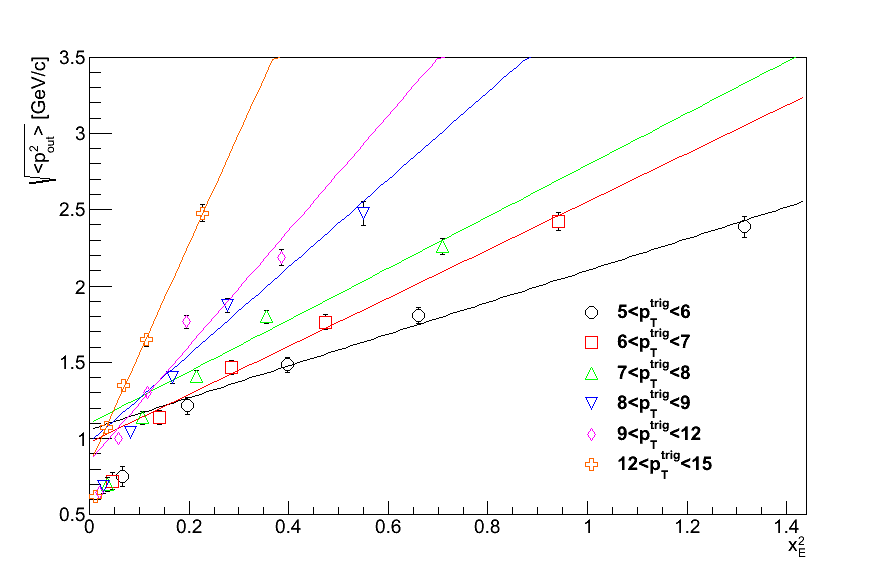}
	\caption{The values of \rmspout are shown as a function of $x_E^2$, as defined in the text. The values exhibit similar qualitative features to Ref.~\cite{Angelis:1980bs}, namely that they are approximately linear with a deviation from linear at small $x_E^2$.}
	\label{fig:pi0_rmspout_xe}
\end{figure}

To cross check the results with other literature, the $\pion$ \rmspout were plotted as a function of the hard scattering variable $x_E=-\ptassoc\cdot\pttrig/|\pttrig|^2$ as defined in Section 1.7 which is approximately $\frac{\ptassoc}{\pttrig}$ for \dphi$\sim\pi$. The qualitative features of the first measurement of \rmspout by the CCOR collaboration in Ref.~\cite{Angelis:1980bs} are reproduced in this analysis; Fig.~\ref{fig:pi0_rmspout_xe} shows that the values are a positively sloped function of $x_E^2$ that is approximately linear with a deviation from linear at small $x_E^2$. Additionally the slope of of this linear dependence increases with $\pttrig$. Quantitatively this result is consistent with Refs.~\cite{Angelis:1980bs,ppg095} as the value of $k_T$ should be larger in \sqs=~510 GeV, leading to larger values of \rmspout.

Due to the significantly larger data set available in this run, the \rmspout values were constructed in finer $\pttrig\otimes\ptassoc$ bins. Examples of the \pion-\h and \photon-\h \rmspout values are shown as a function of both \ptassoc and \pttrig in Figs.~\ref{fig:run13_rmspout_vspttrig} and~\ref{fig:run13_rmspout_vsptassoc}. All of the values can be found in the supplemental material of Ref.~\cite{ppg195}. Figure~\ref{fig:run13_rmspout_vsptassoc} shows that \rmspout has the expected positive dependence on \ptassoc. This is expected from both the geometrical definition of \pout as well as the definition of the fit function used to extract \rmspout in Eq.~\ref{eq:rmspout}. The \rmspout values from \photon-\h correlations are, in general, larger than the corresponding \rmspout values from \pion-\h. As a function of \pttrig, shown in Fig.~\ref{fig:run13_rmspout_vspttrig}, the quantity \rmspout decreases with \pttrig. This indicates that as the hard scale increases, the RMS of \pout is decreasing for a given \ptassoc. Again, when comparing the \photon-\h correlations with the \pion-\h correlations, the \photon-\h \rmspout values are in general larger than their \pion-\h counterparts. This could also be interpreted as a consequence of the values decreasing with the hard scale; since the \photon-\h correlations probe smaller hard scales, they would in turn be expected to be larger than the \pion-\h \rmspout values at a similar \pttrig. It is also interesting to note that the \photon-\h \rmspout values are very strongly a function of \pttrig, while the \pion-\h correlations are only slightly a function of \pttrig. \par

\begin{figure}[thb]
	\centering
	\includegraphics[width=0.7\textwidth]{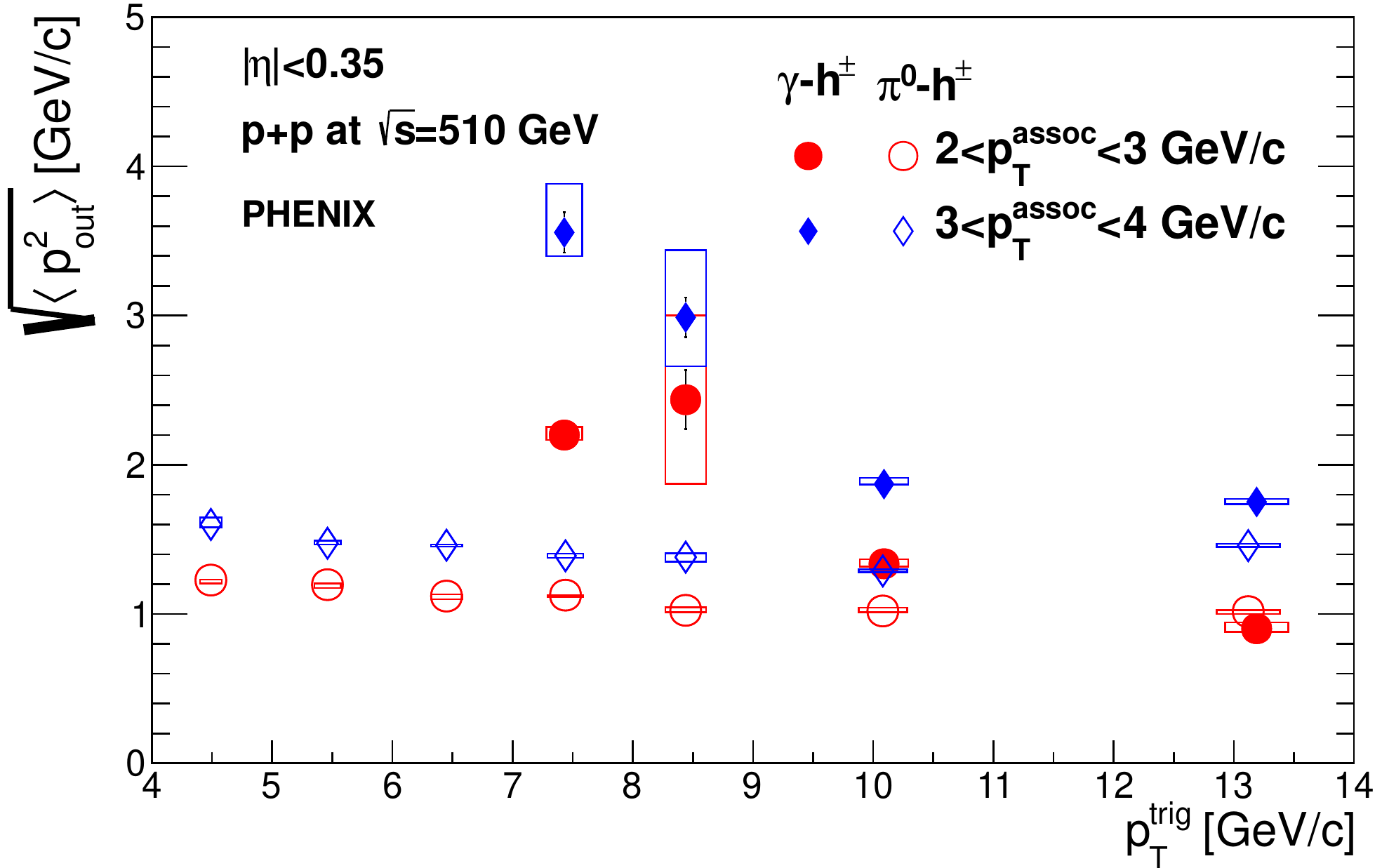}
	\caption{The quantity \rmspout is shown for two example \ptassoc bins as a function of \pttrig. The filled points are for direct photon-hadron correlations while the open points are for \pion-\h correlations.}
	\label{fig:run13_rmspout_vspttrig}
\end{figure}

\begin{figure}[thb]
	\centering
	\includegraphics[width=0.7\textwidth]{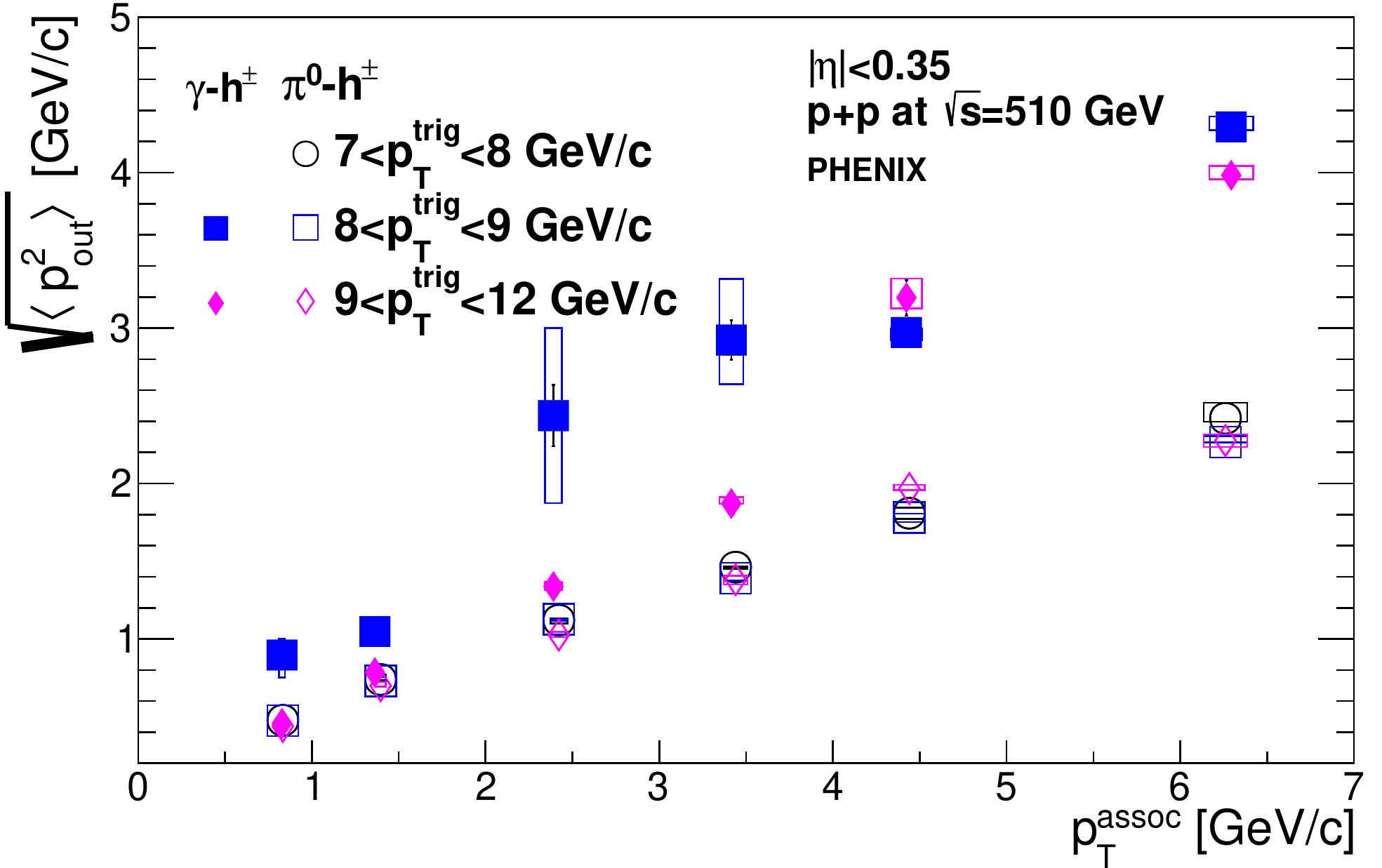}
	\caption{The quantity \rmspout is shown for several \pttrig bins as a function of \ptassoc. The filled points are for direct photon-hadron correlations while the open points are for \pion-\h correlations.}
	\label{fig:run13_rmspout_vsptassoc}
\end{figure}

Since \rmspout is dependent on both \rmskt and \rmsjt~\cite{ppg029}, it is necessary to also measure \rmsjt to identify the fragmentation component of the values of \rmspout. This is useful anyway as a nonperturbative momentum width sensitive only to fragmentation can be studied in addition to those sensitive to both initial-state \kt and final-state \jt. While not explicitly relevant for factorization breaking studies, it is necessary to measure \rmsjt so that any potential fragmentation dependence in \rmspout may be studied.

\section{\rmsjt Determination}

The quantities \rmspout, \rmskt, and \rmsjt are related via the following equation as derived in Ref.~\cite{ppg029}:

\begin{equation}\label{eq:rmspout_ktjt}
\frac{\langle z_T\rangle \rmskt}{\hat{x}_h}  = \frac{1}{x_h}\sqrt{\langle p_{\rm out}^{2}\rangle - \langle j_{T_{y}}^2\rangle(1+x_h^2)}\,.
\end{equation}

\noindent In this equation all quantities on the left side of the equality are partonic values, and all quantities on the right can be measured from the correlation functions. Here, $\langle z_T\rangle=\pttrig/\hat{p}_T^{\rm trig}$ and $x_h = \langle\ptassoc\rangle/\langle\pttrig\rangle$, and any quantity with a hat indicates partonic level quantities. This gives a clear relation between the initial and final state partonic transverse momenta and the quantity \pout. Thus, to completely determine the relation such that the partonic quantities could be calculated, the experimental value of \rmsjt should be determined.

The value of \rmsjt was determined by measuring the widths of Gaussian fits to the near-side of the \pion-\h correlation functions, similarly to Ref.~\cite{ppg029}. Examples of the fits are shown on the near-side \pion-\h peaks in Fig.~\ref{fig:isodppiptys}. The values of \rmsjt were calculated with the following equation:

\begin{equation}\label{eq:rmsjt}
\rmsjt = \sqrt{2\langle j_{T_{y}}^2\rangle}\simeq\sqrt{2}\frac{\pttrig\ptassoc}{\sqrt{p_T^{\rm trig^2}+p_T^{\rm assoc^2}}}~\sigma_N\,.
\end{equation}

\begin{figure}[thb]
	\centering
	\includegraphics[width=0.7\textwidth]{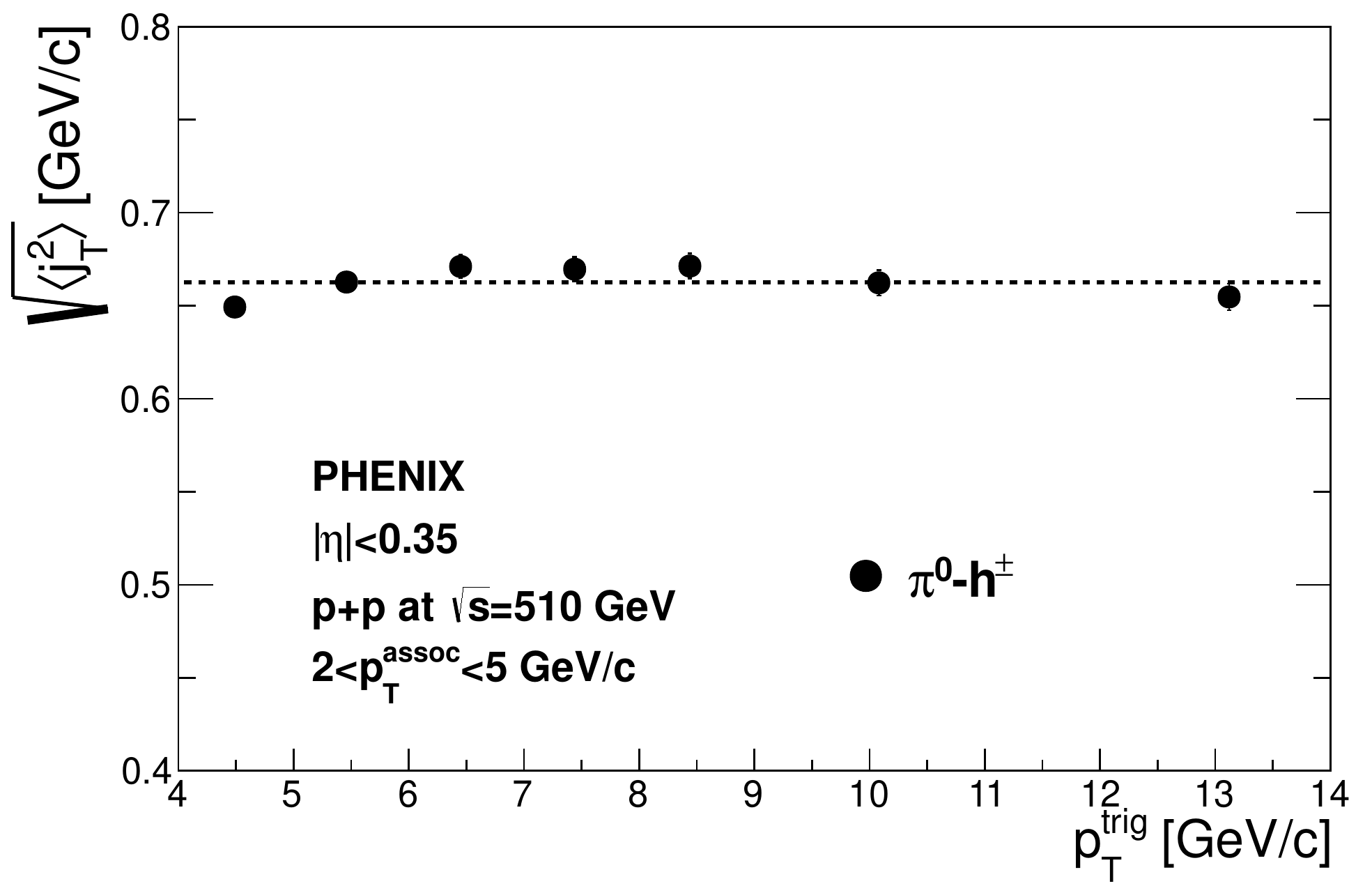}
	\caption{The quantity \rmsjt is shown as a function of \pttrig in \sqs=~510 GeV \pp collisions. The line shows a constant fit to the data, which is used to determine the \pttrig and \ptassoc averaged \rmsjt value.}
	\label{fig:rmsjt}
\end{figure}

Note that in deriving this equality it is assumed that the fragmentation component $j_{T_y}$ is the same in both the $x$ and $y$ directions and that it is sampled from the same Gaussian distribution of \rmsjt on both the away and near sides. Here $\sigma_N$ is the near-side Gaussian width of the correlation functions. Only bins where $\ptassoc>2$ GeV/$c$ were used to satisfy the assumption that $\ptassoc\gg\sqrt{2}j_T$ which was made in the derivation of Eq.~\ref{eq:rmsjt}. Each $\pttrig$ bin was fit to a constant and was averaged over $\ptassoc$ as previous measurements have shown \rmsjt to be approximately constant with \sqs and \pttrig~\cite{ppg029,ppg089, Angelis:1980bs}. Figure~\ref{fig:rmsjt} shows the measurements as a function of \pttrig, where a constant fit was used to determine the average value of $\rmsjt = 0.662\pm0.003$(stat)$\pm0.012$(sys) \gevc. The systematic uncertainty is due to both the momentum resolution of the detector as well as approximations made to determine Eq.~\ref{eq:rmsjt}. Recent ATLAS results show a similar fragmentation variable over a significantly larger range of hundreds of GeV/$c$ in jet \pt, and show that the average transverse momentum with respect to the jet axis rises slowly with the \pt of the jet over this larger \pt range~\cite{Aad:2011sc}.

While \rmspout and \rmsjt are useful quantities, they are extracted from the \dphi correlations and thus they may lack the necessary sensitivity to TMD factorization breaking effects since they contain perturbative contributions in addition to nonperturbative contributions. Since the observable \pout clearly distinguishes the nonperturbative and perturbative regimes, momentum widths from these per-trigger yields will contain more specific information since only nonperturbative contributions will affect the widths.

\section{Gaussian Widths}

\begin{figure}[thb]
	\centering
	\includegraphics[width=0.7\textwidth]{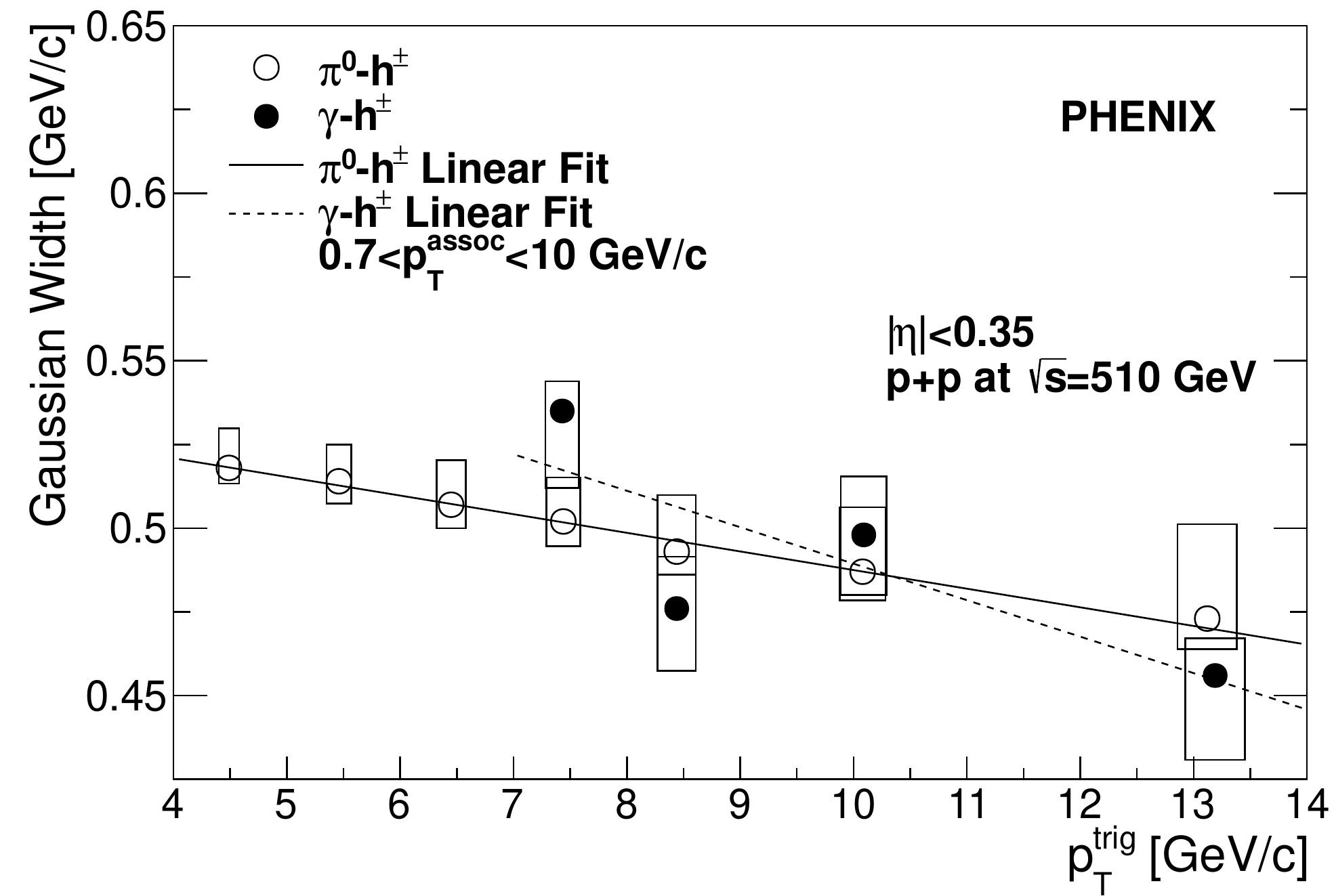}
	\caption{The measured Gaussian widths of \pout are shown as a function of \pttrig for \sqs=~510 GeV \pp collisions.}
	\label{fig:gausswidths_pp510}
\end{figure}

The Gaussian widths of \pout are extracted from fits to the \pout per-trigger yields. The fits are performed in the nonperturbative region and are shown as a function of \pttrig in Fig.~\ref{fig:gausswidths_pp510} for both \pion-\h and direct $\gamma$-\h correlations. Systematic uncertainties on the widths are estimated by altering the fit region by $\pm$0.15 \gev in \pout and taking the absolute value of the difference of the resulting widths. As the systematic uncertainties dominate the uncertainty in the widths, the uncertainties in Fig.~\ref{fig:gausswidths_pp510} are the statistical and systematic uncertainties combined in quadrature. Similarly to the values of \rmspout, the direct photons and \pion both show that the widths decrease with \pttrig. Linear fits to the two sets of widths give slopes of $-0.0055\pm$0.0018(stat)$\pm$0.0010(sys) for \pion mesons and $-0.0109\pm$0.0039(stat)$\pm$0.0016(sys) for direct photons. Systematic uncertainties on the slopes were conservatively estimated by evaluating the fit when the points were placed at the limits given by the systematic uncertainties, and then taking the difference of the slopes.

It should be noted that the magnitude of the slope of the Gaussian widths changes if the minimum \ptassoc cut is increased, but that the slope always remains negative. Integrating over the full range of $0.7<\ptassoc<10$ \gev allowed by the PHENIX detector gives the smallest magnitude slope, thus it is the most conservative measurement for comparing to CSS evolution. For example, the slope of the Gaussian widths of \pout as a function of \pttrig for $1.2<\ptassoc<10$ GeV/c was determined to be $-0.012\pm$0.003(stat)$\pm$0.001(sys) for \pion-\h correlations and $-0.023\pm$0.007(stat)$\pm$0.003(sys) for direct photon-\h correlations. The same behavior can be seen in the values of \rmspout in Fig.~\ref{fig:run13_rmspout_vspttrig} and in the supplemental material of Ref.~\cite{ppg195}.

\section{\zt Correction to Estimate $p_T^{\rm jet}$}

The \pion-hadron correlations contain an implicit dependence on the near-side fragmentation function not present in the direct photon-hadron correlations since the direct photons emerge directly from the hard scattering. To explore this dependence, \pythia 6.4~\cite{Sjostrand:2006za} hard scattered QCD events were analyzed to determine the average $\langle z_T\rangle =\langle\pttrig/\hat{p}_T^{\rm trig}\rangle$ of a \pion where the hat refers to the partonic \pt. \zt was determined for the various \pttrig$\otimes$\ptassoc bins so that the jet \pt could be estimated from the \pion \pt to make a more apples-to-apples comparison between the \pion-\h and $\gamma$-\h correlations. 

\pythia was run at \sqs=~510 GeV with a Gaussian intrinsic \kt value of 3.2 \gev, corresponding to the expectation for partonic $k_T$ from Ref.~\cite{ppg029}. Only standard hard scattering QCD events were run in \pythia, and leading \pion mesons with $4<\pt<15$ \gev were analyzed in the PHENIX pseudorapidity acceptance $|\eta|<0.35$. A single charged hadron was required to be in the away-side direction of the \pion to create a more accurate representation of the kinematics probed in the data. Once these requirements were met the \pion ancestry was traced back to the partonic hard scattering and the momentum fraction $z_T$ was calculated in bins of \pttrig and \ptassoc. Note that quark and gluon jets were treated equivalently and no attempt was made to differentiate between the two types of jets. This is not a significant cause for concern though, since at RHIC energies there is an approximately equal fraction of quark and gluon jets produced in the \pion \pt ranges observed. To find the average $z_T$ the double differential binned values were averaged via their statistical weights.

Results from the \pythia study are shown below in Figs.~\ref{fig:zt1}~-~\ref{fig:ztlast}. The values are intuitive, namely that \zt rises with \pttrig. This also is consistent with the idea of ``trigger bias'' briefly discussed in Ref.~\cite{ppg029} and above; the leading \pion trigger particle carries a large fraction of the near-side jet momentum and this is a result of choosing the highest \pt particle in the event and labeling it as the near-side trigger particle. The scale of \zt also decreases with increasing \ptassoc, as in these scenarios the leading \pion might not be the highest \pt particle in the event anymore. This is a consequence of the PHENIX triggering apparatus; the high energy trigger is an electromagnetic trigger and thus limits the high \pt particles that may be triggered on in data.

\begin{figure}[thb]
	\includegraphics[width=0.5\textwidth]{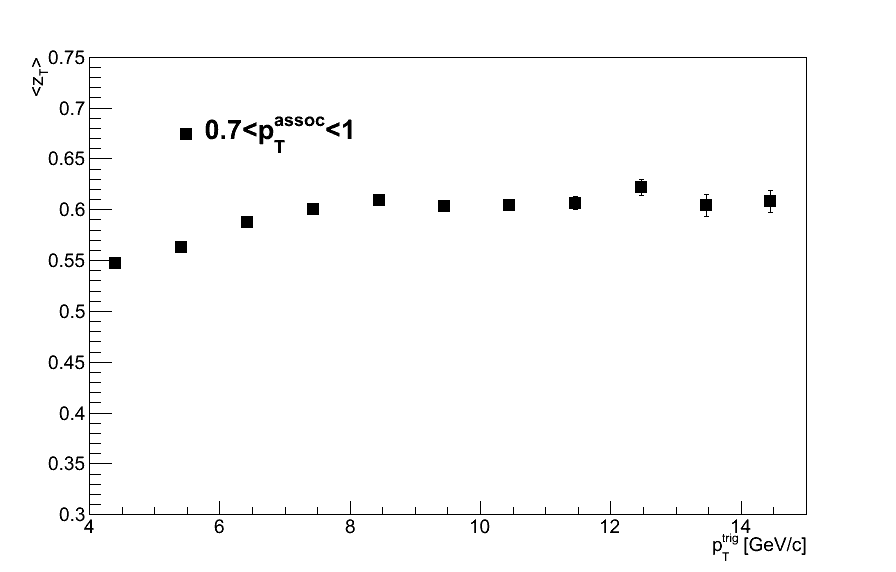}
	\includegraphics[width=0.5\textwidth]{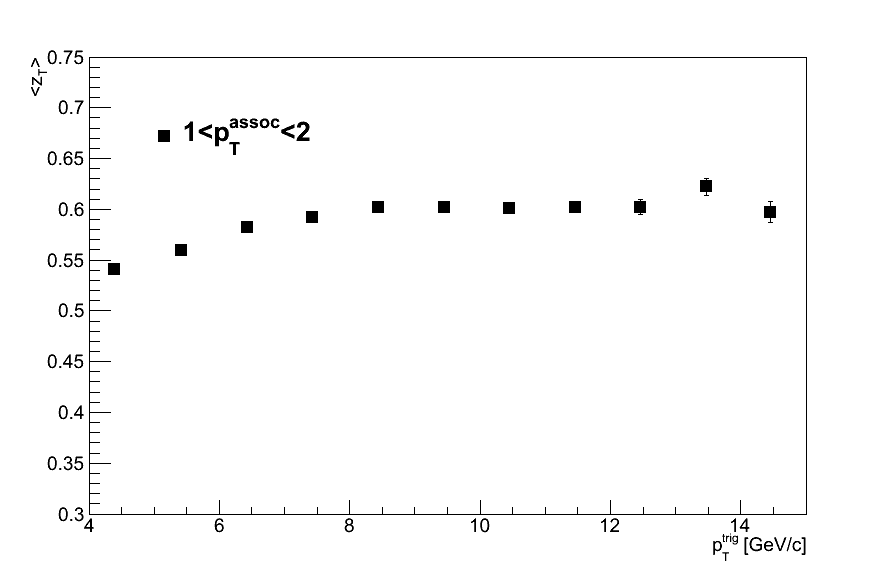}
	\caption{The quantity \zt is shown as a function of \pttrig for $0.7<\ptassoc<2$ \gevc.}
	\label{fig:zt1}
\end{figure}

\begin{figure}[thb]
	\includegraphics[width=0.5\textwidth]{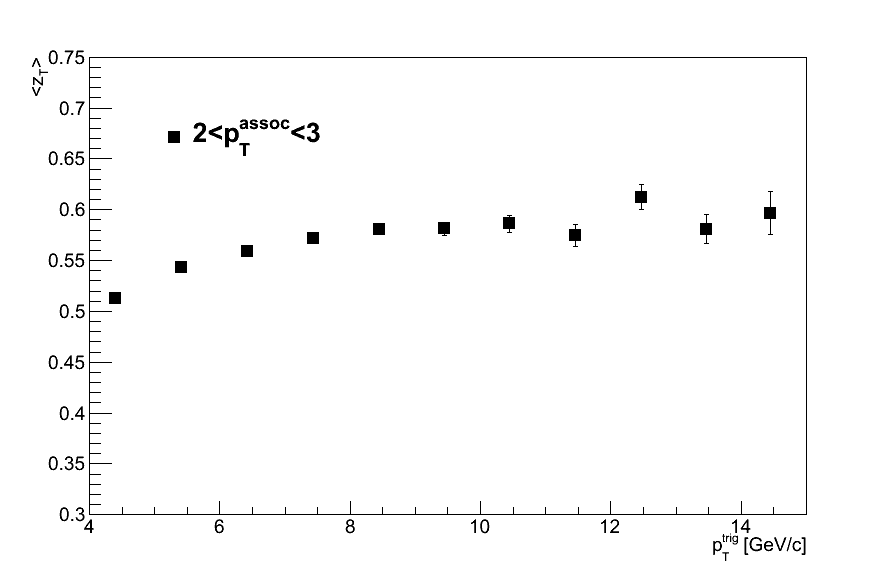}
	\includegraphics[width=0.5\textwidth]{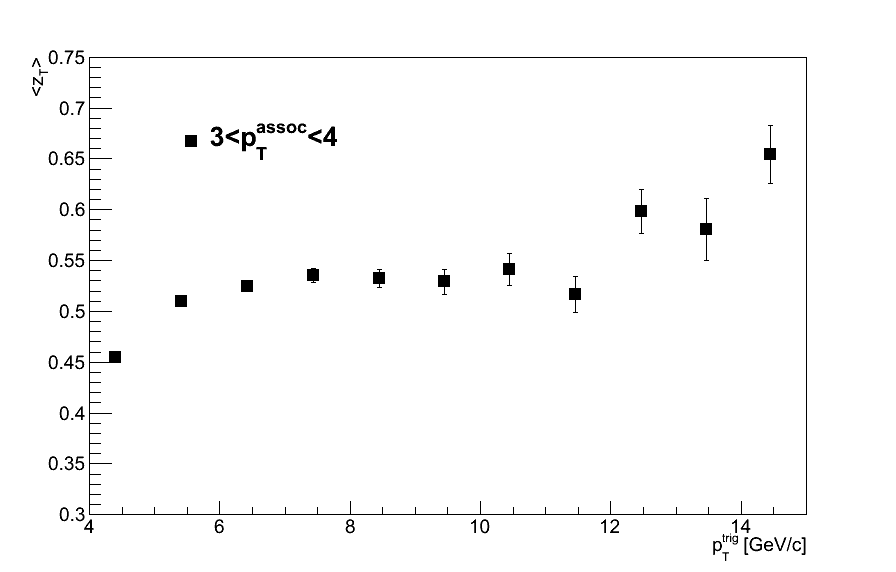}
	\caption{The quantity \zt is shown as a function of \pttrig for $2<\ptassoc<4$ \gevc.}
	\label{fig:zt2}
\end{figure}

\begin{figure}[thb]
	\includegraphics[width=0.5\textwidth]{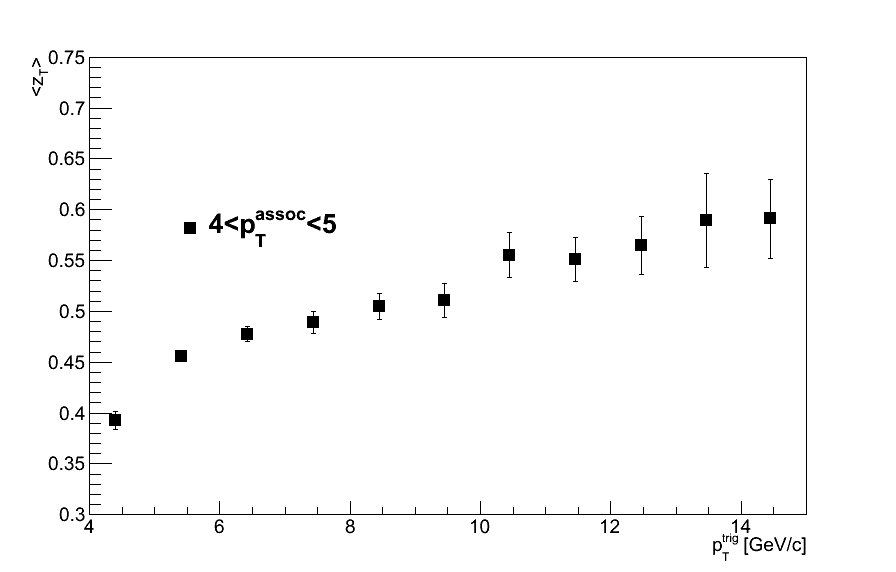}
	\includegraphics[width=0.5\textwidth]{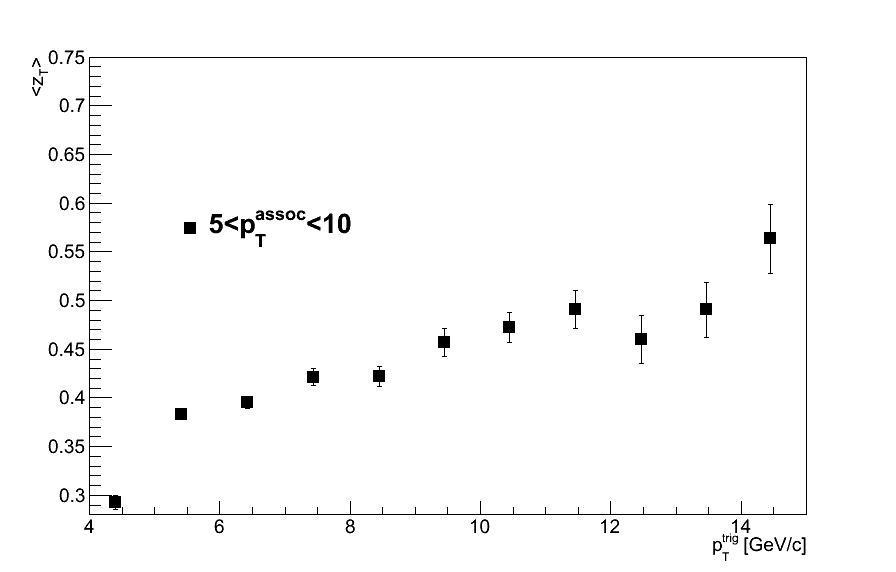}
	\caption{The quantity \zt is shown as a function of \pttrig for $4<\ptassoc<10$ \gevc.}
	\label{fig:ztlast}
\end{figure}

\begin{figure}[thb]
	\centering
	\includegraphics[width=0.7\textwidth]{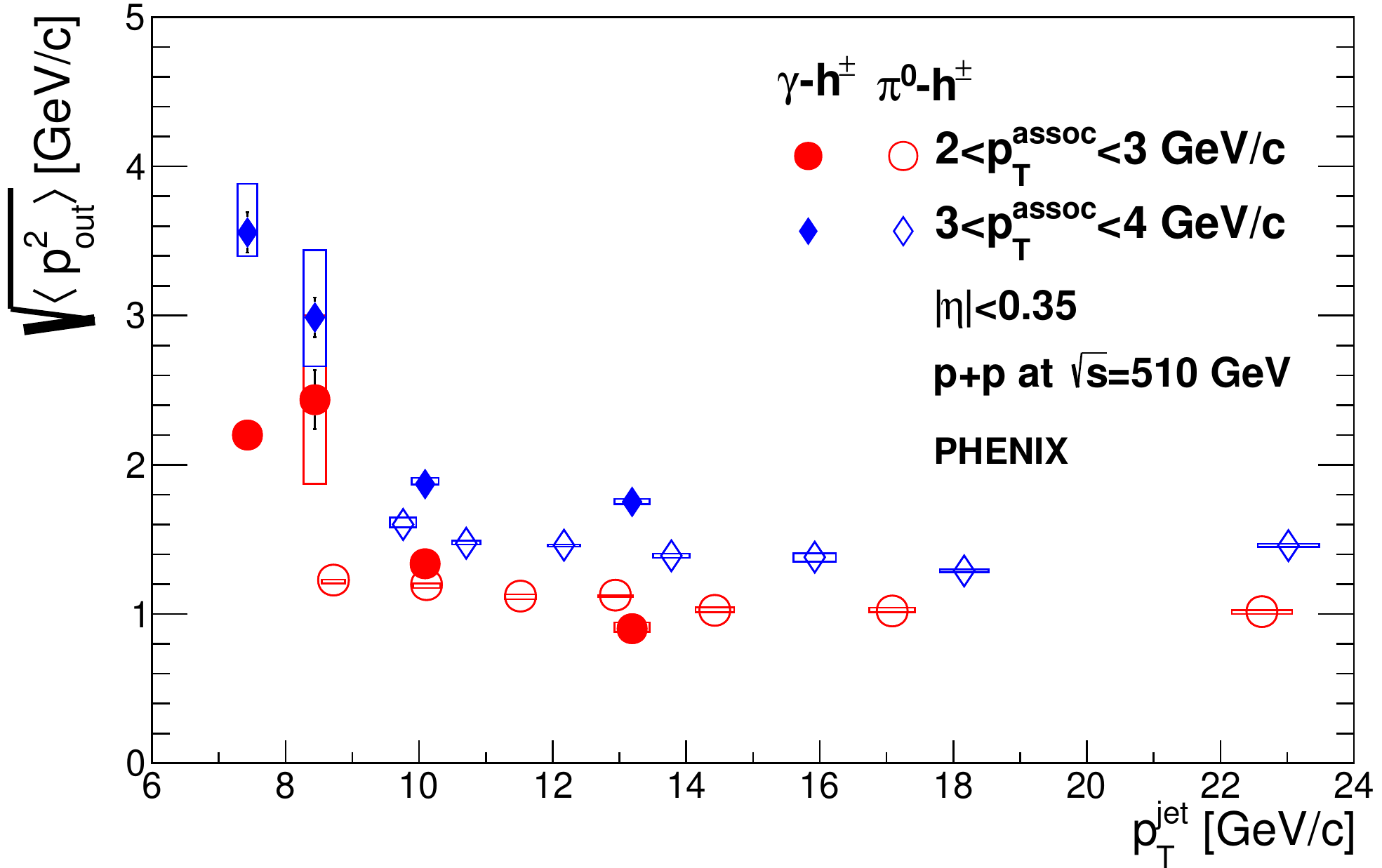}
	\caption{The quantity \rmspout is shown as a function of $p_T^{\rm jet}$ for \pion-hadron and $\gamma$-hadron correlations, where $p_T^{\rm jet}$ is defined as discussed in the text.}
	\label{fig:run13_rmspout_vsptjet}
	
\end{figure}

\begin{figure}[thb]
	\centering
	\includegraphics[width=0.7\textwidth]{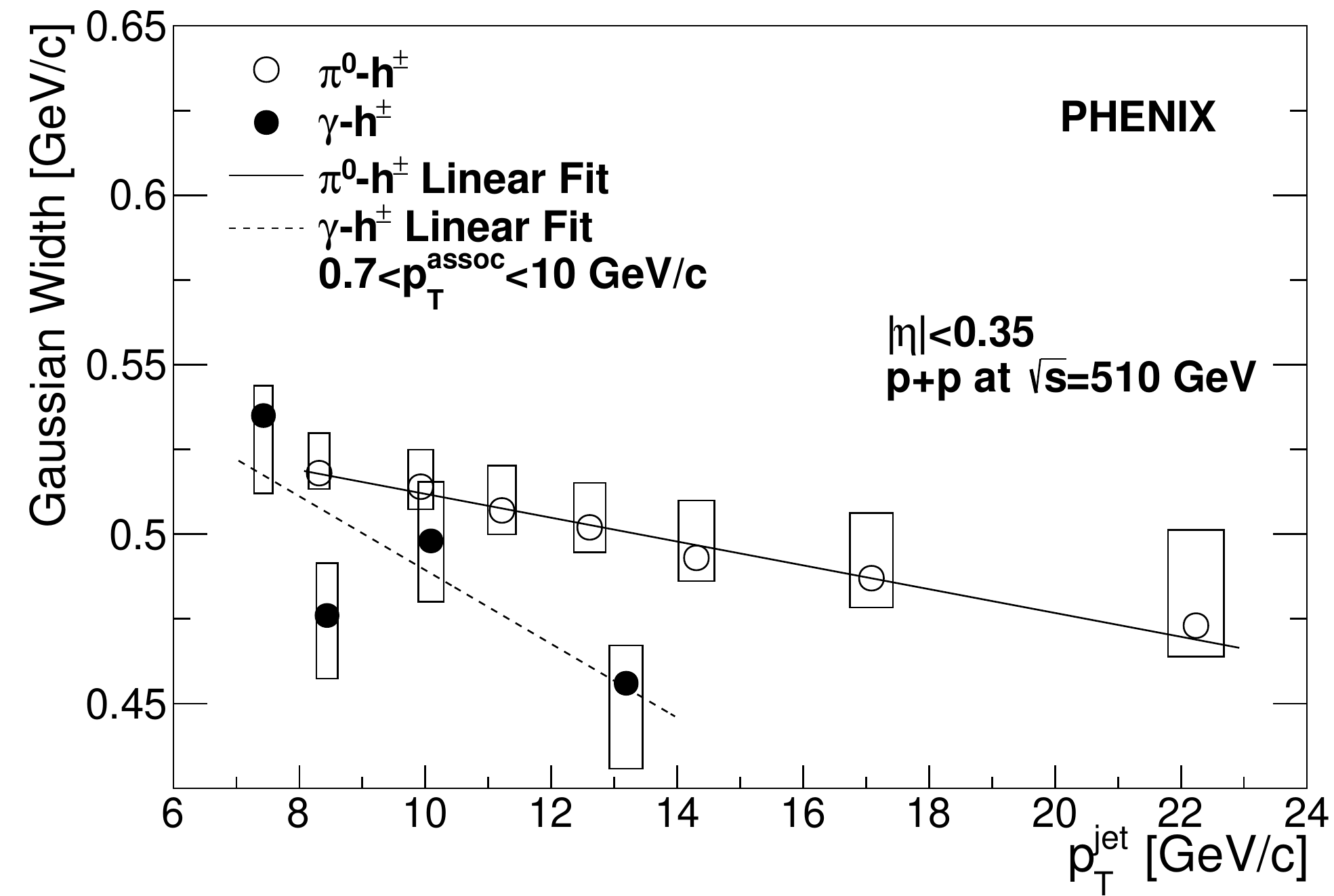}
	\caption{The Gaussian widths of \pout are shown as a function of $p_T^{\rm jet}$ for \pion-hadron and $\gamma$-hadron correlations, where $p_T^{\rm jet}$ is defined as discussed in the text.}
	\label{fig:run13_gausswidth_zt}
\end{figure}

Using these values of \zt, the Gaussian widths and \rmspout values for \pion-hadron correlations were altered to make a more apples-to-apples comparison between the \pion-hadron and $\gamma$-hadron correlations. The \zt correction amounts to shifting the \pion \pt values by a factor of two since the \zt values are roughly constant about $\sim$0.5 across \pttrig and \ptassoc. After the \zt shift is applied, the values of \rmspout appear to not form a continuous function in the two \ptassoc bins shown in Fig.~\ref{fig:run13_rmspout_vsptjet}. Here the quantity $p_T^{\rm jet}$ is just the \pt of the direct photon or the \zt corrected \pt of the leading \pion. The \zt corrected Gaussian widths of \pout are shown in Fig.~\ref{fig:run13_gausswidth_zt}, where $p_T^{\rm jet}$ is defined similarly. These clearly do not form a continuous function; although some care should be taken in the interpretation of these figures since the quantity $p_T^{\rm jet}$ is not the actual jet \pt but rather just an approximation based on \pythia simulations.

\chapter{\sqs=~200 GeV \pp and \pa Results}
\label{chap:paresults}

The previous section detailed the results of the \sqs=~510 GeV analysis and revealed intriguing yet inconclusive results. The decreasing widths of the nonperturbative momentum widths may have pointed to effects from factorization breaking; however, more comparisons needed to be performed before definitive conclusions could be drawn. The presentation of the \sqs=~200 GeV data was altered slightly from the \sqs=~510 GeV results after fruitful discussions with Prof. John Collins of Pennsylvania State University, one of the first theorists to study factorization breaking within a TMD framework~\cite{Collins:2007nk,Collins:2007jp}. In these discussions, Dr. Collins suggested that the effect of the decreasing widths seen in Ref.~\cite{ppg195} may be a result of decreasing the average away-side momentum fraction $z$ of the hadron with respect to the jet as the trigger \pt, and thus hard scale, is increased. Since the correlations were binned in a fixed \ptassoc bin, the dominant $z$ probed, if $\hat{p}_T^{\rm trig}=\hat{p}_T^{\rm assoc}$, would be approximately $0.7/13=0.05$ for the highest \pttrig bin and $0.7/7.4=0.1$ for the lowest \pttrig bin. To account for this, the correlation functions can be binned in the variable $\xe$ which was first presented in the Introduction. As a reminder, the quantity is defined as

\begin{equation}
\xe = -\frac{\pttrig\cdot\ptassoc}{|\pttrig|^2} = -\frac{|\ptassoc|}{|\pttrig|}\cos\dphi\,,
\end{equation}
where in the nearly back-to-back region such that $\cos\dphi\approx-1$ this quantity becomes $|\ptassoc|/|\pttrig|$. At midrapidity this quantity is very similar to the momentum fraction $z$ of the away-side hadron under the assumption $|\pttrig|\approx\hat{p}_T^{\rm assoc}$. This assumption will be explored further; however, it provides a first approximation to study the Gaussian widths such that the away-side hadrons are scaled similarly across different hard scales. This additionally gives the opportunity to study the \pout distributions multidifferentially in both \pout and \xe, which correspond to a transverse and longitudinal momentum component perpendicular and antiparallel to the trigger particle \pt. The previous results at \sqs=~510 GeV were integrated over the full range of \ptassoc, thus any dependence on \ptassoc could not be observed due to the kinematic restrictions on $\pout=\ptassoc\sin\dphi$ as a function of \ptassoc.

\section{\pp Results}

\subsection{\dphi Correlation Functions}

\begin{figure}[tbh]
	\centering
	\includegraphics[width=0.7\textwidth]{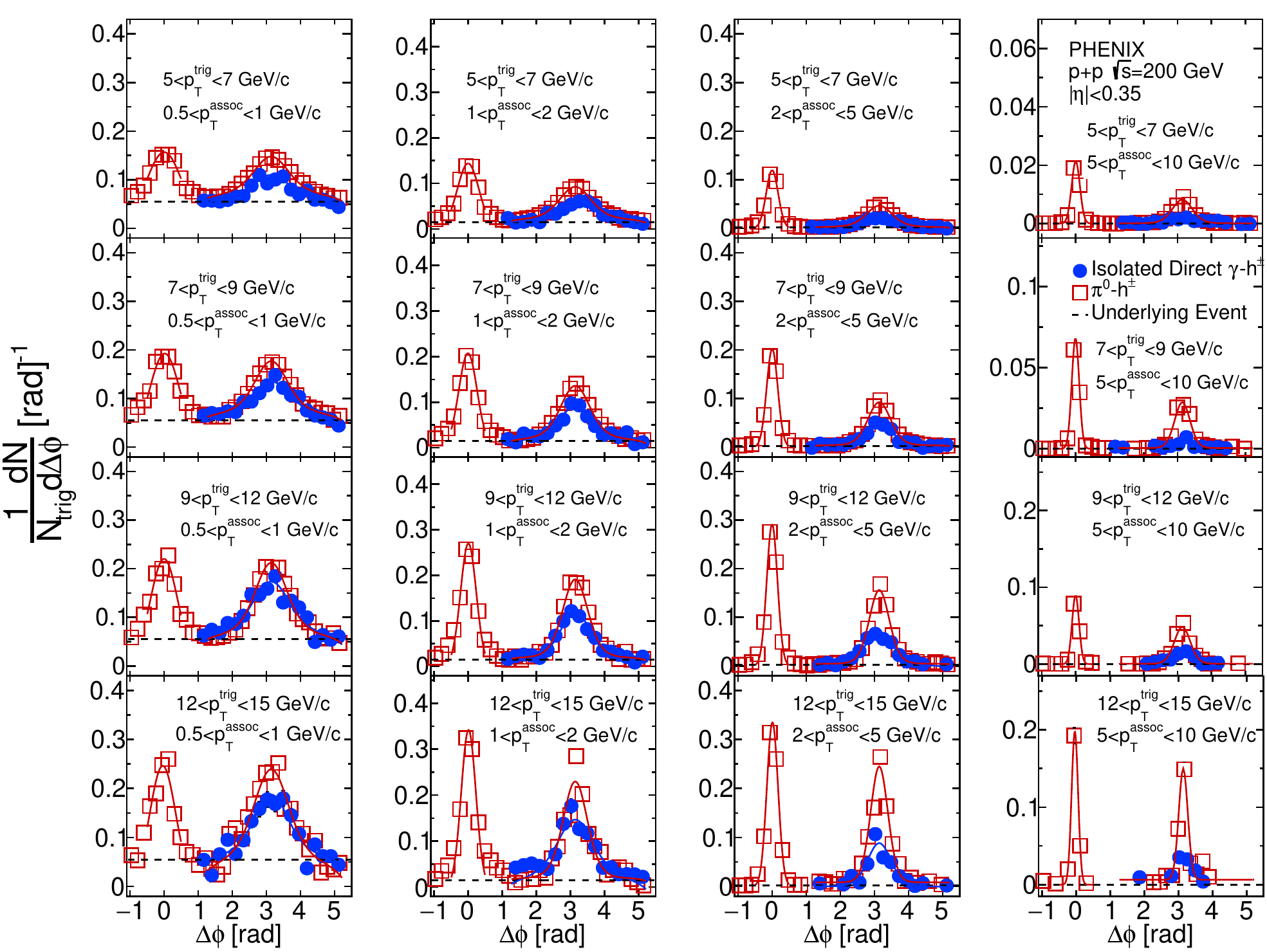}
	\caption{The correlation functions in \dphi space for \pp collisions at \sqs=~200 GeV are shown in several $\pttrig\otimes\ptassoc$ bins. The red open squares show the dihadron correlations and the blue filled circles show the isolated direct photon-hadron correlations.}
	\label{fig:dphi_corr_200gev}
\end{figure}

The correlations as a function of \dphi show the back-to-back jet structure in several $\pttrig\otimes\ptassoc$ bins in Fig.~\ref{fig:dphi_corr_200gev}. The lesser statistical precision of these correlations compared to the \sqs=~510 GeV correlations in Fig.~\ref{fig:isodppiptys} is very clear; for example, there are no counts in certain regions of \dphi at high \pt. Not only is there less integrated luminosity at \sqs=~200 GeV, but a high \pt event is less likely at the smaller center-of-mass energy simply due to the available energy in the collision. In particular, the bin sizes in \pttrig and \ptassoc are wider to accommodate the smaller statistical sample. While the \dphi correlation functions visually show the correlations, the \pout distributions are more relevant for comparisons to the previous data so that the nonperturbative and perturbative contributions can be separated. However, data tables with the \dphi correlation functions can be found in the supplemental material of Ref.~\cite{ppg217}.

\begin{figure}[tbh]
	\centering
	\includegraphics[width=0.6\textwidth]{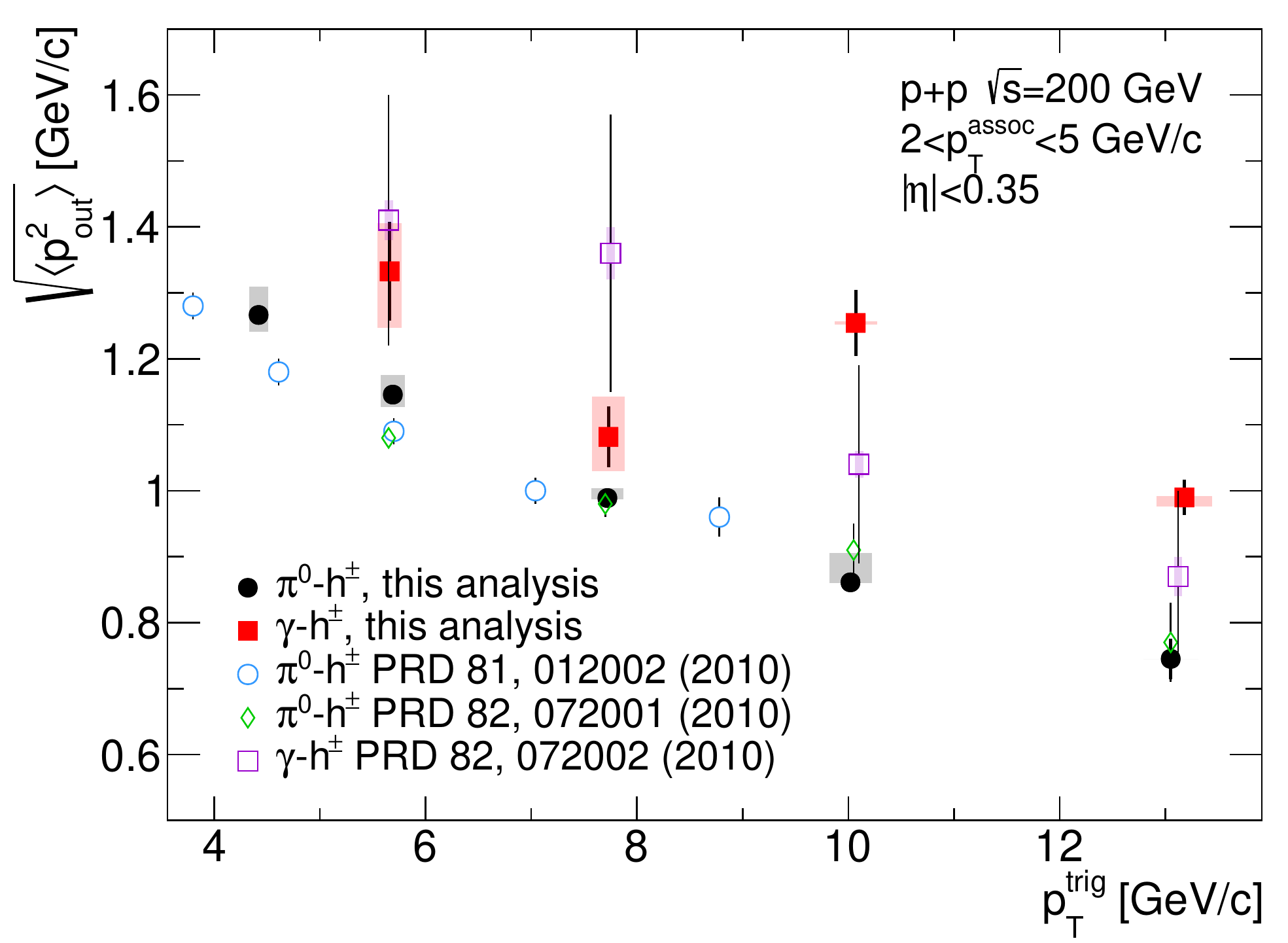}
	\caption{The quantity \rmspout is shown from this analysis and compared to previous PHENIX publications for both dihadron and direct photon-hadron correlations.}
	\label{fig:rmspout_comparison}
\end{figure}

To show consistency with the previous \sqs=~200 GeV publication, the \rmspout is extracted from the away-side \dphi distributions with a similar method to Ref.~\cite{ppg195} and described in Chapter 4. Since the away-side jet width is the physics quantity of interest, and since the \rmspout is the only quantity the previous publication reported, this is the best comparison to make. The results shown in Fig.~\ref{fig:rmspout_comparison} demonstrate good agreement within uncertainties for both the dihadron and direct photon-hadron correlations. At lower \pttrig, there appears to be some non-statistical deviation from the dihadron results of this analysis and the previous analysis. To comment on this, the previous two analyses, Ref.~\cite{ppg095} and Ref.~\cite{ppg089}, were from the exact same data set; therefore, they should yield the exact same results in the overlapping \pttrig region since the data is 100\% correlated. Secondly, the previous analyses did not report any systematic uncertainties on their values of \rmspout; in the analyses presented here we have found that the systematic uncertainty from the fit is in general larger than the corresponding statistical uncertainty from the fit, especially at low \pttrig. Therefore, there is good reason to expect that if systematic uncertainties had been assigned in the previous analyses, the results would be in better agreement.

\subsection{\pout Distributions}
\begin{figure}[tbh]
	\centering
	\vspace{-0.2cm}
	\includegraphics[width=0.6\textwidth]{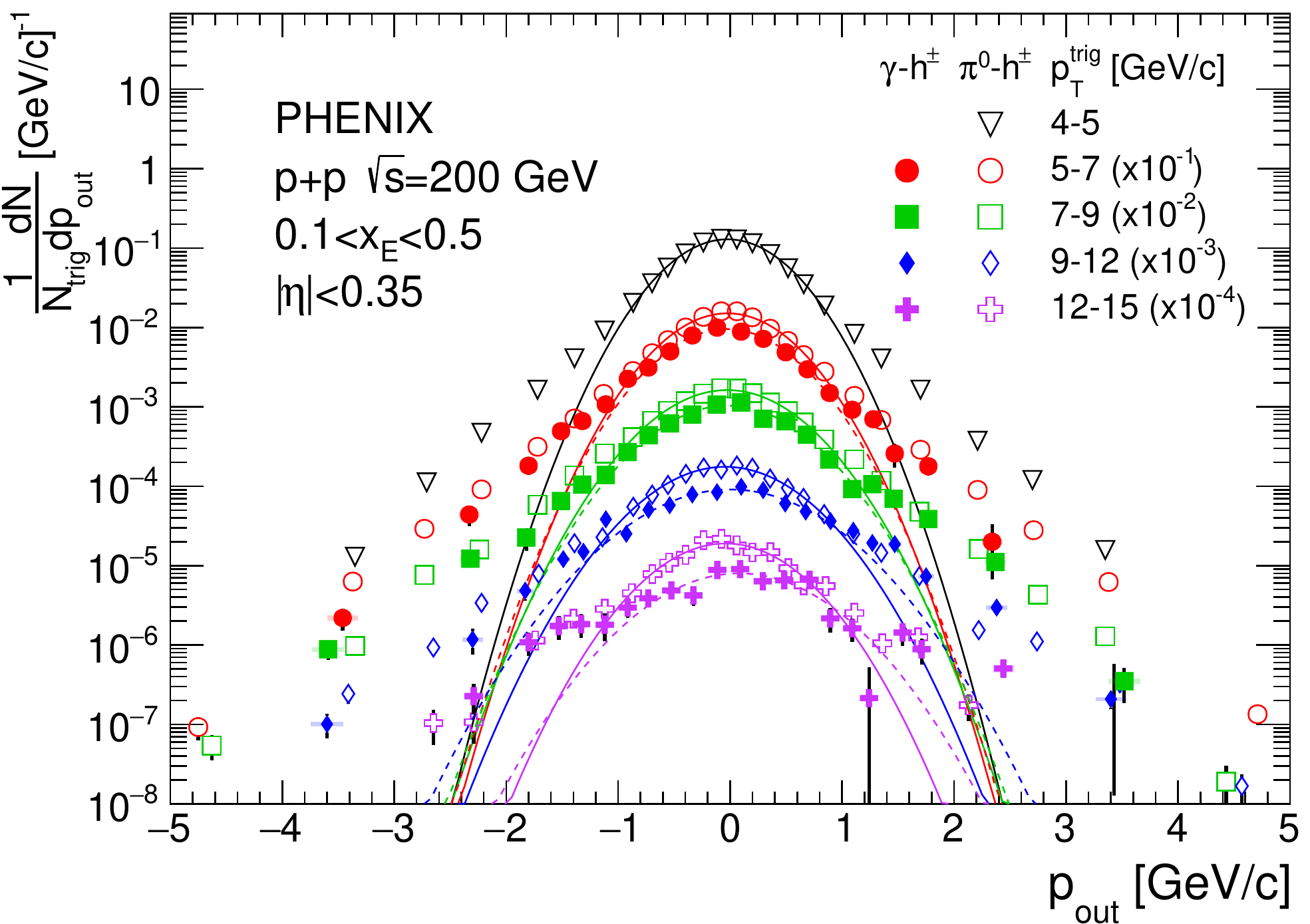}
	\caption{The \pout distributions are shown in several \pttrig bins for a fixed \xe bin between $0.1<\xe<0.5$. The open points are dihadron correlations, while the closed points are direct photon-hadron correlations.}
	\label{fig:pout_dist_fxnpttrig}

\end{figure}

Since the \pout distributions distinguish between the nonperturbative and perturbative contributions, they have sensitivity to potential TMD factorization breaking effects. As discussed previously, the \pout distributions were constructed in a fixed \xe bin so that the various correlation functions at different \pttrig could be compared with similarly scaled away-side momentum fractions. The \pout distributions for $0.1<\xe<0.5$ are shown in several \pttrig bins in Fig.~\ref{fig:pout_dist_fxnpttrig}. Similar behavior between the \sqs=~200 GeV and 510 GeV results, due to kinematics, can be seen; in particular, the direct photon-hadron correlations consistently have smaller yields in the nearly back-to-back region of \pout$\sim0$ whereas the yields are similar at large \pout between direct photon-hadron and dihadron where the NLO gluon radiation contributions begin to dominate. Note that there is no direct photon-hadron correlation for $4<\pttrig<5$ \gevc as this is below the range where PHENIX can reliably measure direct photons using isolation cones and the EMCal. Other methods have been used, especially by the heavy ion community, to measure direct photons at smaller \pt; however, these methods are not explored here since in any case a well defined hard scale must be present for a TMD framework to be applicable. 

One somewhat new feature of the \pout distributions when binned in this way is that the perturbatively generated large \pout correlations get relatively harder as the \pttrig is increased; for example, at large $\pout\sim3$ \gevc the spectra have a broader perturbative distribution as the \pttrig is increased. This is the general expectation from CSS evolution, since at the larger hard scales perturbative gluon radiation is more likely and this results in a larger acoplanarity between the two-particle pair. It is also worth noting that in the previous analysis Kaplan fits were shown with the data, while here there are no Kaplan fits applied. This is because at the smaller center-of-mass energy it is energetically less probable for large \pout correlations to occur; for this reason, the distributions actually fall off more quickly than a power law at very large \pout values. This is also dependent on the \dphi constraint placed, since by definition large \pout correlations are at $\dphi\sim\pi/2$. There is reason to expect this behavior as dijet cross sections as a function of \dphi display a similar trend~\cite{Abazov:2004hm}. In this reference, NLO calculations describe the large \dphi correlations reasonably well, and it is clear by eye that at very large \dphi they do not follow an exact power law shape.

One immediate benefit to binning the \pout distributions in \xe rather than a fixed \ptassoc range is that the distributions can be studied multidifferentially as a function of both the trigger \pt and associated hadron \pt. Due to the definition of $\pout=\ptassoc\sin\dphi$, \pout is limited kinematically by the range of \ptassoc one looks at. For this reason, if the \pout distributions were constructed in a bin of, for example, $1<\ptassoc<2$ \gevc, the \pout distributions would necessarily be 0 at values of \pout~$>$~2 \gevc; thus, the transition from perturbative to nonperturbative behavior may not be clear. However, when binned in \xe, the full range of \pout can be probed for a fixed range of \ptassoc. For example, for $5<\pttrig<7$ \gevc and $0.1<\xe<0.3$, in the nearly back-to-back region \ptassoc is thus limited to approximately $0.5<\ptassoc<2.1$ \gevc. Nonetheless, the full range of \pout can be probed since a particular hadron could have a large \pout, or a large \ptassoc, but a small \xe if it is at a very wide angle from $\dphi=\pi$. In the back-to-back region, the \ptassoc bin becomes more constrained for a given \pttrig bin since $\xe\approx\ptassoc/\pttrig$ when $\dphi\sim\pi$. Tables of the \pout distributions can be found in the supplemental material of Ref.~\cite{ppg217}.

\begin{figure}[tbh]
	\centering
	\includegraphics[width=0.6\textwidth]{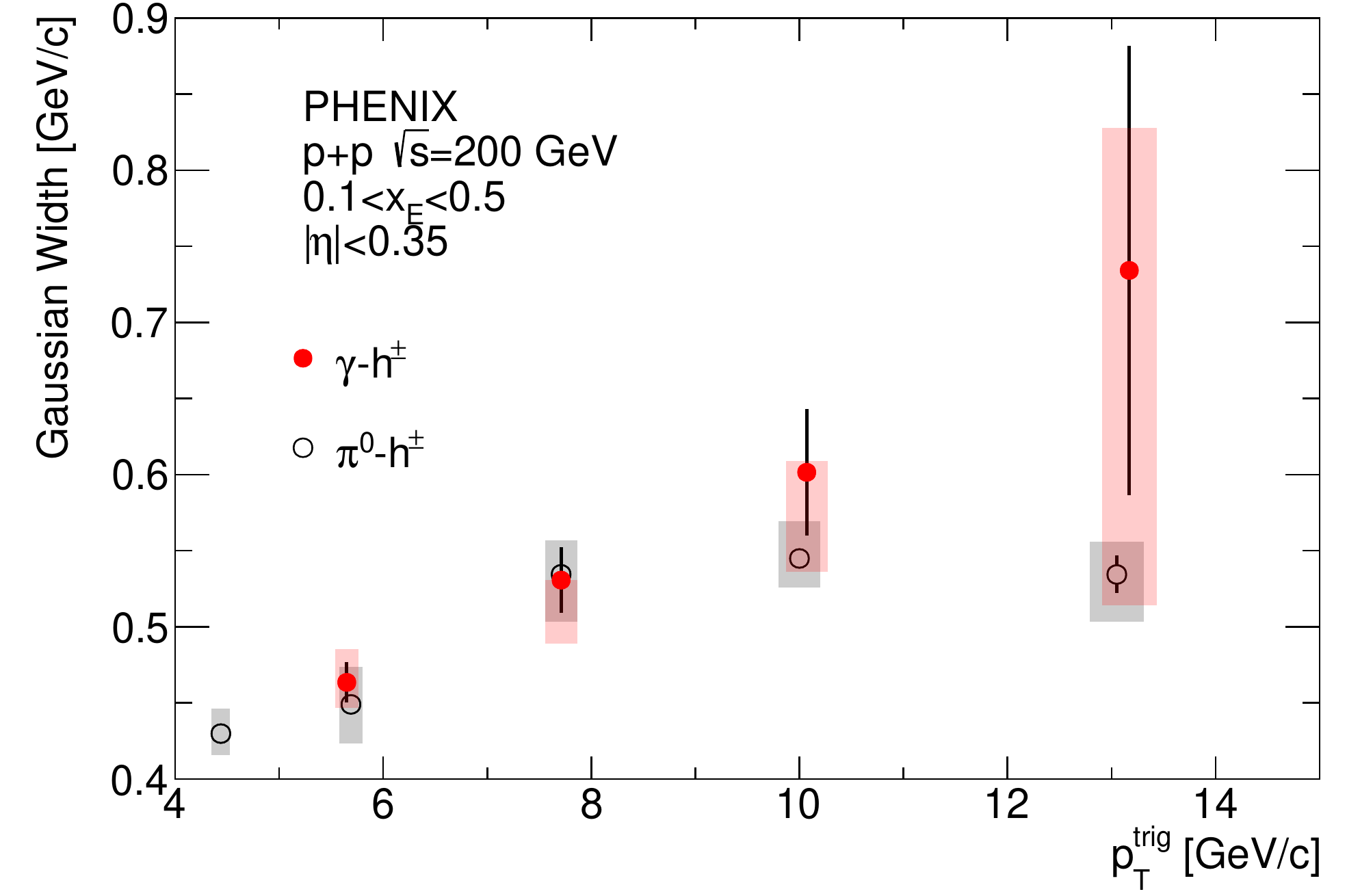}
	\caption{The Gaussian widths of \pout are shown for dihadron and direct photon-hadron correlations in \sqs=~200 GeV \pp collisions for a fixed bin of \xe.}
	\label{fig:gausswidths_200gev_pttrig} 

\end{figure} 

To quantify the evolution of the nonperturbative widths with the interaction hard scale, the \pout distributions were fit with Gaussian functions in the small \pout region similarly to the \sqs=~510 GeV analysis as described in Chapter 4. However, when the distributions are in a fixed bin of \xe, the fit region that best characterizes the nonperturbative behavior strongly depends on the \pttrig bin; for example, the distributions are still Gaussian at $\pout\sim1.5$ for $12<\pttrig<15$ \gevc, while the distributions are no longer Gaussian at this value of \pout for $4<\pttrig<5$ \gevc. This can be seen by eye to some degree by looking directly at the \pout distributions for a fixed \pout and scanning down the various \pttrig bins. The extracted Gaussian widths are shown in Fig.~\ref{fig:gausswidths_200gev_pttrig} for both dihadron and direct photon-hadron correlations. Systematic uncertainties on the widths are evaluated by adjusting the fit region by $\pm$0.2 \gevc and taking the absolute difference of the resulting Gaussian width. The figure clearly indicates that when the \xe bin is fixed rather than \ptassoc, the nonperturbative momentum widths increase with \pttrig. This is in contrast to Ref.~\cite{ppg195}; therefore, this shows that the average momentum fraction $z$ of the away-side hadron was decreasing in the previous analysis and was the cause of the decrease of the widths as a function of \pttrig. When accounting for the momentum fraction of the away-side hadron with the quantity \xe, the behavior is qualitatively similar to that from Drell-Yan measurements where TMD factorization is expected to hold.

\begin{figure}[tbh]
	\centering
	\includegraphics[width=0.6\textwidth]{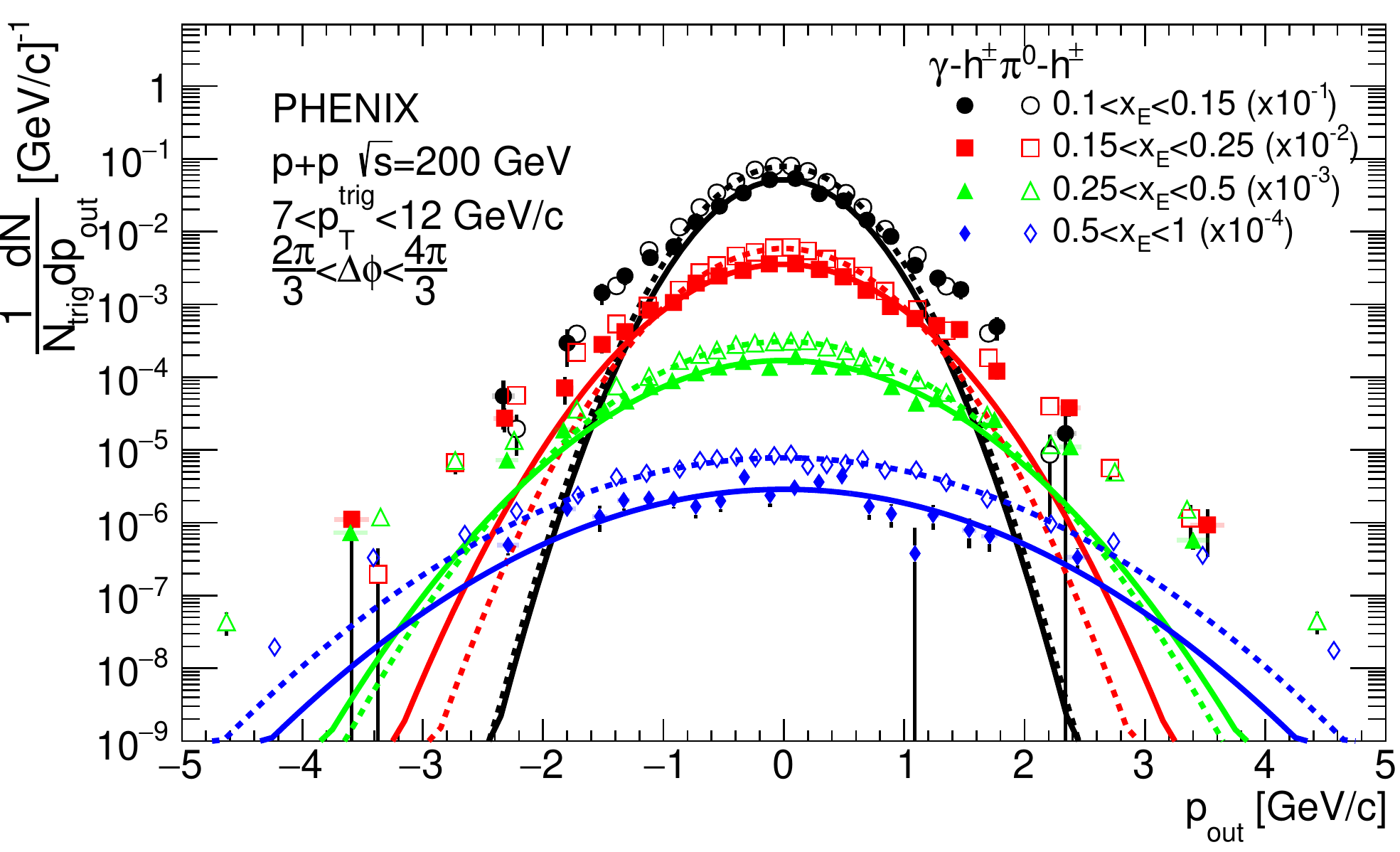}
	\caption{The \pout distributions are shown integrated over $7<\pttrig<12$ \gevc in several bins of \xe for both dihadron and direct photon-hadron correlations.}
	\label{fig:pout_fxn_xe}

\end{figure}

\begin{figure}[tbh]
	\centering
	\includegraphics[width=0.6\textwidth]{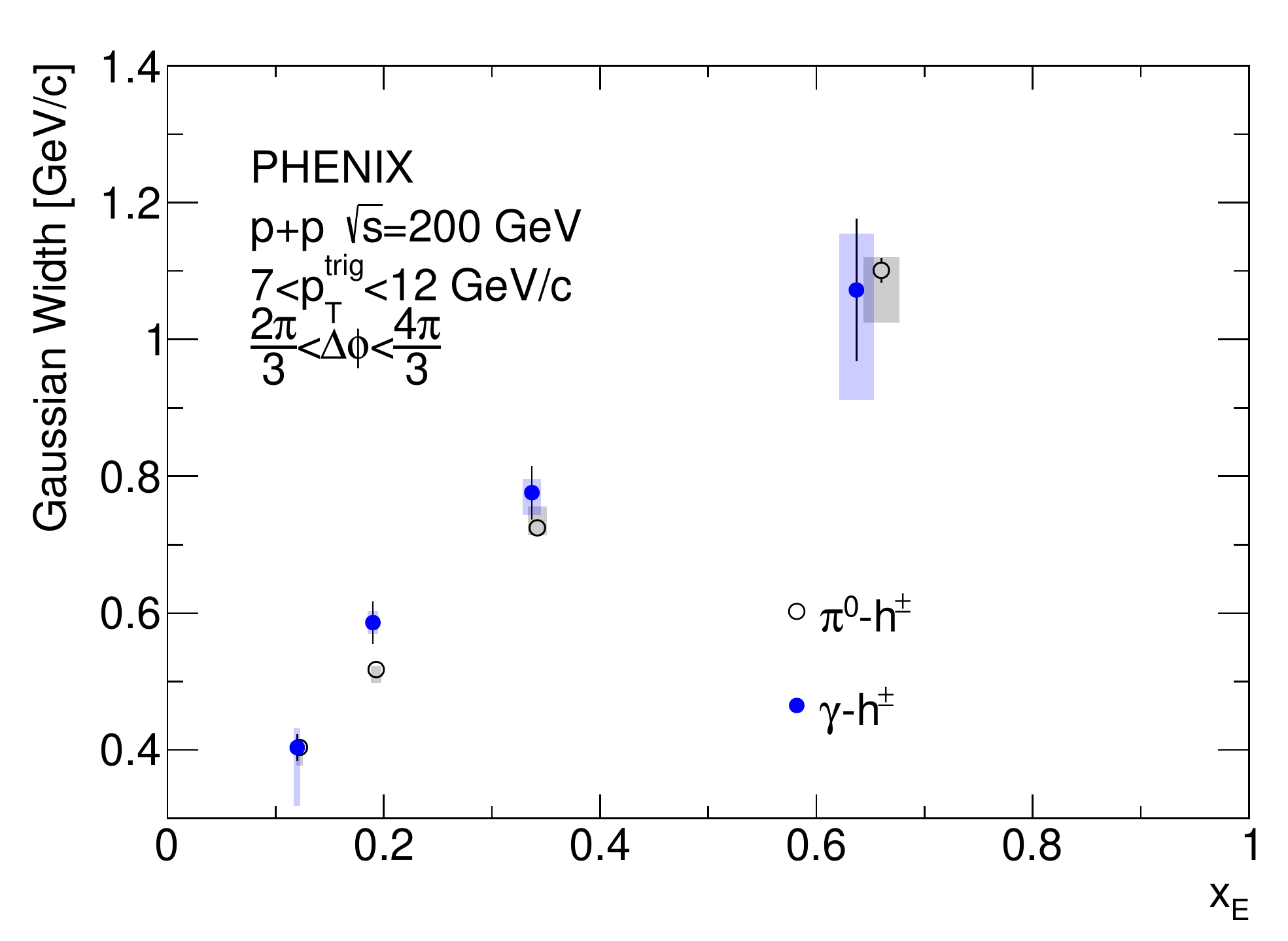}
	\caption{The Gaussian widths are extracted from Fig.~\ref{fig:pout_fxn_xe} and shown as a function of \xe for both dihadron and direct photon-hadron correlations in \pp collisions at \sqs=~200 GeV.}
	\label{fig:gausswidths_fxn_xe}
\end{figure}

To study the dependence of the nonperturbative momentum widths as a function of \xe, the \pout distributions can be constructed in bins of \xe and integrated over a region of \pttrig. Unfortunately, there is not the necessary statistical precision to measure the distributions simultaneously as a function of \pttrig and \xe. To have sufficient statistical precision but with a well defined hard scale, the distributions were constructed integrated over $7<\pttrig<12$ \gevc. This bin was also chosen for direct comparisons to the previous \sqs=~510 GeV results, which will be discussed later. The \pout distributions are shown in Fig.~\ref{fig:pout_fxn_xe} where now each marker indicates a particular \xe bin for both dihadron and direct photon-hadron correlations. It is important to point out that although the binning is limited to \xe=~1, this is not because there is some kinematic limit at \xe=~1; in fact, for both direct photon-hadron and dihadron correlations, pairs with $\xe>1$ exist. However, at these values, the statistical precision becomes very poor and thus the two-dimensional structure cannot be probed with any reliable accuracy. As a function of \xe, the nonperturbative structure is still clearly defined from the perturbative structure; however, the range which encapsulates the nonperturbative dynamics greatly depends on the \xe bin probed which can be clearly seen by the validity of the Gaussian fit over a \pout range of several \gevc in the largest \xe bin. The nonperturbative widths are extracted from the fits to the \pout distributions as a function of \xe, and are shown in Fig.~\ref{fig:gausswidths_fxn_xe}. The nonperturbative widths show a similar qualitative behavior as a function of \xe to the widths as a function of \pttrig; they are also consistent between dihadron and direct photon-hadron correlations.

\begin{figure}[tbh]
	\centering
	\includegraphics[width=0.49\textwidth]{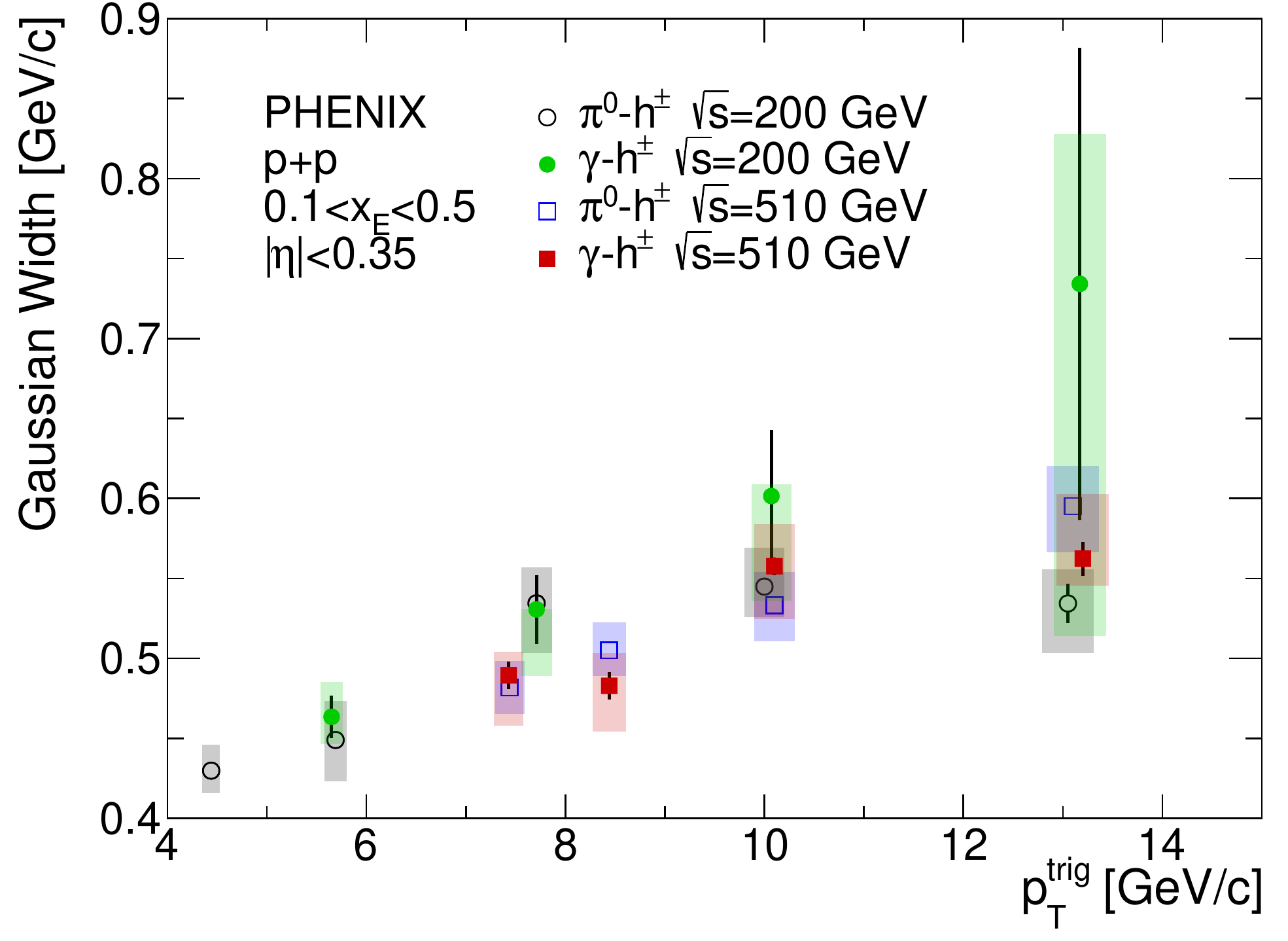}
	\includegraphics[width=0.49\textwidth]{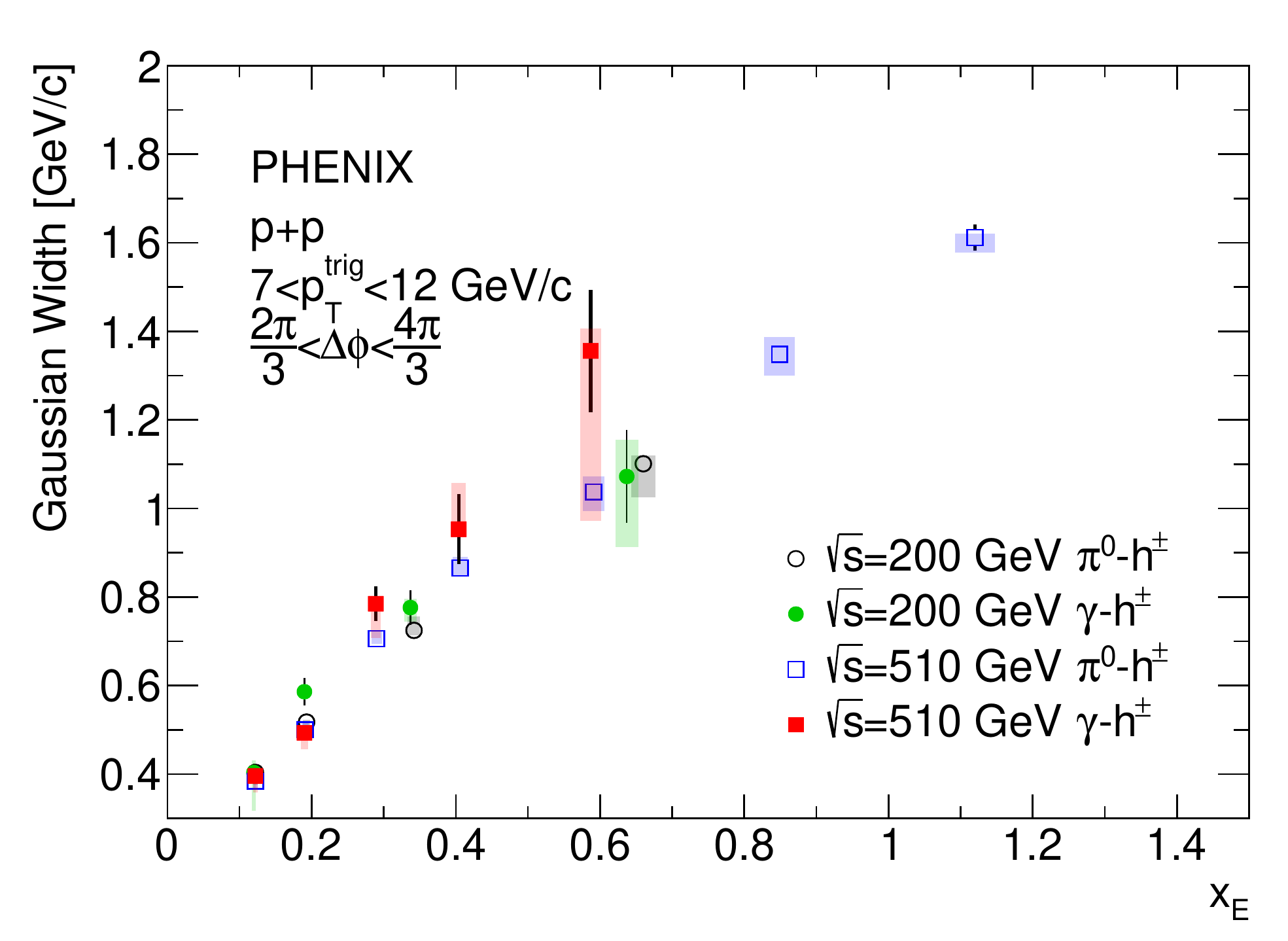}
	\caption{The Gaussian widths in a fixed \xe bin as a function of \pttrig (left) and fixed \pttrig bin as a function of \xe (right) are shown for both \sqs=~200 and 510 GeV \pp collisions.}
	\label{fig:gausswidths_sqs}
\end{figure}

One of the motivations for measuring the correlations in both \sqs=~200 and 510 GeV is that the correlations on average probe different values of the initial partonic momentum fractions $x$ at the two different center-of-mass energies; therefore, the correlations can be studied over a larger range of partonic kinematic variables. For central pseudorapidities, the situation is further simplified in that in general $x_1\sim x_2$ from each incoming proton. To compare the results, the \sqs=~510 GeV data was rebinned in a fixed bin of \xe corresponding to the bin in the \sqs=~200 GeV data in Fig.~\ref{fig:gausswidths_200gev_pttrig}. Only the correlations with $\pttrig>7$ \gevc were reanalyzed in the \sqs=~510 GeV data; this is because the analysis was limited to associated hadrons with $\ptassoc>0.7$ \gevc. For the minimum \xe cut of 0.1 to be unbiased, the \pttrig must be greater than 7 \gevc only in the \sqs=~510 GeV data. The Gaussian widths are shown for both center-of-mass energies and both dihadron and direct photon-hadron correlations in Fig.~\ref{fig:gausswidths_sqs} as a function of \pttrig and \xe. Interestingly, the Gaussian widths show little dependence on the center-of-mass energy as a function of both \xe and \pttrig. Similar conclusions have been drawn by the STAR collaboration at RHIC, where polarized TMD observables are consistent within uncertainties between \sqs=~200 and 500 GeV~\cite{Adamczyk:2017wld,Adamczyk:2017ynk}.

\begin{figure}[tbh]
	\centering
	\includegraphics[width=0.6\textwidth]{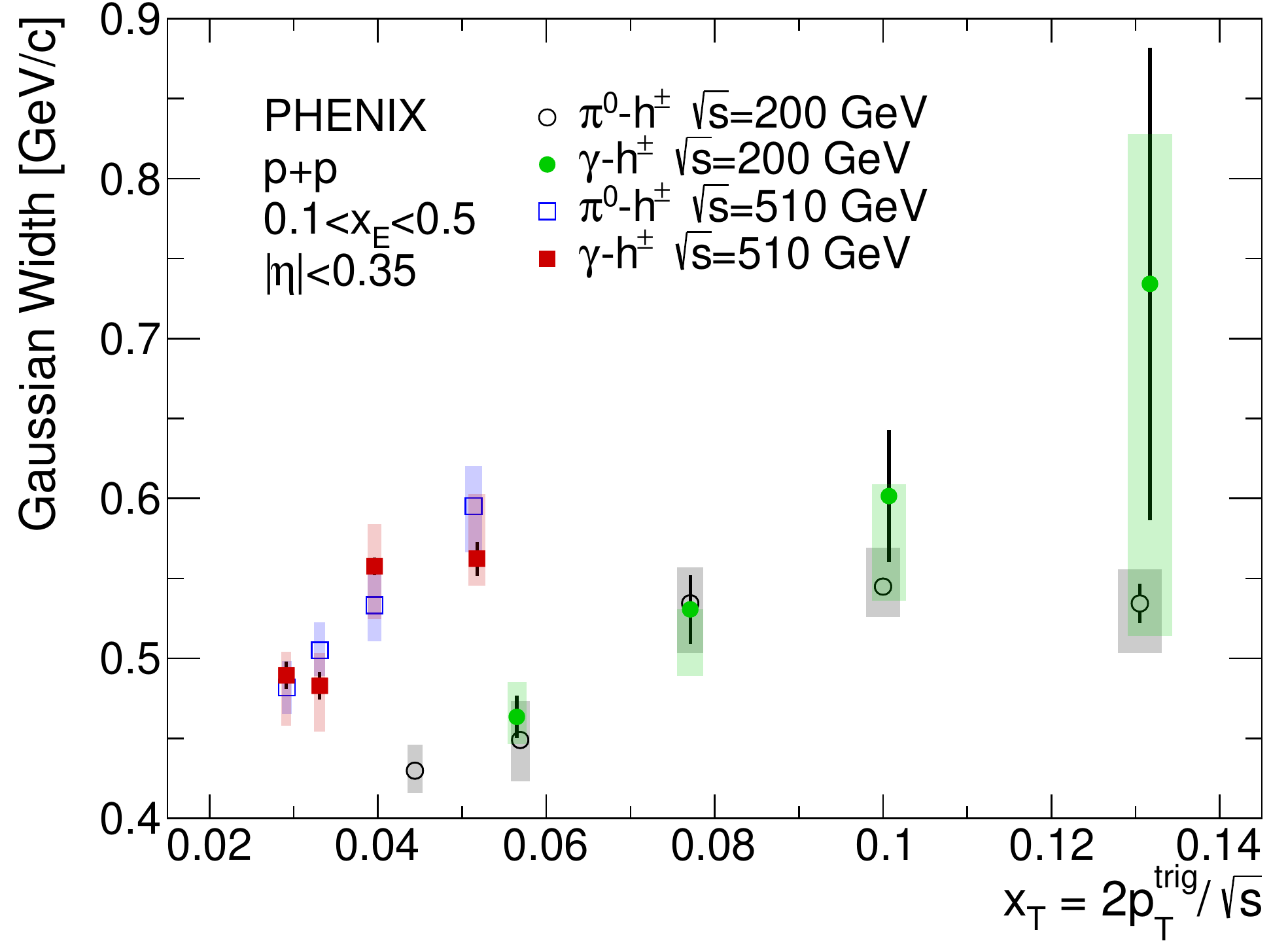}
	\caption{The Gaussian widths are shown as a function of \xt=~2~\pttrig/\sqs for both dihadron and direct photon-hadron correlations.}
	\label{fig:gausswidths_xt}
\end{figure}

To probe the nonperturbative widths as a function of the partonic momentum fraction, the widths can be plotted as a function of $x_T=2\pttrig/\sqs$ rather than \pttrig. While it has no exact relation to $x$, \xt provides a proxy for studying the nonperturbative functions as a function of the partonic momentum fraction. The nonperturbative momentum widths are shown as a function of \xt for the two center-of-mass energies in Fig.~\ref{fig:gausswidths_xt}. The Gaussian widths clearly do not scale with \xt; however, they appear to show qualitatively similar behavior to DY interactions in that the widths are larger at larger center-of-mass energies for a fixed value of \xt. This behavior as a function of $\sqrt{\tau}=x_1x_2$ and \sqs can be observed in TMD momentum widths measured from DY data (see e.g.~\cite{Ito:1980ev}). However, it is interesting to point out that in DY at a similar hard scale nonperturbative momentum widths clearly rise with \sqs, while in the measurements shown here the nonperturbative momentum widths are consistent with each other as a function of \pttrig and \sqs. This may be due to effects from factorization breaking; however, it may also be related to the fact that the \qsq in the correlations is actually the invariant mass of the dijet or photon-jet pair and not \pttrig. 

To summarize, the \sqs=~200 GeV \pp collision results have led to two important conclusions. The first conclusion is that the decreasing nonperturbative momentum widths measured in Chapter 4 were due to the comparison of different momentum fraction $z$ away-side hadrons across the different \pttrig bins. This is concluded from Fig.~\ref{fig:gausswidths_200gev_pttrig} which show that the momentum widths increase with \pttrig for a fixed bin of \xe, which is a proxy for the away-side momentum fraction $z$. Secondly, the comparison between the nonperturbative momentum widths across center-of-mass energies in Fig.~\ref{fig:gausswidths_xt} shows qualitatively similar behavior to DY nonperturbative momentum widths. Therefore, further phenomenological studies will be necessary to fully understand factorization breaking effects quantitatively. Thus, more data should be analyzed to provide additional constraints to phenomenological studies.

\section{\pa Results}\label{pa_results}

While the theoretical prediction for TMD factorization breaking was made for \pp collisions, factorization is predicted to be broken in back-to-back angular correlations in \pal or \pau collisions since these are also hadronic collisions. Since the underlying physical phenomenon which leads to factorization breaking is soft gluon exchanges between the partons involved in the hard scattering and remnants, it might naively be expected that TMD factorization breaking effects may be even larger since there are many nucleons, and thus additional partons and potentially stronger color fields, in the nucleus which may exchange gluons with the partonic hard scattering. However, other nuclear effects can contribute in \pa collisions that are not necessarily present in \pp collisions. 

There are several nuclear effects in \pa collisions that could contribute to the nonperturbative transverse momentum width in addition to any possible factorization breaking effects. Firstly, in \pa collisions nuclear PDFs (nPDFs) must now be considered. nPDFs are known to be phenomenologically different from the naive expectation of superpositions of nucleons~\cite{Eskola:2016oht}, and the knowledge of nPDFs is currently only at the collinear level. In particular, the nuclear modification depends strongly on $x$, and the so-called shadowing and anti-shadowing regions (see e.g. Refs.~\cite{Geesaman:1995yd,Hen:2016kwk}) may lead to different effects in a TMD framework. Therefore, the possibility exists that nPDFs contain additional TMD $k_T$ broadening from interactions of partons with other nucleons in the nucleus;  the nuclear $k_T$ has been found to increase with the atomic number~\cite{Corcoran:1990vq}. Similarly, there could be TMD fragmentation effects from hard scattered partons interacting with the nuclear environment that may lead to broadened $j_T$ compared to \pp collisions. 

In addition to TMD effects, there are several other effects that are more frequently studied in high energy nucleus-nucleus collisions. For example, long-range pseudorapidity correlations could contribute, manifesting themselves in terms of $\cos2\dphi$ and $\cos3\dphi$ modulated amplitudes in a Fourier decomposition of the two-particle correlation function; these have been measured to be nonzero in \pa collisions~\cite{Aidala:2016vgl,ATLAS_pPb_collectivity,CMS_pPb_collectivity,Aidala:2017ajz}. There may also be contributions from various energy loss mechanisms in the nucleus. Energy loss within nuclear media was first studied several decades ago~\cite{Baier:2000mf,Baier:1996sk} in both ``cold'' and ``hot'' nuclear matter, where the distinction refers to a system where a ``hot'' QGP is or is not formed. In this sense, multiple scattering of partons within a nuclear medium could lead to energy loss in several ways: partons could be elastically scattered such that there is angular broadening but the total momentum remains the same, partons could also lose energy via inelastic collisions in the nucleus, and there could be radiative energy loss from the parton in the nuclear medium due to gluon bremsstrahlung. There could additionally be multiple partonic hard scatterings that contribute; however, multiple partonic interactions can also occur in \pp collisions. 

There is additionally the so-called ``Cronin'' effect, which simply refers to the empirically observed enhancement of the inclusive hadron spectra in \pa collisions when compared to \pp collisions at intermediate \pt of approximately $2<\pt<7$ \gevc~\cite{Adler:2006wg,Cronin:1974zm}. This was first attributed to multiple scattering of partons in the nuclear medium; however, this interpretation is too simple and recent identified particle results have shown that additional physical interpretation is necessary~\cite{Adare:2013esx}. It is likely that the physical origin of the enhancement is somehow related to the various effects discussed above. For example, the measurement of larger \kt in nuclear collisions was in a similar kinematic region to the measured Cronin enhancement~\cite{Corcoran:1990vq}. Since this effect refers to the modification of inclusive spectra, correlations of dihadrons may provide additional information about the physical origin of this enhancement in the single particle \pt spectra.

\subsection{\dphi Correlation Functions}

\begin{figure}[tbh]
	\centering
	\includegraphics[width=0.8\textwidth]{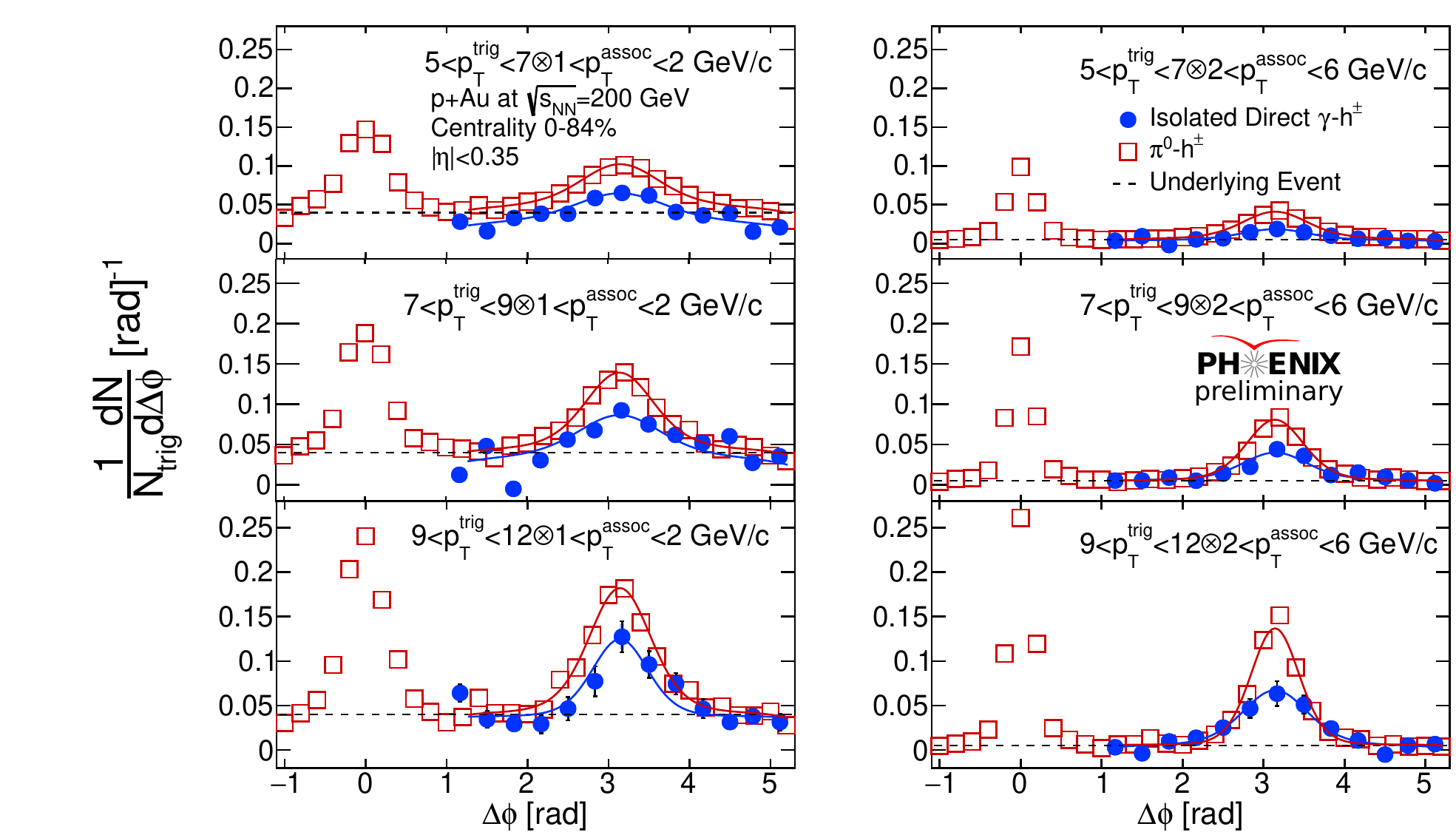}
	\caption{Dihadron and direct photon-hadron correlations are shown in \pau collisions at \sqsn=~200 GeV in several $\pttrig\otimes\ptassoc$ bins.}
	\label{fig:pau_dphis}
\end{figure}

\begin{figure}[tbh]
	\centering
	\includegraphics[width=0.7\textwidth]{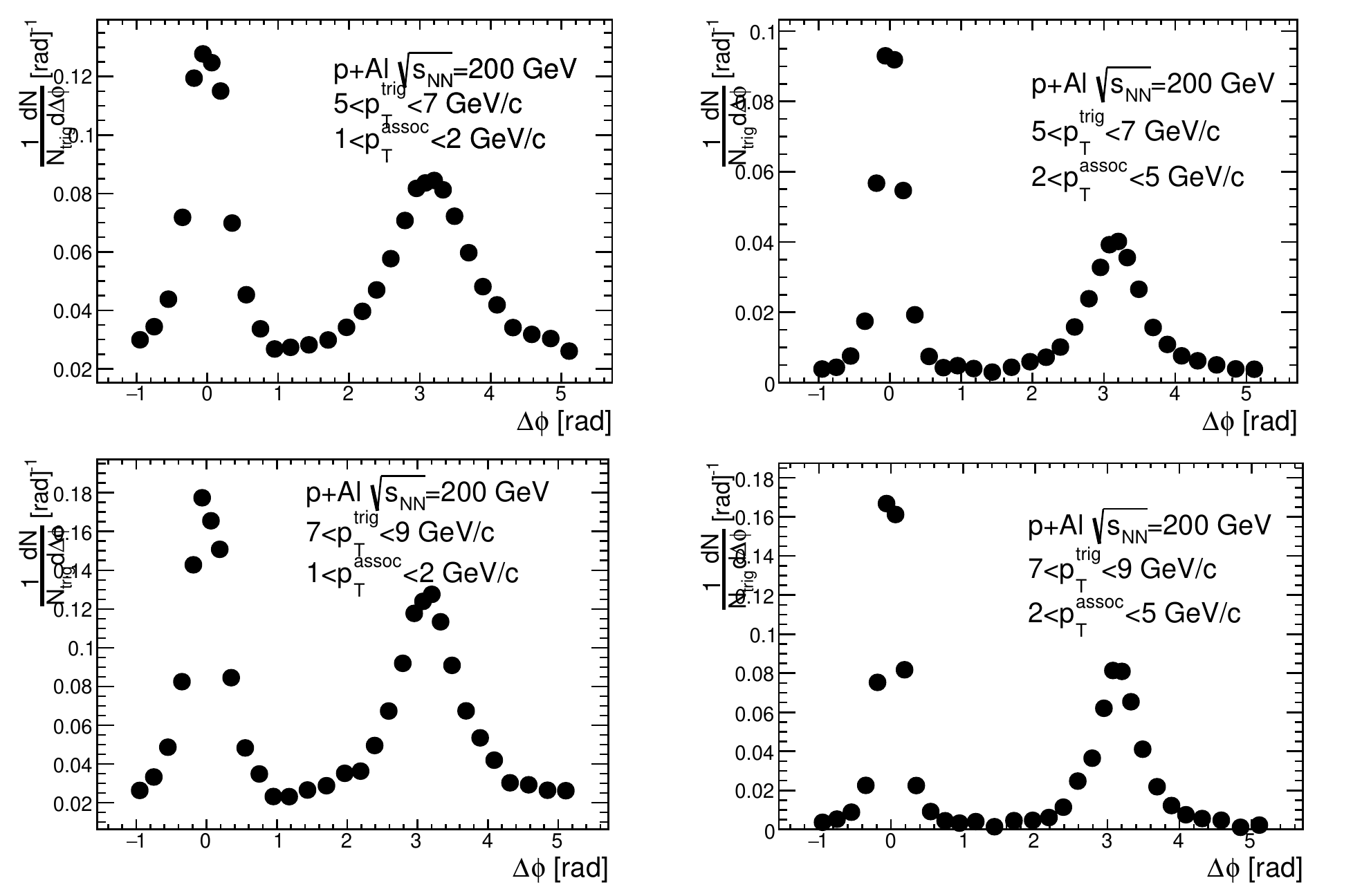}
	\caption{Several examples of dihadron correlations are shown in \pal collisions at \sqsn=~200 GeV.}
	\label{fig:pal_dphis}
\end{figure}

The \dphi correlations for the \pau dihadron and direct photon-hadron correlations are shown in Fig.~\ref{fig:pau_dphis} for a small set of $\pttrig\otimes\ptassoc$ bins. The statistical precision of the direct photon-hadron \pau data is clearly limited; the bin sizes in \dphi are wider than the previous analyses. However, the dihadron correlations show good statistical precision over the full range of \dphi and $\pttrig\otimes\ptassoc$. The underlying event contribution in \pau collisions is clearly much larger than in \pp collisions; the signal to underlying event background in the direct photon-hadron correlations is quite small whereas in the dihadron correlations the signal to background is roughly 50\% in the away-side peak region. The statistical precision of the \pal data set allowed only for dihadron correlations to be measured, several of which are shown in Fig.~\ref{fig:pal_dphis}. Again the effect of the underlying event from uncorrelated hadrons can be seen; for example, comparing the two top left panels of Figs.~\ref{fig:pau_dphis} and~\ref{fig:pal_dphis} the underlying event in \pau is roughly 0.05 charged hadrons per trigger, while in \pal the value is roughly 0.025.

\begin{figure}[tbh]
	\centering
	\includegraphics[width=0.6\textwidth]{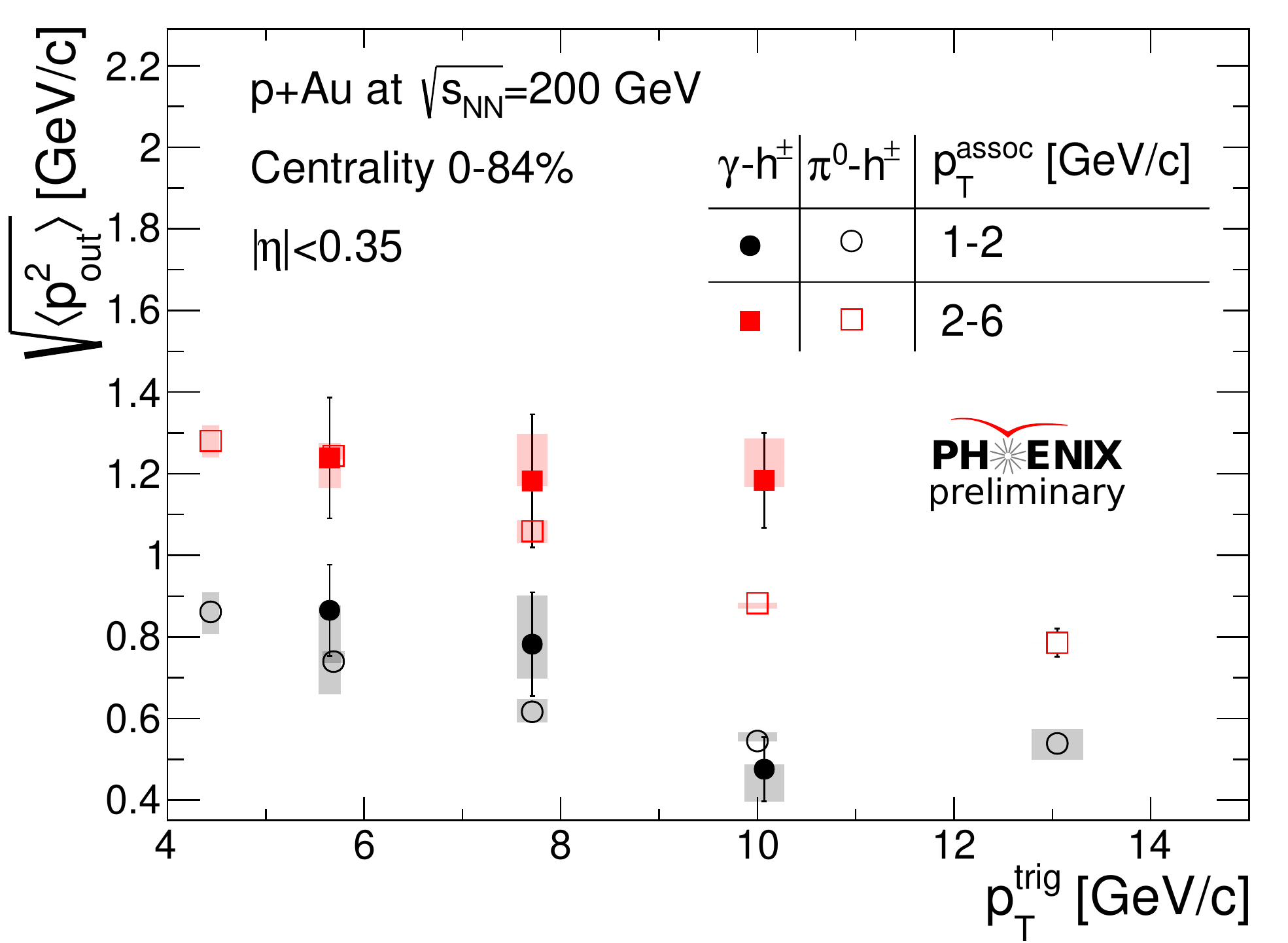}
	\caption{The quantity \rmspout is shown for several \ptassoc bins in \pau collisions at \sqsn=~200 GeV. Open points are dihadron correlations, while filled points are direct photon-hadron correlations.}
	\label{fig:pau_rmspout} 
\end{figure}

\begin{figure}[tbh]
	\centering
	\includegraphics[width=0.6\textwidth]{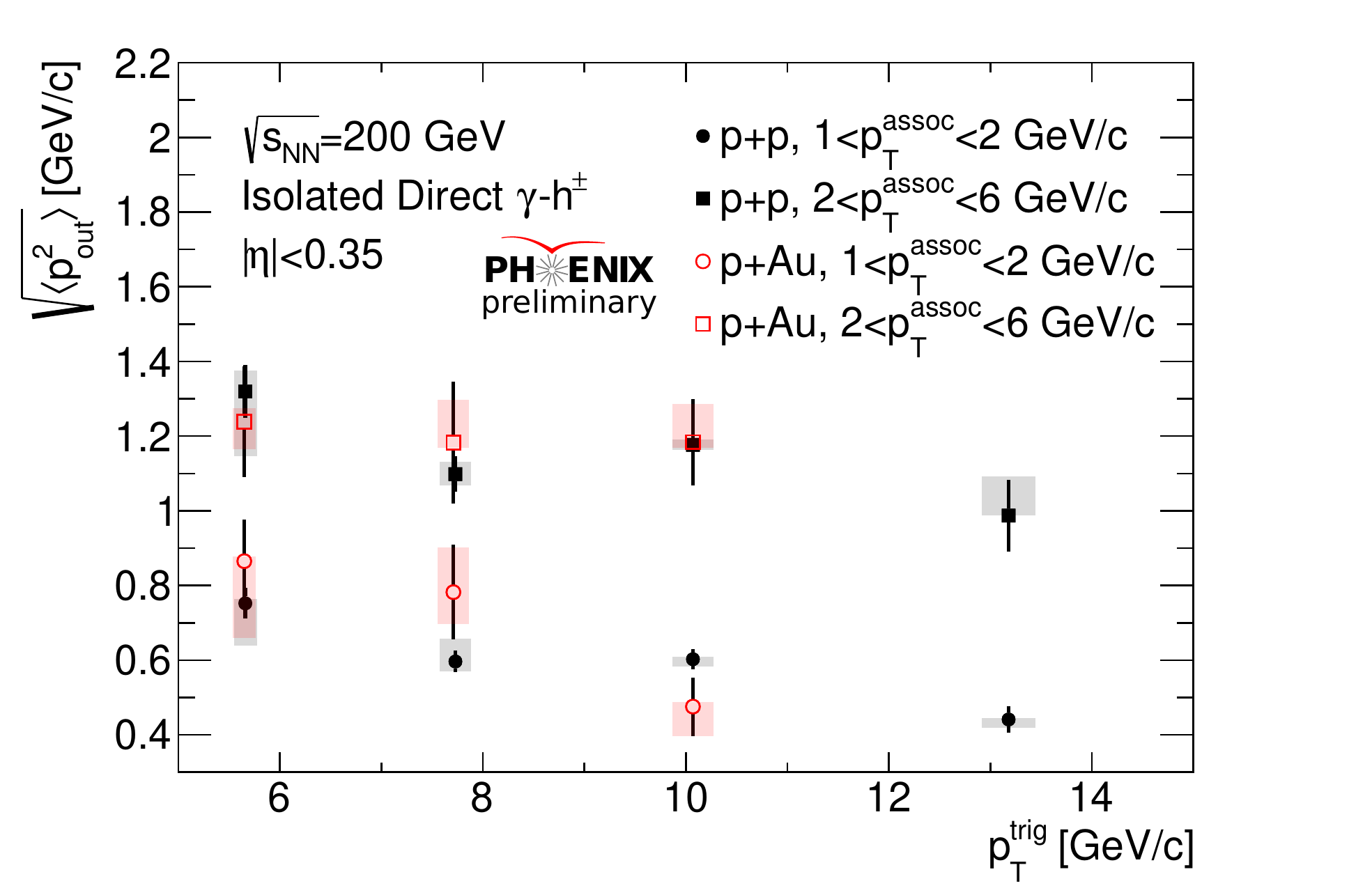}
	\caption{The \pp and \pau direct photon-hadron \rmspout values are shown in several \ptassoc bins. They are consistent within uncertainties.}
	\label{fig:pp_pau_dp_rmspout}

\end{figure}

The \rmspout is extracted from the \dphi correlations similarly to the previous \pp analyses, and the \pau results are shown in Fig.~\ref{fig:pau_rmspout}. The values are mostly consistent between direct photon-hadron and dihadron correlations; however, this may be due to the poor statistical precision of the direct photon-hadron correlations as based on previous measurements the direct photon-hadron \rmspout values are expected to be larger than the corresponding dihadron values. This in turn gives less precision on the away-side jet width. Nonetheless, the direct photon-hadron \rmspout values between \pp and \pau at \sqsn=~200 GeV are shown in Fig.~\ref{fig:pp_pau_dp_rmspout} in two \ptassoc bins. The values are consistent within uncertainties between \pau and \pp collisions, which will be an important consideration when the dihadron \rmspout values are compared between \pa and \pp.

\begin{figure}[tbh]
	\centering
	\includegraphics[width=0.6\textwidth]{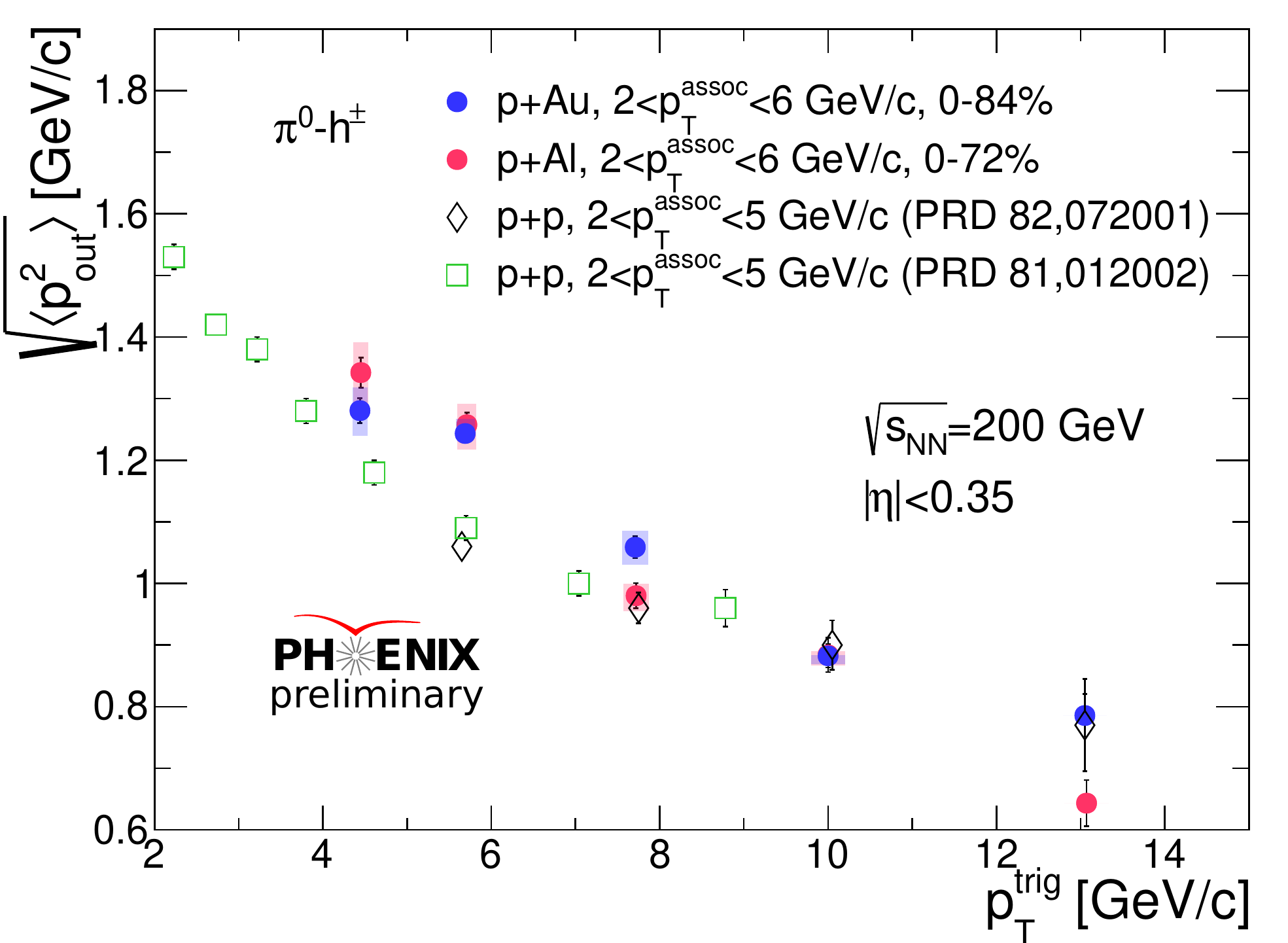}
	\caption{The measured \pau and \pal \rmspout values are shown as compared to the published \pp values at \sqsn=~200 GeV.}
	\label{fig:pa_rmspout_dihadrons}
\end{figure}

The comparison between the dihadron \rmspout values in \pa collisions and a wide range of published \pp values at intermediate \ptassoc is shown in Fig.~\ref{fig:pa_rmspout_dihadrons}. The plot indicates that there is an enhancement of the \rmspout values in \pau and \pal when compared to \pp collisions in the $4<\pttrig<7$ \gevc range. Interestingly this is the same region where the inclusive \pion \pt spectra show an enhancement when compared to \pp collisions, and thus these correlations may provide additional indications as to the physical mechanism for the Cronin peak. The enhancement in the \rmspout values being related to the inclusive \pt spectrum enhancement is also supported by the fact that the associated hadrons are in the range $2<\ptassoc<6$ \gevc, which is also in the range of the Cronin peak. Thus, both the leading trigger \pion and the associated hadrons are likely experiencing whatever the physical mechanism is that leads to the Cronin peak. Again, as two-particle correlations provide more information than inclusive hadron spectra, these results may provide insights into the physical origin of the enhancement in the inclusive hadron spectrum.

\subsection{\pout Distributions}

\begin{figure}[tbh]
	\centering
	\includegraphics[width=0.6\textwidth]{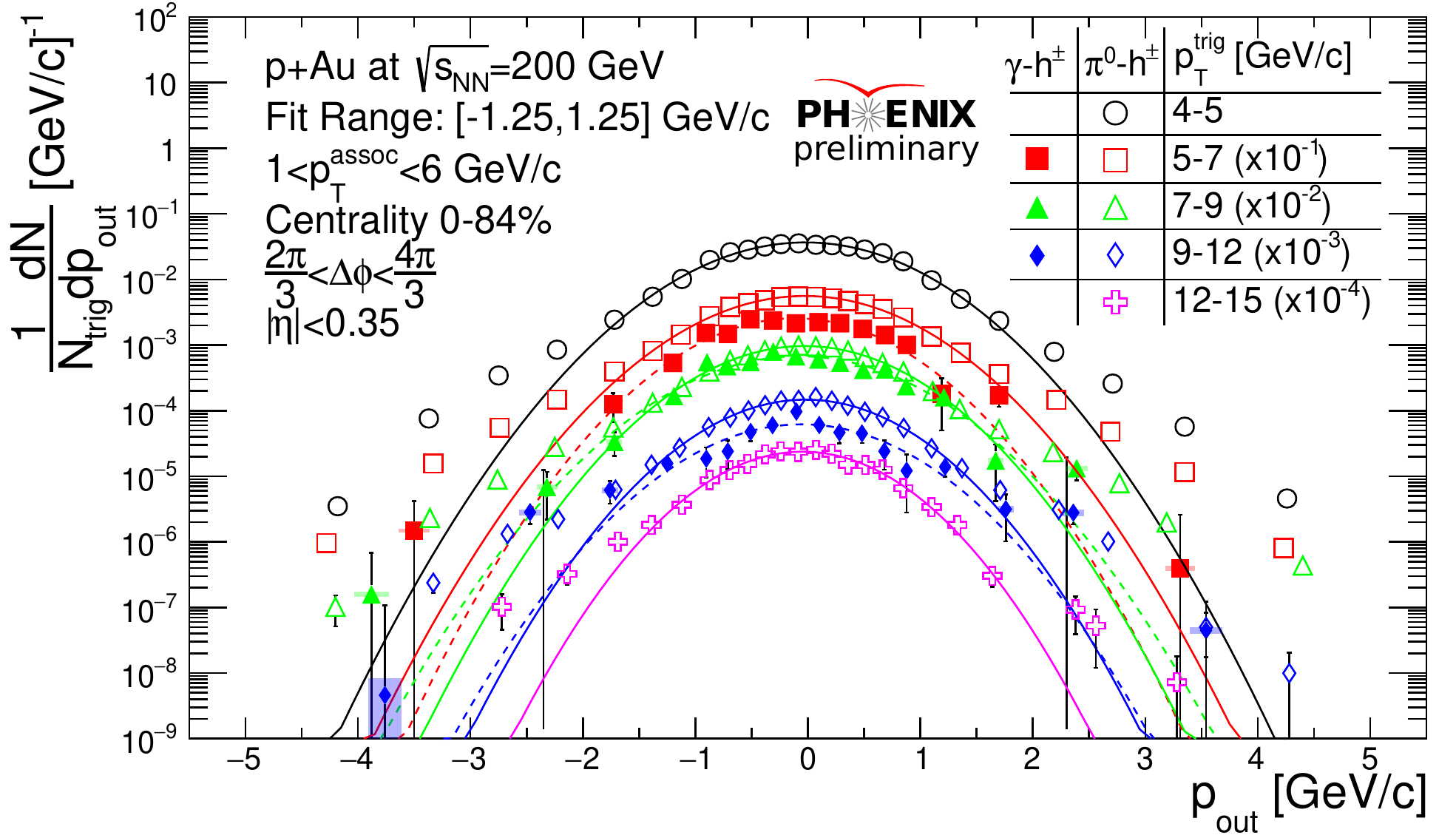}
	\caption{The \pout distributions in \pau collisions are shown in a fixed \ptassoc bin for direct photon-hadron and dihadron correlations in several \pttrig bins. Gaussian fits are also shown in the small \pout region.}
	\label{fig:pau_pout_dist}
\end{figure}

The \pout distributions for dihadron and direct photon-hadron correlations in \pau collisions are shown in Fig.~\ref{fig:pau_pout_dist}. The distributions in \pau also show the transition from nonperturbative structure at small \pout to perturbative structure at large \pout. In the \pau collisions, the small \ptassoc region of $0.5<\ptassoc<1$ \gevc was dominated by background in the direct photon-hadron events, so this motivated the lower limit of \ptassoc=~1 \gevc. While these \pout distributions are shown in a fixed \ptassoc bin following the analysis of Ref.~\cite{ppg195}, the distributions as a function of \xe will be shown and studied following these results.

\begin{figure}[tbh]
	\centering
	\includegraphics[width=0.49\textwidth]{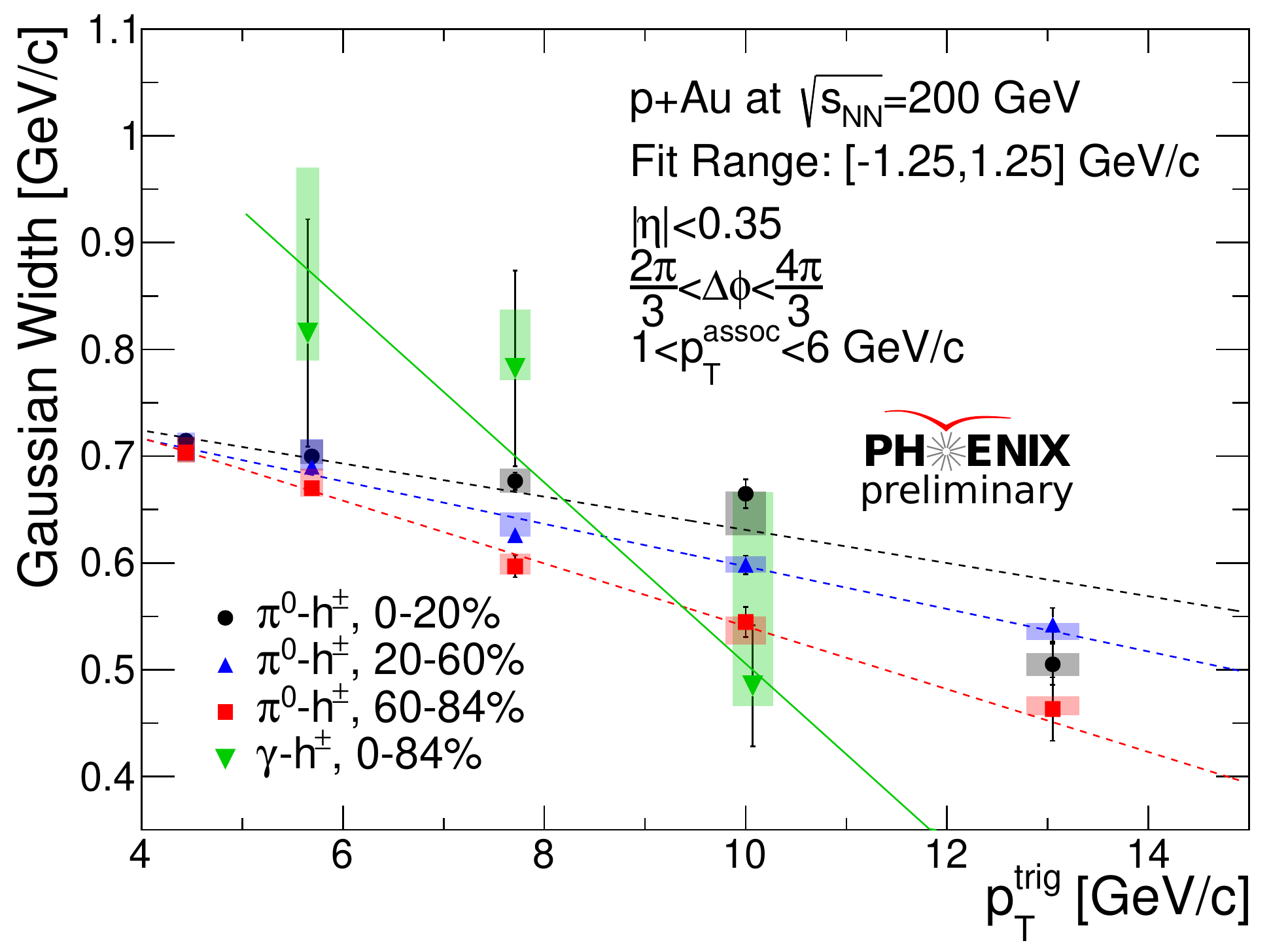}
	\includegraphics[width=0.49\textwidth]{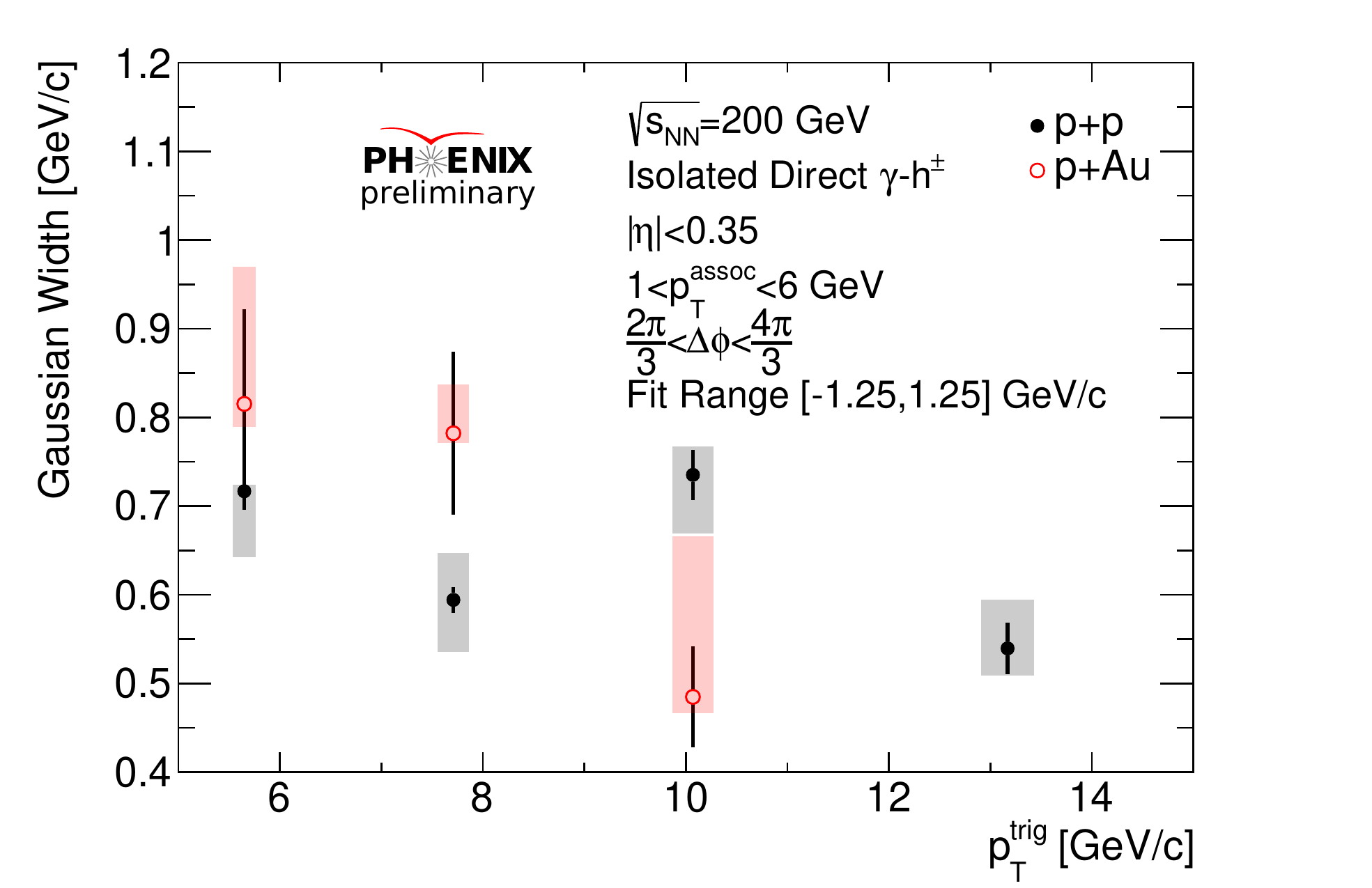}
	\caption{The Gaussian widths of \pout in \pau collisions are shown for dihadron and direct photon-hadron correlations (left). The dihadron points are shown as a function of centrality as well, where 0-20\% (60-84\%) indicates larger (smaller) final-state multiplicities. The direct photon-hadron Gaussian widths are compared in \pau and \pp and are consistent within uncertainties (right).}
	\label{fig:pau_gausswidths}
\end{figure}

The Gaussian widths are extracted from the fits for both types of correlations and the widths are shown in Fig.~\ref{fig:pau_gausswidths}. Note that the dihadron correlations have the statistical precision to be studied as a function of centrality, which is a proxy for the final-state multiplicity in \pa collisions; these \pout distributions are not explicitly shown in Fig.~\ref{fig:pau_pout_dist}, however they show similar features to the centrality integrated case. The widths exhibit similar behavior to the \sqs=~510 GeV analysis; in particular, the direct photon-hadron correlations are systematically larger than the dihadron correlations due to the smaller hard scale probed. However, it is clear from the figure that the direct photon-hadron correlations are significantly statistically limited. The dihadron correlations show a clear centrality dependence, where more central events have in general larger Gaussian widths as a function of \pttrig. The widths are shown with linear fits to highlight this trend. The right panel of Fig.~\ref{fig:pau_gausswidths} shows the direct photon-hadron Gaussian widths in \pau and \pp; within rather large uncertainties, they are consistent.

The observed centrality dependence in the dihadron correlations will be explored further; however, it is important to point out the advantages of the dihadron correlations over the direct photon-hadron correlations in \pau. Direct photon-hadron correlations are highly valued because of the simplicity of photon reconstruction as well as the fact that the photon is largely unmodified by any QCD interactions; since it does not couple via the strong force, it will emerge mostly unmodified from the initial partonic hard scattering. In the case of \pa collisions, dihadron correlations may be more desirable because any effect from QCD interactions is expected to be very small. For example, DY measurements in \pa collisions have placed upper limits on cold nuclear matter energy loss of order $\mathcal{O}(100)$ MeV$^2$~\cite{Vasilev:1999fa}. Thus, dihadron correlations may actually be preferred to direct photon-hadron correlations because of the additional QCD interactions both hadrons undergo when traversing any kind of nuclear medium. In addition to the much higher cross section, both the near and away side hadrons will interact with the medium which may give better sensitivity to nuclear effects when compared to direct photon-hadron correlations where only the away-side hadron interacts strongly.

\subsection{Centrality Dependence}

\begin{figure}[tbh]
	\centering
	\includegraphics[width=0.49\textwidth]{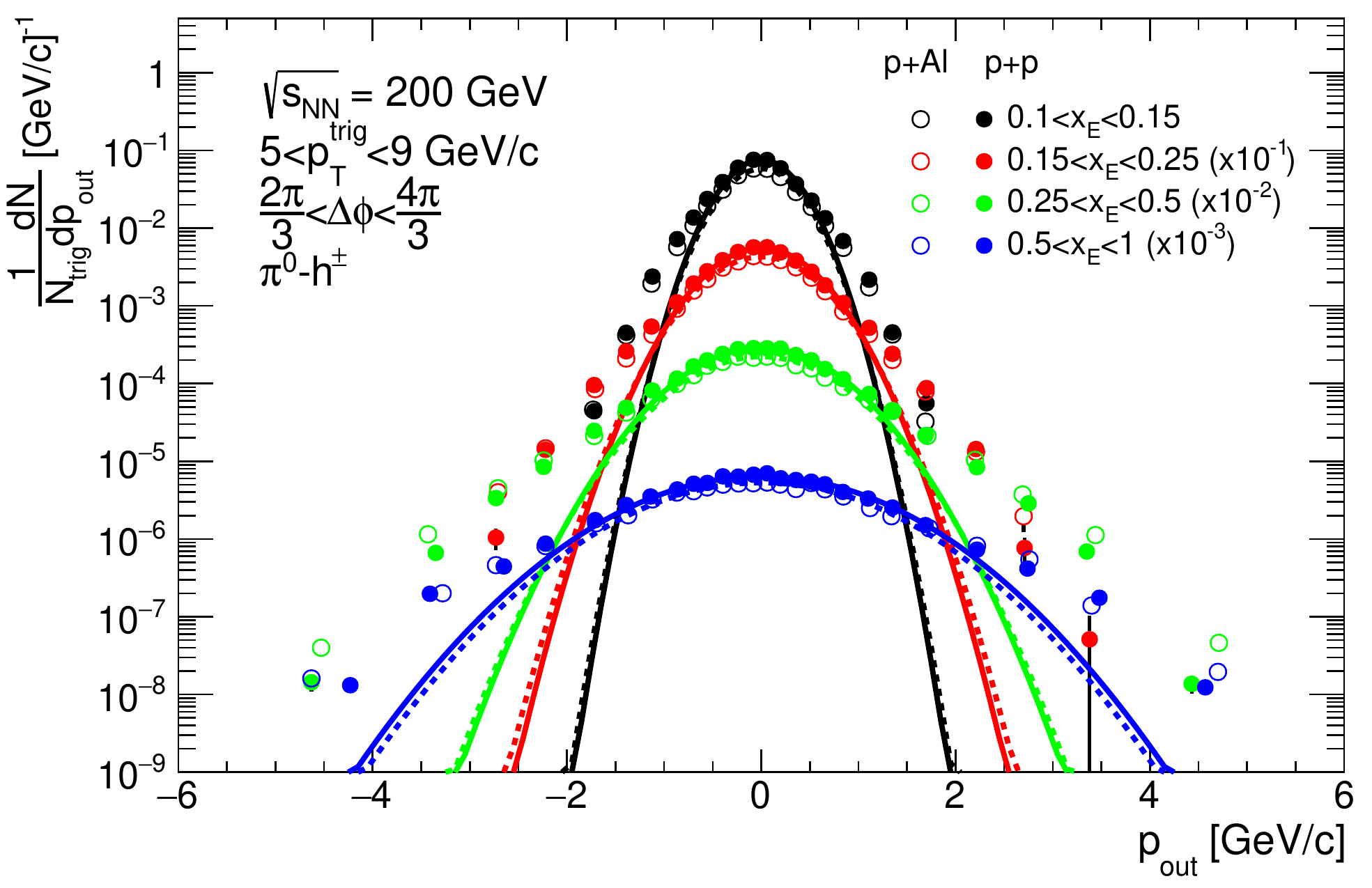}
	\includegraphics[width=0.49\textwidth]{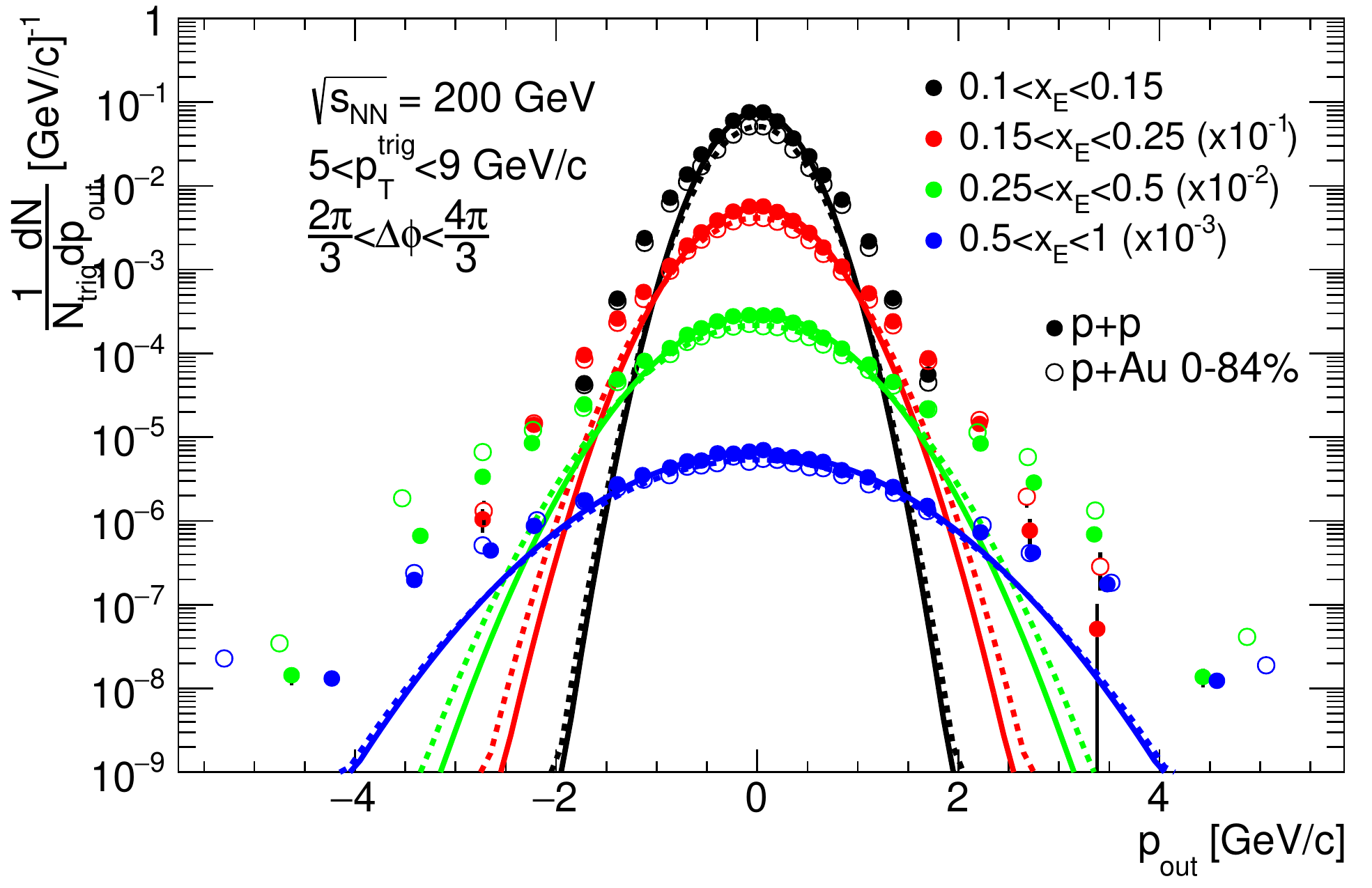}
	\caption{The \pout distributions in \pal (left) and \pau (right) collisions are shown as a function of \xe compared to the \pp distributions.}
	\label{fig:pout_pa_awayside}
\end{figure}

To explore the centrality dependence in the dihadron correlations, the \pout distributions in \pa collisions were constructed in bins of \xe and thus compared to the \pout distributions in \pp collisions. This provides a baseline where nuclear effects are not present; it additionally allows for the \pout widths to be studied as a function of \pttrig and \ptassoc simultaneously as discussed previously. The \pout distributions in \pau and \pal on the away-side are shown in Fig.~\ref{fig:pout_pa_awayside} in bins of \xe for a fixed \pttrig range. Since the largest effects when comparing to \pp collisions were found in the \rmspout distributions in Fig.~\ref{fig:pa_rmspout_dihadrons} at a \pttrig in the range $5<\pttrig<9$ \gevc, this range was used to construct the \pout distributions. A similar centrality dependence was observed in the Gaussian widths in Fig.~\ref{fig:pau_gausswidths} for a fixed \ptassoc bin. The distributions are fit to a Gaussian function at small \pout as shown in the figure and similarly to previous \pout distributions.

The Gaussian widths of \pout in \pa vs. \pp collisions can be used to explore several of the nuclear effects that may arise in \pa collisions as discussed at the beginning of Section~\ref{pa_results}. Initial-state and final-state TMD broadening in \pa collisions can be explored by comparing the near-side widths in \pa and \pp and the away-side widths, respectively, in \pa and \pp collisions. This is because the away-side widths are sensitive to both soft \kt and \jt, while the near-side widths are only sensitive to soft \jt since the hadrons are contained within the same jet. Comparing the values between \pa and \pp may indicate if there is additional \kt or \jt in \pa collisions when compared to \pp collisions. The \pout distributions can also be used to study radiative energy loss mechanisms in a nucleus, as outlined in Ref.~\cite{Tannenbaum:2017afg}. In this publication, the quantity \rmspout was used to extract the quantity $\langle\qhat L\rangle$ in Au+Au collisions; \qhat is the radiative energy loss per-unit length $L$ in the QGP in the case of Au+Au collisions. Reference~\cite{Tannenbaum:2017afg} extracted $\langle\qhat L\rangle$ via the following equation

\begin{equation}\label{eq:qhat}
	\langle\qhat L\rangle/2 = \left[\frac{\hat{x}_h}{\langle z_T\rangle}\right]^2\left[\frac{\langle p_{\rm out}^2\rangle_{AA}-\langle p_{\rm out}^2\rangle_{pp}}{x_h^2}\right]\,,
\end{equation}
where $x_h=\ptassoc/\pttrig$ and $\langle z_T\rangle=\pttrig/\hat{p}_T^{\rm trig}$. While this equation uses the quantity $\langle p_{\rm out}^2\rangle$, the Gaussian widths of \pout could be used to isolate the nonperturbative contributions. Additionally, this quantity can be extracted directly from the functional form of the \pout distribution instead of from a complicated fit function to the away-side \dphi distributions. Furthermore, in \pa collisions, this quantity could also be used to investigate energy loss in a nucleus where a QGP is not expected to form. This could provide a baseline for energy loss in a nucleus for inputs to hot nuclear matter energy loss in a QGP; it may additionally provide constraints on what kind of medium is formed in \pa collisions since there is currently significant debate within the QCD research community about whether or not a QGP is formed in \pa and \pp collisions. 

To first identify whether or not $\qhat L$ is nonzero in \pa collisions, it should be determined if $\hat{x}_h$ and $z_T$ of the \pion are the same between \pa and \pp collisions. There is good reason to expect that \zt is the same, since the inclusive \pion spectra are largely the same between \pa and \pp collisions. This is demonstrated in Ref.~\cite{Novitzky:2017wdt}, which shows the observable $R_{pA}$. This observable is simply the cross section of \pion production in $p$+Au collisions divided by the cross section in \pp collisions, scaled by the number of nucleon-nucleon collisions to account for the additional nucleons in the nucleus. Since the ratio is not significantly different from unity within uncertainties of approximately 10\%, this indicates that the inclusive \pt spectra follow a similar power law shape. This can also be seen by explicitly fitting the cross sections with a power law, which shows that the cross sections have a power law dependence of approximately $n=-8.3$.

\begin{figure}[tbh]	
	\centering
	\includegraphics[width=0.49\textwidth]{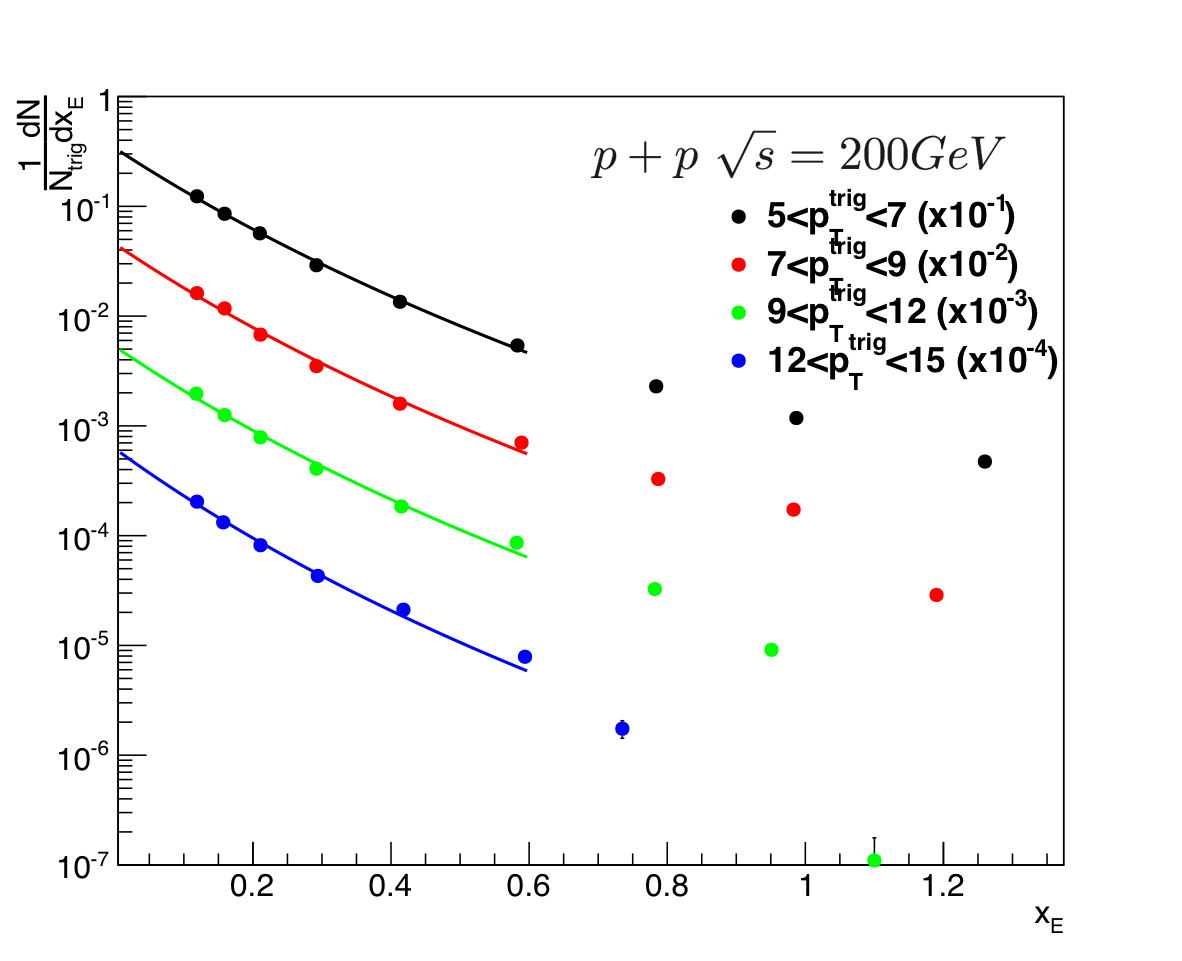}
	\includegraphics[width=0.49\textwidth]{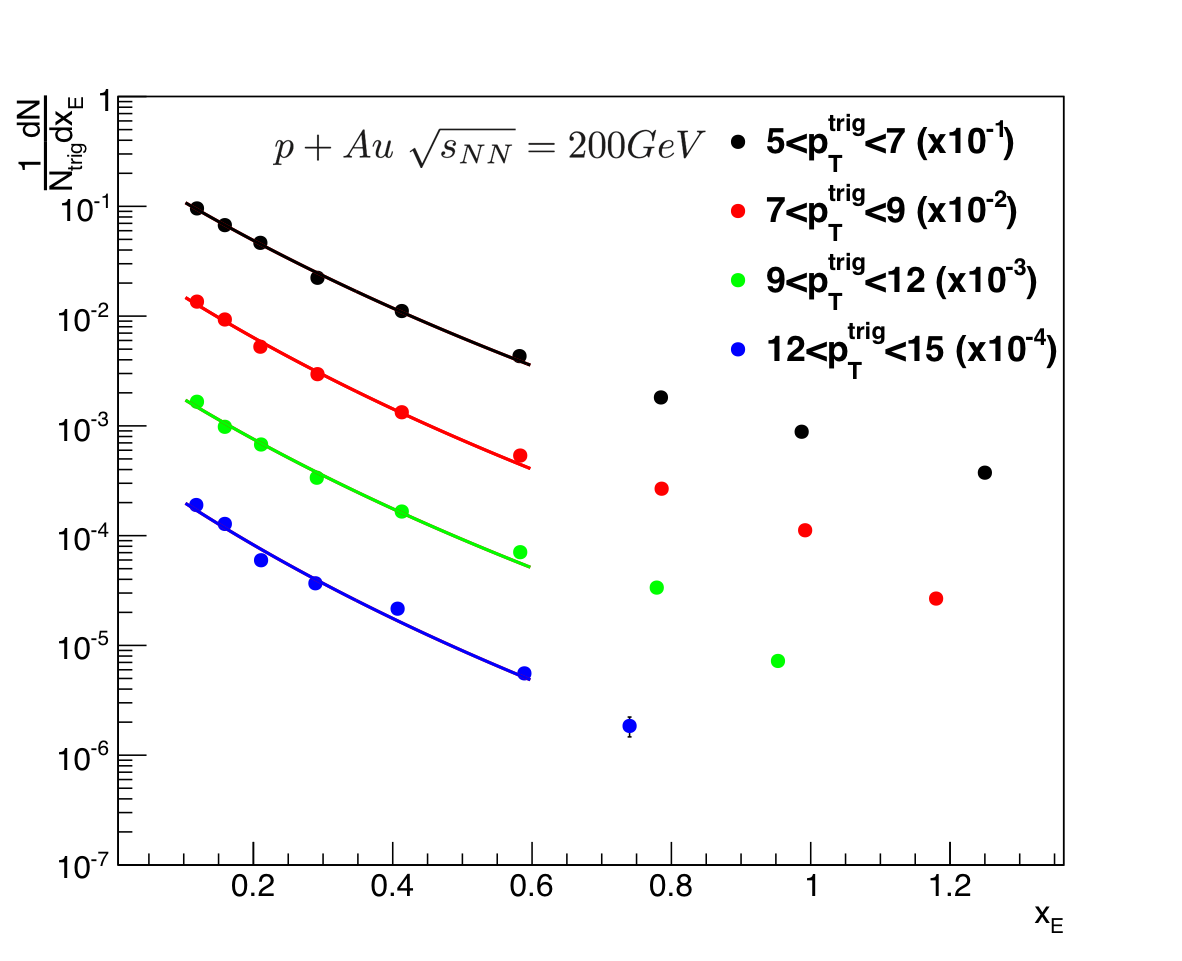}
	\caption{The \xe per-trigger yields are shown in \pp (left) and \pau (right) collisions at \sqsn=~200 GeV. The fits are described in the text and are performed over a similar range of \xe to previous publications~\cite{ppg095}.}
	\label{fig:xe_ptys} 
\end{figure}

\begin{figure}[tbh]	
	\centering
	\includegraphics[width=0.49\textwidth]{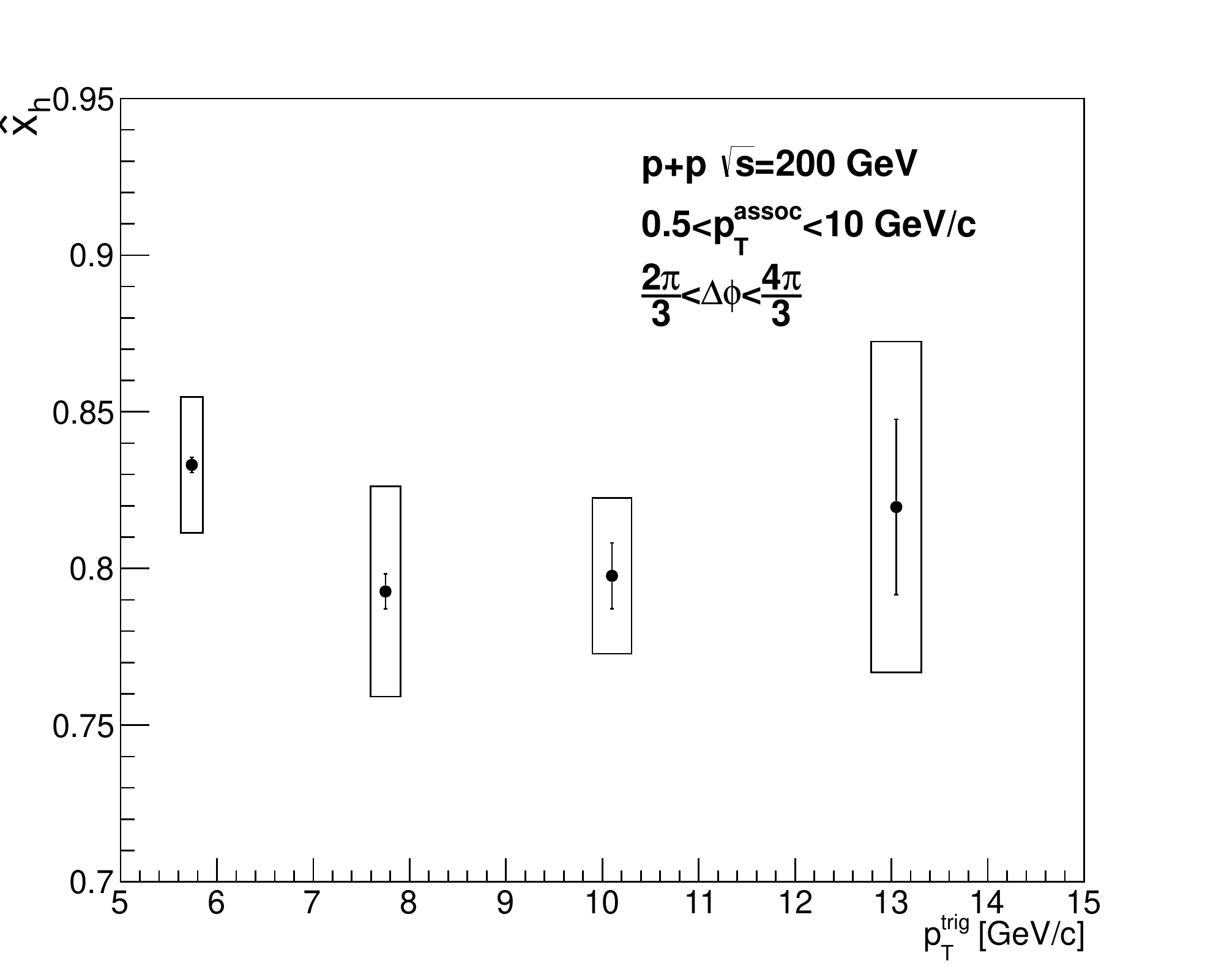}
	\includegraphics[width=0.49\textwidth]{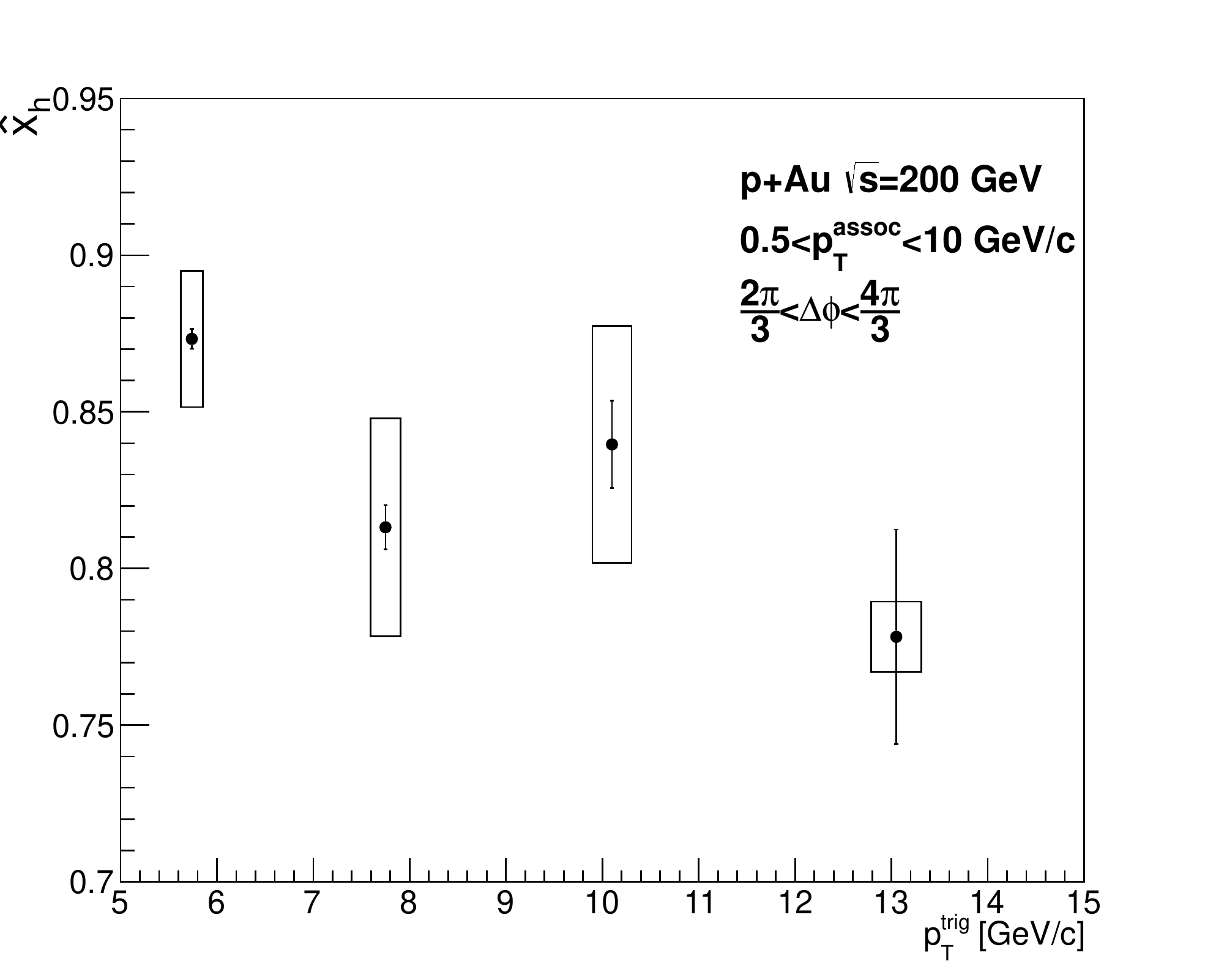}
	\caption{The extracted values of $\hat{x}_h$ from \pp (left) and \pau (right) collisions are shown as a function of \pttrig. The values compare well between the two different collision systems.}
	\label{fig:xhhat}
\end{figure}

To determine if $\hat{x}_h$ is the same between \pa and \pp, the method used in Refs.~\cite{ppg029,ppg095} to extract $\hat{x}_h$ directly from the correlations can be used. The \xe per-trigger yields are constructed from the data and fit to a modified Hagedorn function of the form
\begin{equation}
\frac{dN}{dx_E}\approx N(n-1)\frac{1}{\hat{x}_h}\frac{1}{(1+\frac{\xe}{\hat{x}_h})^n}\,,
\end{equation}
where $N$ and $\hat{x}_h$ are free parameters, and $n$ is the power law value from a fit to the inclusive \pion spectra. Since the $R_{pA}$ is consistent with unity above \pt~=1 \gevc, this means $n$ is approximately the same in \pa and \pp; nevertheless, it was empirically determined to be 8.3 for both from fits to the actual cross sections as briefly described above. The fits are performed to the \xe per-trigger yields in similar \xe regions to Ref.~\cite{ppg095}, shown in Fig.~\ref{fig:xe_ptys}, and the resulting $\hat{x}_h$ values are shown in \pp and \pau collisions in Fig~\ref{fig:xhhat}. The systematic uncertainties are estimated by adjusting the fit region as a function of \xe. The $\hat{x}_h$ values are consistent within uncertainties between \pp and \pau collisions. Therefore, if there is a difference in the \pout Gaussian widths between \pa and \pp collisions, there must be a nonzero $\qhat L$ in \pa collisions since both \zt and $\hat{x}_h$ are consistent between \pp and \pa collisions. This would apply for both direct photon-hadron correlations and dihadron correlations; however, the statistical precision of the measurement will be important as nuclear effects are expected to be small~\cite{Vasilev:1999fa}.

\begin{figure}[tbh]
	\centering
	\includegraphics[width=0.49\textwidth]{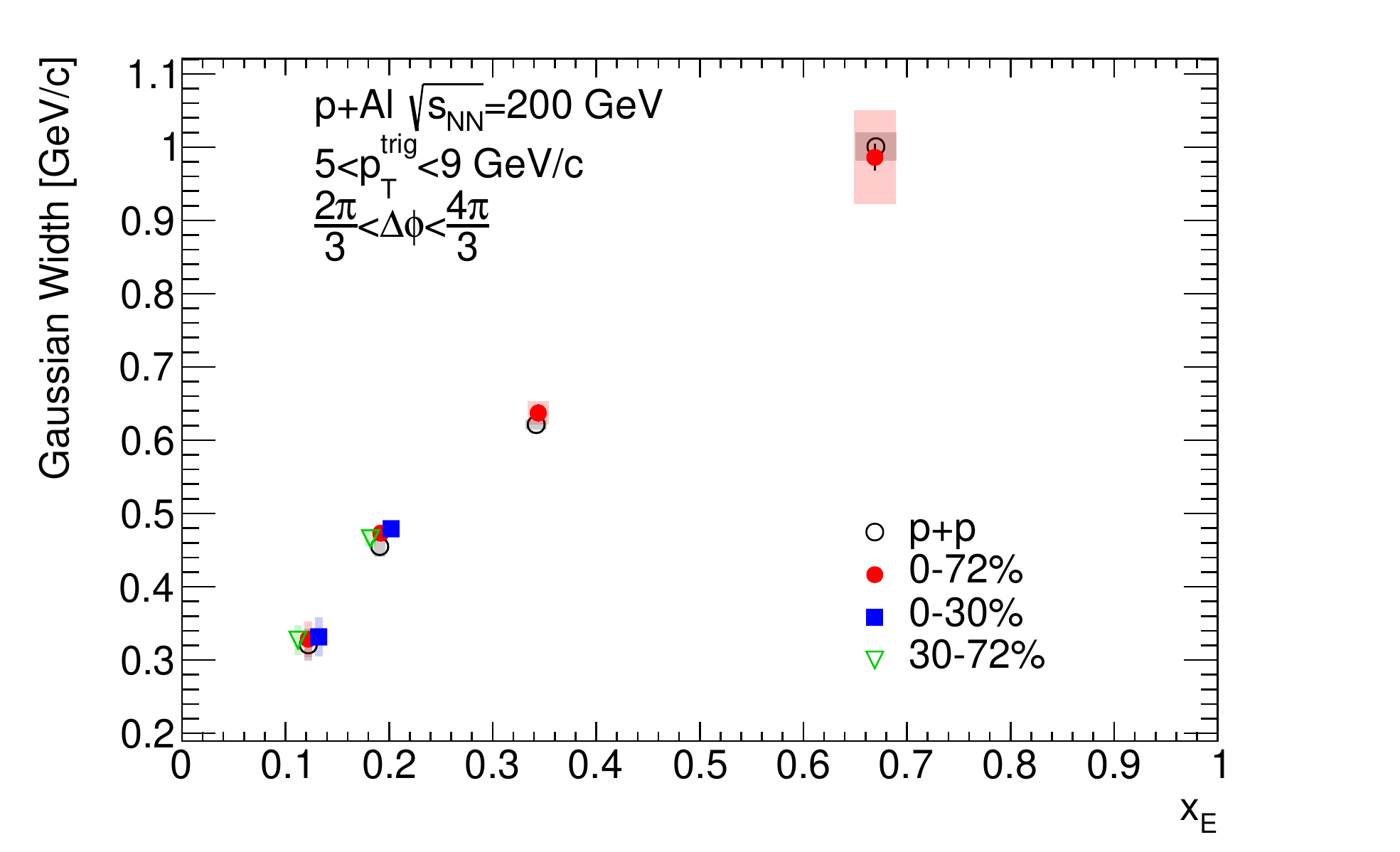}
	\includegraphics[width=0.49\textwidth]{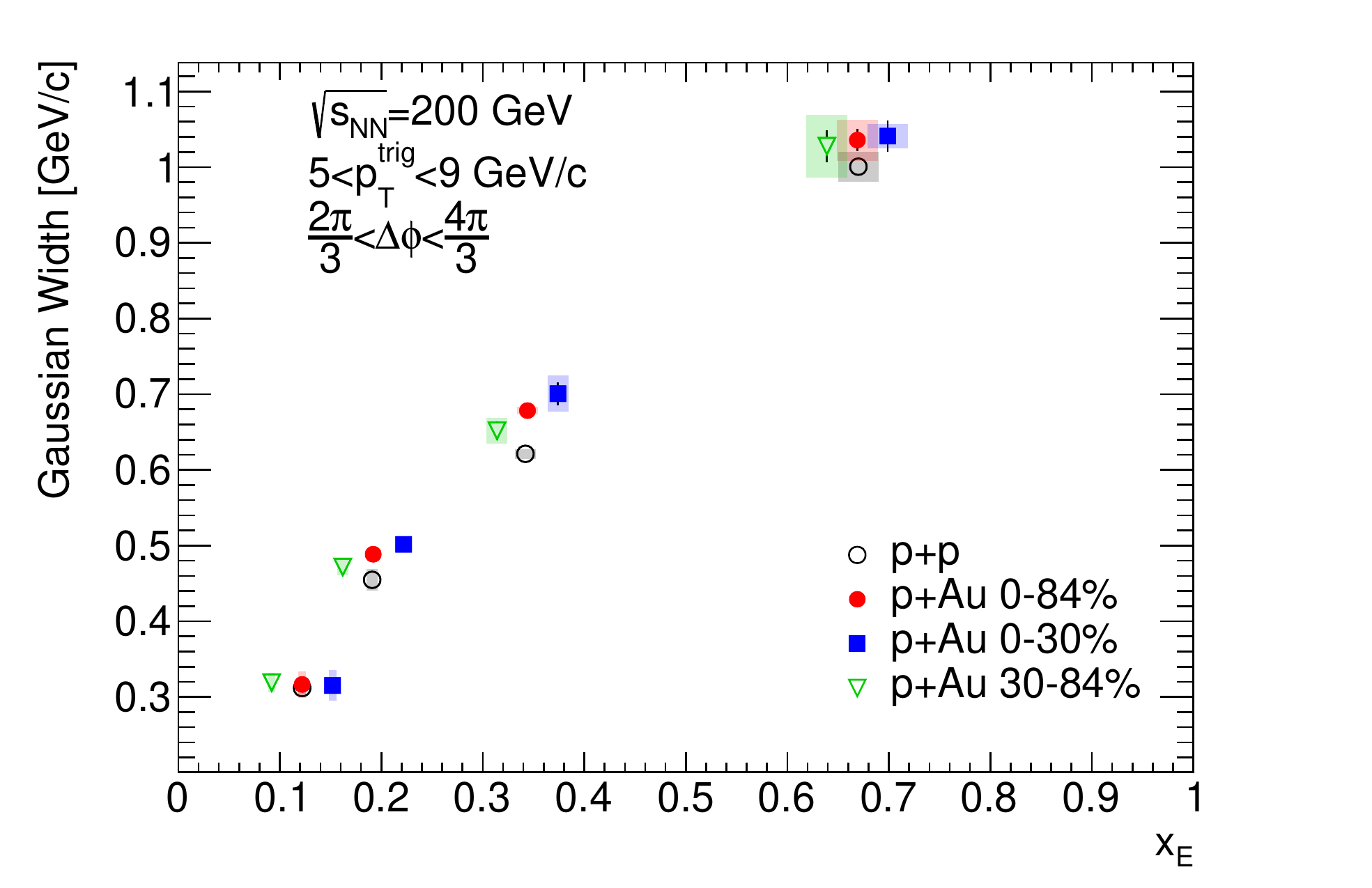}
	\caption{The away-side Gaussian widths of \pout in \pion-\h correlations are shown in \pal (left) and \pau (right) collisions as a function of \xe, in addition to the corresponding \pp widths.}
	\label{fig:gausswidths_pa_pp}
\end{figure}

\begin{figure}[tbh]
	\centering
	\includegraphics[width=0.49\textwidth]{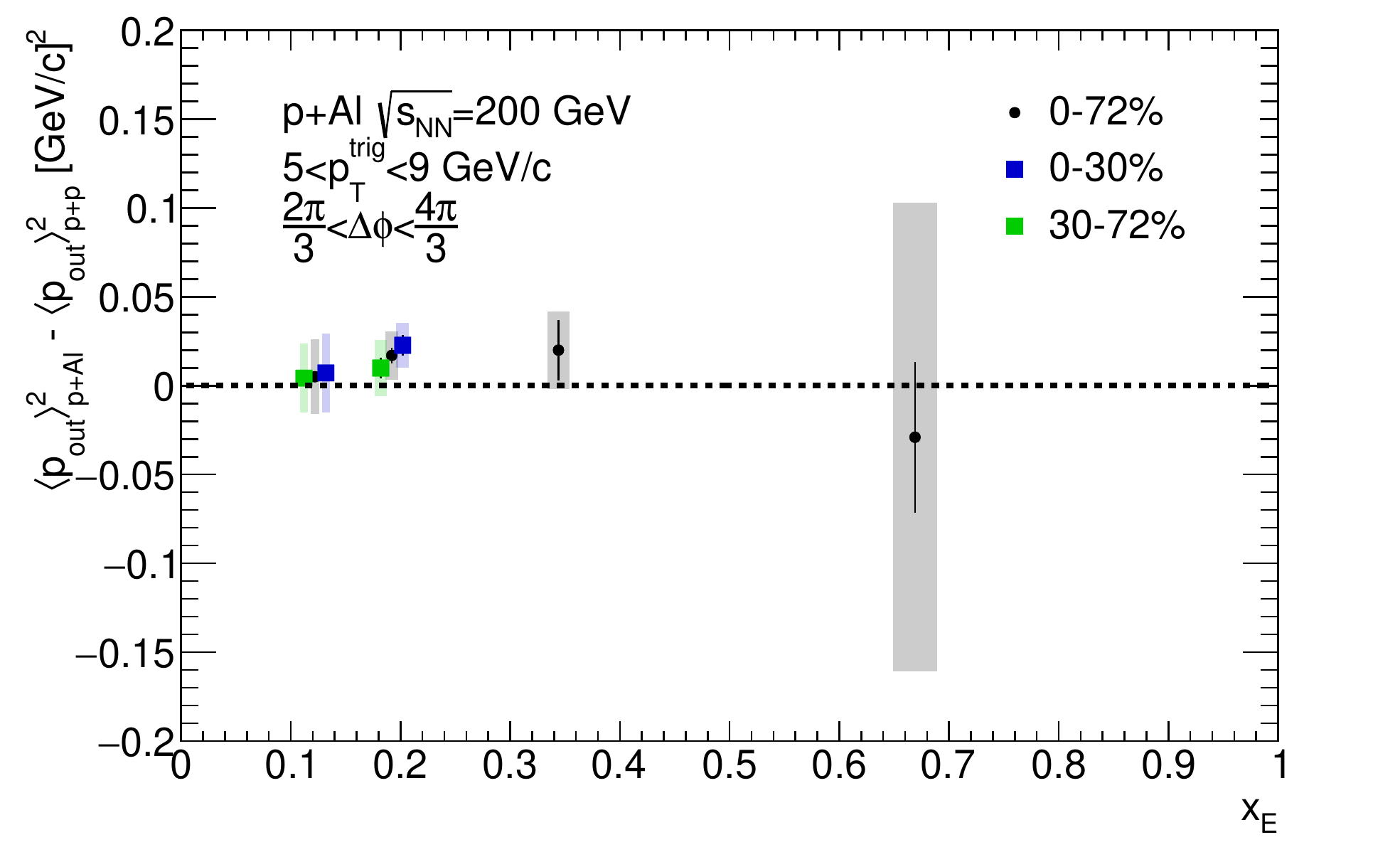}
	\includegraphics[width=0.49\textwidth]{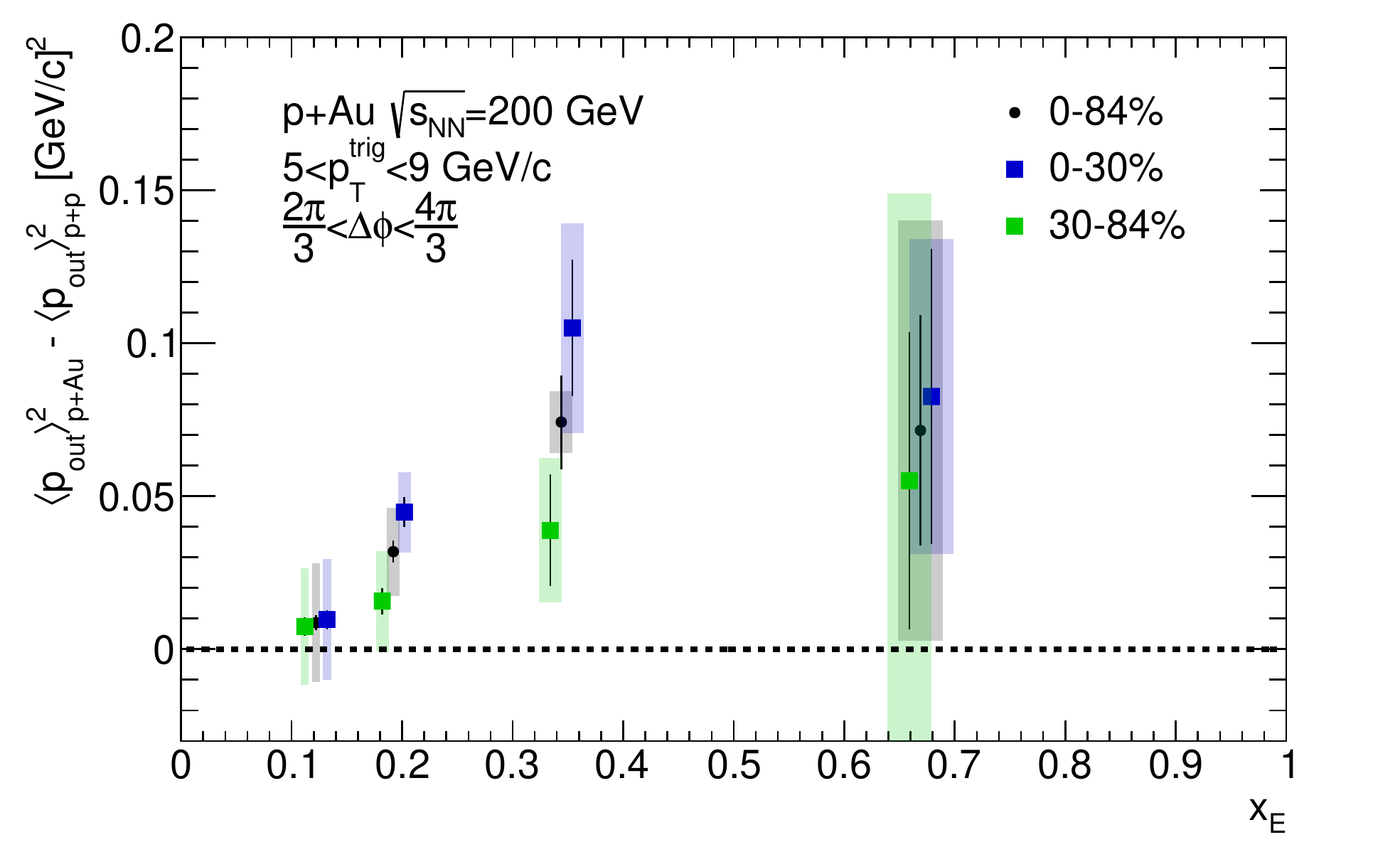}
	\caption{The difference of the away-side squared Gaussian widths of \pout in \pion-\h correlations are shown in \pal (left) and \pau (right) collisions as a function of \xe.}
	\label{fig:pout_sqdiff}
\end{figure}

With the information presented, we are thus equipped to study both the effects of potential TMD initial- and final-state broadening and radiative energy loss in a nucleus using the \pout widths between \pp and \pa collisions. The Gaussian widths of \pout in both \pp and \pa collisions are extracted from, for example, Fig.~\ref{fig:pout_pa_awayside} and shown in Fig.~\ref{fig:gausswidths_pa_pp} in several centrality bins, where the statistical precision of the data permits. The central and peripheral bins are offset slightly in \xe for visibility sake. There are several bins where there is a slight difference between the \pa and \pp away-side widths. To quantify this difference similarly to the form of Eq.~\ref{eq:qhat}, the difference between the \pa and \pp squared widths was determined. To carefully label this, since the Gaussian width determines the average \pout in the nearly back-to-back region, the quantity is labeled $\langle p_{\rm out}\rangle^2_{pA}-\langle p_{\rm out}\rangle^2_{pp}$ in an analogous way to Eq.~\ref{eq:qhat}. However, the distinction is that the Gaussian width is the average width squared, while the $\langle p_{\rm out}^2\rangle$ is the average squared width. The squared differences in both \pal and \pau collisions are shown in Fig.~\ref{fig:pout_sqdiff}; the figures show that at certain values of \xe there is a significant nonzero difference and thus a nonzero value of \qhat based on the analysis Ref.~\cite{Tannenbaum:2017afg}.

\begin{figure}[tbh]
	\centering
	\includegraphics[width=0.49\textwidth]{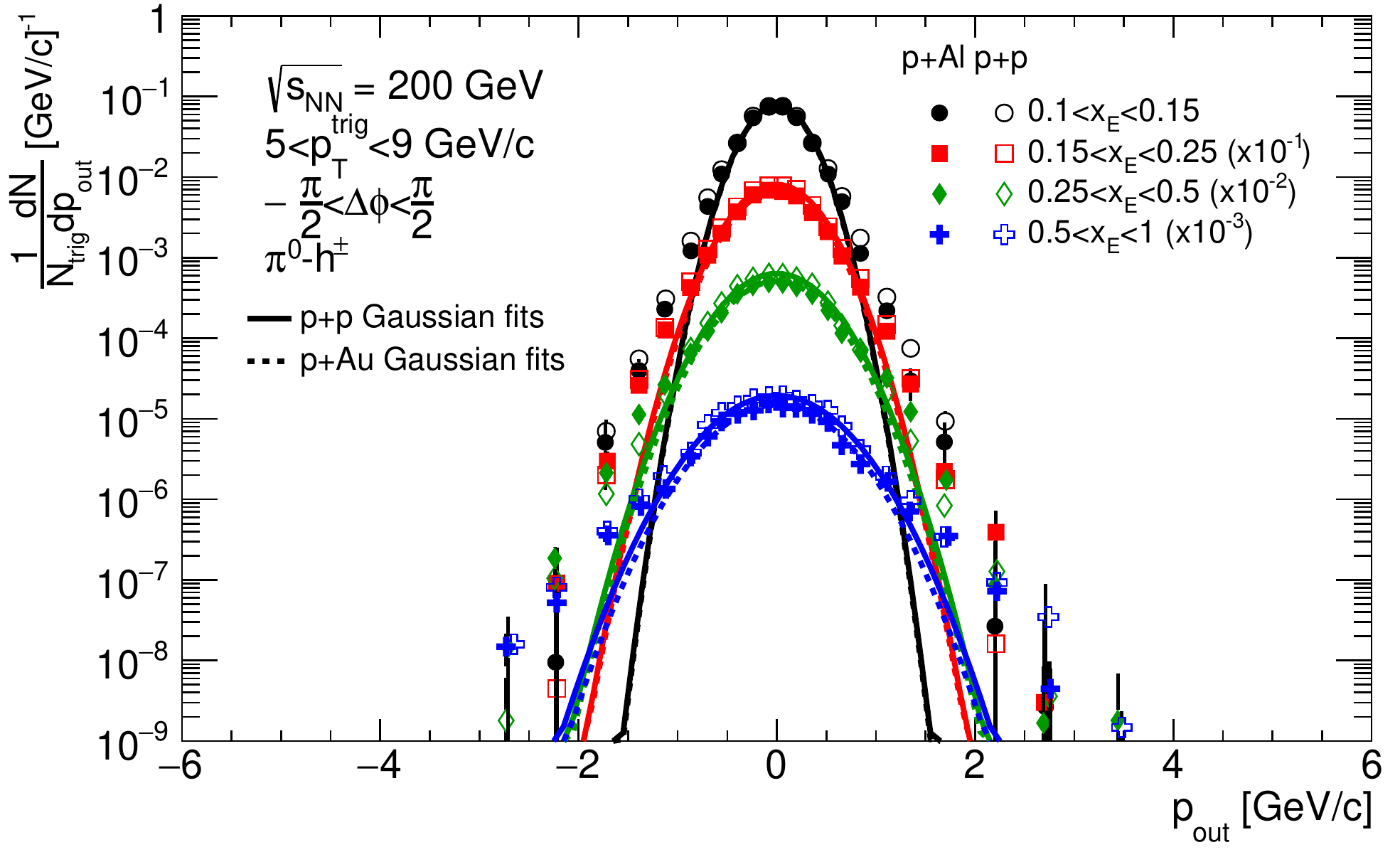}
	\includegraphics[width=0.49\textwidth]{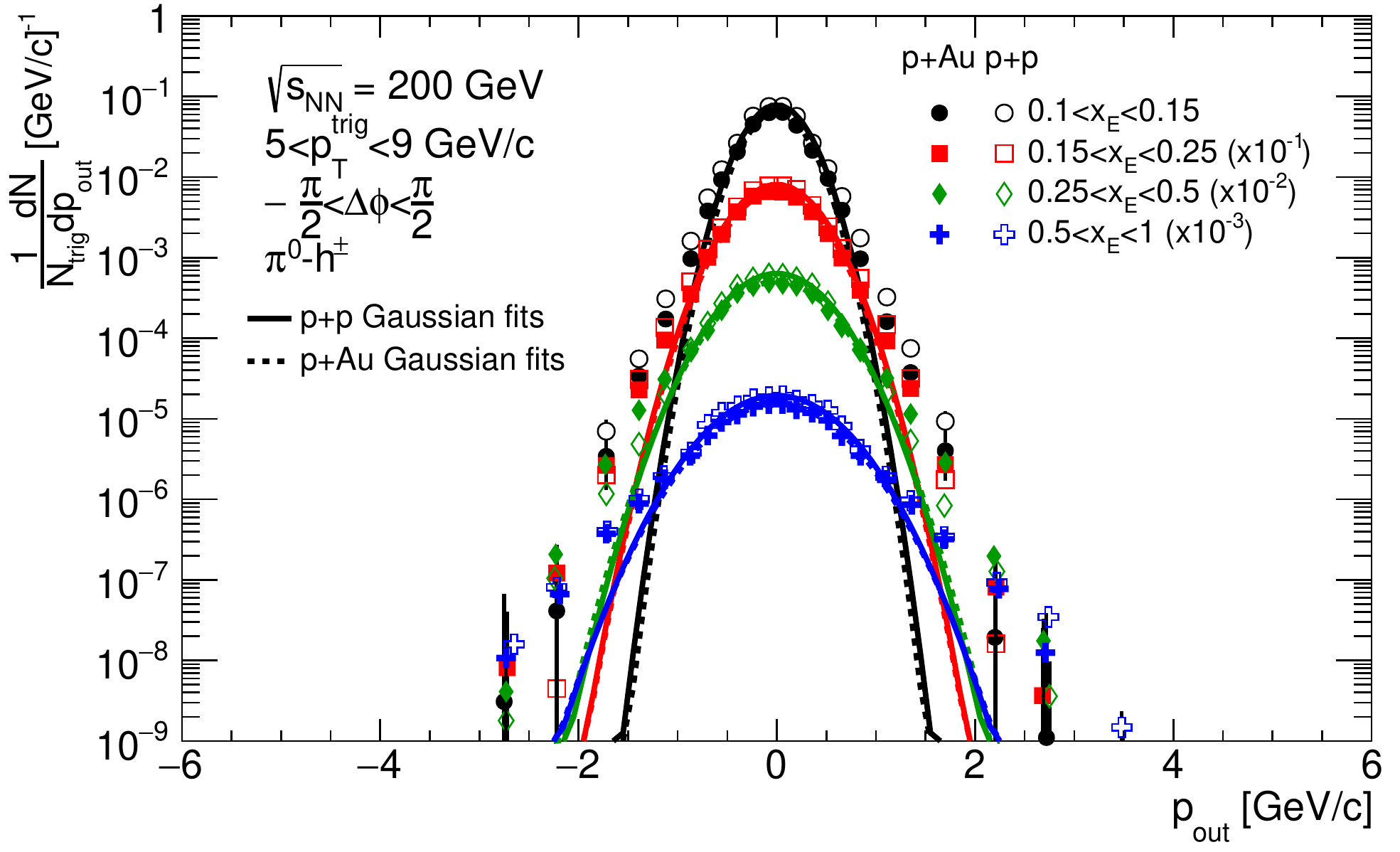}
	\caption{The near-side \pout distributions are shown in \pal and \pp (left) and \pau and \pp (right) collisions in several \xe bins.}
	\label{fig:nearside_pout}
\end{figure}

To explore if these nonzero width differences are potentially due to initial-state or final-state radiation effects, the near-side width differences can also be constructed analogously in \pal and \pau vs. \pp collisions. The near-side widths are only sensitive to final-state radiation since the hadrons in a single jet come from the same parton and are thus not sensitive to \kt. These are extracted from the near-side \pout distributions, shown in Fig.~\ref{fig:nearside_pout}. In these distributions, the underlying event was statistically subtracted in a similar way to the away-side; the per-trigger yields as a function of \dphi were fit to a Gaussian+constant background function, and the underlying event was statistically subtracted from the acceptance and efficiency corrected yield. The near-side \pout distributions have a distinctly different shape than the away-side distributions; while the nonperturbative and perturbative regions are still identifiable, the yields fall towards zero much faster at large \pout on the near sides than the away sides. This is because the near-side per-trigger yields are only sensitive to \jt and are thus not as smeared out in \dphi or \pout. This could be ascertained from any of the \dphi correlations plots, for example in Fig.~\ref{fig:pau_dphis}, which show that the near-side distributions for $\dphi<\pi/2$ are nearly 100\% background except for the region $\sim\pm$~0.4 radians about \dphi=~0. Nevertheless, there is still a clear Gaussian region which fails at large \pout, and this can be quantified similarly to the away-side widths.

\begin{figure}[tbh]
	\centering
	\includegraphics[width=0.49\textwidth]{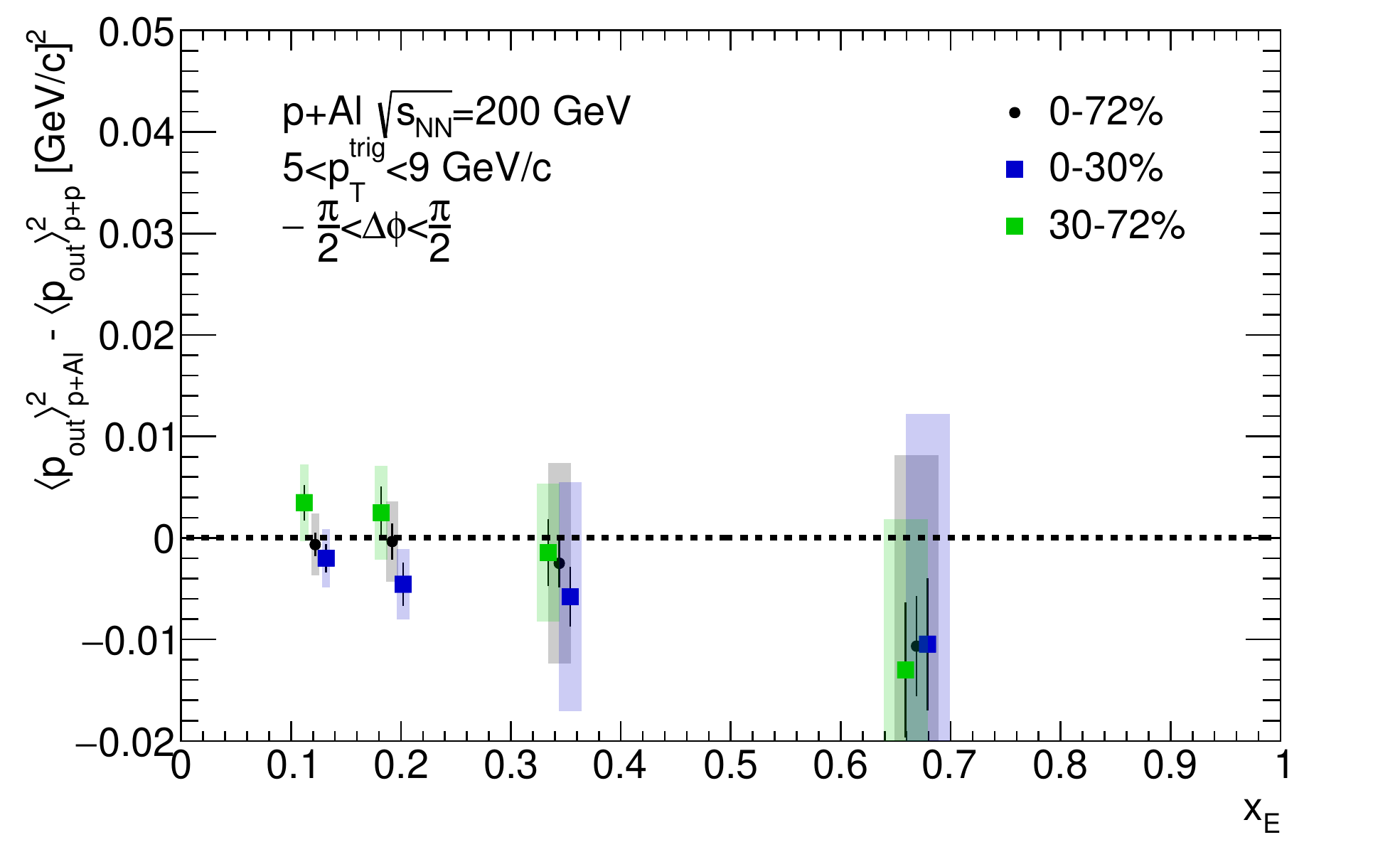}
	\includegraphics[width=0.49\textwidth]{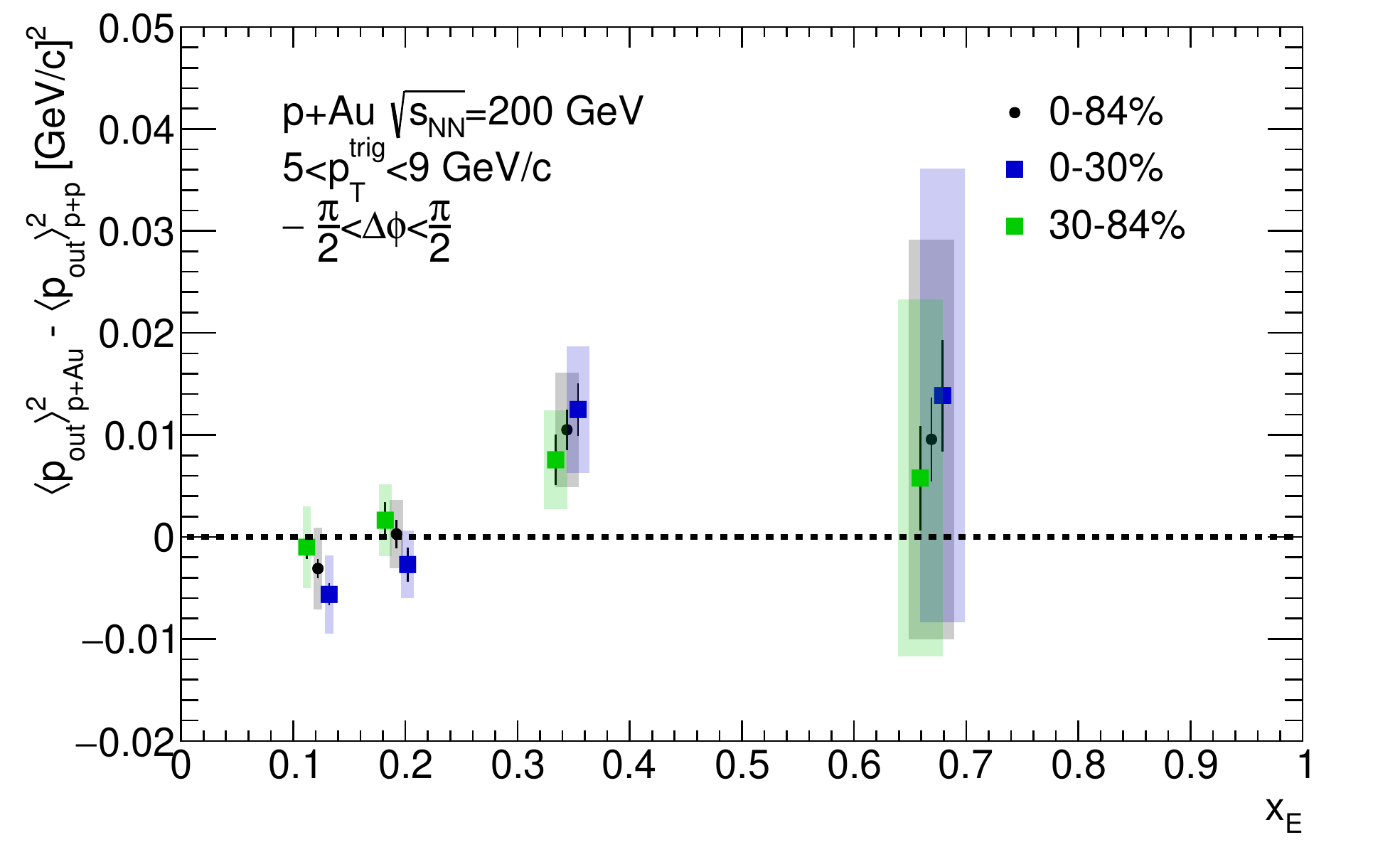}
	\caption{The difference of the near-side squared Gaussian widths of \pout in \pion-\h correlations are shown in \pal (left) and \pau (right) collisions as a function of \xe.}
	\label{fig:nearside_width_diffs}
\end{figure}

The Gaussian widths are extracted from the fits and the squared width differences of the near-side for both \pal and \pau compared to \pp are shown in Fig.~\ref{fig:nearside_width_diffs}. The differences are consistent with 0 across all \xe within uncertainties. The $0.25<\xe<0.5$ bin in \pau data is slightly nonzero; however, within uncertainties the values are slightly less than $2\sigma$ from 0 so it is not statistically significant. Additionally there appears to be no centrality dependence, and in particular there is no systematic trend between the different centrality bins. Similar results from the ALICE collaboration have found that the near-side widths in dihadron events between $p$+Pb and \pp are consistent~\cite{Viinikainen:2016zbl}; ATLAS has also studied fragmentation with full jet reconstruction in $p$+Pb collisions and has found no significant difference between $p$+Pb and \pp fragmentation functions~\cite{Aaboud:2017tke}. This gives the indication that the transverse momentum broadening seen in the away-side widths is not due to final-state fragmentation modification, assuming that the parton fragments into hadrons outside the nucleus. In fact, these results, in addition to those previously mentioned, may suggest that parton fragmentation occurs outside of the nuclear medium since it might be expected that nuclear interactions would modify these distributions.

\begin{figure}[tbh]
	\centering
	\includegraphics[width=0.7\textwidth]{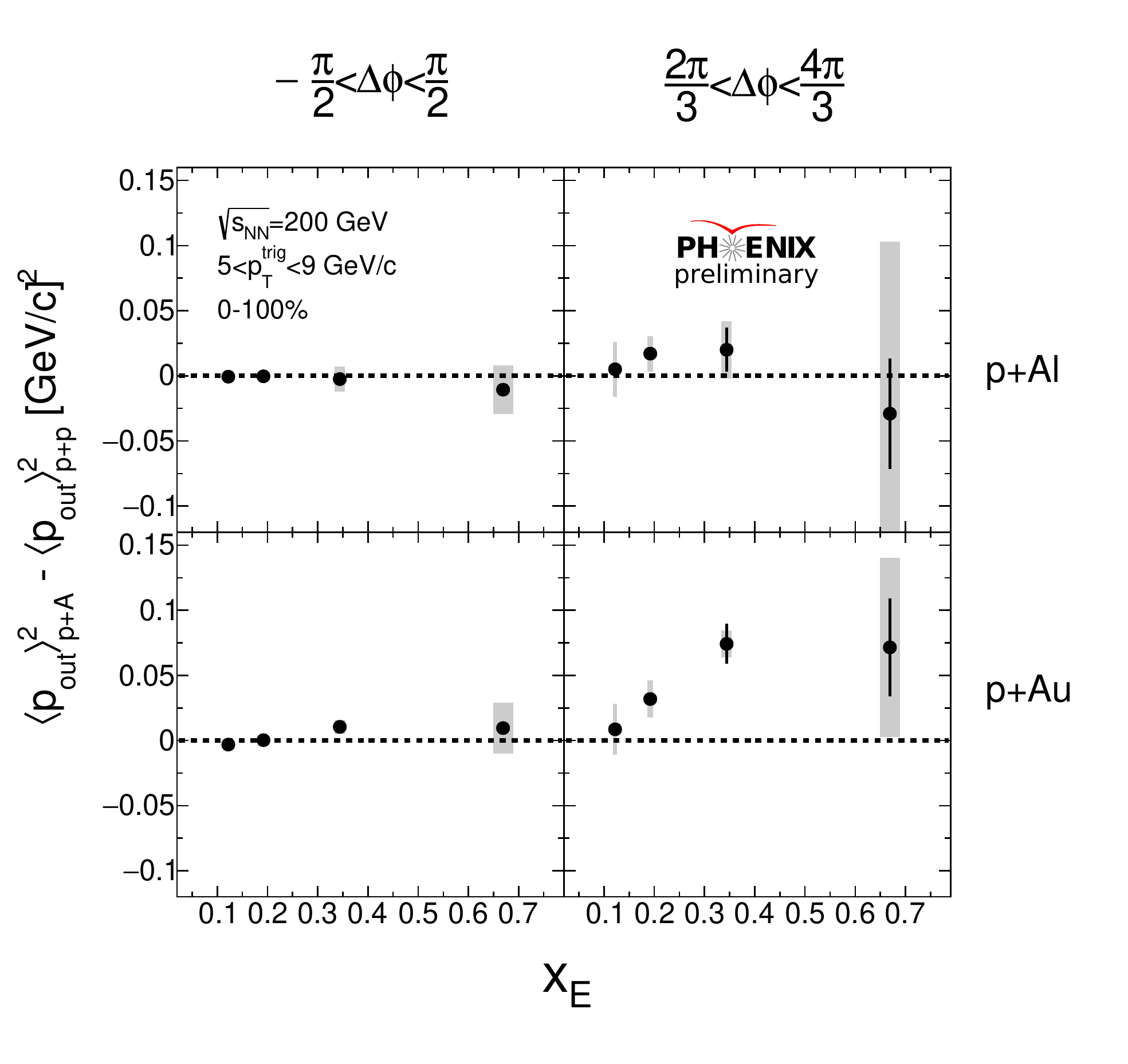}
	\caption{A four panel summary figure that shows the near and away side width differences in \pal or \pau and \pp collisions, highlighting the nonzero effect on the away-side.}
	\label{fig:fourpanel}
\end{figure}

Figure~\ref{fig:fourpanel} shows a summary plot of the squared width differences between \pal, \pau, and \pp. The left column shows the near-side, while the right column shows the away side. The top row shows the \pal results, while the bottom row shows the \pau results. This four panel figure highlights the differences seen; the near-side widths are consistent with 0, however there is statistically significant modification on the away side. Thus, we may conclude that final-state \jt modification does not contribute to the observed away-side azimuthal broadening.

\begin{figure}[tbh]
	\centering
	\includegraphics[width=0.49\textwidth]{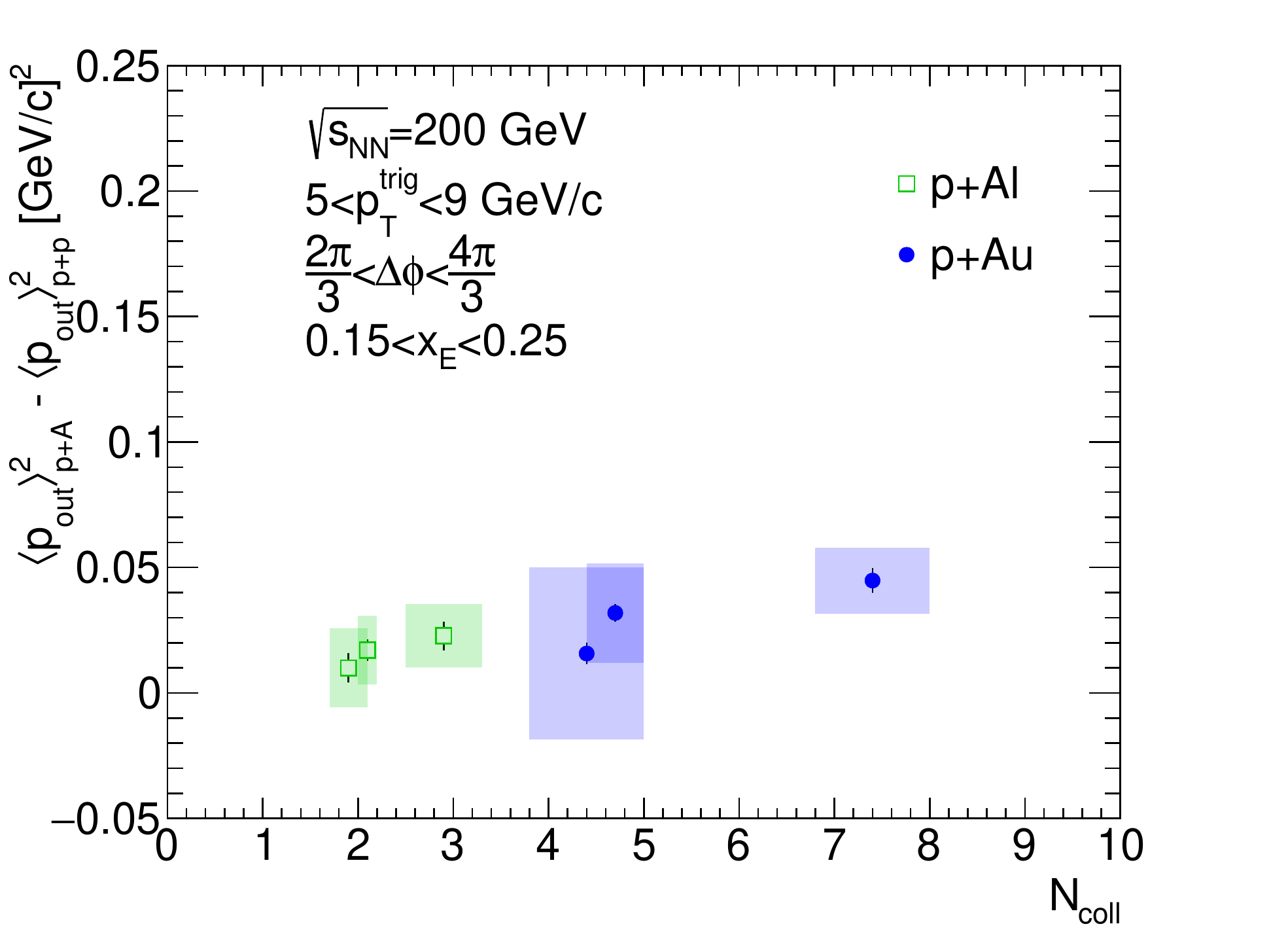}
	\includegraphics[width=0.49\textwidth]{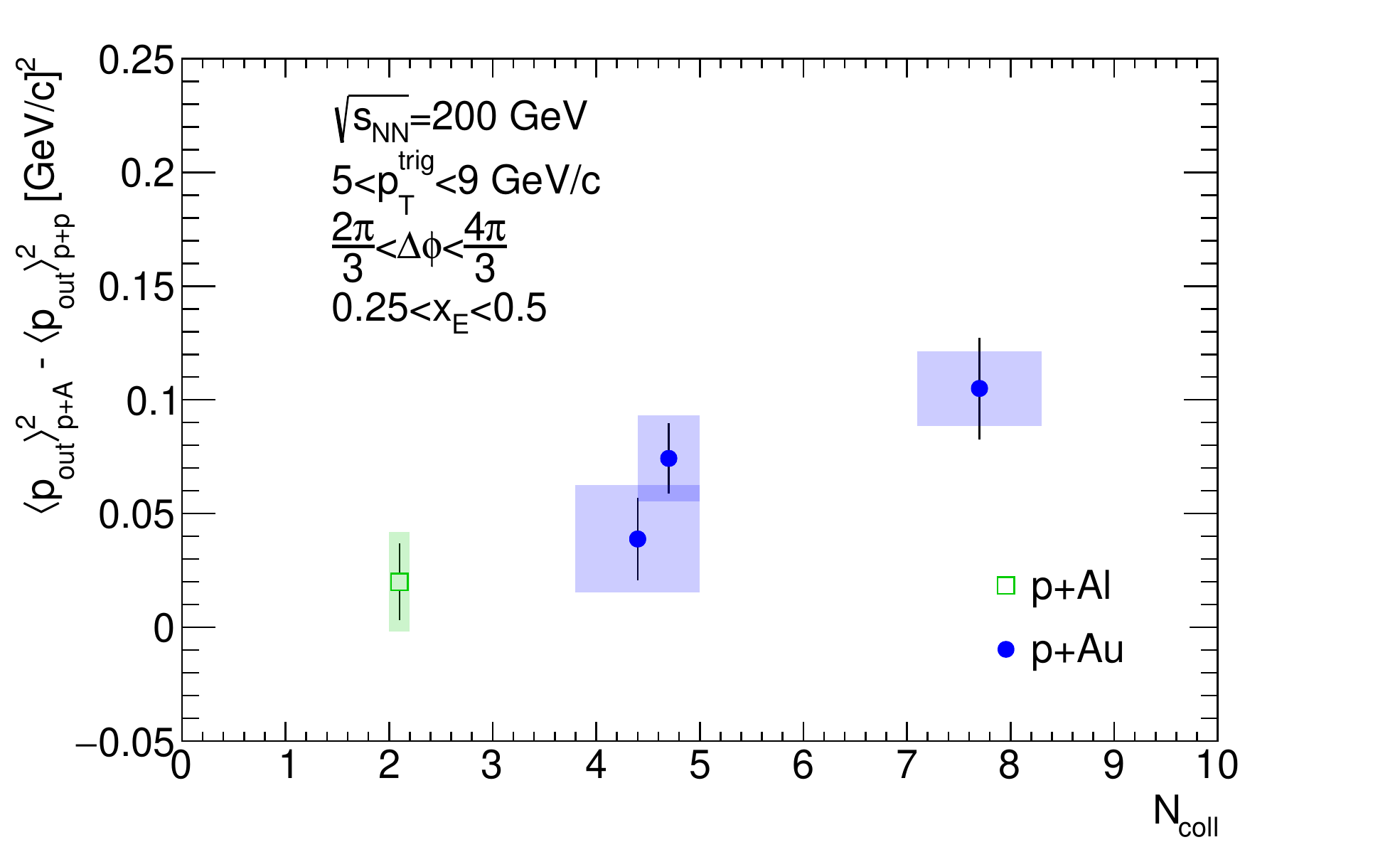}
	\caption{The squared width differences are shown as a function of \ncoll for $0.15<\xe<0.25$ (left) and $0.25<\xe<0.5$ (right). The centrality integrated points are shown in addition to the centrality binned points for completeness, where the middle \ncoll value corresponds to the centrality integrated data in a given data set.}
	\label{fig:widthsvsncoll}
\end{figure}

To study the centrality dependence further, the width differences can be plotted as a function of \ncoll, the number of nucleon-nucleon collisions in the event. The values of \ncoll are determined for a given centrality via Monte Carlo Glauber model simulations as described in Ref.~\cite{dAu_cent_determ}, and to some extent \ncoll can be intuitively thought of as a proxy for the path length that the partons traverse in the interaction. In particular, the \xe bins of interest are those where there are clear nonzero effects in Fig.~\ref{fig:fourpanel}; therefore, the moderate \xe bins within $0.15<\xe<0.5$ are studied further. Figure~\ref{fig:widthsvsncoll} shows the squared width differences as a function of \ncoll in two \xe bins. The data appear to be positively sloped with \ncoll, indicating that as the path length that a hard scattered parton traverses gets larger, the transverse momentum broadening increases.

\begin{figure}[tbh]
	\centering
	\includegraphics[width=0.6\textwidth]{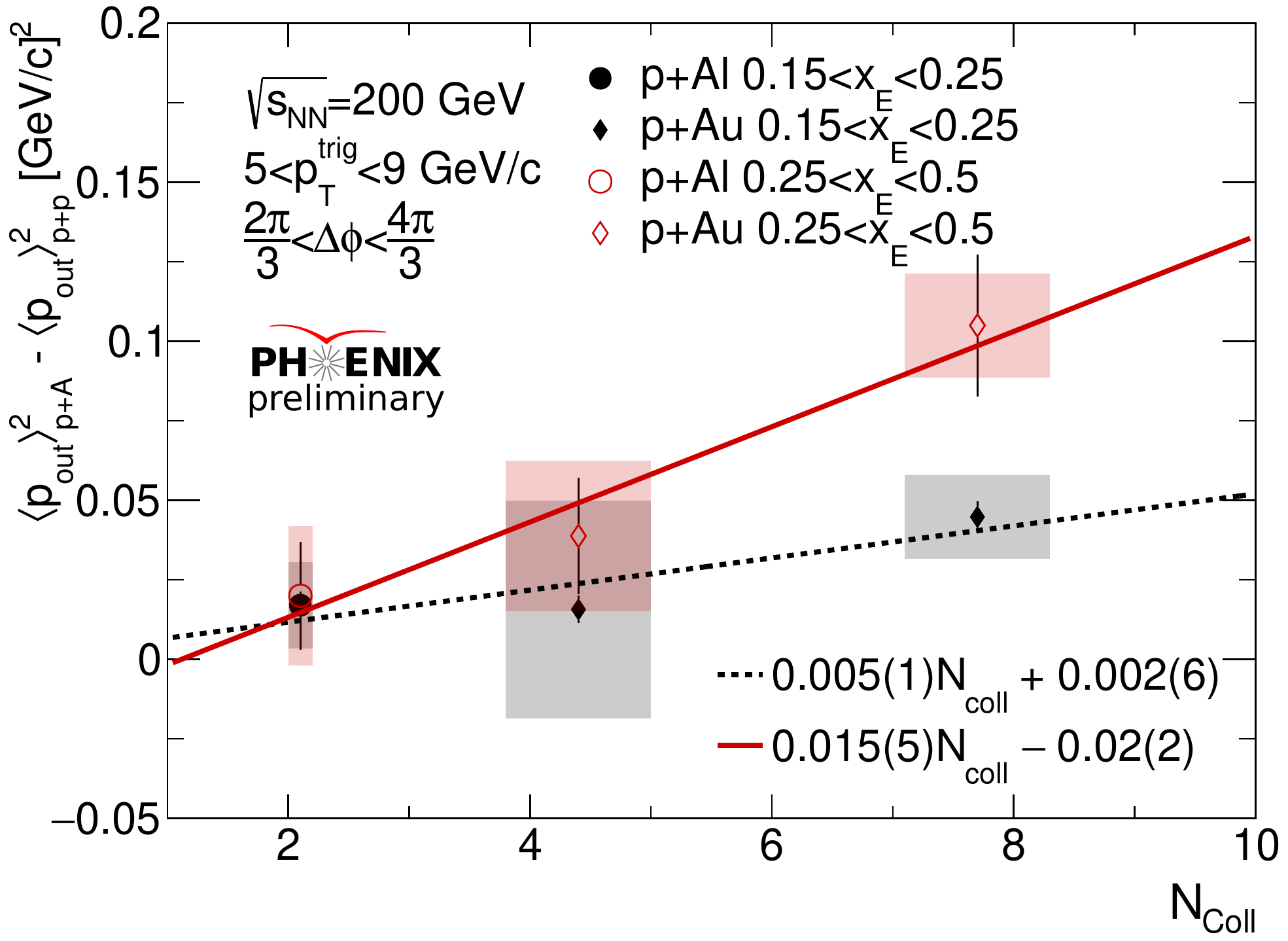}
	\caption{The squared width differences are shown with linear fits in two \xe bins. The fits indicate a positive dependence of the width differences between \pa and \pp with \ncoll, which can be seen in the fit parameters in the lower right of the figure.}
	\label{fig:widthsvsncollfit}
\end{figure}

\begin{figure}[tbh]
	\centering
	\includegraphics[width=0.6\textwidth]{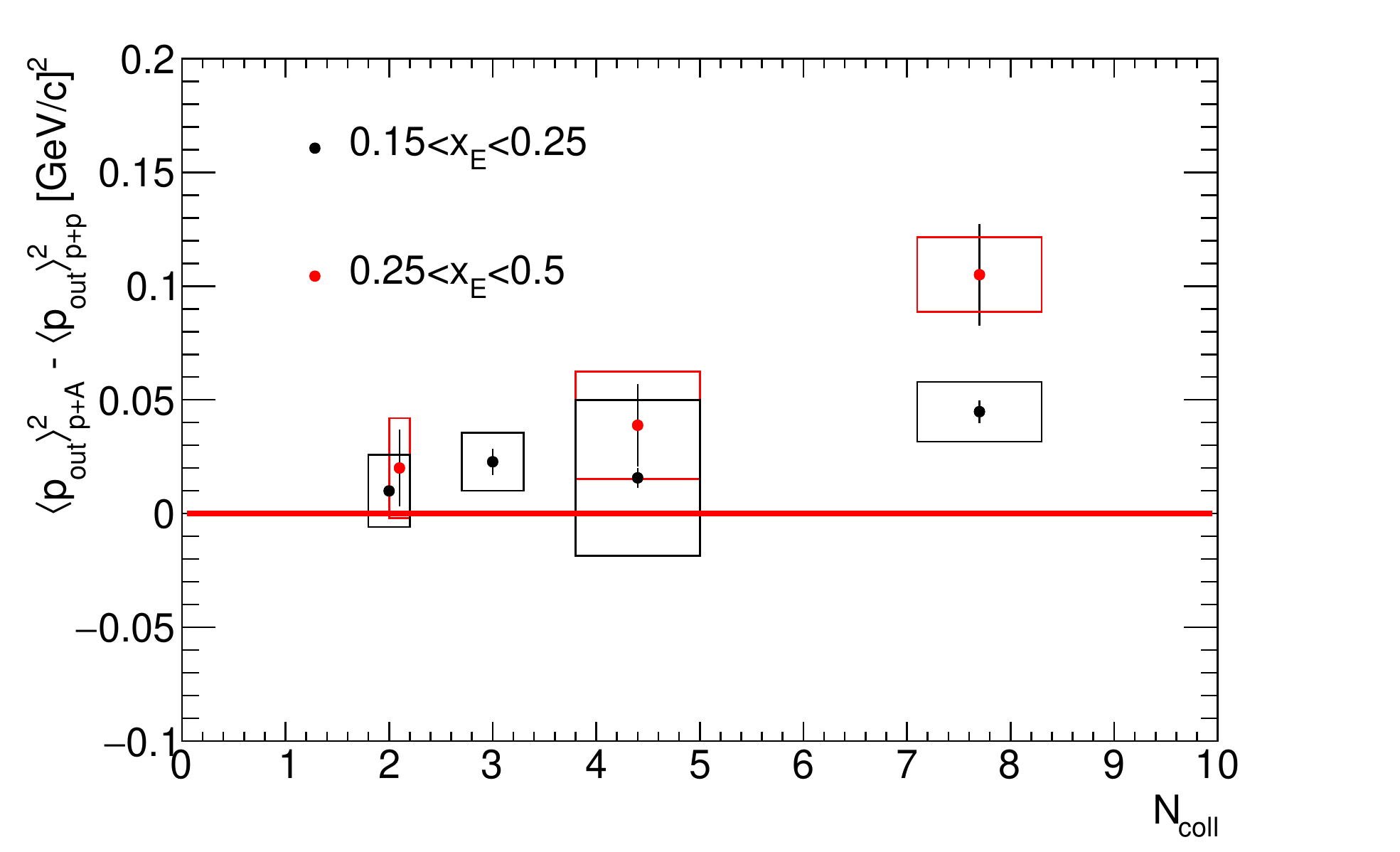}
	\caption{The data are shown with a constant function forced to be 0. When this constraint is placed, the $\chi^2$ per number of degree of freedom becomes 11.8/2 (16.5/2) for the smaller (larger) \xe bin, as discussed in the text.}
	\label{fig:fitforce0}
\end{figure}

To test the dependence of the squared width differences, the data was fit with a linear function in each \xe bin. The results of the fit are shown in Fig.~\ref{fig:widthsvsncollfit}, where the fits indicate that there is a positive dependence of the squared width differences on \ncoll. The fits were performed with only the independent data points; for example, the 0-30\% \pau data is clearly a subset of the 0-84\% \pau data. For these reasons, the centrality integrated data was excluded from the fits when there was enough statistical precision to bin the data in separate centrality bins. Note that the $\chi^2$ per number of degrees of freedom is less than 1 for each of these fits due to the large systematic uncertainties with respect to the magnitude of the data. To test whether or not the data are compatible with a squared width difference of 0, the data was fit to a constant function forced to be 0. The result is shown in Fig~\ref{fig:fitforce0}; in this case, the fit $\chi^2$ per number of degree of freedom becomes 11.8/2 for the $0.15<\xe<0.25$ bin and 16.5/2 for the $0.25<\xe<0.5$ bin. This indicates that the data is significantly less compatible with zero transverse momentum broadening in \pa compared to \pp on the away side.

\subsubsection{Potential Long Range $\Delta\eta$ Contributions}

Since many different underlying physical phenomena could be contributing to the measured transverse momentum broadening in \pa compared to \pp collisions, it would be ideal if one could systematically test whether each physical effect is or is not contributing. While this is not possible for some effects, for example it would not be possible to experimentally differentiate between \pt broadening due to elastic or inelastic collisions while the parton traverses the nucleus, there are some studies which can be performed to rule out certain contributions. For example, the lack of any near-side broadening indicates that there are not significant TMD fragmentation effects, and that the final-state radiation effects are not modified significantly by the nuclear target. Another study can be performed to test any contribution from collective behavior due to multi-particle final-state correlations. Nonzero second and third harmonic Fourier amplitudes, referred to as \vtwo and \vthree, are thought to arise due to hydrodynamic and collective behavior in hadronic collisions, and are generally associated with multi-particle correlations. These quantities are defined as a cosine modulated background as a function of \dphi; therefore, since \vtwo and \vthree have been measured to be nonzero in \pa collisions~\cite{ATLAS_pPb_collectivity,Aidala:2017ajz}, it is possible to systematically test if they are contributing to the transverse momentum broadening in \pa collisions.

To perform this study, the cosine modulated background from \vtwo and \vthree must be explicitly removed from the \pout per-trigger yields. This is performed similarly to the underlying event subtraction, following the methods typically used in nucleus-nucleus collisions where this background is large and must be removed~\cite{ppg090}
\begin{equation}\label{eq:v2v3subtraction}
\frac{dN}{d\pout} = C(\pout,\xe,\pttrig) - b_0(1+2v_2^{\pion} v_2^{\h}\cos(2\dphi)+3v_3^{\pion} v_3^{\h}\cos(3\dphi))\,,
\end{equation} 
where $b_0$ is the zero-yield-at-minimum underlying event level and $v_n^{\pion}$ and $v_n^{\h}$ are the measured \vtwo and \vthree values for \pion mesons and nonidentified charged hadrons, respectively. In the cases where there were no measurements of \vtwo or \vthree in \pa collisions, the equivalent measurements were taken from 0-20\% central Au+Au collisions. This is in no way intended to be realistic; in fact, it is known to be a gross overestimation of the actual values. This could be ascertained from, for example, comparing the measured \vtwo and \vthree values in \pau and Au+Au collisions~\cite{Aidala:2016vgl,Adare:2011tg}; the larger values are due to the elliptic shape of the overlapping area in a Au+Au collision when the impact parameter is nonzero in comparison to the random shape of the overlapping area in a \pau collision. However, the values will provide an incredibly conservative upper limit to any contribution from \vtwo and \vthree, and thus if there is a nonzero contribution it will be more clear using these unrealistically large values of \vtwo and \vthree. 

\begin{figure}[tbh]
	\centering
	\includegraphics[width=0.49\textwidth]{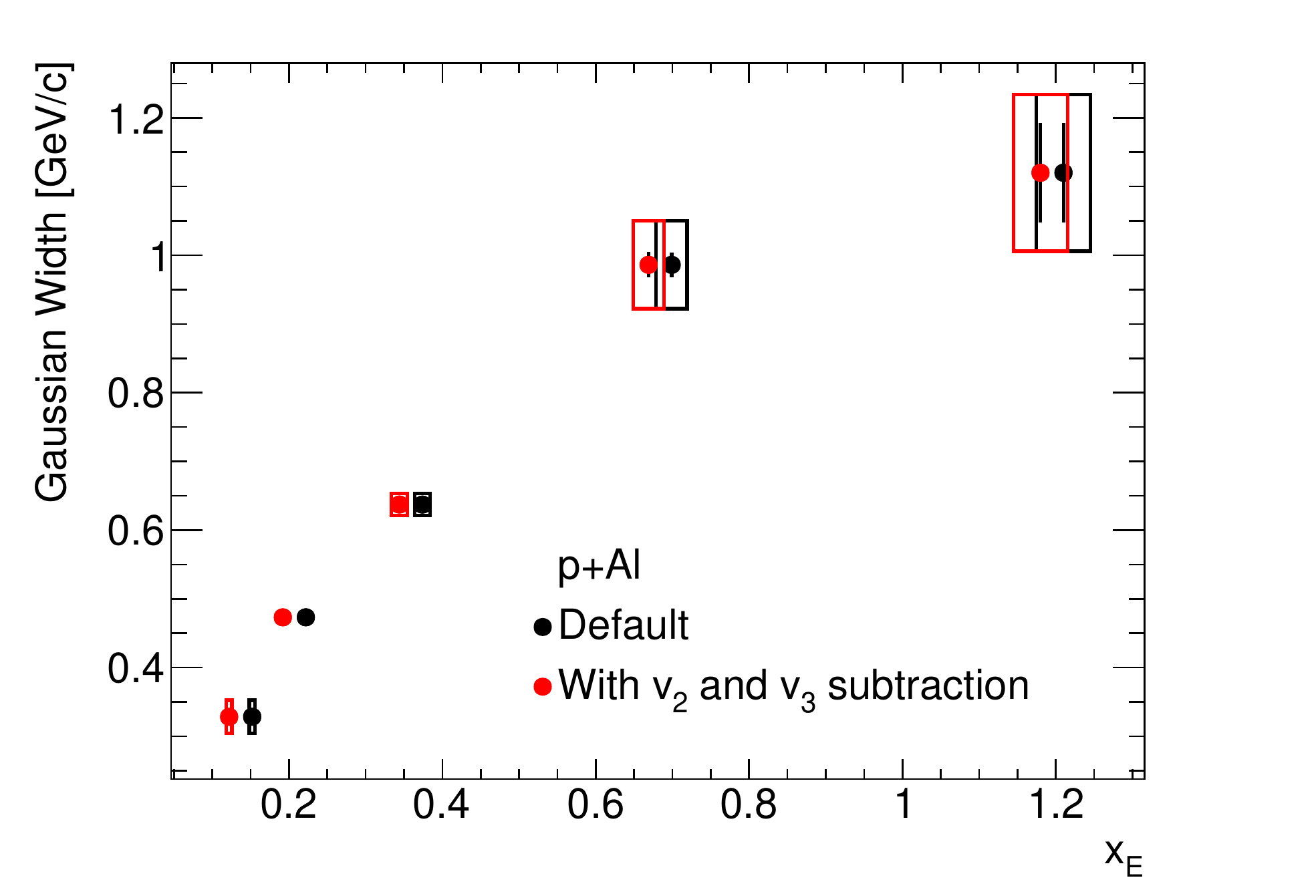}
	\includegraphics[width=0.49\textwidth]{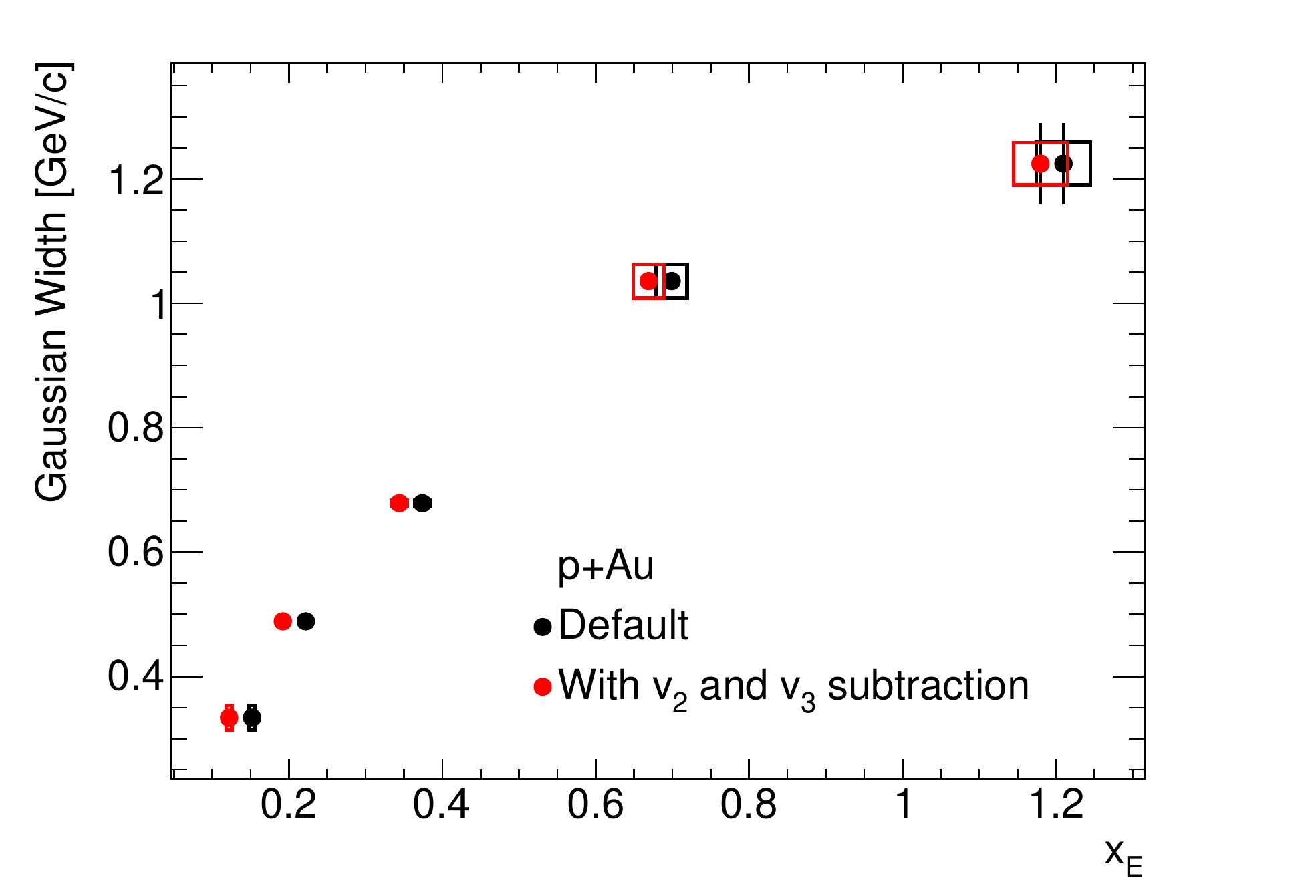}
	\caption{The Gaussian widths are extracted from \pout distributions with and without \vtwo and \vthree modulated background contributions for \pal (left) and \pau (right) collisions.}
	\label{fig:modulated}
\end{figure}

\begin{figure}[tbh]
	\centering
	\includegraphics[width=0.6\textwidth]{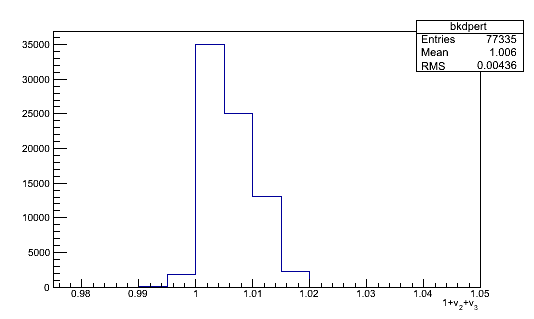}
	\caption{The term in parentheses in Eq.~\ref{eq:background} is shown, which shows the percent contribution from the modulated background compared to the underlying event contribution unity. This indicates that the contribution from any modulated background in \pa collisions is at most 2\%, however the average contribution is less than 1\%.}
	\label{fig:backgroundcontribution}
\end{figure}

The correlation functions were constructed with the modulated background subtraction, and the Gaussian widths were extracted and compared to the widths without the modulated background subtraction. The result is shown in Fig.~\ref{fig:modulated}, which shows that there is no difference between the Gaussian widths with and without the modulated background subtraction. At first, this seems unreasonable given the large Au+Au values of \vtwo and \vthree used; however, this can be rationalized with the following. In the small $\pout\approx0$ region, $\dphi\approx\pi$, thus the modulated background contribution can be approximately reduced to
\begin{equation}\label{eq:background}
b_0(1+2v_2^{\pion}v_2^{\h}-3v_3^{\pion}v_3^{\h})
\end{equation}
since $\cos(2\pi)=1$ and $\cos(3\pi)=-1$. The maximum values of \vtwo used in the study are $v_2^{\pion}v_2^{\h}=0.075\times0.14=0.0105$. The \vthree contribution is even smaller since the \vthree values are smaller; at their maximum values  $v_3^{\pion}v_3^{\h}=0.05\times0.01=0.0005$. Therefore, the modulated background contributes, at most, approximately $2\times0.0105-3\times0.0005=0.02$; compared to the first term in the modulated subtraction of Eq.~\ref{eq:background}, 1, this amounts to a 2\% effect. As a reminder, this is already an incredibly conservative upper limit since the \vtwo and \vthree values were taken from Au+Au collisions, and could be expected to be roughly 3 times larger than similar measurements in \pa collisions at RHIC energies. To confirm this, the quantity in parentheses in Eq.~\ref{eq:background} was plotted and is shown in Fig.~\ref{fig:backgroundcontribution}; the figure confirms that the modulated background on average contributes less than 1\% with the unreasonably large \vtwo and \vthree values. Therefore, this study has shown that any modulated background from hydrodynamic flow can be safely assumed to be negligible and thus not contributing to the observed transverse momentum broadening in \pa collisions.

\subsection{Summary of \pa Results}

Since the \pa results have many different possible effects contributing, some summarizing remarks should be made based on the results presented. The original motivation for studying \pa collisions was to potentially measure processes that may be more sensitive to factorization breaking effects. Since there are more nucleons and thus colored objects in a \pa collision, it could be naively expected that the magnitude of factorization breaking effects might be stronger. However, there are other physical phenomena that may come into play in addition to effects from factorization breaking.

The main result is that the nonperturbative away-side jet widths in \pa collisions are slightly broadened compared to those in \pp collisions at the same center-of-mass energy. This broadening is clearly dependent on \ncoll, the number of nucleon-nucleon interactions in the collision, which suggests that there is a nuclear medium path length dependence to the broadening. The relation to factorization breaking is in the soft gluon exchanges; since the Gaussian widths of \pout are broadened they may be undergoing more soft gluon exchanges due to the higher number of colored objects in the event. For example, factorization violating soft color exchanges have been argued to be responsible for the suppression of $\psi(2S)$ mesons compared to $\psi(1S)$ in \pa collisions~\cite{Ma:2017rsu}. The results also indicate that in certain bins of \xe the Gaussian widths increase at a faster rate in central collisions as opposed to peripheral \pa and \pp collisions. Perturbative QCD calculations in \pp collisions which assume factorization could ultimately be compared to the \pp measurements, and then based on the rate of change of the widths in \pa collisions conclusions about the magnitude of factorization breaking effects in \pa collisions may be possible.

\begin{figure}[tbh]
	\centering
	\includegraphics[width=0.7\textwidth]{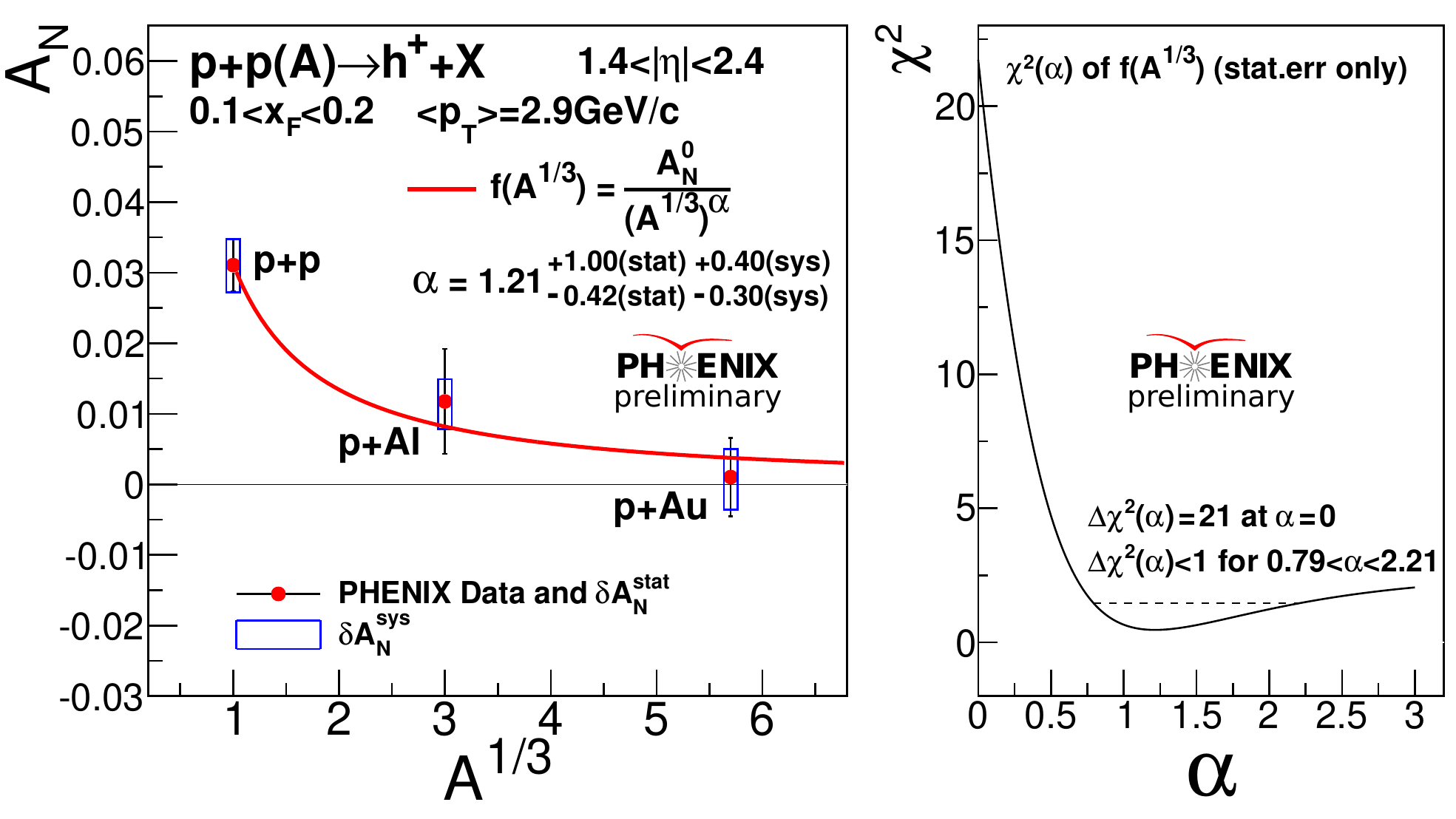}
	\caption{The transverse single spin asymmetry of nonidentified charged hadrons is shown as a function of the atomic number A. The asymmetry is suppressed in larger nuclei, as shown by the $\chi^2$ per degree of freedom in the right panel which indicates that it is statistically unlikely that there is no dependence on the atomic number A.}
	\label{fig:pa_an}

\end{figure}

As previously discussed, there are other physical phenomena that are known to be present in \pa collisions. In particular, the present results have ruled out both final-state TMD fragmentation modification in \pa collisions and collective multi-particle correlations as a contributor to the away-side width broadening. Nevertheless, it is still possible that energy loss mechanisms could be playing a role in the broadened widths. Interestingly, the interactions that lead to factorization breaking, soft gluon exchanges, could be manifesting themselves as radiative energy loss of a parton in a nuclear medium. It is plausible that the physical mechanism which leads to energy loss may be related to factorization breaking in this sense. However, nPDFs may contribute to the broadening as well; since nPDFs are still only roughly determined at the collinear level, it is possible that additional partonic \kt in nuclei may contribute~\cite{Corcoran:1990vq}. This additional \kt could lead to the away-side width broadening. It is also plausible that this could be a function of the impact parameter of the collision; for example, partons within nucleons at the center of the nucleus may experience additional \kt compared to partons at the edge of a nucleus due to the larger number of possible QCD interactions. There is also the possibility that multiple semi-hard partonic scatterings occur in these collisions, giving rise to a broadening of the jet widths. Interestingly, the transverse single spin asymmetry of charged hadrons in the polarized process $p\uparrow+p\rightarrow\h+X$ is found to have a strong nuclear dependence, shown in Fig.~\ref{fig:pa_an}. This may also indicate the effect of additional soft gluon exchanges that, in the case of the asymmetry, tend to reduce any anisotropy due to the spin-momentum correlation. Similar effects could lead to the \pt broadening observed in the \pout distributions.

The kinematic region where the away-side nonperturbative jet widths are broadened in \pa collisions is exactly the region where the Cronin peak is observed. The intermediate \pt range where the Cronin peak is found is roughly $2<\pt<7$ \gevc. The away-side jet width modification is observed in the kinematic region $5<\pttrig<9\otimes1<\ptassoc<4$. Both of these ranges are approximately within the kinematic region where the inclusive hadron \pt spectrum is enhanced in \pa compared to \pp collisions, and since correlations inherently give more information about the event, the observed broadening may be able to provide insights into the physical mechanism which leads to the single particle enhancement. It is entirely possible that this enhancement is related to the physical mechanisms described above; however, this has not yet been rigorously shown. Nonetheless, it is necessary to continue providing new measurements with different observables to provide more information to ultimately understand the underlying physics in \pa collisions.

\chapter{Discussion}
\label{chap:discussion}

To understand qualitative and quantitative effects from factorization breaking, it is necessary to compare observables from processes where factorization is predicted to hold and where it is predicted to break. The results presented in Chapter 4 and 5 are the first measurements which are sensitive to predicted TMD factorization breaking effects; thus world measurements of TMD sensitive observables in SIDIS and DY where factorization is predicted to hold must be compiled and compared to those presented here. With the recent global effort to increase available data sensitive to TMD observables, it has become possible to make more robust comparisons between the SIDIS, DY, and \pp(\pa)$\rightarrow h_1+h_2+X$ or \pp(\pa)$\rightarrow\gamma+h+X$ processes.

\section{Discussion of Measured Results}

While the results of Ref.~\cite{ppg195} were intriguing, the analysis in Ref.~\cite{ppg217} has shown that the decreasing momentum widths as a function of the hard scale was a result of a fragmentation effect; as the hard scale was increased for a fixed range of \ptassoc the average longitudinal momentum fraction of the away-side hadron with respect to the hard scale decreased. When the distributions are binned in the modified fragmentation variable \xe instead of \ptassoc, the widths increase with the hard scale, which is qualitatively similar to measurements in DY. Future estimates of the magnitude of factorization breaking effects will rely on calculations to compare the rate of evolution of the nonperturbative widths between, for example, DY and the results presented here. Nonetheless, there are several important observations to consider concerning the results presented here.

There have been several previous results from RHIC which studied the behavior of \rmspout as a function of \pttrig and \ptassoc~\cite{ppg029,ppg089,ppg095}, but there is an important distinction to be made between this quantity and the Gaussian widths of the \pout distributions presented here. The quantity \rmspout is extracted from fits to the entire away-side jet, which means that, while dominated by nonperturbative contributions in the nearly back-to-back region, it also contains perturbative contributions away from \dphi$\sim\pi$. Observables sensitive to TMD factorization breaking effects need to be sensitive only to nonperturbative contributions since it is specifically predicted to arise from soft gluon exchanges within a TMD framework; the Gaussian widths of \pout are by definition only sensitive to nonperturbative $k_T$ and $j_T$ in the nearly back-to-back region. This subtle difference is important to note, as the modified sensitivity to nonperturbative contributions of the two different observables may be important for interpreting quantitative effects from factorization breaking.

Moreover, since the widths are in the nonperturbative nearly back-to-back region, the quantity \xe is actually a much closer approximation to $z$ than might at first be expected given its definition. The largest smearing effect then comes from the approximation that the near- and far-side jets have the same but oppositely pointing \pt if one assumes no contributions from nonperturbative \kt. This is of course a LO approximation as it has been shown that even in direct photon-jet production the direct photon and jet are imbalanced in \pt in the back-to-back region~\cite{Sirunyan:2017qhf}. However, without full jet reconstruction, the quantity \xe suffices to approximate the away-side hadron momentum with respect to the jet. 

\begin{figure}[tbh]
	\centering
	\includegraphics[width=0.49\textwidth]{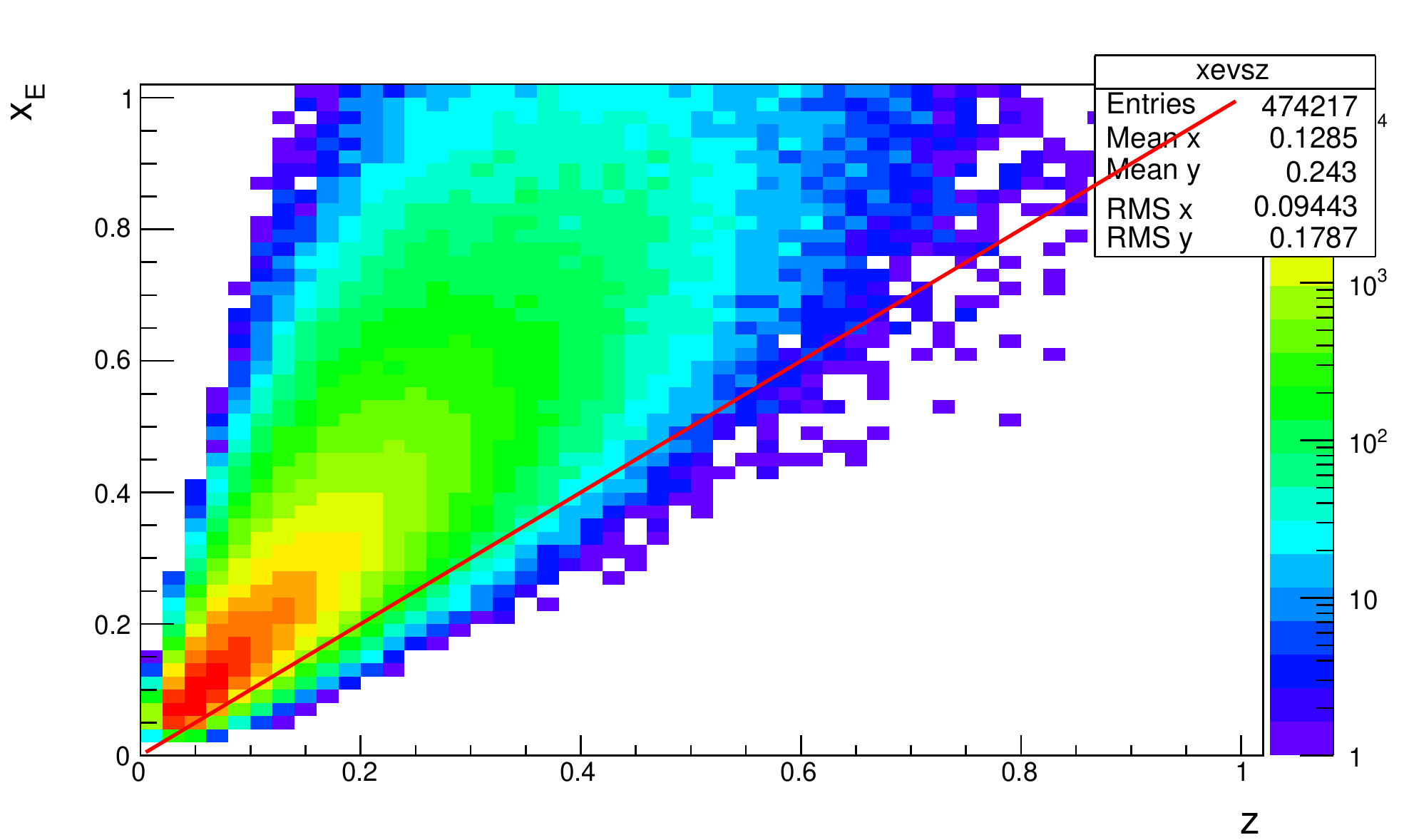}
	\includegraphics[width=0.49\textwidth]{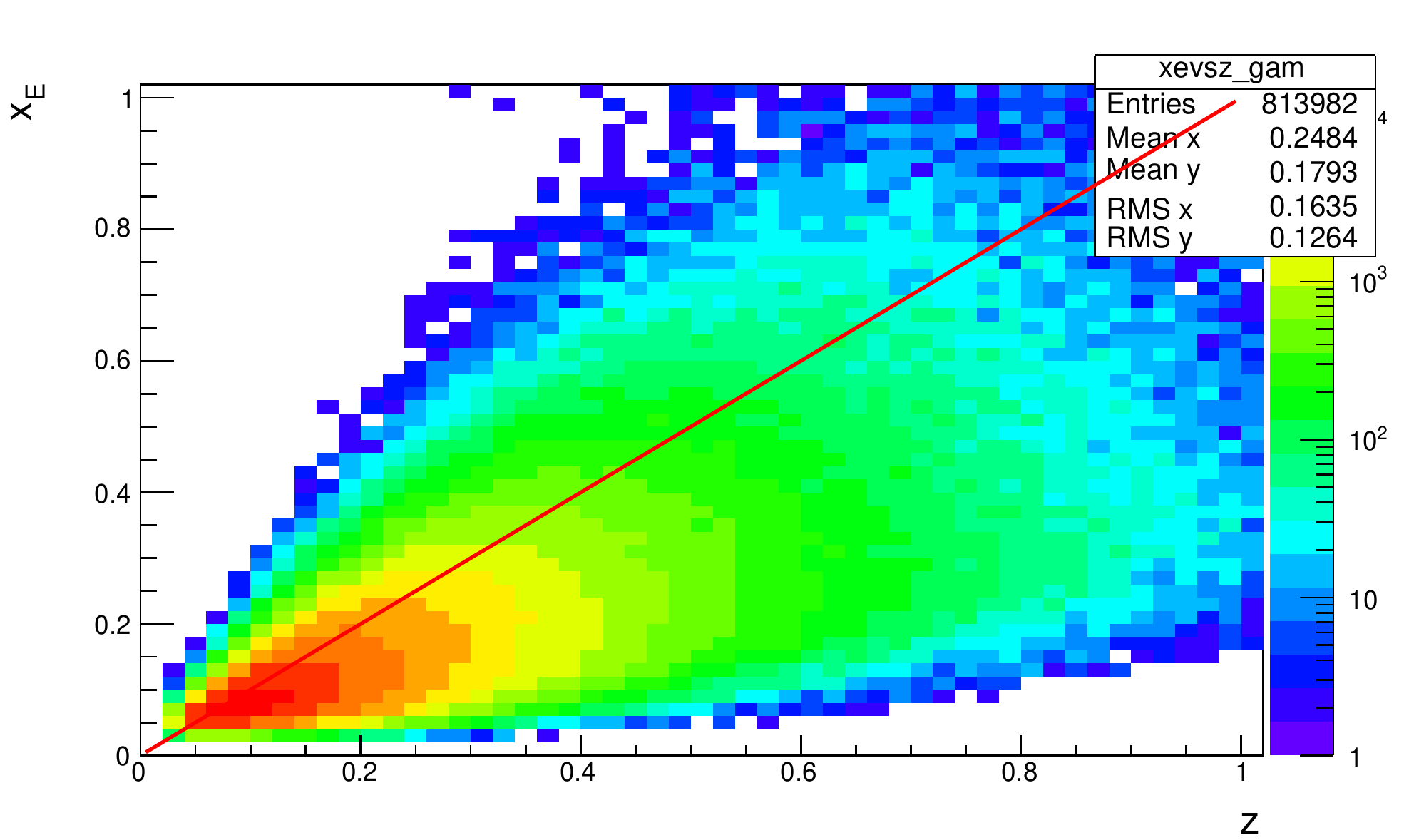}

	\caption{The correlation between \xe and $z$ is shown as determined in \sqs=~200 GeV \pythia Monte Carlo simulations, for $\pion$-\h (left) and $\gamma$-\h (right) correlations.}
	\label{fig:xevszcorrelation}
\end{figure}

Nonetheless, to study the correlation between $z$ and \xe further, a \pythia study was performed for both \pion-\h and $\gamma$-\h correlations to quantitatively determine how closely they are correlated. Correlations were collected in \sqs=~200 GeV \pp collisions in the PHENIX pseudorapidity acceptance, similarly to previous \pythia studies. The value of the away-side hadron $z$ was determined by matching the hadron to the hard scattered parton closest in kinematic phase space defined by $0.3>\Delta R=\sqrt{\Delta\phi^2+\Delta\eta^2}$. This is the best way to match partons to particular hadrons, since the way that \pythia performs the fragmentation is complicated and there is not necessarily a one-to-one correspondence between partons and hadrons. The correlation between \xe and $z$ for \pion-\h and $\gamma$-\h correlations is shown in the left and right panel, respectively, of Fig.~\ref{fig:xevszcorrelation}. The plots also show a red line corresponding to \xe=~$z$ to highlight the strength of the correlation. The \pythia study shows that on average, $\xe>z$ for \pion-\h correlations and $\xe<z$ for $\gamma-\h$ correlations. This can be explained due to the fact that $\hat{x}_h\approx0.9$ and \zt$\approx$~0.6 at RHIC center-of-mass energies. 

\begin{figure}[tbh]
\centering
\includegraphics[width=0.6\textwidth]{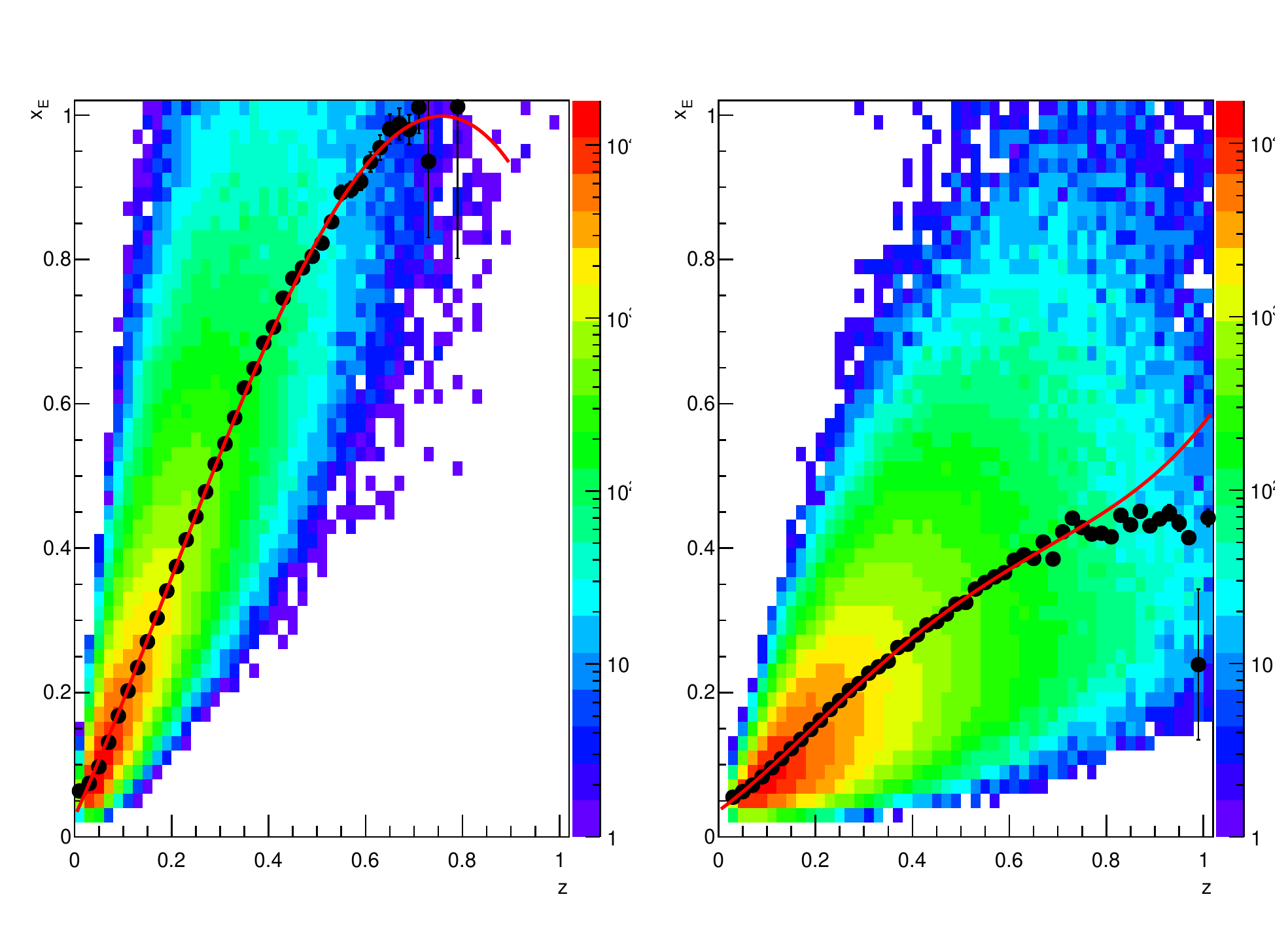}
\caption{The correlation between \xe and $z$ for \pion-\h (left) and $\gamma$-\h (right) is shown with Gaussian fits applied to each slice in $z$. The mean of the Gaussian is shown as the black points, and a third order polynomial is fit to the points to form a continuous function.}
\label{fig:xezcorrfits}
\end{figure}

\begin{figure}[tbh]
	\centering
	\includegraphics[width=0.6\textwidth]{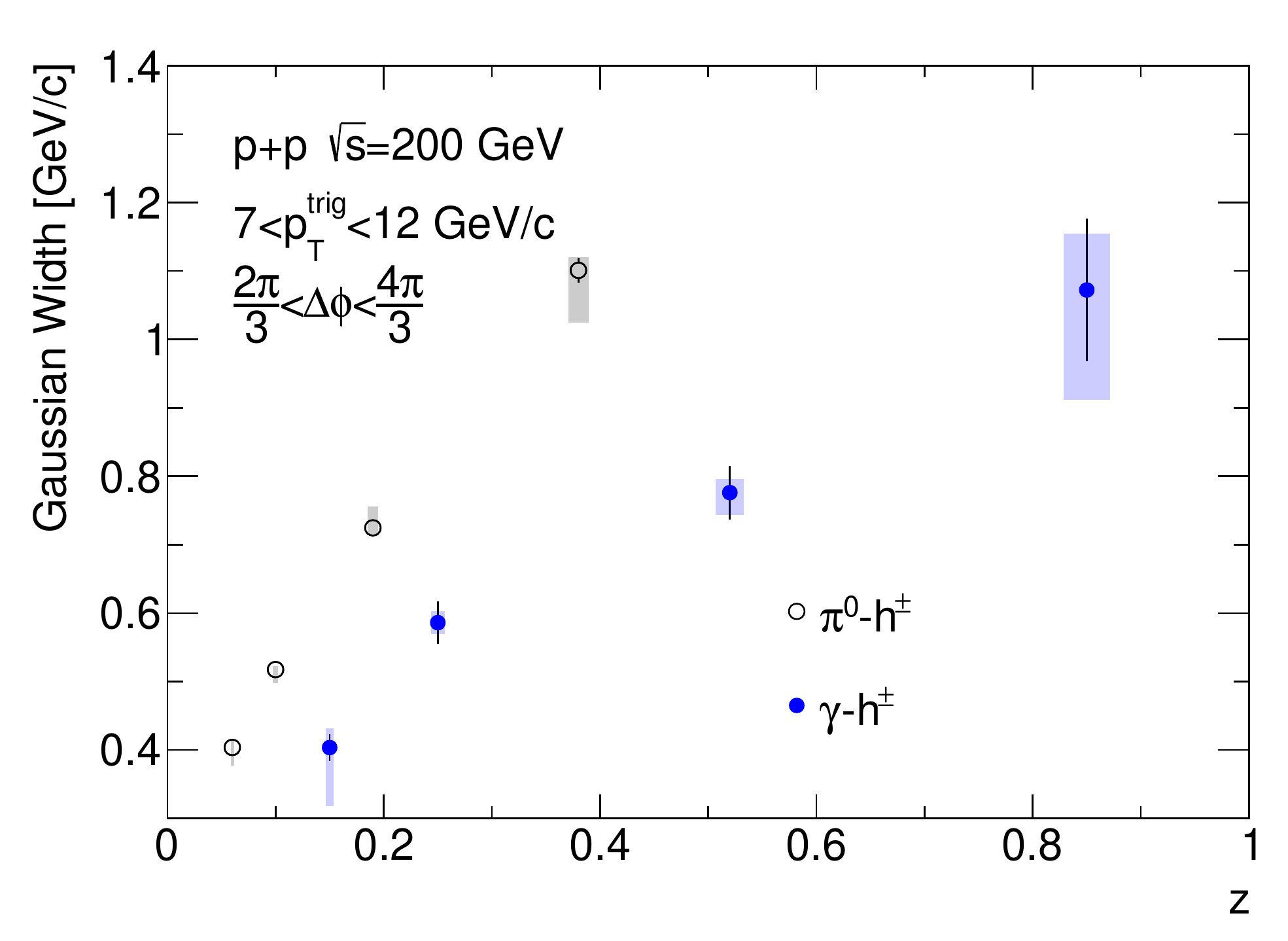}
	\caption{The \sqs=~200 GeV \pp Gaussian widths of \pout are shown as a function of $z$, where $z$ was determined with \pythia Monte Carlo simulations as described in the text.}
	\label{fig:widthsfxnz}
\end{figure}

To construct a plot comparing the \sqs=~200 GeV Gaussian widths as a function of the \pythia determined $z$, the correlations were fit with Gaussian functions to determine the mean value between $z$ and \xe. The mean of the Gaussians were fit with a third order polynomial to smooth out the relation, as shown in Fig.~\ref{fig:xezcorrfits}. With this function, the Gaussian widths as a function of \xe were shifted to be shown as a function of $z$ with the caveat that this correlation was determined with \pythia Monte Carlo and could be fragmentation model dependent. Since the correlation between $\xe$ and $z$ becomes highly non-Gaussian at large $z$ for the $\gamma$-\h correlations, the correction is only applied to the smaller \xe bins for these particular correlations. Due to the difference in \xe-$z$ correlation between \pion-\h and $\gamma$-\h, the Gaussian widths as a function of away-side $z$ are quite different between the two trigger types as shown in Fig.~\ref{fig:widthsfxnz}. This could be due to an extra component of soft \jt contributing to the \pion-\h correlations compared to the $\gamma$-\h correlations. Another contributing factor could be that \pion triggers probe a mix of gluon and quark jets, while the direct photons probe a significantly higher fraction of quark jets than gluon jets.

\begin{figure}[tbh]
	\centering
	\includegraphics[width=0.49\textwidth]{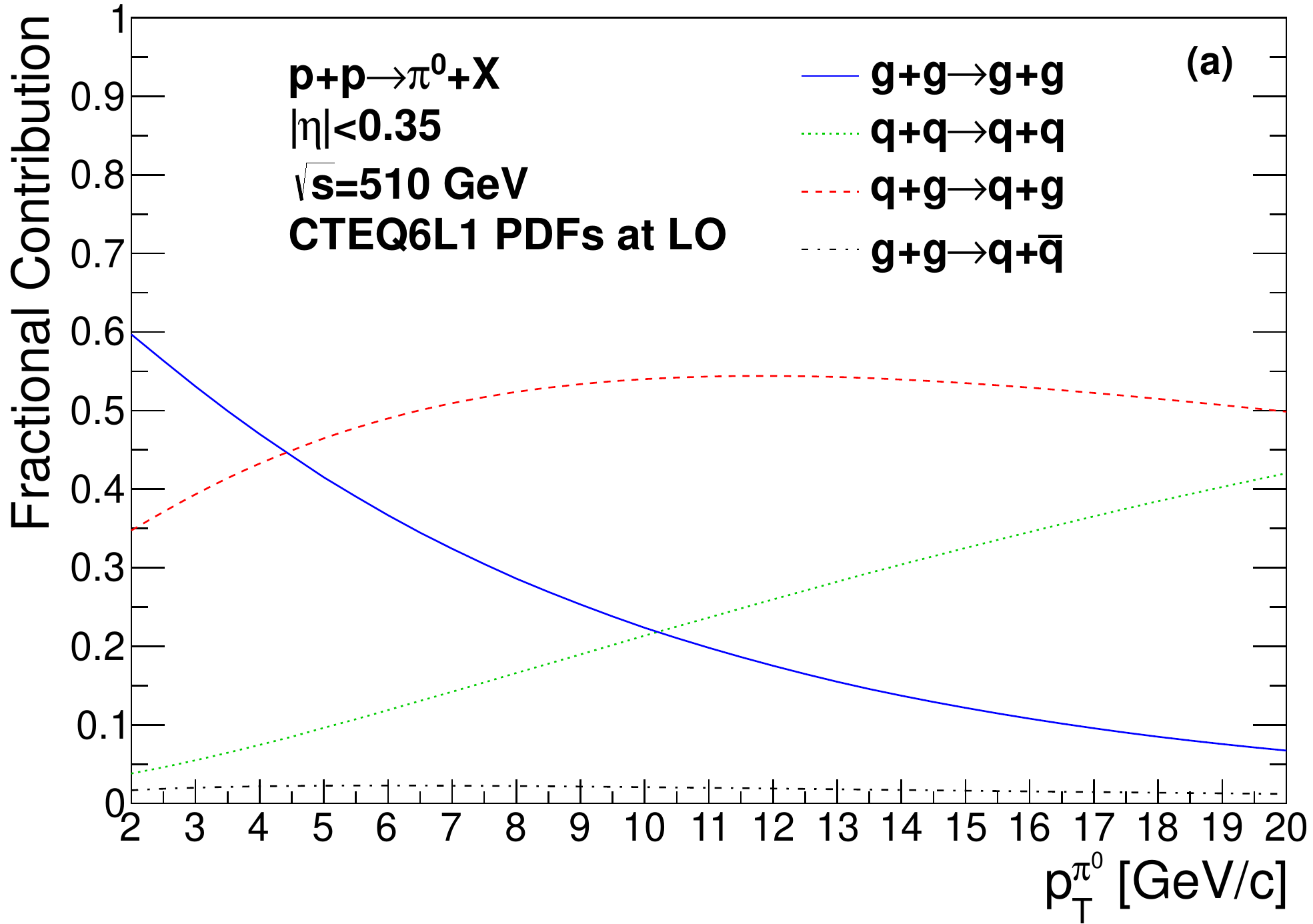}
	\includegraphics[width=0.49\textwidth]{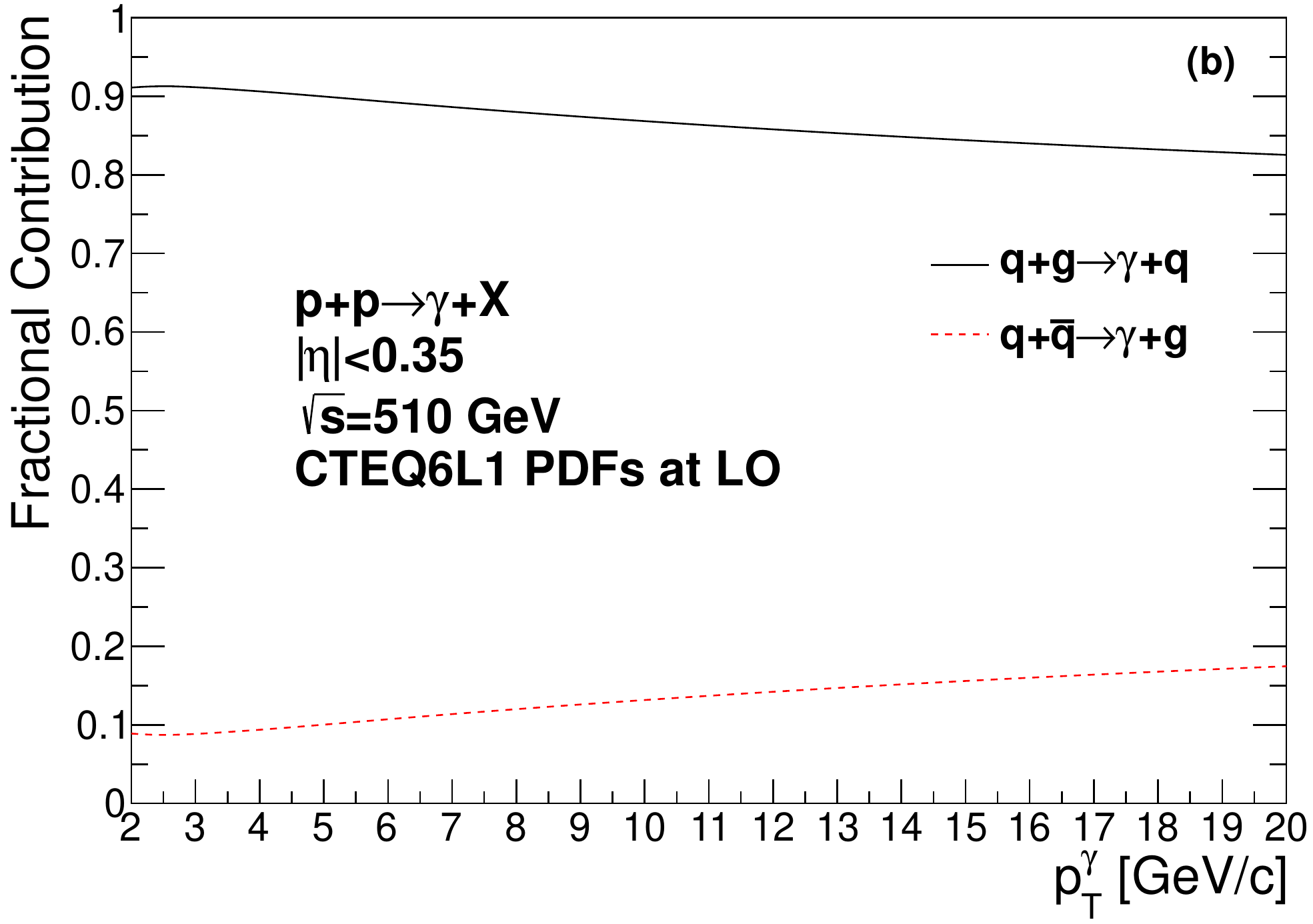}
	
	\caption{The leading order partonic contributions to inclusive \pion and direct photon production at \sqs=~510 GeV and midrapidity are shown on the left and right, respectively. Note that the process $q\bar{q}\rightarrow gg$ is not drawn in the left panel because its contribution is less than one percent in this \pt range.}
	\label{fig:partonic_mixes510}
\end{figure}

A benefit to measuring both \pion-hadron and direct photon-hadron correlations is that the processes probe different partonic hard scattering fractions. Figure~\ref{fig:partonic_mixes510} shows the mix of LO diagrams calculated within pQCD that inclusive-\pion and direct photon triggers probe at \sqs=~510 GeV. The CTEQ6L1 PDFs~\cite{Pumplin:2002vw} were used for the calculations in addition to the DSS14 FFs~\cite{deFlorian:2014xna} for the \pion fragmentation. The same diagrams for \sqs=~200 GeV collisions are shown in Ref.~\cite{ppg095} and are very similar to the \sqs=~510 GeV fractional contributions. From the diagrams it is clear that direct photon-hadron production is dominated by quark gluon Compton scattering such that the away-side hadron will be produced by a quark jet roughly 85\% of the time, whereas \pion triggers probe a significant mix of $qg$, $qq$ and $gg$ scatterings. Interestingly, the results show that, within uncertainties, there is little difference between the direct photon and \pion triggered away-side jet widths as a function of \xe, \pttrig, and \sqs, which could point to a similar nonperturbative fragmentation mechanism between quark and gluon jets. However, studies between light quark and gluon with robust jet finding and jet constituent tracking have shown differences between the fragmentation patterns of the two fundamentally different partons~\cite{Alexander:1995bk,Aad:2016oit}.

The results presented here will have to be compared to calculations before any quantitative statements can be made about factorization breaking effects. However, it is interesting to point out that the large inclusive hadron transverse single spin asymmetries measured in \pp collisions at forward rapidities do not appear to follow standard perturbative evolution. In charged pion production the measured asymmetries change very little from \sqs=~4.9 GeV to \sqs=~62.4 GeV~\cite{Aidala:2012mv}, and more recent measurements show that this holds true even up to \sqs=~200 and 500 GeV~\cite{HEPPELMANN:2013ewa,Dilks:2016ufy}. The asymmetries are predicted to fall off towards 0 at large \pt, however they have been measured to be nearly 8\% up to \pt=8 \gevc~\cite{Dilks:2016ufy}. Comparing future calculations to the data presented here to determine how well, if at all, they follow predictions from standard perturbative evolution.

Moreover, it is also interesting to point out that in processes where factorization breaking should play a role, or processes where there are hadrons in both the initial and final states, there seems to be little change of observables over a large range of center-of-mass energy. The results presented here are all consistent with each other at similar \pttrig but very different \sqs. Similarly, the transverse single spin asymmetries have been measured to be largely constant as a function of \sqs. Recent measurements from RHIC of Collins-like jet asymmetries~\cite{Adamczyk:2017wld} and transverse spin dependent dihadron correlations~\cite{Adamczyk:2017ynk} also show little to no change of observables which have hadrons in the initial and final states over a large range of center-of-mass energy. However, TMD observables in SIDIS and DY show significant changes over large ranges of center-of-mass energies; this will be explored further later in the discussion. While this is not necessarily evidence of factorization breaking effects, these observations could point to a mechanism which is present in $p+p\rightarrow$hadron measurements that is not present in SIDIS or DY interactions.

Since TMD evolution effects have yet to be well constrained, the data presented here at two different center-of-mass energies may also provide input to future nonperturbative global fits for TMD evolution calculations, assuming factorization breaking effects are small. While TMD evolution is only applicable to how observables change with \qsq, the different center-of-mass energies will provide data to constrain the relation between $x$ and \kt. World measurements have provided differing inputs to the question of how TMD observables behave with \sqs and \qsq; for example, the Collins-like FF and transversity TMD PDF show little change from \sqs=~200 to 510 GeV~\cite{Adamczyk:2017wld,Adamczyk:2017ynk}. Note that these measurements, in addition to those presented here, have a clearly defined, large hard scale well above the soft scale necessary to be treated within a TMD framework. Global measurements in SIDIS, at a relatively low $Q^2$ of 1-12 GeV$^2$/$c^2$, and high mass DY and $Z$ boson data have shown disagreements in the predicted size of TMD evolution effects~\cite{Collins:2014jpa}. The recent data provided by both the STAR and PHENIX collaborations may help ameliorate these problems, as the range of $Q^2$ probed is between the DY and $Z$ boson scales.

\section{Other World Measurements}

Many worldwide measurements have been made of processes predicted to break factorization or processes predicted to factorize; thus, it is useful to compile these measurements and compare to those presented here which were the first to explicitly search for TMD factorization breaking effects. Ultimately the quantitative magnitude of any factorization breaking effects will have to be determined by comparisons between, for example, the DY and dihadron processes; therefore, this compilation of results can serve as a starting point for future global phenomenological studies.

\subsection{Processes Predicted to Factorize}
\subsubsection{Drell-Yan}

Most of the world DY data that is also sensitive to a small transverse momentum scale exists from fixed target experiments at Fermilab. Phenomenological studies have already shown that DY and $Z$ boson TMD observables follow perturbative expectations from CSS evolution~\cite{Landry:2002ix,Konychev:2005iy,Schweitzer:2010tt}; however, it is also useful to compile world data and perform fits to extract observables that may be more directly comparable to those measured in dihadron and direct photon-hadron correlations. For example, the measured quantities of $\langle\pt\rangle$ that are often reported in DY analyses (see e.g.~\cite{Moreno:1990sf}) contain both perturbative and nonperturbative contributions since the average is taken over the entire \pt spectrum. For this reason it would be ideal to also have an observable in DY that is only sensitive to the nonperturbative contributions, similarly to the Gaussian widths of \pout in dihadron and direct photon-hadron correlations. \par

\begin{figure}[tbh]
	\centering
	\includegraphics[width=0.49\textwidth]{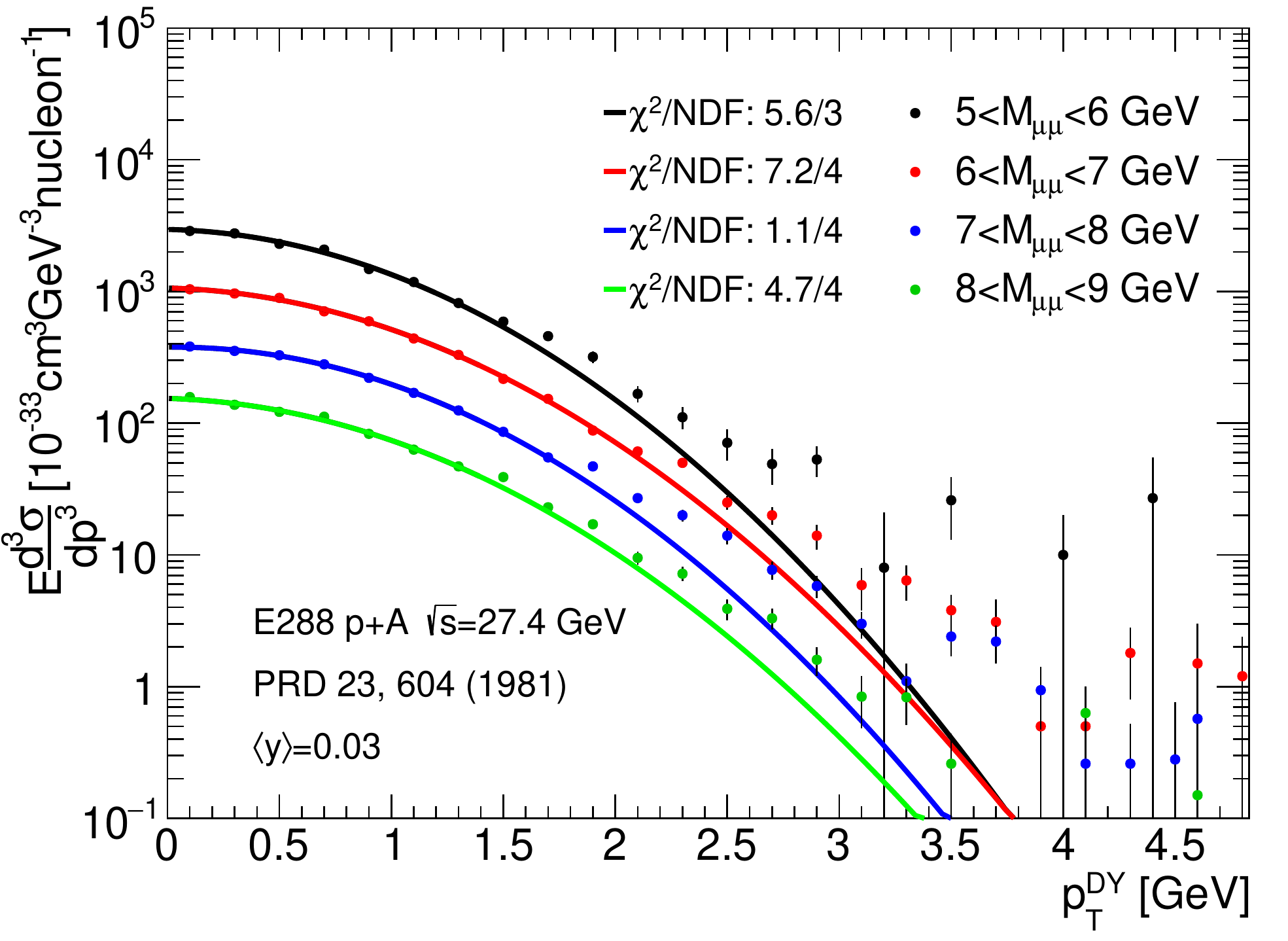}
	\includegraphics[width=0.49\textwidth]{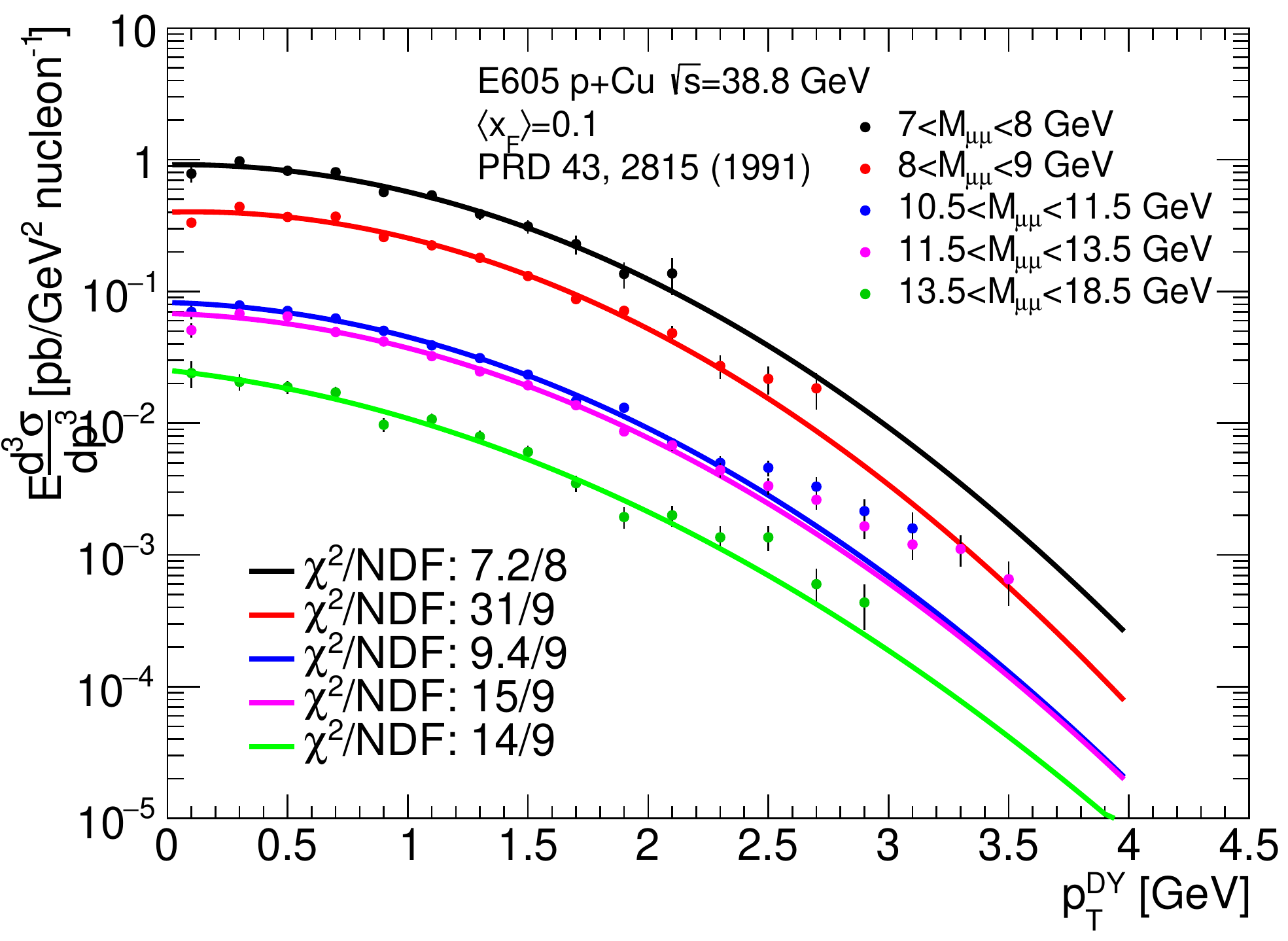}\\
	\includegraphics[width=0.49\textwidth]{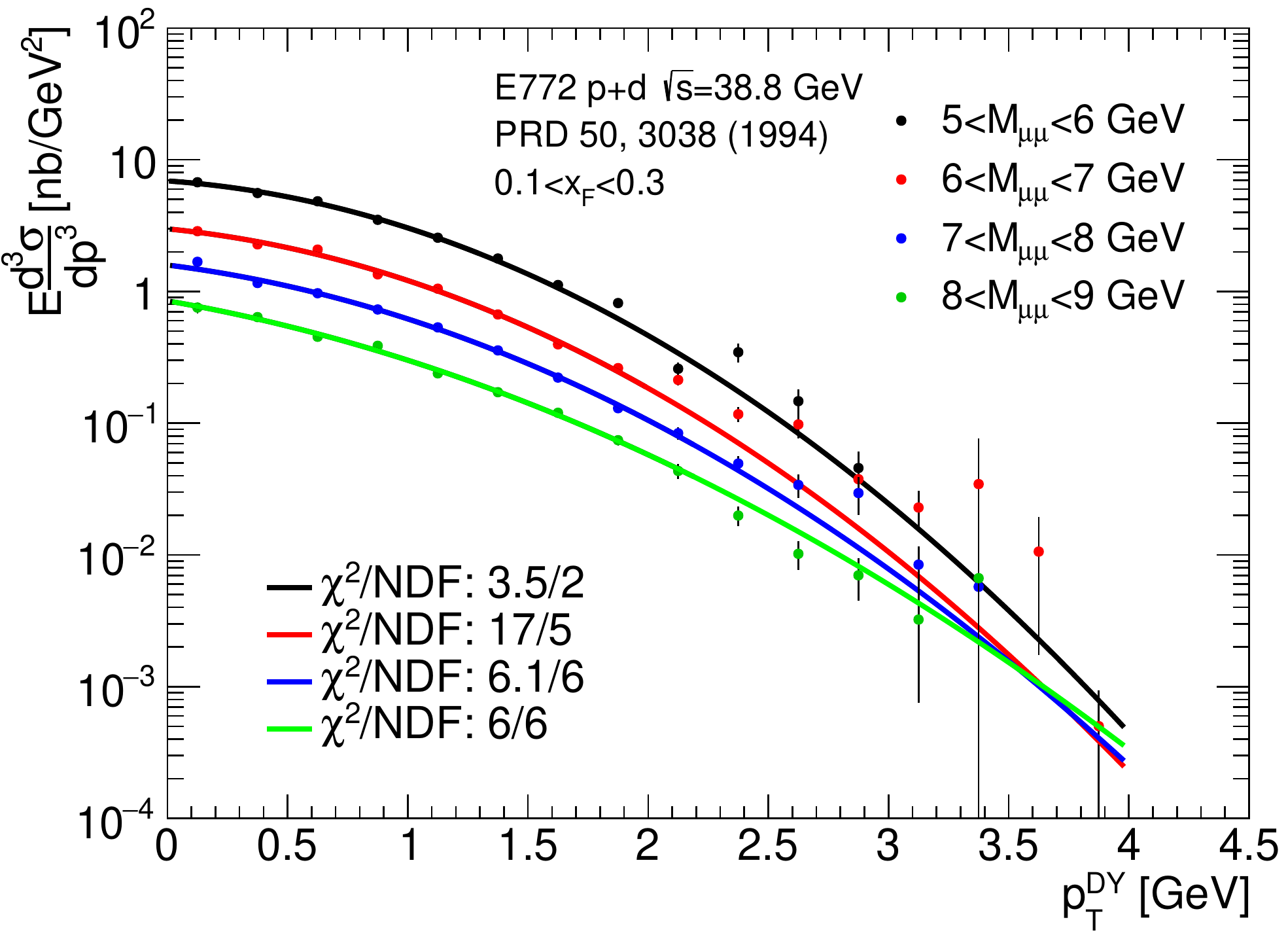}
	\includegraphics[width=0.49\textwidth]{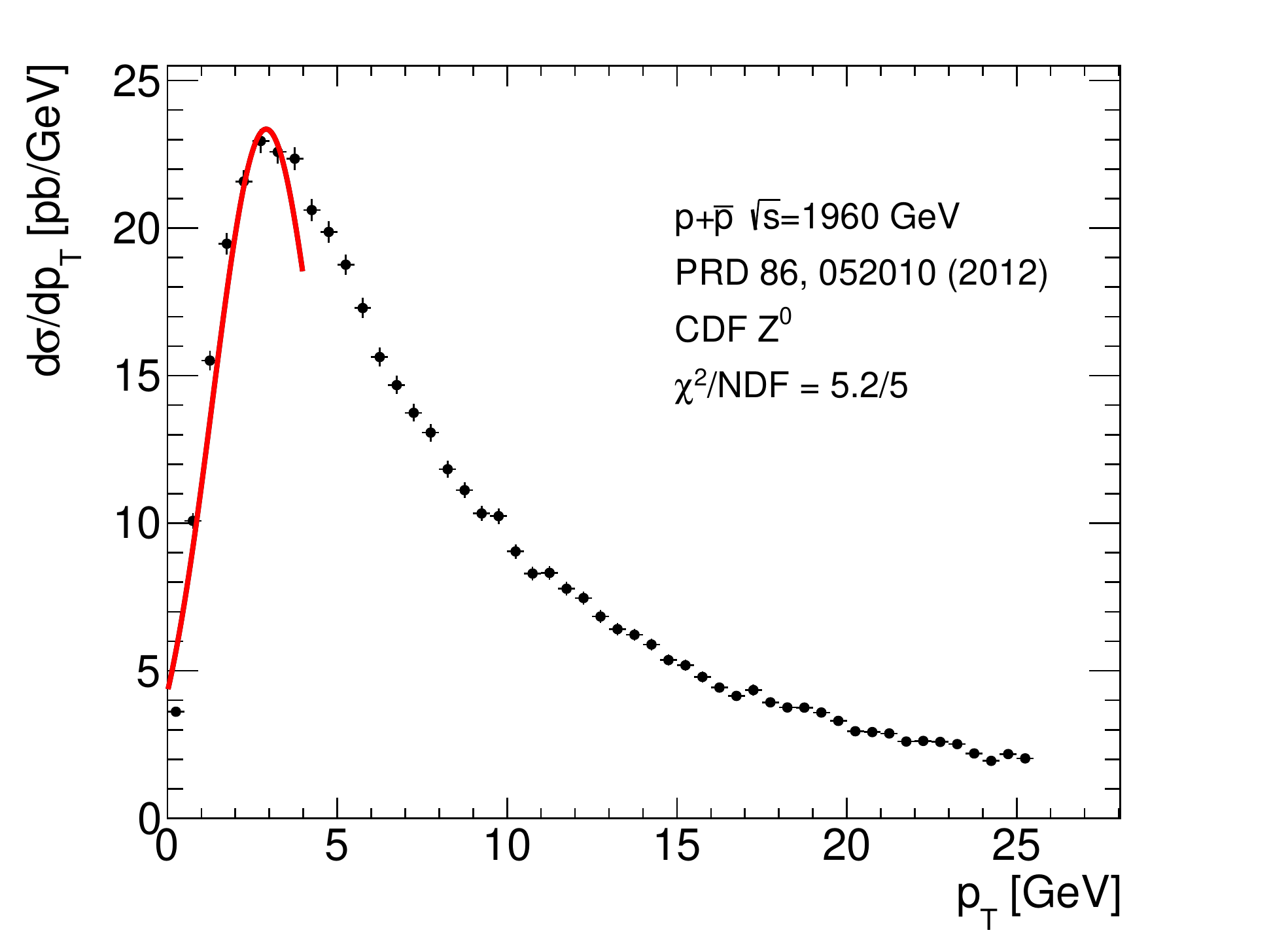}
	\caption{Published DY and $Z$ boson cross sections are shown with Gaussian fits at small \pt as described in the text.}
	\label{fig:dy_xsec_fits}

\end{figure}

\begin{figure}[tbh]
	\centering
	\includegraphics[width=0.6\textwidth]{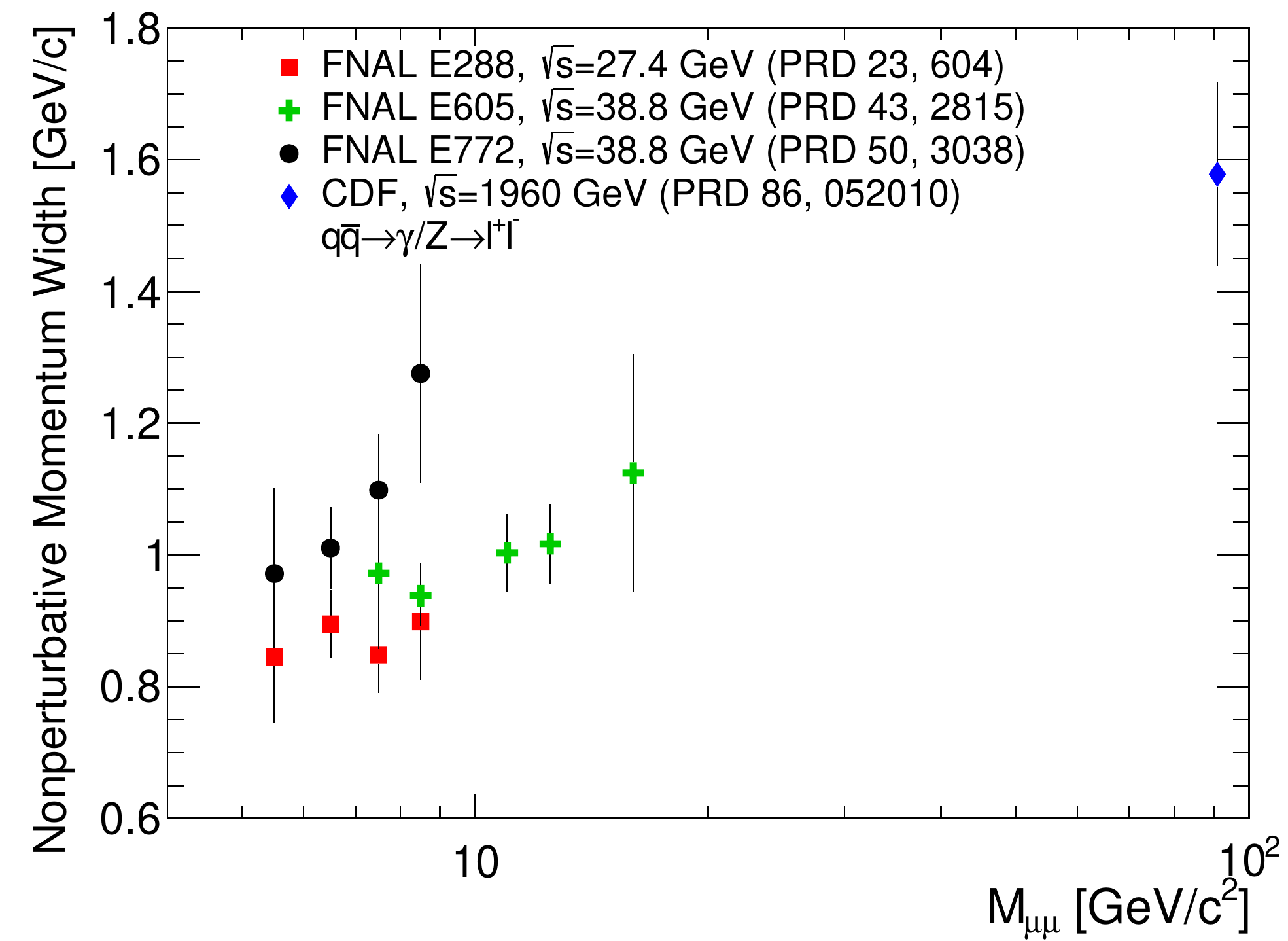}
	\caption{The nonperturbative Gaussian widths are shown as a function of $M_{\mu\mu}$ for the DY and $Z$ boson \pt distributions.}
	\label{fig:dy_gauss_widths}
\end{figure}

To make the most precise comparisons to data with similar partonic quantities, only data was chosen that was as close as possible to the kinematic phase space probed by the dihadron and direct photon-hadron correlations. Data was taken from Refs.~\cite{Ito:1980ev,Moreno:1990sf, McGaughey:1994dx} in the smallest possible $x_F$ bin, which is the closest possible rapidity region to $x_F\sim0$ that is covered by the PHENIX central arm spectrometers. The $Z$ boson cross section measurement was also analyzed from the CDF collaboration~\cite{Aaltonen:2012fi}, which is the most precise measurement to date in the small \pt region that is sensitive to the initial-state \kt of the colliding $q\bar{q}$ pair. The cross sections were fit with Gaussian functions in the small \pt region, as shown in the panels of Fig.~\ref{fig:dy_xsec_fits}. Similarly to the \pout distributions in \pp$\rightarrow$ dihadrons, at large \pt the cross section becomes perturbative and the Gaussian fits no longer describe the data. It should be noted that the Gaussian fit for the $Z$ boson distribution is not the most precise at high \pt, however this fit range gave the best $\chi^2$ per degree of freedom of 5.2/5. The Gaussian widths were extracted from the fits and are displayed as a function of $M_{\mu\mu}$ in Fig.~\ref{fig:dy_gauss_widths}. The widths clearly show a systematic positive trend with both $M_{\mu\mu}$ and \sqs, however the uncertainties from the fit are in general large. This is partly due to the statistical accuracy of the data as well as the fact that the Gaussian width is only constrained by one half of the actual functional form since $p_T$ is necessarily positive.

\begin{figure}[tbh]
	\centering
	\includegraphics[width=0.6\textwidth]{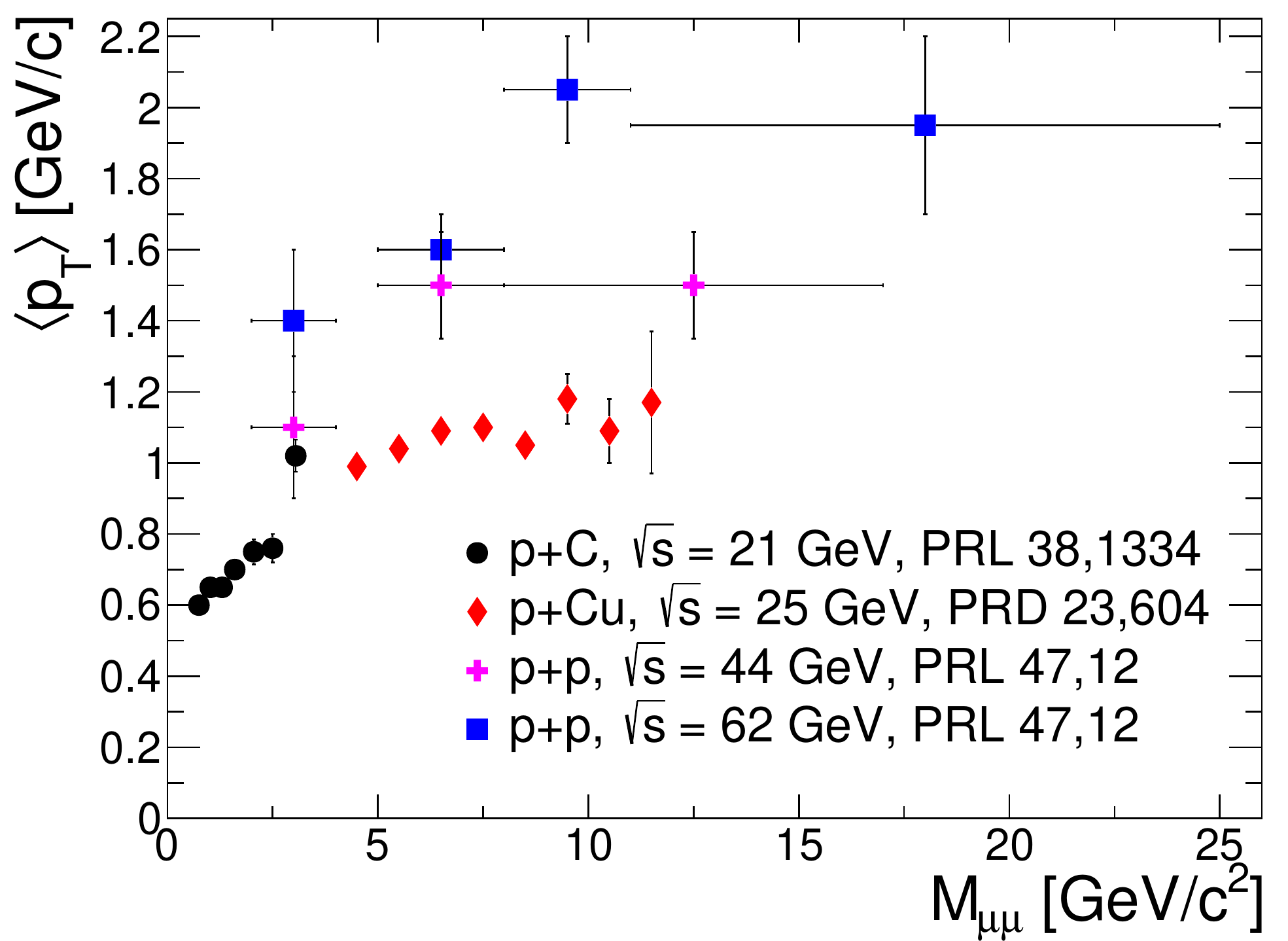}
	\caption{The quantity $\langle\pt\rangle$ is shown for DY events over a wide range of $M_{\mu\mu}$ and \sqs, where the data is taken from the references noted on the figure. This quantity behaves similar to the Gaussian fits to the DY \pt spectra, however it contains both perturbative and nonperturbative contributions since the average is performed over the entire \pt range.}
	\label{fig:dy_avg_pt}
\end{figure}

To compile additional DY data, the average \pt of DY dimuon pairs was also collected over a range of invariant masses and center-of-mass energies. While not directly comparable to the Gaussian width of \pout as $\langle\pt\rangle$ contains both perturbative and nonperturbative contributions, the quantity can still be treated in a TMD framework and may be useful for comparisons to, for example, the RMS of \pout which contains both perturbative and nonperturbative quantities. Figure~\ref{fig:dy_avg_pt} shows a collection of $\langle\pt\rangle$ measurements in DY events over a wide range of $M_{\mu\mu}$. The general characteristics of CSS evolution are met, namely that the $\langle\pt\rangle$ increases with both \sqs and $M_{\mu\mu}$.

While DY is a useful benchmark process to compare to dihadron and direct photon-hadron correlations since they are both processes from hadronic collisions, SIDIS is a useful comparison because there are both initial- and final-state PDFs that are relevant for the scattering process. However, DY and SIDIS should display similar qualitative behavior since factorization is predicted to hold in both processes. Thus, they each provide different but complementary comparisons to dihadron and direct photon-hadron correlations. 

\subsubsection{Semi-Inclusive Deep-Inelastic-Scattering}

Recent data from the COMPASS experiment will likely provide the greatest constraints on the unpolarized TMD functions due to their extremely large data sets; COMPASS has published results from unpolarized SIDIS where the transverse momentum of an outgoing hadron is measured~\cite{Adolph:2013stb,Aghasyan:2017ctw}. In particular the most recent measurement in Ref.~\cite{Aghasyan:2017ctw} is multi-differential in all LO kinematic variables \qsq, $x$, $z$ and \pt. Similarly to the DY and $Z$ boson data, the SIDIS TMD data can be fit to functions in the nonperturbative region and the nonperturbative behavior can be extracted and characterized.

\begin{figure}[tbh]
	\centering
	\includegraphics[width=0.8\textwidth]{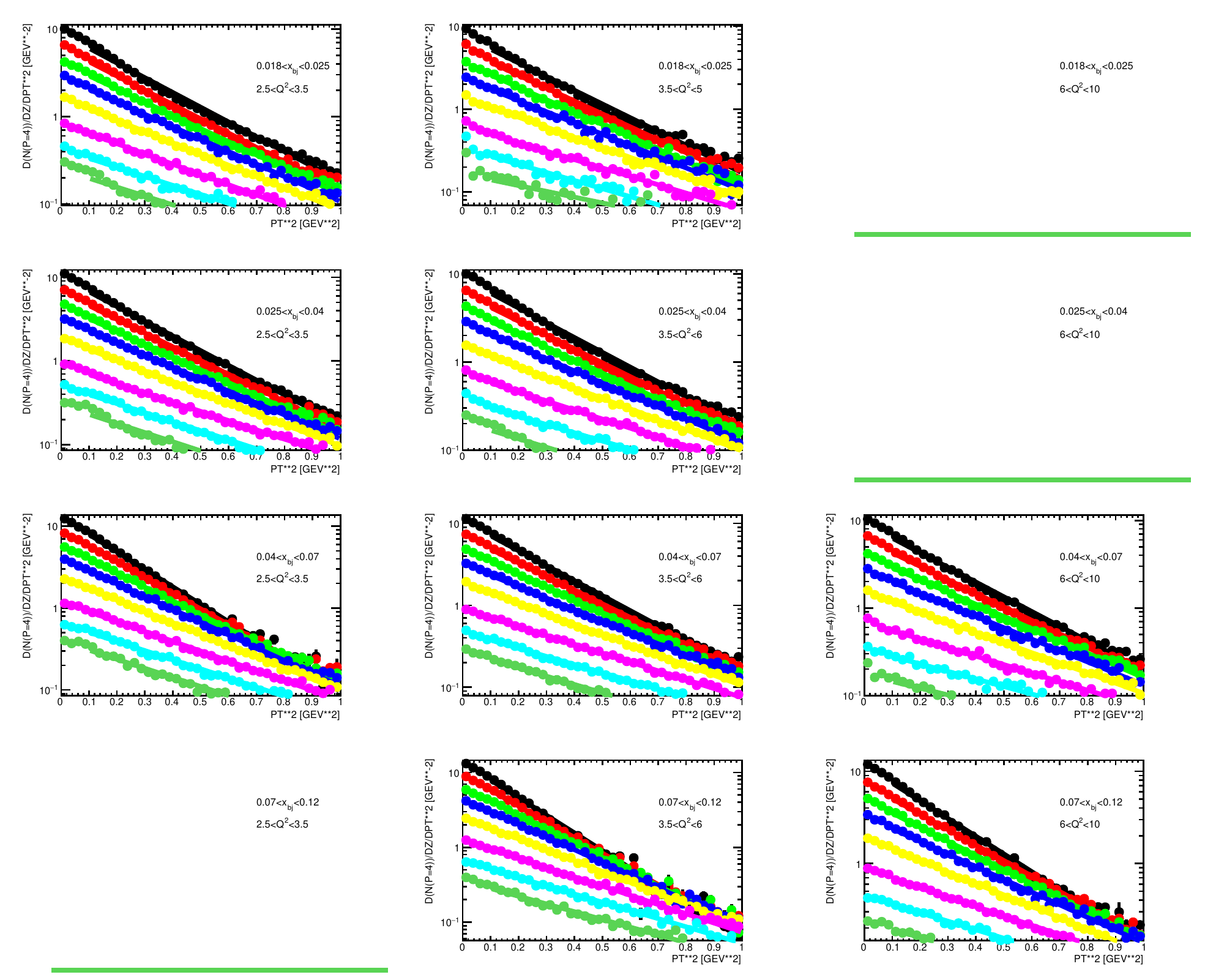}
	\caption{The measured hadron multiplicities from COMPASS are shown as a function of \pt, $z$, $x$, and \qsq. The data is taken from Ref.~\cite{Adolph:2013stb}. Several panels are blank because no data exists in these kinematic bins.}
	\label{fig:compass_multiplicities}
\end{figure}

\begin{figure}[tbh]
	\centering
	\includegraphics[width=0.7\textwidth]{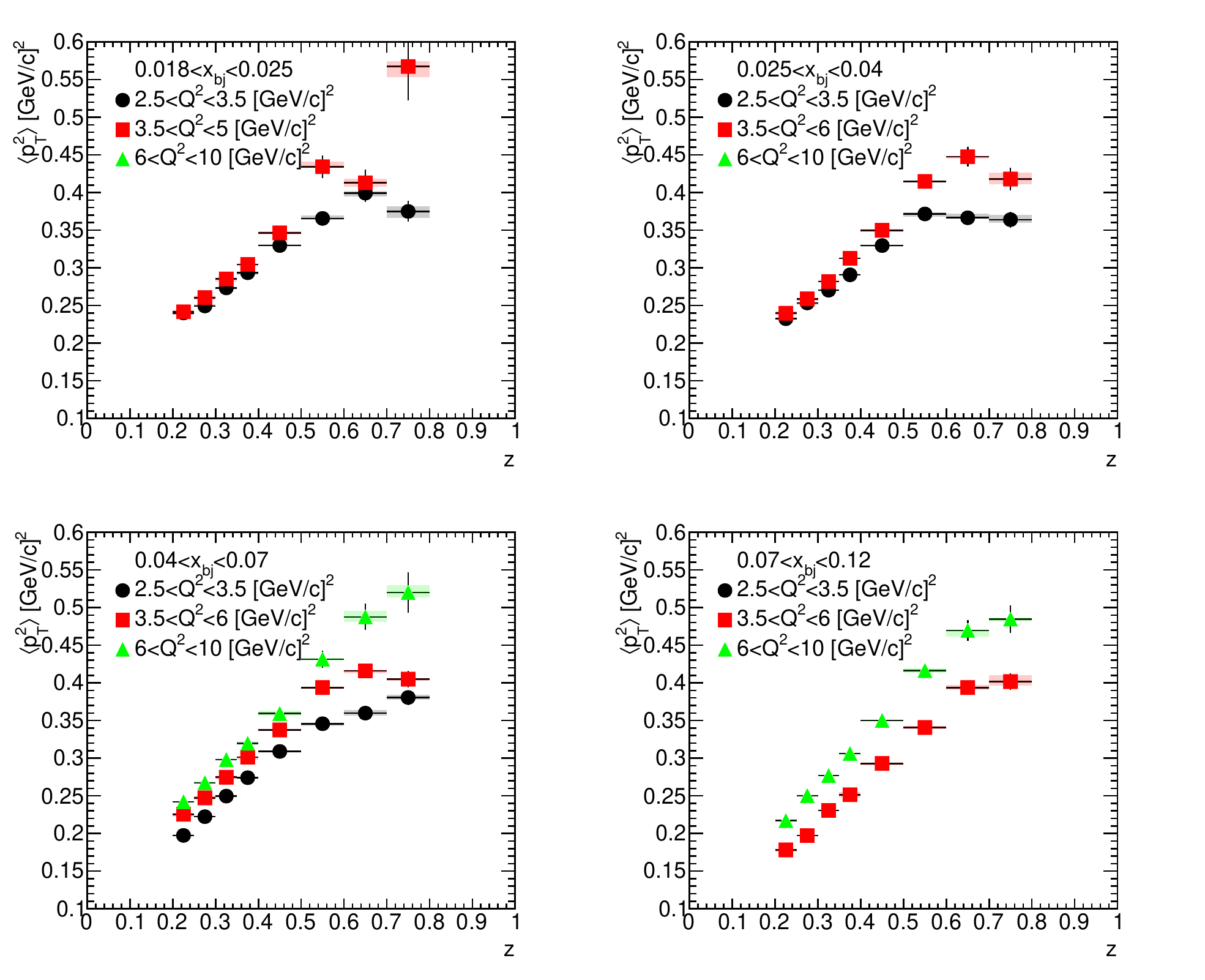}
	\caption{The quantity $\langle p_T^2\rangle$ is shown as a function of $z$ for several bins of \qsq and $x$, as extracted from fits to the data in Fig.~\ref{fig:compass_multiplicities}.}
	\label{fig:compass_widths}
\end{figure}

The analysis performed here uses the data from Ref.~\cite{Adolph:2013stb}; at the time of analysis the newer and more statistically robust data points from Ref.~\cite{Aghasyan:2017ctw} were not yet publicly available from the COMPASS collaboration. The hadron multiplicities are shown as a function of \qsq, $p_T^2$, and $z$ in Fig.~\ref{fig:compass_multiplicities}. The multiplicities are fit to an exponential function in the small $p_T^2$ region and the fits are also drawn in the figure; however, they cannot be seen due to the large number of data points and the scale required in the plot to fit each $z$ bin. Nonetheless, it is clear that the distributions deviate from an exponential function at large $p_T^2$. Several of the panels are blank and have no data; this is simply because no data existed in the publication for that particular $Q^2$ and $x$ kinematic bin. From the exponential fit, the quantity $\langle p_T^2\rangle$ is extracted and plotted as a function of \qsq and $z$ in Fig.~\ref{fig:compass_widths}. Systematic uncertainties on the values are estimated by adjusting the fit range by $\pm$0.05 GeV$^2/c^2$ in $p_T^2$, since again the exact transition from nonperturbative to perturbative behavior is not explicitly defined. The widths show that the qualitative expectation from CSS evolution is met; namely that there is a systematic increase in $\langle p_T^2\rangle$ as a function of \qsq for a given bin of $z$ at fixed $x$.

\begin{figure}[tbh]
	\centering
	\includegraphics[width=0.49\textwidth]{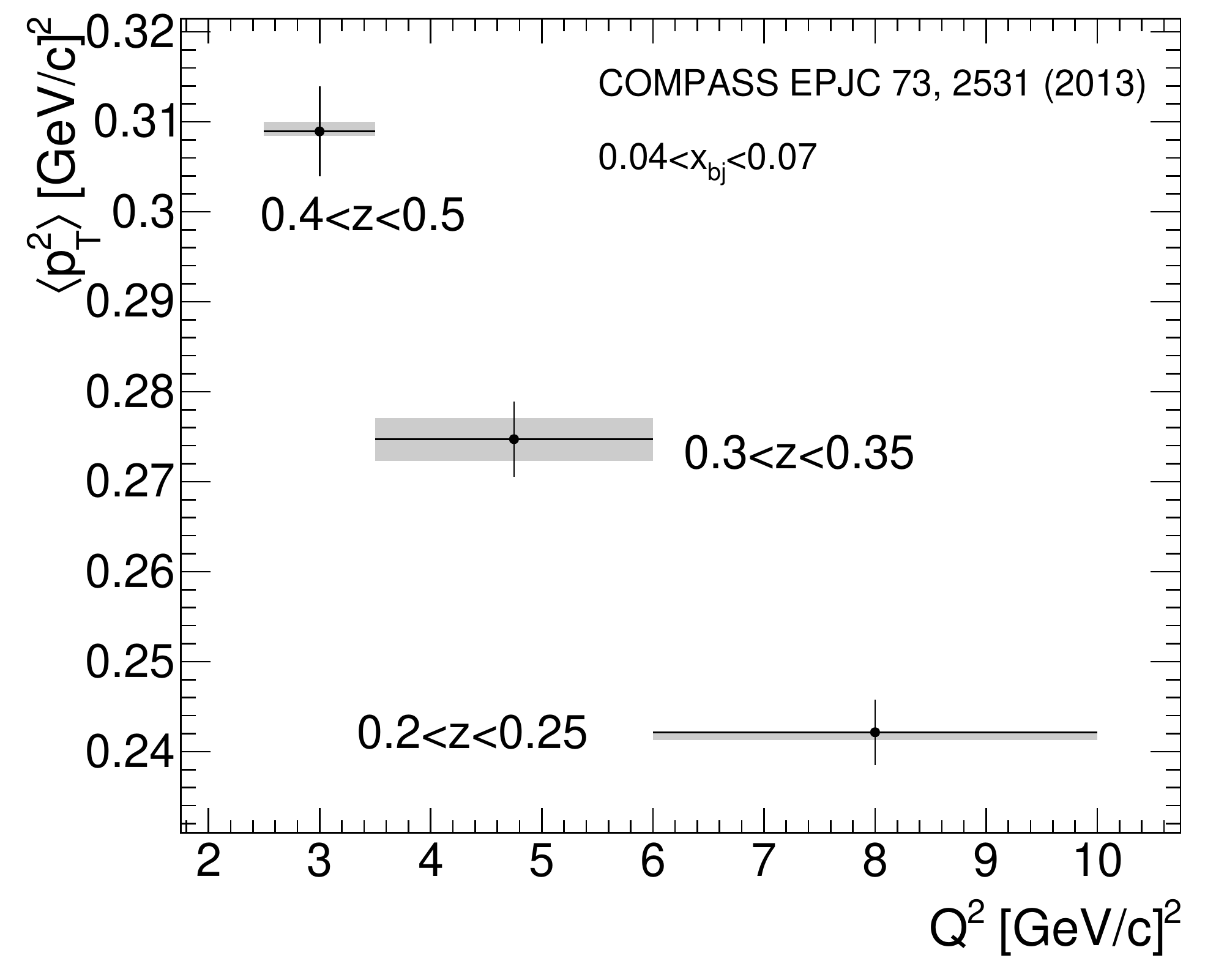}
	\includegraphics[width=0.49\textwidth]{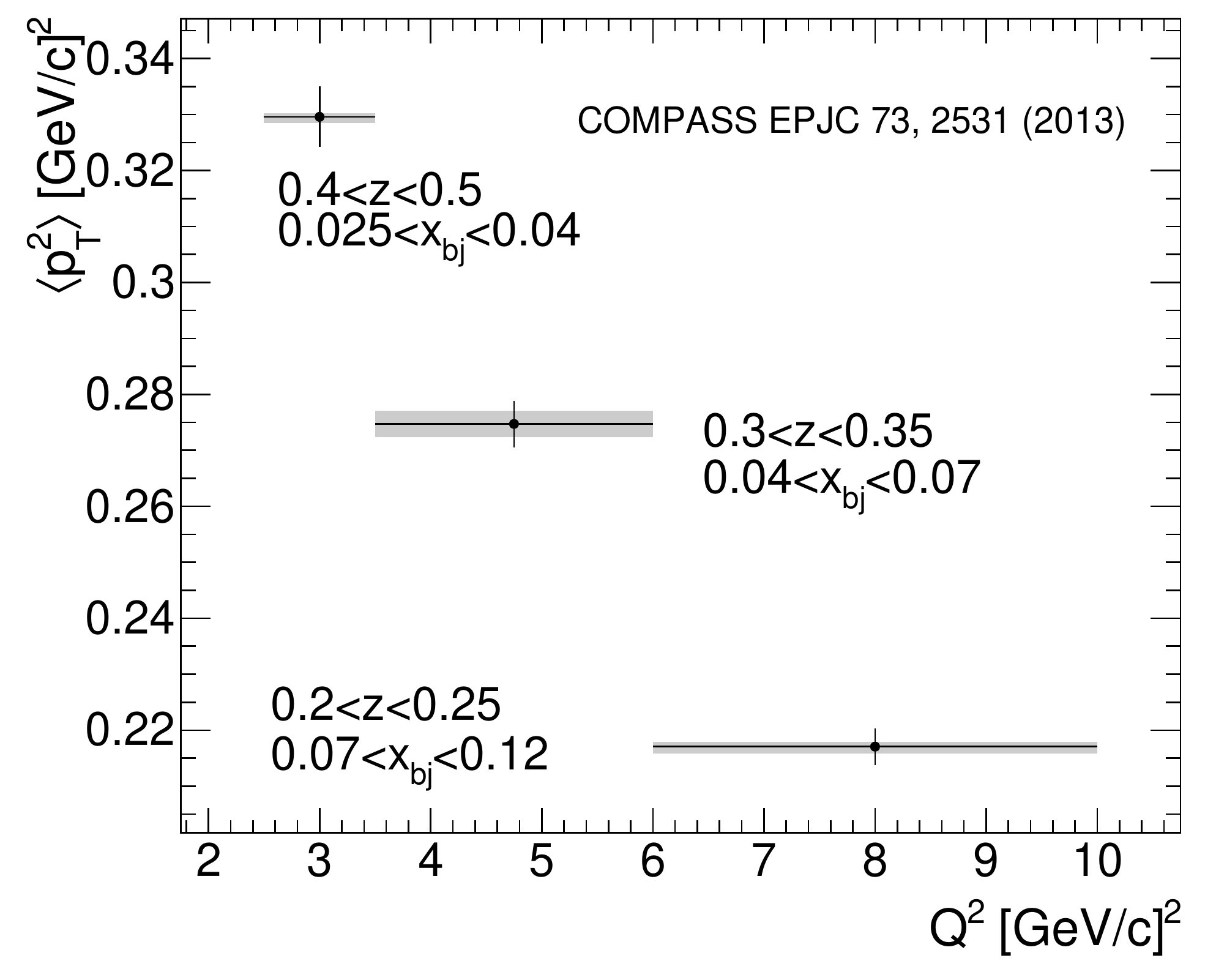}	
	\caption{The $\langle p_T^2\rangle$ is shown as a function of \qsq for varying $z$ bins as indicated in each figure. The left figure shows the quantity for a fixed bin of $x$, while the right figure shows the quantity for a varying bin of $x$ as indicated.}
	\label{fig:compass_widths2}
\end{figure}

If the decreasing widths from the \sqs=~510 GeV analysis seen in \pp$\rightarrow$~dihadrons or photon-hadrons are truly a fragmentation effect due to the average $z$ that is probed, the same behavior should be able to be seen in SIDIS since both initial and final state transverse momentum can be probed. Moreover, the data from COMPASS can be studied differentially in all of the partonic kinematic variables of interest since they can be measured at LO in SIDIS interactions. The $\langle p_T^2\rangle$ is shown as a function of \qsq for sequentially decreasing bins of $z$ in Fig.~\ref{fig:compass_widths2}. The quantity decreases with \qsq, indicating that the result from \pp$\rightarrow$~dihadrons or photon-hadrons occurs because the average $z$ is decreasing as \pttrig is increased. To make the most qualitatively similar comparison to the results in Ref.~\cite{ppg195}, the right panel of Fig.~\ref{fig:compass_widths2} shows the $\langle p_T^2\rangle$ as a function of \qsq for decreasing bins of $z$ and increasing bins of $x$, as would be the case in \pp collisions. Increasing the value of $x$ in conjunction with \qsq does not change the qualitative conclusion that can be drawn from the data.

\subsection{Processes Predicted to Break Factorization}

While the dihadron and direct photon-hadron correlations presented here were the first measurements explicitly searching for effects from factorization breaking, there are many worldwide measurements that may be sensitive to factorization breaking despite this not being the original intent for the measurement. For example, dihadron correlations are a proxy for dijet correlations; dijets satisfy all the necessary conditions for factorization breaking when the jets are nearly back-to-back and a nonperturbative scale can be resolved. Compiling complementary measurements that may exhibit factorization breaking effects may also be useful for comparison to the results presented here. 

\subsubsection{Dijets and Dihadrons}

Dihadron and dijet correlations have been used for decades to probe various physics effects; because there is an underlying correlation between the two jets they are a useful probe for a variety of physics effects. However, the prediction of factorization breaking within a TMD framework in this process only came about approximately 10 years ago; therefore, many of the previous measurements were focused on measuring azimuthal correlations rather than momentum space correlations. The use of \pout as a momentum space correlation which has the resolution to clearly identify a perturbative and nonperturbative region is what sets the results presented here apart from previous measurements; however, the previous measurements can still be used to probe potential factorization breaking effects. 

\begin{figure}[tbh]
	\centering
	\includegraphics[width=0.6\textwidth]{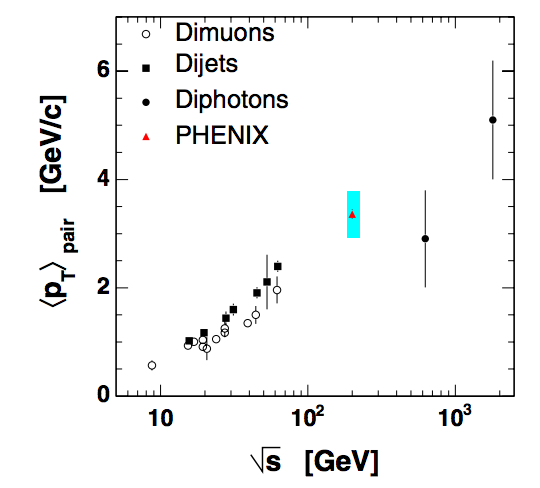}
	\caption{A compilation of measurements of the initial-state partonic $\langle\pt\rangle_{\rm pair}$ in various processes, taken from Ref.~\cite{ppg029}.}
	\label{fig:kt_compilation}
\end{figure}

For example, the evolution of the initial-state \kt in processes predicted to break factorization has been studied in previous PHENIX dihadron and direct photon-hadron analyses. Figure~\ref{fig:kt_compilation} shows a compilation of partonic initial-state \kt measurements from Ref.~\cite{ppg029}, converted from \kt to the total $\langle\pt\rangle_{\rm pair}$ of the partons. In this particular figure, processes predicted to break factorization are compared to those where factorization is predicted to hold as a function of \sqs. In particular the data could be used for comparing to perturbative calculations; for example, the dijet and dihadron measurements appear to have additional \pt when compared to dimuons at the same \sqs.   

\begin{figure}[tbh]
	\centering
	\includegraphics[width=0.6\textwidth]{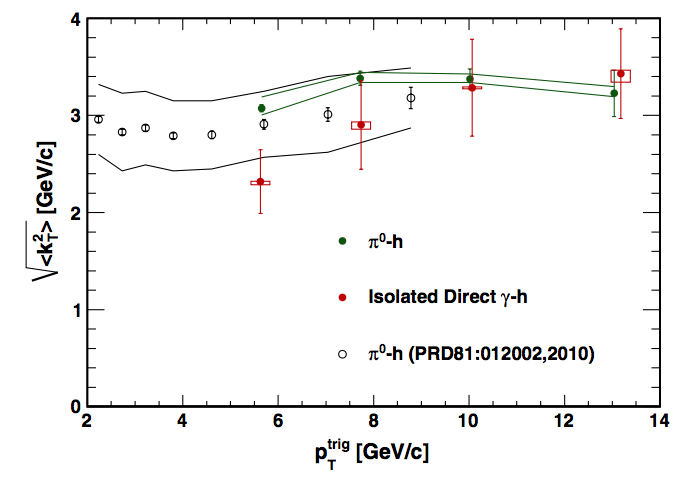}
	\caption{Measurements of \rmskt as a function of \pttrig are shown for dihadron and direct photon-hadron correlations~\cite{ppg095}.}
	\label{fig:kt_ppg095}
\end{figure}

Similarly, the quantity \rmskt was extracted in Refs.~\cite{ppg029,ppg095} and compared between \pion and direct photon triggers. Figure~\ref{fig:kt_ppg095} shows the \rmskt as measured in \pion-hadron and direct photon-hadron correlations over a wide range of \pttrig accessible by the PHENIX experiment. These measurements show that there is little dependence on \rmskt as a function of \pttrig, which is very different from the behavior seen in, for example, DY $\langle\pt\rangle$ measurements as a function of $M_{\mu\mu}$. There is some indication that the \rmskt rises in direct photon-hadron correlations, however the uncertainties are still quite large. This may indicate that full jet measurements are necessary in order to completely reconstruct the hard scale, or the invariant mass, of the dijet or direct photon-jet pair. 

Most dijet measurements have focused on the decorrelation of the associated jet with respect to the trigger jet at large angles of \dphi. For example, Refs.~\cite{Abazov:2004hm,Khachatryan:2011zj,daCosta:2011ni} study the azimuthal distribution of correlated dijet pairs at both Tevatron and LHC center-of-mass energies. In these references, the focus has been on understanding the NLO and NNLO behaviors at very large angles of $\dphi\sim\pi/2$. Additionally, dijet measurements at large \pt and \sqs usually suffer from poor resolution at $\dphi\sim\pi$, hence the nonperturbative behavior in \dphi cannot be studied clearly. NLO calculations diverge at this region of phase space as well, which is shown in Fig. 3 of Ref.~\cite{Khachatryan:2011zj} where NLO calculations are not shown at $\dphi\sim\pi$. 

\begin{figure}[tbh]
	\centering
	\includegraphics[width=0.8\textwidth]{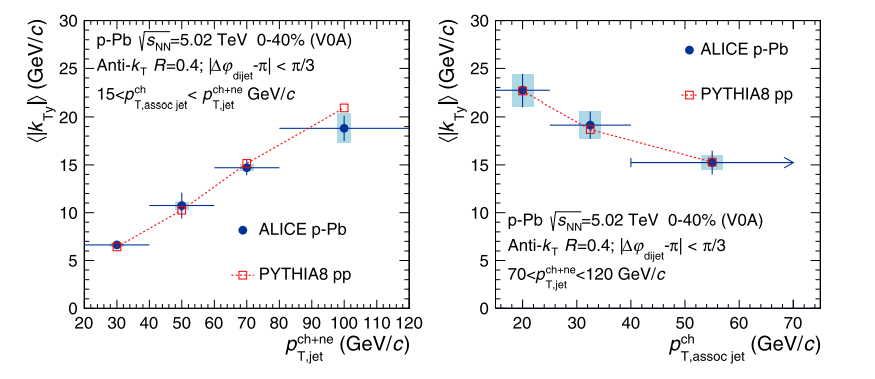}
	\caption{Mean of the $|k_{T_y}|$ distributions as a function of the leading jet transverse momentum (left) and the associated jet transverse momentum (right) in $p$+Pb collisions~\cite{Adam:2015xea}. The $k_T$ values are compared to \pythia \pp simulations.}
	\label{fig:kt_ppb}
\end{figure}

A recent measurement of the dijet channel in $p$+Pb collisions has measured the acoplanarity in momentum space of the dijet pair at LHC energies~\cite{Adam:2015xea}. Figure~\ref{fig:kt_ppb} shows the average $k_{T_y}$ of the dijet pair as a function of the leading jet transverse momentum (left panel) and the associated jet transverse momentum (right panel). This measurement was motivated by studying potential energy loss mechanisms in cold nuclear matter and thus providing a benchmark for energy loss in the QGP. However, the fact that a momentum space observable was used shows that new observables are being tested to try and better understand various QCD mechanisms. In particular, this measurement could be used to compare to phenomenological calculations given that the dijet pair is required to have $|\dphi-\pi|<\pi/3$ and the jet transverse momenta are not hundreds of \gevc, thus the measurements may have sensitivity to the underlying nonperturbative dynamics.

\subsubsection{\jpsi and $\Upsilon$ Mesons}

\begin{figure}[tbh]
	\centering
	\includegraphics[width=0.49\textwidth]{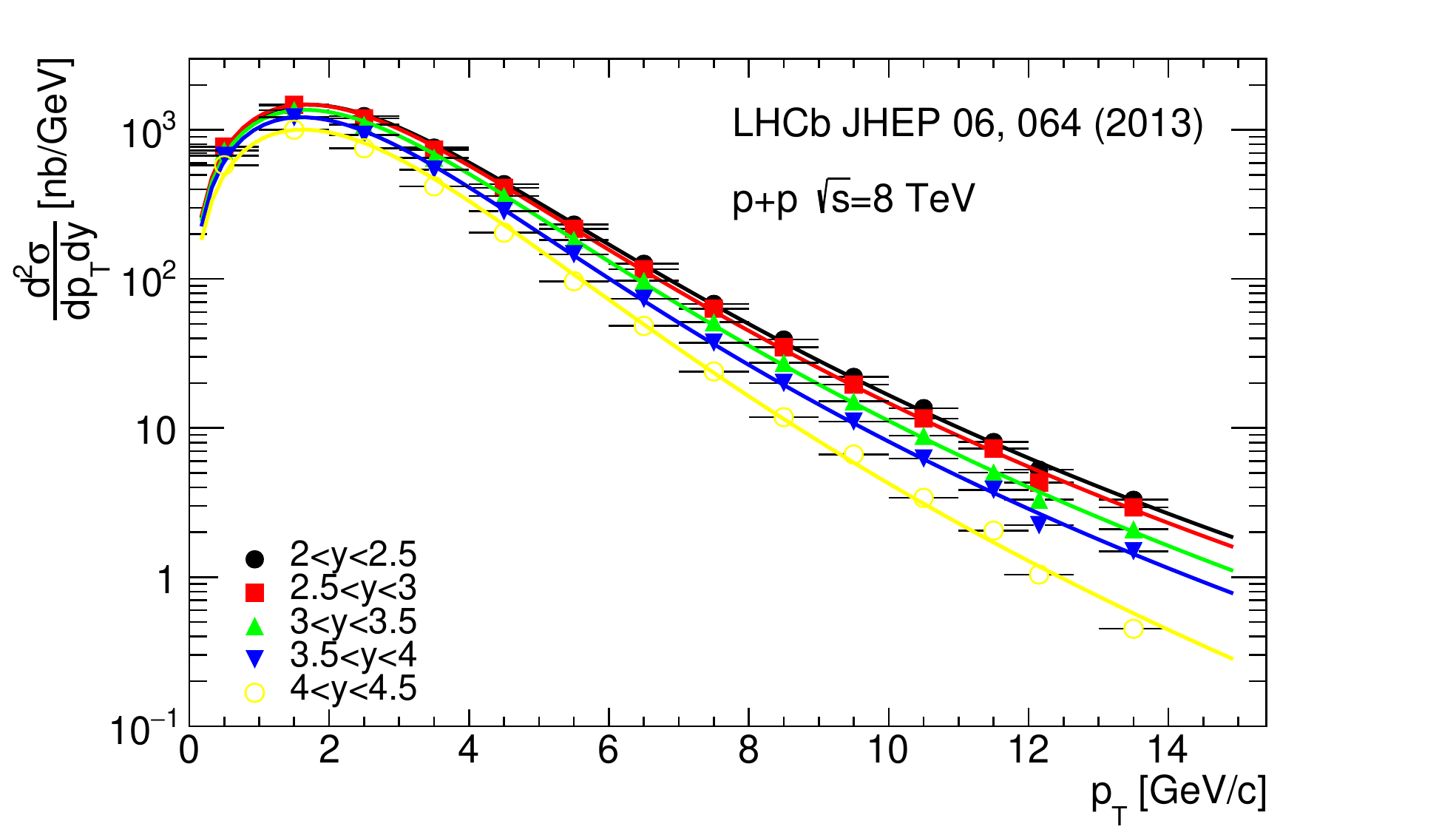}
	\includegraphics[width=0.49\textwidth]{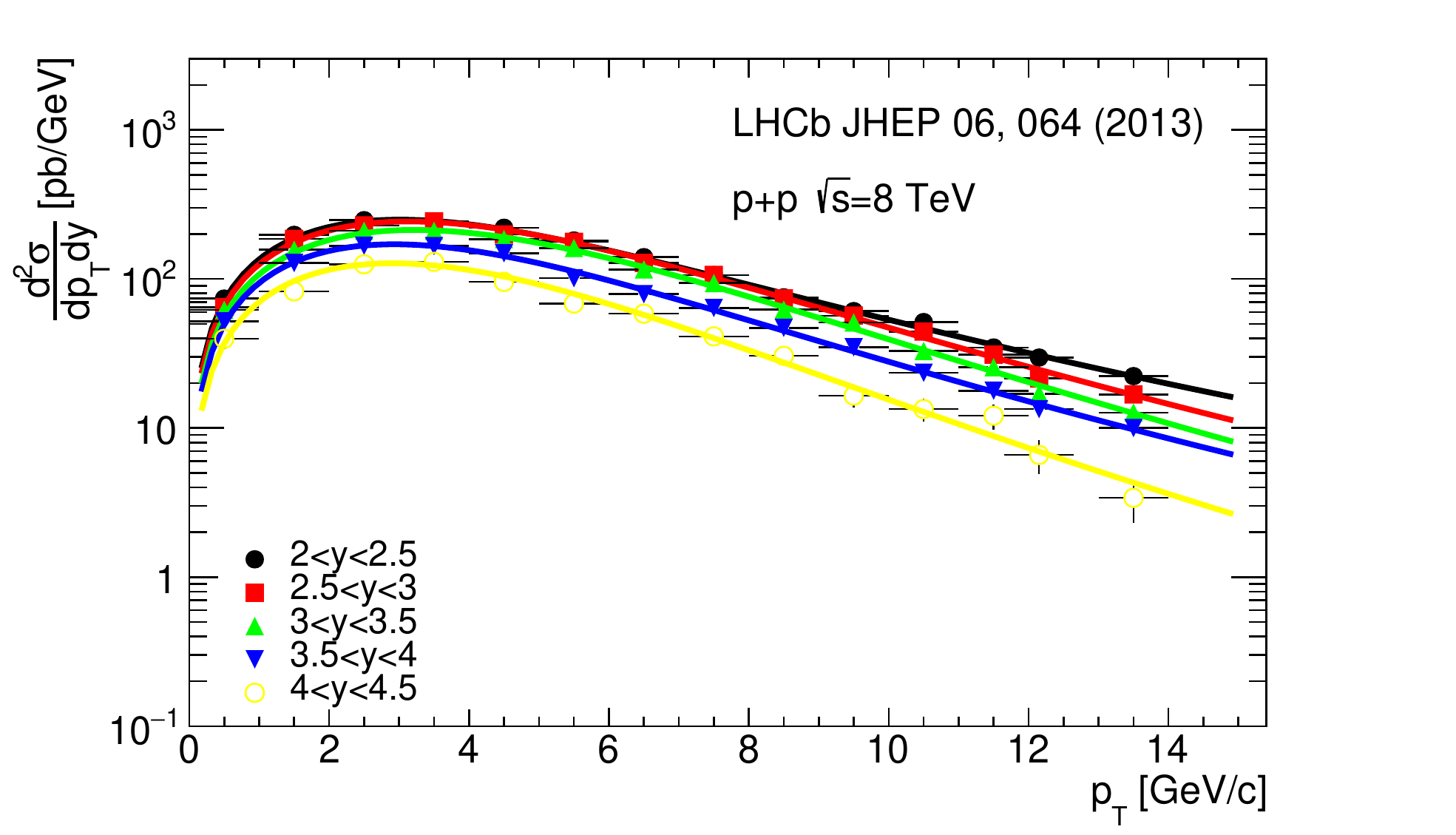}
	\caption{Measured prompt \jpsi (left) and $\Upsilon(1S)$ meson cross sections from Ref.~\cite{Aaij:2013yaa} are shown with Kaplan fits.}
	\label{fig:jpsixsecs}

\end{figure}

Since \jpsi and \ups mesons have colored partons in the final-states, they should in principle be sensitive to factorization breaking effects when a TMD framework is applicable. This may also depend on whether or not they are produced in a color singlet or color octet state~\cite{Dunnen:2014eta}; nonetheless there is significant data from the LHC now which may allow for the cross sections to be studied as a function of the invariant mass of the quarkonium pair. Similarly to the idea for direct photon-hadron and dihadron correlations, differences from expectations of CSS evolution could be observed in the evolution of these widths when compared to processes where factorization is predicted to hold. While the ATLAS and CMS experiments do not have the \pt resolution required to measure \jpsi or \ups cross sections at small \pt where a TMD framework is applicable, the LHCb experiment has already published these cross sections with reasonably small \pt resolution at forward rapidities~\cite{Aaij:2011jh,Aaij:2013yaa,Aaij:2015awa}. Several of these cross sections are shown in Fig.~\ref{fig:jpsixsecs}.

\begin{figure}[tbh]

	\centering
	\includegraphics[width=0.7\textwidth]{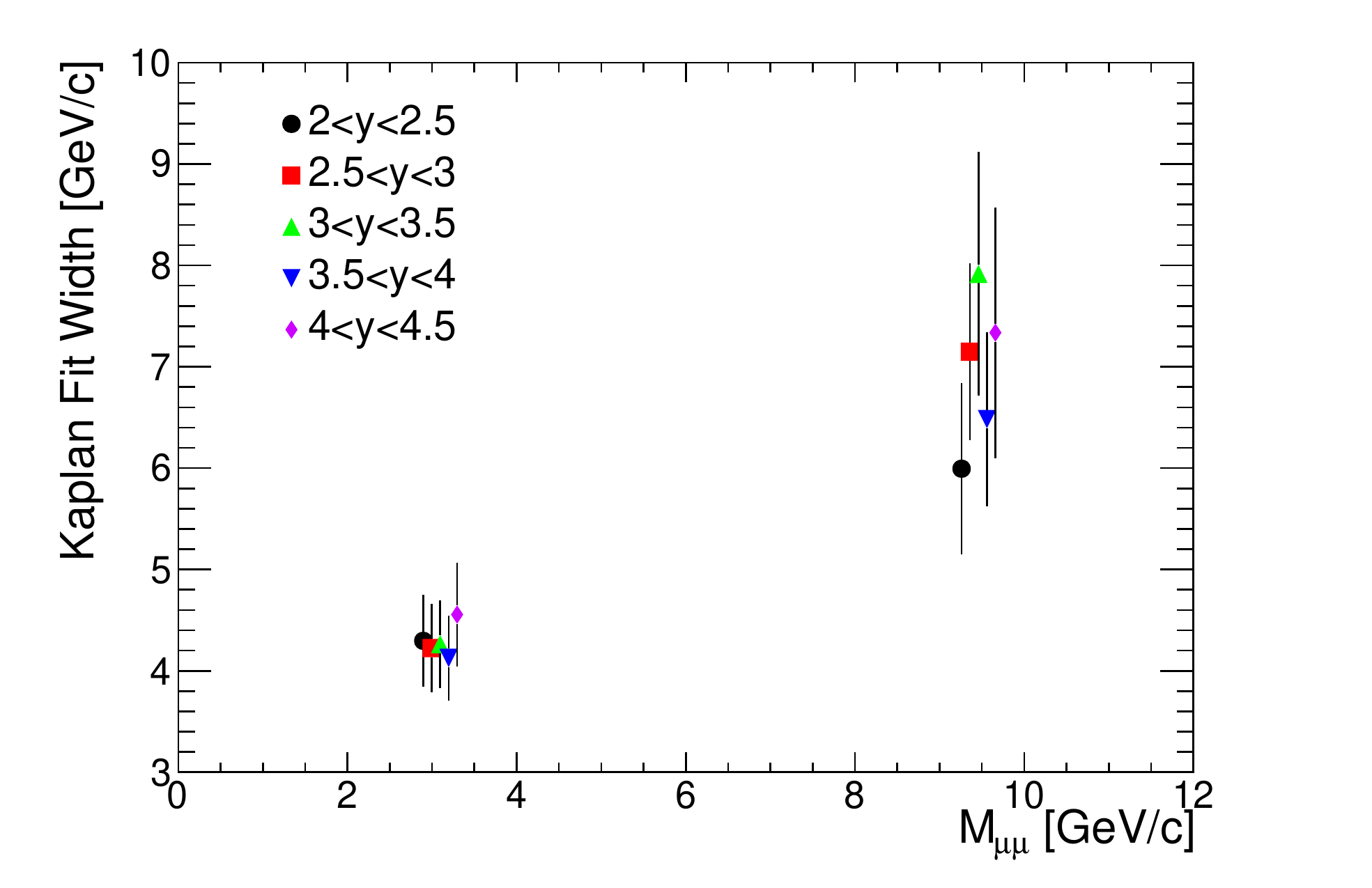}
	\caption{The extracted width parameter from the Kaplan fits shown in Fig.~\ref{fig:jpsixsecs} is shown as a function of the invariant mass of the \jpsi and $\Upsilon$. The points are offset in $M_{\mu\mu}$ for visibility.}
	\label{fig:jpsi_widths}
\end{figure}

The cross sections were taken from Ref.~\cite{Aaij:2013yaa} and fit to Kaplan functions to accurately describe both the small and large \pt nonperturbative and perturbative behavior, respectively. These fits are shown in Fig.~\ref{fig:jpsixsecs} for several different rapidity regions that the LHCb experiment published. Comparing the figures, similar behavior to the \pout correlations as a function of \xe can be seen in that the perturbative region of the $\Upsilon(1S)$ cross section has a noticeably harder spectrum than that of the \jpsi. The widths are extracted from the fits and are shown in Fig.~\ref{fig:jpsi_widths}; the points are offset in $M_{\mu\mu}$ for visibility. Each rapidity region shows that the measured widths increase in \jpsi and $\Upsilon$ production, albeit within large uncertainties. There also appears to be no discernible rapidity dependence. These measurements provide an alternative way to study potential factorization breaking effects in a TMD framework where the final-state is a QCD bound state and thus is composed of colored partons. Depending on the production mechanism, the presence or lack thereof of factorization breaking effects might also provide information about the largely unknown \jpsi and $\Upsilon$ production mechanisms; if they are produced in such a state where they are not able to exchange soft gluons with other colored remnants they may not exhibit effects from factorization breaking.

\chapter{Future Measurements}
\label{chap:future}

The work presented here comprises the first ever measurements with the intent to search for effects from processes which are predicted to explicitly break factorization in a TMD framework. There are significantly more measurements that can be made which may be able to shed light on color entanglement effects; many of these measurements are now feasible due to the extraordinarily high luminosity of both RHIC and the LHC in addition to the excellent detector facilities at these accelerator complexes. It should also be noted that it is imperative to make measurements sensitive to color entanglement in the near future at both RHIC and the LHC, as the Electron Ion Collider will not be able to probe these effects directly. Entanglement effects are only present in hadronic collisions where a hadron is present in the final-state, so it is necessary to make measurements now so that observables from the EIC can be interpreted within the context of data from hadronic collisions. This will be especially important for interpreting data from the EIC with regards to color flow in hadronic interactions including SIDIS, which is currently one of the most important questions in proton structure physics~\cite{LRP_2015}. \par

In particular, using jets as a probe for entanglement effects provides several benefits over hadrons. Jets can be used to probe effects that are sensitive to only initial-state parton $k_T$ within the colliding hadrons. This eliminates any fragmentation dependence on the observables and would allow for a cleaner comparison to Drell-Yan momentum widths which are also only sensitive to initial-state $k_T$. Secondly, jets can be used to reconstruct, at leading order, the longitudinal momentum fractions of the colliding partons. This will allow for any possible correlation between $x$ and $k_T$ to be studied at leading order. In addition, the partonic center-of-mass energy can be reconstructed at leading order between dijet and direct photon-jet correlations, thus a full kinematic mapping of the $x$, $k_T$, and $Q^2$ dependence could be measured without any fragmentation contributions. This would also provide cleaner and more direct comparisons to the Drell-Yan momentum widths, where the $Q^2$ dependence can be measured directly from the dilepton pair. While jet measurements have these advantages, hadrons also have the advantages of better angular resolution; this motivates making measurements of both processes for comparisons between both and to take advantage of each of their particular strengths.  

The most ideal channels for probing color entanglement effects are either the $Z^0$-jet or direct photon-jet channel. This channel has already been studied theoretically and it has been shown that certain spin asymmetries due to color entanglement effects may arise in both \pp and \pa collisions~\cite{Rogers:2013zha,Schafer:2014xpa}. While dijet correlations can be used, and are in some ways beneficial since the cross section is significantly larger, \gammajet events are the most ideal due to the number of colored objects present. In dijet events, color can be exchanged between four hard scattered partons, while in \gammajet events the photon is unable to exchange color with any remnants. This allows the color flow to be limited between objects in the final-state, and in some sense is the simplest case for entanglement since at least three colored objects are necessary for factorization to break down. The \gammajet channel is also ideal due to the fact that at leading order, assuming no nonperturbative $k_T$, the photon and jet emerge exactly back-to-back with equal and opposite \pt. This gives a benchmark to compare the photon and jet with each other, and the possibility to understand the role, if any, of fragmentation by comparing \gammajet and $\gamma$-hadron correlations. An additional benefit is that the spatial resolution of photons is essentially only limited by the segmentation of the electromagnetic calorimeter, which gives better resolution on \dphi and \pout so that the \gammajet pair can be treated in a TMD framework. On the other hand jets have significantly worse spatial resolution due to their composite nature, and also come with additional systematic uncertainties that arise from jet reconstruction. For these reasons the \gammajet channel is the most ideal for measuring effects sensitive to factorization breaking.

Performing dijet and \gammajet correlation measurements at both the LHC and RHIC will be crucial in the search for color entanglement effects. In fact, \gammajet and $Z^0$-jet results at midrapidity from the ATLAS and CMS experiments respectively already exist in the literature~\cite{Aaboud:2016sdm,Sirunyan:2017jic}; however, an important point should be noted about results from the LHC. In, for example, Ref.~\cite{Aaboud:2016sdm}, the transverse energy of the photon is required to be at least 100 GeV while the away-side jet is required to have \pt$>$100 \gev at midrapidity. These requirements are largely made to reduce backgrounds at \sqs=~8 TeV; it is additionally motivated by the fact that the LHC detectors were designed to measure extremely high \pt processes. To measure factorization breaking effects, it is necessary that the correlation measurement has sensitivity to the nonperturbative dynamics of the interaction; this requires excellent resolution on the observable \dphi$\sim\pi$ or \pout$\sim$~0. Suppose one measures a 100 GeV photon and a \pt=~80 \gev jet at \dphi=~3.1 radians; this corresponds to $\pout=80\sin(3.1)\approx3.3$ \gev. At \dphi=~3 radians \pout becomes roughly 11 \gev, which is no longer a nonperturbative scale! Since the jet \pt is very large it is necessary to have very fine resolution in \dphi to have sensitivity to the nonperturbative dynamics; unfortunately, acquiring this precision in jet characteristics is nearly impossible with current jet finding algorithms. To ameliorate this problem, one needs to identify other observables that are sensitive to color, such as the jet pull vector, or measure jet correlations at smaller \pt where this extreme resolution in \dphi is not necessary; at the LHC systematic uncertainties associated with jet production at 20-50 \gev become very large. This motivates making jet correlation measurements at RHIC, where the detectors are designed to measure jets of order \pt~$\approx$~20-50 \gev; therefore, sensitivity to the nonperturbative dynamics will be possible while also suppressing systematic uncertainties in jet measurements.

\begin{figure}[tbh]
	\centering
	\includegraphics[width=0.85\textwidth]{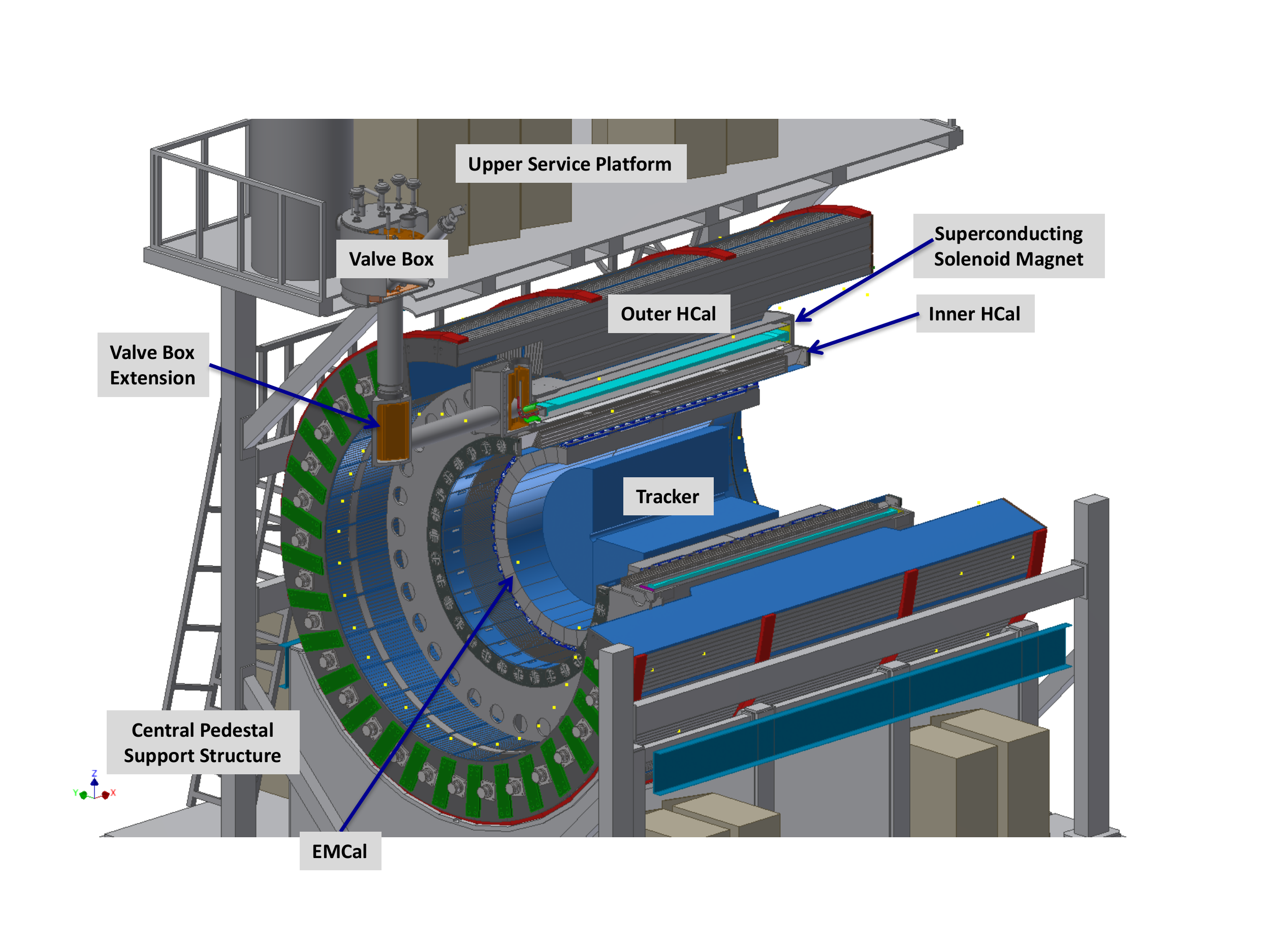}
	\caption{An engineering drawing of the proposed sPHENIX barrel detector shows the various subdetectors. The electromagnetic and hadronic calorimeters are the most important subsystems for high \pt jet measurements.}
	\label{fig:sphenix_detector}
\end{figure}

\section{sPHENIX}

The sPHENIX experiment is a new proposed experiment which will be housed in the current experimental hall of the PHENIX experiment, which completed data taking in 2016, at RHIC~\cite{Adare:2015kwa}. sPHENIX is designed to measure jets, jet correlations, and bottomonium states in \pp, \pau, and Au+Au collisions at RHIC. In particular, the experiment is principally motivated by probing the QGP at different temperature and length scales than experiments at the LHC to fully characterize the nature of the strongly coupled plasma. While the flagship sPHENIX measurements including characterizing the nature of the QGP with dijet and photon-jet measurements in heavy ion collisions, the detector must also be able to make the same measurements in \pp collisions to provide baseline measurements without nuclear modification for the heavy ion collisions. For these reasons sPHENIX will be an ideal facility to probe color entanglement effects in both \pp and $p$+Au collisions with dijet and direct photon-jet correlations.  \par

\subsection{sPHENIX Detector}

The sPHENIX detector will be the first dedicated high-rate jet detector at the RHIC accelerator complex. The proposed physics measurements have driven the requirements for the detector components of sPHENIX; the experiment must be able to precisely measure jets with energies of 10-70 GeV for a range of jet cone sizes in both \pp and Au+Au collisions. This requires that the detector have a large acceptance and excellent triggering capabilities so that each produced high \pt jet process is recorded; additionally, the detector components must have excellent resolution to be able to accurately measure the jet properties. To make $\gamma$-jet measurements the detector must have an electromagnetic calorimeter that has high rejection triggering capabilities and a modest energy resolution, since these are rare high \pt processes. To satisfy these requirements, the sPHENIX detector has been proposed, as shown in Fig.~\ref{fig:sphenix_detector}, to have an electromagnetic calorimeter and two hadronic calorimeters, where the inner and outer hadronic calorimeters surround the central solenoidal magnet. sPHENIX also has a rich bottom quark physics program, for which a time projection chamber and various silicon vertex trackers are installed immediately surrounding the beam pipe. The tracking detectors will also be essential for fragmentation studies, in particular for understanding the role of fragmentation, if any, in factorization breaking effects.  \par

\begin{figure}[tbh]
	\centering
	\includegraphics[width=0.7\textwidth]{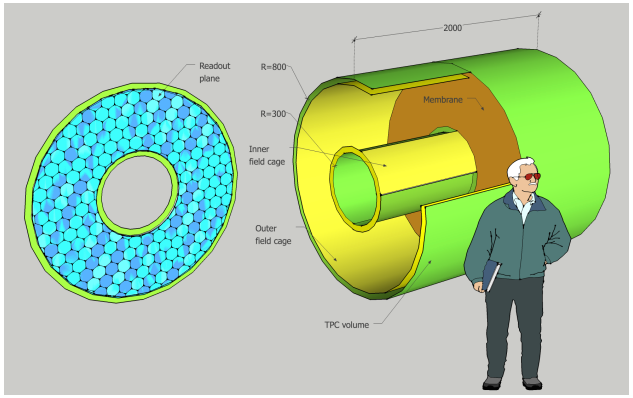}
	\caption{A schematic diagram shows the main elements of the proposed sPHENIX time projection chamber.}
	\label{fig:sphenix_tpc_layout}
\end{figure}

\subsubsection{Tracking}

The sPHENIX tracking performance is primarily driven by the goals to measure the three upsilon states, $\Upsilon$(1S), $\Upsilon$(2S), and $\Upsilon$(3S), and fragmentation function measurements in Au+Au collisions. Given the difficulty of tracking in high multiplicity heavy nucleus collisions, the proposed tracking performance will necessarily be excellent in lower multiplicity \pp collisions. To measure the $\Upsilon(nS)$ states in Au+Au collisions, the tracker requires a momentum resolution of roughly 1.2\% in the range of 4-8 \gev. For fragmentation studies at low and high $z$, this requirement can be extended to require $\Delta p/p\simeq$~0.2\%$p~(\gevc)$ . For low $z$ tracks it is imperative that the apparatus be able to measure the jet and the track, which might be  highly displaced from the actual jet axis since it has a low momentum fraction of the total jet. This (among other measurements) reinforces the requirement that sPHENIX must have a large and uniform acceptance.

\begin{figure}[tbh]
	\centering
	\includegraphics[width=0.7\textwidth]{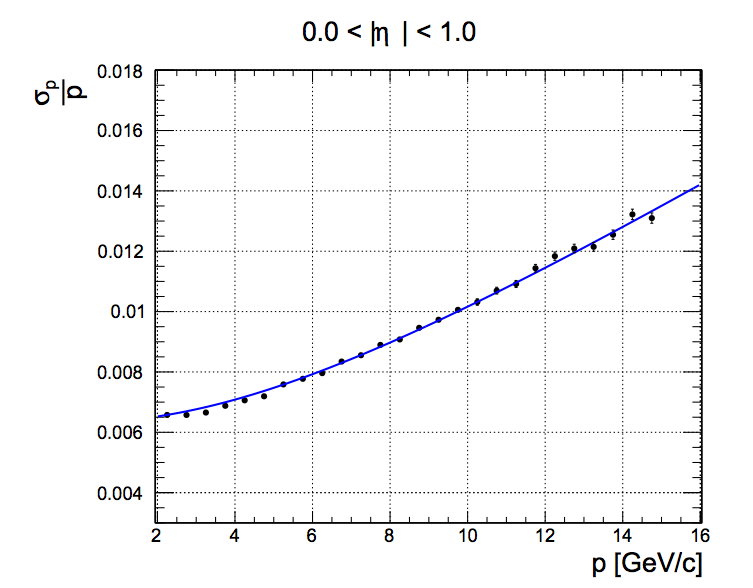}
	\caption{A GEANT4~\cite{Agostinelli:2002hh} simulation of the sPHENIX time projection chamber shows that the momentum resolution requirements for measurements of fragmentation functions will be met.}
	\label{fig:sphenix_tpc_sim}
\end{figure}

To meet these requirements, the sPHENIX detector will utilize a cylindrical double-sided time projection chamber (TPC) which has been used in several other experiments around the world. The TPC is similar to the ALICE TPC which has been successfully implemented at the LHC~\cite{Alme:2010ke}. Figure~\ref{fig:sphenix_tpc_layout} shows the generic layout of the proposed TPC. The TPC will have a central membrane electrode which divides it into two mirror-symmetric volumes surrounding the interaction point. In each volume the readout plane will be located on the endcap inner surface; primary ionization will drift towards the endcaps by setting the readout planes to ground potential while the central membrane will be set to a high bias voltage setting. Simulations of a realistic detector modeled in GEANT4~\cite{Agostinelli:2002hh} show that the momentum resolution of tracks, shown in Fig.~\ref{fig:sphenix_tpc_sim}, will meet the requirements for upsilon and fragmentation function measurements in both \pp and Au+Au collisions.

\subsubsection{Electromagnetic Calorimeter}

While the tracking system is important for fragmentation studies, the most important subsystems for jet measurements are the electromagnetic and hadronic calorimeters. Directly surrounding the tracking system is the electromagnetic calorimeter (EMCal), which is essential for both the calorimetric jet reconstruction as well as the identification of direct photons. The physics requirements for the EMCal are quite modest as its primary purpose is for the measurement of electrons from $\Upsilon$(nS) decays and hard probes from, for example, direct photon scatterings. Since both of these measurements will be at high \pt, one requirement is that the EMCal have a resolution of approximately $\Delta E/E\leq$~15\%/$\sqrt{E}$ (\gevc). The EMCal must have a large acceptance so that a large percentage of rare probes can be triggered on and measured. Additionally it must be carefully designed as the calorimeter must fit inside the solenoid magnet as well as allow enough space for the tracking system to sit inside of it. This means that the EMCal must occupy minimal radial space and that the readout work within a magnetic field. \par

The technology chosen for the EMCal uses an absorber consisting of a matrix of tungsten powder and epoxy with embedded scintillating fibers. This so called SPACAL design has been implemented in several other experiments, for example in Ref.~\cite{Leverington:2008zz}. The readout utilizes silicon photomultipliers, which provide high gain and require minimal space in addition to functioning within a magnetic field. The towers are the first SPACAL towers which, at high rapidity, will be projective in both azimuth and pseudorapidity. At small rapidities the towers are 1D projective in only azimuth. The choice of 2D projectivity, in both azimuth and rapidity, at large rapidities is largely because electrons from $\Upsilon$ decays enter the individual towers at larger angles at high rapidity, thus for towers that are only projective in azimuth the resulting electromagnetic shower develops over several towers. This increases the chances of underlying event activity being included in the cluster, and thus increases the background for the $\Upsilon$ measurements.  

\begin{figure}[thb]
	\centering
	\includegraphics[width=0.7\textwidth]{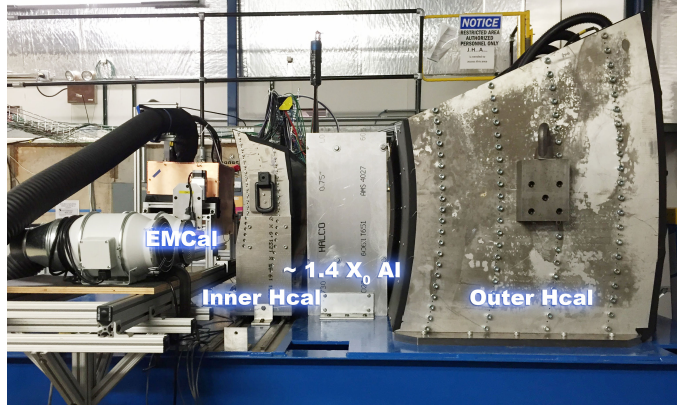}
	\caption{The calorimeters in the 2016 Fermilab T-1044 test beam experiment are shown. The beam enters perpendicularly from the left side of the page.}
	\label{fig:sphenix_2016_testbeam}

\end{figure}

\begin{figure}[thb]
	\centering
	\includegraphics[width=0.85\textwidth]{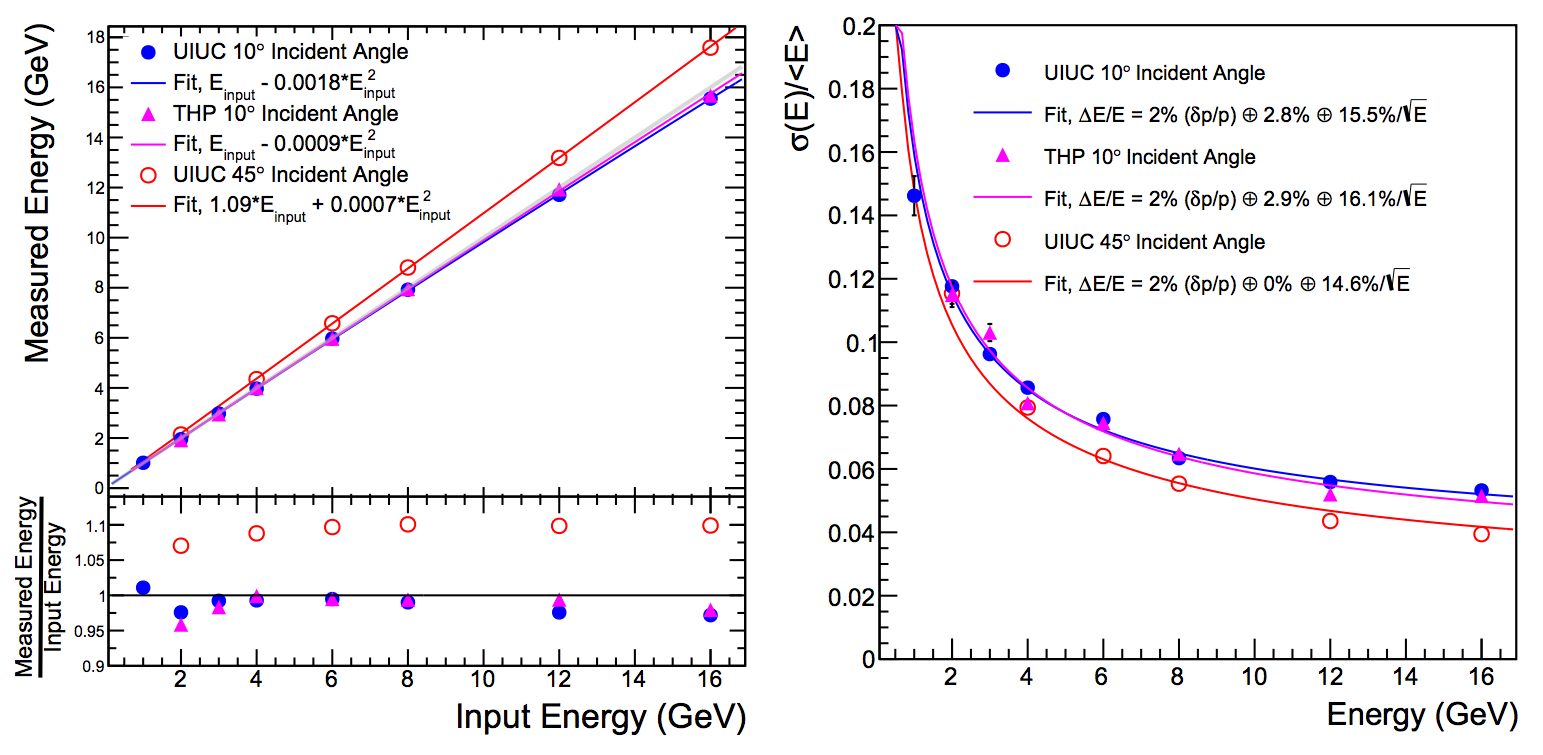}
	\caption{The linearity (left) and resolution (right) of the 1D projective SPACAL towers are shown with corresponding simulation curves. The measured resolution meets the sPHENIX physics requirements.}
	\label{fig:sphenix_testbeam_res_2016}
\end{figure}

Both simulations and prototype beam tests have been essential for characterizing and understanding the proposed detector. In particular the verification that GEANT4 simulations of prototype modules have matched actual test beam analysis is crucial for understanding how the detector will respond in heavy nucleus collision environments since these cannot be explicitly tested in test beam scenarios. The first T-1044 test beam experiment with a full sPHENIX calorimetry prototype was performed in 2016 at the Fermilab Test Beam Facility and was intended to emulate the central rapidity sPHENIX calorimetry. Figure~\ref{fig:sphenix_2016_testbeam} shows a picture of the actual test beam setup, where the beam enters perpendicularly from the left side of the page. An aluminum cryostat is inserted between the inner and outer hadronic calorimeters to simulate the solenoidal magnet. Hodoscopes are placed upstream from the EMCal so that the beam location can be precisely identified. The energy resolution of the 1D projective SPACAL towers as measured in the 2016 test beam is shown in Fig.~\ref{fig:sphenix_testbeam_res_2016}~\cite{Aidala:2017rvg}, and meets the sPHENIX physics requirement of a small constant term with a less than 15\% stochastic term. GEANT4 simulations match the measured resolution quite well, indicating that the detector description within GEANT will be suitable for understanding backgrounds that cannot be measured in test beam situations.

\begin{figure}[thb]
	\centering
	\includegraphics[width=0.6\textwidth]{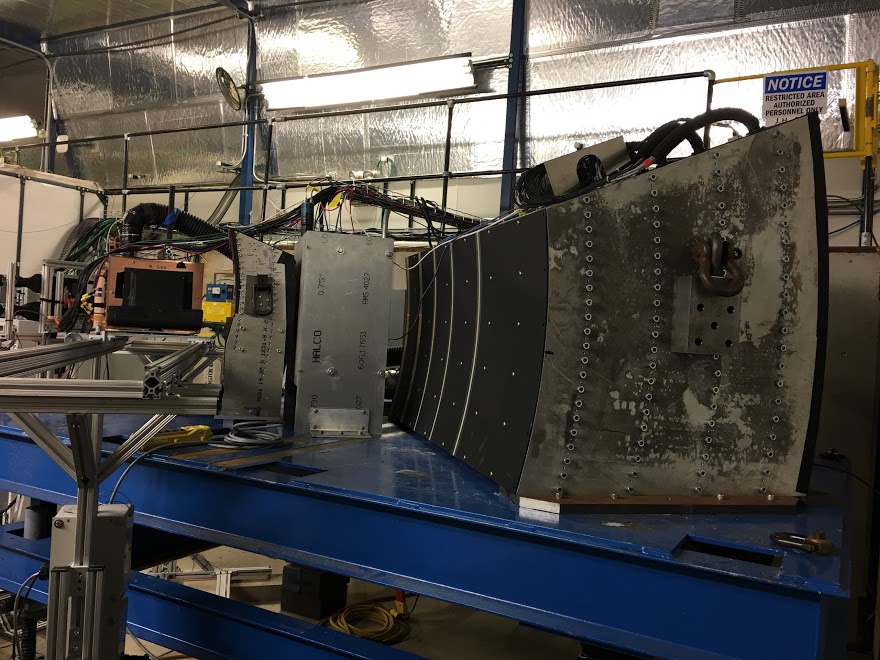}
	\caption{The calorimeters in the 2017 Fermilab T-1044 test beam experiment are shown. The beam comes out of the page; the picture is taken from behind the outer hadronic calorimeter.}
	\label{fig:sphenix_2017_testbeam}
\end{figure}

\begin{figure}[thb]
	\centering
	\includegraphics[width=0.49\textwidth]{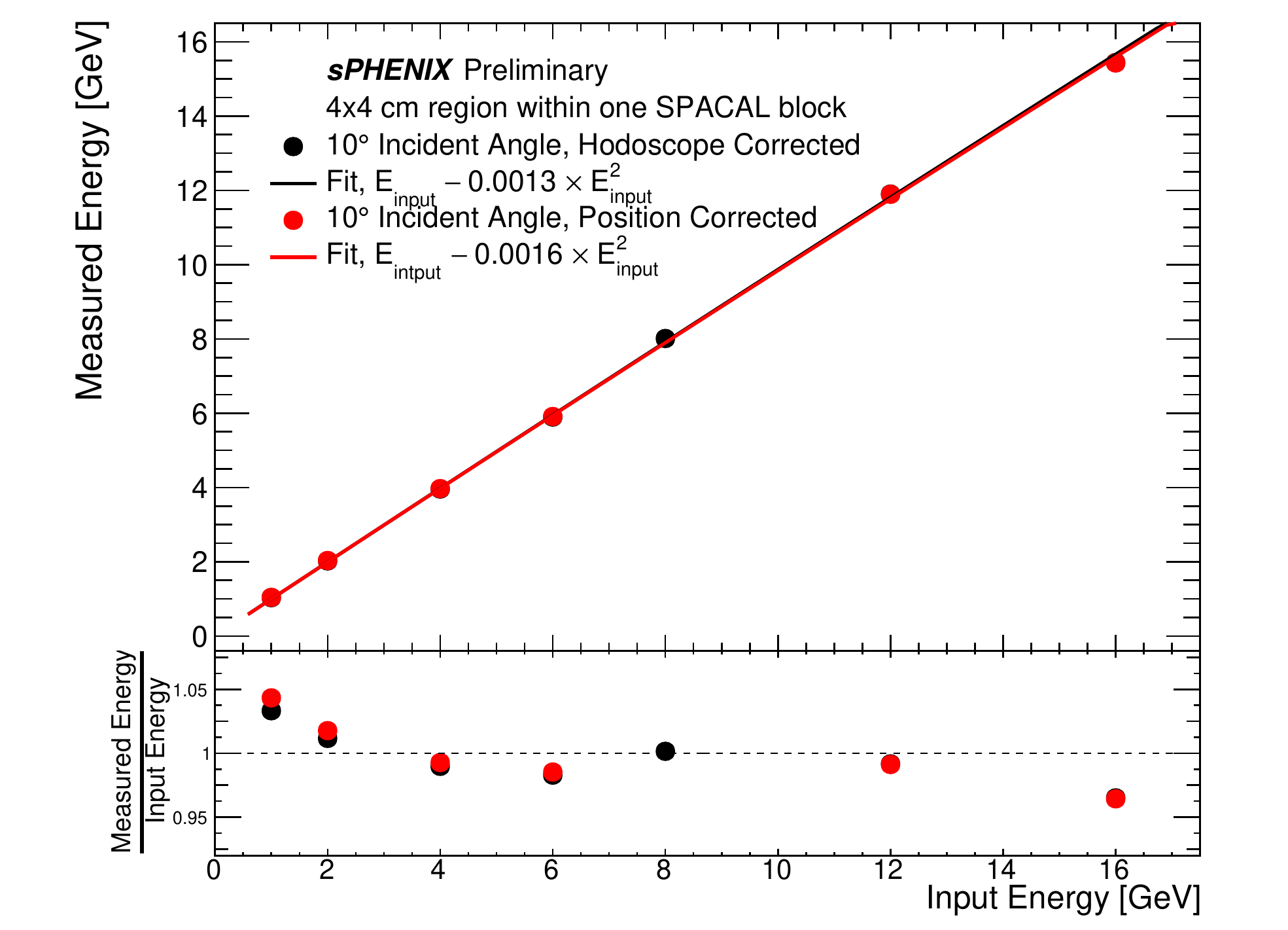}
	\includegraphics[width=0.49\textwidth]{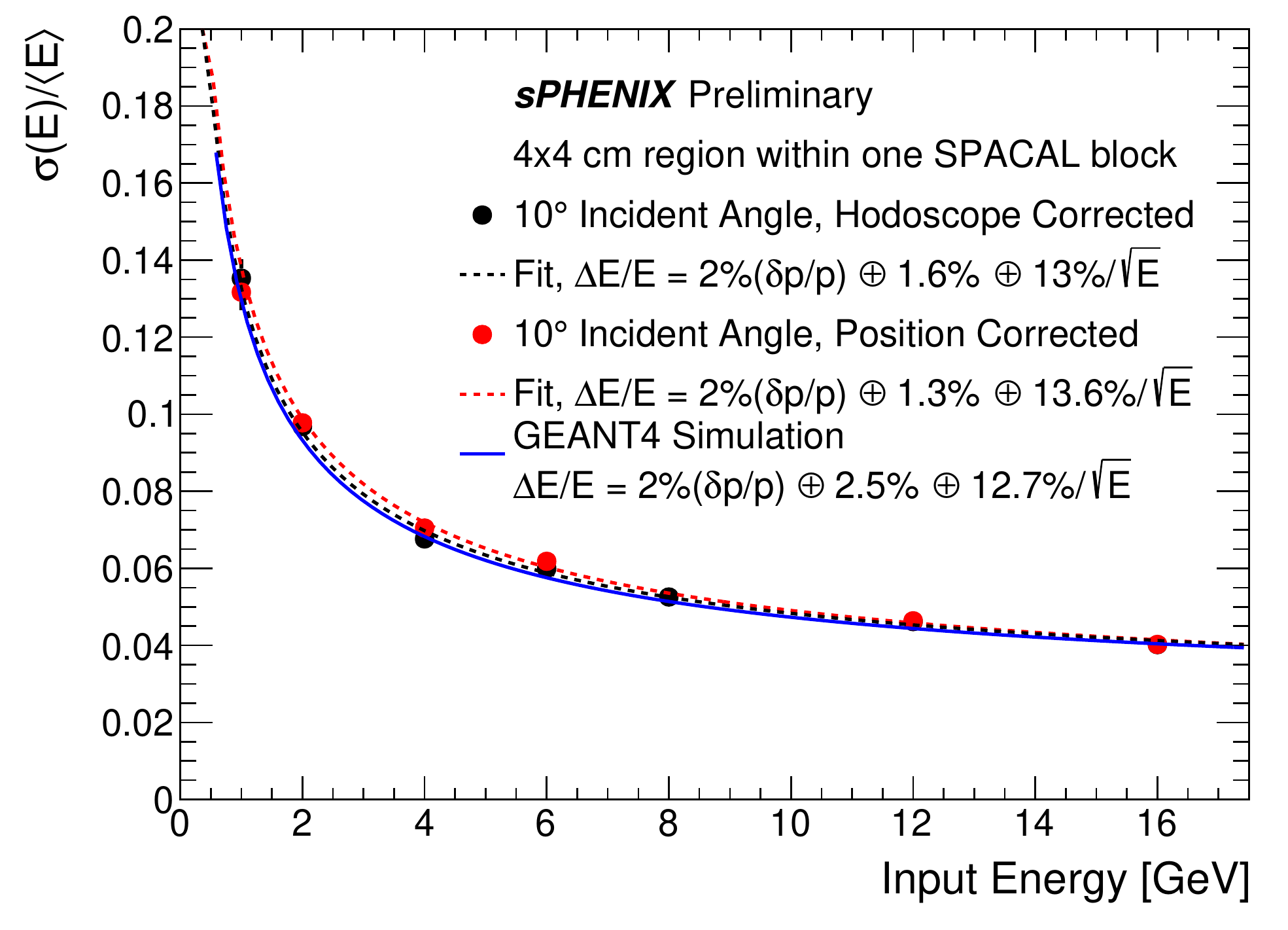}

	\caption{The linearity (left) and resolution (right) of the 2D projective SPACAL towers are shown with corresponding simulation curves. The measured resolution when the beam is required to be centered on a tower exceeds the sPHENIX physics requirements.}
	\label{fig:sphenix_testbeam_res_2017}
\end{figure}

The next installment of the T-1044 experiment was performed in 2017 at the Fermilab Test Beam Facility and featured the large rapidity $\eta\sim$0.9 calorimetry. Figure~\ref{fig:sphenix_2017_testbeam} shows a picture of the high rapidity setup, where here the beam is coming out of the page and the picture is taken from behind the outer hadronic calorimeter. The EMCal used in this test beam includes the first ever 2D projective SPACAL blocks, so in particular this test beam was very important for characterizing the towers as well as learning the best methods to build them. Preliminary results show the linearity and resolution of the 2D SPACAL towers in Fig.~\ref{fig:sphenix_testbeam_res_2017}~\cite{2017_emcal_analysis}. The simulation matches the data well and the resolution meets the sPHENIX EMCal physics requirements.

\begin{figure}[thb]
	\centering
	\includegraphics[width=0.9\textwidth]{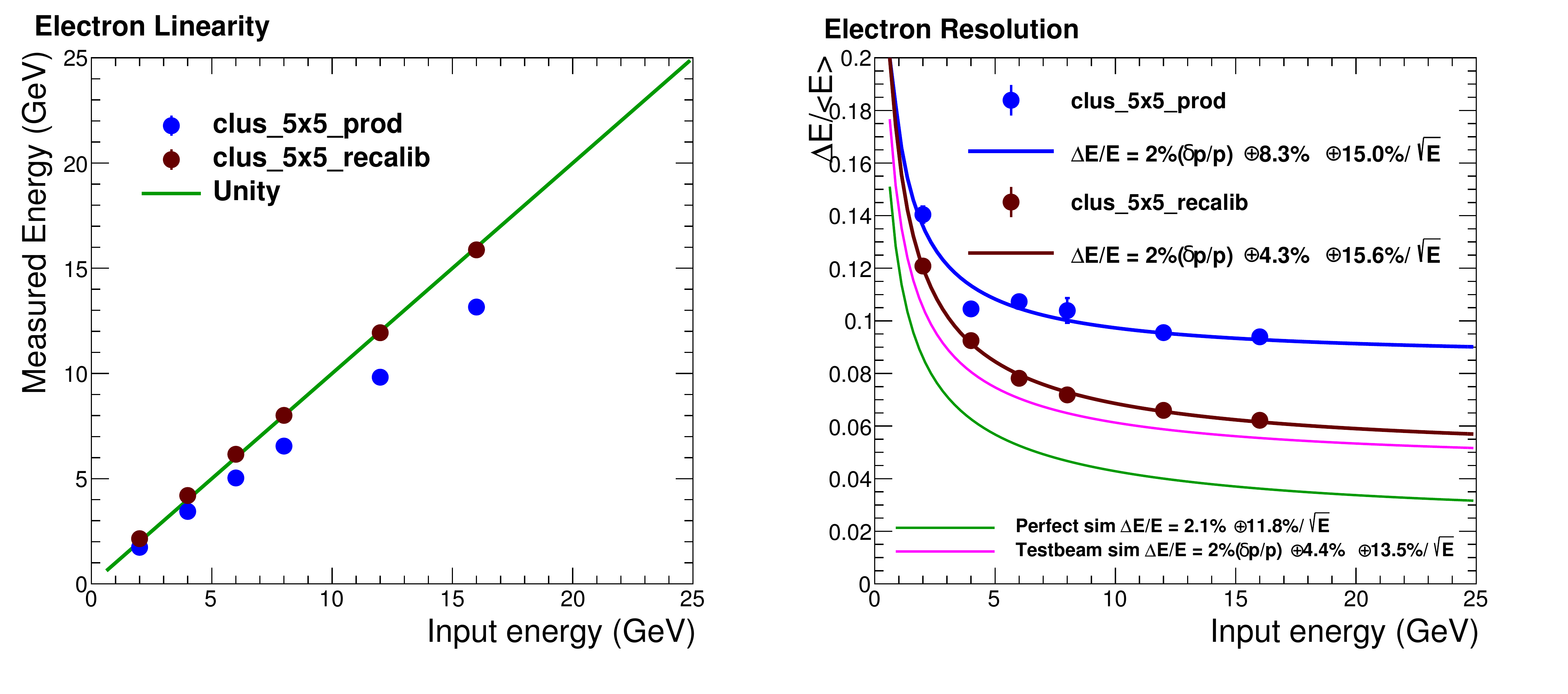}
	\caption{The linearity (left) and resolution (right) of the 2D projective SPACAL towers across the entire face of the calorimeter are shown with corresponding simulation curves. The simulation does not match the data, largely because of significant energy loss in the block boundaries. These block boundaries will be explored in greater depth in the 2018 test beam analysis.}
	\label{fig:sphenix_testbeam_res_2017_thirdjointscan}
\end{figure}

However, in these results the beam was required to be centered on a tower; when this requirement is relaxed the resolution and linearity are shown in Fig.~\ref{fig:sphenix_testbeam_res_2017_thirdjointscan}. The brown points in Fig.~\ref{fig:sphenix_testbeam_res_2017_thirdjointscan} show that the resolution degrades considerably, and that the simulation no longer matches the test beam data. This is due to the block boundaries of this particular prototype. Since these were the first 2D projective SPACAL towers ever produced, the methods that were used to combine blocks of towers together were still being refined. A second prototype has been constructed utilizing the knowledge gained from the first prototype, which is already known to have significantly better block boundaries. Another test beam with 2D projective towers in the spring of 2018 will test this new prototype and have more conclusive results of the resolution across the entire face of the 2D projective SPACAL blocks~\cite{2017_emcal_analysis}. Nonetheless, the preliminary results show encouraging progress towards the sPHENIX physics goals being met.

\subsubsection{Hadronic Calorimeters}

The hadronic calorimeters (HCals) are essential for measuring well calibrated and meaningful jets, since jets include both electromagnetic and hadronic components. The physics requirements for the HCals are largely motivated by jet measurements in heavy ion collisions. The HCals have a resolution requirement of $\Delta E/E = $100\%/$\sqrt{E}$ (\gevc) since the jets that will be measured will be at high \pt with respect to the beam energy. In order to measure and trigger on jets the HCal needs to have a large uniform acceptance, similar to the EMCal. To minimize systematic uncertainties associated with energy that is missed by the detector, the HCal is required to absorb greater than 95\% of the incident hadronic energy; therefore the required calorimeter depth should be 5.5 nuclear interaction lengths. These requirements are imposed since the largest uncertainty for jet finding in heavy ion collisions is from the background subtraction of the underlying event; in \pp collisions this is not a substantial issue and thus these requirements will also be suitable for high \pt jet measurements. \par

The calorimeter is constructed of absorber plates tilted from the radial direction to provide more uniform sampling in the azimuthal plane. Plastic scintillators with embedded wavelength shifting fibers are placed in between the absorber plates and are read out with silicon photomultipliers. The tilt angle of the plates is chosen so that a track which emerges from the interaction point traverses at least four scintillator tiles. This is intentional in order to meet the requirement that the total HCal system absorbs at least 95\% of the incident hadronic energy. Both the inner and outer HCals are constructed in this fashion, with some minor changes for the construction of each based on the radial space they occupy to meet the 5.5 interaction length requirement. \par

\begin{figure}[tbh]
	\centering
	\includegraphics[width=0.9\textwidth]{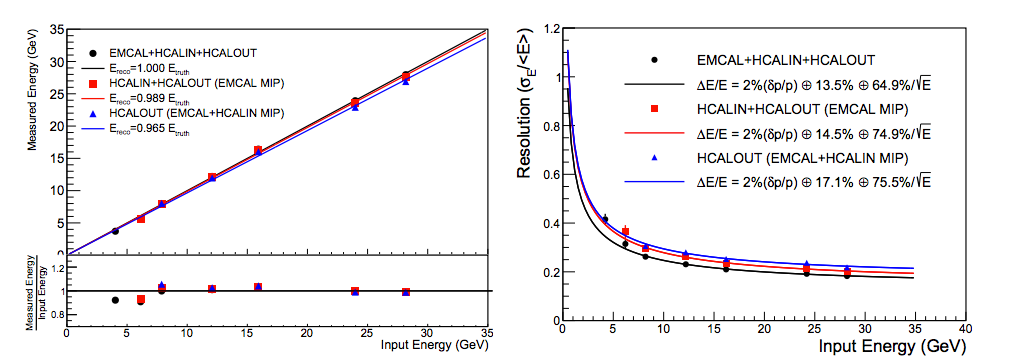}
	\caption{The hadronic calorimeter linearity (left) and resolution (right) are shown for the central rapidity sPHENIX system~\cite{Aidala:2017rvg}. The linearity and resolution are shown for all three calorimeters, only the inner and outer HCal, and the outer HCal alone.}
	\label{fig:hcal_2016_results}

\end{figure}

\begin{figure}[tbh]
	\centering
	\includegraphics[width=0.49\textwidth]{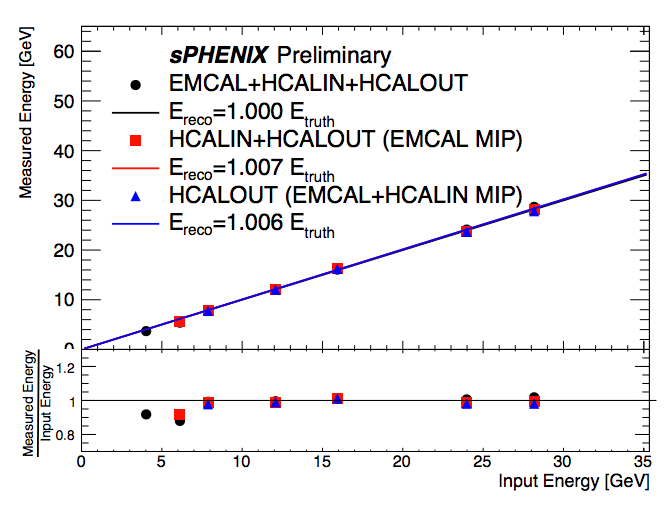}
	\includegraphics[width=0.49\textwidth]{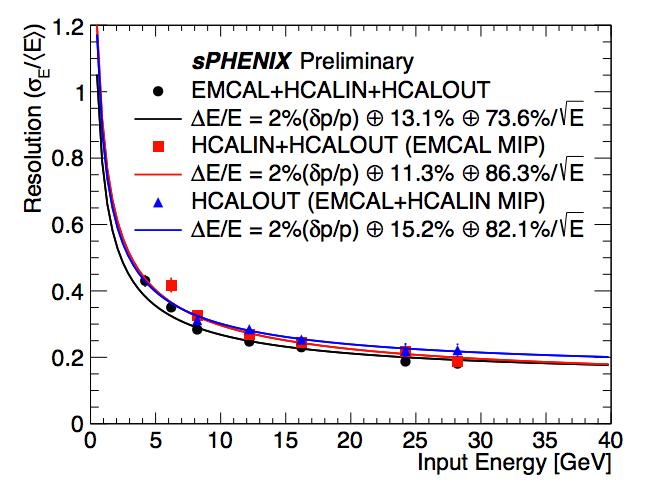}
	\caption{The hadronic calorimeter linearity (left) and resolution (right) are shown for the high rapidity sPHENIX system. The linearity and resolution are shown for all three calorimeters, only the inner and outer HCal, and the outer HCal alone.}
	\label{fig:hcal_2017_results}

\end{figure}

The inner and outer HCals were both tested in the 2016 and 2017 test beams~\cite{Aidala:2017rvg, 2017_hcal_analysis}. In each test beam the HCal system is evaluated as a standalone detector as well as with the EMCal in front of the system to determine the hadron resolution for the entire calorimeter system. The central rapidity results are shown in Fig.~\ref{fig:hcal_2016_results}. The hadronic calorimeter system shows excellent linearity, with deviations at lower energies due to difficulties from distinguishing the hadron peak from the minimum ionizing particle (MIP) peak~\cite{Aidala:2017rvg}. The measurements meet and exceed the sPHENIX requirement of hadronic calorimetry which has a stochastic term less than 100\%/$\sqrt{E}$ (\gevc). The 2017 test beam was the first to test the high rapidity HCal configurations, although the actual calorimeter segments are largely the same as the central rapidity version. The linearity and resolution of the 2017 test beam are shown in Fig.~\ref{fig:hcal_2017_results}, where again the results meet and exceed the sPHENIX requirements for the hadronic calorimetry. \par

%
%
%
%
%
%
%
%

\subsection{Simulation Study and Estimated \gammajet Yields}

To investigate the statistical precision with which sPHENIX will be able to measure \gammajet events, a simulation study was performed using a complete GEANT4~\cite{Agostinelli:2002hh} description of the detector. Performing this simulation study will determine the quantitative precision with which sPHENIX will be able to measure this channel and potential sensitivity to factorization breaking effects. \pythia8~\cite{Sjostrand:2006za, Sjostrand:2007gs} was used to generate \gammajet events, and all jets were reconstructed with the anti-$k_T$ algorithm using the Fastjet package~\cite{Cacciari:2011ma,Cacciari:2005hq}.

\pythia \pp simulations were performed at \sqs=~200 GeV with the CTEQ6L PDF set and all prompt photon processes. A requirement that the hard scale of the interaction be greater than 6 \gev was set so that enough statistics could be accumulated. Due to the jet resolution of the detector, an offline jet \pt cut of greater than 8 \gevc was set, which to some degree sets the scale of the direct photon. The only trigger requirement in the \pythia event was that a high \pt photon of at least 10 \gev was within the sPHENIX acceptance of $|\eta|<1$; this way the sPHENIX high energy EMCal trigger could be emulated. This also makes the events as kinematically unbiased as possible.

\subsubsection{Acceptance and Efficiency}

\begin{figure}[tbh]
	\centering
	\includegraphics[width=0.49\textwidth]{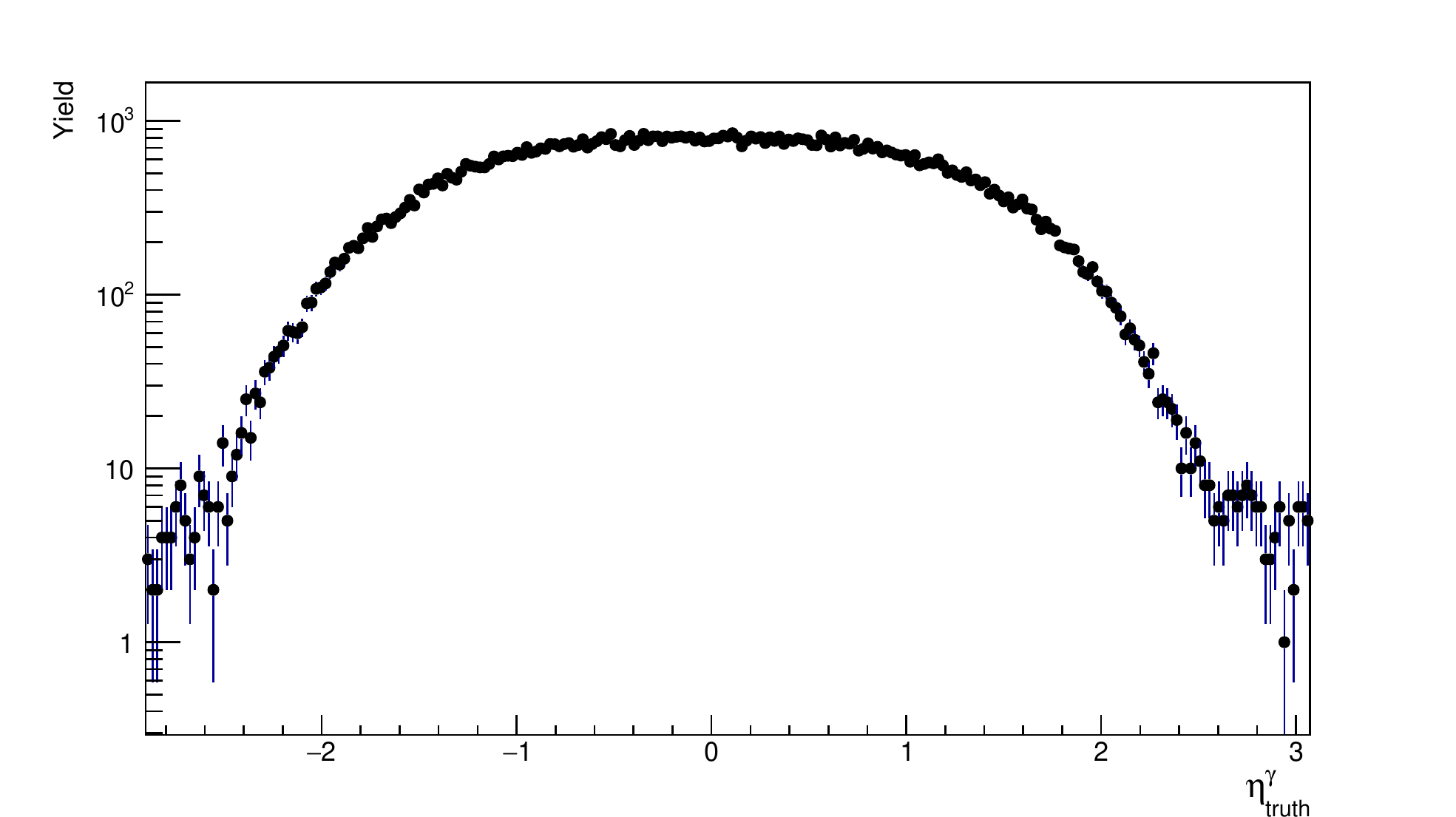}
	\includegraphics[width=0.49\textwidth]{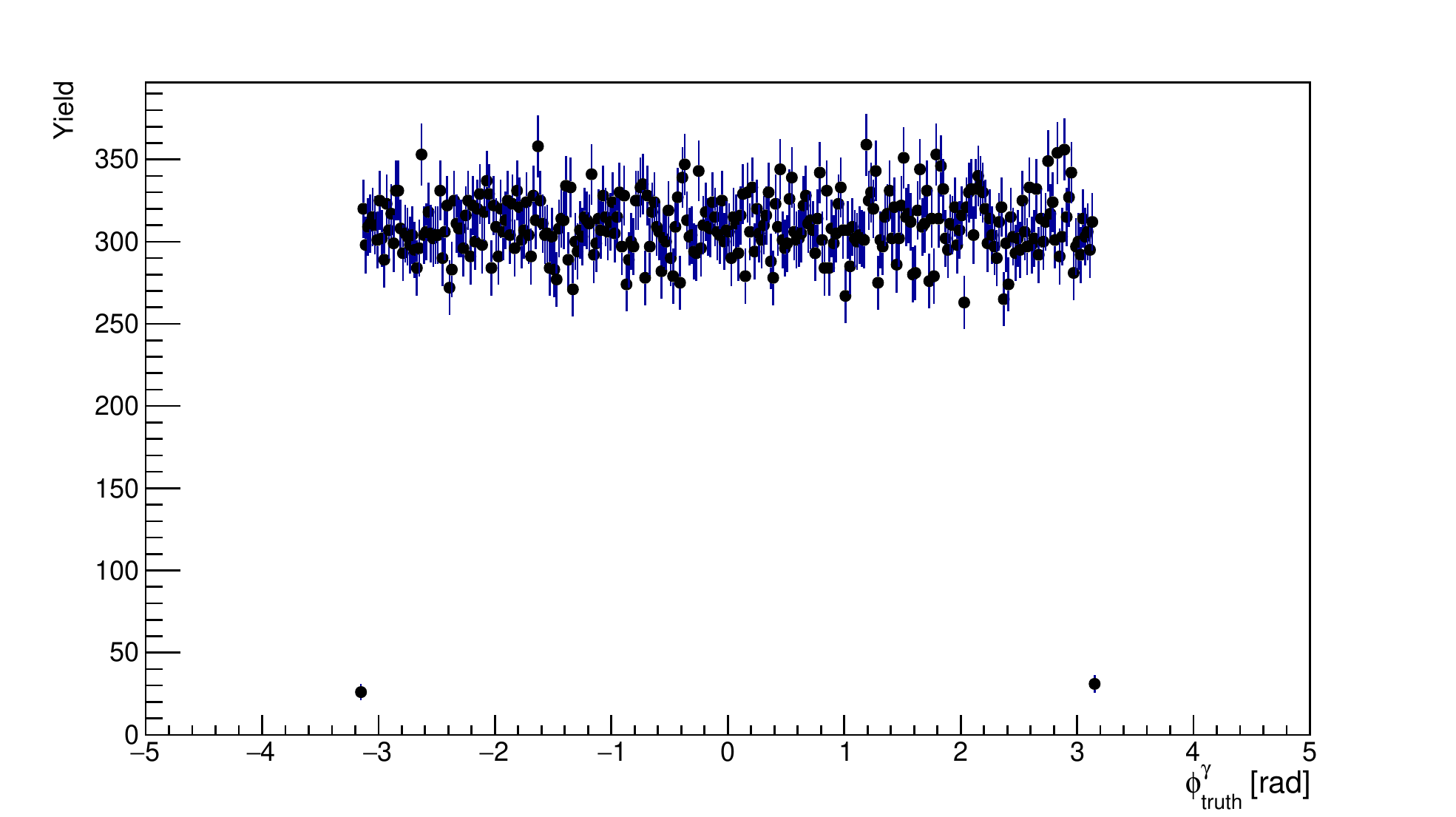}
	\includegraphics[width=0.5\textwidth]{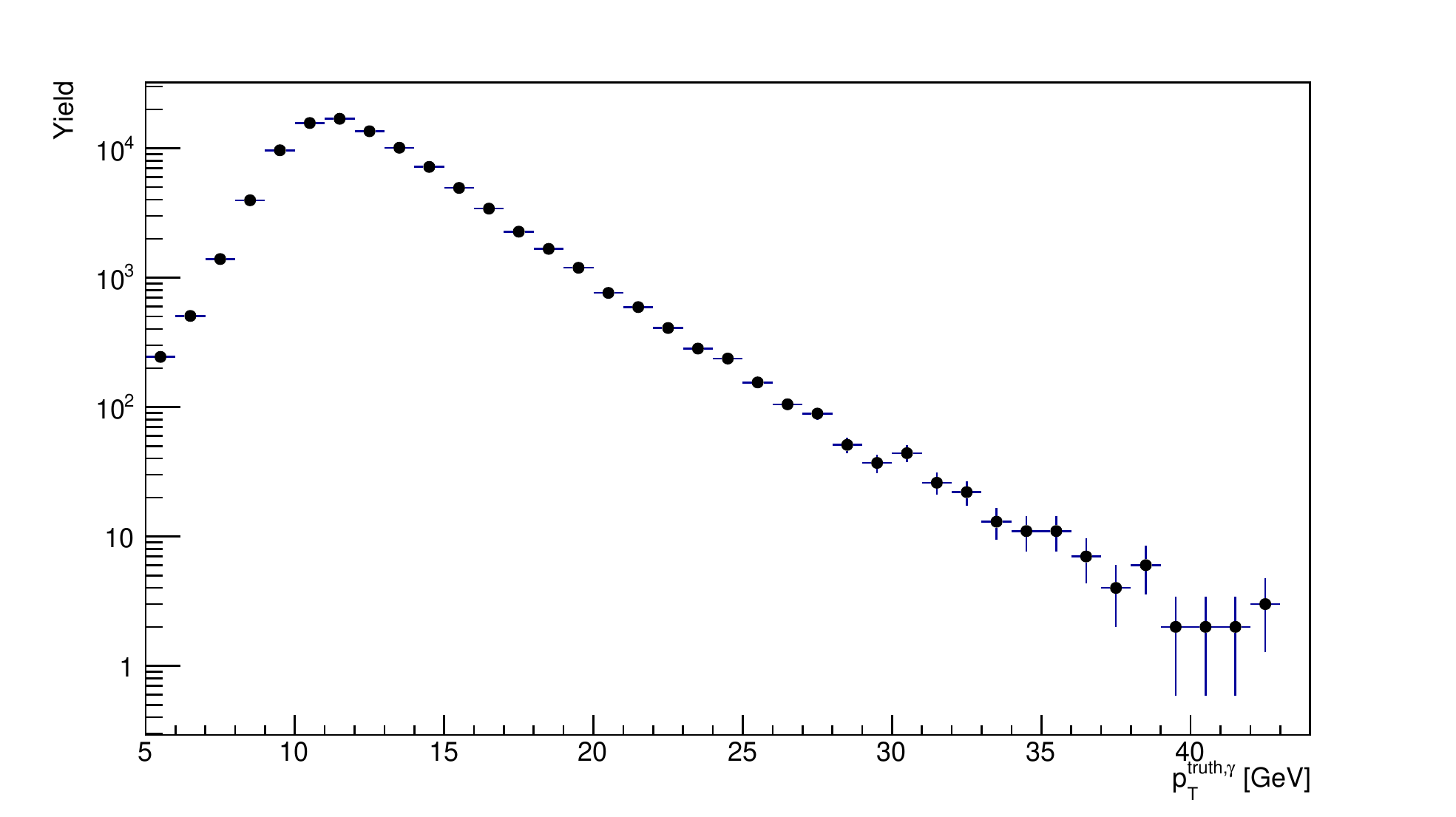}
	\caption{The truth photon distributions for the \pythia acceptance and efficiency study are shown as a function of $\eta$ (top left), $\phi$ (top right), and \pt (bottom).}
	\label{fig:truth_photon_dist}
\end{figure}

\begin{figure}[tbh]
	\centering
	\includegraphics[width=0.49\textwidth]{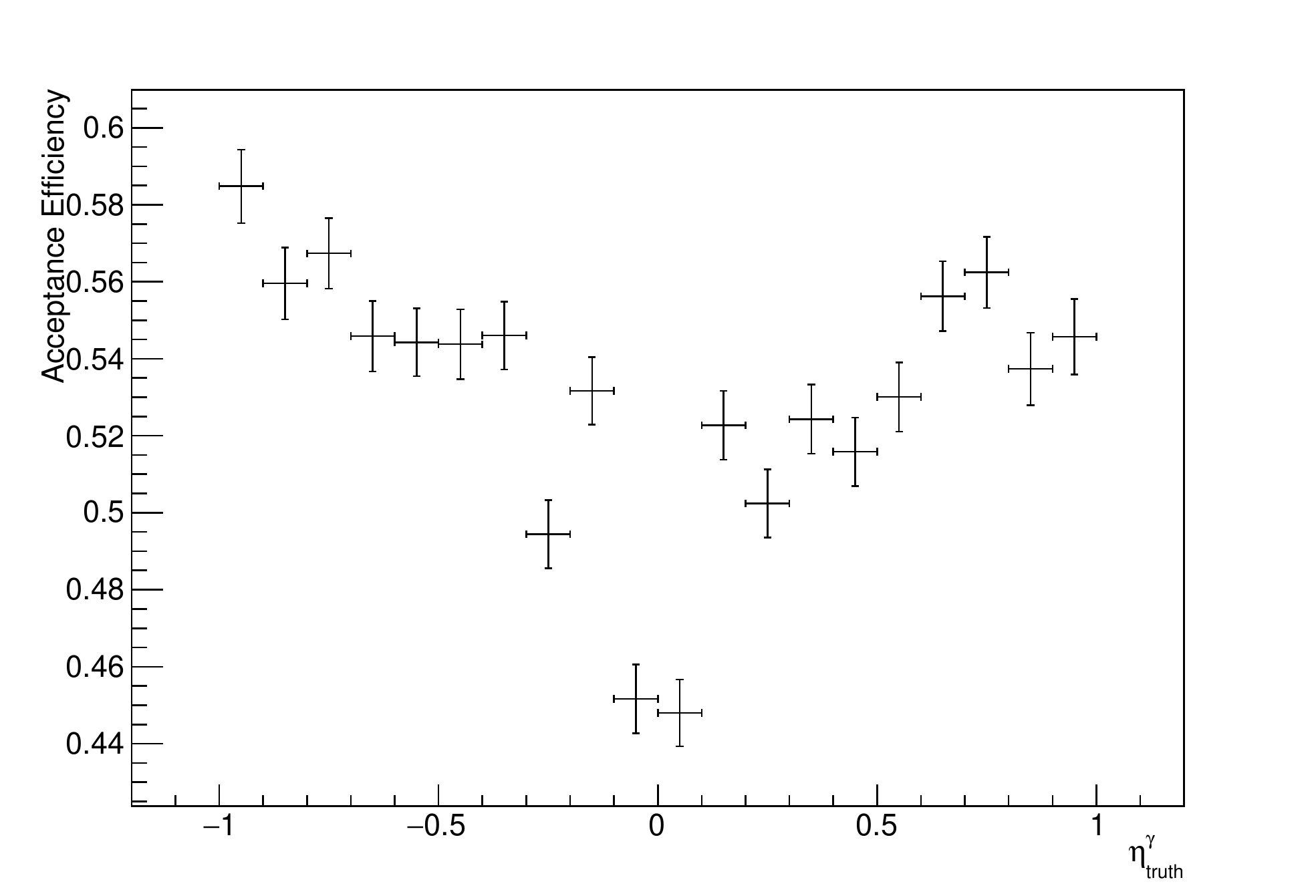}
	\includegraphics[width=0.49\textwidth]{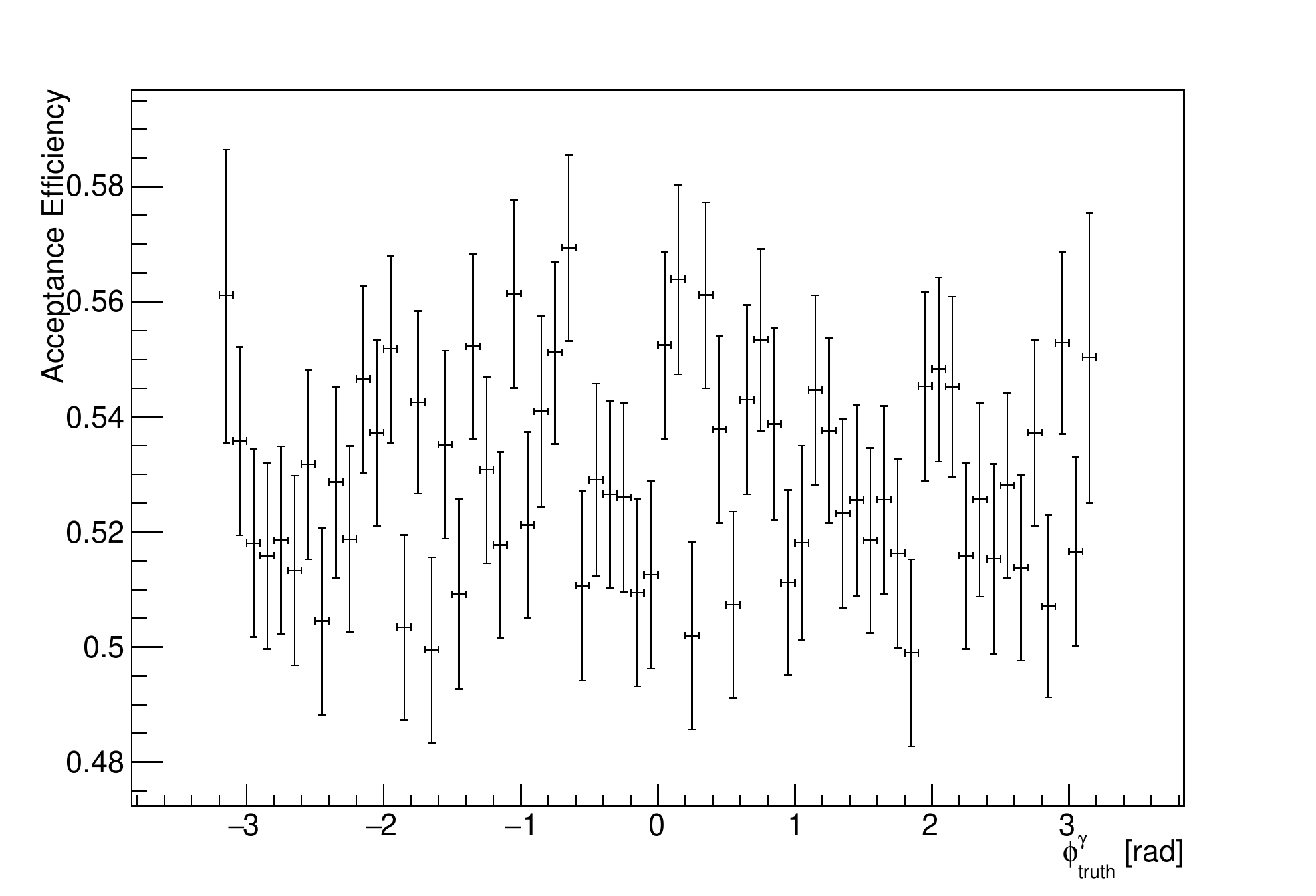}
	\includegraphics[width=0.5\textwidth]{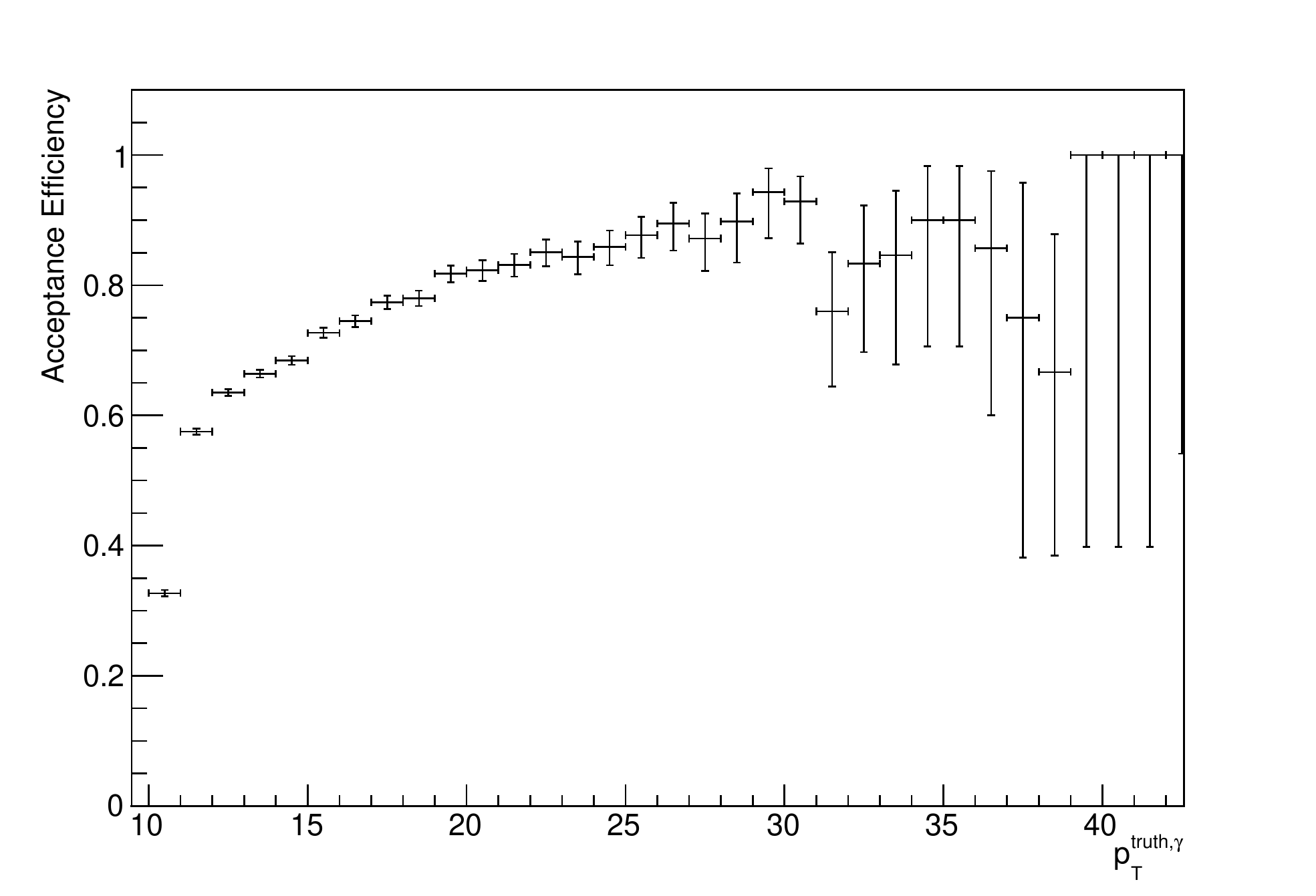}
	\caption{The \gammajet efficiencies as defined above are shown as a function of the truth photon $\eta$ (top left), $\phi$ (top right), and \pt (bottom).}
	\label{fig:gammajet_eff_sphenix}
\end{figure}

To quantify the efficiency of the detector, an acceptance and efficiency study was performed. To determine the acceptance and efficiency, \gammajet events were thrown in \pythia without any \pythia level trigger requirement. The only requirement was that the partonic hard scale be set to at least 10 \gev to accumulate enough statistics in the kinematic region of interest. The truth photon distributions are shown in Fig.~\ref{fig:truth_photon_dist} as a function of the photon azimuth, pseudorapidity, and \pt. To construct the efficiencies, the \pythia events were processed through the GEANT4 description of the sPHENIX detector. The efficiency was determined by dividing the number of reconstructed \gammajet pairs by the number of truth photons which were thrown in $|\eta|<1$. This restriction is necessary as the yields will be calculated only for $|\eta|<1$, not for a full rapidity integrated cross section. The efficiency values are shown as a function of the truth photon \pt, $\eta$, and $\phi$ in Fig.~\ref{fig:gammajet_eff_sphenix}. The efficiency is flat as a function of azimuth due to the uniform azimuthal coverage. There is a significant dip in the efficiency at $\eta\sim0$ because the events were thrown with $z_{vtx}=0$ cm, and there is a physical boundary in the EMCal at exactly $\eta=0$ which joins the two hemispheres of the detector. This boundary causes the photon detection to be significantly worse. The efficiency rises with the photon \pt as one might expect, since by definition at higher \pt the photon-jet pair are at more central rapidities and are thus more likely to be reconstructed.

These efficiencies should be applied to a perturbatively calculated cross section, scaled by the appropriate luminosity projections, to determine the resulting expected yields. Unfortunately such a calculation does not exist for RHIC energies as the current RHIC experiments have not been able to measure this channel, so theorists likely skipped over the particular calculation. PHENIX has published direct photon cross sections at \sqs=~200 GeV~\cite{Adare:2012yt}. Therefore projected yields given a total integrated luminosity can be determined for direct photons, and a ``\gammajet'' efficiency can be applied to this which is determined from \pythia. This efficiency is just the probability of reconstructing a \gammajet pair given a reconstructed direct photon. \par

\begin{figure}[tbh]
	\centering
	\includegraphics[width=0.7\textwidth]{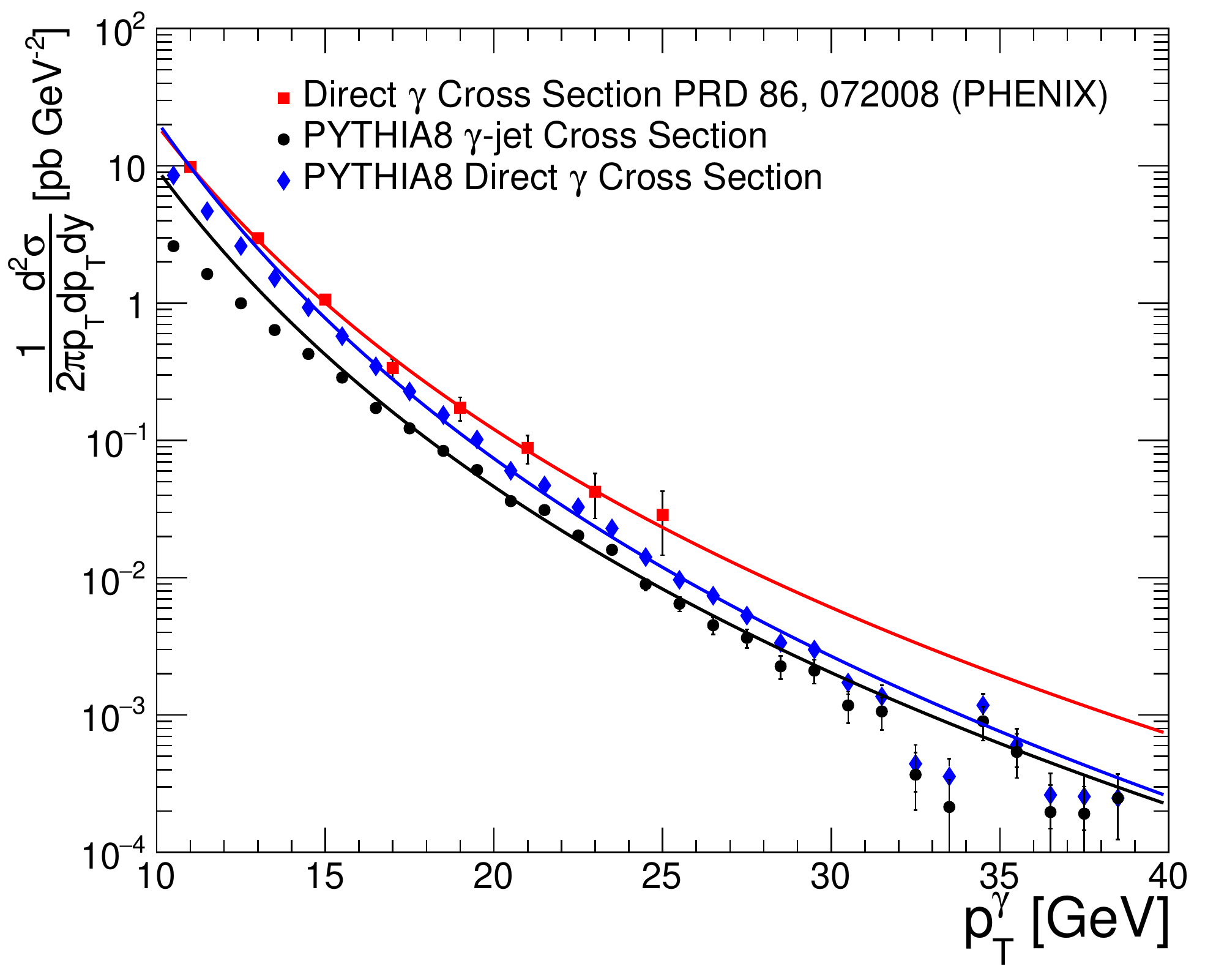}
	\caption{The direct photon cross sections as measured by PHENIX~\cite{Adare:2012yt} and calculated in \pythia are shown with the \gammajet cross section from PYTHIA.}
	\label{fig:photon_cross_sections}
\end{figure}

\begin{figure}[tbh]
	\centering
	\includegraphics[width=0.7\textwidth]{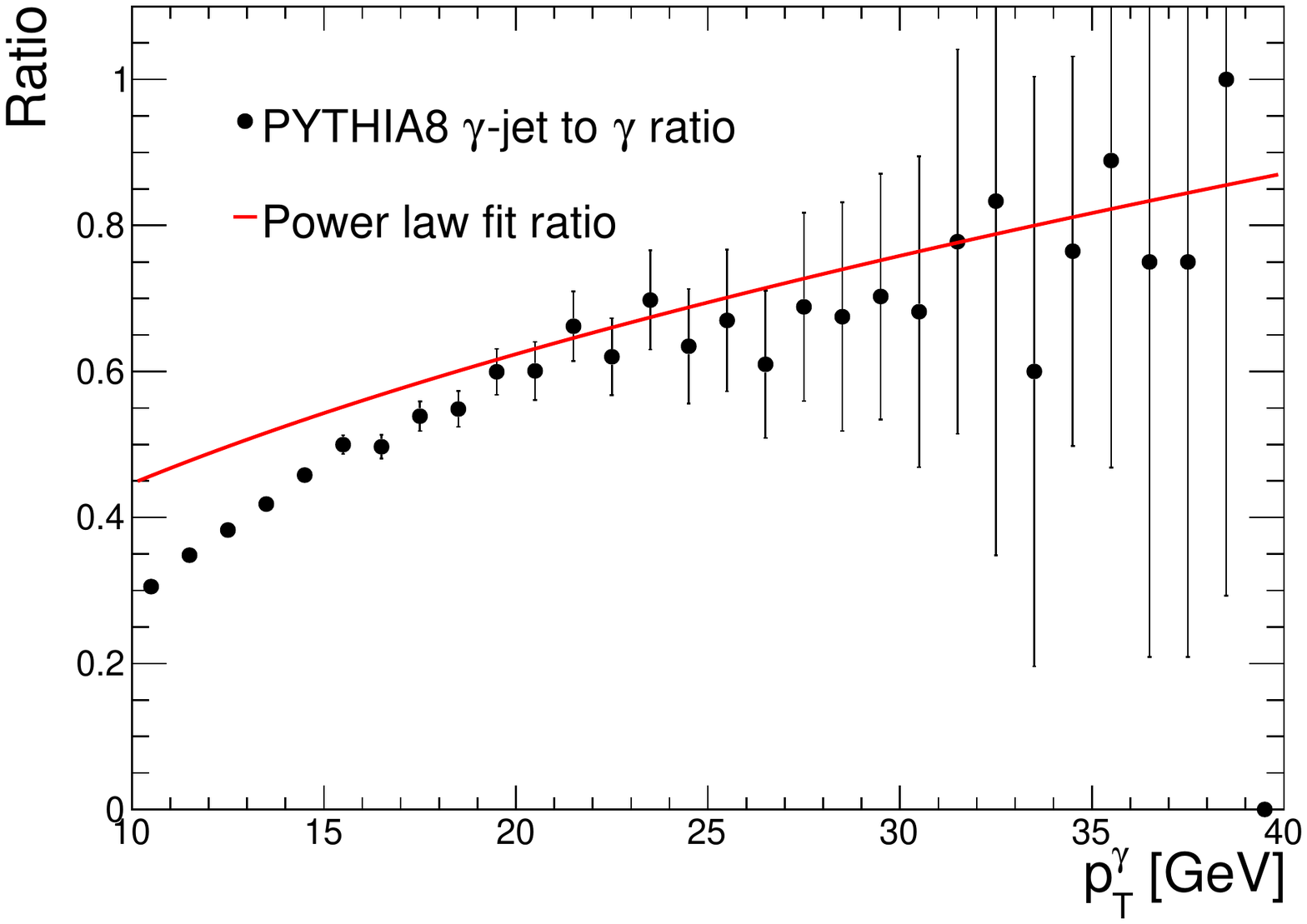}
	\caption{The ratio of \gammajet to direct photons as determined from \pythia cross section calculations is shown. The points represent the actual ratio of the cross sections, while the line represents the ratio of power law fits to the cross sections.}
	\label{fig:gammajet_eff}
\end{figure}

The cross section of direct photons and \gammajet was determined from \pythia, and this ratio was applied as a \gammajet efficiency factor to the statistical estimates of the yields determined from the direct photon cross section. Figure~\ref{fig:photon_cross_sections} shows the cross sections for direct photons and \gammajet as determined from a PYTHIA simulation together with the measured PHENIX inclusive direct photon cross section from Ref.~\cite{Adare:2012yt}. The ratio of the \pythia \gammajet to direct photon cross section was taken as the \gammajet efficiency factor, and is shown in Fig.~\ref{fig:gammajet_eff}. The points show the ratio of the actual cross section points in Fig.~\ref{fig:photon_cross_sections}, while the line shows the ratio of power law fits to the cross sections. The line is taken as the efficiency since, as one can see from Fig.~\ref{fig:photon_cross_sections}, the cross sections deviate from power law behavior at smaller \pt. This is not physical and is simply a result of the simulation being generated with the requirement that the partonic hard scale be at least 10 \gev within $|\eta|<$1.

\begin{figure}[tbh]
	\centering
	\includegraphics[width=0.7\textwidth]{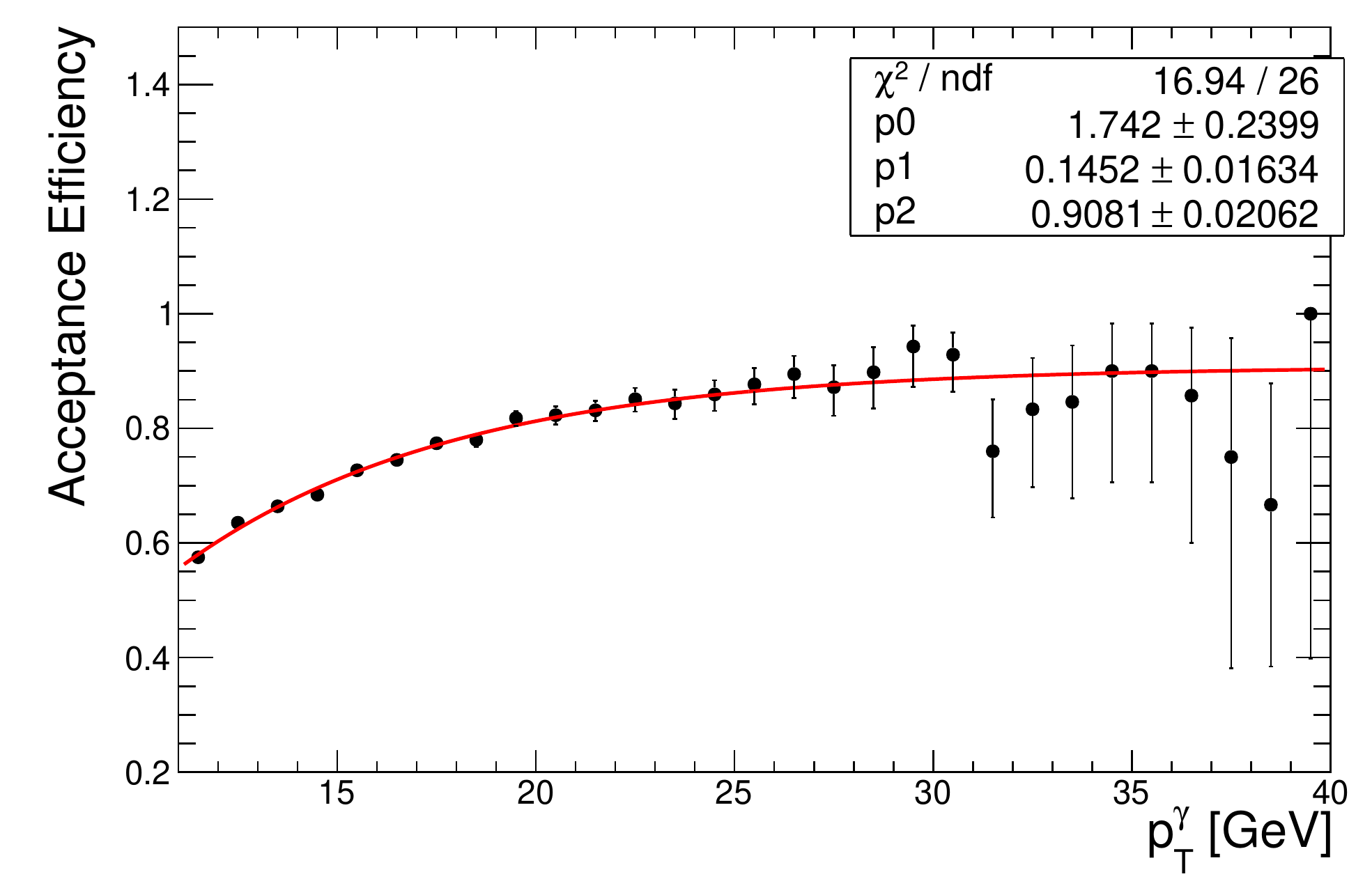}
	\caption{The direct photon efficiencies were fit with a saturated exponential to model the functional form and to ultimately apply to the direct photon cross section yields.}
	\label{fig:gammajet_eff_satexp}
\end{figure}

\begin{figure}[tbh]
	\centering
	\includegraphics[width=0.49\textwidth]{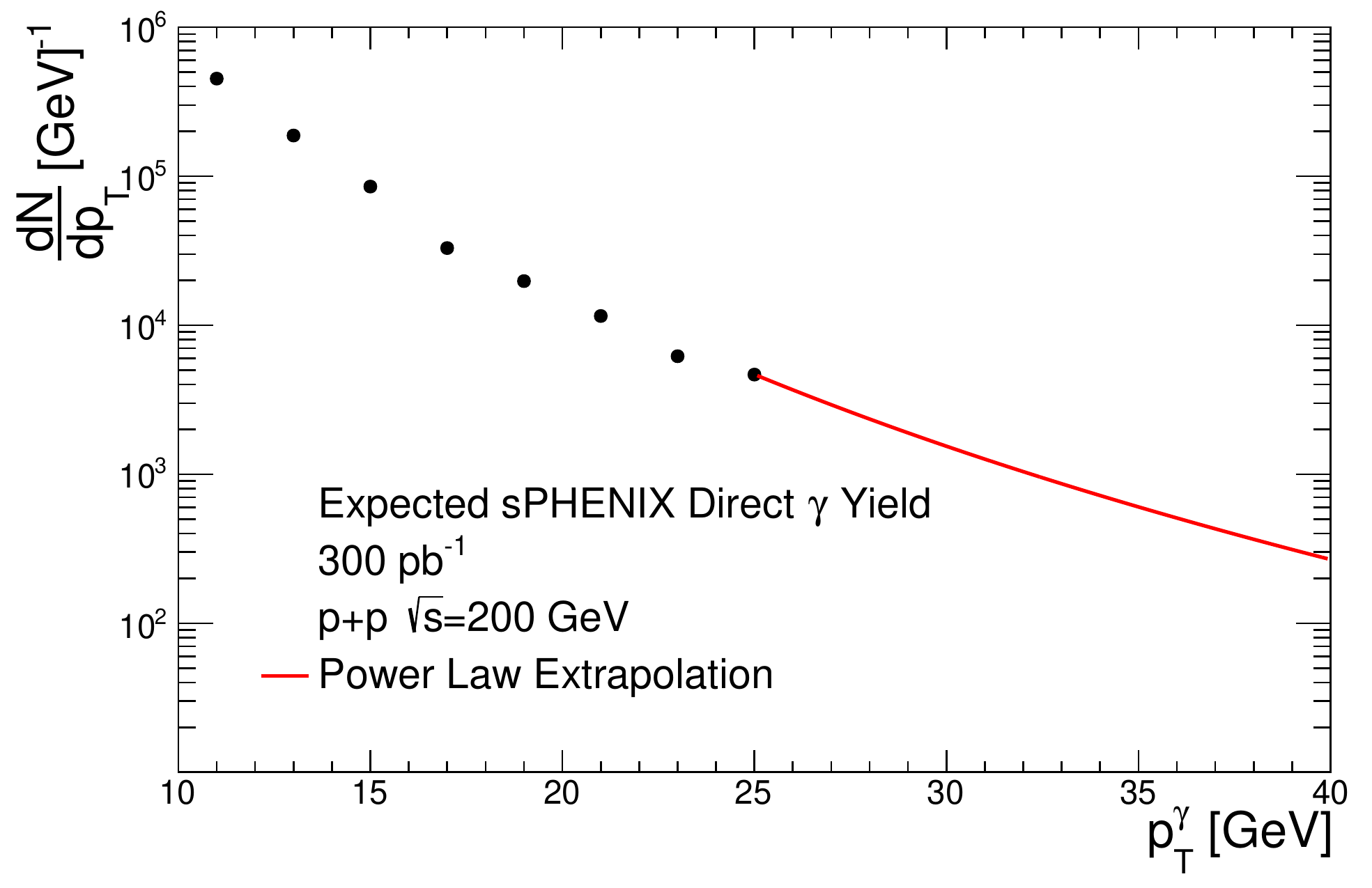}
	\includegraphics[width=0.49\textwidth]{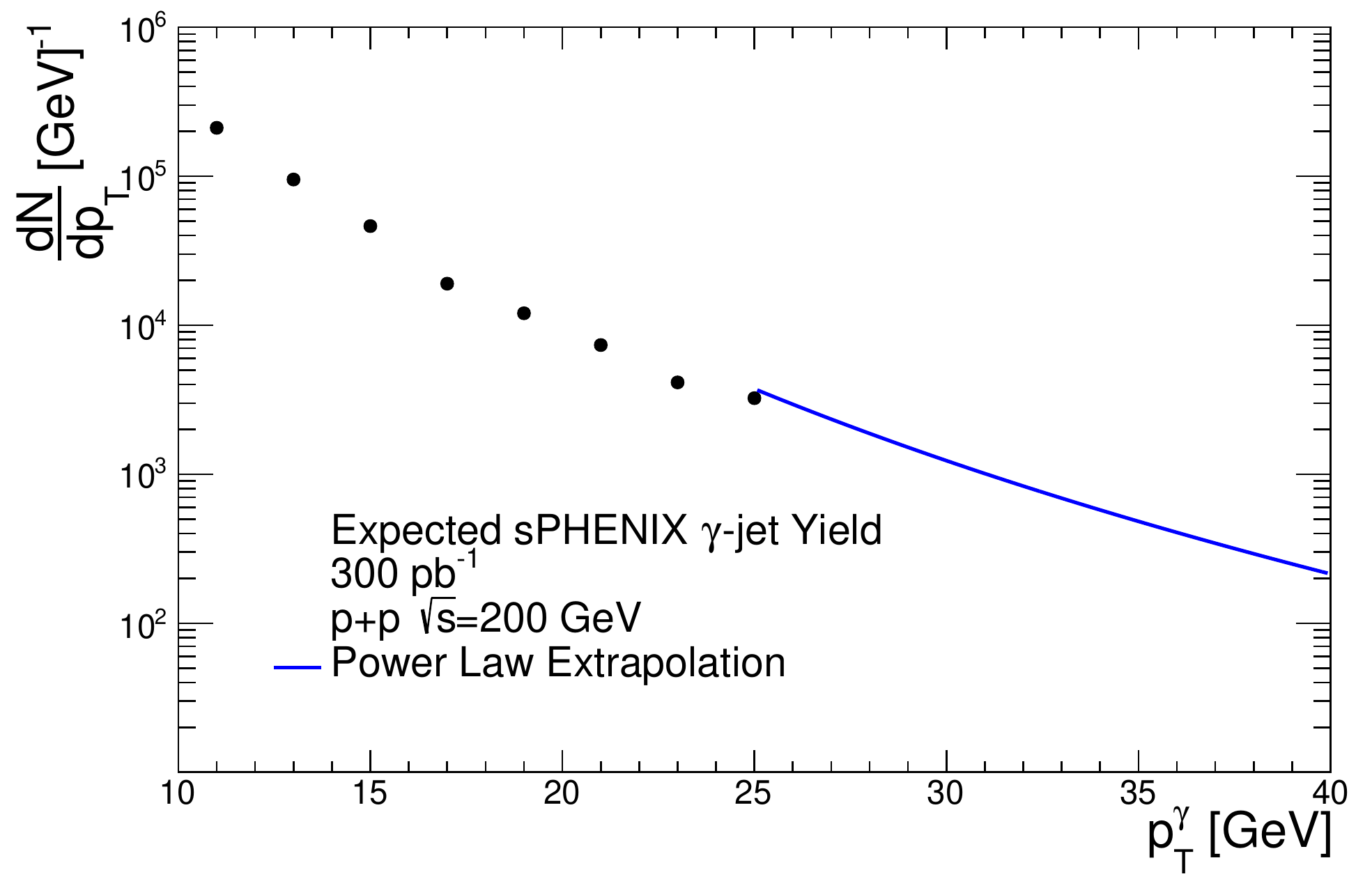}
	\caption{The direct photon and \gammajet projected yields are shown for integrated luminosities of 300 pb$^{-1}$. The power law extrapolations are determined by applying the efficiencies to a power law fit to the direct photon data from Ref.~\cite{Adare:2012yt}.}
	\label{fig:proj_phot_yields}
\end{figure}

With the appropriate efficiencies determined, a statistical estimate for \gammajet yields can be calculated. The measured direct photon cross section from Ref.~\cite{Adare:2012yt} was converted to a yield as a function of \pt by multiplying by the appropriate factors as well as the 300 pb$^{-1}$ integrated \pp luminosity projection from the RHIC cold QCD plan~\cite{Aschenauer:2016our}. The efficiency as a function of \pt was fit to a saturated exponential to model the functional form, seen in Fig.~\ref{fig:gammajet_eff_satexp}. This was applied to the cross sections and the resulting yields are shown in the left panel of Fig.~\ref{fig:proj_phot_yields}, where the power law extrapolation is just the power law fit from the published data multiplied by the saturation term from the acceptance and efficiency fit. The right panel of Fig.~\ref{fig:proj_phot_yields} shows the projected \gammajet yields, which are obtained by applying the $\gamma/\gammajet$ efficiency factor to the left panel of Fig.~\ref{fig:proj_phot_yields}.

To make statistical projections for the actual physical observables of \gammajet correlations, the resulting projected yields were binned as a function of the correlated observables $\dphi$ and \pout. The yields were estimated as per-trigger yields for photons with at least \pt=~10 \gev and jets with at least \pt=~8 \gev, and the quantities are binned as reconstructed values. No attempt was made to do an unfolding procedure since these are just statistical projections; any unfolding will introduce systematic uncertainties in the measurement and will be determined when actual data is being recorded.

\begin{figure}[tbh]
 \centering
 	\includegraphics[width=0.7\textwidth]{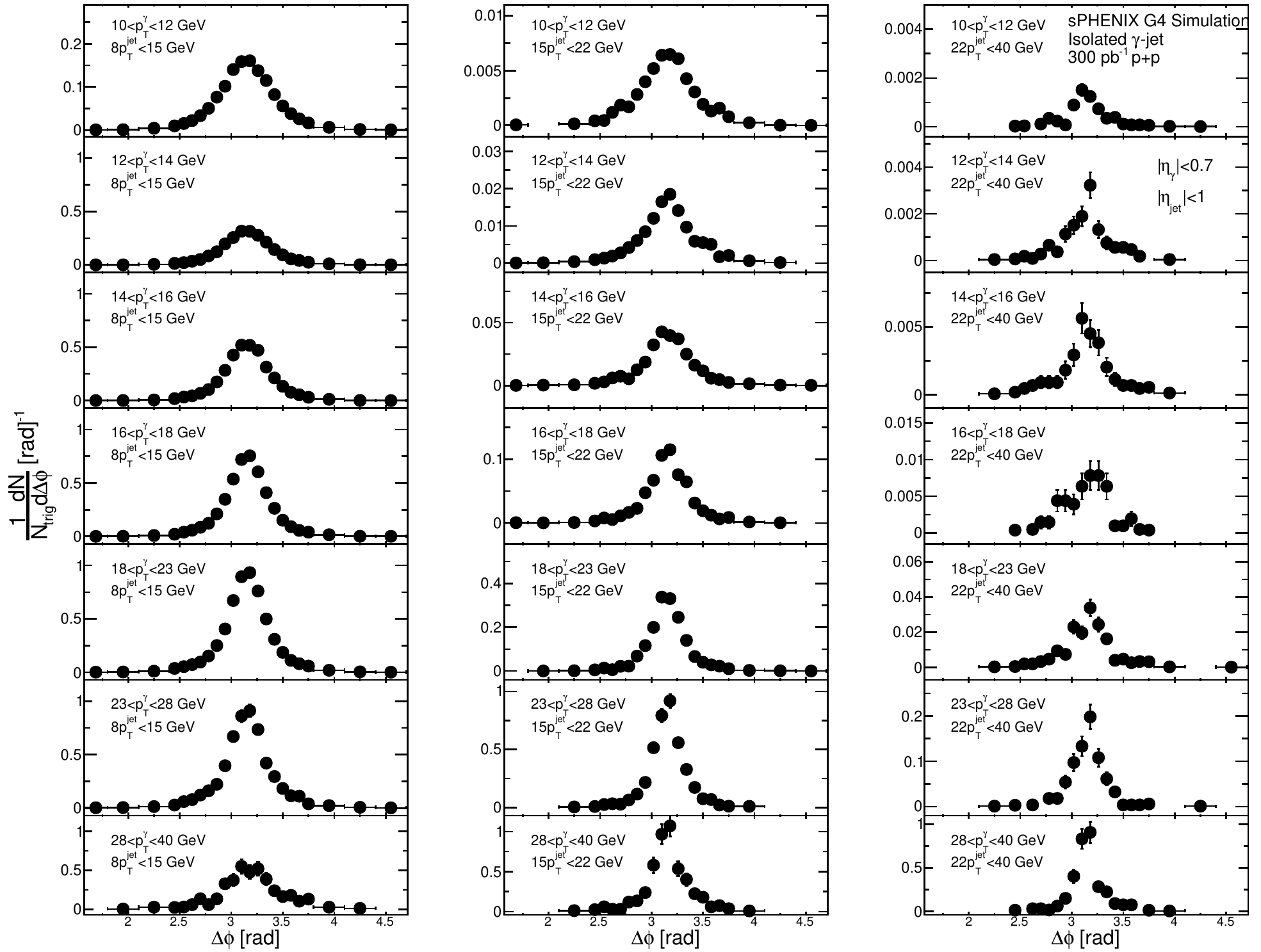}
	\caption{The statistical projections for the \gammajet \dphi distributions at sPHENIX are shown for several bins of $p_T^\gamma$ and $p_T^{\rm jet}$.}
	\label{fig:gammajet_dphis}
\end{figure}

\begin{figure}[tbh]
	\centering
	\includegraphics[width=0.8\textwidth]{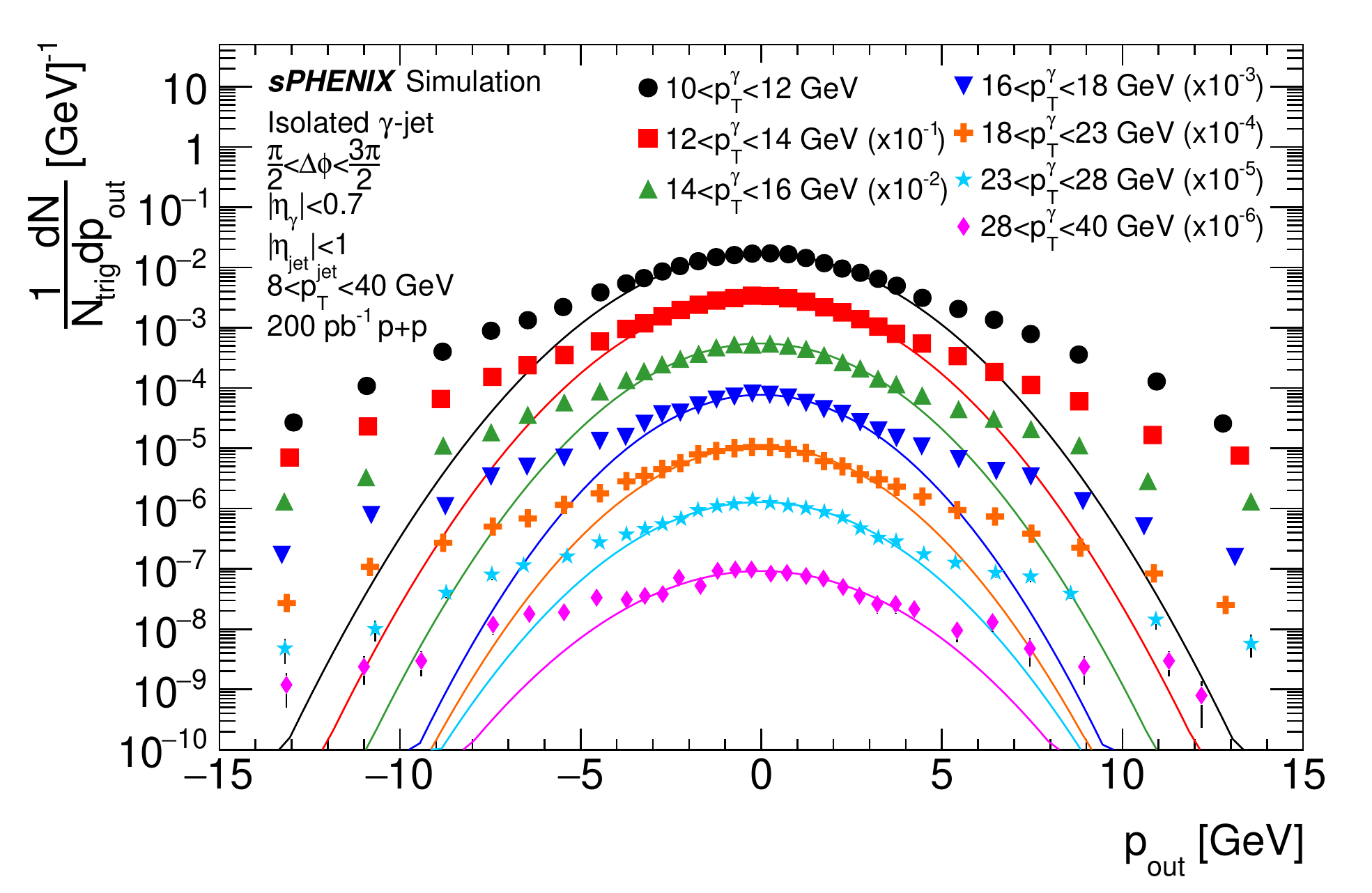}
	\caption{The statistical projections for the \gammajet \pout distributions at sPHENIX are shown for several bins of $p_T^\gamma$.}
	\label{fig:gammajet_pouts}
\end{figure}

Figure~\ref{fig:gammajet_dphis} shows the projected statistical uncertainties on \dphi correlations between \gammajet pairs as a function of the \photon and jet \pt. The \dphi bin widths at \dphi$\sim\pi$ were estimated based on a study of the jet azimuthal angle resolution which was found to be approximately 0.08 radians. The study shows that there will be a significant number of \gammajet correlations to measure at sPHENIX. Additionally, in \dphi space, the observables will have sensitivity to the nonperturbative physics as the resolution of \dphi is small enough to clearly see a transition from Gaussian to power-law behavior. 

Nonetheless, the \pout distributions have a greater sensitivity to the nonperturbative physics; these projections are shown in Fig.~\ref{fig:gammajet_pouts}. Here the yields are shown with 200 pb$^{-1}$ of integrated luminosity. This is the actual projected luminosity by the sPHENIX experiment and the projections are more up-to-date than those in the RHIC cold QCD plan. Even with the reduction of integrated luminosity between the two projections there will be more than enough statistical precision to measure the \pout distributions accurately. Additionally the resolution of the sPHENIX detector is precise enough to be able to distinguish between the nonperturbative and perturbative regions of the \gammajet correlations. This is again largely due to the kinematic region with which sPHENIX will be able to probe \gammajet correlations at RHIC; sPHENIX is designed to measure lower \pt jets than experiments at the LHC and this resolution allows for the nonperturbative structure to be identified.

This short simulation study has shown that the sPHENIX detector will be able to probe the ``golden channel'' for factorization breaking measurements, direct photon-jet, with excellent statistical precision in the kinematic region of interest. This measurement must be performed before the future EIC is turned on, as factorization breaking is only predicted in hadronic collisions. In particular it is important to perform the measurement at RHIC due to the kinematic region that can be probed; however, there are unique measurements at the LHC that may also be able to probe factorization breaking effects.

\section{New Observables at the LHC}

The collider experiments at the LHC benefit from the significantly higher luminosity that the LHC is able to deliver. For processes with small cross sections, this is particularly important to accumulate the necessary statistical precision to make a meaningful measurement. Both ATLAS and CMS have already performed analyses which probe color coherence effects~\cite{Chatrchyan:2013fha,Aaboud:2016sdm}; color coherence appears to be qualitatively similar to color entanglement effects as it describes effective regions where gluons destructively interfere. These interferences result in regions where products from gluon radiations are more likely to be measured, and these analyses determine these regions by measuring the angular distribution of jets with respect to a nearly back-to-back dijet system. Color coherence measurements should also be performed at sPHENIX as a test of their relation to color entanglement effects. This will allow for a \sqs dependence to be studied, as measurements have additionally been made at the Tevatron~\cite{Abe:1994nj,Abbott:1997bk}.

\begin{figure}[tbh]
	\centering
	\includegraphics[width=0.4\textwidth]{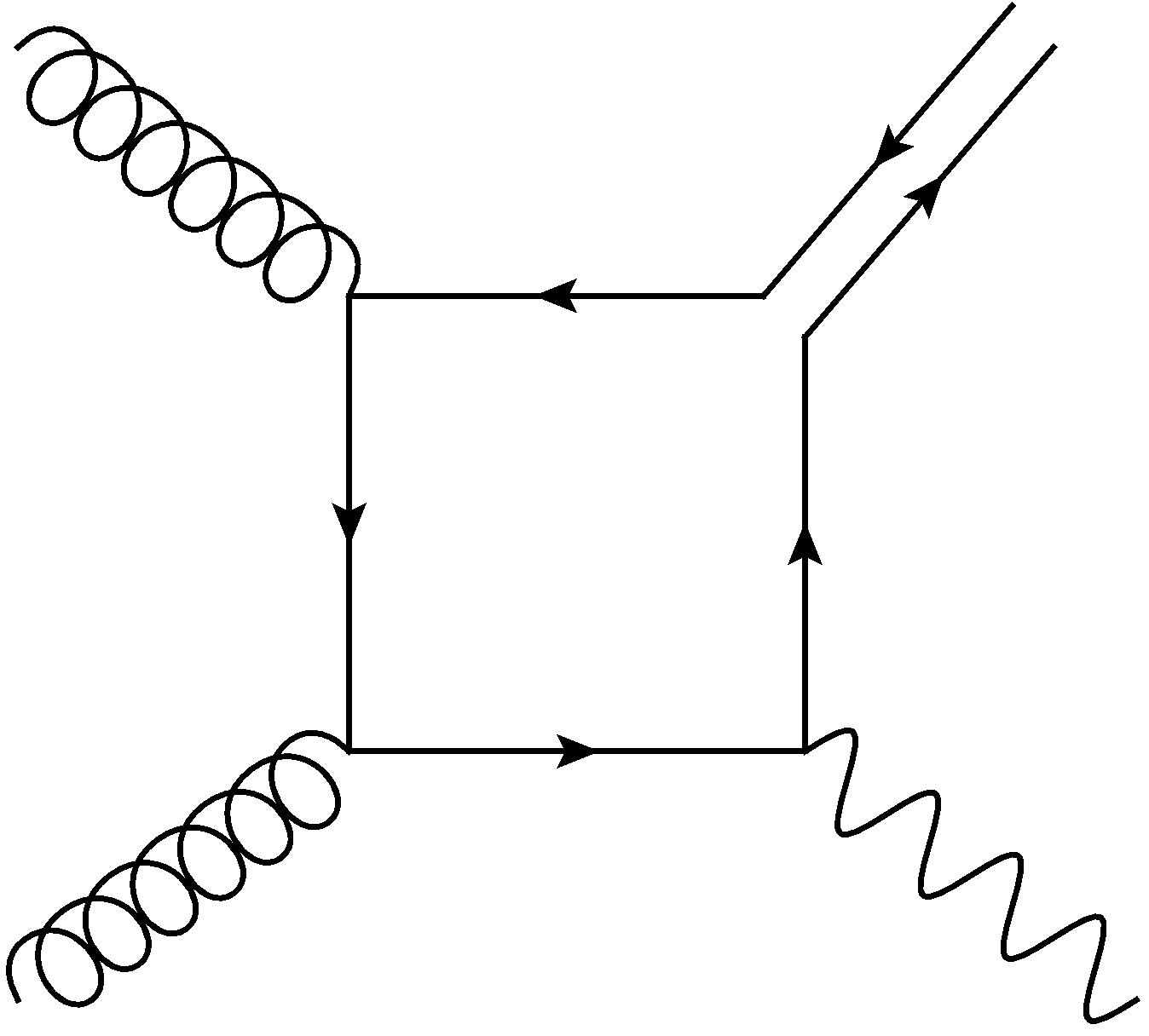}\hspace*{1cm}
	\includegraphics[width=0.4\textwidth]{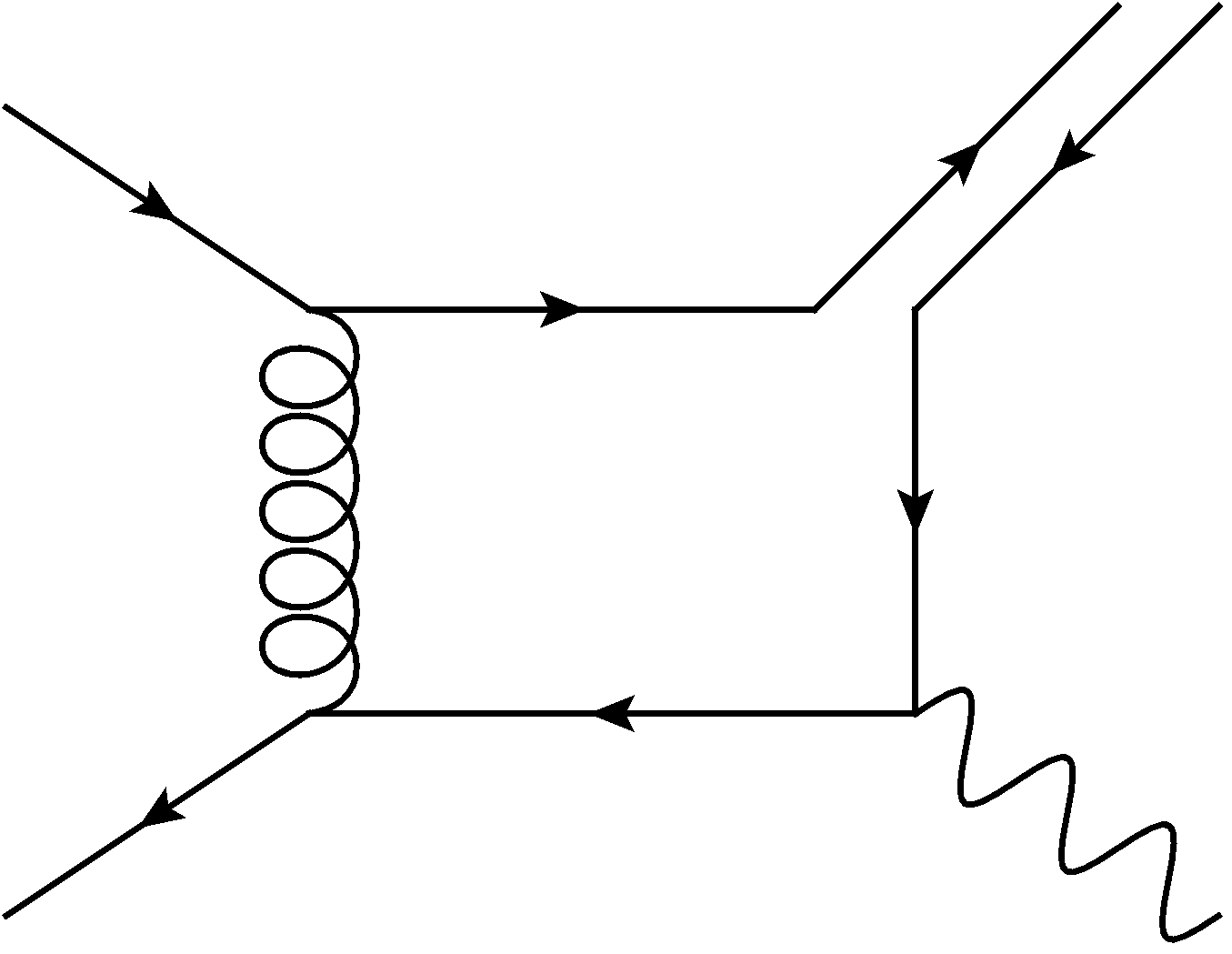}
	\caption{Two Feynman diagrams are shown for direct photon-quarkonium production in \pp collisions at LHC center-of-mass energies. The gluon-gluon fusion process (left) dominates.}
	\label{fig:isophot_quarkonia}
\end{figure}

\begin{figure}[tbh]
	\centering
	\includegraphics[width=0.8\textwidth]{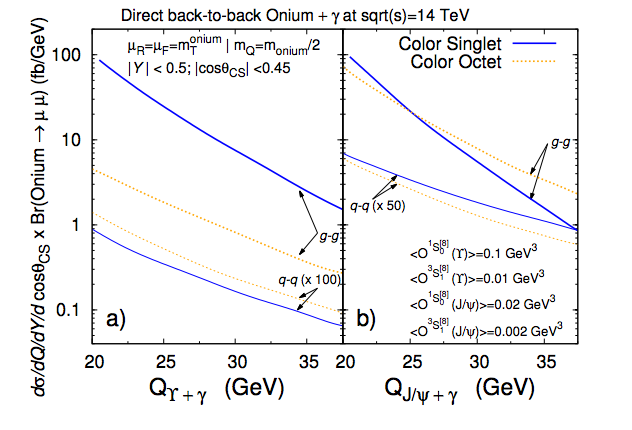}
	\caption{Predictions are shown for isolated photon-upsilon (left) and isolated photon-\jpsi (right) cross sections in both the color octet and color singlet models~\cite{Dunnen:2014eta}.}
	\label{fig:onium_rates}
\end{figure}

However, a completely new measurement could be performed only at the LHC experiments which would probe several different physical effects related to both TMD PDFs and potential factorization breaking. The proposed channel to measure is a nearly back-to-back isolated photon and (either) \jpsi or $\Upsilon$ state. The Feynman diagram for such a process is shown in Fig.~\ref{fig:isophot_quarkonia} and is a NLO process. Nevertheless, with the significantly larger center-of-mass energies and luminosities accessible by the LHC, Ref.~\cite{Dunnen:2014eta} has shown that a measurement of this process is feasible at both ATLAS and CMS. A measurement in this channel would probe several important, and largely unknown, gluon distribution functions. This channel additionally has the potential to identify effects from color in the production of heavy quarkonium and also could probe factorization breaking effects because of these color effects. Predictions for the cross sections in the color singlet and octet channels are shown in Fig.~\ref{fig:onium_rates} for both $\gamma-\Upsilon$ and $\gamma-\jpsi$. The predictions show that the gluon-gluon channel dominates at hard scales accessible at the LHC. \par

This process is very similar to the ones presented in this thesis, except that rather than an away-side hadron an away-side heavy quarkonium state is measured. Thus the applicability of a TMD framework is trivial; a hard scale is well defined by the large \pt of the final-state particles or the large invariant mass of the two-particle pair and a soft transverse momentum scale is observed when they are nearly back-to-back. This is especially important since the CMS and ATLAS experiments do not have sub GeV resolution on \pt dependent cross sections and thus they cannot be treated in a TMD framework. As a point of reference, $Z$ boson cross sections require a \pt resolution of less than 0.5 GeV to fully resolve the nonperturbative TMD structure~\cite{Aaltonen:2012fi}. TMD evolution effects can also be observed by adjusting the hard scale of the process, similarly to what has been presented in this document.

Factorization breaking effects are relevant as there are QCD bound states in both the initial and final states. However, if the heavy-quark pair is produced in a colorless state at short distances as in the color singlet model~\cite{Chang:1979nn, Baier:1981uk,Baier:1983va}, it is possible that factorization breaking effects would be minimal. It is also necessary that the quarkonium state is not accompanied by other colored partons that it may be interacting with after the bound state is formed. Previous studies have shown that the color octet contributions to the process $\gamma-\mathcal{Q}$ are likely smaller than in the inclusive quarkonium case~\cite{Li:2008ym,Kim:1996bb,Brambilla:2010cs}; however, a measurement of this process would give important information to the validity of this statement. If effects due to factorization breaking could be measured, this would indicate whether or not the heavy quarkonium state is produced more often in a color octet state state or if this is truly negligible in this process as suggested by Fig.~\ref{fig:onium_rates}. On the other hand, a measurement of the cross section could indicate whether or not color octet or singlet states are dominant, and thus, whether or not factorization breaking effects are present. If the measured cross section is more compatible with color octet contributions, this would indicate that the process may be violating TMD factorization. \par

If the processes are more compatible with a color singlet cross section, this may imply that the magnitude of TMD factorization breaking  effects should be small; however, there is still a significant amount of parton structure physics that could be identified from the observable. At the very minimum, the measurement would constrain unpolarized gluon TMD PDFs since the dominant partonic process that leads to a $\gamma-\mathcal{Q}$ state is gluon-gluon fusion. The unpolarized gluon TMD PDF is largely unconstrained due to the fact that it is a NLO process in SIDIS, therefore there is little statistically precise data to constrain the functions; ultimately the EIC will provide the best constraints due to its high luminosity. There are \pt dependent \jpsi cross sections from, for example, the PHENIX and LHCb collaborations~\cite{Adare:2012qf,Aaij:2013yaa}; however, the LHCb results do not have the necessary resolution at small \pt to constrain the gluon TMDs. There are additionally the complications of potential factorization breaking effects in these processes. Nevertheless, the $\gamma-\mathcal{Q}$ channel would ultimately provide information on both factorization breaking effects and unpolarized gluon TMDs. \par

There has also been recent interest in the linearly polarized gluon PDF (see e.g.~\cite{Boer:2017xpy}), as it is also unconstrained and is necessary to accurately describe gluon fusion processes. This includes, for example, Higgs boson production via gluon fusion processes. To describe the Higgs transverse momentum distribution accurately, the unpolarized and linearly polarized distributions are necessary as at small \pt the Higgs \pt will be largely due to the small transverse momentum of the gluons which fused. The linearly polarized gluon distribution can easily be measured by constructing an azimuthal asymmetry from the $\phi$ integrated cross section which leads to the unpolarized gluon distribution.  Therefore, assuming that enough integrated luminosity is collected, the linearly polarized distribution can be constrained trivially once the cross section is measured. Figure~\ref{fig:onium_rates} shows that the cross section is roughly 100 fb/GeV at $Q$=20 GeV for \sqs=14 TeV; ATLAS has recently published results from \sqs=13 TeV collisions with an integrated luminosity of 36 fb$^{-1}$ (see e.g.~\cite{Aaboud:2017buh}) which should be enough integrated luminosity to measure this process. Depending on the actual number of events collected the azimuthally binned cross section may be able to be measured; however, in several years the LHC will enter the high luminosity era which is projected to collect on the order of hundreds of fb$^{-1}$ of data and thus these observables can be measured before the EIC which is, at best, 10 years away. \par

In summary, the process $\pp\rightarrow\gamma+\mathcal{Q}+X$ is a promising avenue to search for several different nonperturbative QCD physics effects. When the $\mathcal{Q}$ and $\gamma$ are nearly back-to-back, the process can be treated within a TMD framework, and thus, transverse parton dynamics can be accessed. In particular the process would provide insight into whether or not the heavy quarkonium state is produced in a color singlet or color octet state, which would give important information on the production mechanisms of \jpsi or $\Upsilon$ states. From this it may be possible to infer the magnitude of factorization breaking effects in this process; if the cross section is more compatible with color octet production this would indicate that factorization breaking effects need to be considered. Additionally, a measurement of the cross section would parasitically provide information on the unpolarized gluon TMD PDF, as well as the linearly polarized gluon PDF assuming that enough data was collected to measure the cross section as a function of azimuth.

\chapter{Conclusion}
\label{chap:conclusion}

QCD research has entered a new era which is focused on quantitative measurements that have the ability to illuminate fundamental aspects about non-Abelian gauge invariant quantum field theories. This is largely due to the incredible modern day facilities with which experimentalists can make measurements, in addition to the significant theoretical work that is now exploring the role of various correlations within the nucleon. Both theoretical and experimental physicists have revealed complicated yet exciting new phenomena in QCD that are beginning to account for the complex nature of composite bound states in a strongly confined theory. In particular, the role of color has come to the forefront of both theoretical calculations and experimental measurements across many subfields of QCD. Theoretically, in the cases of modified universality and factorization breaking, the role of color exchanges with spectators of the interaction has become clearer. Experimentally, multidifferential measurements, in addition to the improved methods with which measurements are made have shown measurable effects from color flow unique to QCD interactions. \par

In particular, the focus of this work was searching for effects due to TMD factorization breaking, which results from complex color flows in hadronic collisions in dihadron and direct photon-hadron angular correlations. These color flows connect the spectators and hard interaction, resulting in an inability to uniquely define individual TMD PDFs and TMD FFs. This behavior is specifically a consequence of QCD as a non-Abelian gauge invariant quantum field theory. Dihadron correlations are relatively straightforward to measure compared to direct photon-hadron correlations and can be determined via correlation functions that account for the acceptance and efficiency of the PHENIX spectrometer. Direct photon-hadron correlations require an additional statistical subtraction to remove the decay photon-hadron background that is still present even after an isolation cut as well as removing tagged decay background. To probe effects sensitive to nonperturbative transverse momentum, the signed quantity \pout was used which characterizes the transverse momentum with respect to the near-side trigger particle; additionally the underlying event can be statistically removed to better isolate away-side hadrons associated with the hard scattered parton.

Final results presented include the \dphi distributions which show the qualitative and quantitative visual features expected from dihadron and direct photon-hadron angular correlations. On the other hand, the away-side \pout distributions show the yields in momentum space and display a two component distribution where a nonperturbative region transitions to a perturbatively behaved region. Nonperturbative momentum widths in these processes are measured experimentally for the first time which are sensitive to only the soft \kt and \jt present in the nearly back-to-back region. While the measurements of these nonperturbative widths do not show any notable qualitative factorization breaking effects, they represent the first measurements that can be rigorously interpreted within a TMD framework and thus compared to future calculations assuming factorization holds. With the surge of worldwide data that has been reported recently, theoretical global fits that more accurately constrain the TMD PDFs and TMD FFs should be possible in the future. Therefore, calculations can be compared to the measurements shown here and will ideally be able to shed light on the magnitude of factorization breaking effects since it is well established within the theoretical community that factorization is broken in these processes.

Additionally, this work has laid the foundation for searching for future factorization breaking effects in the direct photon-jet channel. The demonstration of measurements in the direct photon-hadron channel is an important first step towards direct photon-jet measurements; in addition, the simulation studies of the photon-jet channel at sPHENIX have shown that ample data will be collected to measure an observable predicted to break factorization which can be treated in a TMD framework. While measurements at the LHC have already demonstrated that this channel is sensitive to color coherence effects~\cite{Aaboud:2016sdm}, effects from color coherence have not yet been rigorously treated within a TMD framework and the data have only been compared to model dependent Monte Carlo simulations. While this is useful, ultimately QCD as the theory of the strong force must be able to explain these phenomena, thus comparisons to rigorous calculations made directly within the theory are important for understanding the magnitude of color flow effects.

Looking ahead to new observables, there are significant opportunities to study factorization breaking effects in future RHIC running periods as well as with the high luminosity data from the LHC. Nearly back-to-back direct photon-quarkonium production is a potential probe for measurable color flow effects. Significant effort has been placed on advancing jet grooming algorithms, in particular in the beyond-the-Standard-Model community, to isolate jet components that are truly from the hard scattered parton while excluding components due to soft radiation. Jet groomed observables have already been suggested to be more sensitive to factorization violating effects within SCET~\cite{Schwartz:2018obd}. The same techniques could be used in reverse; one could instead remove the jet components from the hard scattered parton and then measure ``soft radiation'' jets where the constituents are only from soft radiation. A potential observable could be measuring the invariant mass of ``anti-groomed'' jets, where this refers to the jets that have been anti-groomed to exclude the hard scattered parton constituents. Additional channels include spin asymmetries that may arise in photon-jet production in both \pp and \pa collisions~\cite{Zhou:2017mpw,Rogers:2013zha,Schafer:2014xpa}; it will be necessary to measure these processes at RHIC as it is the world's only polarized proton-proton collider, and in particular at sPHENIX as the only dedicated high rate jet detector at RHIC.

Ultimately, QCD research is seeing a push towards understanding more global observables which can be sensitive to overall color correlations. Rather than considering single inclusive particle production, global observables such as two, four, and six particle correlations, for example in Refs.~\cite{Aidala:2017ajz,CMS_pp_collectivity3}, can probe novel collective phenomena in hadronic collisions. These particular measurements are also observed across large pseudorapidity regions of $\Delta\eta>2$, where long distance correlations across hadrons are expected to arise from either the initial state or immediately following the hard interaction~\cite{Dumitru_pp_ridge}. Not surprisingly, the color coherence effects measured by the CMS collaboration are found to be stronger in the forward rapidity region than at central rapidity~\cite{Chatrchyan:2013fha}, where color correlations across hadrons are expected to arise based on TMD factorization breaking in dijet processes. With the solid angle coverage by modern detectors, global measurements rather than inclusive measurements will continue to probe novel long distance correlations.

Closely related to the study of global observables, the study of multipartonic interactions will be another important avenue for studying color correlations in future QCD research. Rather than considering simply the $2\rightarrow2$ hard process, there has been a greater push to understand hadronic collisions where multiple partons interact, producing several semi-hard interactions. While these have primarily been studied in the context of Monte Carlo event generators or models of QCD (see e.g. Refs.~\cite{Ortiz:2013yxa,Blok:2017pui}), there have also been rigorous theoretical QCD studies of double parton scattering in the context of multipartonic interactions~\cite{Rogers:2009ke,Blok:2013bpa}. It is perhaps not a coincidence that many of the QCD phenomena that are becoming prominent areas of study involve multi-scale problems, for example in the case of multipartonic interactions~\cite{Blok:2013bpa} or within the TMD framework in the case of factorization breaking~\cite{Rogers:2010dm}. Additionally, the consideration of the proton as a complex bound state has led to many of these new predictions; moving from inclusive observables to global observables, which are necessarily sensitive to the many partons that interact in a collision, has had a profound effect on advancing the dialogue on how nucleons and nuclei are treated. The results presented in this thesis will continue to advance this dialogue by providing measurements with which future calculations can constrain effects from color correlations.

While the work presented here was focused on studying TMD factorization breaking effects from color correlations in dihadron and direct photon-hadron angular correlations, the role of color within QCD is becoming a more prominent field of study across various subfields of QCD in general. Several examples include the prediction of factorization breaking and the modified universality of certain TMD PDFs~\cite{Collins:2002kn,Rogers:2010dm}, $J/\psi$ and $\psi^\prime$ suppression in \pa collisions~\cite{Ma:2017rsu}, color interference effects in long-range pseudorapidity correlations in \pp and \pa collisions~\cite{Blok:2017pui}, as well as in color coherence measurements~\cite{Chatrchyan:2013fha,Aaboud:2016sdm}. In some ways, historically separate subfields of QCD studying phenomena in \pp, \pa, and A+A collisions are converging towards measurements and an understanding of the fundamental properties that make QCD unique and that should apply in all hadronic systems. The next decade will bring a deeper understanding of these properties in hadronic collisions; this is a crucial time period in the history of QCD research as many of these properties in hadronic collisions will need to be measured and understood so that the data from the Electron Ion Collider can be interpreted within the context of our knowledge from hadronic collisions. The eventual turn-on of the Electron Ion Collider will signal the next great step in the precision measurement era of QCD research that has already begun.


\startappendices
 \appendix{Derivation of Equation for \rgammaprime}
 \label{app:rgammaprime_derivation}
 
This appendix shows the derivation of the equation used to calculate $\rgammaprime$ from the most fundamental definition of the quantity. We start from the definition of $\rgammaprime=N^{iso}_{inc}/N^{iso}_{dec}$ as well as $\rgamma=N_{inc}/N_{dec}$. We must evaluate $N_{dec}^{iso}$ with some other method, since we can't just measure isolated decay photons as we don't \textit{a priori} know which isolated photons are decay and which are direct. We have as definitions

\[
N_{dec} \equiv N^{iso}_{dec}+N^{tag}_{dec}+N^{niso}_{dec}\,,
\]
and thus
\begin{equation}\label{eq:ndec}
N_{dec}^{iso} = N_{dec}-N^{tag}_{dec}-N_{dec}^{niso}\,.
\end{equation}
Here, and throughout this appendix, niso refers to not isolated. The quantities are defined as follows: $N^{iso}_{dec}$ is the number of isolated decay photons, $N_{dec}^{tag}$ is the number of photons tagged as decay, and $N^{niso}_{dec}$ is the number of non isolated decay photons. The sum of all of these is clearly the total sample of decay photons. The problem here is that we don't know both $N_{dec}^{niso}$ and $N_{dec}^{iso}$, so we have two unknowns. To proceed we find isolated $\pion$s in the data and then use the fact that the photon isolation efficiency depends only on the parent $p_T$. We can then use this to map the isolation efficiency from the parent mesons to the daughter photon $p_T$ using the Monte Carlo decay probability functions that estimate the decay per trigger yields with
\begin{equation}\label{eq:fracptag}
\frac{N_{dec}^{iso}(p_T^\gamma)}{N_{dec}^{niso}(p_T^\gamma)} = \frac{\sum P^{tag}(p_T^\pi,p_T^\gamma)\otimes N_\pi^{iso}(p_T^\pi)}{\sum P^{tag}(p_T^\pi,p_T^\gamma)\otimes N_\pi^{niso}(p_T^\pi)}\,.
\end{equation}
The same principle is used to determine the isolated decay per trigger yields in the full statistical subtraction equation. For convenience, I will write the convolution of the Monte Carlo mapping function with the number of iso (niso) $\pion$s as $\mathcal{P}(N_{parent})$. We then solve equations~\ref{eq:ndec} and~\ref{eq:fracptag} to eliminate $N_{dec}^{niso}$ so that we can use $N^{iso}_{dec}$ in the original $\rgammaprime$ equation. Solving the equations gives 

\begin{equation}
N_{dec}^{iso} = \frac{N_{dec}-N_{dec}^{tag}}{(1+\frac{\prob^{niso}}{\prob^{iso}})}\,.
\end{equation}
Now we have an expression for $N_{dec}^{iso}$, so plug this into the expression for $\rgammaprime$ gives
\begin{equation}
\rgammaprime = \frac{N_{inc}^{iso}}{N_{dec}^{iso}} = \frac{N_{inc}^{iso}(1+\frac{\probniso}{\probiso})}{N_{dec}-N_{dec}^{tag}}\,.
\end{equation}
This is our expression for $\rgammaprime$. Using algebraic manipulations we can write this expression in terms of quantities we measure, so that everything in this expression is from data except the use of the Monte Carlo mapping functions. These functions are just used to map the isolated and non isolated $\pion$s to their daughter counter parts. Of course, it is impossible to measure any quantity of $\rgamma$ or $\rgammaprime$ without some help from Monte Carlo to identify the decay component background in full. Continuing on the denominator can be reduced as

\begin{equation}\label{eq:rgammaprimedecfact}
\rgammaprime = \frac{N_{inc}^{iso}(1+\frac{\probniso}{\probiso})}{N_{dec}(1-\frac{N_{dec}^{tag}}{N_{dec}})}\,.
\end{equation}
We can now write the quantity $\frac{N^{tag}_{dec}}{N_{dec}}$ as the tagging efficiency $\epsilon_{tag}$. We calculate this by

\begin{equation}\label{eq:tageffeq}
\frac{N^{tag}_{dec}}{N_{dec}} = \frac{N^{tag}_{dec}}{N_{inc}}\frac{N_{inc}}{N_{dec}} = \frac{N^{tag}_{dec}}{N_{inc}}\rgamma\,.
\end{equation}
Now we can continue with $\rgammaprime$. We can write $N_{inc} \equiv N^{iso}_{inc}+N^{niso}_{inc}+N^{tag}_{dec}$. Again this relation is a definition and is just the number of inclusive photons one can detect. Solving for $N^{iso}_{inc}$ allows us to substitute this quantity into equation~\ref{eq:rgammaprimedecfact} to give

\begin{equation}
\rgammaprime = \frac{N_{inc}-N^{tag}_{dec}-N^{niso}_{inc}}{N_{dec}}\frac{(1+\frac{\probniso}{\probiso})}{(1-\epsilon^{tag})}\,.
\end{equation}
We can then use $\rgamma=\frac{N_{inc}}{N_{dec}}$ to substitute in for the $N_{dec}$ in the denominator, i.e. $1/N_{dec}=\rgamma/ N_{inc}$. This gives

\begin{equation}
\rgammaprime = \frac{N_{inc}-N^{tag}_{dec}-N^{niso}_{dec}}{N_{inc}}\rgamma\frac{(1+\frac{\probniso}{\probiso})}{(1-\epsilon^{tag})}\,.
\end{equation}
The first fraction is now, again, something we can calculate from data by just counting the photons which pass the various cuts. The numerator $N_{inc}-N^{tag}_{dec}-N^{niso}_{dec}$ is simply the number of photons that pass the tagging and isolation cuts. The denominator is just the inclusive set of all photons. For convenience, define and rewrite this fraction as $\alpha$, i.e. $\alpha\equiv\frac{N_{inc}-N^{tag}_{dec}-N^{niso}_{dec}}{N_{inc}}$. In conclusion, we have

\begin{equation}\label{eq:rprimeeqnoepsiso}
\rgammaprime = \alpha\rgamma\frac{(1+\frac{\probniso}{\probiso})}{(1-\epsilon^{tag})}\,.
\end{equation}
Everything in this equation is calculated from counting photons, with the mapping function $\prob$ used to map the isolated $\pion$ background to isolated decay photon background that we cannot measure and need Monte Carlo to evaluate. To get the exact equation used in Ref.~\cite{ppg095}, some more algebraic manipulation to the $1+\frac{\probniso}{\probiso}$ term is needed, which finally gives

\begin{equation}\label{eq:finalrprime}
\rgammaprime = \frac{\alpha\rgamma}{(1-\epsilon^{tag})(1-\epsilon^{iso})}\,,
\end{equation}
where the so called ``isolation efficiency" is given by

\begin{equation}
\epsilon^{iso} = (1+\frac{\probiso}{\probniso})^{-1}\,.
\end{equation}
Therefore, equation~\ref{eq:finalrprime} was determined directly from the definition of $\rgammaprime$.

The derivation to put equation~\ref{eq:rprimeeqnoepsiso} into the final form, with $\epsilon^{iso}$ in the denominator, is shown here. To recall, we are starting from equation~\ref{eq:rprimeeqnoepsiso}. So we need to get $(1+\frac{\probniso}{\probiso})$ into the denominator to look like $(1-\epsilon^{iso})$. The derivation to do this is shown below.

\[
1+\frac{\probniso}{\probiso}
\]
\[
=\frac{\probniso+\probiso}{\probiso}
\]
\[
= \frac{\probniso+\probiso}{\probniso}\cdot\frac{\probniso}{\probiso}
\]
\[
= \frac{\frac{\probniso+\probiso}{\probniso}}{\frac{\probiso}{\probniso}}
\]
\[
= \frac{1+\frac{\probiso}{\probniso}}{\frac{\probiso}{\probniso}}
\]
\[
=\left(\frac{\frac{\probiso}{\probniso}}{1+\frac{\probiso}{\probniso}}\right)^{-1}
\]
\[
=\left(\frac{1+\frac{\probiso}{\probniso}}{1+\frac{\probiso}{\probniso}} - \frac{1}{1+\frac{\probiso}{\probniso}}    \right)^{-1}
\]
\[
=\left(1-\frac{1}{1+\frac{\probiso}{\probniso}}\right)^{-1}
\]
\[
=\left(1-\left(1+\frac{\probiso}{\probniso}\right)^{-1}\right)^{-1}
\]
Recall that $\epsilon^{iso}$ is defined as $\left(1+\frac{\probiso}{\probniso}\right)^{-1}$, so we can just substitute this in to get
\[
=\left(1-\epsilon^{iso}\right)^{-1}\,.
\]

Therefore it is shown that 
\[
1+\frac{\probniso}{\probiso}=\left(1-\epsilon^{iso}\right)^{-1}\,,
\]
and we can write the equation for $\rgammaprime$ as is found in Ref.~\cite{ppg095}
\begin{equation}
\rgammaprime = \frac{\alpha\rgamma}{(1-\epsilon^{tag})(1-\epsilon^{iso})}\,.
\end{equation}

 \begin{singlespace} 
  \bibliography{thesis_references}   
  \bibliographystyle{ieeetr}
 \end{singlespace}


\end{document}